ORGANISATION EUROPÉENNE POUR LA RECHERCHE NUCLÉAIRE
# CERN EUROPEAN ORGANIZATION FOR NUCLEAR RESEARCH

# THE CLIC POTENTIAL FOR NEW PHYSICS

## Corresponding editors

J. de Blas [1,2], R. Franceschini [3,4], F. Riva [5], P. Roloff [6], U. Schnoor [6], M. Spannowsky [7], J. D. Wells [8], A. Wulzer [1,6,9] and J. Zupan [10]

[1]Dipartimento di Fisica e Astronomia "Galileo Galilei", Università di Padova, Padova, Italy
[2]INFN, Sezione di Padova, Padova, Italy
[3]Dipartimento di Matematica e Fisica, Università degli Studi Roma Tre, Rome, Italy
[4]INFN, Sezione di Roma Tre, Rome, Italy
[5]Départment de Physique Théorique, Université de Genève, Genève, Switzerland
[6]CERN, Geneva, Switzerland
[7]Institute for Particle Physics Phenomenology, Department of Physics, Durham University, UK
[8]Leinweber Center for Theoretical Physics, Physics Department, University of Michigan, USA
[9]Theoretical Particle Physics Laboratory (LPTP), EPFL, Lausanne, Switzerland
[10]Department of Physics, University of Cincinnati, Cincinnati, Ohio, USA

## Abstract

The Compact Linear Collider (CLIC) is a mature option for the future of high energy physics. It combines the benefits of the clean environment of $e^+e^-$ colliders with operation at high centre-of-mass energies, allowing to probe scales beyond the reach of the Large Hadron Collider (LHC) for many scenarios of new physics. This places the CLIC project at a privileged spot in between the precision and energy frontiers, with capabilities that will significantly extend knowledge on both fronts at the end of the LHC era. In this report we review and revisit the potential of CLIC to search, directly and indirectly, for physics beyond the Standard Model.

**Contributors:** S. Alipour-Fard [1], W. Altmannshofer [2], A. Azatov [3,4], D. Azevedo [5,6], J. Baglio [7], M. Bauer [8], F. Bishara [9,10], J.-J. Blaising [11], S. Brass [12], D. Buttazzo [13], Z. Chacko [14,15], N. Craig [1], Y. Cui [16], D. Dercks [9,17], P.S.Bhupal Dev [18], L. Di Luzio [8,13,19], S. Di Vita [20], G. Durieux [9,21], J. Fan [22], P. Ferreira [5,23], C. Frugiuele [24], E. Fuchs [24], I. García [25,26], M. Ghezzi [7,27], A. Greljo [26], R. Gröber [8,28], C. Grojean [9,28], J. Gu [29], R. Hunter [30], A. Joglekar [16], J. Kalinowski [31], W. Kilian [12], C. Kilic [32], W. Kotlarski [33], M. Kucharczyk [34], E. Leogrande [26], L. Linssen [26], D. Liu [35], Z. Liu [14,15], D. M. Lombardo [36], I. Low [35,37], O. Matsedonskyi [24], D. Marzocca [4], K. Mimasu [38], A. Mitov [39], M. Mitra [40], R.N. Mohapatra [14], G. Moortgat-Pick [9,17], M. Mühlleitner [41], S. Najjari [42], M. Nardecchia [4,26], M. Neubert [29,43], J. M. No [44], G. Panico [9,45,46,47], L. Panizzi [48,49], A. Paul [9,28], M. Perelló [25], G. Perez [24], A. D. Plascencia [8], G. M. Pruna [50], D. Redigolo [24,51,52], M. Reece [53], J. Reuter [9], M. Riembau [36], T. Robens [54,55], A. Robson [26,56], K. Rolbiecki [31], A. Sailer [26], K. Sakurai [31], F. Sala [9], R. Santos [5,23], M. Schlaffer [24], S. Y. Shim [57], B. Shuve [16,58], R. Simoniello [26,59], D. Sokołowska [31,60], R. Ström [26], T. M. P. Tait [61], A. Tesi [46], A. Thamm [26], N. van der Kolk [62], T. Vantalon [9], C. B. Verhaaren [63], M. Vos [25], N. Watson [64], C. Weiland [8,65], A. Winter [64], J. Wittbrodt [9], T. Wojton [34], B. Xu [39], Z. Yin [37], A. F. Żarnecki [31], C. Zhang [66], Y. Zhang [18]

[1]Department of Physics, University of California, Santa Barbara, CA, USA
[2]Santa Cruz Institute for Particle Physics, Santa Cruz, CA, USA
[3]SISSA, Trieste, Italy
[4]INFN, Sezione di Trieste, Trieste, Italy
[5]Centro de Física Teórica e Computacional, Universidade de Lisboa, Campo Grande, Lisboa, Portugal
[6]LIP, Departamento de Física, Universidade do Minho, Braga, Portugal
[7]Institute of Theoretical Physics, Eberhard Karls Universität Tübingen, Germany
[8]Institute for Particle Physics Phenomenology, Department of Physics, Durham University, UK
[9]DESY, Hamburg, Germany
[10]Rudolf Peierls Centre for Theoretical Physics, University of Oxford, UK
[11]Laboratoire d'Annecy-le-Vieux de Physique des Particules, France
[12]Department of Physics, University of Siegen, Siegen, Germany
[13]INFN, Sezione di Pisa, Pisa, Italy
[14]Maryland Center for Fundamental Physics, Department of Physics, University of Maryland, USA
[15]Theoretical Physics Department, Fermi National Accelerator Laboratory, Batavia, IL, USA
[16]Department of Physics and Astronomy, University of California, Riverside, CA, USA
[17]II. Institute f. Theo. Physics, University of Hamburg, Hamburg, Germany
[18]Dept. of Physics and McDonnell Center for the Space Sciences, Washington University, St.Louis, USA
[19]Dipartimento di Fisica dell'Università di Pisa, Italy
[20]INFN, Sezione di Milano, Milano, Italy
[21]Physics Department, Technion — Israel Institute of Technology, Haifa, Israel
[22]Physics Department, Brown University, Providence, RI, USA
[23]Instituto Superior de Engenharia de Lisboa, Instituto Politécnico de Lisboa, Lisboa, Portugal
[24]Weizmann Institute of Science, Rehovot, Israel
[25]IFIC (UV/CSIC) Valencia, Spain
[26]CERN, Geneva, Switzerland
[27]Paul Scherrer Institut, Villigen PSI, Switzerland
[28]Institut für Physik, Humboldt-Universität zu Berlin, Berlin, Germany
[29]PRISMA Cluster of Excellence, Institut für Physik, Johannes Gutenberg-Universität, Mainz, Germany
[30]Department of Physics, University of Warwick, Coventry, UK
[31]Faculty of Physics, University of Warsaw, Warsaw, Poland
[32]Theory Group, Department of Physics, The University of Texas at Austin, Austin, TX, USA
[33]Institut für Kern- und Teilchenphysik, TU Dresden, Dresden, Germany

[34]The Henryk Niewodniczanski Institute of Nuclear Physics, Polish Academy of Sciences, Poland
[35]High Energy Physics Division, Argonne National Laboratory, Argonne, IL, USA
[36]Départment de Physique Théorique, Université de Genève, Genève, Switzerland
[37]Department of Physics and Astronomy, Northwestern University, Evanston, IL, USA
[38]Centre for Cosmology, Particle Physics and Phenomenology, UCLouvain, Belgium
[39]Cavendish Laboratory, University of Cambridge, Cambridge, UK
[40]Institute of Physics (IOP), Sachivalaya Marg, Bhubaneswar, India
[41]Institute for Theoretical Physics, Karlsruhe Institute of Technology, Karlsruhe, Germany
[42]Theoretische Natuurkunde and IIHE/ELEM, Vrije Universiteit Brussel, Brussels, Belgium
[43]Department of Physics and LEPP, Cornell University, Ithaca, NY, USA
[44]Departamento de Física Teórica and Instituto de Física Teórica, IFT-UAM/CSIC, Madrid, Spain
[45]Dipartimento di Fisica e Astronomia Università di Firenze, Sesto Fiorentino, Italy
[46]INFN, Sezione di Firenze, Italy
[47]IFAE and BIST, Universitat Autònoma de Barcelona, Bellaterra, Barcelona, Spain
[48]Uppsala University, Uppsala, Sweden
[49]School of Physics and Astronomy, University of Southampton, Southampton, UK
[50]INFN, Laboratori Nazionali di Frascati, Frascati, Italy
[51]Raymond and Beverly Sackler School of Physics and Astronomy, Tel-Aviv University, Tel-Aviv, Israel
[52]Institute for Advanced Study, Princeton, NJ, USA
[53]Department of Physics, Harvard University, Cambridge, MA, USA
[54]MTA-DE Particle Physics Research Group, University of Debrecen, Debrecen, Hungary
[55]Theoretical Physics Division, Rudjer Boskovic Institute, Zagreb, Croatia
[56]University of Glasgow, Glasgow, UK
[57]Department of Physics, Konkuk University, Seoul, Korea
[58]Harvey Mudd College, 301 Platt Blvd, Claremont, CA, USA
[59]Johannes Gutenberg-Universität Mainz, Mainz, Germany
[60]International Institute of Physics, Universidade Federal do Rio Grande do Norte, Brazil
[61]Department of Physics and Astronomy, University of California, Irvine, CA, USA
[62]Max-Planck-Institut für Physik, Munich, Germany
[63]Center for Quantum Mathematics and Physics, University of California, Davis, CA, USA
[64]University of Birmingham, Birmingham, UK
[65]PITT PACC, Department of Physics and Astronomy, University of Pittsburgh, PA, USA
[66]Institute of High Energy Physics, Chinese Academy of Sciences, Beijing, China

# Contents

# Executive summary

CLIC is a high-energy $e^+e^-$ collider that enables precision studies that qualitatively improve our knowledge and probe our understanding of the next generation questions of particle physics that have arisen from what we have learned at the LHC and other facilities around the world. CLIC can reach unprecedented precision in the properties and interactions of the Higgs boson, top quark and electroweak gauge bosons in the high energy realm. It can discover new states that are inaccessible at any other experiment, including the potential to discover Dark Matter and possibly give some experimental insight to cosmological questions such as (in)stability of the vacuum or the origin of the baryon asymmetry. The science program ranges from the "guaranteed physics" of precision studies of the Standard Model effective theory (SMEFT) beyond the capacity of any other facility running or proposed, to the "prospect physics" of directly producing new states or witnessing new interactions otherwise not allowed in the Standard Model, balancing a full guaranteed-plus-risk portfolio of physics capabilities.

Throughout this report many detailed studies are described that show the capability of CLIC to improve our understanding in numerous domains of particle physics, ranging from Higgs boson properties to flavour dynamics. Below, we give short paragraph descriptions of a variety of these capabilities. These "key findings" are broadly stated so as to be brief and not too technically narrow, but the reader is invited to follow the references at the end of each finding to read the details of the analyses and context behind these statements. Within these statements the terms CLIC-1, CLIC-2, and CLIC-3 are used to refer to the three different stages of the CLIC collider as defined in Table 1. Roughly stated, CLIC-1 is a Higgs and top quark factory at 350/380 GeV centre-of-mass energy, CLIC-2 is an intermediate high-energy stage at 1.5 TeV centre-of-mass energy, and CLIC-3 is the highest energy stage at 3 TeV centre-of-mass energy.

### Value of multi-TeV stages

The unique capability of CLIC of advancing to multi-TeV centre-of-mass energies brings extraordinary analysis advantages: 1) Single and double Higgs production increases substantially through vector boson fusion, enabling unprecedented precision on single Higgs couplings and the trilinear Higgs coupling. 2) Higher-dimensional effective operators contributing to $f\bar{f}$, $WW$, $ZH$ and $t\bar{t}$ production are significantly more activated at higher energies; this vastly improves the indirect sensitivities to motivated scenarios of physics beyond the Standard Model (BSM), such as composite Higgs/top, $Z'$, Dark Matter in loops, etc. And, 3) the discovery capacity increases to see BSM states through direct production and decay, surpassing by far the LHC sensitivity in several cases. (See Chapter 2 for a detailed discussion of points 1) and 2). The benefit of the high energy stages for CLIC direct searches is extensively documented in all the other chapters.)

### Precision single Higgs couplings

CLIC-1 alone already enables precision frontier determinations well beyond HL-LHC capabilities of single Higgs couplings to $WW$, $ZZ$, $c\bar{c}$ and $b\bar{b}$. Moreover, an absolute measurement of Higgs production in association with a $Z$ boson is made possible at CLIC due to the known kinematic constraints in the $e^+e^-$ colliding environment. This allows an absolute determination of the Higgs couplings, as opposed to the ratios accessible at the LHC. Finally, there is a per mille sensitivity to a universal overall rescaling of the Higgs boson couplings to other SM states, which is also significantly more sensitive than HL-LHC capabilities and relevant for several BSM scenarios. Later CLIC stages at higher energy and higher luminosity substantially improve sensitivities even further. (See Figure 1 and Sections 2.1, 2.2.1 and Sections 4.2 and 6.1 for more discussion.)

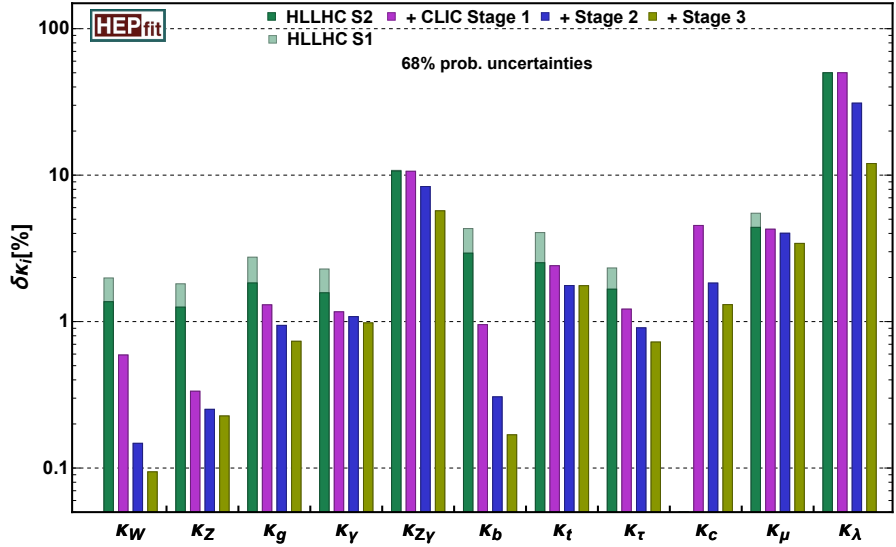

Figure 1: CLIC sensitivity to the different Higgs boson couplings (combined with the HL-LHC projections). See Sections 2.1 and 2.2.1.

**Higgs self coupling**

CLIC-2 and 3 will allow an exclusive measurement of the Higgs trilinear coupling with a precision at the $10\%$ level. This precision does not deteriorate in a global analysis thanks to the accurate single Higgs measurements performed at all CLIC stages. (See Figure 1 and Section 2.2.1. The impact of this measurement for models of electroweak baryogenesis is discussed in Section 4.2 and Chapter 6 and illustrated in Figure 4.)

**Precision electroweak analysis**

In addition to the renormalizable operators of the SM, the SMEFT contains all the non-renormalizable interactions that are consistent with the SM particle content and symmetries. CLIC enables precision searches for the existence of these non-renormalizable operators in the electroweak sector at coupling scales of many tens of TeV, generally far surpassing the sensitivities that can be achieved at HL-LHC. (See Chapter 2 and in particular Section 2.9.)

**Precision top quark analysis**

Study of $e^+e^- \to t\bar{t}$ at all CLIC stages allows precision searches for the SMEFT operators related to top physics significantly beyond the sensitivity of the HL-LHC. In some cases the sensitivity at CLIC-2 (CLIC-3) to top-dependent operators is one (two) orders of magnitude better than HL-LHC. (See Chapter 2 and in particular Section 2.9.) Moreover, the accurate measurement of the top quark mass at CLIC will help clarify the Standard Model vacuum (in)stability issue.

**Global sensitivity to BSM effects in the SMEFT**

The complementarity between low and high-energy measurements of different observables in the electroweak, Higgs and top sectors allows CLIC to explore multiple directions in the SMEFT parameter space simultaneously. CLIC not only significantly increases the precision in those directions that will be tested by the HL-LHC, but also gives access to others that can hardly be accessed at hadron colliders, e.g. new physics contributing to light quark Yukawa couplings. The CLIC potential to test SMEFT interactions under the (theoretically well motivated) assumptions of universal new physics is shown in Figure 2. (See Section 2.9.)

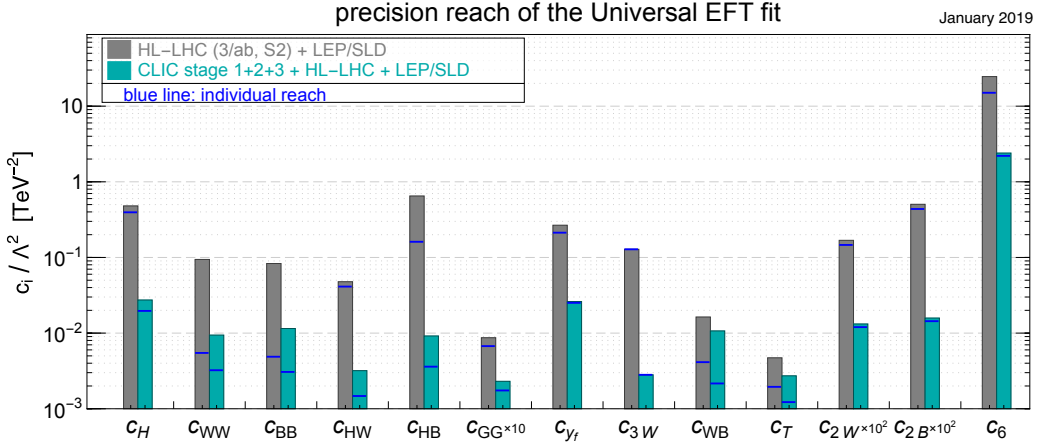

Figure 2: Comparison of the HL-LHC and CLIC reach from a global fit to universal new physics in the context of the SMEFT.

### Higgs and top compositeness

CLIC will *discover* Higgs and top compositeness if the compositeness scale is below 8 TeV. Scales up to 40 TeV can be discovered, in particularly favorable conditions, for large composite sector couplings $g_\star \simeq 8$. For comparison, the model-independent projected HL-LHC *exclusion* reach is only of 3 TeV. For $g_\star \simeq 8$, the HL-LHC reach is of 7 TeV. (See Section 2.10 and Figure 3.)

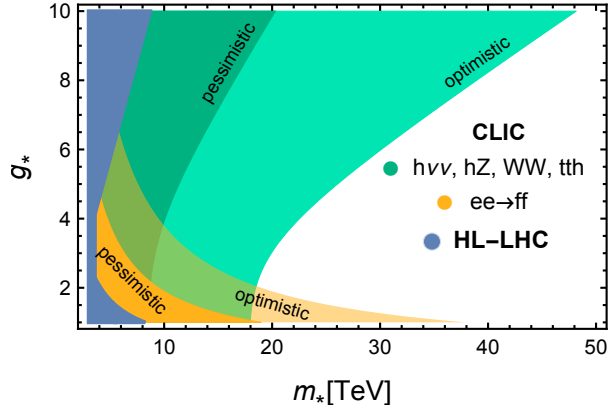

Figure 3: Discovery ($5\sigma$) reach on Composite Higgs at CLIC, and projected exclusion ($2\sigma$) at the HL-LHC, from Section 2.10.

### Baryogenesis

In models of electroweak baryogenesis new scalar particles are proposed to facilitate a strong first-order phase transition during which the electroweak symmetry is broken. We find that CLIC can exhaustively probe large classes of such models by combining precision measurements of the Higgs couplings to gauge bosons, of the Higgs self coupling, and direct searches for new scalar particles. CLIC significantly outperforms the HL-LHC in all these three ways of testing extended scalar sectors, which are frequently considered an ingredient of models for electroweak baryogenesis. (See Figure 4 and Section 6.1, as well as Section 4.2.) A radically different approach is "WIMP baryogenesis", a scenario where the baryon asymmetry is generated via the baryon number violating decays of TeV-scale long-lived particles. The favorable experimental conditions of CLIC allow to probe unexplored regions of the mass-lifetime parameter space of this model. (See Section 6.2.)

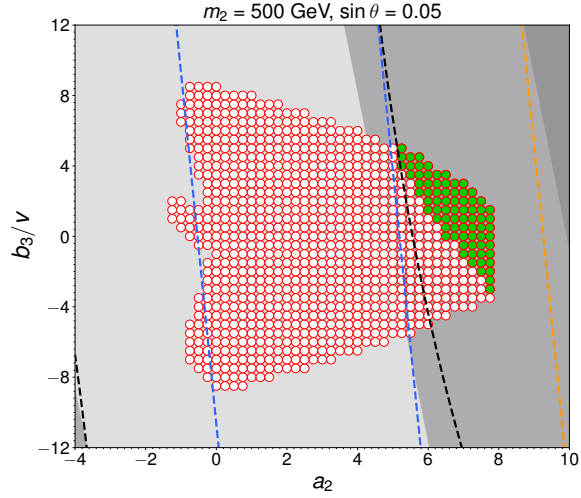

Figure 4: In green, the points compatible with electroweak baryogenesis, for $m_2 = 500\,\text{GeV}$ and $\sin\theta = 0.05$, in the model discussed in Section 6.1.3. These could all be tested both by the Higgs self coupling measurement and by direct searches, indicated by the black and the blue dashed lines respectively.

### Direct discoveries of new particles

Various BSM theories, such as supersymmetry, have substantial parameter space that yields no new discoveries at the LHC, but can be discovered through direct production at high-energy CLIC. This occurs when the new BSM states have highly compressed mass differences, rendering them invisible at the LHC, or when the only interactions allowed by the new BSM states are through electroweak and/or Higgs boson interactions, rendering their rates too small to discern from the large LHC backgrounds. Examples presented in this document range from supersymmetry and extended Higgs sectors potentially related with electroweak baryogenesis, to Dark Matter, neutrino mass models and feebly interacting particles.

### Extra Higgs boson searches

CLIC is ideally suited to discover and study heavy additional Higgs bosons (either new singlets or new doublets) that couple to the SM Higgs boson via its $|H|^2$ portal. Indirect and direct sensitivities on exotic Higgs bosons are typically substantially better than HL-LHC capabilities. (See Figure 5 and Section 2.1 and 4.2 for more discussion.)

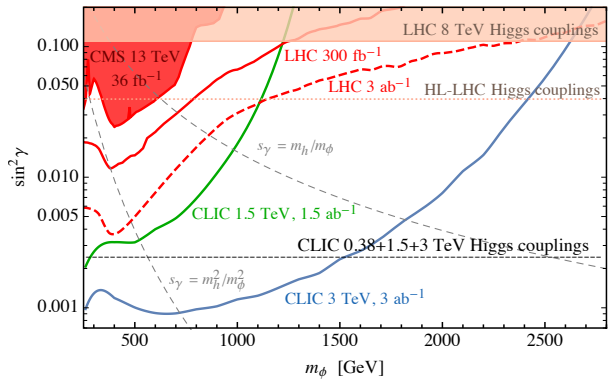

Figure 5: Reach on heavy scalar singlets, from Section 4.2.

### Studying new physics and new scales

If new physics is discovered at the LHC and/or CLIC, then the experimental environmenat at CLIC would provide the opportunity to study new states with great precision. These analyses could answer questions pertaining to the precise nature of the discovered new states and help point to yet new mass scales for the future. (See Section 4.4 for more discussion.)

### Dark matter searches

The relatively simple kinematic properties of the incoming $e^+e^-$ beam collisions and the relatively low rate of outgoing background at CLIC enables unprecedented searches for Dark Matter created in the laboratory, reaching sensitivities in parameter space interesting for cosmology and well beyond LHC capabilities. In particular, CLIC has sensitivity to the thermal Higgsino by stub tracks and to Minimal EW charged matter by its indirect radiative effects. (See Figure 6, Chapter 5 and in particular Sections 5.2 and 5.3 for more discussion.)

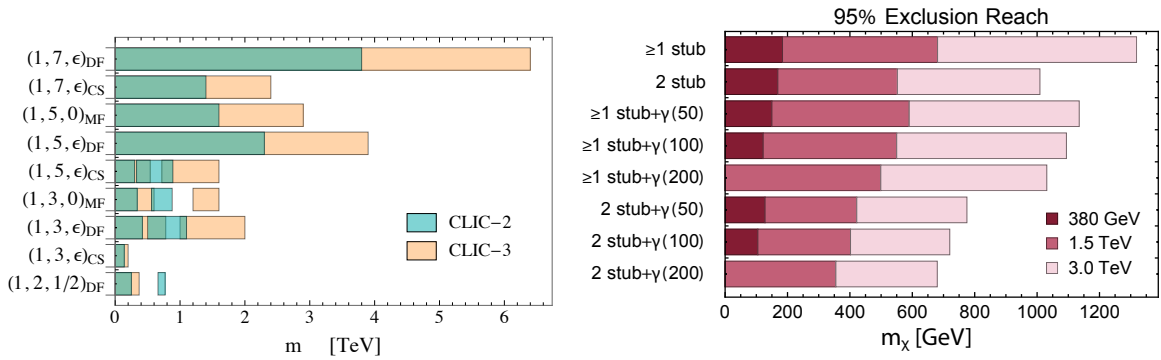

Figure 6: Left: DM in loops, from Section 5.3. Right: Higgsino reach from stub tracks, from Section 5.2.

### Lepton and flavour violation

Lepton-number violating and top quark flavour-changing neutral current interactions can be generated by SMEFT operators whose effects grow in importance with energy. These can be probed at the CLIC high-energy stages at levels far exceeding what can be achieved at the LHC (See Chapter 3 for more discussion.)

### Neutrino properties

Several mechanisms for the breaking of lepton number can be probed at CLIC both in direct searches and precision physics. CLIC is capable to probe directly weakly charged states involved in the generation of neutrino masses e.g. in Type-2 see-saw model and in gauge-extended models. It can also probe new heavy neutrinos and other states responsible for the breaking of lepton number by precision studies of leptonic two-body final states as well as WWH final states. (See Chapter 7 for more discussion.)

### Hidden sector searches

The clean $e^+e^-$ collision environment offers a clear chance to investigate rare and subtle signals from feebly coupled new physics and generic hidden sectors beyond the Standard Model. Displaced signals from long-lived particles are a very typical signature of these scenarios and CLIC enjoys a unique vantage point to look at these signals both in Higgs boson decays and in more general production of long-lived states that may be linked, for instance, to the Naturalness Problem or to the generation of the baryon asymmetry of the Universe. (See Section 6.2 and Chapter 8 for more discussion.)

### Exotic Higgs boson decays

Higgs decays directly into exotic states, such as axion-like particles or hidden sector particles, can be searched for at CLIC at levels far more sensitive than the LHC or any other facility for large regions of the parameter space. (See Chapter 8 for more discussion.)

### Impact of the charm Yukawa measurement on flavour models

Models of new physics at the weak scale typically require some mechanism of flavour protection, but can still induce sizable deviations in the Yukawa couplings to light quarks, like the charm. The charm Yukawa coupling will not be accessible to the HL-LHC but it could be measured at CLIC with $1\%$ precision. The impact of this measurement on several benchmark models of new physics is shown in Figure 7. (See Section 3.5.1.)

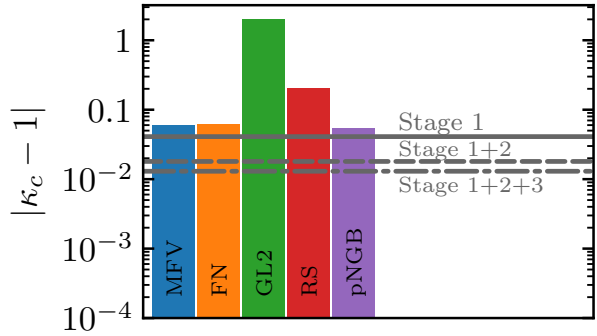

Figure 7: Deviations of the charm Yukawa coupling, $\kappa_c \equiv y_c/y_c^{\rm SM}$, in several flavour models consistent with current bound, confronted with the sensitivity expected at the different CLIC stages. See Section 3.5.1 for details.

# 1 Introduction

The Standard Model (SM) can be extrapolated up to the Planck scale without technical inconsistencies. Since we cannot probe this scale directly or indirectly, future experiments will not have access to a regime where the standard theory is guaranteed to break down and New Physics is guaranteed to show up. This is pretty unusual in the long history of high energy physics. Our field only dealt until now with "incomplete enough" theories, which required to be extended at relatively low energies. For instance the lack of high-energy consistency of the massive vector boson theory required the existence of a sector responsible for ElectroWeak Symmetry Breaking (EWSB) below around 3 TeV. This led to the famous "Higgs no-lose theorem", according to which EWSB physics (the Higgs boson, or something else) was guaranteed to be in the energy range probed by the LHC. No similar argument can be made for future colliders.

On the other hand, the quest for physics Beyond the SM (BSM) is even more pressing now than in the past. The results of the LHC and other experiments (prominently, the Dark Matter Direct Detection ones) have not answered any of the long-standing questions on the microscopic origin of the SM and of its parameters (e.g., the Naturalness Problem), nor those related with the lack of a SM explanation for observed phenomena (e.g., Dark Matter). On the contrary they made the problems harder, by putting strong pressure on many of the solutions we had hypothesized for them. The positive implication is that BSM physics is most likely something that has not yet been thought of. Consequently, by discovering it we will make a major breakthrough in the understanding of fundamental interactions. The other side of the coin is that we cannot define the goals and quantify the performances of future colliders in terms of a few benchmark models or scenarios. The goal is to explore the landscape of fundamental physics as broadly as possible, taking the many different proposed solutions to the SM problems as guidance, but also trying to be ready for the Unexpected as much as possible. This is the general ideology that underlies the present Report on the physics potential of the CLIC project.

CLIC is a mature option for the future of high energy physics. The technological feasibility of the collider is established and the detector design is at a very advanced stage. These aspects are documented in Refs. [1] and [2], respectively (see also [3–6]). The latter documents, together with the present one, are summarized in Ref. [7]. CLIC is an $e^+ e^-$ collider that operates in three stages, at 380 GeV, 1.5 and 3 TeV centre-of-mass energies, respectively. Stage 1 also foresees a short run around 350 GeV in order to study the top quark production threshold. While electron beams will be polarized at $\pm 80\%$ at all stages, positron polarization is not part of the CLIC baseline. The integrated luminosity delivered at each stage, and the fraction of it that is collected for each polarization configuration is reported in Table 1. Notice that the luminosities are significantly larger than what was originally foreseen in Ref. [8]. The choices of centre-of-mass energy, luminosity, and divisions of run-time for the different polarizations result from the optimization of the performance for top and Higgs measurements. The table also reports the luminosity of $e^+ e^-$ collisions occurring at an energy above $90\%$ and $99\%$ of the nominal collider energy. This is only a fraction of the total luminosity because of beam-beam interactions. The staging scenario presented in Table 1 has been assumed as baseline for most of the studies included in the Report. While minor departures are present in some analyses, these do not affect the conclusions. Similar considerations apply to the level of detail of the simulations employed for our studies. While detector and beam effects are not always included, they have been taken into account where particularly relevant.

The present Report reviews the potential of the CLIC project to probe BSM physics. It results from the activity of the "Physics Potential" Working Group within the CLICdp collaboration, but it summarizes a global community effort and collects material from different sources. It mostly consists of invited contributions that review, adapt, and in some cases significantly extend, recent papers on CLIC physics. A number of these papers were initiated in the context of the working group activities. Part of the material instead appears here for the first time. The Report obviously also builds upon previous CLICdp summary documents, and in particular on those presenting the CLIC potential for measurements in the Higgs [9] and top quark [10] sectors. Outlining the BSM implications of those results (supplemented in some case with recent updates) is one of our purposes. Our report has common aspects with older reviews of the CLIC physics potential [11], but also radical differences. In this respect it is worth noting that those documents were written in a different historical context than ours. LHC was expected to discover BSM particles, and therefore a lot of emphasis was given to CLIC capability to characterize the properties of those particles and to unveil the underlying BSM theory. While this aspect is still present (see Section 4.4), most of our Report assumes instead that no discovery is made at the LHC and at the High-Luminosity (HL-) LHC. This change of attitude towards the most probable LHC outcome, together with the much more accurate and realistic projections that are now available for Higgs and top quark physics at CLIC (and at the HL-LHC), strongly motivates our fresh look at the CLIC physics potential. Needless to say, the CLIC ability to characterize new particles will strike back in the event of an LHC discovery.

Table 1: The new baseline CLIC staging scenario [12]

|  | Stage 1 | | Stage 2 | Stage 3 |
|---|---|---|---|---|
| Nominal Energy ($\sqrt{s}$) | 380 GeV | 350 GeV | 1.5 TeV | 3 TeV |
| Integrated Luminosity [ab$^{-1}$] | 0.9 | 0.1 | 2.5 | 5.0 |
| Lumi. > 90% of $\sqrt{s}$ [ab$^{-1}$] | 0.81 | 0.09 | 1.6 | 2.85 |
| Lumi. > 99% of $\sqrt{s}$ [ab$^{-1}$] | 0.54 | 0.06 | 0.95 | 1.7 |
| Beam Polarizations | $P_{e^-} = -(+)80\%$ | | $P_{e^-} = -(+)80\%$ | $P_{e^-} = -(+)80\%$ |
| Lumi. Fraction by Polarization | 1/2 (1/2) | | 4/5 (1/5) | 4/5 (1/5) |

The Report forms a coherent body, and all its chapters are highly interconnected. However, the approach we follow in Chapters 2 and (part of) 3 is conceptually different from the other chapters. Latter chapters start from a given BSM scenario and discuss how to probe it at CLIC. Early chapters start from measurements that will be possible at CLIC and discuss their BSM implications. The language of the SM Effective Field Theory (SMEFT) is extensively employed in this context. The SMEFT is arguably the most minimal SM extension, because it contains no new degrees of freedom but it just extends the $d = 4$ SM Lagrangian to include interaction operators of energy dimension $d > 4$. Clearly, BSM particles are implicitly assumed to exist, to be sufficiently light and strongly-coupled to generate large enough EFT operators and in turn to give visible effects. Still they must be sufficiently heavy not to be produced directly such that the EFT expansion is justified. In the SMEFT framework, any BSM theory that respects the assumptions above is encapsulated in the values of the Wilson coefficients of the $d > 4$ EFT operators. All these theories are thus probed at once in a model-independent fashion. Because of the agnostic attitude we are obliged to take on New Physics, as previously outlined, the SMEFT is the ideal benchmark scenario to illustrate the CLIC potential for the indirect exploration of new physics and to compare its performances with the ones of other projects.

Chapter 2 is entirely devoted to the SMEFT and to the CLIC potential to explore it by measurements in the Higgs, top and EW sector. We will see in Section 2.9 that CLIC can probe the SMEFT much more precisely than the HL-LHC and that it can do it "globally", i.e. in all directions of the Wilson coefficients parameter space. This results from the interplay among different measurements performed at the different CLIC stages, which are found to be highly synergetic. Notice that the high-energy stages, which are unique of CLIC among all proposed lepton colliders, are found to be crucial for the precision program. Their importance stems from two facts. First, Higgs production cross-sections in the Vector Boson Fusion channels are enhanced at high energy, and the integrated luminosity is larger. Hence, the determination of all the couplings of the Higgs is significantly improved at Stage 2 and 3 with respect to Stage 1. Furthermore, the Higgs trilinear coupling can be measured only at high energies (see Section 2.2.1). Second, the EFT new physics effects on $2 \to 2$ scattering processes, relative to the SM, are enhanced at high energy. Hence even a moderate percent-level accuracy on these processes, due to their relatively small cross-section, can probe some of the EFT operators much better than very accurate

Higgs measurements. As a concrete illustration of the BSM power of these probes, in Section 2.10 we will discuss the impact of CLIC measurements on Higgs and top Compositeness. We will see that CLIC will unavoidably *discover* Higgs and top compositeness, if present, up to a compositeness scale of 8 TeV. A discovery might occur even for a compositeness scale of 40 TeV in particularly favourable conditions. These figures are well above what the HL-LHC can *exclude*.

Simplifying assumptions are made in Chapter 2 on the flavour structure of the underlying BSM theory, and in turn on the structure of the EFT operators. Relaxing these assumptions requires care, in order to avoid the severe bounds on flavour-breaking new physics from low-energy measurements, but allows us to probe BSM scenarios that address the SM flavour puzzle. CLIC physics in this context is described in Chapter 3. Among the topics discussed in this chapter, the possibility of directly probing the occurrence of BSM Flavour-Changing-Neutral-Current (FCNC) processes at high energy deserves a special mention, because it is a unique possibility offered by the CLIC project. Exploiting once again the high-energy enhancement of the EFT operators, FCNC operators with the structure $(\bar{e}e)(\bar{e}\tau)$ (Section 3.1.1) and $(\bar{e}e)(\bar{t}q)$ (Section 3.1.2) are probed better with this strategy than by exotic $\tau$ and top decays, respectively. Other topics include light quark Yukawa couplings determinations, the search for neutrino mass-generation models and for direct signatures of lepto-quarks that might be responsible for the lepton flavour universality violation anomaly in $b$-quark physics.

The rest of the Report aims at quantifying the CLIC reach on specific BSM scenarios which address some of the questions the SM does not answer. These include most prominently the origin of the scale of weak interactions, the nature of Dark Matter, and the origin of the asymmetry between baryons and anti-baryons in the Universe. Questions also surround the origin of measured parameters of the SM that appear to have peculiar values, for instance the mass of neutrinos and of all other fermions. Many models that solve these issues have been formulated over the last decades. The phenomenology of each of these models can give us an idea of the relevant experimental exploration necessary to investigate the big question they aim to address. However, it might also mislead us, making us focus on signatures that are peculiar to the specific model at hand, rather than to the general problem. Some amount of model-independence is needed also in this context. Therefore in what follows we will organize our discussion in terms of the guiding principles behind the formulation of the specific models. In fact, exploration of BSM physics at future colliders can be considered thorough and exhaustive only if it provides a test, and possibly a conclusive one, of general scenarios addressing the SM issues, and not just of specific models.

In Chapter 4 we study the origin and the structure of the EWSB sector. In this context, we already mentioned that CLIC can make a big jump ahead in probing the composite nature of the Higgs through indirect searches. Progress is possible also on some aspects of Supersymmetry (SUSY) searches. In particular it is worth outlining that the clean experimental environment and the available high centre-of-mass gives CLIC the unique opportunity to probe "Natural" (i.e., light) SUSY spectra that would have evaded LHC detection. The chapter also investigates extended Higgs sectors that are not necessarily motivated by SUSY. For instance the "Higgs plus singlet" scenario discussed in Sections 4.2 (and 6.1) might emerge from the Non-Minimal Supersymmetric SM (NMSSM), from Twin Higgs models and from some implementations of the recent "Relaxion" proposal for the solution of the Naturalness Problem. It is also connected with models of EW Baryogenesis (EWBG). As CLIC will operate in a physics landscape that is currently partially unknown, we also consider in this chapter possible investigations of the properties of yet undiscovered particles, whose hints, or discovery, may happen at the LHC or at the HL-LHC. In particular we evaluate the possible tests of the MSSM prediction of the Higgs boson mass and tests of the properties of new particles that appear in composite Higgs models of the "Twin" type.

Dark Matter (DM) is another very important topic, which we discuss in Chapter 5. CLIC can make progress in this area by probing DM candidates that are charged under the SM EW interactions, realizing the Weakly Interacting Massive Particle (WIMP) miracle in its most appealing form.[1] Clearly the WIMP is not the only option for DM, however it remains a compelling possibility that is not ruled out by Direct Detection experiments. Particularly motivated candidates in this class are the Higgsino and Minimal DM, which we study in Sections 5.2 and 5.3 by exploiting direct searches by stub-tracks and indirectly by loop effects, respectively. We will see that CLIC can discover the Higgsino for a mass of 1.1 TeV, which is the one it must have in order to be responsible for the observed DM density. CLIC can be conclusive also on other relevant and less standard DM scenarios discussed in the rest of the chapter.

---
[1]Namely, by exploiting the SM EW force, and not a new interaction, to obtain the relic abundance by thermal freeze-out.

Electroweak baryogenesis is the possibility of generating the asymmetry between baryons and anti-baryons in the Universe during the EW phase transition. It requires, among other things, a strong first order phase transition, unlike the one that occurs in the SM Higgs thermal potential. Given that the EW phase transition occurs at temperatures around 100 GeV, it is rather surprising that this possibility is still viable. Namely it is surprising that BSM models where the EW phase transition is of strong first order are not experimentally excluded in spite of the fact that they radically modify the Higgs potential at relatively low temperatures. This signals that our experimental knowledge of the Higgs sector is limited, and motivates further investigation. In the first part of Chapter 6 we will see how CLIC can probe these models indirectly, by measuring Higgs couplings among which the trilinear coupling, but also directly by searching for the states responsible for the modifications of the Higgs potential. The second part deals with an alternative mechanism for TeV-scale Baryogenesis, which employs new EW-scale particles that decay out of thermal equilibrium via B- and CP-violating interactions. This is the so-called "WIMP Baryogenesis" scenario, which predicts long-lived particles and can be probed at CLIC by displaced vertices.

CLIC is ideal to probe TeV-scale models for the origin of neutrino masses. In Chapter 7 we consider a number of these models and show the CLIC sensitivity through direct particle production, e.g. of mediators of Type-2 see-saw mechanism discussed in Section 7.2, scalars from extended models of the Left-Right type discussed in Section 7.4, and through precision measurements of SM processes such as $e^+e^- \to W^+W^-H$ discussed in Section 7.1 for models of the inverse see-saw type or $e^+e^- \to \ell^+\ell^-$ that can be mediated by exotic doubly charged SU(2) singlet scalars that are discussed in Section 7.3. Overall CLIC has sensitivity to TeV-scale degrees of freedom that may be responsible for the generation of neutrino masses and can show up in a wide range of collider signals.

Finally, in Chapter 8 we outline the CLIC sensitivity to particles that interact with the SM much more feebly than the weak interaction. Benchmark scenarios are long-lived particles coupled to the Higgs boson and Axion-like particles. The quiet experimental environment of CLIC allows them to be efficiently detected even if they decay relatively close to the interaction point. The high available energy allows CLIC to produce these particles in a wide range of masses.

We emphasized at the beginning of this chapter that it is impossible, given the current theoretical situation, to identify uniquely the new physics benchmark scenarios to be studied in order to assess the potential of future collider projects. Consequently the choice of subjects we made in the present Report could be criticized. The choice of the topics and the presentation unavoidably reflects the present-day perception of the community and, to a lesser extent, our personal taste. However we tried to be as inclusive as possible, and we paid special attention to topics and benchmark models that are currently being investigated for other future colliders. This should allow for a straightforward comparison.

## 2 Standard Model effective field theory

In this section we describe the potential of CLIC to search indirectly for physics beyond the SM. This mode of exploration focusses on heavy BSM dynamics, associated with a mass scale beyond the CLIC direct energy reach, and it exploits the fact that such dynamics can still have an impact on processes at smaller energy, via virtual effects. In this context the well-established framework of effective field theories allows to systematically parametrise BSM effects and how they modify SM processes. The leading such effects can be captured by dimension-6 operators

$$\mathcal{L}_{\text{eff}} = \mathcal{L}_{\text{SM}} + \frac{1}{\Lambda^2} \sum_i c_i \mathcal{O}_i + \cdots \qquad (1)$$

for dimensionless coefficients $c_i$ and a common suppression scale $\Lambda$; throughout this report we will measure $c_i/\Lambda^2$ in units of TeV$^{-2}$, equivalent to taking $\Lambda = 1$TeV. We will generically refer to these dimension-6 operators as BSM effects. Table 2 proposes a set of operators relevant for the present analysis. This set is *redundant*, in the sense that different combinations of operators might lead to the same physical effect. In practical applications we will always be interested in identifying non-redundant subsets of operators that, moreover, shall also be well motivated from a BSM perspective. Throughout this chapter we will therefore project into different *minimal* (non-redundant) bases, in such a way as to simplify as much as possible the BSM description of individual physical processes at CLIC, and we will choose different well-motivated BSM scenarios to focus our EFT analysis.

Table 2: A list of dimension-6 SMEFT operators used in this chapter, defined for one family only; operators suppressed in the minimal flavour violation assumption [13] have been neglected (in particular dipole-type operators). Operators specific to top physics will be discussed in Section 2.7, and other four-fermion operators will be discussed in Section 2.6. Some combinations are redundant and can be eliminated as described in the text. We define $\overset{\leftrightarrow}{D}{}^a \equiv \sigma^a \overset{\rightarrow}{D} - \overset{\leftarrow}{D} \sigma^a$.

| Higgs-Only Operators | | |
|---|---|---|
| $\mathcal{O}_H = \frac{1}{2}(\partial^\mu |H|^2)^2$ | $\mathcal{O}_6 = \lambda |H|^6$ | |
| $\mathcal{O}_{y_u} = y_u |H|^2 \bar{Q} \widetilde{H} u$ | $\mathcal{O}_{y_d} = y_d |H|^2 \bar{Q} H d$ | $\mathcal{O}_{y_e} = y_e |H|^2 \bar{L} H e$ |
| $\mathcal{O}_{BB} = g'^2 |H|^2 B_{\mu\nu} B^{\mu\nu}$ | $\mathcal{O}_{GG} = g_s^2 |H|^2 G^A_{\mu\nu} G^{A\mu\nu}$ | $\mathcal{O}_{WW} = g^2 |H|^2 W^a_{\mu\nu} W^{a\mu\nu}$ |
| Universal Operators | | |
| $\mathcal{O}_T = \frac{1}{2}(H^\dagger \overset{\leftrightarrow}{D}_\mu H)^2$ | $\mathcal{O}_{HD} = (H^\dagger D^\mu H)^*(H^\dagger D_\mu H)$ | |
| $\mathcal{O}_W = \frac{ig}{2}(H^\dagger \overset{\leftrightarrow}{D}{}^a_\mu H) D_\nu W^{a\mu\nu}$ | $\mathcal{O}_B = \frac{ig'}{2}(H^\dagger \overset{\leftrightarrow}{D}_\mu H) \partial_\nu B^{\mu\nu}$ | $\mathcal{O}_{WB} = gg'(H^\dagger \sigma^a H) W^a_{\mu\nu} B^{\mu\nu}$ |
| $\mathcal{O}_{HW} = ig(D^\mu H)^\dagger \sigma^a (D^\nu H) W^a_{\mu\nu}$ | $\mathcal{O}_{HB} = ig'(D^\mu H)^\dagger (D^\nu H) B_{\mu\nu}$ | |
| $\mathcal{O}_{3W} = \frac{1}{3!} g \epsilon_{abc} W^{a\,\nu}_\mu W^b_{\nu\rho} W^{c\,\rho\mu}$ | $\mathcal{O}_{2B} = \frac{1}{2}(\partial_\rho B_{\mu\nu})^2$ | $\mathcal{O}_{2W} = \frac{1}{2}\left(D_\rho W^a_{\mu\nu}\right)^2$ |
| and $\mathcal{O}_H, \mathcal{O}_6, \mathcal{O}_{BB}, \mathcal{O}_{WW}, \mathcal{O}_{GG}, \mathcal{O}_y = \sum_\psi \mathcal{O}_{y_\psi}$ | | |
| Non-Universal Operators that modify $Z/W$ couplings to fermions | | |
| $\mathcal{O}_{HL} = (iH^\dagger \overset{\leftrightarrow}{D}_\mu H)(\bar{L} \gamma^\mu L)$ | $\mathcal{O}^{(3)}_{HL} = (iH^\dagger \overset{\leftrightarrow}{D}{}^a_\mu H)(\bar{L} \sigma^a \gamma^\mu L)$ | $\mathcal{O}_{He} = (iH^\dagger \overset{\leftrightarrow}{D}_\mu H)(\bar{e} \gamma^\mu e)$ |
| $\mathcal{O}_{HQ} = (iH^\dagger \overset{\leftrightarrow}{D}_\mu H)(\bar{Q} \gamma^\mu Q)$ | $\mathcal{O}^{(3)}_{HQ} = (iH^\dagger \overset{\leftrightarrow}{D}{}^a_\mu H)(\bar{Q} \sigma^a \gamma^\mu Q)$ | |
| $\mathcal{O}_{Hu} = (iH^\dagger \overset{\leftrightarrow}{D}_\mu H)(\bar{u} \gamma^\mu u)$ | $\mathcal{O}_{Hd} = (iH^\dagger \overset{\leftrightarrow}{D}_\mu H)(\bar{d} \gamma^\mu d)$ | |
| CP-odd operators | | |
| $\mathcal{O}_{H\widetilde{W}} = (H^\dagger H) \widetilde{W}^a_{\mu\nu} W^{a\mu\nu}$ | $\mathcal{O}_{H\widetilde{B}} = (H^\dagger H) \widetilde{B}_{\mu\nu} B^{\mu\nu}$ | $\mathcal{O}_{\widetilde{W}B} = (H^\dagger \sigma^a H) \widetilde{W}^a_{\mu\nu} B^{\mu\nu}$ |
| | $\mathcal{O}_{3\widetilde{W}} = \frac{1}{3!} g \epsilon_{abc} W^{a\,\nu}_\mu W^b_{\nu\rho} \widetilde{W}^{c\,\rho\mu}$ | |

These operators induce two types of effects: some that are proportional to the SM amplitudes and some that produce genuinely new amplitudes. The former are the reason indirect searches are often presented as SM Precision Tests, and they are better accessed by high-luminosity experiments in kinematic regions where the SM is the largest, in such a way as to minimize statistical errors. Examples of this are Higgs couplings measurements. In the context of CLIC, *high-luminosity tests* have been already discussed in some detail in Ref. [8, 9], and we will update some of the past studies, presented in modern EFT language, in Section 2.1 and 2.2.

Effects associated with new amplitudes might instead be better tested in regions where the SM is relatively small. In this class, particularly interesting for CLIC are BSM effects that grow with energy: *high-energy* tests constitute a dedicated program focussed on the 3 TeV CLIC operational mode. At dimension-6 level we find effects that grow quadratically with the energy, implying a quadratic energy-growth in the sensitivity. This can be contrasted with high-intensity effects, whose sensitivity increases only with the square root of the integrated luminosity, and eventually saturates as systematics become comparable. Notice that most high-intensity tests can also be performed at 380 GeV; for instance measurements of Higgs couplings benefit enormously from the CLIC Stage 3 run, because of the enhanced VBF cross section and because of the larger integrated luminosity. In this sense, the high-intensity/high-energy separation should be thought merely as a qualitative label to classify different processes.

In the majority of this chapter we focus on processes and effects that fall in the high-energy category, that is $2 \to 2$ or $2 \to 3$ processes. In particular we discuss $e^+e^- \to W^+W^-, ZH, \psi\bar{\psi}$ or $\bar{t}th$ in Sections 2.3, 2.4, 2.6 and 2.7 respectively, while Section 2.5 studies multiboson processes. At the same time, we will emphasise the power of some precise high-intensity tests to access EFT parameters. For instance, Section 2.8 includes loop effects from modified top Yukawas into the $h \to gg$ branching rate, and Section 2.2.1 provides an alternative way of studying the Higgs self-coupling via its loop-induced effects in many high-intensity probes. We will combine the information from different runs and discuss the various high/low-energy connection, including a comparison with other machines running on the $Z$-pole.

The picture that will emerge is that the simple processes studied in this chapter are enough to test with high precision all the operators from Eq. (1) that can be tested at lepton colliders: with a few exceptions the study of more complex processes would not add relevant information to this endeavour. This information will be made quantitative in Section 2.9 where a global EFT fit is performed. Beside impressive reach on individual operators, it is shown that comparable results hold also when many operators are simultaneously present.

The generic EFT language used in this chapter provides a systematic tool to parametrise BSM effects searchable in precision tests. Yet, concrete BSM scenarios will not generate all operators, but rather a subset, with perhaps large hierarchies between the coefficients (expressing for instance the fact that some operators arise at loop level, while some already at tree level). In such a subset, the contributions to different observables will exhibit correlations in addition to those already discussed in the context of the global EFT study, thus providing a different perspective on CLIC reach. A notable example is that of universal theories, that take into account only modifications to the SM bosonic interactions, and are well motivated from a BSM point of view (see e.g. [14]); the operators they generate are singled out in Table 2. Most composite Higgs (CH) models [15, 16] belong to this class. CH models represent the prototype scenario for large EFT effects [17], since the underlying strong coupling implies some large coefficients $c_i$, for a fixed New Physics scale. Seen in a different way, given that we assume the scale of new physics is larger than the direct CLIC reach, CH models are more likely to have visible effects at CLIC than their weakly coupled counterpart (such as the minimal supersymmetric models). For this reason we discuss an interpretation in terms of CH models in Section 2.10. The strong coupling/weak coupling dichotomy is also at the heart of the issue of EFT validity, which questions the truncation of Eq. (1) at the leading order. In our discussion, rather than pointing out that there can be theories where the EFT truncation breaks down, we rely in this chapter on the fact that it is possible to find theories (based perhaps on strong coupling) where the scale $\Lambda$ is above the direct CLIC reach (see [18]). In cases where the new physics is instead within CLIC kinematic reach, we assume that direct searches (discussed later in this report) provide a better search strategy.

Unless otherwise stated, the contributions in this chapter are based on the CLIC baseline scenario from Table 1, summarized in Table 3 (the electron beam polarized as $P_{e^-} = \pm 80\%$, while no polarizations is envisaged for the positron beam $P_{e^+} = 0\%$).

Table 3: CLIC Baseline scenarios considered in this study (from Table 1). In some contributions, for comparison, we include a scenario running without beam polarization ("CLIC Unpolarized").

| Scenario | | CLIC Baseline | | CLIC Unpolarized |
|---|---|---|---|---|
| $(P_{e^-}, P_{e^+})$ | | $(80\%, 0\%)$ | $(-80\%, 0\%)$ | $(0\%, 0\%)$ |
| $\sqrt{s}$ | | $L = \int \mathcal{L} dt$ | | |
| Stage 1 | 380 GeV | 500 fb$^{-1}$ | 500 fb$^{-1}$ | 1000 fb$^{-1}$ |
| Stage 2 | 1500 GeV | 500 fb$^{-1}$ | 2000 fb$^{-1}$ | 2500 fb$^{-1}$ |
| Stage 3 | 3000 GeV | 1000 fb$^{-1}$ | 4000 fb$^{-1}$ | 5000 fb$^{-1}$ |

## 2.1 Higgs couplings

The operators from the top panel of Table 2 have the form $|H|^2 \times \mathcal{L}_{\text{SM}}$, with $\mathcal{L}_{\text{SM}}$ denoting operators in the SM Lagrangian, and it is easy to see [19, 20] that, at tree level, they modify *only* processes which involve at least one physical Higgs, i.e. they cannot be measured in the vacuum $\langle H \rangle = v/\sqrt{2}$. In the context of Higgs physics they imply small modifications $\propto v^2/\Lambda^2$ of the Higgs couplings to other SM fields, with respect to the SM value. These have been parametrised in CLIC studies as rescalings of the SM rates, for instance $\kappa_{HWW}^2 = \Gamma_{HWW}/\Gamma_{HWW}^{\text{SM}}$ ($\Gamma$ the partial width, and $\Gamma_i^{\text{SM}}$ its SM expectation) captures deviations of the Higgs coupling to $W$ bosons *assuming* the same Lorentz structure as that of the SM, i.e. providing an overall energy-independent factor.[2]

The estimated reach on the several $\kappa$ values from different CLIC runs [12] is reported in Table 4 ($\kappa_\lambda$, which is associated with modifications of the Higgs self-coupling, will be the subject of a dedicated Section 2.2.1).[3] To better understand these results it is useful to recall the energy dependence of Higgs production cross section in the different channels (see Figure 30); in particular the energy-growth of the cross section in the vector boson fusion channels plays an important role for the improved sensitivity from Stage 3. In this section we discuss the impact of a Higgs-couplings analysis performed in terms of $\kappa$'s, for BSM effects in terms of the coefficients of EFT operators. In this section we discuss the impact of a Higgs-couplings analysis performed in terms of $\kappa$'s, for BSM effects in terms of the coefficients of EFT operators.

Operators in the top panel of Table 2 are the primary target of Higgs couplings measurements. There are in fact more combinations of operators that modify processes with the Higgs and we can divide them in four categories:

*i)* Operators that are redundant and generate effects that are identical to those from the top panel of Table 2;

---

[2] In the EFT framework the only light particles are those of the SM, so that the total Higgs width $\Gamma_H$ is fixed by the sum of those in the individual channels. This scenario is referred in previous reports as *model dependent fit*, and is parametrized by $\kappa_i$ factors, as opposed to the *model-independent fit*, associated with $g_i$ factors where $\Gamma_H$ is an additional free parameter.

[3] For comparison, estimates for the HL-LHC are also shown in the table. These were obtained by combining the ATLAS and the CMS projected sensitivities for Higgs signal strength measurements, presented in Ref. [21]. Since $H \to Z\gamma$ projections from CMS were not available, we assume the same precision as ATLAS. Furthermore, correlations between the two sets of projections were not available at the time this fit was performed and were therefore neglected. We show the results for the 2 scenarios of systematics denoted as S1 and S2 in the previous reference.

Table 4: Expected reach on $\kappa_i$ parameters from a global fit of Higgs couplings, from Ref. [12]. The last column shows estimates for the HL-LHC in two scenarios for the extrapolated systematics, see Footnote 3 for details.

|  | Stage 1 | Stage 1+2 | Stage 1+2+3 | HL-LHC S1 (S2) |
|---|---|---|---|---|
| $\kappa_{\text{HZZ}}$ | 0.4 % | 0.3 % | 0.2 % | 1.8 (1.3) % |
| $\kappa_{\text{HWW}}$ | 0.8 % | 0.2 % | 0.1 % | 2.0 (1.4) % |
| $\kappa_{\text{Hbb}}$ | 1.3 % | 0.3 % | 0.2 % | 4.3 (2.9) % |
| $\kappa_{\text{Hcc}}$ | 4.1 % | 1.8 % | 1.3 % | – |
| $\kappa_{\text{H}\tau\tau}$ | 2.7 % | 1.2 % | 0.9 % | 2.3 (1.7) % |
| $\kappa_{\text{H}\mu\mu}$ | – | 12.1 % | 5.6 % | 5.5 (4.4) % |
| $\kappa_{\text{Htt}}$ | – | 2.9 % | 2.9 % | 4.1 (2.5) % |
| $\kappa_{\text{Hgg}}$ | 2.1 % | 1.2 % | 0.9 % | 2.8 (1.8) % |
| $\kappa_{\text{H}\gamma\gamma}$ | – | 4.8 % | 2.3 % | 2.3 (1.6) % |
| $\kappa_{\text{HZ}\gamma}$ | – | 13.3 % | 6.6 % | 11 (11) % |

*ii)* Operators with different Lorentz structure that contribute to Higgs couplings in an energy-dependent way;

*iii)* Operators that are very tightly constrained from other measurements and are not expected to have a substantial impact on Higgs physics.

*iv)* Operators that enter at loop-level in Higgs processes.

Let us discuss these in turn. Table 2 represents a redundant set of operators, meaning that two different combinations might lead to exactly the same physical effect. These redundancies can be eliminated, using integration by parts and field redefinitions, which in practice eliminate any combination of operators proportional to the SM equations of motion. These imply relations between the operators of Table 2; the most important ones being ($Y$ denotes here the hypercharge)

$$\mathcal{O}_{HB} = \mathcal{O}_B - \frac{1}{4}\mathcal{O}_{BB} - \frac{1}{4}\mathcal{O}_{WB}, \qquad \mathcal{O}_{HW} = \mathcal{O}_W - \frac{1}{4}\mathcal{O}_{WW} - \frac{1}{4}\mathcal{O}_{WB} \qquad (2)$$

$$\mathcal{O}_B = \frac{g'^2}{2}\sum_\psi Y_\psi \mathcal{O}_{H\psi} - \frac{g'^2}{2}\mathcal{O}_T, \qquad \mathcal{O}_T = \mathcal{O}_H - 2\mathcal{O}_{HD} \qquad (3)$$

$$\mathcal{O}_W = \frac{g^2}{2}\Big[(\mathcal{O}_{y_u} + \mathcal{O}_{y_d} + \mathcal{O}_{y_e} + \text{h.c.}) - 3\mathcal{O}_H + 4\mathcal{O}_6 + \frac{1}{2}\sum_{\psi_L}\mathcal{O}_{H\psi_L}^{(3)}\Big], \qquad (4)$$

and similar expressions for $\mathcal{O}_{2W}$ and $\mathcal{O}_{2B}$ in terms of the products of $SU(2)$ and $U(1)$ SM currents. Eqs. (2-4) can be used to define minimal, non-redundant operator bases; for instance, in the context of Higgs physics, the operators $\mathcal{O}_H, \mathcal{O}_W, \mathcal{O}_B, \mathcal{O}_{HW}, \mathcal{O}_{HB}$ are retained at the expense of $\mathcal{O}_{HD}, \mathcal{O}_{WW}, \mathcal{O}_{WB}, \mathcal{O}_{HL}^{(3)}, \mathcal{O}_{HL}$ in what is known as the SILH basis [17], while in the opposite case we refer to the Warsaw basis [22].[4] So, Eqs. (2-4) allow us to remove contributions from operators belonging to class *i)*, in favour of operators appearing on the left-hand side of Table 2, in particular $4(\mathcal{O}_W - \mathcal{O}_B - \mathcal{O}_{HW} + \mathcal{O}_{HB}) \to \mathcal{O}_{WW} - \mathcal{O}_{BB}$.

Operators from class *ii)*, that generate energy-dependent effects, can not be put in one-to-one correspondence with the $\kappa$'s, which implicitly assume the same structure as the SM. For instance, operators like $\mathcal{O}_W$ or $\mathcal{O}_{He}$ induce a centre-of-mass energy growth in the $\bar{e}e \to Zh$ rate with respect to the SM, while $\kappa_{\text{HZZ}}$ only captures a constant departure from the SM. These energy growing effects will be the subject of dedicated sections in this document and omitting them from the analysis of this section would simplify it substantially, without quantitatively compromising its results. The combination of Higgs couplings and other high-energy measurments will be the subject of Section 2.9.

---
[4]In addition, the SILH basis gives preference to the operators $\mathcal{O}_{2W}$ and $\mathcal{O}_{2B}$, which are more easily found in universal BSM theories, while the Warsaw basis swaps them in terms of four-fermions operators.

Table 5: Higgs couplings 68% C.L. reach on the dimension-6 operator coefficients $c_i/\Lambda^2$ [TeV$^{-2}$] in the top panel of Table 2, as translated from the $\kappa$ fit presented in Table 4 (see text for details and the assumptions going into this translation). Numbers without brackets: global fit with all operators, then marginalised; number in brackets: global fit limited to the operators from the top and bottom panel respectively. Combines the three CLIC stages.

| 68% reach on tree-level operators | | | | | | |
|---|---|---|---|---|---|---|
| $c_H$ | $c_6$ | $c_{y_t}$ | $c_{y_c}$ | $c_{y_b}$ | $c_{y_\tau}$ | $c_{y_\mu}$ |
| 0.033 [0.032] | 1.7 [1.7] | 0.48 [0.13] | 0.21 [0.21] | 0.030 [0.029] | 0.14 [0.14] | 0.92 [0.92] |

| 68% reach on loop-level operators | | |
|---|---|---|
| $c_{BB}$ | $c_{WW}$ | $c_{GG}$ |
| 0.018 [0.017] | 0.017 [0.016] | 0.0012 [0.0002] |

In a similar category enter operators that, beside modifying Higgs physics, also modify other very precisely measured observables, such as those used to determine the SM input parameters, or those associated with $Z$ pole physics relevant for LEP - class *iii)*. These precise measurements, already available, are strong enough that these operators cannot modify Higgs physics in a way that is relevant for the discussion in this section. Nevertheless, CLIC can improve on the measurements of some of these operators via its high-energy reach, see Section 2.9.1.

We have so far focussed on tree-level effects where the relation between observables and EFT effects is rather transparent. At loop level many more operators can contribute to a given observable. We have in mind tiny BSM effects, which become even smaller when loop-suppressed; so generally loop effects can be ignored. Yet, there are situations where an operator contributes at tree-level to a poorly measured observable, and at loop level to a very precisely measured one [23–26]. In this case the reach from the latter might be better. Notable examples, relevant for this chapter, are modifications of the top Yukawa, and modifications of the Higgs self-coupling that enter respectively in the $h \to gg, \gamma\gamma, Z\gamma$ rates and in $\bar{e}e \to Zh$, as well as CP-odd effects (bottom panel of Table 2) that enter the computations for electric dipole moments, which are constrained with extremely high precision [27]. We study this in detail in Sections 2.8 and 2.2.1 but ignore loop-effects in the rest of this section.

Focusing on tree-level effects, and the operators in the top panel of Table 2, we can translate the reach on Higgs couplings ($\kappa$s) into EFT. The operators $\mathcal{O}_H$ and $\mathcal{O}_{y_\psi}$ (for the different fermions species $\psi$) modify processes that are tree-level in the SM, and have a simple translation in terms of $\kappa$s:

$$\kappa_{HWW} = 1 - \hat{c}_H/2, \quad \kappa_{HZZ} = 1 - \hat{c}_H/2, \quad \kappa_{H\psi\psi} = 1 - (\hat{c}_H/2 + \hat{c}_\psi), \quad \kappa_\lambda = 1 + \hat{c}_6 - 3\hat{c}_H/2, \quad (5)$$

where we use a hat to denote

$$\hat{c}_i = c_i \frac{v^2}{\Lambda^2}. \quad (6)$$

These operators are also typically generated together in models that account for modified Higgs dynamics, such as composite Higgs models [17]. Notice that, under the assumptions we have spelled out, the modification of the Higgs couplings to $Z$ and to $W$ bosons are identical. Their 68% CL reach, as extracted from Higgs couplings, is reported in the top panel of Table 5.

On the other hand $\kappa_{Hgg}$, $\kappa_{H\gamma\gamma}$ and $\kappa_{HZ\gamma}$ are loop-level generated in the SM. Their expression in terms of EFT operator coefficients can be found, for instance, in Ref. [28]; what suffices for the present

Table 6: Left: Projected statistical precision (68% C.L.) of the $\kappa$ parameter Eq. (8) and $\hat{c}_H$ for the three CLIC stages. Middle: Same for the $g$ and $\Gamma_H$ parameters defined in the text. Right: Precision on $\Gamma_H$ alone.

|            | $\Delta\kappa$ | $|\hat{c}_H|$ |            | $\Delta g$ | $\Delta\Gamma_H$ |            | $\Delta\Gamma_H$ |
|------------|--------|--------|------------|-------|-------|------------|-------|
| Stage 1    | 0.22%  | 0.0011 | Stage 1    | 0.58% | 2.3%  | Stage 1    | 0.47% |
| Stage 1+2  | 0.10%  | 0.0005 | Stage 1+2  | 0.57% | 2.3%  | Stage 1+2  | 0.20% |
| Stage 1+2+3| 0.06%  | 0.0003 | Stage 1+2+3| 0.57% | 2.3%  | Stage 1+2+3| 0.13% |

purpose is that, at tree level in BSM,

$$\kappa_{Hgg} \leftrightarrow \hat{c}_H, \hat{c}_{GG}, \quad \kappa_{H\gamma\gamma} \leftrightarrow \hat{c}_H, \hat{c}_{BB}, \hat{c}_{WW}, \quad \kappa_{HZ\gamma} \leftrightarrow \hat{c}_H, \hat{c}_{BB}, \hat{c}_{WW}. \tag{7}$$

These operators are also typically generated at loop-level if additional BSM particles couple to the Higgs and to the SM gauge bosons. Their reach from studies of Higgs couplings is reported in the bottom panel of Table 5.

**Universal coupling scale parameter fit**

As mentioned in the introduction, specific BSM scenarios might exhibit special correlations between the coefficients of the operators they generate. An interesting example is the one where only $\mathcal{O}_H$ appears in the EFT, as could be the case in models where the Higgs sector is extended by a scalar singlet (with a small self-coupling), see Section 4.2. Then dedicated Higgs coupling fits are performed with the assumption that the Higgs boson couplings to all SM particles scale in the same manner, defining

$$\kappa^2 = \Gamma_i / \Gamma_i^{\text{SM}} \text{ with } i = \text{HZZ}, \text{HWW}, \text{Hbb}, \text{Hcc}, \text{H}\tau\tau, \text{H}\mu\mu, \text{Htt}, \text{Hgg}, \text{H}\gamma\gamma \text{ and } \text{HZ}\gamma, \tag{8}$$

where, in terms of the coefficient of $\mathcal{O}_H$,

$$\kappa = 1 - \hat{c}_H/2. \tag{9}$$

The implementation of the $\chi^2$ minimisation follows the model-dependent approach described in Ref. [9]. Correlations of the experimental projections are included where relevant. The expected statistical precisions of the $\kappa$ parameter for three CLIC energy stages are summarised in the left panel of Table 6. Systematic uncertainties can not be ignored given the expected statistical accuracies, especially at high energy. The most precise input to the fit is the rate of H $\to$ b$\bar{\text{b}}$ events in WW fusion at multi-TeV energies. It is expected that the systematic uncertainty of this measurement can be controlled with similar precision compared to its statistical uncertainty [9]. The other fitted Higgs observables have statistical precisions on the percent level and it is generally expected that the systematic uncertainties are small compared to the statistical errors. All measurements used as input to the fit are affected by the knowledge of the total luminosity, which is fully correlated for all measurements at a given energy stage. With the luminosities envisaged for CLIC [29, 30], it is expected that this impact on $\kappa$ will be small compared to the statistical uncertainty for the first CLIC stage and on the per mille level for the higher-energy stages. While a full study of all sources of systematic uncertainties requires more knowledge of the technical implementation of the detector than is currently available, it seems possible to control the systematic uncertainty on $\kappa$ to a level not largely exceeding the expected statistical precisions even for the high-energy stages of CLIC.

**Simplified Higgs fits including the total width**

It is worth considering an additional scenario, which departs from our original EFT assumptions, in which the Higgs boson has additional, non-SM, decays. This scenario cannot be captured by our parametrizations above, but is easily addressed by adding the total Higgs width, $\Gamma_H$, as a second fit parameter in addition to the universal coupling scale parameter here referred to as $g$ (the analog of $\kappa$ in the previous paragraph). Analogously to the model-independent fit described in Ref. [9], the total cross section for the $e^+e^- \to ZH$ process obtained using the recoil method is directly proportional to $g^2$. This provides sensitivity to $\Gamma_H$ from a global fit to the measurements of individual Higgs decay modes in ZH and $WW$ fusion events.

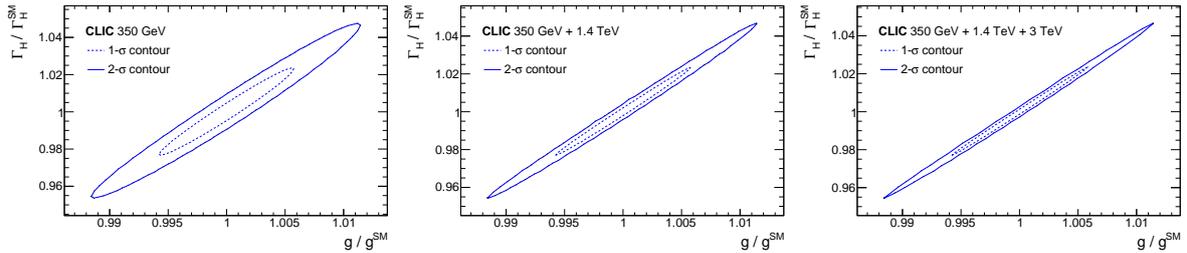

Figure 8: 1- and 2-$\sigma$ contours from the two-parameter Higgs fit for the three CLIC energy stages.

The middle panel of Table 6 gives the expected statistical precisions of the $g$ and $\Gamma_H$ parameters. The accuracy of disentangling both parameters is limited by the measurement of the total ZH cross section at the first CLIC stage and hence only improves marginally when including the higher energy stages; this is manifest in the contour plots of $g$ versus $\Gamma_H$ as shown in Figure 8. The systematic uncertainties are expected to be small compared to the expected statistical precisions for this two-parameter fit.

If all Standard Model couplings of the Higgs boson are fixed to their default values, the precision on the total Higgs width improves considerably. The result of such a fit is shown in the right panel of Table 6. In contrast to the two parameter fit, the width is not limited by the ZH measurement at the first CLIC stage and its projected precision improves with energy.

## 2.2 Determination of the Higgs trilinear self-coupling

In this section we perform a detailed analysis of measurements that aim at identifying effects from the operator $\mathcal{O}_6$, which modifies the SM triple Higgs coupling. In Section 2.2.1 we present a parton-level analysis that includes all possible BSM contributions (in the form of the dimension-6 operators from Table 2). Yet, this analysis is not optimised and neglects detector effects, as well as ISR and beamstrahlung. For this reason, we present in Section 2.2.2 a more detailed study that focusses on the effects of $c_6$ (or $\kappa_\lambda$) only and should be thought as an illustration of how much additional reach can be gained by a dedicated study. Section 2.2.3 is dedicated to discuss other interesting BSM effects that enter in di-Higgs processes.[5]

### 2.2.1 Global perspective on the Higgs self-coupling[6]

**High-energy**

The optimal way to measure the Higgs trilinear self coupling at high-energy lepton colliders is through the exploitation of Higgs pair production processes, whose cross section is affected by the Higgs self coupling at tree level. An electron-positron collider like CLIC offers two main di-Higgs production modes [9], namely double Higgsstrahlung ($e^+e^- \to Zhh$) and vector boson fusion ($e^+e^- \to \nu\bar{\nu}hh$), see Figure 9 for representative diagrams. The cross section for the two channels has different scaling as a function of the centre-of-mass energy of the collider (see Figure 10). Double Higgsstrahlung reaches a maximum not far from threshold (at $\sqrt{s} \sim 500\,\text{GeV}$) and then decreases due to the $s$-channel $Z$ boson propagator. On the other hand, the vector boson fusion cross section benefits from a $t$-channel logarithmic enhancement and grows with the collider energy. Double Higgsstrahlung and vector boson fusion cross sections are equal at a centre-of-mass energy of around 1 TeV. In this section we focus on CLIC Stages 2 and 3 and we perform simulations using MadGraph [31].

---

[5]These three sections, utilise a CLIC running scenario that differs slightly from that of Table 3: Stage 1 runs at 350 GeV, while Stage 2 runs at $\sqrt{s} = 1.4$ TeV.

[6]Based on a contribution by S. Di Vita, G. Durieux, C. Grojean, J. Gu, Z. Liu, G. Panico, M. Riembau, T. Vantalon.

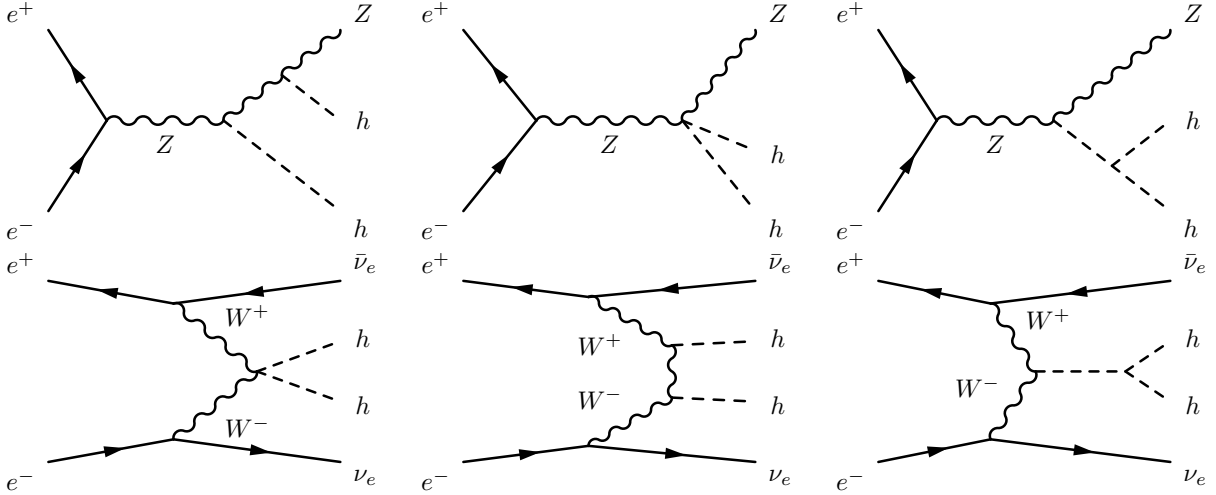

Figure 9: Illustrative diagrams contributing to the di-Higgs boson production at lepton colliders

The dependence of both di-Higgs production cross sections on the value of the trilinear Higgs self coupling weakens with the centre-of-mass energy. At $\sqrt{s} = 1.4$ TeV, this dependence is already relatively weak for the double Higgsstrahlung cross section. It is significantly larger in vector boson fusion. The right panel in Figure 10 shows how the total cross section of the two Higgs pair-production channels depends on the trilinear Higgs self coupling. The result is presented as a function of

$$\delta\kappa_\lambda = \kappa_\lambda - 1 = \hat{c}_6 - \frac{3}{2}\hat{c}_H \qquad (10)$$

which denotes the correction to the Higgs self coupling normalized to its SM value, here given in terms of the dimension-6 operator of Table 2.

The right panel of Figure 10 shows an interesting complementarity between the two Higgs pair production channels. Due to a positive interference, the $Zhh$ cross section grows for $\delta\kappa_\lambda > 0$, so that it can more easily constrain positive deviations in the trilinear Higgs self coupling, but is mostly insensitive to negative deviations. On the contrary, $\nu\bar{\nu}hh$ production is more sensitive to negative shifts of the trilinear coupling that increase the cross section. Notice moreover that the vector-boson-fusion cross section reproduces the SM one also for $\delta\kappa_\lambda \sim 1$, therefore such large positive deviations can not be tested with the $\nu\bar{\nu}hh$ inclusive rate. So, although the $Zhh$ sensitivity is weaker than the $\nu\bar{\nu}hh$ one, the former can still be useful to probe values $\delta\kappa_\lambda \sim 1$. We stress that the above considerations are valid in the case in which the true value of the Higgs trilinear self coupling is close to the SM one (i.e. $\delta\kappa_\lambda \simeq 0$). In the presence of sizeable deviations the sensitivity can become significantly different.

We find that, after combining both vector boson fusion and double Higgsstrahlung channels, CLIC Stages 2 and 3 are sufficient to exclude the second fit minimum at $\delta\kappa_\lambda \sim 1$ at 95%C.L. . Another possibility to lift the degenerate minima is to consider the information on the invariant mass spectrum of the two Higgs bosons, $m_{hh}$, since it offers an excellent discrimination power thanks to the large sensitivity to modifications of the Higgs trilinear coupling [32]. Large positive values of $\delta\kappa_\lambda$ lead to a spectrum with a sharp peak close to threshold followed by a steep fall off. A simple cut-and-count analysis with a few bins is thus sufficient to distinguish this distribution from the SM one [33]. Here we present a simplified version of the analysis in Section 2.2.2, where the $m_{hh}$ distribution is splitted in 5 bins.

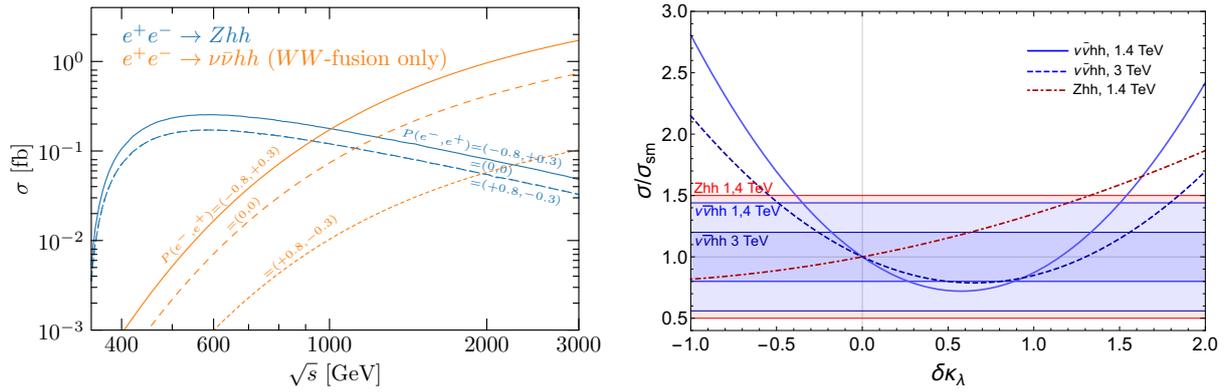

Figure 10: **Left:** Cross section of the main di-Higgs production modes in a lepton collider as a function of the centre-of-mass energy. **Right:** Dependence of the signal strengths on the trilinear coupling of the Higgs for different centre-of-mass energies. The horizontal bands show expected sensitivities.

As can be seen from the results in Table 7, differential information in vector boson fusion di-Higgs production at $\sqrt{s} = 3\,\text{TeV}$ allows one to constrain $\delta\kappa_\lambda$ to the range $[-0.11, 0.13]$ at the $\Delta\chi^2 = 1$ level. This result should be compared with the $[-0.13, 0.16] \cup [1.13, 1.42]$ constraint that is achievable with inclusive cross section measurements only.

Table 7: Exclusive constraints on $\delta\kappa_\lambda$ deriving from the measurements of $Zhh$ and $\nu\bar{\nu}hh$ cross sections, with all other parameters fixed to their standard-model values. A differential $m_{hh}$ measurement in weak boson fusion di-Higgs production at $\sqrt{s} = 3\,\text{TeV}$ is additionally considered in the last two rows.

|  | $\Delta\chi^2 = 1$ | $\Delta\chi^2 = 4$ |
|---|---|---|
| CLIC Stage 2 | $[-0.22,\ 0.48]$ | $[-0.40,\ 1.05]$ |
| CLIC Stage 3 | $[-0.13,\ 0.16] \cup [1.13,\ 1.42]$ | $[-0.24,\ 0.42] \cup [0.87,\ 1.53]$ |
| CLIC Stage 2+3 | $[-0.12,\ 0.14]$ | $[-0.21,\ 0.35]$ |
| 5 bins in $\nu\bar{\nu}hh$ | $[-0.11,\ 0.13]$ | $[-0.21,\ 0.29]$ |

**Low-energy and global fit**

Let us now consider the impact of the low-energy CLIC Stage 1 run. Such a run leads to very small double-Higgs-production rates, making these channels irrelevant for determining the Higgs trilinear self coupling. As an alternative, one could exploit high precision measurements of single-Higgs-production processes, which are affected by deviations in the trilinear Higgs self coupling at the one-loop level [36].

Interestingly, single-Higgs processes show a good sensitivity to the Higgs self coupling, thanks to the very high precision with which they can be measured at a lepton collider. In the left panel of Figure 11 we show, in dashed pink, how an exclusive fit to the Higgs self coupling using single-Higgs processes can achieve an $\mathcal{O}(1)$ sensitivity on the Higgs trilinear, surpassing the HL-LHC projections (dotted blue lines). It is important to stress that this result holds in the case in which one performs an exclusive study of the trilinear dependence, assuming that all single-Higgs couplings take exactly their SM values. In most new physics scenarios, however, deviations in the Higgs potential are generically accompanied by modifications in other Higgs couplings. It is therefore essential to assess the robustness of the previous observation within a global fit that includes the relevant set of Higgs coupling deformations. Following Refs. [33, 37, 38] (see also Section 2.9 in this report), we perform a global fit featuring 13 effective operators that parametrize the relevant deviations from SM Higgs couplings:

$$\{\mathcal{O}_{gg}, \mathcal{O}_{WW}, \mathcal{O}_{BB}, \mathcal{O}_{HW}, \mathcal{O}_{HB}, \mathcal{O}_6, \mathcal{O}_H, \mathcal{O}_{y_t}, \mathcal{O}_{y_b}, \mathcal{O}_{y_c}, \mathcal{O}_{y_\tau}, \mathcal{O}_{y_\mu}, \mathcal{O}_{3W}\}\,. \tag{11}$$

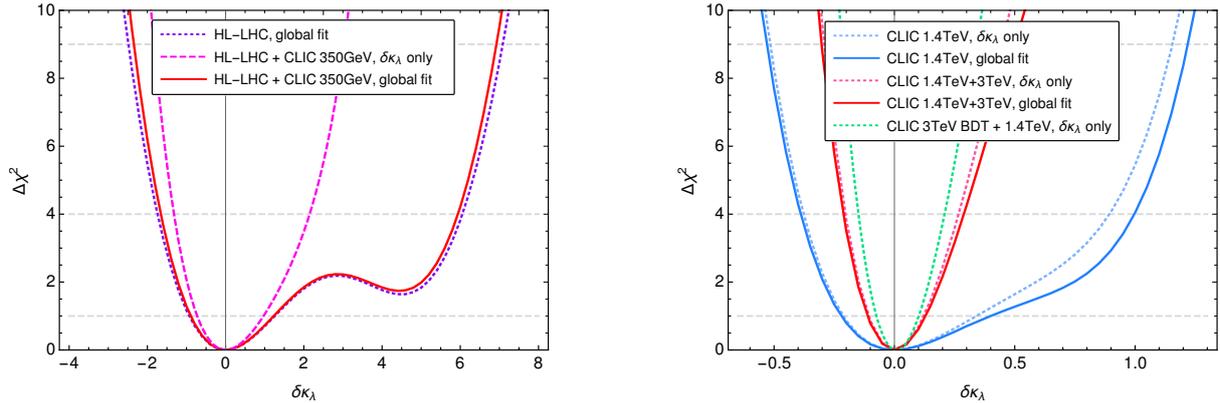

Figure 11: Chi-squared fit on the anomalous Higgs self coupling $\delta\kappa_\lambda$. **Left:** In dashed pink we show an exclusive fit on $\delta\kappa_\lambda$ using single Higgs data only for a 350 GeV run with 0.5 ab$^{-1}$ of integrated luminosity. In solid red, we profile over the rest of EFT parameters, after combination with the HL-LHC likelihood, shown in dotted blue for reference. **Right:** In blue, the chi-squared resulting from the differential $\nu\bar{\nu}hh$, $Zhh$ and single Higgs measurements at $\sqrt{s} = 1.4$ TeV, in combination with HL-LHC measurements [32, 34, 35]. In red, combined with a 3 TeV run. In dotted, the exclusive fit to $\delta\kappa_\lambda$, while in solid lines we show the result from a global fit.

Table 8: Exclusive and global constraints on $\delta\kappa_\lambda$ after the CLIC Stage 2 and 3 runs, and in combination with HL-LHC measurements. This combination explains the improvement compared to the constraints reported in Table 7.

|  | 68 %C.L. | 95%C.L. |
| --- | --- | --- |
| CLIC Stage 2, exclusive | $[-0.21, 0.34]$ | $[-0.38, 0.89]$ |
| CLIC Stage 2, global | $[-0.22, 0.40]$ | $[-0.39, 1.00]$ |
| CLIC Stage 2+3, exclusive | $[-0.11, 0.12]$ | $[-0.20, 0.27]$ |
| CLIC Stage 2+3, global | $[-0.11, 0.13]$ | $[-0.21, 0.29]$ |

We find that a Stage 1 run alone leaves a very strong correlation among deviations in the Higgs trilinear $\delta\kappa_\lambda$ and modifications of the $hZZ$ coupling. As a consequence a global fit does not set meaningful constraints on the trilinear Higgs self coupling. The flat direction can be partially resolved by a combination with HL-LHC measurements. The solid red curve shows the obtained chi-squared in the left panel of Figure 11. One can see that a low-luminosity Stage 1 run can only marginally improve the HL-LHC constraints.

As discussed before, the centre-of-mass energies of CLIC Stages 2 and 3, give access to double Higgs production processes, sensitive to the trilinear Higgs self coupling at tree level. They are however also affected by modifications in other Higgs couplings, and it is important to test the impact these have on the exclusive self coupling constraints reported in Table 7. We performed a study comparing the projections of an exclusive fit to the Higgs trilinear self coupling with a global fit where all the other parameters are profiled over, for a CLIC Stage 2 run alone or in combination with Stage 3 one. We included in the fit $\nu\bar{\nu}hh$ production with a differential analysis including 4 bins in the $m_{hh}$ distribution. The inclusive $Zhh$ cross section and the $\delta\kappa_\lambda$ dependence of the single-Higgs processes. We report the results in Table 8 and in the right panel of Figure 11.

The CLIC Stage 2 constraints on the Higgs self coupling are dominated by the differential $\nu\bar{\nu}hh$ measurement. Moreover, the constraints on the other Higgs couplings obtained from the single-Higgs observables are strong enough to close any possible flat direction. As a consequence the exclusive study gives similar constraints as the global fit. This shows that a possible low-energy CLIC Stage 1 run is not strictly necessary to avoid correlations and obtain a robust determination of the trilinear Higgs self coupling. The CLIC Stage 3 run will drastically increase the sensitivity on the Higgs self coupling with respect to CLIC Stage 1. This is due to the increase in statistics to get access to detailed differential distributions.

**Two parameter fit**

These results will be discussed from a global fit perspective, including all operators and more processes, in Section 2.9. An important special case that is worth discussing in detail is that in which only the effects of $\mathcal{O}_H$ and $\mathcal{O}_6$ are taken into account, corresponding to new physics effects only in the Higgs sector, for instance in generic models of Higgs and a singlet, see Section 4.2.[7] The sensitivity to this scenario is reported in Figure 12, where we compare it (solid vs dashed) with the reach to the same operators when also all other operators are present, and then marginalised.

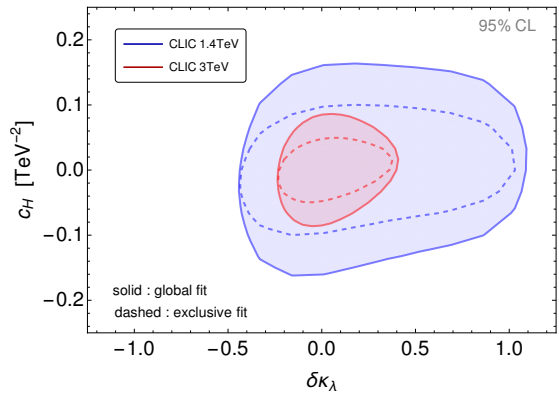

Figure 12: Reach on the coefficients of the operators $\mathcal{O}_H$ and $\mathcal{O}_6$, reported in terms of $c_H$ and $\delta\kappa_\lambda$ (see Eq. (10)). Solid lines include marginalization of all other operators, dashed lines has other operators set to zero.

### 2.2.2 Full simulation study of experimental prospects[8]

In this section we focus on the prospects for the extraction of the trilinear Higgs self-coupling in double Higgs production ($\kappa_\lambda$ only), which have been evaluated with full detector simulation of the CLIC_ILD [5] detector model in the process of vector boson fusion Higgs production ($e^+e^- \to hh\nu_e\bar{\nu}_e$) at 3 TeV. The analysis is based on Ref. [39] and [9] and is summarized here.

The analysis is optimized for the signal of $e^+e^- \to hh\nu_e\bar{\nu}_e$ in the decay channel $hh \to b\bar{b}b\bar{b}$. This channel benefits from the clean environment in electron-positron collisions, the good jet energy resolution of the detector, and the good heavy flavor tagging performance. This study uses a luminosity of $\mathcal{L} = 5000\,\text{fb}^{-1}$ at the 3 TeV energy stage of CLIC with an electron beam polarization of -80 %(+80 %) for 80 %(20 %) of the run [12], here called the "4:1 polarization scheme". With this mix of polarizations, the signal cross section is enhanced by a factor of 1.48. The same factor is applied to all backgrounds as a conservative estimate.

Events containing isolated leptons or hadronic $\tau$ lepton candidates are rejected. The remaining events are required to pass exclusive clustering in four jets using the $k_t$ algorithm with a jet size parameter of $R = 0.7$. Flavour tagging is based on the LCFIPLUS package [40]. The sum of the $b$-tag values in the $b\bar{b}b\bar{b}$ candidate events is required to be $\sum b-\text{tag} > 2.3$. These cuts define the pre-selection. The four jets are then grouped into two Higgs candidates by minimizing the absolute difference between the resulting di-jet masses $|m_{ij} - m_{kl}|$ among the possible combinations. A multivariate analysis based on Boosted Decision Trees is performed using the pre-selected events. The algorithm is trained for the $hh \to b\bar{b}b\bar{b}$ signal based on observables of the two Higgs candidates (angular distance and invariant masses), flavor tagging information (b-tag weights and c-tag weights), the invariant mass of the $b\bar{b}b\bar{b}$ system, the total missing transverse momentum, the number of photons, as well as the maximum absolute pseudorapidity of the four jets. A cut is applied on the resulting BDT response. For the cross section measurement, the optimal cut value is chosen to maximize the significance ("tight BDT"). The cut criterion is redefined for the template fit to allow higher statistics ("loose BDT").

---
[7]Differently from the discussion around Eq. (9), the analysis of this section applies also in models where the singlet-self coupling is large.

[8]Based on a contribution by U. Schnoor, P. Roloff, R. Simoniello, B. Xu.

Table 9 presents the resulting event yields in the loose and tight BDT regions. Considering the full process $e^+e^- \to hh\nu_e\bar{\nu}_e$ with all decay modes as the signal, the precision on the measurement is $\Delta\lambda/\lambda = 7.4\,\%$.

Table 9: Selection efficiencies as well as expected number of events after the tight and loose BDT selections for $\mathcal{L} = 5\,\text{ab}^{-1}$ with the 4:1 polarization scheme for the $hh\nu_e\bar{\nu}_e$ process and the main backgrounds at $\sqrt{s} = 3\,\text{TeV}$.

| Process | $\sigma$/fb | $\epsilon_{\text{loose BDT}}$ | $N_{\text{loose BDT}}$ | $\epsilon_{\text{tight BDT}}$ | $N_{\text{tight BDT}}$ |
|---|---|---|---|---|---|
| $e^+e^- \to hh\nu_e\bar{\nu}_e$ | 0.59 | 17.6 % | 766 | 8.43 % | 367 |
| only $hh \to b\bar{b}b\bar{b}$ | 0.19 | 39.8 % | 559 | 17.8 % | 250 |
| only $hh \to$ other | 0.40 | 6.99 % | 207 | 3.95 % | 117 |
| $e^+e^- \to q\bar{q}q\bar{q}$ | 547 | 0.0065 % | 259 | 0.00033 % | 13 |
| $e^+e^- \to q\bar{q}q\bar{q}\nu\bar{\nu}$ | 72 | 0.17 % | 876 | 0.017 % | 90 |
| $e^+e^- \to q\bar{q}q\bar{q}\ell\bar{\nu}$ | 107 | 0.053 % | 421 | 0.0029 % | 23 |
| $e^+e^- \to q\bar{q}h\nu\bar{\nu}$ | 4.7 | 3.8 % | 1171 | 0.56 % | 174 |
| $e^\pm\gamma \to \nu q\bar{q}q\bar{q}$ | 523 | 0.023 % | 821 | 0.0014 % | 52 |
| $e^\pm\gamma \to qqh\nu$ | 116 | 0.12 % | 979 | 0.0026 % | 21 |

A template fit based on full detector simulation of the samples with different values of the trilinear Higgs self-coupling $\lambda$ is used to determine the prospective precision of the extraction of $\lambda$ in the case that the SM is observed. For this, the deviations from the SM case are quantified by a $\chi^2$ measure. Pseudoexperiments are used to account for the statistical limitation in real experiments.

If only the cross section measurement of $hh\nu\bar{\nu}$ at 3 TeV is taken into account for the extraction of the trilinear coupling, the ambiguity demonstrated in Figure 10(right) results in two combined confidence intervals for $\delta\kappa_\lambda$: $[-0.10, +0.12] \cup [1.40, 1.61]$ obtained in the tight BDT selection. This can be combined with the prospective measurement of the cross section of double Higgsstrahlung $Zhh$ production, for which the second stage of CLIC is most suitable. In lack of a full simulation study, we assume with guidance from CLIC prospects for the $Zh$ measurement [9] that the event selection for $Zhh$ production at 1.4 TeV can be performed with a signal efficiency of 50 % and is background free as heavy flavor tagging requirements and the invariant mass separation between $W$, $Z$, and $h$ suppress any possible contributions from multiboson final states (cf. Section 2.5). Combining both cross section measurements resolves the ambiguity and results in the confidence interval $[-0.10, +0.11]$ for $\delta\kappa_\lambda$.

As in the previous paragraph, (cf. Table 7), the sensitivity of the invariant mass of the Higgs boson pair, reconstructed from the four identified $b$ jets, is exploited to improve the precision. For this we use the loose BDT selection to increase statistics. The invariant mass $M(hh)$ is combined with the response of the multivariate analysis ("BDT score") to define kinematic bins for the fit procedure. The resulting distribution is shown in Figure 14. With the templates based on these kinematic bins, the 68 % C.L. constraints improve to $[-0.07, +0.12]$. Combining this with the $Zhh$ cross section measurement as defined above, we obtain $[-0.07, +0.11]$ as constraints on $\delta\kappa_\lambda$. The constraints on $\delta\kappa_\lambda$ are summarized

Table 10: Constraints on $\delta\kappa_\lambda$ obtained in the full detector simulation study using a multivariate analysis for selection. The constraint from cross section only is obtained in a tight selection optimized for cross section precision. The constraints based on differential distributions are derived in a looser selection.

| Constraints for $\delta\kappa_\lambda$ based on | $\Delta\chi^2 = 1$ |
| --- | --- |
| $hh\nu\bar{\nu}$ cross section only (3 TeV) | $[-0.10, +0.12] \cup [1.40, 1.61]$ |
| $hh\nu\bar{\nu}$ (3 TeV) and $Zhh$ (1.4 TeV) cross section | $[-0.10, +0.11]$ |
| $hh\nu\bar{\nu}$ differential (3 TeV) | $[-0.07, +0.12]$ |
| $hh\nu\bar{\nu}$ differential (3 TeV) and $Zhh$ cross section (1.4 TeV) | $[-0.07, +0.11]$ |

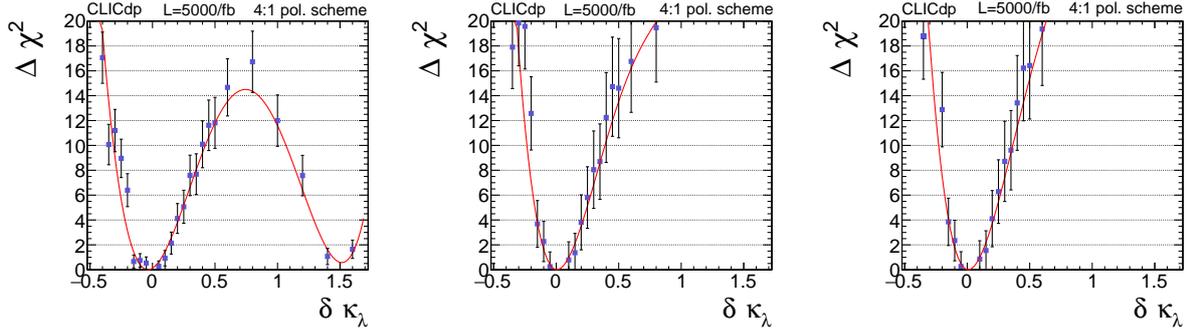

Figure 13: Nominal $\Delta\chi^2$ distributions of the templates for different values of $\kappa\lambda$ using (Left) only cross section information for the $hh\nu\bar{\nu}$ process at 3 TeV in the tight BDT selection, (Center) the differential distribution comprising the BDT score and $M(hh)$ in $hh\nu\bar{\nu}$ at 3 TeV in the loose BDT selection, and (Right) the combination of the differential distribution in $hh\nu\bar{\nu}$ at 3 TeV in the loose BDT selection with the cross section measurement of $Zhh$ at 1.4 TeV.

in Table 10 and the corresponding $\chi^2$ curves are shown in Figure 13.

In conclusion, the measurement of the trilinear Higgs self-coupling makes use of differential distributions sensitive to modifications of the $hhh$ vertex in VBF double Higgs production at 3 TeV as well as the combination with a cross-section measurement of $Zhh$ production at 1.4 TeV. The resulting constraints on the trilinear Higgs self-coupling modification $\delta\kappa_\lambda$ that can be reached with the CLIC high-energy stages, assuming no other BSM effects, are $[-0.07, +0.11]$.

### 2.2.3 Precision measurements of HVV and HHVV Couplings[9]

In this paragraph we motivate precision measurements on the tensor structures of one Higgs couplings with two electroweak gauge bosons (HVV) and two Higgses couplings with two electroweak gauge bosons (HHVV) in CLIC. This motivation is based on models of composite Higgs which, as discussed in the introduction, are the primary target of precision tests; yet it is not limited to the dimension-6 EFT effects discussed so far, hence the different approach taken in this section. There exist special relations between HVV and HHVV couplings in composite Higgs models that are universal, independent of the symmetry breaking pattern invoked in a particular model. These "universal relations" are controlled by a single input parameter, the decay constant $f$ of the pseudo-Nambu-Goldstone Higgs boson. Testing the universal relations requires measuring the tensor structures of HVV and HHVV couplings to high precision. In particular, HHVV interactions remains as one of the few untested predictions of the Standard Model Higgs boson, which can be probed through the double Higgs production in the vector boson fusion and associate production channels at CLIC. Below we summarize the main results. The phenomenological details and theoretical foundation can be found in Refs. [41–44].

---

[9]Based on a contribution by D. Liu, I. Low and Z. Yin.

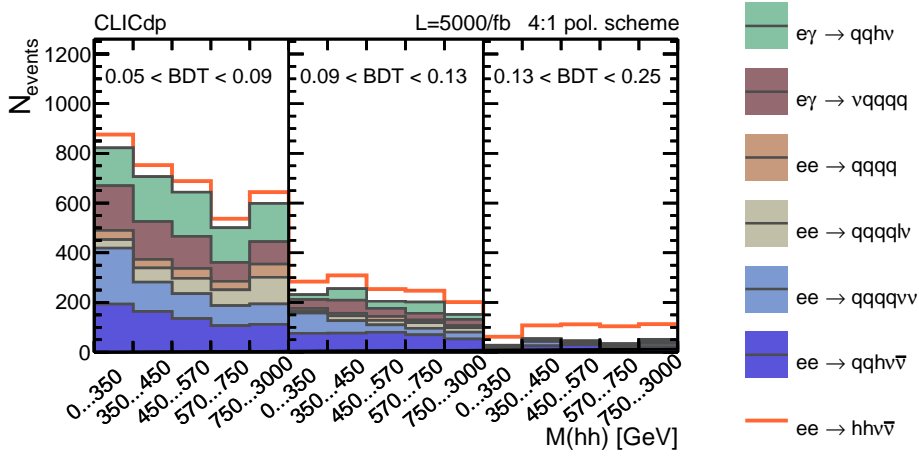

Figure 14: Distribution of the di-Higgs invariant mass $M(hh)$ in three bins of the BDT score in the loose BDT selection. These kinematic bins are used for the templates of different values of $\kappa_\lambda$ in the fits with differential distribution. The orange line marked as $ee \to hh\nu\bar{\nu}$ corresponds to the SM case ($\kappa_\lambda = 1$).

At the leading two-derivative order, the HVV and HHVV couplings in composite Higgs models in the unitary gauge is given by the following simple expression:

$$\mathcal{L}^{(2)} = \frac{1}{2}\partial_\mu h \partial^\mu h + \frac{g^2 f^2}{4}\sin^2(\theta + h/f)\left(W_\mu^+ W^{-\mu} + \frac{1}{2\cos^2\theta_W}Z_\mu Z^\mu\right), \tag{12}$$

where $v = 246$ GeV, $f$ is the decay constant of the composite Higgs boson and $\sin\theta = v/f$. This result is independent of the symmetry breaking pattern of the strong composite sector in the UV, apart from the overall normalization of $f$, which does depend on the UV model.

At the four-derivative level, we parameterize the HVV and HHVV couplings as follows:

$$\mathcal{L}^{(4)} = \sum_i \frac{m_W^2}{m_\rho^2}\left(C_i^h \mathcal{I}_i^h + C_i^{2h} \mathcal{I}_i^{2h}\right), \tag{13}$$

where the definition of the operators $\mathcal{I}_i^h$ and $\mathcal{I}_i^{2h}$ are presented in Table 11 and Table 12. On the other hand, $C_i^h$ and $C_i^{2h}$ are Wilson coefficients which depend on six unknowns ($\theta, c_3, c_4^\pm, c_5^\pm$) in composite Higgs models and on four unknowns ($c_W, c_B, c_{HW}, c_{HB}$) in the Standard Model Effective Field Theory (SMEFT). In the above $m_\rho = g_\rho f$ is the typical mass scale of the new composite resonances. The different Lorentz structures lead to different angular distributions in the decay products and, therefore, can be measured accordingly. At the LHC Run 1, testing the tensor structure of HVV couplings was among the top priorities and gave confidence to the Higgs nature of the 125 GeV resonance. (See, for example, Ref. [45].) A similar program for HHVV coupling is currently lacking and should be pursued at CLIC.

In general, we have two different Lorentz structure in the HVV couplings:

$$\frac{h}{v}V_{1\,\mu}\mathcal{D}^{\mu\nu}V_{2\,\nu}\,, \qquad \frac{h}{v}V_{1\,\mu\nu}V_2^{\mu\nu}\,, \tag{14}$$

where $\mathcal{D}^{\mu\nu} = \partial^\mu\partial^\nu - \eta^{\mu\nu}\partial^2$ and $V_{1,2} \in \{W, Z, \gamma\}$ with electric charge conservation implicitly indicated. For HHVV couplings we have:

$$\frac{h^2}{v^2}V_{1\,\mu}\mathcal{D}^{\mu\nu}V_{2\,\nu}\,, \quad \frac{h^2}{v^2}V_{1\,\mu\nu}V_2^{\mu\nu}\,, \quad \frac{\partial_\mu h \partial_\nu h}{v^2}V_1^\mu V_2^\nu\,, \quad \frac{\partial_\mu h \partial^\mu h}{v^2}V_1^\mu V_{2\,\mu}\,. \tag{15}$$

The ultimate goal then will be to measure these different tensor structures at CLIC.

Table 11: Single Higgs coupling coefficients $C_i^h$ for the non-linearity case (NL) and the purely dimension-6 contributions (D6) in SMEFT. Here $c_w, t_w$ and $c_\theta$ denote $\cos\theta_W, \tan\theta_W$ and $\cos\theta$, respectively, where $\theta_W$ is the weak mixing angle. $\mathcal{D}^{\mu\nu}$ denotes $\partial^\mu \partial^\nu - \eta^{\mu\nu}\partial^2$. Hermitian conjugate of an operator is implied when necessary.

| $\mathcal{I}_i^h$ | $C_i^h$ (NL) | $C_i^h$ (D6) |
|---|---|---|
| (1) $\frac{h}{v}Z_\mu \mathcal{D}^{\mu\nu} Z_\nu$ | $\frac{4c_{2w}}{c_w^2}(-2c_3 + c_4^-) + \frac{4}{c_w^2}c_4^+ c_\theta$ | $2(c_W + c_{HW}) + 2t_w^2(c_B + c_{HB})$ |
| (2) $\frac{h}{v}Z_{\mu\nu}Z^{\mu\nu}$ | $-\frac{2c_{2w}}{c_w^2}(c_4^- + 2c_5^-) - \frac{2}{c_w^2}(c_4^+ - 2c_5^+)c_\theta$ | $-(c_{HW} + t_w^2 c_{HB})$ |
| (3) $\frac{h}{v}Z_\mu \mathcal{D}^{\mu\nu} A_\nu$ | $8(-2c_3 + c_4^-)t_w$ | $2t_w(c_W + c_{HW} - c_B - c_{HB})$ |
| (4) $\frac{h}{v}Z_{\mu\nu}A^{\mu\nu}$ | $-4(c_4^- + 2c_5^-)t_w$ | $-t_w(c_{HW} - c_{HB})$ |
| (5) $\frac{h}{v}W_\mu^+ \mathcal{D}^{\mu\nu} W_\nu^-$ | $4(-2c_3 + c_4^-) + 4c_4^+ c_\theta$ | $2(c_W + c_{HW})$ |
| (6) $\frac{h}{v}W_{\mu\nu}^+ W^{-\mu\nu}$ | $-4(c_4^- + 2c_5^-) - 4(c_4^+ - 2c_5^+)c_\theta$ | $-2c_{HW}$ |

Table 12: The coupling coefficients $C_i^{2h}$ involve two Higgs bosons for universal nonlinearity case (NL) and the dimension-six case in SMEFT (D6). A cross ($\times$) means there is no contribution at the order we considered. Notice $C_i^{2h} = C_i^h/2$ for SMEFT at the dimension-6 level. $c_{2\theta}$ and $s_\theta$ denote $\cos 2\theta$ and $\sin\theta$, respectively.

| $\mathcal{I}_i^{2h}$ | $C_i^{2h}$ (NL) | $C_i^{2h}$ (D6) |
|---|---|---|
| (1) $\frac{h^2}{v^2}Z_\mu \mathcal{D}^{\mu\nu} Z_\nu$ | $\frac{2c_{2w}}{c_w^2}(-2c_3 + c_4^-)c_\theta + \frac{2}{c_w^2}c_4^+ c_{2\theta}$ | $\frac{1}{2}C_1^h$ |
| (2) $\frac{h^2}{v^2}Z_{\mu\nu}Z^{\mu\nu}$ | $-\frac{c_{2w}}{c_w^2}(c_4^- + 2c_5^-)c_\theta - \frac{1}{c_w^2}(c_4^+ - 2c_5^+)c_{2\theta}$ | $\frac{1}{2}C_2^h$ |
| (3) $\frac{h^2}{v^2}Z_\mu \mathcal{D}^{\mu\nu} A_\nu$ | $4t_w(-2c_3 + c_4^-)c_\theta$ | $\frac{1}{2}C_3^h$ |
| (4) $\frac{h^2}{v^2}Z_{\mu\nu}A^{\mu\nu}$ | $-2t_w(c_4^- + 2c_5^-)c_\theta$ | $\frac{1}{2}C_4^h$ |
| (5) $\frac{h^2}{v^2}W_\mu^+ \mathcal{D}^{\mu\nu} W_\nu^-$ | $2(-2c_3 + c_4^-)c_\theta + 2c_4^+ c_{2\theta}$ | $\frac{1}{2}C_5^h$ |
| (6) $\frac{h^2}{v^2}W_{\mu\nu}^+ W^{-\mu\nu}$ | $-2(c_4^- + 2c_5^-)c_\theta - 2(c_4^+ - 2c_5^+)c_{2\theta}$ | $\frac{1}{2}C_6^h$ |
| (7) $\frac{(\partial_\nu h)^2}{v^2}Z_\mu Z^\mu$ | $\frac{8}{c_w^2}c_1 s_\theta^2$ | $\times$ |
| (8) $\frac{\partial_\mu h \partial_\nu h}{v^2} Z^\mu Z^\nu$ | $\frac{8}{c_w^2}c_2 s_\theta^2$ | $\times$ |
| (9) $\frac{(\partial_\nu h)^2}{v^2}W_\mu^+ W^{-\mu}$ | $16 c_1 s_\theta^2$ | $\times$ |
| (10) $\frac{\partial^\mu h \partial^\nu h}{v^2} W_\mu^+ W_\nu^-$ | $16 c_2 s_\theta^2$ | $\times$ |

From Table 11 and Table 12, we can extract relations among $C_i^h$ and $C_i^{2h}$ that only depend on the $\theta$. We call them "universal relations" as they represent universal predictions of a composite Higgs boson, whose nonlinear interactions are dictated by the underlying shift symmetries acting on the four components of the Higgs doublet [41–44]. Some examples of universal relations involving both HVV and HHVV couplings are:

$$\frac{C_3^{2h}}{C_3^h} = \frac{C_4^{2h}}{C_4^h} = \frac{1}{2}\cos\theta, \tag{16}$$

$$\frac{C_5^{2h} - C_3^{2h}/2t_w}{C_5^h - C_3^h/2t_w} = \frac{C_6^{2h} - C_4^{2h}/t_w}{C_6^h - C_4^h/t_w} = \frac{\cos 2\theta}{2\cos\theta} \approx \frac{1}{2}\left(1 - \frac{3}{2}\xi\right), \tag{17}$$

$$\frac{s_{2w} C_1^{2h} - c_{2w} C_3^{2h}}{s_{2w} C_1^h - c_{2w} C_3^h} = \frac{s_{2w} C_2^{2h} - c_{2w} C_4^{2h}}{s_{2w} C_2^h - c_{2w} C_4^h} = \frac{\cos 2\theta}{2\cos\theta} \approx \frac{1}{2}\left(1 - \frac{3}{2}\xi\right). \tag{18}$$

These relations depend on one single parameter $\theta$ or, equivalently, $\xi = v^2/f^2$. In other words, they can be used to over-constrain the parameter $f$. If the 125 GeV Higgs boson indeed arises as a pseudo-

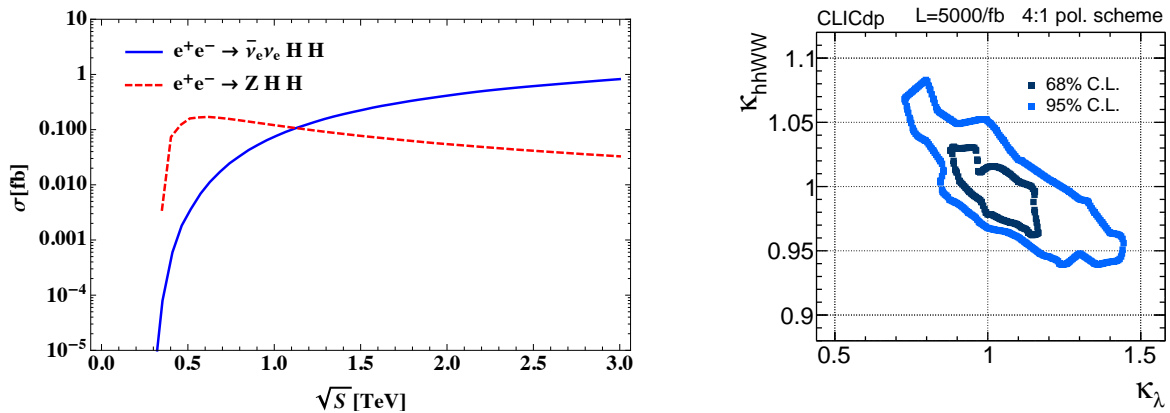

Figure 15: (Left) $WW$ fusion and associate production of double Higgs bosons at CLIC. (Right) Two-dimensional constraints at 68 % C.L. of the trilinear Higgs coupling $\kappa_\lambda$ and the quartic coupling $\kappa_{hhWW}$ based on the differential distribution in $HH\nu_e\bar\nu_e$ production at 3 TeV and the measurement of the $Zhh$ cross section at 1.4 TeV, see Section 2.2.2 and Ref. [39].

Nambu-Goldstone boson, the decay constant $f$ as measured from the different universal relations must be consistent with one another.

In order to test the universal relations, it is necessary to measure the tensor structures of HHVV couplings. This is where CLIC could have an advantage over other lepton colliders, given its 3 TeV centre-of-mass energy. At CLIC, $C_i^h$ can be measured from single Higgs decays into four leptons in a fashion similar to the analysis performed in Ref. [45], while measurements on $C_i^{2h}$ would have to rely on double Higgs production in the $WW$ fusion channel and the associated production with a $Z$ boson. The production topologies are displayed in the second and fourth diagrams of Figure 9.

In Figure 15 we show the double Higgs production rate in the $WW$ fusion channel and the associated production channel in an $e^+e^-$ collider as a function of the centre-of-mass energy $\sqrt{S}$. For a CM energy below roughly 1.2 TeV, the associate production dominates over the $WW$ fusion. However, in the high energy regime, the WW fusion production rate rises continually, reaching 0.9 fb for $\sqrt{S} = 3$ TeV. As a comparison, at $\sqrt{S} = 1$ TeV the associate production rate is about 0.1 fb. This demonstrates that a 3 TeV CLIC is a unique machine to probe the HHVV coupling structure. In the right panel of Figure 15 we show the 68 and 95 % C.L. reach on $\kappa_{hhWW} = g_{hhWW}/g_{hhWW}^{SM}$ and $\kappa_\lambda$ as discussed in Section 2.2.2.

## 2.3 Probing the Higgs with $ZH$ angular observables[10]

In this section, we investigate the prospects for CLIC to constrain or observe anomalous contributions in the Higgs-strahlung production process $e^+e^- \to hZ$, extending the analysis of Ref. [46] and including beam polarization effects following Ref. [37]. We parameterize deviations from Standard Model predictions in the framework of the dimension-6 Standard Model EFT (SMEFT), which takes the schematic form of Eq. (1). Only a subset of possible operators in the dimension-6 SMEFT contribute to $e^+e^- \to hZ$. Some of these operators induce effects that grow with energy, and are well constrained by rate measurements. However only one combination of operators (see Ref. [47]) survives in the high-energy regime. For this reason, it becomes interesting to also exploit the information hidden in angular observables. One of the goals of this section is to assess the comparative reach and complementarity of rate versus angular measurements.

We will illustrate the relationship between SMEFT operators and angular observables in terms of a minimal operator basis given in Ref. [48]. The relevant operators in this basis are listed in Table 2. The effects of these operators on both rate measurements and angular distributions in $e^+e^- \to hZ$ are most transparently conveyed by the following linear combinations [48]:

$$
\begin{aligned}
\alpha_{ZZ}^{(1)} &= -2c_H - \frac{1}{2}\hat{\delta}_{G_F} + \frac{1}{4}c_{HD}, \\
\alpha_{ZZ} &= c_w^2 c_{WW}/g^2 + s_w^2 c_{BB}/g'^2 + s_w c_w c_{WB}/(gg'), \\
\alpha_{Z\widetilde{Z}} &= c_w^2 c_{W\widetilde{W}}/g^2 + s_w^2 c_{B\widetilde{B}}/g'^2 + s_w c_w c_{\widetilde{W}B}/(gg'), \\
\alpha_{AZ} &= 2s_w c_w (c_{WW} g^2 - c_{BB}/g'^2) + (s_w^2 - c_w^2) c_{WB}/(gg'), \\
\alpha_{A\widetilde{Z}} &= 2s_w c_w (c_{W\widetilde{W}}/g^2 - c_{B\widetilde{B}}/g'^2) + (s_w^2 - c_w^2) c_{\widetilde{W}B}/(gg'), \\
\alpha_{H\ell}^V &= c_{He} + \left(c_{HL} + c_{HL}^{(3)}\right), \\
\alpha_{H\ell}^A &= c_{He} - \left(c_{HL} + c_{HL}^{(3)}\right), \\
\delta g_V &= -\widehat{\alpha}_{H\ell}^V + \frac{\widehat{c}_{HD}}{4} + \frac{\widehat{\delta}_{G_F}}{2} + \frac{4s_w^2}{c_{2w}}\left[\frac{\widehat{c}_{HD}}{4} + \frac{c_w}{s_w}\widehat{c}_{WB} + \frac{\widehat{\delta}_{G_F}}{2}\right], \\
\delta g_A &= -\widehat{\alpha}_{H\ell}^A - \frac{\widehat{c}_{HD}}{4} - \frac{\widehat{\delta}_{G_F}}{2},
\end{aligned} \quad (19)
$$

where $\delta_{G_F} = -c_{4L} + 2c_{HL}^{(3)}$, with $c_{4L}$ the coefficient of the operator

$$\mathcal{O}_{4L} = (\bar{L}\gamma_\mu L)(\bar{L}\gamma^\mu L) \quad (20)$$

that enters because its shift of the input parameter $G_F$ from measurements of the muon decay rate (this effect can be eliminated by using $\alpha_{em}, m_W, m_Z$ as input parameters [49]). These relate (after appropriate normalization) to the coefficients of the respective broken-phase interactions

$$
\begin{aligned}
\mathcal{L}_{\text{eff}} \supset{} & m_Z^2(1+\hat{\alpha}_{ZZ}^{(1)})\frac{h}{v}Z_\mu Z^\mu + \hat{\alpha}_{ZZ}\frac{h}{v}Z_{\mu\nu}Z^{\mu\nu} + \hat{\alpha}_{Z\widetilde{Z}}\frac{h}{v}Z_{\mu\nu}\widetilde{Z}^{\mu\nu} + \hat{\alpha}_{AZ}\frac{h}{v}Z_{\mu\nu}A^{\mu\nu} + \hat{\alpha}_{A\widetilde{Z}}\frac{h}{v}Z_{\mu\nu}\widetilde{A}^{\mu\nu} \\
& + \frac{h}{v}Z_\mu\bar{\ell}\gamma^\mu(c_V + c_A\gamma_5)\ell + \frac{m_Z}{2v}Z_\mu\bar{\ell}\gamma^\mu(1-4s_w^2+\delta g_V - \gamma_5 - \delta g_A\gamma_5)\ell - g_{\text{em}}Q_\ell A_\mu\bar{\ell}\gamma^\mu\ell
\end{aligned} \quad (21)
$$

where $1/v \equiv (\sqrt{2}G_F)^{1/2}$, and hatted quantities are defined in Eq. (6). We will present projected constraints directly in terms of the $\widehat{\alpha}$ and $\delta g$ coefficients, making it straightforward to work out implications in various SMEFT operator bases by constructing the generalization of Eq. (19).

---

[10]Based on a contribution by N. Craig, J. Gu, Z. Liu.

There are six independent form factors in the $e^+e^- \to hZ$ amplitude, leading to six possible independent angular observables, each of which will be sensitive to different linear combinations of the above dimensionless parameters. A convenient basis of angular observables was specified in Ref. [48], namely

$$
\begin{aligned}
\mathcal{A}_{\theta_1} &= \frac{1}{\sigma}\int_{-1}^{1} d\cos\theta_1\, \text{sgn}(\cos(2\theta_1))\frac{d\sigma}{d\cos\theta_1} \\
\mathcal{A}_\phi^{(1)} &= \frac{1}{\sigma}\int_0^{2\pi} d\phi\, \text{sgn}(\sin\phi)\frac{d\sigma}{d\phi} & (22) \\
\mathcal{A}_\phi^{(2)} &= \frac{1}{\sigma}\int_0^{2\pi} d\phi\, \text{sgn}(\sin(2\phi))\frac{d\sigma}{d\phi} & (23) \\
\mathcal{A}_\phi^{(3)} &= \frac{1}{\sigma}\int_0^{2\pi} d\phi\, \text{sgn}(\cos\phi)\frac{d\sigma}{d\phi} & (24) \\
\mathcal{A}_\phi^{(4)} &= \frac{1}{\sigma}\int_0^{2\pi} d\phi\, \text{sgn}(\cos(2\phi))\frac{d\sigma}{d\phi} & (25)
\end{aligned}
$$

$$\mathcal{A}_{c\theta_1,c\theta_2} = \frac{1}{\sigma} \int_{-1}^{1} d\cos\theta_1 \, \text{sgn}(\cos\theta_1) \int_{-1}^{1} d\cos\theta_2 \, \text{sgn}(\cos\theta_2) \frac{d^2\sigma}{d\cos\theta_1 d\cos\theta_2} \qquad (26)$$

where $\text{sgn}(\pm|x|) = \pm 1$ and the observables are all normalized to the total cross section $\sigma$. These angular observables depend linearly on the dimensionless coefficients defined in Eq. (19), and so may be used to place constraints on SMEFT coefficients complementary to those offered by rate measurements alone.

In order to obtain preliminary projections for CLIC sensitivity to deviations in $e^+e^- \to hZ$ angular observables, we consider the process $e^+e^- \to hZ$ with $Z \to \mu^+\mu^-/e^+e^-$ and $h \to b\bar{b}$ at the three CLIC energy stages (1ab$^{-1}$ at $\sqrt{s} = 380$ GeV, 2.5ab$^{-1}$ at $\sqrt{s} = 1.5$ TeV, and 5ab$^{-1}$ at $\sqrt{s} = 3$ TeV). Beam polarization may significantly improve discrimination of angular observables, and so we assume baseline polarization at the $\pm 80\%$ level for the electron beam, with 50% of the data collected for each of the two polarization configurations at $\sqrt{s} = 380$ GeV. At $\sqrt{s} = 1.5, 3$ TeV we assume 67% data collection with $P(e^-) = -80\%$ and 33% data collection with $P(e^-) = +80\%$, anticipating run conditions optimizing the $HH\nu\nu$ cross section at these energies.

The choice of processes for our study is conservative, in the sense that we only make use of 7% of $Z$ decays to ensure a relatively background-free analysis. A significant increase in statistics is achievable by including hadronic decays of the $Z$ and potentially also additional Higgs decay modes – indeed, the hadronic recoil mass analysis provides the highest statistical precision for a rate measurement of $e^+e^- \to hZ$ – but this requires a detailed understanding of background contributions to angular observables that is beyond the scope of this analysis.

In order to forecast CLIC sensitivity to angular observables, we assume experimental results are SM-like and perform a simple $\chi^2$ analysis along the lines of Ref. [46], illustrating respective sensitivity purely from rate measurements, angular observables, and their combination. In particular, the $\chi^2$ from rate measurements, angular measurements, and the combination thereof are defined as

$$\chi^2_{\text{rate}} = \frac{(X_{\text{NP}} - X_{\text{SM}})^2}{\sigma_X^2}, \qquad (27)$$

$$\chi^2_{\text{angles}} = \sum_i \frac{(\mathcal{A}^i_{\text{NP}} - \mathcal{A}^i_{\text{SM}})^2}{\sigma_{\mathcal{A}^i}^2}, \qquad (28)$$

$$\chi^2_{\text{total}} = \chi^2_{\text{rate}} + \chi^2_{\text{angles}}, \qquad (29)$$

where $X$ denotes the cross section, to disambiguate from the standard deviation $\sigma$. Here $X_{\text{SM}}$ and $\mathcal{A}^i_{\text{SM}}$ are (assumed SM-like) measured values, $X_{\text{NP}}$ and $\mathcal{A}^i_{\text{NP}}$ are the predictions in the presence of dimension-6 operators, which we approximate at linear order in the EFT coefficients. In both cases the angular variables $\mathcal{A}^i$ are summed over $\mathcal{A}_{\theta_1}, \mathcal{A}^{(1)}_\phi, \mathcal{A}^{(2)}_\phi, \mathcal{A}^{(3)}_\phi, \mathcal{A}^{(4)}_\phi$ and $\mathcal{A}_{c\theta_1,c\theta_2}$. The errors $\sigma_X$ and $\sigma_{\mathcal{A}^i}$ denote the one-sigma statistical uncertainties for the rate and angular observables, respectively. The absolute statistical uncertainty of an angular observable $A$ is given by [46]

$$\sigma_A = \sqrt{\frac{1 - \bar{A}^2}{N}}, \qquad (30)$$

where $\bar{A}$ is the expectation value of $A$, and $N$ is the number of events. For the CLIC runs under consideration, the SM expectation of $\mathcal{A}_{\theta_1}$ is within the range $[-0.6, -0.8]$. The SM expectations of the other angular observables are either equal or very close to zero, and $\sigma_A \approx 1/\sqrt{N}$ provides a very good approximation. Our analysis neglects possible correlations among experimental measurements of the observables, which are expected to be small.

In what follows, we assume 50% signal efficiency and negligible background at each of the three CLIC centre-of-mass energies for a recoil mass selection of $e^+e^- \to hZ$ with $Z \to \ell^+\ell^-$ and $h \to b\bar{b}$. Following Ref. [46], realistic signal cuts are not expected to alter the response of our asymmetry observables to dimension-6 operators in the regime of interest. Sensitivity to angular observables may be further impacted by instrumental uncertainties such as beam energy resolution, initial-state radiation, and particle reconstruction energy resolution, but these are all expected to be small corrections. Theoretical uncertainties will also play a significant role; here we anticipate precision at the sub-percent level and neglect these uncertainties in our forecasting.

Table 13: One-sigma uncertainties for individual coefficients, with the assumption that all other coefficients are zero. The results are shown for each of the three runs at CLIC as well as the combination of them. For each run, the first row shows the constraints from the rate measurements only, the second row shows the constraints from measurements of angular observables only, and the third row shows the combined constraints from both rate and angular measurements. A $\infty$ is shown if no constraint could be derived within our procedure.

| | $\widehat{\alpha}_{ZZ}$ | $\widehat{\alpha}_{ZZ}^{(1)}$ | $\widehat{\alpha}_{H\ell}^V$ | $\widehat{\alpha}_{H\ell}^A$ | $\widehat{\alpha}_{AZ}$ | $\delta g_V$ | $\delta g_A$ | $\widehat{\alpha}_{Z\widetilde{Z}}$ | $\widehat{\alpha}_{A\widetilde{Z}}$ |
|---|---|---|---|---|---|---|---|---|---|
| | \multicolumn{9}{c}{380 GeV, 1 ab$^{-1}$, $P(e^-, e^+) = (\mp 0.8, 0)$ [50%/50%]} |
| rate | 0.0013 | 0.011 | 0.00085 | 0.00068 | 0.0021 | 0.014 | 0.0056 | $\infty$ | $\infty$ |
| angles | 0.0059 | $\infty$ | 0.024 | 0.39 | 0.0090 | 0.023 | 0.36 | 0.013 | 0.017 |
| total | 0.0013 | 0.011 | 0.00085 | 0.00068 | 0.0020 | 0.012 | 0.0056 | 0.013 | 0.017 |
| | \multicolumn{9}{c}{1.5 TeV, 2.5 ab$^{-1}$, $P(e^-, e^+) = (\mp 0.8, 0)$ [67%/33%]} |
| rate | 0.0026 | 0.030 | 0.00014 | 0.00011 | 0.0039 | 0.037 | 0.015 | $\infty$ | $\infty$ |
| angles | 0.0045 | $\infty$ | 0.013 | 0.20 | 0.0061 | 0.18 | 2.8 | 0.0098 | 0.010 |
| total | 0.0022 | 0.030 | 0.00014 | 0.00011 | 0.0033 | 0.036 | 0.015 | 0.0098 | 0.010 |
| | \multicolumn{9}{c}{3 TeV, 5 ab$^{-1}$, $P(e^-, e^+) = (\mp 0.8, 0)$ [67%/33%]} |
| rate | 0.0036 | 0.043 | 0.000050 | 0.000041 | 0.0054 | 0.053 | 0.022 | $\infty$ | $\infty$ |
| angles | 0.0049 | $\infty$ | 0.0088 | 0.14 | 0.0059 | 0.49 | 7.9 | 0.0072 | 0.0072 |
| total | 0.0029 | 0.043 | 0.000050 | 0.000041 | 0.0040 | 0.052 | 0.022 | 0.0072 | 0.0072 |
| | \multicolumn{9}{c}{Combined} |
| rate | 0.0011 | 0.010 | 0.000047 | 0.000038 | 0.0017 | 0.013 | 0.0051 | $\infty$ | $\infty$ |
| angles | 0.0029 | $\infty$ | 0.0069 | 0.11 | 0.0038 | 0.022 | 0.36 | 0.0053 | 0.0056 |
| total | 0.0010 | 0.010 | 0.000047 | 0.000038 | 0.0016 | 0.011 | 0.0051 | 0.0053 | 0.0056 |

From this analysis, we can project the sensitivity of CLIC measurements of angular observables to new physics scenarios. Here we focus on model-independent constraints on terms in the dimension-6 SMEFT, parameterized by the coefficients defined in Eq. (19), by computing the dependence of the $hZ$ production rate and angular observables on these coefficients at each polarization and centre-of-mass energy. As there are 9 such coefficients but only 7 constraints coming from the rate and angular measurements at fixed energy and polarization, it is not possible to independently constrain all coefficients without further assumptions. To this end, we first consider the constraints placed by rate and angular measurements on individual coefficients considered in isolation, the results for which are shown at each energy stage (and in combination) in Table 13, and will be discussed in what follows.

**Rate measurements**

At the level of constraining individual coefficients, rate measurements typically provide the strongest constraints, particularly at $\sqrt{s} = 380$ GeV where CLIC parameters are optimized for the $e^+e^- \to hZ$ rate measurement. The exception being contributions from the CP-violating coefficients $\alpha_{Z\widetilde{Z}}$ and $\alpha_{A\widetilde{Z}}$, which are not accessible to the rate measurement at leading order.[11] Some operators have particularly strong energy dependence (most notably the $hV\ell\ell$ contact interactions), leading to exquisite sensitivity by $\sqrt{s} = 3$ TeV despite the small cross section. However, rate measurement alone only constrains one linear combination of coefficients at a given centre-of-mass energy, though some polarization-sensitive coefficients may be further constrained by rate measurements at opposite polarizations, as we will discuss further. For this reason the primary strength of the angular observables lies in breaking these degeneracies.

Rate measurements with polarized beams are much more sensitive to $\widehat{\alpha}_{AZ}$ than those with unpolarized beams, due to an accidental cancellation in the unpolarized case that reduces the sensitivity of $e^+e^- \to hZ$ to the $hZ\gamma$ vertex. This makes angular measurements of $\widehat{\alpha}_{AZ}$ at CLIC complementary to proposed future circular $e^+e^-$ colliders. See Ref. [37] for further discussion.

Rate measurements at a single centre-of-mass energy can resolve the degeneracies of some polarization-sensitive operator coefficients with $\widehat{\alpha}_{ZZ}^{(1)}$ since the rate measurement now entails two constraints from the two polarizations. In particular, the dependence of $\sigma(hZ)$ on $\widehat{\alpha}_{H\ell}^V$, $\widehat{\alpha}_{AZ}$, and $\delta g_V$ is sensitive to beam polarizations because the SM $e^+e^-Z$ vertex is dominated by the axial coupling ($|g_A| \gg |g_V|$ in the term $Z_\mu \bar{l}\gamma^\mu(g_V - g_A\gamma_5)l$). Both $\widehat{\alpha}_{H\ell}^V$ and $\delta g_V$ generate vector-like couplings, while $\widehat{\alpha}_{AZ}$ contributes to the diagram with an $s$-channel photon for which the electron-photon coupling is also vector-like. The interference terms between the SM $e^+e^- \to hZ$ amplitude and the ones generated by these coefficients are thus dominantly axial-like and change signs under a flip of the beam polarization. For these operator coefficients, even rate measurements at a single centre-of-mass energy can partially resolve the degeneracy with $\widehat{\alpha}_{ZZ}^{(1)}$.

**Angular measurements**

To illustrate the potential of angular observables to break degeneracies among different coefficients, we next consider constraints on pairs of coefficients. To the extent that $e^+e^- \to hZ$ rate measurements provide the strongest probe of $\widehat{\alpha}_{ZZ}^{(1)}$ (which parameterizes additional contributions to the SM-like coupling $hZ_\mu Z^\mu$), we focus on pairing $\widehat{\alpha}_{ZZ}^{(1)}$ with one of the other Wilson coefficients in Eq. (19), setting the rest to zero. These constraints are illustrated in Figure 16.

To a certain extent, rate measurements at different centre-of-mass energies already begin to gently disambiguate different coefficients, as the rate measurements depend on slightly different linear combinations of coefficients at each energy. However, angular measurements still provide significant discriminating power.

While the improvement in discrimination provided by the angular observables is marginal in many cases (particularly when polarization effects are taken into account), there are notable exceptions, namely $\widehat{\alpha}_{ZZ}, \widehat{\alpha}_{Z\widetilde{Z}}$, and $\widehat{\alpha}_{A\widetilde{Z}}$. While the latter two are likely better constrained by direct probes of CP violation, discrimination between $\widehat{\alpha}_{ZZ}$ and $\widehat{\alpha}_{ZZ}^{(1)}$ is particularly promising. In this case, the sensitivity of the rate measurement to both $\widehat{\alpha}_{ZZ}$ and $\widehat{\alpha}_{ZZ}^{(1)}$ does not depend on the polarization, so the rate measurement at one energy only constrains a fixed combination of them, and changes of centre-of-mass energy do not provide substantial improvement.

CLIC provides an unprecedented potential to explore the structure of the Higgs sector and the physics of electroweak symmetry breaking. Here we have characterized the sensitivity of CLIC to rate and angular observables in $e^+e^- \to hZ$ measurements at a variety of centre-of-mass energies, including polarization effects, and translated this sensitivity into constraints on linear combinations of operator coefficients in the dimension-6 Standard Model EFT. Our projections for bounds on the $\widehat{\alpha}$ and $\delta g$ dimensionless couplings may be readily related to coefficients in different SMEFT operator bases.

---

[11]Notice that these might be more strongly constrained by direct tests of CP violation.

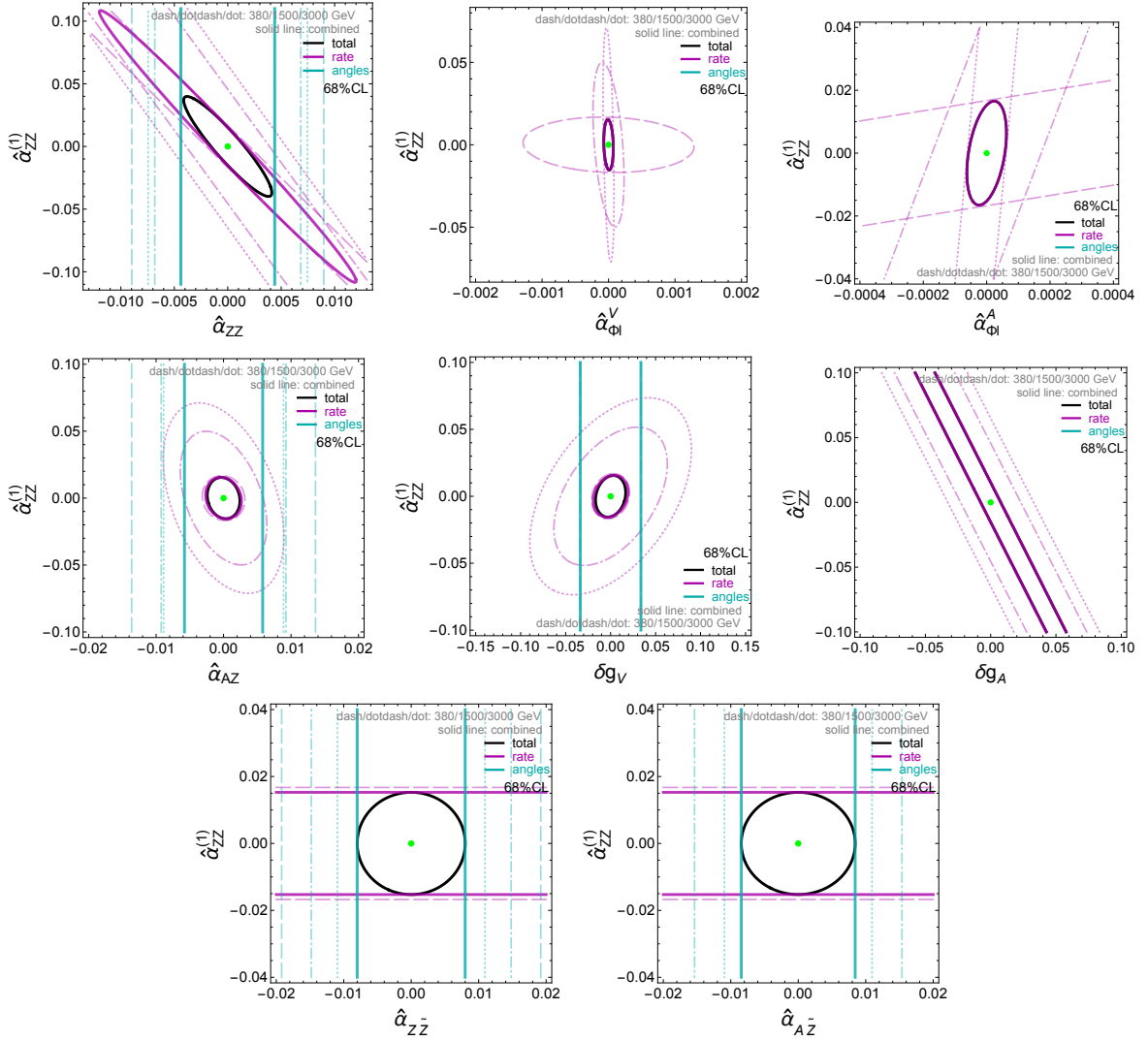

Figure 16: Expected constraints from the $e^+e^- \to hZ \to \ell^+\ell^- b\bar{b}$ process at CLIC, assuming SM-like results. Each plot shows the 68% confidence level (C.L.) contour for $\widehat{\alpha}_{ZZ}^{(1)}$ and one of the other Wilson coefficients in Eq. (19), with the rest set to zero. The purple (cyan) contours show the constraints from the rate (angular) measurements only, and the black contours show the combined constraints from both rate and angular measurements. The dashed, dot-dashed and dotted contours correspond to the measurements from the 380 GeV, 1.5 TeV and 3 TeV runs, respectively. The solid contours correspond to the combination of the three runs. The green dot at $(0,0)$ indicates the SM prediction.

Broadly speaking, the use of angular observables in $e^+e^- \to hZ$ measurements at CLIC both improves statistics in combination with rate measurements and contributes valuable discriminating power between different operator coefficients that are otherwise degenerate in rate measurements (though the availability of different beam polarizations and centre-of-mass energies at CLIC improves the discrimination of rate measurements compared to proposed circular $e^+e^-$ colliders). In terms of discriminating power, angular measurements at CLIC are particularly effective in differentiating between various tensor structures in the coupling of the Higgs to $Z$ bosons, namely the SM-like coupling $hZ_\mu Z^\mu$ and $hZ_{\mu\nu}Z^{\mu\nu}$.

## 2.4 Diboson $W^+W^-$ processes as new physics probes[12]

Diboson processes are rich in information about the SM structure: the physics of the longitudinal polarizations is related, in the high energy limit, to that of the Higgs boson, while that of the transverse polarizations is unique and controlled by the SM electroweak symmetry. From an experiment perspective, the challenge is to isolate these qualitatively different physics, and transform them into precise BSM probes. This is the goal of this section.

At high-energy, the amplitudes become simple objects, almost entirely determined by the helicity of the initial and final states, and dimensional analysis [47]. In particular, if we are interested in effects that grow quadratically with the energy (as relevant for dimension-6 operators) there is only one amplitude for a given external state configuration (see however Ref. [50]). For this reason, the analysis of high-energy effects can be focussed on individual operators; the results can be easily transferred into the context of a global fit.

**BSM in the transverse polarizations**

This part focusses on searching for New Physics in the transverse polarizations of vectors, reflected in the EFT via the operator $\mathcal{O}_{3W}$ and its CP-odd counterpart $\mathcal{O}_{3\tilde{W}}$ [51] (see Table 2 for the definition of operators). This is related to anomalous trilinear gauge couplings (TGCs) parameters $\lambda_\gamma = \lambda_Z$ of Ref. [52] via $c_{3W}/\Lambda^2 = -\lambda_\gamma/m_W^2$. These effects can be tested in high-energy $W^+W^-$ processes. Unfortunately SM and BSM exhibit different $W$ helicity structures (see Ref. [53] for a recent discussion), so that the two amplitudes do not interfere in inclusive measurements: an important drawback of traditional analyses in the context of a precision program. We will study here how differential distribution measurements of the azimuthal angles of the $W$ boson decay planes do bear the interference information (see Ref. [54]).

Analyses of diboson processes are often presented as measurements of anomalous trilinear gauge couplings (TGCs), associated with the parameters $\lambda_\gamma$, $g_1^Z$ and $\kappa_\gamma$ of Ref. [52]. These are in correspondence with dimension-6 operators through the Wilson coefficients. In this note we will focus on the first one $\lambda_\gamma$. The relationships between this parameter and the Wilson coefficient of the dimension six Lagrangian read

$$c_{3W} = -\frac{\Lambda^2}{m_W^2}\lambda_\gamma. \tag{31}$$

Where the BSM scale is assumed to be $\Lambda \sim 1\text{TeV}$. Our results concerns the reach on the parameter $\lambda_\gamma$ achieved by studying the transverse part of diboson production amplitudes.

$\mathcal{O}_{3W}$ produces, at tree-level and at high energy, dominantly $++$ or $--$ helicities in the $WW$ final states, with amplitudes $\mathcal{A}_{BSM}^{++} = \mathcal{A}_{BSM}^{--}$. These amplitudes do not interfere, in inclusive $2 \to 2$ scattering, with the SM amplitude $\mathcal{A}_{SM}$ [53]. SM processes have, in the high-energy and classical limits, dominantly $+-$, $-+$ or $00$ helicity. The latter is however smaller and has little impact on this part of the analysis. Nevertheless, the amplitude for $e^+e^- \to W^+W^- \to 4f$ decays into fermions can in principle interfere. This interference is proportional to a function of the azimuthal angles of the decay planes of the fermion/anti-fermion originating from the $W^+$ and $W^-$ respectively. In this note we focus on the distribution with respect to the azimuthal angle $\phi$ of the plane defined by the decay products of one of the two $W$ bosons, relative to the scattering plane, as illustrated in Figure 17. We remain inclusive about the other $W$, which can then be thought as a state of well defined helicity. The interference term, between the transverse-transverse amplitudes, reads

$$I^{WW} \propto \mathcal{A}_{++}^{\text{BSM}}\left[\mathcal{A}_{-+}^{\text{SM}}+\mathcal{A}_{+-}^{\text{SM}}\right]\cos 2\varphi\,, \tag{32}$$

see also Ref. [54] for more details. Interference vanishes when integrated over $\varphi$, reproducing the non-interference result.

---
[12]Based on a contribution by D. Lombardo, F. Riva, P. Roloff.

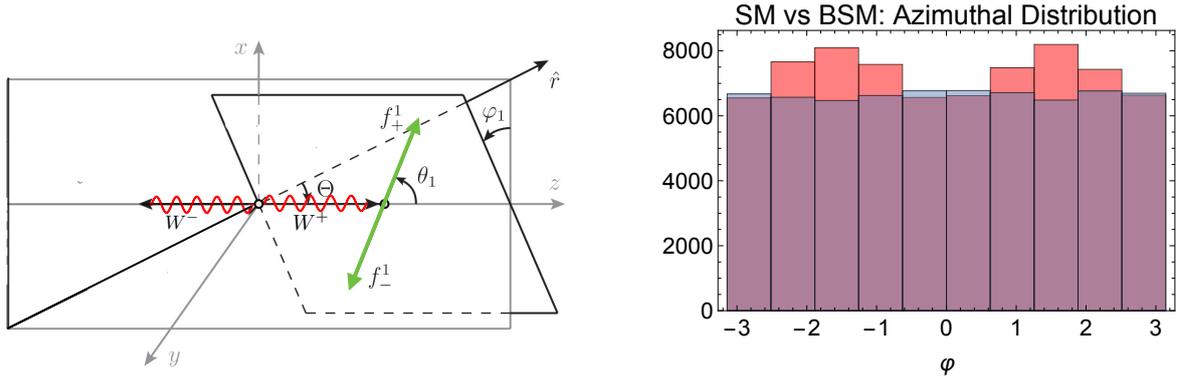

Figure 17: (Left) Definitions of the polar angle $\Theta$ and azimuthal angle $\varphi$. (Right) Azimuthal differential distribution, in arbitrary units, for the semileptonic case with $0 < \cos\Theta < 0.5$ (in blue the SM, in red the BSM with coupling value $c_{3W} = 0.294$); other cuts as in the text.

The azimuthal angle $\varphi$ is measured choosing one of the two fermions produced from the decay. For instance, in Figure 17 it is defined making reference to the outgoing fermion of positive helicity ($f_+$ in the figure). When the $W$ is decaying *hadronically*, selecting one of the two fermions is not possible,[13] implying an ambiguity

$$\varphi_h \leftrightarrow \varphi_h + \pi. \tag{33}$$

Distributions of the form Eq. (32) are insensitive to this ambiguity. This is not true for distributions of the form $\sin 2\varphi$, such as those originating from CP odd $\mathcal{O}_{3\widetilde{W}}$ effects, which vanish because of the ambiguity and become almost impossible to see.

For *leptonically* decaying $W$ bosons the situation is different: here, one could measure the $\varphi$ angle always making reference to the charged lepton, but the decaying plane is defined only if the neutrino momentum is also known. This can be reconstructed from the kinematics if only one neutrino is present in the event; however the resulting kinematic equation is quadratic, and has two solutions, which differ in their longitudinal momentum component. At hadron machines, where the initial longitudinal momentum is unknown, it is impossible to single out which of these two solutions corresponds to the real one. At lepton colliders this could in principle be possible, because the total initial centre-of-mass energy is known. However, in practice, initial state radiation (ISR) and beam-strahlung largely spoil the exact knowledge of the initial energy collision, making the leptonic-collider case similar to the one of LHC: the initial momentum on the beam-pipe direction is not known with sufficient precision. This leads eventually to an ambiguity similar to Eq. (33)

$$\varphi_l \leftrightarrow \pi - \varphi_l. \tag{34}$$

Moreover, in the *fully hadronic* channel there is an additional ambiguity due to the impossibility of distinguishing $W^+$ and the $W^-$, resulting in

$$\Theta_h \to \Theta_h + \pi. \tag{35}$$

In our analysis we randomly select one of the fermions and one of the $W$ bosons to define the $\varphi$ and $\Theta$ angles (the latter is analogous to a charge-inclusive analysis).

The *semileptonic* channel $\nu l \bar{q} q$ does not suffer from a $\Theta$ ambiguity. For simplicity we work with $l = \mu^-$ and multiply by 4 the luminosity to account for $l = \mu^+$ and $l = e^\pm$ (in principle the analysis with $l = e^\pm$ is slightly more complicated because of a $t$-channel diagram not present in the muon channel; due to its different kinematics this effects can be however efficiently singled out).

---

[13] In this context it would be interesting to study decays including charm quarks; we leave this for the future.

At CLIC, ISR and Brehmsstrahlung broaden the beam energy-spectrum, implying that a non-negligible portion of events has energy much smaller than the nominal energy $E^{nom} = 380, 1500, 3000$ GeV. At smaller energy, non-trivial azimuthal distributions can be observed also in the SM alone due to interference, e.g. of the $+-$ helicity with the $+0$ helicity (the latter suppressed by one power of $m_W/E$). Given that at smaller energy the cross section is in fact larger, this effect is amplified, and it becomes difficult to recognise a distribution of the form Eq. (32). For this reason, in our analysis, we include a selection cut on the energy of the events, namely

$$\sqrt{s} > 2600, 1300, 330 \,\text{GeV} \qquad (36)$$

for the 3 TeV, 1.5 TeV and 380 GeV runs respectively.

Knowledge of the SM amplitude can guide us through the most appropriate choice of cuts and binning in the kinematic variables.
In the high-energy limit, the tree-level SM amplitudes for the parton-level $2 \to 2$ process read

$$\mathcal{A}_{\text{SM}}^{-+} = g^2 \sin \Theta \qquad \mathcal{A}_{\text{SM}}^{+-} = -2g^2 \cos^4 \frac{\Theta}{2} \csc \Theta \qquad \mathcal{A}_{\text{SM}}^{00} = -\frac{1}{2}(g^2 + g'^2) \sin \Theta, \qquad (37)$$

where $\Theta$ is the polar angle, corresponding to the angle between the incoming electron and the outgoing $W^-$, see Figure 17. The BSM transverse amplitude is instead

$$\mathcal{A}_{++}^{\text{BSM}} = \mathcal{A}_{--}^{\text{BSM}} \approx c_{3W} 6e\sqrt{2} \frac{M_{W\gamma}^2}{\Lambda^2} \sin \Theta \,. \qquad (38)$$

The important lessons here are:

– In the backward region $\cos \Theta \approx -1$ both SM and BSM vanish, so that this region is not favorable
– In the forward region $\cos \Theta \approx +1$ the BSM vanishes and the SM explodes because of the $t$-channel neutrino pole; the interference term is in fact finite. The signal over sqrt-background vanishes in the backward point, but increases rapidly ($\sim \Theta^{3/2}$) as we approach the central region, so that even this backward region can have interesting information.
– In the central region $\cos \Theta \approx 0$ the BSM amplitude has its maximum, and the SM switches from being dominated by the $+-$ to being dominated by $-+$. Most importantly, since the latter SM amplitudes have opposite sign (see Eq. (37)): the overall SM amplitude changes sign!

In light of these remarks, we understand that the most important region for our analysis will be $\cos \Theta \sim 0$; however, it is important to separate the analysis (or implement an asymmetry) for

$$\cos \Theta < 0 \quad \text{and} \quad \cos \Theta > 0; \qquad (39)$$

because of the opposite SM amplitude sign, the sum of the interference terms from these distinct regions tends to cancel, and this is the reason why being inclusive in the $W$ bosons charge (as for the fully hadronic case where we find the ambiguity Eq. (35)), generates a partial cancellation which weakens the signal in the central region. From this discussion, our choice of binning, is

- *Fully hadronic* channel: we take 6 bins to separate the peripheral region from the central one, and, in the latter one, to highlight the zone where the cancellation in the SM amplitudes is milder,

$$\cos \Theta \in [-1, -0.5, -0.2, 0, 0.2, 0.5, 1]. \qquad (40)$$

- *Semileptonic* channel: we consider 4 bins in polar angle

$$\cos \Theta \in [-1, -0.5, 0, 0.5, 1]. \qquad (41)$$

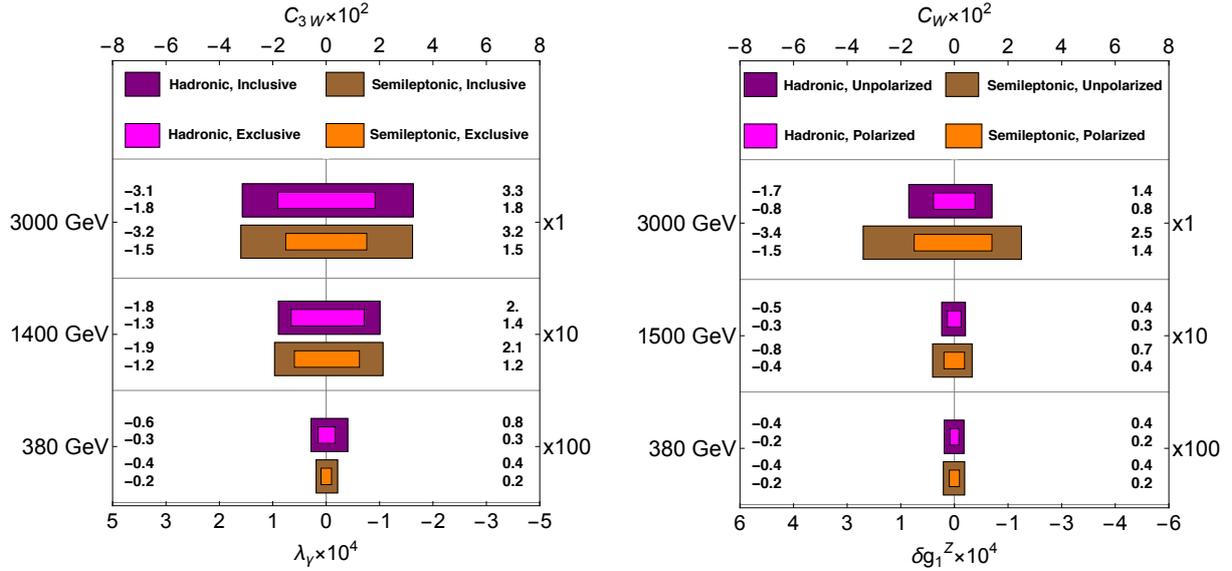

Figure 18: (Left) $1\sigma$ sensitivity on the Wilson coefficient $c_{3W}$ (or equivalently $\lambda_\gamma$), for $\Lambda = 1$ TeV from different channels. (Right) same for $c_W$ (or equivalently $\delta g_1 Z$). Numbers refer to the size of bars, for 1% systematic errors.

In addition we bin $\varphi \in [-\pi, \pi]$ in 10 bins and compare the performance of a binned and unbinned analysis (i.e. exclusive/inclusive in the azimuthal angle). There are in principle two angles we can choose from: the one defined by the hadronically decaying $W$ and the one defined by the leptonically decaying one; the results are practically identical and we present the analysis for the latter. Our study is based on simulations generated with WHIZARD [55], where detector effects are approximated by smearing the energy of jets on a normal distribution with a 4% resolution. The right panel of Figure 17 shows an example of the azimuthal distribution of interest[14]. The analysis includes a series of other small acceptance cuts: a polar angle cut $10 < \Theta_{l,j} < 170$ for jets and leptons, a lower cut in energy for all particles $E > 10$ GeV, a cut in the polar angle of $W$ bosons to avoid the very forward region, $-0.95 < \cos(\Theta) < 0.95$, a cut in the invariant mass of jets $M_{jj} > 10$ GeV, to avoid events with jets from virtual photons in the fully hadronic case.

Table 14: **One sigma sensitivity on $c_{3W} \times 10^2$** for both the hadronic and semileptonic channels and for different systematic uncertainties, run conditions and analysis strategy. The first columns denotes inclusive analysis in the hadronic angle $\varphi_h$, the second exclusive.

| **Semileptonic Channel** | | Inclusive $\varphi$ | Exclusive $\varphi$ | **Fully Hadronic Channel** | | Inclusive $\varphi$ | Exclusive $\varphi$ |
|---|---|---|---|---|---|---|---|
| $\delta_{syst} = 3\%$ | Stage 1+2+3 | [-2.79, 3.22] | [-1.13, 1.16] | $\delta_{syst} = 3\%$ | Stage 1+2+3 | [-3.40, 3.54] | [-1.99, 2.01] |
| | Stage 1+2 | [-10.1, 11.0] | [-3.38, 3.42] | | Stage 1+2 | [-11.8, 13.0] | [-6.52, 6.90] |
| | Stage 1 | [-61.5, 89.1] | [-23.8, 25.2] | | Stage 1 | [-82.0, 173] | [-37.4, 43.9] |
| $\delta_{syst} = 1\%$ | Stage 1+2+3 | [-2.64, 3.07] | [-1.07, 1.09] | $\delta_{syst} = 1\%$ | Stage 1+2+3 | [-3.14, 3.26] | [-1.81, 1.83] |
| | Stage 1+2 | [-9.06, 10.1] | [-3.12, 3.16] | | Stage 1+2 | [-10.3, 11.5] | [-5.67, 5.92] |
| | Stage 1 | [-42.7, 53.6] | [-20.01, 21.0] | | Stage 1 | [-56.8, 80.8] | [-30.4, 34.4] |

We compare two scenarios with optimistic 1% and pessimistic 3% systematic uncertainty $\delta_{syst}$ in all bins, in addition to statistical uncertainty. We also assume a 50% signal acceptance. The results are summarized in Table 14 and Figure 18. As expected, in both the channels, the analysis binned in the azimuthal angle, which takes into account the interference effects, gives additional sensitivity, and the reach is dominated by the central bins.

---

[14]The SM is almost flat, and exhibits a tiny $\cos(\varphi)$ behaviour, due to the above-mentioned SM-SM interference between different helicity amplitudes.

**BSM in the longitudinal polarizations**

According to the equivalence theorem, BSM modifications of the Higgs sector imply modifications of the amplitude $e^+e^- \to W_L^+ W_L^-$, involving the longitudinal polarisations of vectors for a thorough discussion. As a matter of fact, the effects that enter in this amplitude, are the same that enter $e^+e^- \to Zh$, discussed in the previous section, see Ref. [47].

Here the obstacle to a precise measurement is the fact that, in the SM, the cross-section is dominated by the transverse components. This is true both for the total cross section, which is dominated by forward scattering of transverse bosons, but also in the central region, where the longitudinal component is maximal. For right-handed polarized electrons, however, the transverse contribution vanishes in the high-energy limit.

Table 15: **One sigma sensitivity on $c_W \times 10^2$** for both the hadronic and semileptonic channels and for different systematic uncertainties, and different polarization setups as defined in Table 3 (polarized refers to the standard baseline scenarios).

| **Semileptonic Channel** | | Unpolarized | Polarized | **Fully Hadronic Channel** | | Unpolarized | Polarized |
|---|---|---|---|---|---|---|---|
| | Stage 1+2+3 | [-3.43, 2.50] | [-1.52, 1.45] | | Stage 1+2+3 | [-1.98, 1.62] | [-0.96, 0.94] |
| $\delta_{syst} = 3\%$ | Stage 1+2 | [-8.80, 7.04] | [-4.3, 4.11] | $\delta_{syst} = 3\%$ | Stage 1+2 | [-5.95, 5.08] | [-3.17, 3.15] |
| | Stage 1 | [-56.8, 51.1] | [-22.8, 23.2] | | Stage 1 | [-62.6, 56.4] | [-23.4, 24.1] |
| | Stage 1+2+3 | [-3.64, 2.60] | [-1.61, 1.53] | | Stage 1+2+3 | [-1.71, 1.44] | [-0.78, 0.77] |
| $\delta_{syst} = 1\%$ | Stage 1+2 | [-8.09, 6.60] | [-3.91, 3.80] | $\delta_{syst} = 1\%$ | Stage 1+2 | [-4.69, 4.14] | [-2.57, 2.53] |
| | Stage 1 | [-41.2, 38.1] | [-19.0, 19.10] | | Stage 1 | [-38.0, 35.6] | [-16.7, 17] |

In this ending paragraph we compare the reach in a simple analysis with and without polarization. Since CLIC polarization is not 100%, we keep separated analyses in 4 bins in the polar angle, in order to be able to single out the forward bin, where remnant transversely polarized vectors still provide an important background. The analysis of this section therefore parallels that of the previous section, with only the exception of the azimuthal angle, which plays no particular role here, and we treat inclusively. The results are summarized in Table 15 and in the right panel of Figure 18.

## 2.5 Multiboson processes[15]

The CLIC collider with a staged energy range between 380 GeV and 3 TeV allows for simultaneous on-shell production of more than two massive electroweak bosons $V = W, Z, H$:

$$e^-e^+ \to VV, VVV, VVVV, \dots \tag{42}$$

For instance, the kinematical production thresholds for triboson production range from 252 GeV ($WWZ$) to 331 GeV ($ZHH$), and four-boson final states become kinematically accessible beyond 322 GeV ($4W$). Given the Feynman diagrams for the amplitudes, we get direct access to a set of elementary bosonic interactions of the type $V^* \to nV$. Likewise, multiple massive electroweak bosons are produced in $W^+W^-$ fusion processes (VBF),

$$e^-e^+ \to \nu_e \bar{\nu}_e + VV, VVV, VVVV, \dots \tag{43}$$

Two-boson final states $VV$ have been studied in great detail. The prospects for isolating beyond the SM (BSM) contributions from future CLIC data in vector-boson pair and associated production $W^+W^-$, $ZZ$, and $ZH$, are considered in various chapters elsewhere in this document.

---
[15]*Based on a contribution by S. Brass, W. Kilian, S. Y. Shim and J. Reuter.*

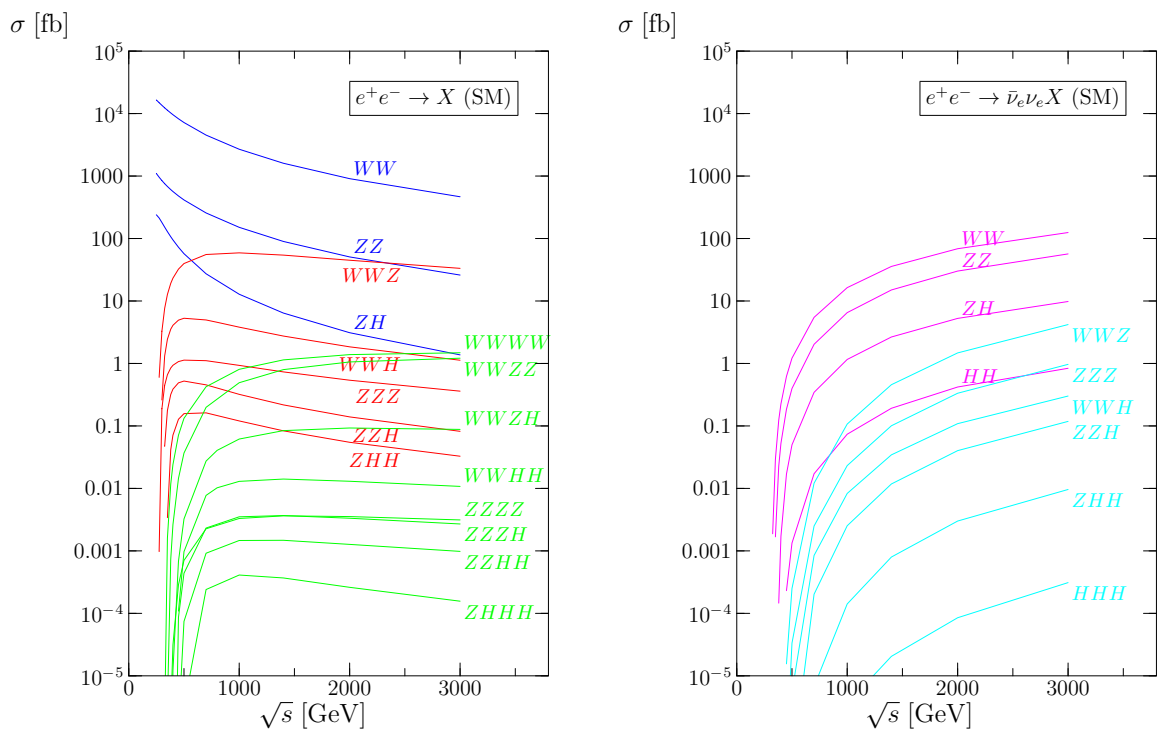

Figure 19: Total cross section for multi-boson production in $e^+e^-$ annihilation in the SM, standard (left panel) and vector boson fusion (right panel). Leading order, no beam polarization, no ISR or beamstrahlung corrections included (for VBF we require $M_{\bar{\nu}_e \nu_e} > 150$ GeV).

In this section, we give an overview of the phenomenology of processes that involve multiple vector bosons, both in annihilation and in VBF. Multi-boson and VBF final states, in the SM and at sub-TeV energy, naturally have lower event rates than di-boson production. Nevertheless, they may become an additional and important probe of new physics in the electroweak sector.

**Multi-boson rates in the SM**

We have computed the SM total cross sections for all processes of type $e^-e^+ \to nV$ ($n = 2, 3, 4$) and $e^-e^+ \to \nu_e \bar{\nu}_e + nV$ ($n = 2, 3$).[16] Figure 19 displays the result. We observe that while the di-boson final states are clearly dominant both in annihilation and in VBF, various higher-order final states are also accessible. At the highest energy ($\sqrt{s} = 3$ TeV), the rates for corresponding final states in annihilation and in VBF become comparable in magnitude.

The cross section for $e^-e^+ \to W^+W^-Z$ as the most important tri-boson process reaches about 50 fb. Beyond the maximum at about 800 GeV, the rate decreases rather slowly with energy and overtakes the $ZZ$ di-boson process. Other processes with cross sections above 1 fb include the tri-boson final states $W^+W^-H$ and, marginally, $ZZZ$. The quadruple-boson final states $W^+W^-W^+W^-$ and $W^+W^-ZZ$ also enter this range. Regarding VBF, the $W^+W^-$, $ZZ$, $ZH$, and $W^+W^-Z$ final states have cross sections that rise above 1 fb. Given integrated luminosity rates of a few ab$^{-1}$ and the full coverage of the detector and analysis, distributions should become accessible to detailed studies.

In the range between 1 ab and 1 fb, we find a set of additional final states that include more $Z$ and $H$ bosons in place of $W^\pm$. We expect observable event rates above $\sim 10$ ab (see below), although efficiencies and background severely limit the capacities for precision studies. These processes compete with other multi-boson final states that involve even more $W^\pm$ emission, not shown in the plot, as well as with QCD jet radiation in the continuum. The set includes processes that depend on the triple Higgs coupling: $ZHH$, and $W^+W^-HH$ in annihilation, and likewise $HH$ and $ZHH$ in VBF.

---

[16]The numerical computations have been performed at leading order using WHIZARD [55], with cross checks using MadGraph5 [31] with the model [56].

Table 16: Estimates for the number of events that can be analyzed in the hadronic decay channels of various multi-boson final states. We quote SM results for $e^+e^-$ annihilation ($N_a$) and vector-boson fusion ($N_{\text{VBF}}$) processes. We assume an integrated luminosity of Table 3, and a detection efficiency of 50 %.

|  | Stage 1 | | Stage 2 | | Stage 3 | |
| --- | --- | --- | --- | --- | --- | --- |
|  | $N_a$ | $N_{\text{VBF}}$ | $N_a$ | $N_{\text{VBF}}$ | $N_a$ | $N_{\text{VBF}}$ |
| $WWZ$ | 3500 | 0 | 25000 | 210 | 30000 | 3800 |
| $WWH$ | 470 | 0 | 920 | 11 | 750 | 200 |
| $ZZZ$ | 210 | 0 | 460 | 62 | 450 | 1200 |
| $ZZH$ | 57 | 0 | 98 | 5 | 74 | 110 |
| $ZHH$ | 5 | 0 | 29 | 0 | 23 | 6 |
| $HHH$ | – | 0 | – | 0 | – | 0 |
| $WWWW$ | | 1 | | 290 | | 740 |
| $WWZZ$ | | 0 | | 280 | | 870 |
| $WWZH$ | | 0 | | 23 | | 48 |
| $WWHH$ | | 0 | | 2 | | 4 |
| $ZZZZ$ | | 0 | | 1 | | 3 |
| $ZZZH$ | | 0 | | 1 | | 2 |
| $ZZHH$ | | 0 | | 0 | | 0 |
| $ZHHH$ | | 0 | | 0 | | 0 |
| $HHHH$ | | – | | – | | – |

Finally, the plots show rates for triple Higgs production, $ZHHH$ in annihilation and $HHH$ in VBF, always below 1 ab and thus unlikely to be detected if the SM is correct. As stated before, in this range there are further processes such as quadruple production in VBF that we do not include here.

The plots indicate that for annihilation processes, the decrease with energy is less pronounced if more bosons are produced. Likewise, VBF processes with higher multiplicity rise faster with energy than the quasi-elastic $2 \to 2$ processes. This is easily explained since in the SM, extra bosons can be regarded as real radiative corrections. Radiated particles, in the total cross section, come with additional logarithms of the energy. We expect that the angular and momentum distributions of the extra radiated particles exhibit the singular behavior of splitting processes, cut off by the finite mass values.

To relate these bare cross sections to the expected sensitivity at the CLIC collider, for an order-of-magnitude estimate, we may adopt the following assumptions: (1) $W$ bosons are measured in the hadronic decay channel (BR = 0.67); $Z$ bosons are measured in all visible decay channels (BR = 0.8); Higgs bosons are measured in the $b\bar{b}$ decay channel (BR = 0.6). (2) All final states are detected with uniform efficiency $\epsilon = 0.5$. With these assumptions we may evaluate the expected number of events in the SM for all final states, as displayed in Table 16. These can be used to estimate the BSM reach of these precision probes.

This estimate does not take into account any details of the detector, the dependency of the acceptance on the final state, or the probability of misidentification of $W$ vs. $Z$ in the hadronic decay channel. On the other hand, an actual experimental analysis would make use of kinematical information, it is not restricted to the total cross section as the only observable. Within the SM, we expect extra radiated bosons to be collimated towards the forward region. By contrast, contributions beyond the SM are gen-erically expected to prefer the central region. In what follows we compute rates for multiboson processes in the SM EFT, using the operators defined in Table 2. We study the impact of anomalous contributions to bosonic interaction vertices as indicated in Table 17.

Table 17: Direct anomalous contributions of dimension-six operators to 3- and 4-boson vertices.

|  | $\mathcal{O}_6$ | $\mathcal{O}_H$ | $\mathcal{O}_T$ | $\mathcal{O}_W$ | $\mathcal{O}_B$ | $\mathcal{O}_{HW}$ | $\mathcal{O}_{HB}$ | $\mathcal{O}_{BB}$ | $\mathcal{O}_{3W}$ |
|---|---|---|---|---|---|---|---|---|---|
| $W^+W^-Z$ |  |  |  | * |  | * | * |  | * |
| $W^+W^-\gamma$ |  |  |  | * |  | * | * |  | * |
| $W^+W^-H$ |  |  |  | * |  | * |  |  |  |
| $ZZH$ |  |  | * | * | * | * | * | * |  |
| $Z\gamma H$ |  |  |  | * | * | * | * | * |  |
| $\gamma\gamma H$ |  |  |  |  |  |  |  | * |  |
| $HHH$ | * | * |  |  |  |  |  |  |  |
| $W^+W^-W^+W^-$ |  |  |  | * |  | * |  |  | * |
| $W^+W^-ZZ$ |  |  |  | * |  | * |  |  | * |
| $W^+W^-Z\gamma$ |  |  |  | * |  | * |  |  | * |
| $W^+W^-\gamma\gamma$ |  |  |  | * |  |  |  |  | * |
| $W^+W^-ZH$ |  |  |  | * |  | * | * |  |  |
| $W^+W^-\gamma H$ |  |  |  | * |  | * | * |  |  |
| $W^+W^-HH$ |  |  |  | * |  | * |  |  |  |
| $ZZHH$ |  |  | * | * | * | * | * | * |  |
| $Z\gamma HH$ |  |  |  | * | * | * | * | * |  |
| $\gamma\gamma HH$ |  |  |  |  |  |  |  | * |  |
| $HHHH$ | * | * |  |  |  |  |  |  |  |

**Higgs self-interactions**

In Figure 20 we show the energy dependence of cross sections for the processes with multiple Higgs production, including the $ZH$ curve for comparison. We display the results for the SM and for the nonzero coefficient values[17] $c_6 = \pm 0.5/\lambda \text{TeV}^{-2}$, corresponding to a 30% effect on $\kappa_\lambda$. We verify the fact that this coefficient can be measured both at low energy using $ZHH$, and at high energy using $\nu\bar{\nu}HH$. The two measurements are complementary in the sense that the term linear in $c_6$ has opposite sign, see Section 2.2.1 for a detailed analysis. The plot also indicates that additional precision could be gained by analyzing the $W^+W^-HH$ and $\nu\bar{\nu}ZHH$ final states, depending on energy. The two triple-Higgs processes shown in the plot are likely below observable levels, but they exhibit a much stronger dependence on $c_6$ than the pair-production processes. Setting a limit on either of those may thus further constrain the allowed parameter space. None of the anomalous contributions increases with energy, but collecting data at different energies improves the accuracy by the different set of final states that becomes accessible.

**Higgs-gauge interactions**

The operators $\mathcal{O}_W$, $\mathcal{O}_B$, $\mathcal{O}_{HW}$, $\mathcal{O}_{HB}$, and $\mathcal{O}_{BB}$ from Table 2 collectively describe non-SM interactions of the Higgs-doublet field with gauge bosons, inducing anomalous triple and quartic gauge interactions via the Higgs vacuum expectation value, and anomalous couplings of the Higgs to transverse gauge bosons. Multi-boson production at CLIC can provide independent information that eventually allows us to disentangle all five parameters with good precision.

---

[17]Here the factor $\lambda$ has been factored out from the definition in Table 2.

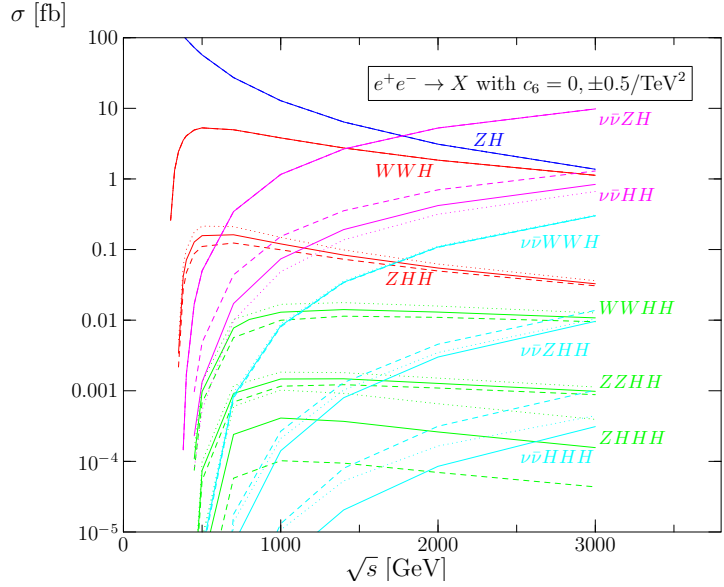

Figure 20: Cross sections for multi-Higgs production in $e^+e^-$ annihiliation. Solid lines: SM; dashed/dotted: $c_6 = \pm(0.5/\lambda)\text{TeV}^2$.

In the following, we consider only $\mathcal{O}_{HW}$ for a specific example. In Figure 21 we show the energy dependence of the coefficients $\sigma^{(0,1,2)}$ from an expansion of the cross section of the form $\sigma(c) = \sigma^{(0)} + \sigma^{(1)}c + \sigma^{(2)}c^2$, the three dominant processes $e^-e^+ \to W^+W^-$, $W^+W^-Z$, and $ZH$. We observe that the term linear in $c_{HW}$ (dashed), i.e., the interference of the SM contribution with the new interaction, is roughly constant with energy for the diboson processes, and rising with energy for the triboson process. This behavior is due to the presence of field derivatives in the interaction which translate into energy-momentum factors and cancel the $1/s$ suppression of the SM processes. The quadratic term (dotted) increases with energy, for all three processes. In fact, due to the $SU(2)_L$ symmetry which becomes manifest at high energy, this term coincides for the $W^+W^-$ and $ZH$ final states. Considering all five operators of this type, we obtain a correlated pattern. The overall energy dependence of the deviation is similar to $c_{HW}$ in all cases, but the contribution to the individual processes differs between operators. The staged CLIC run proposal allows for measurements of all accessible processes at distinct energy values. This is essential for disentangling the various effects that can be present.

**Gauge self-interactions**

At the dimension-six level of the SMEFT, there is one operator that modifies the gauge self-interactions only, $\mathcal{O}_{3W}$. This term affects the strength and Lorentz structure of the triple and quartic gauge self-couplings, independent of the gauge coupling to matter (i.e., fermions and the Higgs doublet). Due to field derivatives acting on gauge fields, the induced deviation from the SM rises with energy. The anomalous contributions predominantly concerns the transverse polarization modes of the gauge bosons.

In Figure 21, we display the constant, linear, and quadratic prefactors $\sigma^{(0,1,2)}$ for the processes $e^-e^+ \to W^+W^-$ and $W^+W^-Z$ and the operator $\mathcal{O}_{3W}$. For the diboson process, we verify the well-known fact that due to a mismatch in helicity, the interference term decreases with energy. By contrast, for the triboson process, the interference is unsuppressed and slightly increases with energy. The quadratic term $\sigma^{(2)}$ is rather large for both processes, and it increases with energy.

The results presented above indicate that multi-boson processes may significantly contribute to the res-olution power of the CLIC collider. A substantial number of new processes becomes accessible to direct detection. Selecting final states, multi-boson interactions can be studied individually. The numerical results of this preliminary study provide an order-of-magnitude estimate for the level and kind of new-physics contributions that should be detectable by a careful analysis of CLIC multi-boson data. A dedicated study will include initial-state radiation and beamstrahlung effects in the calculation, and account for final-state branching fractions, detection efficiencies, and mis-identification probabilities. In analogy to the analysis of di-boson final states, we expect kinematical distributions and angular observables to further improve the sensitivity, similarly to Section 2.4.

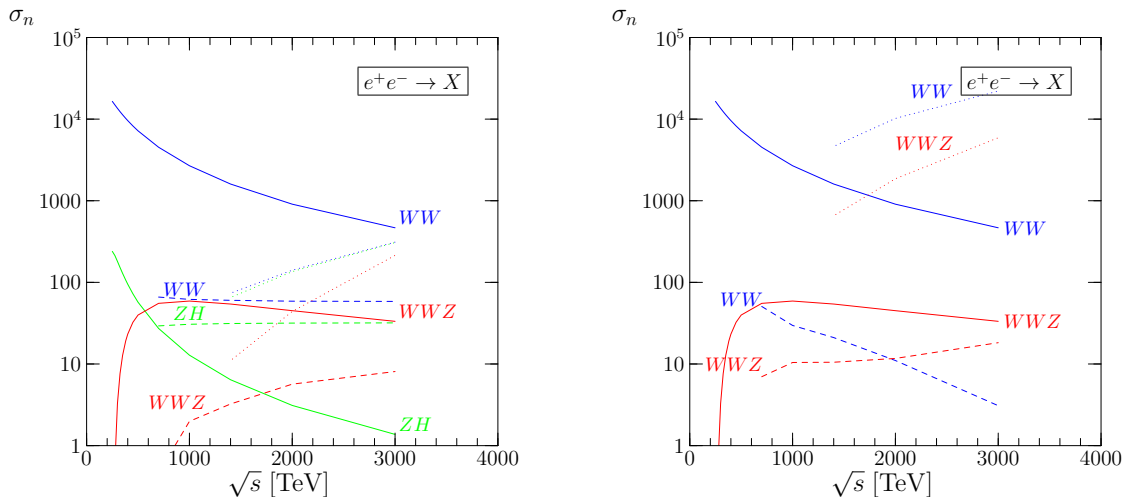

Figure 21: Multi-boson production in $e^+e^-$ annihilation: dependence on non-zero $c_{HW}$ (left panel) and non-zero $c_{3W}$ (right panel). Solid lines: $\sigma^{(0)}$ = SM contribution, in fb. Dashed: $\sigma^{(1)}$ = interference, in fb TeV$^2$. Dotted: $\sigma^{(2)}$ = quadratic term, in fb TeV$^4$.

## 2.6 Sensitivity to universal theories via $e^+e^- \to \psi\bar{\psi}$ [18]

Two to two fermion reactions are one of the most basic processes in particle physics. At $e^+e^-$ colliders, and within the Standard Model (SM), such reactions are mediated only by electroweak interactions. In particular, the study of $e^+e^- \to \psi\bar{\psi}$ at the $Z$-pole at LEP/SLD was crucial in the confirmation of the SM description of the properties of the neutral current with a precision at the per mille level. Being sensitive to effects modifying the structure of the electroweak sector makes these processes relevant in testing many different types of new physics effects. In this section we study the potential of the high-energy measurements possible at CLIC in constraining new physics using $e^+e^- \to \psi\bar{\psi}$ within the context of general "universal" theories. These type of theories include SM extensions with new vector bosons mixing with the SM ones [57–60], models of vector compositeness [51], etc. Using the formalism of the SM effective field theories (SMEFT), and truncating the effective expansion at dimension 6, universal new physics effects in $e^+e^- \to \psi\bar{\psi}$ can be described by the well-known "oblique parameters" $S$, $T$, $W$ and $Y$ [61, 62]. While the first two, $S$ and $T$, induce effects that are constant with the energy, the relative importance of the $W$ and $Y$ effects grows with the energy, thus making a high-energy lepton collider such as CLIC the ideal setup to test such types of oblique new physics.

In the what follows we briefly describe the different types of new physics that can modify $e^+e^- \to \psi\bar{\psi}$ within the context of the dimension-6 SMEFT, and the connection with the oblique parameters. We then describe in the different observables included in the analysis and the estimation of their uncertainties

---

[18]Based on a contribution by J. de Blas and A. Wulzer.

at CLIC. We finally present the projected sensitivity to universal new physics.

**New physics in $e^+e^- \to \psi\bar\psi$**

Around the $Z$ pole, due to the effect of the resonance, $e^+e^- \to \psi\bar\psi$ is sensitive almost exclusively to new physics modifications of the SM neutral current interactions. Within the context of the dimension-6 SMEFT, and using the notation introduced in Table 2, direct contributions to such properties are given by the operators

$$\mathcal{O}_{H\psi}, \quad \mathcal{O}_{H\psi}^{(1)} \quad \text{and} \quad \mathcal{O}_{H\psi}^{(3)}, \tag{44}$$

which modify the neutral and charged current couplings, as well as by

$$\mathcal{O}_{WB} \quad \text{and} \quad \mathcal{O}_{HD}, \tag{45}$$

which modify the gauge boson propagators.[19] At energies above or below the resonance measurements of $e^+e^- \to \psi\bar\psi$ become sensitive to heavy new physics generating effective four-fermion interactions at low energies. The differential cross section for $e^+e^- \to \psi\bar\psi$, including the contributions from the four-fermion interactions, and neglecting fermion masses[20] can be written as [63, 64]

$$\frac{1}{N_\psi}\frac{4s}{\alpha^2}\frac{d\sigma}{d\Omega}\left(e^+e^- \to \psi\bar\psi\right) = \left[\left|\mathcal{M}_{LR}^{ee}(t)\right|^2 + \left|\mathcal{M}_{RL}^{ee}(t)\right|^2\right]\frac{s^2}{t^2}\delta_{e\psi} +$$
$$+ \left[\left|\mathcal{M}_{LR}^{e\psi}(s)\right|^2 + \left|\mathcal{M}_{RL}^{e\psi}(s)\right|^2\right]\frac{t^2}{s^2} + \tag{46}$$
$$+ \left[\left|\mathcal{M}_{LL}^{e\psi}(s)\right|^2 + \left|\mathcal{M}_{RR}^{e\psi}(s)\right|^2\right]\frac{u^2}{s^2}.$$

where $\alpha$ is the QED coupling constant, $N_\psi$ the number of colors for each fermion $\psi$, $s = 4E_\text{beam}^2$, $t = -\frac{1}{2}s(1-\cos\theta)$ and $s+t+u=0$. Whereas the helicity amplitudes read

$$\mathcal{M}_{\alpha\beta}^{ee}(t) = 1 + \frac{g_\alpha^e g_\beta^e}{\sin^2\theta_W \cos^2\theta_W}\frac{t}{t-M_Z^2} + \frac{t}{4\pi\alpha}\frac{\mathcal{A}_{\alpha\beta}^f}{\Lambda^2}, \quad (\alpha \neq \beta),$$

$$\mathcal{M}_{\alpha\beta}^{e\psi}(s) = -Q_\psi + \frac{g_\alpha^e g_\beta^\psi}{\sin^2\theta_W \cos^2\theta_W}\frac{s}{s-M_Z^2+iM_Z\Gamma_Z} + \frac{s}{4\pi\alpha}\frac{\mathcal{A}_{\alpha\beta}^\psi}{\Lambda^2}, \quad (\alpha \neq \beta),$$

$$\mathcal{M}_{\alpha\alpha}^{e\psi}(s) = -Q_\psi + \frac{g_\alpha^e g_\alpha^\psi}{\sin^2\theta_W \cos^2\theta_W}\left[\frac{s}{s-M_Z^2+iM_Z\Gamma_Z} + \frac{s}{t-M_Z^2}\delta_{e\psi}\right] + \frac{s}{t}\delta_{e\psi} + (1+\delta_{e\psi})\frac{s}{4\pi\alpha}\frac{\mathcal{A}_{\alpha\alpha}^\psi}{\Lambda^2},$$

with $Q_\psi$ the electric charge of $\psi$, $\theta_W$ the weak mixing angle and the indices $\alpha, \beta = L, R$. The neutral current couplings, $g_{L(R)}^\psi \equiv g_{L(R)}^{\psi\text{SM}} + \delta g_{L(R)}^\psi$, take into account the new physics contributions from Eqs. (44) and (45). The effects of the four-fermion operators are encoded in the coefficients $\mathcal{A}_{\alpha\beta}^\psi$:

$$\mathcal{A}_{LL}^\ell = \frac{(c_{LL})_{11ii} + (c_{LL})_{1ii1}}{1+\delta_{e\ell}}, \qquad \mathcal{A}_{LR}^\ell = (c_{Le})_{1ii1},$$
$$\mathcal{A}_{LL}^u = \sum_{k,l} V_{ik}\left(c_{LQ}^{(1)} - c_{LQ}^{(3)}\right)_{11kl} V_{li}^\dagger, \qquad \mathcal{A}_{LR}^u = (c_{Lu})_{1ii1},$$
$$\mathcal{A}_{LL}^d = \left(c_{LQ}^{(1)} + c_{LQ}^{(3)}\right)_{11ii}, \qquad \mathcal{A}_{LR}^d = (c_{Ld})_{1ii1},$$

$$\mathcal{A}_{RR}^\ell = \frac{2(c_{ee})_{11ii}}{1+\delta_{e\ell}}, \qquad \mathcal{A}_{RL}^\ell = (c_{Le})_{i11i},$$
$$\mathcal{A}_{RR}^u = (c_{eu})_{11ii}, \quad \text{and} \quad \mathcal{A}_{RL}^u = \sum_{kl} V_{ik}(c_{Qe})_{k11l} V_{li}^\dagger,$$
$$\mathcal{A}_{RR}^d = (c_{ed})_{11ii}, \qquad \mathcal{A}_{RL}^d = (c_{Qe})_{i11i},$$

---

[19] If using $\alpha$, $M_Z$ and $G_F$ as inputs of the SM, as we do here, there are extra indirect contributions from the four-lepton operator $(\bar{L}_L^i \gamma_\mu L_L^j)(\bar{L}_L^k \gamma^\mu L_L^l)$, as this modifies the amplitude for $\mu$ decay, which is used to extract the value of $G_F$.

[20] The effects of the top quark mass are included in our calculations of the new physics corrections to the $e^+e^- \to t\bar{t}$ process.

where $V$ is the CKM matrix and $i$ stands for any given flavor. The factors $1+\delta_{e\psi}$ have been introduced for convenience in order to take into account the different contributions to the Lagrangian for $e^+e^- \to \ell^+\ell^-$, depending on whether $\ell$ is or not an electron. The notation for the coefficients of the four-fermion operators, $c_i$, entering in the different $\mathcal{A}^\psi_{\alpha\beta}$ is introduced in Table 18.

Table 18: Notation for the different four-fermion operators appearing in the $\mathcal{A}^\psi_{\alpha\beta}$ coefficients, classified according to the chiralities of the fermion multiplets entering in the operator.

| LLLL | RRRR | LLRR |
|---|---|---|
| $\mathcal{O}_{LL} = (\overline{L}^i_L \gamma_\mu L^j_L)(\overline{L}^k_L \gamma^\mu L^l_L)$ | $\mathcal{O}_{ee} = (\overline{e}^i_R \gamma_\mu e^j_R)(\overline{e}^k_R \gamma^\mu e^l_R)$ | $\mathcal{O}_{Le} = (\overline{L}^i_L \gamma_\mu L^j_L)(\overline{e}^k_R \gamma^\mu e^l_R)$ |
| $\mathcal{O}^{(1)}_{LQ} = (\overline{L}^i_L \gamma_\mu L^j_L)(\overline{Q}^k_L \gamma^\mu Q^l_L)$ | $\mathcal{O}_{eu} = (\overline{e}^i_R \gamma_\mu e^j_R)(\overline{u}^k_R \gamma^\mu u^l_R)$ | $\mathcal{O}_{Lu} = (\overline{L}^i_L \gamma_\mu L^j_L)(\overline{u}^k_R \gamma^\mu u^l_R)$ |
| $\mathcal{O}^{(3)}_{LQ} = (\overline{L}^i_L \gamma_\mu \sigma_a L^j_L)(\overline{Q}^k_L \gamma^\mu \sigma_a Q^l_L)$ | $\mathcal{O}_{ed} = (\overline{e}^i_R \gamma_\mu e^j_R)(\overline{d}^k_R \gamma^\mu d^l_R)$ | $\mathcal{O}_{Ld} = (\overline{L}^i_L \gamma_\mu L^j_L)(\overline{d}^k_R \gamma^\mu d^l_R)$ |
| | | $\mathcal{O}_{Qe} = (\overline{Q}^i_L \gamma_\mu Q^j_L)(\overline{e}^k_R \gamma^\mu e^l_R)$ |

As is apparent from the helicity amplitudes above, the relative contributions to the cross section from the four fermion interactions grows with the centre-of-mass energy. Therefore the access to large energies gives a very useful handle for performing precision tests of such effects. At hadron colliders like the LHC, this energy dependence can compensate for the lack of absolute experimental accuracy. On the other hand, the access to high energy collisions together with the much cleaner environment of lepton colliders makes CLIC a much better option to prove such effects.

For the case of universal new physics, and truncating the EFT expansion at dimension-6, the effects can be described by the following Lagrangian [62],[21]

$$\Delta \mathcal{L}_{\text{Universal}} = \frac{S}{16\pi v^2} \mathcal{O}_{WB} - \frac{2\alpha T}{v^2} \mathcal{O}_{HD} - \frac{Y}{2M_W^2} \mathcal{O}_{2B} - \frac{W}{2M_W^2} \mathcal{O}_{2W}, \qquad (47)$$

where the dimension-six operators $\mathcal{O}_{2B}$ and $\mathcal{O}_{2W}$ are given in Table 2.

Using the Bianchi identities and a perturbative field redefinition it is easy to see that, from the point of view of $e^+e^- \to \psi\bar\psi$, $\mathcal{O}_{2B}$ and $\mathcal{O}_{2W}$ can be traded by the "square" of the SM hypercharge and weak isospin currents. These contribute, in particular, to several types of four-fermion operators, which induce the effects that grow with energy in $e^+e^- \to \psi\bar\psi$ mentioned above. The effects of $S$ and $T$, however, only enter in the modifications of the effective electroweak couplings, and therefore the new physics amplitudes induced by these parameters have the same energy dependence as the SM ones.

The current status on the knowledge of these oblique parameters is controlled by fit to the electroweak precision observables (EWPO) for the case of the $S$ and $T$ parameters [65, 66]. Prior to the LHC, the leading constraints on the $W$ and $Y$ parameters came from the off-pole measurements of $e^+e^- \to \psi\bar\psi$ taken at LEP2 [67]. From the analysis of Ref. [68],

$$\begin{pmatrix} S \\ T \\ W \\ Y \end{pmatrix} = \begin{pmatrix} -0.10 \pm 0.13 \\ 0.02 \pm 0.08 \\ (-0.1 \pm 0.6) \times 10^{-3} \\ (-1.2 \pm 0.9) \times 10^{-3} \end{pmatrix}, \quad \rho = \begin{pmatrix} 1 & & & \\ 0.86 & 1 & & \\ -0.12 & -0.06 & 1 & \\ 0.70 & 0.39 & -0.49 & 1 \end{pmatrix} \qquad (48)$$

which we will use in the comparison with the CLIC projections. As shown in Refs. [69, 70], however, the bounds on $W$ and $Y$ from the Drell-Yan process at the LHC 8 TeV run are already more constraining than those from LEP2, and will be significantly improved at the HL-LHC.

---

[21]The normalization of the Lagrangian has been chosen such that the definition of the $S$ and $T$ parameters matches the original one in Ref. [61].

$e^+e^- \to \psi\bar{\psi}$ **processes**

For this study we use the following observables measured in difermion production:

1. The differential distribution of the number of events $dN_{ev}/d\cos\theta$ in $e^+e^- \to \ell^+\ell^-$, $\ell = e, \mu, \tau$. We assume 100% efficiency in the reconstruction and identification of the electrons and muons. For the $\tau^+\tau^-$ channel we consider only hadronic decays. We assume an overall efficiency of 50% in the reconstruction of the $\tau$'s and a 3% fake rate from jets.

2. The differential distribution of the number of events $dN_{ev}/d\cos\theta$ in $e^+e^- \to c\bar{c}, b\bar{b}$. We assume 80% tag efficiency for $b$ quarks and 10% (1%) mistag rate of $c$ quarks ($u, d, s$ quarks) as $b$ jets. For the charm we use a 50% tag efficiency and fake rates of 10% and 2% from bottom and $u, d, s$ quarks, respectively. Two $b$ ($c$) tags are required for each event. Note that reconstructing the full $\cos\theta$ distribution implicitly assumes the charge of the final state $b$ and $c$-hadrons can be measured. We explicitly checked that an analysis blind to the $b$ and $c$ charges, i.e. using the $|\cos\theta|$ distributions, does not have a significant impact on the limits obtained in the global $e^+e^- \to \psi\bar{\psi}$ fit presented below.

3. For the $t\bar{t}$ final state we follow the CLIC study at 1500 and 3000 GeV centre-of-mass energies presented in Ref. [10]. We include the total cross section and forward-backward distributions in $e^+e^- \to t\bar{t}$ as observables in our analysis.

In all cases the polar scattering angle $\theta$ is defined as the angle of the outgoing fermion with respect to the electron direction in the rest frame of the difermion system. We assume the data are binned in 20 bins of $\cos\theta$ in the whole range $[-1, 1]$. We only consider events within an estimated CLIC detector acceptance for $8° < \theta < 172°$.

The sensitivity to the previous observables has been estimated for the CLIC energies and luminosities detailed in Table 3; here we consider both the baseline scenario with polarized beams, and an unpolarized scenario for comparison.

For the calculation of the statistical uncertainties of the different processes we used event samples generated with the WHIZARD Monte Carlo software [55, 71]. We include in the simulations the effects of both initial state radiation (ISR) and beamstrahlung. The statistical uncertainties were obtained assuming a SM signal only.

At large energies, the effects of ISR and beamstrahlung can induce a significant energy loss, thus reducing the effective energy entering in the hard process, $\sqrt{s'}$, with respect to the nominal center of mass energy $\sqrt{s}$. The measured processes, whose observed cross section is then given by the convolution of the cross section of the hard process with the luminosity spectrum, thus involve events distributed over a wide range of energies. To isolate the high-energy effects of the signal, we select events close to the nominal collision energy by imposing a cut on the difermion invariant mass $M_{ff} \geq 0.85\sqrt{s}$. We therefore consider only events with $M_{ff}$ above 325/1300/2600 GeV for $\sqrt{s} = 380/1500/3000$ GeV. The impact of these beam effects is shown in Figure 22. In the left panel we present the distribution of events for the $e^+e^- \to e^+e^-$ process at different energies, as a function of the $e^+e^-$ invariant mass. The impact of the beam effects on the angular differential distributions and the importance of keeping high energy events to maximize the sensitivity to new physics is illustrated in the right panel of the figure. There we show the normalized distributions of the number of events at $\sqrt{s} = 3000$ GeV for the case of ($P_{e^-} = -80\%, P_{e^+} = 0\%$) beams, for the SM and the contributions from the $W$ and $Y$ parameters (solid, dashed and dotted lines, respectively). We compare the $M_{ff}$ cut introduced above, with another one at 50% of the nominal centre-of-mass energy, illustrating the loss in sensitivity to the effects that grow with energy.

For illustration of the statistics we are dealing with, we show in Figure 23 the distributions of events obtained for the $e^+e^-$, $\mu^+\mu^-$, $b\bar{b}$ and $c\bar{c}$ channels for the $\sqrt{s} = 3000$ GeV run, which, as we will see, dominates in the constraints on $W$ and $Y$.

For the $e^+e^- \to t\bar{t}$ process we use the experimental sensitivity reported by the dedicated analysis presented in Ref. [10] for the total cross section and forward-backward asymmetries. We only include the information about the 1500 and 3000 GeV stages (the 380 GeV stage will be shown later to be of little importance in the analysis of the $W$ an $Y$ constraints). The numbers in that analysis were computed assuming only polarized beams and different luminosities than those in Table 3.[22] We rescale the statistical uncertainties accordingly. Given that the preselection and selection efficiencies vary very little across polarizations, we apply their average to obtain the statistical uncertainties in the case with unpolarized beams.

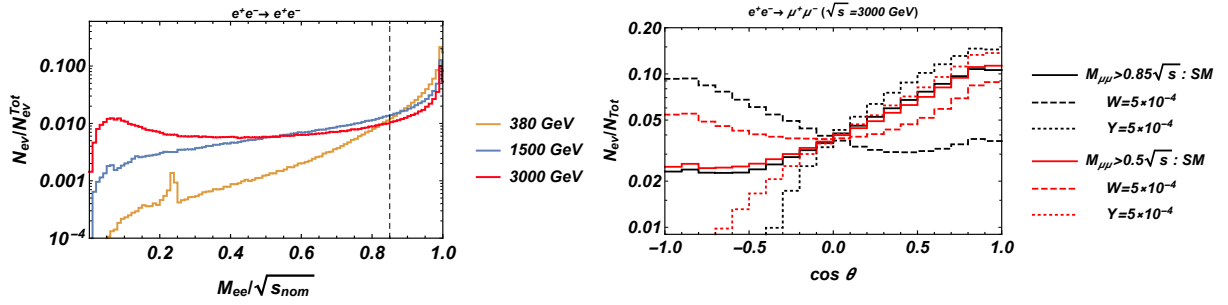

Figure 22: (Left) Distribution of $e^+e^- \to e^+e^-$ events as a function of the $e^+e^-$ invariant mass, taking into account the CLIC luminosity spectrum. (The dashed line indicates the cut applied in the analysis.) (Right) Effect of the convolution with the luminosity spectrum for the differential cross section in $e^+e^- \to \mu^+\mu^-$ for CLIC at 3000 GeV.

On top of the statistical uncertainties we also consider in our analysis the impact of systematic errors. For the $e^+e^- \to t\bar{t}$ channel we follow again the results in Ref. [10]. Combining the most relevant systematic uncertainties coming from the variation of the normalization and the modeling of the background, the systematics errors at 1500 GeV were found to be in the range $\sim 1-3\%$ for both the total cross section and the forward-backward asymmetry. We also assume such systematics for the 3000 GeV run. Since these uncertainties are associated mainly with background effects, they are expected to be much smaller for the other di-fermion processes. To asses the impact of systematic uncertainties in the light fermion channels we study 3 different scenarios:

- A *conservative* case with a global systematic uncertainty of $\delta_{sys,ff} = 1\%$ for $f \neq t$. For the top quark channel we use in this case $\delta^{(C)}_{sys,tt} = 3\%$.
- In a more *realistic* scenario systematic uncertainties in the light fermion channels are expected to be somewhat smaller, and we use $\delta_{sys,ff} = 0.3\%$ for $f \neq t$. For the $t\bar{t}$ observables we assume $\delta^{(O)}_{sys,tt} = 1\%$ as a systematic global error.
- Finally, we also consider an *optimistic* case where systematic effects are assumed to be subdominant and only statistical errors are considered: $\delta_{sys} = 0$.

The effects of new physics effects from the oblique parameters $S$, $T$, $W$ and $Y$ were computed using the same tools and same level of approximation as the SM signal by using reweight methods. Since the form of the new physics amplitude has exactly the same helicity structure as in the SM, the sampling of the phase space of the SM case has sufficient coverage to ensure the validity of this reweighting method.

The reweighted samples, computed for different values of the oblique parameters, are then used to extract semi-analytical expressions for the number of events in each bin of the different distributions. In this process only the leading effects coming from the interference between the NP and SM amplitudes

---

[22]Also, in that study a centre-of-mass energy of 1500 GeV is considered for the second CLIC stage. We assume the results for that energy translate directly into the updated baseline scenario with $\sqrt{s} = 1500$ GeV.

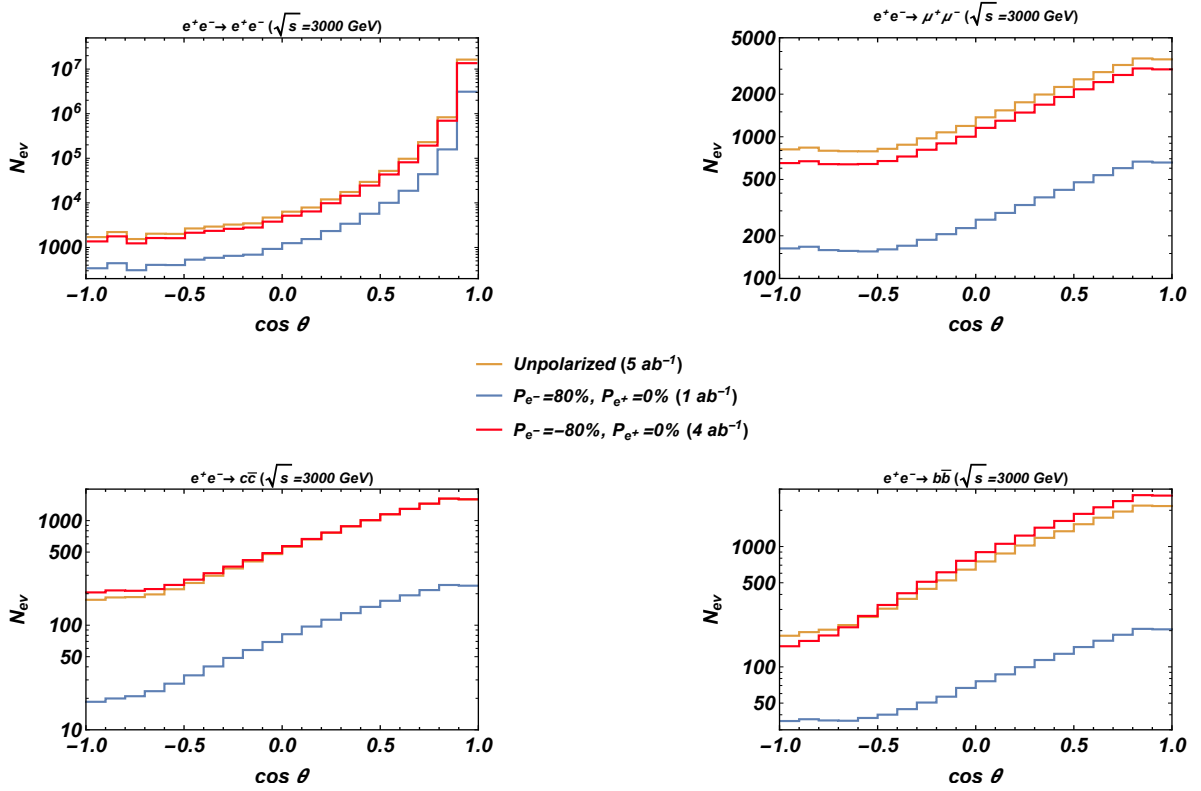

Figure 23: Distribution of the number of events as a function of $\cos\theta$ at $\sqrt{s} = 3000$ GeV for $e^+e^- \to e^+e^-$ (upper-left panel), $e^+e^- \to \mu^+\mu^-$ (upper-right panel), $e^+e^- \to c\bar{c}$ (lower-left panel) and $e^+e^- \to b\bar{b}$ (lower-right panel). For each channel we show the distributions obtained assuming unpolarized beams and $L = 5$ ab$^{-1}$ (yellow line), $(80\%, 0\%)$ polarization with $L = 1$ ab$^{-1}$ (blue line) and $(-80\%, 0\%)$ polarization with $L = 4$ ab$^{-1}$.

are considered, consistently with dim-6 effective Lagrangian expansion:

$$N_i = N_i^{\text{SM}} + \sum_{\alpha=S,T,W,Y} A_\alpha \alpha.$$

In the following we use the previous setup to estimate the CLIC sensitivity to these universal new physics effects.

**Constraints on universal new physics**

Using the different observables, theory predictions and uncertainties estimated as described above, we compute a $\chi^2$ function

$$\chi^2 = \sum_i \frac{(N_i - N_i^{\text{SM}})^2}{N_i^{\text{SM}} + (\delta_{sys,i} N_i^{\text{SM}})^2}, \qquad (49)$$

with $i$ running over the different energies, fermion channels and bins/observables. The results of the global fit including all di-fermion processes and the corresponding projections for the uncertainties of the different oblique parameters are shown in Table 19 and Figure 24. The results are given in the "CLIC Baseline" scenario and also compared with the "Unpolarized" case described Table 3. In Table 19 we report the 68% C.L. Gaussian uncertainties on each parameter, as well as the correlations, from the fit assuming the "realistic" projections for the systematic uncertainties. We also show in the table the results assuming new physics contributes only to one type of oblique parameter.[23] Figure 24 left (right)

---
[23] As mentioned above, not being able to identify the charge of the $b$ and $c$ hadrons only has a small impact in the global

shows the 95% confidence regions (C.R.) in the $S$-$T$ ($W$-$Y$) plane, profiling over the other parameters. The contours assuming the "realistic" systematic uncertainties are compared with the other assumptions discussed above.

Table 19: (Right) 68% C.L. Gaussian errors on the different oblique parameters. In parenthesis, the results assuming the other oblique parameters are set to 0. All numbers obtained from the fit assuming $\delta_{\text{sys}} = 0.3\%$ for the light fermion channels and $\delta_{\text{sys},tt}^{(O)} = 1\%$ systematics for $e^+e^- \to t\bar{t}$. (Left) The correlations between the different oblique parameters from the fit for the CLIC Baseline scenario.

| Scenario $(P_{e^-}, P_{e^+})$ | Current | CLIC Baseline $(\mp 80\%, 0\%)$ | CLIC Unpolarized $(0\%, 0\%)$ |
|---|---|---|---|
| $S$ | 0.13 | 0.09 (0.05) | 0.16 (0.10) |
| $T$ | 0.08 | 0.10 (0.05) | 0.12 (0.07) |
| $W\,[\times 10^6]$ | 600 | 1.7 (1.5) | 3.0 (2.2) |
| $Y\,[\times 10^6]$ | 900 | 2.0 (1.8) | 2.3 (1.7) |

| CLIC Baseline Correlation matrix | | | | |
|---|---|---|---|---|
| $S$ | 1 | | | |
| $T$ | 0.86 | 1 | | |
| $W$ | 0.08 | 0.19 | 1 | |
| $Y$ | 0.10 | 0.05 | -0.41 | 1 |

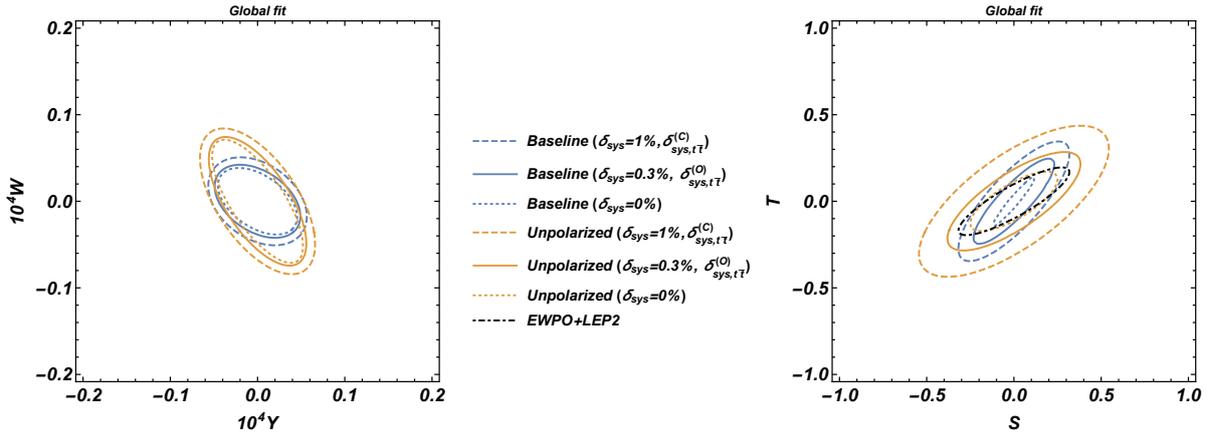

Figure 24: (Left) 95% C.R. in the $W$-$Y$ plane, profiling over $S$ and $T$, for the different CLIC scenarios and assumptions on the systematic errors. (Right) 95% C.R. in the $S$-$T$ plane, profiling over $W$ and $Y$, for the different CLIC scenarios and assumptions on the systematic errors.

The CLIC-only results can be compared with the current constraints on $S$, $T$, $W$ and $Y$ from EWPO and LEP2 measurements from Ref. [68], also shown in Table 19 and Figure 24. As is apparent, the measurements of $e^+e^- \to \psi\bar{\psi}$ at CLIC can only constrain the $S$ and $T$ parameters at a level similar to current EWPO. On the other hand, due to the access to very high energies, the projected sensitivities for the $W$ and $Y$ parameters are not only several orders of magnitude better than the LEP2 bounds, but would also greatly improve the projections at the HL-LHC: $\delta_{68\%}W_{\text{HL-LHC}} \sim 0.4 \times 10^{-4}$, $\delta_{68\%}Y_{\text{HL-LHC}} \sim 0.6 \times 10^{-4}$ [70]. For this reason, in what follows we focus the discussion on these 2 parameters. The results in Figure 24 also show that reducing the systematic errors below the $\sim 0.1\%$ level does not have any significant impact in the results for $W$ and $Y$, as the uncertainties become statistics dominated.

---

bounds in Table 19. For completeness, in a $b$ and $c$ charge-blind analysis we find the following 68% C.L. uncertainties for the CLIC-Baseline scenario: $\delta_{68\%}S = 0.11$, $\delta_{68\%}T = 0.11$, $\delta_{68\%}W = 1.8 \times 10^{-6}$ and $\delta_{68\%}Y = 2.1 \times 10^{-6}$. The individual bounds obtained assuming the other oblique parameters are set to zero remain unchanged with respect to Table 19.

The effects of polarization are only sizable along the direction $W \approx -Y$. The impact of polarization is however much more pronounced in the constraints set by each individual difermion channel, as shown in Figure 25, and it is only washed out in the global fit due to the complementarity between the different channels. From the figure it is also apparent that the constraints from the top quark channel, which is subject to larger systematics and whose statistics is more affected by the different selection efficiencies, are fairly irrelevant in the global fit. Finally, as shown in the left panel of Figure 26, and it is expected from the energy dependence of the new physics contributions, the bounds on $W$ and $Y$ are dominated by the 3 TeV run.

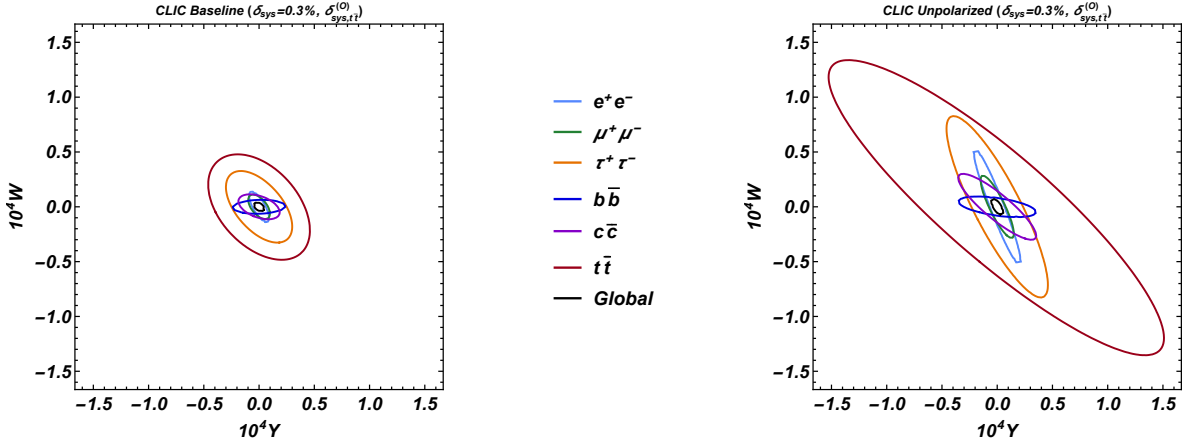

Figure 25: (Left) 95% C.R. in the $W$-$Y$ plane, profiling over $S$ and $T$, for the different final fermion states, assuming the CLIC Baseline scenario. (Right). The same in the scenario assuming CLIC operation with unpolarized beams.

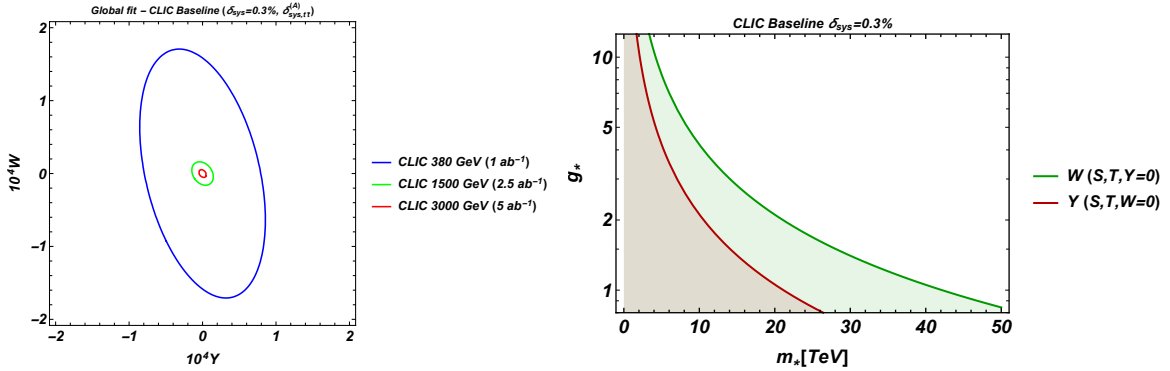

Figure 26: (Left) 95% C.R. in the $W$-$Y$ plane, profiling over $S$ and $T$, including data only from $\sqrt{s} =$ 380, 1500 and 3000 GeV, respectively, and assuming the CLIC Baseline scenario. (Right). 95% C.L. limit in the $g_*$-$m_*$ plane assuming CLIC operation with polarized beams and 0.3% systematics.

The results presented above can be interpreted within more definite scenarios, either via matching of the SMEFT with specific UV completions [72–86] or using power-counting rules for classes of models [17, 51]. For instance, assuming the Higgs originates from a strongly coupled strongly sector characterized by only one coupling $g_*$ and one scale $m_*$,

$$W = 2\frac{g^2}{g_*^2}\frac{M_W^2}{m_*^2}, \quad Y = 2\frac{g'^2}{g_*^2}\frac{M_W^2}{m_*^2}. \tag{50}$$

One can therefore translate the bounds on $W$ and/or $Y$ into exclusion regions in the $g_*$-$m_*$ plane. These are shown in Figure 26 for $\delta_{sys} = 0.3\%$, for the cases where the new physics only generates contributions to one of the 2 parameters, $W$ or $Y$.

## 2.7 Global effective-field-theory analysis of top-quark pair production[24]

The top quark and the Higgs boson are the only SM particles having escaped the precise scrutiny of the previous generation of lepton colliders. In $e^+e^-$ collisions at a centre-of-mass greater than twice the top quark mass, top quarks are dominantly produced in pairs (see Figure 27a). The study of the pair production process provides a precise characterization of the $t\bar{t}Z$ and $t\bar{t}\gamma$ vertices, complementing the study of the strong and charged-current interactions of the top quark at hadron colliders. With a centre-of-mass energy of $\sqrt{s} = 380\,\text{GeV}$, the initial stage of CLIC enables top-quark physics from the onset. The high-energy stages at $1.5$ and $3\,\text{TeV}$ strongly enhance the sensitivity to high-scale new physics that generates four-fermion operators of $e^+e^-t\bar{t}$ field content. They also give access to new processes, such as the single top-quark production of Figure 27b, the associated production of a top-quark pair and a Higgs boson (see Figure 27d), and the vector-boson-fusion production of top quark pairs of Figure 27e (discussed briefly below).

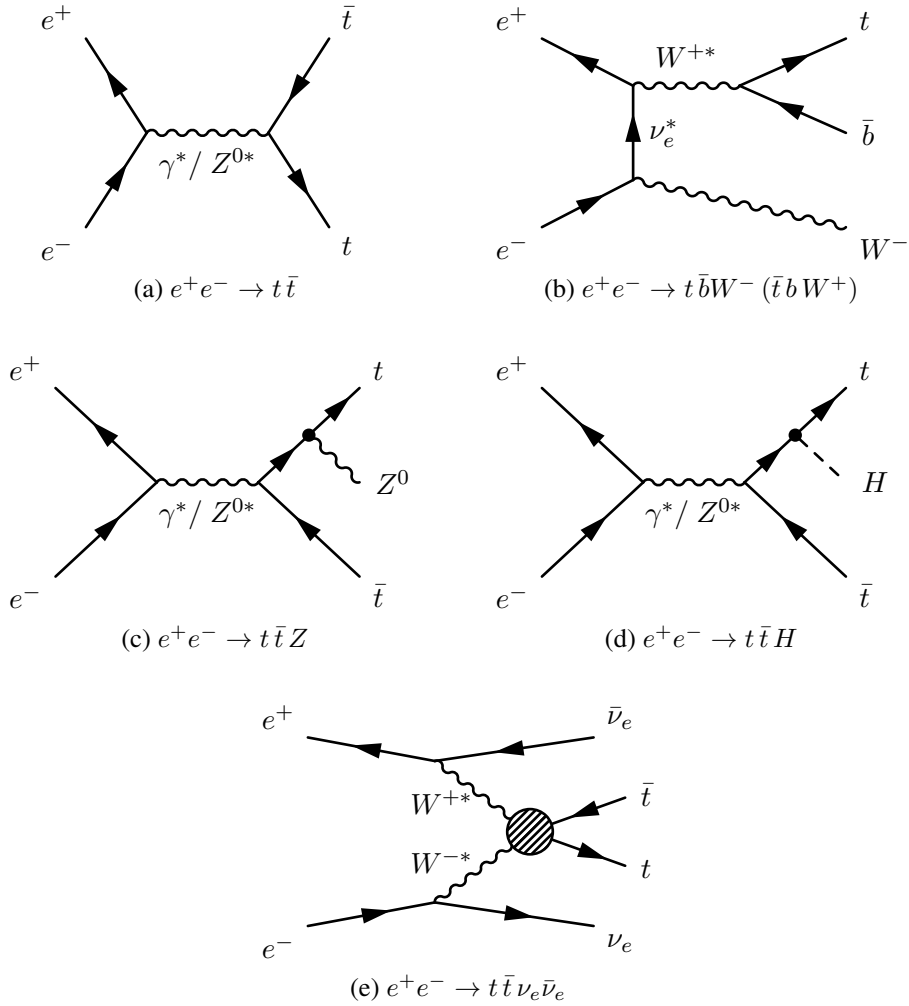

Figure 27: Relevant top-quark production processes: (a) $t\bar{t}$, (b) single-top, (c) $t\bar{t}Z$, (d) $t\bar{t}h$, (e) $t\bar{t}\nu_e\bar{\nu}_e$.

We discuss in this section the indirect sensitivity to physics beyond the SM gained through precision measurements of the top-quark pair production. Various observables are considered in the framework of the effective field theory introduced in Section 2, and the dependence on the centre-of-mass energy and the beam polarization are quantified. All ten dimension-six operators of the Warsaw basis which involve a top quark and interfere with the leading-order SM $e^+e^- \to t\bar{t} \to bW^+\bar{b}W^-$ amplitudes in the vanishing $b$ mass limit, are considered. CP-violating and four-fermion operators are included. Realistic statistical uncertainties on cross section, forward-backward asymmetry and statistically optimal observables measurements in top-quark pair production are estimated with full-simulation studies at three centre-of-mass energies and in two beam polarization configurations. The obtained limits are interpreted in a concrete extension of the SM in Section 2.10.

---

[24]Based on a contribution by G. Durieux, I. García García, M. Perelló Roselló, P. Roloff, R. Ström, M. Vos, N. Watson, A. Winter and C. Zhang.

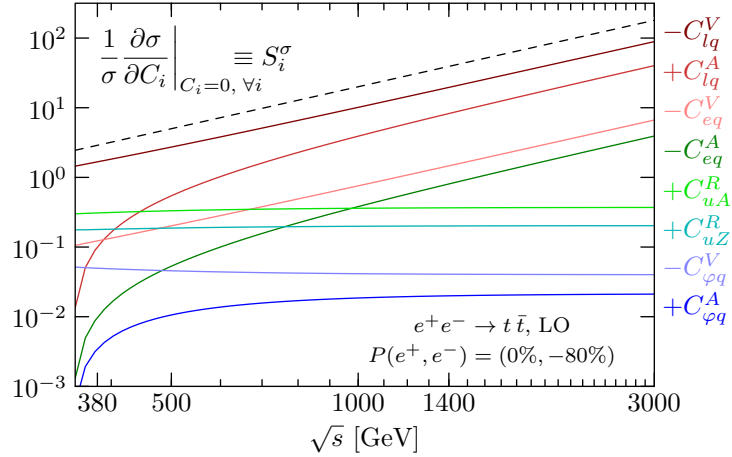

Figure 28: Leading-order sensitivity of the total $e^+e^- \to t\bar{t}$ cross section to various operators, as a function of the centre-of-mass energy, for a mostly left-handed electron beam polarization. That of four-fermion operators grows quadratically (as the dashed line) while that of two-fermion operators becomes constant. Axial-vector combinations of operators suffer from a threshold suppression. Figure taken from Ref. [88].

The sensitivity to the different dimension-six operators displays a diverse dependence on the centre-of-mass energy [87, 88] (see Figure 28. The operator used throughout this section are presented in Table 20. For two-fermion operators that modify the left- and right-handed couplings of the top quark to the $Z$-boson, the sensitivity is constant; the sensitivity of some observables to dipole operators grows linearly; and the sensitivity of four-fermion operators grows quadratically with $\sqrt{s}$. Operation of CLIC at several centre-of-mass energies is therefore a key asset for constraining otherwise degenerate combinations of operator coefficients.

**Observables**

The measurement of the $e^+e^- \to t\bar{t}$ cross-section provides precise constraints on the anomalous couplings of the top quarks to the photon (dipole) and $Z$-boson. In the lepton-plus-jets final state also the forward-backward asymmetry of the top quark is readily available. The combination of the two measurements in two runs with opposite-sign beam polarization [89] allows to disentangle the contributions of the $Z$-boson and photon. Measurements of the top-quark polarization using the charged lepton as polarimeter [88], of spin correlations, and of specific CP-odd observables [90] provide complementary information, that improves the constraint on poorly bounded combinations of operator coefficients.

We define a set of statistically optimal observables [91, 92] on the $e^+e^- \to t\bar{t} \to bW^+\bar{b}W^-$ differential distribution to simultaneously and efficiently constrain all considered directions in the effective-field-theory parameter space [88]. Similar techniques, based on differential information, were previously discussed for top-quark pair production at lepton colliders in Ref. [91, 93–95]. A discrete set of statistically optimal observables is ideally suited to probing multidimensional effective-field-theory parameter spaces. Resting on firm theoretical bases, optimal observables allow for a transparent study of higher-order corrections and systematic uncertainties.

**Selection and reconstruction**

The statistical uncertainties of the measurement of cross section, forward-backward asymmetry and statistically optimal observables are estimated using detailed detector simulations and reconstruction algorithms, and taking into account the impact of the luminosity spectrum. Only the electron- and muon-plus-jets channels are exploited. It is expected that systematic uncertainties could be controlled to the level of statistical ones. A detailed discussion is available in Ref. [10]. In the following, only a brief description is given.

Fully simulated events are reconstructed with the standard Pandora particle flow algorithm [96–98]. Isolated electron and muon candidates are identified by studying the pattern of energy depositions in the calorimeters, impact parameters, and isolation in cone around each input track. The remaining particle-flow objects are clustered into exactly four jets with the VLC algorithm [99] using $R = 1.6$ and $\beta = \gamma = 0.8$. The LCFI algorithm [100] identifies jets that contain $B$-hadrons. The event selection requires exactly one isolated lepton, one $b$-tagged jet according to a strict criterion and a second jet that satisfies a looser cut. The neutrino is reconstructed, up to a twofold ambiguity, using the missing transverse energy measurement, the charged lepton candidate, and the $W$-boson mass constraint. The hadronic $W$-boson candidate is formed by the two jets with the smallest $b$-tag score. Top-quark candidates are constructed by pairing the $W$-boson candidates with the $b$-tagged jets. The combination that minimizes a $\chi^2$ criterion based on the reconstructed values of the $W$-boson and top-quark mass, the top-quark boost and the angle between $W$-boson and $b$-jet, the expected values of the same quantities, and their resolution. The selection procedure yields a signal efficiency of approximately 70%, and a sample purity of greater than 80%. A cut on the $\chi^2$ score is applied for some observables to remove poorly reconstructed events and further reduce the background.

The selection and reconstruction of $t\bar{t}$ events collected during the high-energy stage, at $\sqrt{s} = 1.4$ and 3 TeV, requires a different strategy, specifically designed for the collimated decays of highly boosted top quarks. After removal of the charged-lepton candidate the particle-flow objects are clustered into exactly two jets with the VLC algorithm (with $R = 1.4$ for $\sqrt{s} = 1.4$ TeV, $R = 1$ for $\sqrt{s} = 3$ TeV and $\beta = \gamma = 1$). The hadronic top-quark candidate is tagged using an adaptation of the Johns Hopkins top tagging algorithm [101]. The leptonic top-quark candidate is obtained by adding the momenta of the charged lepton, the neutrino (reconstructed as before) and the remaining jet. The final step of the signal selection is performed using several boosted decision trees trained to reject different background sources. The $e^+e^- \to l\nu_l b\bar{b}q\bar{q}'$ sample contains a sizeable fraction of single-top events. The long tail of the luminosity spectrum leads to events with a centre-of-mass energy that is significantly below the nominal value. Events with $\sqrt{s} < 0.86$ times the nominal centre-of-mass energy, or with $|m_{Wb} - m_t| > 15$ GeV are considered part of the background, as well as events with $\tau$-leptons. The efficiency on the fiducial region is 37-39% at $\sqrt{s} = 1.4$ TeV, and 33-37% at $\sqrt{s} = 3$ TeV.

**Global reach**

Statistical uncertainties estimated with full simulation of the detector response at three centre-of-mass energies and with two beam polarizations serve as input to estimate the global reach of statistically optimal observable measurements on the ten-dimensional effective-field-theory space considered. Note that two centre-of-mass energies are required to disentangle two- and four-fermion operators involving top-quark currents of identical Lorentz structures. See Ref. [88] for further details and discussions of the impacts of the centre-of-mass energy and beam polarization.

Figure 29 provides global one-sigma projections in the reduced parameter space of top-philic scenarios defined in Section 3.5 of Ref. [10].[25] The relevant operator definitions are reminded in Table 20. In-dividual constraints on modifications of the left- and right-handed vector interactions of the top quark to the $Z$ boson ($C_{\varphi Q}^-, C_{\varphi t}$) are one to two orders of magnitude more stringent that the ones presently deriving from measurements of top-quark decay, single-top production and associated $t\bar{t}V$ production at the LHC and Tevatron [102, 103]. Individual constraints on the CP-conserving components of the electroweak dipole of the top quark ($C_{tW}, C_{tB}$) are about three orders of magnitude better than present ones. The improvement expected from the HL-LHC phase on current individual constraints is of at most an order of magnitude. CLIC sensitivity on four-fermion operator coefficients ($C_{lq,B}, C_{lq,W}, C_{lt,B}$) naively translates to scales probed indirectly of at least 20 TeV, beyond its direct reach. In comparison, present constraints on colour octet $q\bar{q}t\bar{t}$ operators from Ref. [103] naively translate to scales probed of the order of few hundred GeV.

---

[25]Note that event selection and reconstruction of top-quark candidates affect the measurements of optimal observables (as

Table 20: Dimension-six operators relevant for top-quark pair production in the top-philic scenario defined in Ref. [10]. Notice that the Hermitian conjugate is added to the Lagrangian for the operators $\mathcal{O}_{\varphi t}$ and $\mathcal{O}_{\varphi q}^-$, in spite of the fact that they are manifestly real. Hence, they effectively appear with an extra factor of 2. No factor of 2 is introduced neither for the 4-fermion operators, nor for any other of the Hermitian operators considered in this report.

| | |
|---|---|
| $\mathcal{O}_{\varphi q}^- = \frac{1}{2}(iH^\dagger \overleftrightarrow{D}_\mu H)(\bar{Q}\gamma^\mu Q) - \frac{1}{2}(iH^\dagger \overleftrightarrow{D}_\mu^a H)(\bar{Q}\sigma^a\gamma^\mu Q)$ | $\mathcal{O}_{\varphi t} = (iH^\dagger \overleftrightarrow{D}_\mu H)(\bar{t}\gamma^\mu t)$ |
| $\mathcal{O}_{tW} = (\bar{Q}\sigma^{\mu\nu}\sigma^a t)\tilde{H}W_{\mu\nu}^a$ | $\mathcal{O}_{tB} = (\bar{Q}\sigma^{\mu\nu}t)\tilde{H}B_{\mu\nu}$ |
| $\mathcal{O}_{lq,B} = (\bar{Q}\gamma_\mu Q)(\bar{e}\gamma^\mu e + \frac{1}{2}\bar{L}\gamma^\mu L)$ | $\mathcal{O}_{lq,W} = (\bar{Q}\sigma^a\gamma_\mu Q)(\bar{L}\sigma^a\gamma^\mu L)$ |
| $\mathcal{O}_{lt,B} = (\bar{t}\gamma_\mu t)(\bar{e}\gamma^\mu e + \frac{1}{2}\bar{L}\gamma^\mu L)$ | |

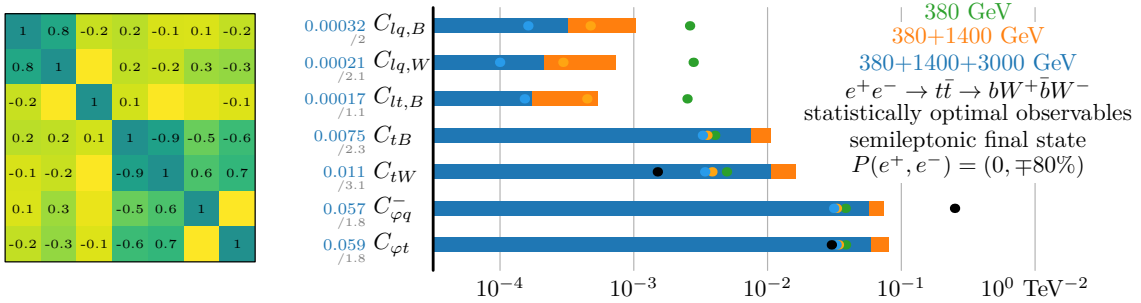

Figure 29: Global EFT reach from statistically optimal observables. Bars (blobs) indicate the global (individual) one-sigma reach on each operator coefficient for the three stages of CLIC. Black dots denote the reach from high-energy measurements of $tth$ and $WW \to tt$. Numerical values are provided for the correlation matrix, the global constraints and their ratio to individual ones for the entire CLIC programme. Systematic uncertainties are not included. Statistical ones account for the backgrounds, luminosity spectrum, acceptance and efficiency estimated with a full simulation of the electron- and muon-plus-jets channels.

In summary, the CLIC physics program offers a privileged window on the interactions of the top quark with neutral electroweak gauge bosons. The sensitivity of top-quark pair production measurements to dimension-six operator coefficients significantly exceeds that of current and future measurements at hadron colliders. The combination of the three CLIC energy stages and the possibility of electron beam polarization yields robust constraints in a global fit of the relevant operator coefficients.

---

used in Figure 29) at a level that exceeds the statistical uncertainty. A combination of in-situ and Monte-Carlo-based correction techniques is therefore required to correct measurements.

**Improved reach from the high energy regime in top processes**

The differential information in the top pair production process, studied in runs at different centre-of-mass energies and beam polarizations, allows the separation of the different operators affecting top-quark pair production. Part of the reason behind this success is that different operators induce a different energy dependence in this process. Four-fermion operators ($\mathcal{O}_{lq,B}$, $\mathcal{O}_{lq,W}$, $\mathcal{O}_{lt,B}$) have interferences with SM amplitudes which grow quadratically with energy and therefore benefit mostly from the different runs. The angular distributions of the top-quark decay products need to be exploited to extract the interferences between the amplitudes featuring different top-quark helicities which grow linearly with energy and involve dipole-type operators $\mathcal{O}_{tW}, \mathcal{O}_{tB}$. Finally operators like $\mathcal{O}_{\varphi Q}^{-}$ and $\mathcal{O}_{\varphi t}$ do not lead to interferences with SM amplitudes which grow with energy and are more efficiently determined at the lower energy where the top-quark pair production rate is higher.

It is therefore interesting to identify other processes in which the operators $\mathcal{O}_{\varphi Q}^{-}$, $\mathcal{O}_{\varphi t}$, $\mathcal{O}_{tW}$ and $\mathcal{O}_{tB}$ induce more rapid energy growths and could benefit from the high-energy operation of CLIC. Dipole-type operators actually lead to amplitude interferences growing quadratically with the centre-of-mass energy in the $e^+e^- \to \bar{t}th, \bar{t}bW, \bar{t}tZ$ processes. In addition, the helicity structure they induce coincides with that of the SM, so that SM and EFT amplitudes interfere in a simple analysis without angular distributions. The processes $\bar{t}tZ$ and $\bar{t}bW$ have larger cross sections. The dominant SM amplitudes however have $t$-channel singularities which limit their phase-space overlap the dipole operator amplitudes. The $\bar{t}th$ process has instead a smaller cross section, but allows for a simpler analysis. We report here the results for the last CLIC phase, with 5 ab$^{-1}$ collected at a centre-of-mass energy of 3 TeV, of a simple analysis in which effects of ISR and brehmstrahlung have been taken into account by a 50% decrease in the effective luminosity. We assume that all decay modes of the $W$-bosons arising from top-quark decays can be reconstructed, and obtain the following one-sigma range:

$$|C_{tW}| < 0.0008(0.0015) \text{ TeV}^{-2} \tag{51}$$

where a 100% (50%) reconstruction efficiency for the entire final state and no (3%) systematic uncertainties are assumed. This individual sensitivity is seen, in Figure 29, to be competitive with the one obtained with statistically optimal observables in top-quark pair production. In a global analysis, the $\bar{t}th$ data could help better constraining some direction in this parameter space in addition to providing a determination of the top-quark Yukawa coupling.

Similarly, the operators $\mathcal{O}_{\varphi Q}^{-}$ and $\mathcal{O}_{\varphi t}$ induce interferences with SM amplitudes which grow maximally with energy in $W^*W^* \to t\bar{t}$. This can be understood from the equivalence theorem, where the longitudinal polarisations of $W$-bosons belong to the Higgs doublet; then $\mathcal{O}_{\varphi Q, \varphi t}$ clearly correspond to contact interactions between longitudinally polarised $W$-bosons and top quarks, implying that the amplitude $A_{WW \to t\bar{t}} \propto E^2/\Lambda^2$. This process has already been studied in Ref. [10], obtaining for a one-by-one fit,[26]

$$|C_{\varphi t}| < 0.02 \ (0.03) \text{ TeV}^{-2} \quad |C_{\varphi Q}^1| < 0.1 \ (0.25) \text{ TeV}^{-2} \quad |C_{\varphi Q}^3| < 0.03 \ (0.05) \text{ TeV}^{-2}. \tag{52}$$

Again, a combination with top-quark pair production measurements could be beneficial.

## 2.8 Determination of the top-quark Yukawa coupling at lower energy[27]

The top-quark Yukawa coupling $y_t$ dominates the renormalization group evolution of the Higgs potential at high energy scales (see, for instance, Refs. [104–106]). Therefore $y_t$ is among the main drivers of SM predictions at very high energies and often dominates in studies of the self-consistency of the SM at very high energies. In addition, top-coupling modifications are expected in various models, see Section 2.10.

---

[26]Note that $C_{\varphi Q}^{-}$ which is probed in top-quark $e^+e^- \to t\bar{t}$ is $C_{\varphi Q}^1 - C_{\varphi Q}^3$. The loosest individual limit which applies on $C_{\varphi Q}^1$ from $WW \to t\bar{t}$ is reported in Figure 29 in comparison with $C_{\varphi Q}^{-}$.

[27]Based on a contribution by S. Boselli, A. Mitov and R. Hunter.

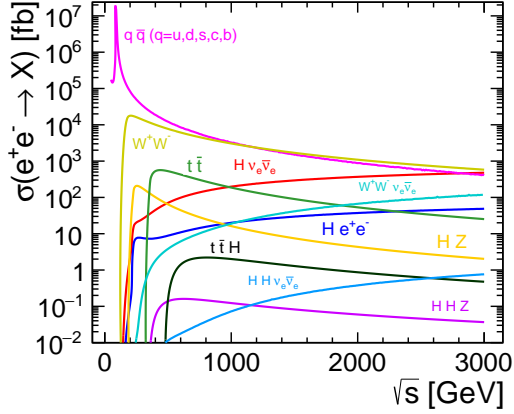

Figure 30: LO cross-section for important Standard Model processes in electron-positron collisions including single Higgs production. The process labeled HZ includes all decays of the Z boson. The effect of Initial-State Radiation is included.

Preliminary studies concerning $t\bar{t}h$ final states suggest [9, 12] that CLIC will be able to measure $y_t$ with a precision of about 3%, contrasted with the expected 5% precision [107] from the HL-LHC and the 1% precision expected at a future 100 TeV hadron collider [108].

In this section we explore a new approach for the determination of the top-quark Yukawa coupling, utilizing loop-induced Higgs production and decay processes. The advantage of such an approach is that it potentially allows for a precise determination of $y_t$ even at c.m. energies below the $t\bar{t}h$ threshold. Furthermore, even above the $t\bar{t}h$ threshold, measurements at different c.m. energies could be combined in order to derive more precise determination of $y_t$ than from $t\bar{t}h$ final states alone. Such indirect approaches are already being pursued in, for example, the determination of the Higgs self-interaction at the LHC [33, 34, 38, 109–111].

In what follows, we will assume that the only New Physics effects come from the operator $\mathcal{O}_{y_t}$, via[28]

$$y_t = y_t^{\text{SM}} + \Delta y_t \quad \text{with} \quad \Delta y_t = -\hat{c}_{y_t}. \tag{53}$$

We stress therefore that our analysis is not model independent but works only if the only modifications are of the form of Eq. (53). Such an assumption is however well-motivated, as discussed in Section 2.10.

The dominant single-Higgs production mechanisms are the $s$-channel Higgstrahlung process $e^+e^- \to hZ$ and the $t$-channel charged vector-boson fusion (VBF) process resulting in $h\nu\bar{\nu}$ final states. The relative importance of these two processes depends on the c.m. energy; the Higgstrahlung process dominates around 240-250 GeV while the VBF cross section takes over around 500 GeV. At even higher energies, the neutral VBF process $e^+e^- \to he^-e^+$ also becomes significant. One should keep in mind, however, that the separation of the various processes is not unambiguous once the $Z$ decays have been taken into account. In particular, Higgstrahlung with $Z$ decaying to neutrinos (electrons) yields the same final-state as the charged (neutral) vector boson fusion processes. In Figure 30 we show the centre-of-mass dependence of the inclusive cross section computed at leading order (LO) for the final states described above, and other states relevant for this analysis. The $h\nu\bar{\nu}$ and $he^+e^-$ channels include the Higgstrahlung contribution; the process labeled $hZ$ on the other hand includes all $Z$ decay modes.

As far as Higgs decays are concerned, the loop-induced processes $h \to gg$ and $h \to \gamma\gamma$ are both sensitive to $y_t$ and will be considered in the following. The Higgs decay to gluons is generated by massive quarks in the loops with the top quark being the dominant contribution. In the $m_t \to \infty$ limit this coupling is known with next-to-next-to-next to leading order (N$^3$LO) QCD accuracy [112]. In contrast, the Higgs decay to photons (as well as $h\gamma$ production) has a dominant contributions from loops involving gauge bosons which results in a reduced sensitivity to $y_t$ compared to $h \to gg$.

---

[28]In general, the top-quark Yukawa coupling is modified also by the operator $\mathcal{O}_H$ as $\Delta y_t = -\frac{\hat{c}_H}{2} - \hat{c}_{y_t}$.

Table 21: The estimated one-sigma uncertainties $\delta_{ij}$ used in Eq. (54) from Ref. [9], scaled to the current CLIC baseline. The $\sqrt{s} = 3$ TeV CLIC $t\bar{t}h$ result is derived by extrapolating the 1.4 TeV one with the corresponding number of events. The process labeled $hZ$ includes selected $Z$ decays, and their content is specific to each analysis referenced in this table.

| $\sqrt{s}$ (GeV) | $\mathcal{L}$ (ab$^{-1}$) | $h \to gg$ | | $h \to \gamma\gamma$ | | $h \to b\bar{b}$ |
|---|---|---|---|---|---|---|
| | | $hZ$ | $\nu\bar{\nu}h$ | $hZ$ | $\nu\bar{\nu}h$ | $t\bar{t}h$ |
| 350 | 1 | 4.3% | 7.2% | - | - | - |
| 1500 | 2.5 | - | 3.2% | - | 9.9% | 5.7% |
| 3000 | 5 | - | 2.2% | - | 4.9% | 7.9% |

In order to constrain $\Delta y_t$ we define a global $\chi^2$ for each run of the future colliders described in the previous section

$$\chi^2(\Delta y_t) = \sum_{i=1}^{N_p} \sum_{j=1}^{N_d} \frac{[\mu_{ij}(\Delta y_t) - 1]^2}{\delta_{ij}^2}, \qquad (54)$$

with $N_p$ and $N_d$ being, respectively, the number of available production and decay channels. The sums in Eq. (54) include only the processes for which $\delta_{ij}$ values are explicitly shown in Table 21. The one-sigma uncertainties $\delta_{ij}$ appearing in Eq. (54) are listed in Table 21. These are based on the information in Refs. [9, 37, 113, 114]. The values in these references have been derived in a realistic framework that accounts for acceptance cuts, background contributions and detector simulation for the reconstruction of the final state. The numbers in the previous references, however, correspond to the old CLIC luminosity setup. The uncertainties in Table 21 have been therefore conservatively scaled to match the updated CLIC baseline [12]. In Ref. [115] the loop-induced production of the Higgs in association with a photon was also considered as a possible handle to the top Yukawa coupling at lepton colliders. At CLIC, however, the very small number of $h\gamma$ events dilutes the importance of this process, and therefore we ignore it in this study.

The degrees of freedom of the $\chi^2$ in Eq. (54) are the signal-strengths $\mu_{ij}$ of all Higgs boson processes which are sensitive to a non-vanishing value of $\Delta y_t$ and which can be measured with a sufficient precision. The signal-strength $\mu_{ij}$ for a generic Higgs production mode $i$ and decay channel $j$ can be written in the narrow-width approximation as

$$\mu_{ij} = \left(\frac{\sigma_i}{\sigma_i^{\text{SM}}}\right) \left(\frac{\Gamma_j}{\Gamma_j^{\text{SM}}}\right) \left(\frac{\Gamma_h}{\Gamma_h^{\text{SM}}}\right)^{-1}. \qquad (55)$$

In Eq. (55) $\Gamma_h$ is the total Higgs width and $\sigma_i$ and $\Gamma_j$ are the corresponding production cross-section and partial decay width. Due to the small number of expected $t\bar{t}h$ and $h\gamma$ events, these two production channels are included with only the dominant $h \to b\bar{b}$ decay mode. The analytic expressions for the $t\bar{t}h$ signal-strengths, as functions of $\Delta y_t$, read

$$\mu_{t\bar{t}h}\begin{pmatrix}\sqrt{s} = 1500 \text{ GeV} \\ \sqrt{s} = 3000 \text{ GeV}\end{pmatrix} = \frac{\sigma_{t\bar{t}h}}{\sigma_{t\bar{t}h}^{\text{SM}}} = 1 + \begin{pmatrix}1.83 \\ 1.71\end{pmatrix} \Delta y_t. \qquad (56)$$

In the calculation of the above expressions we do not include corrections beyond LO. Such higher-order effects have been studied in Ref. [9] for the 1.4 TeV run of CLIC. These corrections result in a relatively small shift in the corresponding coefficient in Eq. (56) from 1.83 to 1.89. In turn, this slightly increases the $y_t$ precision in the $t\bar{t}h$ channel.

Table 22: 68% C.L. boundaries on $\Delta y_t$ for different runs and processes. In the last column we report the results of the global $\chi^2$ analysis described in the text.

| $\sqrt{s}$ (GeV) | $\mathcal{L}$ (ab$^{-1}$) | $h \to gg$ | $h \to \gamma\gamma$ | $t\bar{t}h$ | Total |
|---|---|---|---|---|---|
| 350 | 1 | 2.0% | - | - | 2.0% |
| 1500 | 2.5 | 1.7% | 13% | 3.7% | 1.6% |
| 3000 | 5 | 1.2% | 7.1% | 5.1% | 1.2% |

For the loop-induced Higgs decay processes $h \to gg$ and $h \to \gamma\gamma$ we get

$$\mu_{h \to gg} = \frac{\Gamma_{h \to gg}}{\Gamma^{\text{SM}}_{h \to gg}} = 1 + 2\Delta y_t \,, \tag{57}$$

$$\mu_{h \to \gamma\gamma} = \frac{\Gamma_{h \to \gamma\gamma}}{\Gamma^{\text{SM}}_{h \to \gamma\gamma}} = 1 - 0.56\Delta y_t \,. \tag{58}$$

All computations in this work have been carried out in the $G_\mu$ input scheme with the help of the `Madgraph5_aMC@NLO_v2.6.1` code [31]. Eqs. (56–58) have been derived in the following way: we first compute the corresponding cross-sections and decay widths for a number of different values of $\Delta y_t$ and then fit the resulting expressions for $\mu_{ij}$ with a parabola. Finally, we take its linear approximation for small values of $\Delta y_t$. In deriving $\mu_{h \to gg}$ the bottom quark contribution in the loop has been neglected.

Our main results, namely, the 68% C.L. constraints following from Eq. (54), are displayed in Table 22, from which we conclude that the decay process $h \to gg$ is a strong potential candidate for precise determination of $y_t$, while the potential of $h \to \gamma\gamma$ for a precise determination of $y_t$ is much smaller than $h \to gg$. While not directly competitive with $h \to gg$, this additional $y_t$ sensitivity is on a par with that expected at HL-LHC and may be useful for disentangling Wilson coefficients in a more refined Effective Field Theory (EFT) approach (see Section 2.9 for further details).

The results in Table 22 show that, assuming new physics effects modify *only* the top Yukawa coupling, the first stage of CLIC, running at 350/380 GeV, allows $y_t$ to be determined from purely loop-induced processes with a precision of about 2.0%. At higher CLIC energies the precision in the $y_t$ determination from loop-induced processes is significantly larger than the one expected from the standard $t\bar{t}h$-based approach. Our estimates show that by combining the extraction of $y_t$ from $t\bar{t}h$ with that from loop-induced final states one can reach $y_t$-precision of about 1.2–1.6% at both the $\sqrt{s} = 1.4$ TeV and $\sqrt{s} = 3.0$ TeV CLIC runs. This is 2-3 times better than the precision expected from purely $t\bar{t}h$ final states. Combining the information for all the different channels at all the CLIC energies a final precison on $y_t$ at the level of $0.9\%$ is within the CLIC reach. Finally, while these conclusions may be altered in a global analysis where all possible new physics effects are taken into account, the results illustrate the usefulness of the loop-induced processes in providing an extra handle to $y_t$ at CLIC (see also Ref. [116]). The results in this section have been extrapolated from the original results presented in Ref. [115], and we refer to that publication for more details.

## 2.9 Global effective field theory fit[29]

The focus of this chapter has been so far on particular processes, or particular effective operators. In this section we gather all this information to provide a global perspective on the capabilities of CLIC indirect searches. From a BSM perspective, two broad scenarios emerge as particularly interesting: universal [14, 62], and top-philic new physics.

---
[29]*Based on a contribution by J. Gu.*

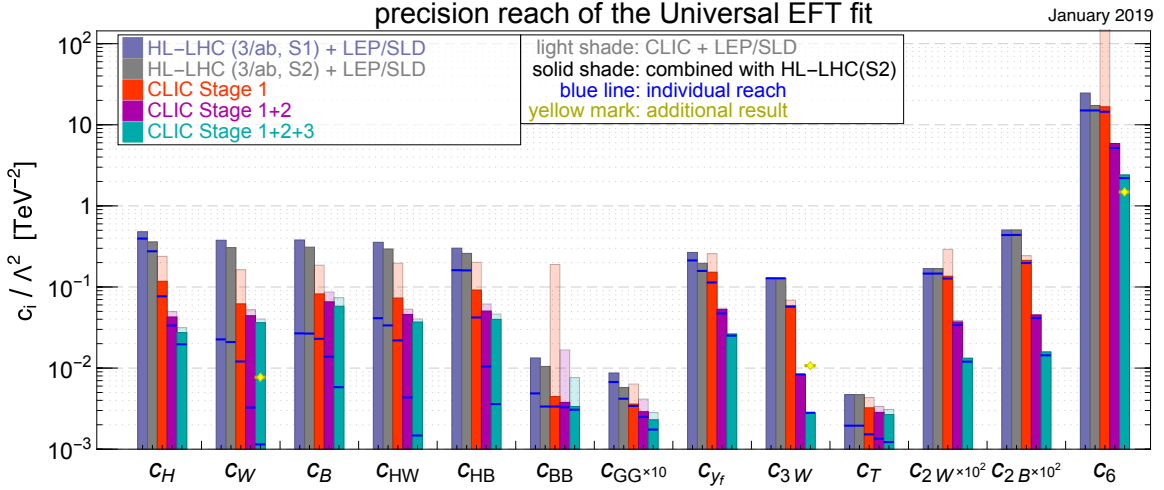

Figure 31: Global fit to the universal operators of Eq. (59). Coloured bands: global fit; blue marks: one-by-one fit; yellow marks: analysis of Section 2.2.2 for $c_6$ and Section 2.4 for $c_{3W}$ and $c_W$. HL-LHC results from Ref [21].

In this section, we will focus on these two scenarios. Of course there is in principle more information (more operators) that can be accessed at CLIC. For instance, we have already seen in Section 2.1, Table 5, that CLIC has an impressive reach on operators modifying couplings of Higgs and the individual fermions $\psi = t, b, c, \tau, \mu$, while in the universal scenario only an overall modification associated with $\mathcal{O}_y$ is considered, and in the top-philic this is complemented with $\mathcal{O}_{y_t}$. Hence the global perspective developed in this section should be thought as a broad and well-motivated BSM interpretation, rather than an exhaustive summary of CLIC capabilities. Yet, the EFT context is also useful as a rather model-independent tool to compare the sensitivity of widely different experiments. For this reason we focus in Section 2.9.1 on operators that allow for a comparison with $Z$-pole observables, as studied at LEP I and envisaged at FCCee and CEPC. Moreover, Section 3 is dedicated to the discussion of a richer BSM flavour structure.

**Universal global fit**

In universal theories, the BSM dynamics only contribute to effective interactions among Higgs and gauge bosons; the effective operators they induce are summarized in the second panel of Table 2. In particular, a non redundant set, obtained via the relations Eqs. (2-4), is given in the SILH basis by

$$\{\mathcal{O}_T, \mathcal{O}_W, \mathcal{O}_B, \mathcal{O}_{HW}, \mathcal{O}_{HB}, \mathcal{O}_{3W}, \mathcal{O}_{2B}, \mathcal{O}_{2W}, \mathcal{O}_H, \mathcal{O}_6, \mathcal{O}_{BB}, \mathcal{O}_{GG}, \mathcal{O}_y\}. \tag{59}$$

In Figure 31 we report the results of a global fit with these operators. We include separate results for the different CLIC stages, and compare with a similar analysis performed in the context of the HL-LHC program [21], supplemented with LEP data [117]. Coloured histograms denote the results of a global fit. The global fit is performed using the forecasts discussed in the previous sections of this document, but employing optimal observables for $WW$ and $ZH$ processes (where a 50% acceptance has been included in the analysis), as discussed in Ref. [37] (see also Section 2.7). In this sense it should be thought as an optimistic scenario in which the experiment capabilities are optimally exploited. In order to asses the extent to which this can be considered realistic, we denote with yellow marks the results of individual analyses discussed in previous sections. In particular, the mark for $c_6$ corresponds to the full detector simulation analysis of Section 2.2.2, while the $c_{3W}$ mark corresponds to Section 2.4. Blue marks correspond instead to an analysis where the operators have been switched on one at a time. Clearly in some cases the discrepancies with the global analysis are striking, implying large correlations between the various measurements; as confirmed in Figure 32.

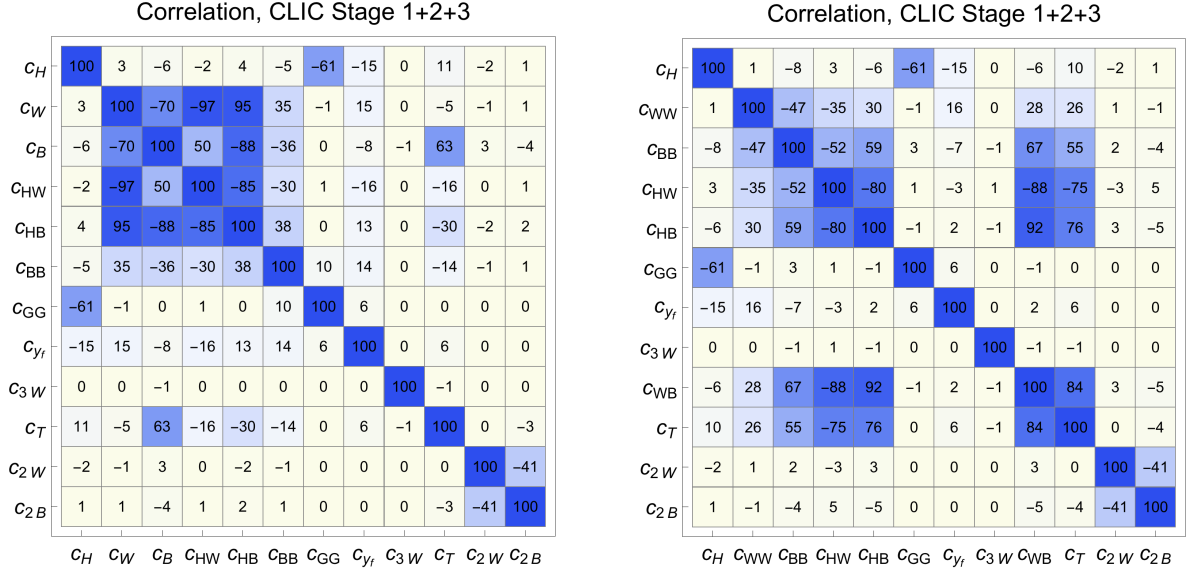

Figure 32: Left (right) pane: correlation matrices associated with the global fit of Figures 31 and 33

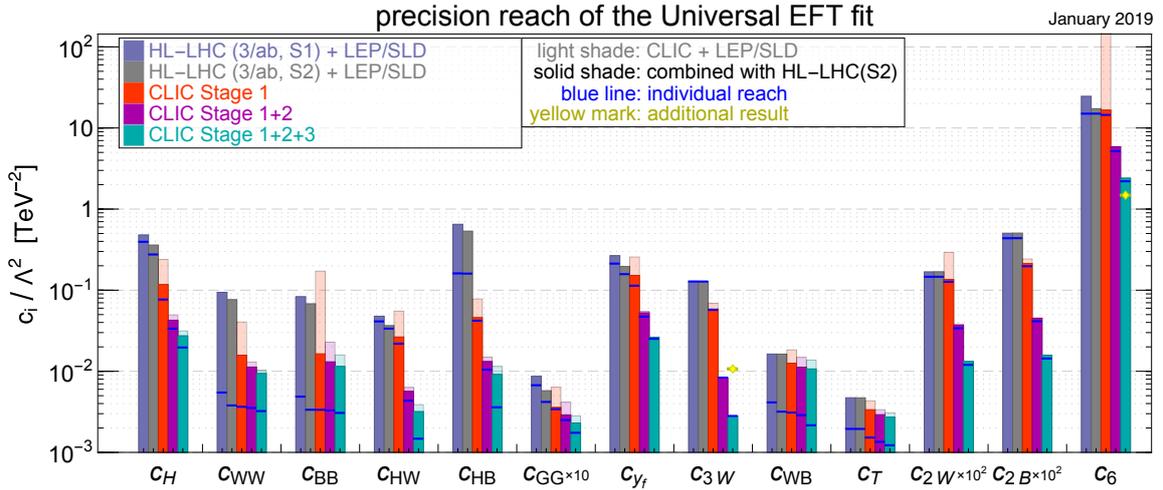

Figure 33: Same as Figure 31, but for the operators in Eq. (60).

We can use the freedom of changing operator basis to something closer to the Warsaw basis [22],

$$\{\mathcal{O}_T, \mathcal{O}_{WB}, \mathcal{O}_{HW}, \mathcal{O}_{HB}, \mathcal{O}_{3W}, \mathcal{O}_{2B}, \mathcal{O}_{2W}, \mathcal{O}_H, \mathcal{O}_6, \mathcal{O}_{BB} \pm \mathcal{O}_{WW}, \mathcal{O}_{GG}, \mathcal{O}_y\}, \quad (60)$$

The results are reported in Figure 33, and the correlation matrix, now almost diagonal, in the right panel of Figure 32. The large correlations can be traced to the choice of operators, that does not align well with measurements at different precision [20, 49, 118].

**Top-philic global fit**

In top-philic scenarios the new physics also couples to the top-quark. To be precise, we are still assuming universal couplings, but allow for *non-universal* couplings in operators where the third family of quarks is involved. This is discussed in the top-dedicated CLIC physics report [10]; the relevant operators, in addition to those of Eq. (59), are

$$\{\mathcal{O}_{\varphi q}^-, \mathcal{O}_{\varphi t}, \mathcal{O}_{tW}, \mathcal{O}_{tB}, \mathcal{O}_{lq,B}, \mathcal{O}_{lq,W}, \mathcal{O}_{lt,B}\}, \tag{61}$$

where we use the notation of Table 20. In this context, the best constraints on the universal operators from Eq. (59) are still obtained from measurements of processes not involving the top quark, so that we can treat them as an independent block and focus here on Eq. (61). Most of these can be constrained by $e^+e^- \to t\bar{t}$ at different energies, complemented with specific processes, such as $WW \to t\bar{t}$ and $e^+e^- \to t\bar{t}h$ at high energy. The results are summarized in Figure 29.

### 2.9.1 Comparison with Z-pole observables

CLIC is inherently a high-energy machine targetting effects that are enhanced at high-energy. This contrasts with the operational mode of circular colliders, which exploit a greater sensitivity at low-energy, in particular on the $Z$-pole. In the context of a global fit, these offer genuinely complementary information. We show this in Figure 34, where we focus on two (universal) effects, $\mathcal{O}_W$ and $\mathcal{O}_B$; their sum is equivalent to the famous $S$ parameter [62], while their difference can only be accessed at high energy colliders. In the left panel we show a global fit, in which also all other operators of Eq. (59) are included in the fit: clearly the fit would benefit from precise measurements on the $Z$-pole. The right panel shows instead a situation where all other operators are set to zero. This can be justified from a BSM perspective because $\mathcal{O}_W$ and $\mathcal{O}_B$ can be generated by the tree-level exchange of resonances, while the other operators relevant for the fit are always generated at loop level. In this limit, we see that the CLIC measurement completely outperforms LEP, and is equivalent in precision with a $\sim O(\text{few} \times 10^{-5})$ precision on the $Z$-pole.

In the non universal context, there can be more contributions to the $Z$-pole observables, one for each coupling of the $Z$-boson to SM fermions. These operators also induce energy-growth in other processes. In particular, modifications of the $Z\bar{e}_R e_R$ and $Z\bar{e}_L e_L$ couplings induce energy growth in $e_{L,R} e_{L,R} \to W^+W^-, ZH$, and they can even be disentangled with runs at different polarizations, as envisaged in the CLIC baseline scenario. This information can be complemented with high-energy data from the HL-LHC [47], to provide a global high-energy perspective on non-universal $Z$-couplings.

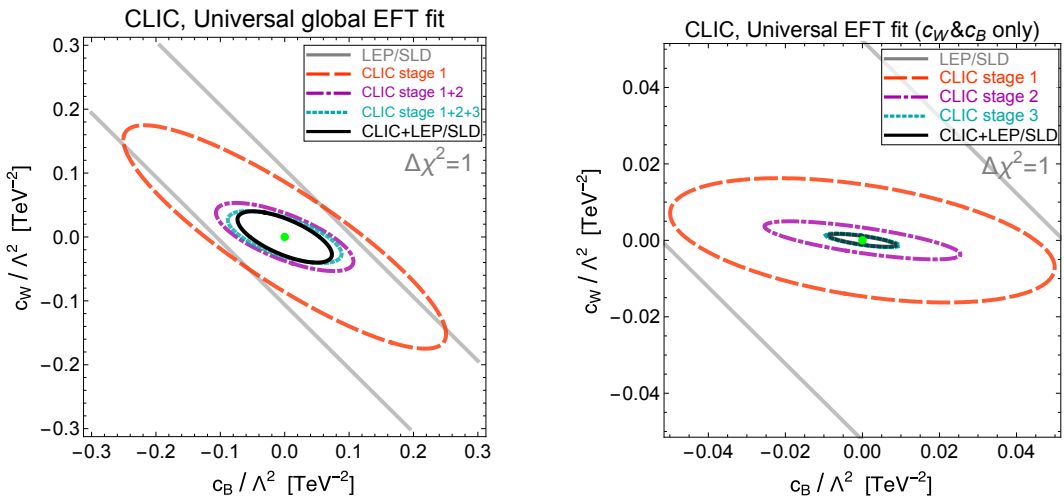

Figure 34: Focus on the operators $\mathcal{O}_W$ and $\mathcal{O}_B$ and comparison with $Z$-pole measurements. Marginalization with respect to all other operators that enter CLIC processes is shown in the left panel, while in the right panel only $\mathcal{O}_W$ and $\mathcal{O}_B$ are included in the fit.

## 2.10 BSM interpretation[30]

We interpret the CLIC sensitivities presented in previous sections in a concrete class of BSM scenarios, namely composite Higgs (CH) models where the Higgs is realized as a pseudo-Nambu-Goldstone boson of an approximate symmetry broken at the confinement scale of a new strong interaction [119] (see Ref. [16] for a recent review). An important feature making this type of model a good example for illustrating the reach of CLIC on new physics (NP) is the potentially large deviations in a wide variety of observables [120]. This remains true even when the mass scale of new states $m_*$ is out of direct reach. For the purpose of this analysis, we assume it is the case at both CLIC and the LHC.

We will use an EFT approach to CH models, faithfully reflecting their most important features while limiting model dependence. This approach differs from the more generic EFT used in previous sections in that we apply a theoretical bias to the coefficients of the effective dimension-six operators. These coefficients now become functions of the typical composite-sector mass $m_*$ and coupling $g_*$ [17], as expected for the Wilson coefficients of the operators which are obtained by integrating out the composite sector characterized by $g_*$ and $m_*$. For the strong sector coupling, one generically expects $1 < g_* \leq 4\pi$, and we assume $m_*$ to lie above 3 TeV, the highest centre-of-mass energy envisioned for CLIC. In the following two sections, we will consider separately flavour-independent (universal) and flavour-dependent NP effects. While the first ones can *a priori* be probed in various fermionic and bosonic final states, the latter ones are mostly restricted to the processes of the top and bottom quark production. Notice, that both universal and non-universal effects can contribute to the same process, however we will find that the strongest constraints on each type of effects come from different measurements. In both cases, the major role for the CLIC searches will often be played by the operators whose interference with SM amplitudes grows with energy and thus benefit from high-energy collisions.

The scaling of the operators with the strong sector parameters is defined by (see e.g. [121])

$$\mathcal{L}^6 = \kappa_i \frac{m_*^4}{g_*^2} \hat{\mathcal{L}}_i^6 \left( \epsilon_\psi \frac{g_* \psi}{m_*^{3/2}}, \frac{g_* H}{m_*}, \frac{\partial_\mu}{m_*}, \frac{g_X X_\mu}{m_*} \right) \tag{62}$$

where $\hat{\mathcal{L}}_i^6$ is a polynomial of its arguments, $\psi$ are fermionic SM fields, $X_\mu$ are SM gauge fields with a gauge coupling $g_X$. $\kappa_i$ are dimensionless coefficients which are expected to be of order one, unless certain symmetry or selection rule suppresses them. The $\epsilon_\psi \leq 1$ mixing parameters measure the degree of compositeness of SM fermions. The latter are expected to be a mixture of the elementary and composite states, with Yukawa interactions given by $y_\psi \simeq \epsilon_{\psi_L} \epsilon_{\psi_R} g_*$.

**Universal effects**

Let us first consider the low-energy effects of the strong sector —called universal in the following— which do not vanish in the limit of absent direct couplings between SM fermions and composite resonances. In the SILH basis [17], they are described by the following operators (omitting CP-violating effects and purely gluonic operators):

$$\begin{aligned}
\mathcal{L}_{universal}^{d=6} &= c_H \frac{g_*^2}{m_*^2} \mathcal{O}_H + c_T \frac{N_c \epsilon_q^4 g_*^4}{(4\pi)^2 m_*^2} \mathcal{O}_T + c_6 \lambda \frac{g_*^2}{m_*^2} \mathcal{O}_6 + \frac{1}{m_*^2} \left[ c_W \mathcal{O}_W + c_B \mathcal{O}_B \right] \\
&+ \frac{g_*^2}{(4\pi)^2 m_*^2} \left[ c_{HW} \mathcal{O}_{HW} + c_{HB} \mathcal{O}_{HB} \right] + \frac{y_t^2}{(4\pi)^2 m_*^2} \left[ c_{BB} \mathcal{O}_{BB} + c_{GG} \mathcal{O}_{GG} \right] \\
&+ \frac{1}{g_*^2 m_*^2} \left[ c_{2W} g^2 \mathcal{O}_{2W} + c_{2B} g'^2 \mathcal{O}_{2B} \right] + c_{3W} \frac{3! g^2}{(4\pi)^2 m_*^2} \mathcal{O}_{3W} \\
&+ c_{y_t} \frac{g_*^2}{m_*^2} \mathcal{O}_{y_t} + c_{y_b} \frac{g_*^2}{m_*^2} \mathcal{O}_{y_b}
\end{aligned} \tag{63}$$

where $\epsilon_q$ stands for the degree of compositeness of the third-generation quark doublet, $\lambda$ is the SM Higgs quartic coupling and $N_c = 3$ is the number of colours. The $c$-coefficients are expected to be of order one.

---

[30] Based on a contribution by O. Matsedonskyi and G. Durieux.

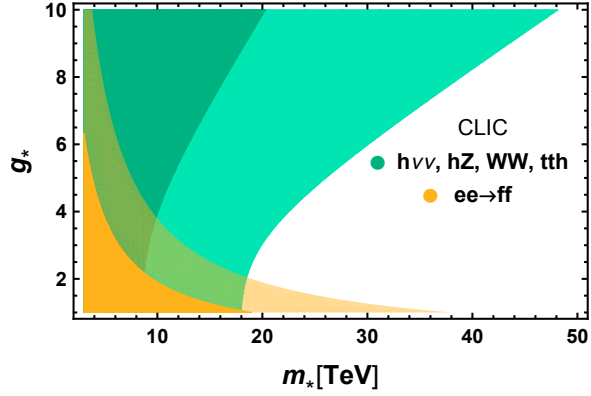

Figure 35: Sensitivity to CH parameter space, in terms of a typical strong sector coupling $g_*$ and a mass $m_*$, using the combined fit from $h\nu\nu$, $hZ$, $WW$ and $tth$ production (in green) and Drell-Yan processes (in orange) at $5\sigma$. The full CLIC programme with polarized beams is considered. For Drell-Yan production the scenario featuring 0.3% of systematic uncertainties is assumed. Two shades of color filling correspond to the strongest and the weakest sensitivities obtained for a factor of 2 increase and decrease of the operator coefficients.

The set (63) contains 12 bosonic operators which is 2 less than the minimal universal set defined in Ref. [14] (neglecting again two purely gluonic operators).

The $\mathcal{O}_W, \mathcal{O}_B, \mathcal{O}_{2W}, \mathcal{O}_{2B}, \mathcal{O}_T$ operators contribute to Drell-Yan production discussed in Section 2.6, as well as to the $t\bar{t}$ production of Section 2.7. The latter however receives larger non-universal contributions, which we discuss next. $\mathcal{O}_T$ and a combination of $\mathcal{O}_W$ and $\mathcal{O}_B$ are already strongly constrained by the LEP data.

The Higgs self-coupling measurements of Section 2.2.1 are a unique probe of $\mathcal{O}_6$, while the other operators contributing to this process are much better probed in other channels. The expected sensitivity is, however, not sufficient to test the typically expected order-one values of $c_6$, given that $m_*/g_*$ is already constrained to be at or above about 800 GeV [122].

Higgs and vector boson production analysed in Sections 2.1, 2.4 and 2.3 are affected by $\mathcal{O}_W, \mathcal{O}_B, \mathcal{O}_{HW}, \mathcal{O}_{HB}, \mathcal{O}_{3W}, \mathcal{O}_{GG}, \mathcal{O}_{BB}$ and $\mathcal{O}_H$. Here one should emphasize that in CH models the dominant contribution to the modification of $hgg$ and $h\gamma\gamma$ interactions comes not from $\mathcal{O}_{GG}$ and $\mathcal{O}_{BB}$, but from $\mathcal{O}_H$ and a non-universal operator $\mathcal{O}_{y_t}$.

Using the projected sensitivities presented in the listed sections, we derive the sensitivities to the strong sector parameters $g_*$ and $m_*$ from the most relevant channels. The results are displayed in Figure 35. The sensitivity of the combined fit to the Higgs and diboson data is dominated by $c_H$, $c_{y_t}$ and $c_{y_b}$ at high $g_*$, and by $c_{W,B}$ at low $g_*$. For each category of measurement, regions probed in pessimistic and optimistic cases are respectively indicated in dark and light colour shades. To derive them we independently vary, in the $[-2, -1/2] \cup [1/2, 2]$ range, the numerical factors up to which the power counting for each operator is satisfied. In the pessimistic case, a point in the $(m_*, g_*)$ plane is considered to be within reach only if it is expected to be probed for any choice of numerical factor within the specified range. In the optimistic case instead, we require the point to be probed for at least one choice of parameters within that range. This procedure aims at covering various possible CH model realizations.

**Top compositeness effects**

The dominant non-universal effects of the strong sector are expected to arise from the sizeable mixings of the top-quark with composite states, required to generate its Yukawa coupling. The latter is given by

$$y_t \simeq \epsilon_q \epsilon_t g_* \qquad (64)$$

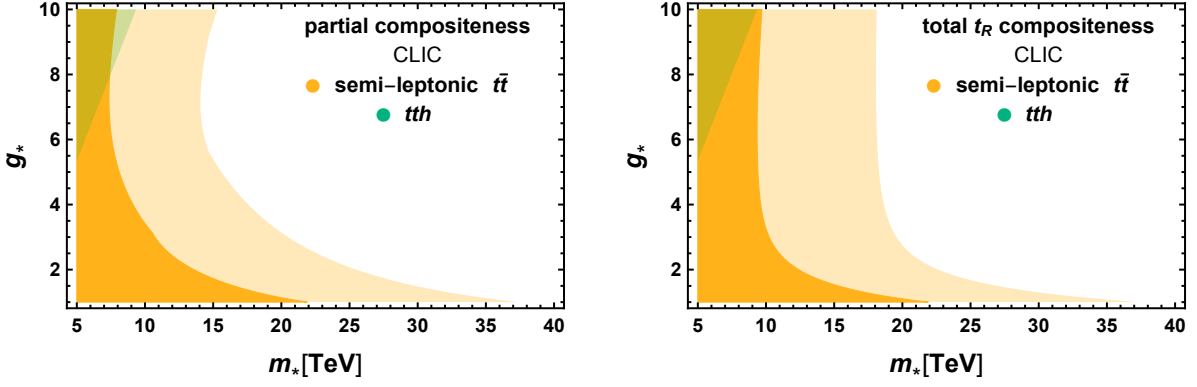

Figure 36: Sensitivity to CH parameter space from the non-universal operators, at $5\sigma$, for equally composite $t_L$ and $t_R$ (left panel) and totally composite $t_R$ (right panel). Orange regions are probed in $t\bar{t}$ production and green in $tth$. Two shades of orange correspond to the strongest and the weakest reach obtained upon varying the operator coefficients from a half to twice the estimates given in Eq. (65).

where $q$ and $t$ in the following refer to the SM third-generation left-handed quark doublet and right-handed singlet, respectively. We consider two representative scenarios: featuring an equal degree of compositeness for both chiralities, $\epsilon_q = \epsilon_t = (y_t/g_*)^{1/2}$, and a totally composite top right [123], $\epsilon_t = 1, \epsilon_q = y_t/g_*$. For a consistent treatment of top-quark compositeness effects, we write down all possible operators involving top quarks and bosons. This redundant set can then be projected onto a basis of independent operators. For operators relevant at CLIC, one obtains the following scaling estimates [10, 124]:

$$\begin{aligned}
\mathcal{L}_{top}^{d=6} &= c_{Hq} \frac{\epsilon_q^2 g_*^2}{m_*^2} \left[ (H^\dagger i \overset{\leftrightarrow}{D}_\mu H)(\bar{q}\gamma^\mu q) - (H^\dagger i \overset{\leftrightarrow}{D}_\mu^I H)(\bar{q}\gamma^\mu \tau^I q) \right] + c_{Ht} \frac{y_t^2}{m_*^2} (H^\dagger i \overset{\leftrightarrow}{D}_\mu H)(\bar{t}\gamma^\mu t) \\
&+ \frac{y_t g_*^2}{(4\pi)^2 m_*^2} \left[ c_{tW} g (\bar{q}\sigma^{\mu\nu} \tau^I u) \tilde{H} W_{\mu\nu}^I + c_{tB} g' (\bar{q}\sigma^{\mu\nu} t) \tilde{H} B_{\mu\nu} \right] \\
&+ \frac{\epsilon_q^2}{m_*^2} \left[ c_{lq}^{(1)} g'^2 (\bar{l}\gamma^\mu l)(\bar{q}_k \gamma^\mu q_l) + c_{lq}^{(3)} g^2 (\bar{l}\gamma^\mu \tau^I l)(\bar{q}\gamma^\mu \tau^I q) + c_{eq} g'^2 (\bar{e}\gamma^\mu e)(\bar{q}\gamma^\mu q) \right] \\
&+ \frac{\epsilon_t^2}{m_*^2} \left[ c_{lt} g'^2 (\bar{l}\gamma^\mu l)(\bar{u}\gamma^\mu u) + c_{et} g'^2 (\bar{e}\gamma^\mu e)(\bar{u}\gamma^\mu u) \right] \\
&+ c_{y_t} \frac{y_t g_*^2}{m_*^2} |H|^2 \bar{q} \tilde{H} t,
\end{aligned} \quad (65)$$

where $l$ and $e$ stand for the first generation SM leptons. The $c$ coefficients are of order one. The contributions to the operator coefficients, proportional to the $\epsilon_{q,t}$ mixings, are always dominant compared to flavour-universal ones, which we have therefore neglected. Notice that processes involving the left-handed bottom quark which appears, together with the top, in the left-handed doublet $q_L$ are also affected by the large top mixings. The operators of Eq. (65) contribute to $\bar{t}t$ and $\bar{b}b$ production, $tth$ coupling modification, and consequently to the $hgg$ and $h\gamma\gamma$ effective couplings. The latter effect on $hgg$ and $h\gamma\gamma$ was already analysed in the previous section and hence we do not discuss it again. With the future experimental sensitivities discussed in Section 2.7, we obtain the very reach of CLIC in terms of $g_*$ and $m_*$, displayed in Figure 36 for the two top compositeness scenarios. As in Figure 35, pessimistic and optimistic probed regions respectively displayed in dark and light colour shades are obtained by varying the numerical factor up to which the power counting for each operator is satisfied in the $[-2, -1/2] \cup [1/2, 2]$ interval.

As is clear from the comparison of Figures 35 and 36, fermion pair production measurements are complementary to that of Higgs and diboson processes. Overall, the strongest sensitivity at high $g_*$ arises from the $\mathcal{O}_H$ operator and extends to $m_* \simeq 48$ TeV, while at low $g_*$, a reach on $m_*$ of up to 35 TeV is driven by the $W$ parameter and non-universal four-fermion operators.

**Summary**

Probing the composite nature of the Higgs and possibly the top quark is a major goal for future collider projects. We see in Figure 35 that CLIC will for sure discover Higgs compositeness if the compositeness scale is below 8 TeV. For large $g_* \simeq 8$ the discovery is possible up to scales of around 40 TeV, in particularly favorable situations. Top compositeness effects may also be discovered, as shown in Figure 36. For comparison, we show in Figure 37 the estimated projected exclusion reach at the HL-LHC. A $g_*$-independent exclusion is possible up to around 3 TeV, and it never exceeds 9 TeV.

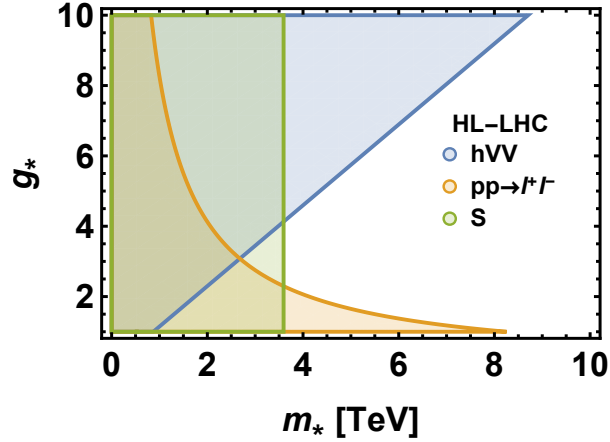

Figure 37: 95% C.L. sensitivity to CH models at the HL-LHC via: 1) modifications on single Higgs couplings ($hVV$); 2) Drell-Yan ($W$ parameter); 3) electroweak precision constraints from LEP/SLD ($S$ parameter). All the effects have been estimated by Eq. (63) for $c$-coefficients equal to one. The projected HL-LHC reach is taken from Refs. [21, 65, 70]. The reach of direct resonant searches is expected to be comparable or inferior [125].

## 3 Flavour physics

In flavour physics there are two outstanding puzzles: (i) why the quark and lepton masses are so hierarchical and have such a peculiar pattern of mixing angles (also called the SM flavour puzzle), and (ii) if there is new physics (NP) at the TeV scale, what is its flavour structure, so that it avoids stringent indirect constraints (the so call NP flavour problem). CLIC offers several unique probes to address the two puzzles. For TeV scale NP that is produced in $e^+e^-$ collisions, one can gain further insight into the origin of flavour by measuring precisely the Higgs and top couplings, as well as by searching for extended Higgs sectors. One can also search for higher NP states through their off-shell contributions to flavour violating processes. In the remainder of this section we detail the opportunities at CLIC for each of these probes.

### 3.1 Direct probes of FCNC point interactions

Heavy new physics can induce, through the exchange of virtual particles, processes that are extremely rare in the SM, such as FCNC in the leptonic sector, or involving the third quark family. If new physics is heavy, above the collider energy, these processes can be described by local operators in an effective field theory. In this regime, the rate for the FCNC processes grows with the energy, and thus the highest energy colliders have the largest lever arm to probe these effects. The high available energy, combined with the low level of background, places CLIC in the ideal position to perform this kind of studies.

#### 3.1.1 Leptonic FCNC operators[31]

In Ref. [126], Lepton Flavour Violating (LFV) operators of the schematic form $(\bar{e}e)(\bar{e}\tau)$ were studied at possible future high energy $e^+e^-$ or $e^-e^-$ colliders (see [127] for an earlier study). It was found that such machines are competitive with searches for rare decays at the low energy precision frontier, such as $\tau \to eee$.

The complete set of dimension 6 four-fermion operators involving three electrons and a tau is [126]

$$\mathcal{L}_{\text{eff}} = V_{LL} \left(\bar{e}\gamma^\mu P_L e\right) \left(\bar{\tau}\gamma_\mu P_L e\right) + V_{RR} \left(\bar{e}\gamma^\mu P_R e\right) \left(\bar{\tau}\gamma_\mu P_R e\right) \\ + V_{LR} \left(\bar{e}\gamma^\mu P_L e\right) \left(\bar{\tau}\gamma_\mu P_R e\right) + V_{RL} \left(\bar{e}\gamma^\mu P_R e\right) \left(\bar{\tau}\gamma_\mu P_L e\right) + \text{h.c.}, \tag{66}$$

where the $V_{ij}$ $(i,j = L, R)$ are complex coefficients of mass dimension $-2$ that parameterize the strength of each interaction, representing the heavy fields that have been integrated out. The current best limits come from Belle searches for $\tau \to eee$ [128] which place limits on the combination of couplings

$$\left(|V_{LR}|^2 + |V_{RL}|^2 + 2|V_{RR}|^2 + 2|V_{LL}|^2\right) \leq 1.63 \times 10^{-16} \text{ GeV}^{-4} = \frac{1}{(8.85 \text{ TeV})^4}. \tag{67}$$

Belle 2, upgraded LHCb and high luminosity ATLAS and CMS could improve on these bounds by one to two orders of magnitude [129–131], resulting in bounds on $(|V_{LR}|^2 + |V_{RL}|^2 + 2|V_{RR}|^2 + 2|V_{LL}|^2)^{-1/4}$ of 20.1 to 35.8 TeV.

At a high energy $e^+e^-$ or $e^-e^-$ lepton collider the $(\bar{e}e)(\bar{e}\tau)$ interaction results in signal events containing an electron and a tau lepton produced back-to-back with energies very close to $\sqrt{s}/2$. While the tau decay results in missing momentum, the ability to uniquely reconstruct the centre-of-mass frame allows the signal to be extracted from the otherwise overwhelming background. The tau decay produces at least one charged particle (typically $e^\pm$, $\mu^\pm$, or $\pi^\pm$), one or more neutrinos, and in some cases, neutral hadrons. The backgrounds from fakes are expected to be negligible, while there are two reducible backgrounds involving real taus and electrons. The electroweak process $e^+e^- \to \tau\nu_\tau e\nu_e$ is the dominant background for $\sqrt{s} \gtrsim 500$ GeV. In addition, the process $e^+e^- \to \tau^+\tau^-$ produces electrons from the decay $\tau \to e\nu_e\nu_\tau$. Such electrons only rarely produce an electron whose energy is $\simeq \sqrt{s}/2$.

---

[31]*Based on a contribution by T. Tait.*

Table 23: Expected number of events for the $e\tau$ signal (assuming $V_{LL} = 1/(10 \text{ TeV})^2$ and $V_{RR} = V_{RL} = V_{LR} = 0$) and background processes, assuming 1 ab$^{-1}$ of data collected at an $e^+e^-$ collider running at $\sqrt{s} = 250, 500, 1000$, and 3000 GeV, before and after the cuts described in the text. Also indicated is $1/\sqrt{V_{95}}$, the expected 95% C.L. limit on $1/\sqrt{V_{LL}}$, assuming no signal is observed.

| Process | 250 GeV | 500 GeV | 1 TeV | 3 TeV |
|---|---|---|---|---|
| | | Before Cuts | | |
| $e\tau$ Signal | 112 | 450 | 1800 | $1.6 \times 10^4$ |
| $e\tau\nu_e\nu_\tau$ | $4.6 \times 10^5$ | $5 \times 10^5$ | $6.6 \times 10^5$ | $1.2 \times 10^6$ |
| $\tau\tau$ | $6.3 \times 10^5$ | $1.5 \times 10^5$ | $3.7 \times 10^4$ | 4200 |
| | | After Cuts | | |
| $e\tau$ Signal | 101 | 405 | 1620 | $1.5 \times 10^4$ |
| $e\tau\nu_e\nu_\tau$ | 9300 | $10^4$ | 5900 | 2480 |
| $\tau\tau$ | 6590 | 1600 | 390 | 44 |
| $1/\sqrt{V_{95}}$ | 8.0 TeV | 11.7 TeV | 18.0 TeV | 34.9 TeV |

Table 23 shows the expected number of signal and background events for a collected data set of 1 ab$^{-1}$, and collider energies $\sqrt{s} = 250, 500, 1000$, and 3000 GeV (for signal we use a reference value consistent with the limits from tau decays, $V_{LL} = 1/(10 \text{ TeV})^2$ and $V_{RR} = V_{RL} = V_{LR} = 0$). The signal to background ratio can be improved by performing the cut on the energy $E_e$ of the most energetic electron in the event, $E_e \geq (1-r)\sqrt{s}/2$, and on the reconstructed centre-of-mass energy of the event, $\sqrt{\bar{s}} \geq (1-r)\sqrt{s}$, where $\bar{s} = (p_e + p_\tau)^2$ with $p_\tau$ reconstructed under signal assumption that all missing momentum is due to the neutrinos from the tau decay, so that $\vec{p}_\tau = \vec{p}_{\text{vis}} + \not{\vec{p}}$. The cut on $\bar{s}$ can be understood to be equivalent to requirement that the reconstructed tau and electron are approximately back-to-back, and is defined to be robust under the presence of visible final state radiation photons.

The resulting number of signal events for $r = 0.1$ are shown in Table 23. Even such rather modest choice for $r$, well below the expected energy resolution for isolated electrons, is sufficient to reduce both backgrounds by $\mathcal{O}(10^2)$. After cuts, the signal to background ratio is 1:10 at the lowest considered energies, and more like 100:1 at the highest. Also shown in Table 23 are the projected 95% C.L. limits on $1/\sqrt{V_{LL}}$ for each collider energy, assuming statistical errors dominate the background estimation and that no excess is observed. At $\sqrt{s} = 250$ GeV, the projected limits are comparable but slightly worse than those currently available from tau decays. These limits steadily improve with collider energy, reaching $\sim 35$ TeV for $\sqrt{s} = 3$ TeV.

Tightening the cut parameter to $r = 0.05$, greater background rejection is possible, though at the cost of larger signal losses ($\sim 30\%$) due to initial state radiation. The net result is a modest gain in sensitivity up to $V_{LL}^{-1/2} \geq \{9.3, 14.2, 20.3, 39.5\}$ TeV for $\sqrt{s} = \{250, 500, 1000, 3000\}$ GeV. Further improvements could be obtained by polarizing the incoming beams, which could reduce the $e\tau\nu_e\nu_\tau$ background, though the cost to the signal would depend on which of the $V_{ij}$ are dominant. This feature might be best exploited to disentangle the chiralities involved once a signal is observed.

At a high energy $e^-e^-$ collider, the signal consists of $e^-e^- \to e^-\tau^-$. The dominant background is $e^-e^- \to e^-\nu_e\tau^-\bar{\nu}_\tau$, and there is no analogue to the $\tau\tau$ background. The same strategy employed above for $e^+e^-$ collisions should be effective at extracting the signal from this background. Given the smaller backgrounds, and assuming 1 ab$^{-1}$ of collected data and $r = 0.1$, we find projected limits $V_{LL}^{-1/2} \geq \{18.6, 21.9, 27.0, 42.2\}$ TeV for $\sqrt{s} = \{250, 500, 1000, 3000\}$ GeV. Even at $\sqrt{s} = 250$ GeV, this is a substantial improvement over the existing constraints from tau decays.

### 3.1.2 Top-quark FCNC operators[32]

In this section we examine top-quark FCNC production, $e^+e^- \to tj$, using effective field theory and derive global constraints on FCNC couplings of the top for different CLIC stages (for TESLA and FCC-ee studies see Refs. [132, 133]). We use statistically optimal observables [91, 92] and include four-fermion operators which were previously overlooked. More details are presented in Ref. [134].

#### 3.1.2.1 Effective field theory

For the dimension-six top-quark EFT we use the conventions of Ref. [135], which are based on the Warsaw basis [22], but with Higgs denoted by $H$ instead of by $\varphi$. The operators relevant at tree-level for $e^+e^- \to tj$ can contain either two quarks and bosons, or two quarks and two leptons. The full list of such operators is

$$
\begin{aligned}
O_{u\varphi}^{(ij)} &= \bar{q}_i u_j \tilde{H} (H^\dagger H), & O_{lq}^{1(ijkl)} &= (\bar{l}_i \gamma^\mu l_j)(\bar{q}_k \gamma^\mu q_l), \\
O_{\varphi q}^{1(ij)} &= (H^\dagger i\overleftrightarrow{D}_\mu H)(\bar{q}_i \gamma^\mu q_j), & O_{lq}^{3(ijkl)} &= (\bar{l}_i \gamma^\mu \tau^I l_j)(\bar{q}_k \gamma^\mu \tau^I q_l), \\
O_{\varphi q}^{3(ij)} &= (H^\dagger i\overleftrightarrow{D}_\mu^I H)(\bar{q}_i \gamma^\mu \tau^I q_j), & O_{lu}^{(ijkl)} &= (\bar{l}_i \gamma^\mu l_j)(\bar{u}_k \gamma^\mu u_l), \\
O_{\varphi u}^{(ij)} &= (H^\dagger i\overleftrightarrow{D}_\mu H)(\bar{u}_i \gamma^\mu u_j), & O_{eq}^{(ijkl)} &= (\bar{e}_i \gamma^\mu e_j)(\bar{q}_k \gamma^\mu q_l), \\
O_{\varphi ud}^{(ij)} &= (\tilde{H}^\dagger i D_\mu H)(\bar{u}_i \gamma^\mu d_j), & O_{eu}^{(ijkl)} &= (\bar{e}_i \gamma^\mu e_j), \\
O_{uW}^{(ij)} &= (\bar{q}_i \sigma^{\mu\nu} \tau^I u_j) \tilde{H} W_{\mu\nu}^I & O_{lequ}^{1(ijkl)} &= (\bar{l}_i e_j)\, \varepsilon\, (\bar{q}_k u_l), \\
O_{dW}^{(ij)} &= (\bar{q}_i \sigma^{\mu\nu} \tau^I d_j) H W_{\mu\nu}^I, & O_{lequ}^{3(ijkl)} &= (\bar{l}_i \sigma^{\mu\nu} e_j)\, \varepsilon\, (\bar{q}_k \sigma_{\mu\nu} u_l), \\
O_{uB}^{(ij)} &= (\bar{q}_i \sigma^{\mu\nu} u_j) \tilde{H} B_{\mu\nu}, & O_{ledq}^{(ijkl)} &= (\bar{l}_i e_j)(\bar{d}_k q_l)(\bar{u}_k \gamma^\mu u_l), \\
O_{uG}^{(ij)} &= (\bar{q}_i \sigma^{\mu\nu} T^A u_j) \tilde{H} G_{\mu\nu}^A.
\end{aligned} \quad (68)
$$

Several of these do not contribute to $e^+e^- \to tj$ at tree level. The operators $O_{\varphi ud}$, $O_{dW}$, $O_{ledq}$ and two combinations of $O_{\varphi q}^{1,3}$, $O_{lq}^{1,3}$ operators only contribute to charged top-quark currents, while $O_{u\varphi}$ and $O_{uG}$ only give rise to interactions with the Higgs or the gluons.

The linear combinations of Wilson coefficients for the operators that do contribute in top-quark FCNCs are thus (cf. also Appendix E of Ref. [135])

$$
\begin{aligned}
c_{\varphi q}^{-[I](3+a)} &\equiv {}^{[\Im]}_{\Re}\{C_{\varphi q}^{1(3a)} - C_{\varphi q}^{3(3a)}\}, \\
c_{\varphi u}^{[I](3+a)} &\equiv {}^{[\Im]}_{\Re}\{C_{\varphi u}^{(3a)}\}, & c_{lq}^{-[I](1,3+a)} &\equiv {}^{[\Im]}_{\Re}\{C_{lq}^{-(113a)}\}, & c_{lequ}^{S[I](1,3a)} &\equiv {}^{[\Im]}_{\Re}\{C_{lequ}^{1(113a)}\}, \\
c_{uA}^{[I](3a)} &\equiv {}^{[\Im]}_{\Re}\{c_W C_{uB}^{(3a)} + s_W C_{uW}^{(3a)}\}, & c_{eq}^{[I](1,3+a)} &\equiv {}^{[\Im]}_{\Re}\{C_{eq}^{(113a)}\}, & c_{lequ}^{S[I](1,a3)} &\equiv {}^{[\Im]}_{\Re}\{C_{lequ}^{1(11a3)}\}, \\
c_{uA}^{[I](a3)} &\equiv {}^{[\Im]}_{\Re}\{c_W C_{uB}^{(a3)} + s_W C_{uW}^{(a3)}\}, & c_{lu}^{[I](1,3+a)} &\equiv {}^{[\Im]}_{\Re}\{C_{lu}^{(113a)}\}, & c_{lequ}^{T[I](1,3a)} &\equiv {}^{[\Im]}_{\Re}\{C_{lequ}^{3(113a)}\}, \\
c_{uZ}^{[I](3a)} &\equiv {}^{[\Im]}_{\Re}\{-s_W C_{uB}^{(3a)} + c_W C_{uW}^{(3a)}\}, & c_{eu}^{[I](1,3+a)} &\equiv {}^{[\Im]}_{\Re}\{C_{eu}^{(113a)}\}, & c_{lequ}^{T[I](1,a3)} &\equiv {}^{[\Im]}_{\Re}\{C_{lequ}^{3(11a3)}\}. \\
c_{uZ}^{[I](a3)} &\equiv {}^{[\Im]}_{\Re}\{-s_W C_{uB}^{(a3)} + c_W C_{uW}^{(a3)}\},
\end{aligned} \quad (69)
$$

The SM contribution to $e^+e^- \to tj$ is negligible, thus the FCNC signal is quadratic in the coefficients of the EFT operators. In total, 56 real EFT degrees of freedom contribute to the $e^+e^- \to tj$ process. Top-up and top-charm FCNCs do not interfere with each other. In the limit of vanishing $m_u$ and $m_c$ masses, operators involving light quarks of different chiralities do not interfere either. Disregarding absorptive parts in the amplitudes and motion-reversal-odd distributions, CP-even and CP-odd contributions also do not interfere. Selecting just 7 real degrees of freedom is therefore already representative of

---
[32]Based on a contribution by G. Durieux.

the full parameter space. For these there are four choices:

$$
\begin{aligned}
& c_{lq}^{-(1,3+a)}, \quad c_{eq}^{(1,3+a)}, \quad c_{\varphi q}^{-(3+a)}, \quad c_{uA}^{(a3)}, \quad c_{uZ}^{(a3)}, \quad c_{lequ}^{S(1,a3)}, \quad c_{lequ}^{T(1,a3)}, \quad \text{and} \quad c_{t\varphi}^{(a3)}, \\
\text{or} \quad & c_{lu}^{(1,3+a)}, \quad c_{eu}^{(1,3+a)}, \quad c_{\varphi u}^{(3+a)}, \quad c_{uA}^{(3a)}, \quad c_{uZ}^{(3a)}, \quad c_{lequ}^{S(1,3a)}, \quad c_{lequ}^{T(1,3a)}, \quad \text{and} \quad c_{t\varphi}^{(3a)}, \\
\text{or} \quad & c_{lq}^{-I(1,3+a)}, \quad c_{eq}^{I(1,3+a)}, \quad c_{\varphi q}^{-I(3+a)}, \quad c_{uA}^{I(a3)}, \quad c_{uZ}^{I(a3)}, \quad c_{lequ}^{SI(1,a3)}, \quad c_{lequ}^{TI(1,a3)}, \quad \text{and} \quad c_{t\varphi}^{I(a3)}, \\
\text{or} \quad & c_{lu}^{I(1,3+a)}, \quad c_{eu}^{I(1,3+a)}, \quad c_{\varphi u}^{I(3+a)}, \quad c_{uA}^{I(3a)}, \quad c_{uZ}^{I(3a)}, \quad c_{lequ}^{SI(1,3a)}, \quad c_{lequ}^{TI(1,3a)}, \quad \text{and} \quad c_{t\varphi}^{I(3a)},
\end{aligned}
\tag{70}
$$

where quark (and lepton) generation indices are enclosed in parentheses. In particular, $a \in \{1, 2\}$ is the generation of the light-quark. We will drop superscript in the following to cover generically each of these four sets, for either light-quark generation. The operators corresponding to $c_{lq,lu}^-$ and $c_{eq,eu}$ coefficients give rise to $tqee$ four-fermion interactions of vector Lorentz structure; $c_{\varphi q,\varphi u}^-$ to $tqZ$ interactions with vector Lorentz structure; $c_{uA}$ and $c_{uZ}$ to dipole $tqA$ and $tqZ$ interactions; $c_{lequ}^S$ and $c_{lequ}^T$ to $tqee$ four-fermion interactions of scalar and tensor Lorentz structures. For completeness we also listed the $c_{t\varphi}$ coefficients which give rise to $tqh$ interactions and do not enter in $e^+e^- \to t\,j$.

The leading order expressions for the $t \to Zj, t \to \gamma j$ and $t \to hj$ branching fractions are given by (see [136] for expressions at NLO in QCD)

$$
\mathrm{BR}(t \to Zj) = \frac{1}{\Gamma_t} \frac{m_t^5}{8\pi\Lambda^4} \left( \frac{\pi\alpha}{s_W^2 c_W^2} \frac{(1-x_Z^2)^2(1+2x_Z^2)}{x_Z^2} |c_{\varphi q,\varphi u}^-|^2 + (1-x_Z^2)^2(2+x_Z^2)|c_{uZ}|^2 \right) \tag{71}
$$
$$
\simeq 0.0016\,|c_{\varphi q,\varphi u}^-|^2 + 0.0048\,|c_{uZ}|^2,
$$

$$
\mathrm{BR}(t \to \gamma j) = \frac{1}{\Gamma_t} \frac{m_t^5}{4\pi\Lambda^4} |c_{uA}|^2 \simeq 0.0081\,|c_{uA}|^2, \tag{72}
$$

$$
\mathrm{BR}(t \to hj) = \frac{1}{\Gamma_t} \frac{m_t^7 G_F |c_{t\varphi}|^2}{4\pi\sqrt{2}\Lambda^4} (1-x_h^2)^2 \simeq 0.00044\,|c_{t\varphi}|^2, \tag{73}
$$

with $\Gamma_t \simeq 1.47\,\mathrm{GeV}$, Eq. (95), $x_i \equiv m_i/m_t$, and $s_W, c_W$ are the sine and cosine of the weak mixing angle. For numerical evaluations the new-physics scale $\Lambda$ was set to $1\,\mathrm{TeV}$ unless otherwise specified.

*3.1.2.2  The CLIC reach on EFT coefficients*

To estimate the top FCNC reach of CLIC from the $e^+e^- \to tj$ process we use statistically optimal observables. Given a phase-space distribution that depends only linearly on a set of small parameters, $C_i$, that are to be constrained,

$$
\frac{d\sigma}{d\Phi} = \frac{d\sigma_0}{d\Phi} + C_i \frac{d\sigma_i}{d\Phi},
$$

the average value of the ratio $n \frac{d\sigma_i}{d\Phi} / \frac{d\sigma_0}{d\Phi}$, with $n$ the number of observed events, is a statistically optimal observable. The covariance matrix for $C_i$ that is obtained using the optimal observables is given, at zeroth order in $C_i$, by the phase-space integral

$$
\mathrm{cov}(C_i, C_j)^{-1} = \epsilon\,\mathcal{L} \int d\Phi \left( \frac{d\sigma_i}{d\Phi} \frac{d\sigma_j}{d\Phi} \Big/ \frac{d\sigma_0}{d\Phi} \right), \tag{74}
$$

where $\mathcal{L}$ is the integrated luminosity and $\epsilon$ the overall efficiency. For $e^+e^- \to tj$ the phases distribution $d\sigma_0/d\Phi$ is the phase-space distribution of the background. Furthermore, for $e^+e^- \to tj$ the unknown coefficients are quadratic in the EFT Wilson coefficients, $C_i = c_k c_l$. They are therefore not independent of each other, e.g., are constrained to be non-negative for $k = l$, and thus the limits following from (74) are not guaranteed to be optimal, though we expect them to still be a good proxy for the CLIC reach. For this we construct for each CLIC run a chi-squared function $\sum_{ij} C_i \,\mathrm{cov}(C_i, C_j)^{-1} C_j$ with $C_i = c_k c_l$ and $C_j = c_m c_n$, and then furthermore take, for practical reasons, $k = l$, $m = n$, with $k, m$ running over seven Wilson coefficients in (70). The LO $e^+e^- \to tj$ matrix elements are computed using the standalone c++ output of MG5_AMC@NLO [31] and the UFO model developed in Refs. [137, 138] (see HTTPS://FEYNRULES.IRMP.UCL.AC.BE/WIKI/TOPFCNC).

Table 24: The 4th column shows 95% C.L. limits on BR($t \to \gamma j$) that follow from bounds on $e^+e^- \to tj$ FCNC production, obtained in Refs. [132, 133] (5th column) for several centre-of-mass energies (1st column) and integrated luminosities (2nd column) with no beam polarization (3rd column). Overall efficiencies $\epsilon$ required to reproduce the bounds in our framework are given in 6th column, while 7th and 8th columns gives the estimates for $\epsilon$ and the branching ratio bounds (proportional to $1/\sqrt{\epsilon}$) using $\epsilon(\sqrt{s}) = (125\,\text{GeV}/\sqrt{s})^{1.92}$, obtained from a power-law fit.

| $\sqrt{s}$ [GeV] | $\mathcal{L}$ [fb$^{-1}$] | $P(e^+, e^-)$ | BR before fit | Ref. | $\epsilon$ before fit | $\epsilon$ after fit | BR after fit |
|---|---|---|---|---|---|---|---|
| 240 | 3000 | (0,0) | $3.70 \times 10^{-5}$ | [133] | 0.30 | 0.30 | $3.7 \times 10^{-5}$ |
| 350 | 3000 | (0,0) | $9.86 \times 10^{-6}$ | [133] | 0.19 | 0.14 | $1.1 \times 10^{-5}$ |
| 500 | 3000 | (0,0) | $6.76 \times 10^{-6}$ | [133] | 0.057 | 0.072 | $6.0 \times 10^{-6}$ |
| 500 | 300 | (0,0) | $2.2 \times 10^{-5}$ | [132] | 0.054 | 0.072 | $1.9 \times 10^{-5}$ |
| 800 | 500 | (0,0) | $7.8 \times 10^{-6}$ | [132] | 0.037 | 0.029 | $8.7 \times 10^{-6}$ |

For the $e^+e^- \to t\bar{u} \to \mu^+\nu_\mu b\,\bar{u}$ signal the main background is due to $e^+e^- \to W^+W^- \to \mu^+\nu_\mu\,\bar{c}\,s$ [132, 133]. We neglect the $b$ mistagging efficiency of light quarks compared to the background from charm quarks. Factors accounting for tagging efficiencies, the charge-conjugate process and the alternative electron channel are folded into an overall efficiency $\epsilon$ when computing the covariance matrix, see Eq. (74). A perfect reconstruction of each particle momentum — neutrino included — is assumed. Their energies are required to be above 10 GeV. A minimal angular separation of 10° is demanded between them, and from the beam. The $\mu^+\nu_\mu$ and $\mu^+\nu_\mu b$ invariant masses are required to be within 15 GeV of $m_W$ and $m_t$, respectively, while the $b\,\bar{u}$ invariant mass is required to be at least 15 GeV away from $m_W$. The last cuts have little impact, though, since the statistically optimal observables already enhance signal-like phase-space regions and suppress background-like ones.

The overall efficiencies $\epsilon$ in Eq. (74) correct for our idealizations. Finite resolutions are expected to decrease the power of optimal observables by smoothing the sharp phase-space features that distinguish signal and background. We determine the $\epsilon$'s for each CLIC run by requiring that they reproduce the limits on BR($t \to \gamma j$) derived in studies of $e^+e^- \to tj$ at centre-of-mass energies of 240, 350, 500 and 800 GeV, without beam polarization [132, 133]. A power law provides a good fit for $\epsilon$ as a function of $\sqrt{s}$ (see Table 24). We use it to determine efficiencies at CLIC centre-of-mass energies and apply them to runs with either $P(e^+, e^-) = (0, \pm 0.8)$ beam polarizations. The fit function yields efficiencies of 0.12, 0.0088 and 0.0023, at centre-of-mass energies of 380, 1500 and 3000 GeV, respectively.

Global constraints are derived for successive runs at centre-of-mass energies of 380 GeV, 1.5 TeV and 3 TeV with integrated luminosities of 500 fb$^{-1}$, 1.5 ab$^{-1}$ and 3 ab$^{-1}$ equally shared between $P(e^+, e^-)$ = $(0, \pm 0.8)$ polarizations. The marginalized 95% C.L. limits in the directions parametrized by the operator coefficients of Eq. (70) are displayed in Figure 38 (dots indicate limits with unpolarized beams, which are 1.1 to 1.6 times weaker). High-energy runs significantly improve the sensitivity to four-fermion operators. The constraints due to the expected CLIC bounds on BR($t \to j\gamma$) and BR($t \to jh$) at 380 GeV, without beam polarization, obtained in Section 3.4.2, are shown with black arrows (BR($h \to b\bar{b}$) = 0.57 is used in the conversion). The decay processes are most powerful at low centre-of-mass energies, where the $t\bar{t}$ production cross section is the highest. In Figure 39 we also show the limits in ($c_{\varphi q, \varphi u}^-$, $c_{uZ}$) and in ($c_{uA}$, $c_{uZ}$) planes, marginalized over the other Wilson coefficients.

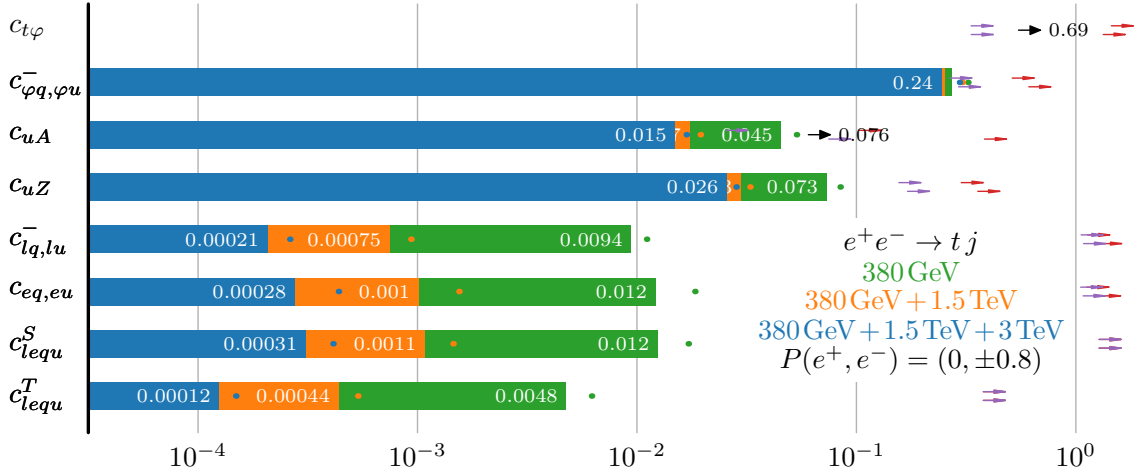

Figure 38: The expected 95% C.L. limits on top-quark FCNC operator coefficients from $e^+e^- \to tj$ production, with top decaying semi-leptonically, for integrated luminosities of $500\,\text{fb}^{-1}$ (green), or in addition $1.5\,\text{ab}^{-1}$ (orange) and $3\,\text{ab}^{-1}$ (blue) at centre-of-mass energies of $380\,\text{GeV}$, $1.5\,\text{TeV}$ and $3\,\text{TeV}$, respectively, and equally shared between $P(e^+, e^-) = (0, \pm 0.8)$ polarizations. The constraints from bounds on $\text{BR}(t \to j\gamma)$ and $\text{BR}(t \to jh)$, Section 3.4.2, are indicated with black arrows. Small dots indicate the limits obtained without beam polarization. Current LHC limits and the projected HL-LHC reach obtained in Ref. [139] are reported as red and purple arrows, respectively. Upper (lower) ones stand for top-up (top-charm) FCNCs.

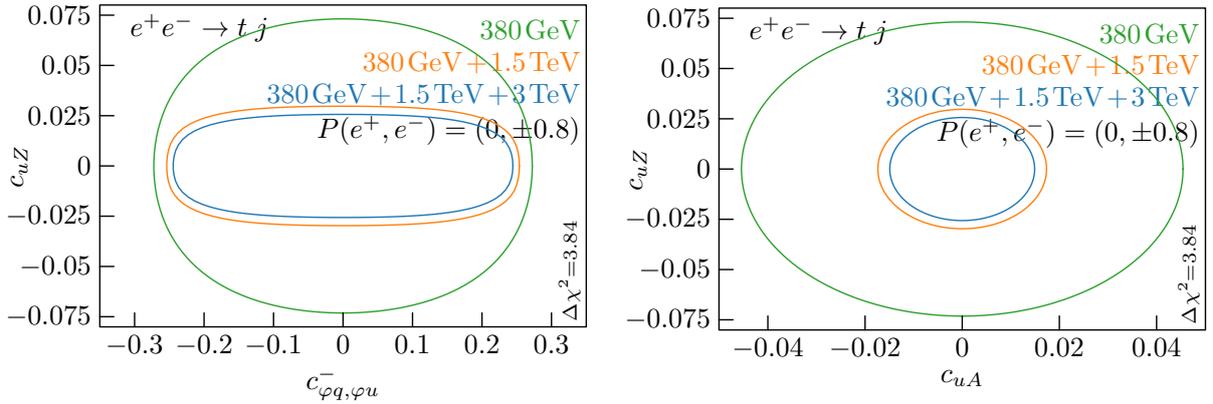

Figure 39: Constraints on top-quark FCNC operator coefficients $(c^-_{\varphi q, \varphi u}, c_{uZ})$ [left panel] and $(c_{uA}, c_{uZ})$ [right panel] that follow from $e^+e^- \to tj$, with top decaying semi-leptonically, marginalized over the other coefficients. Integrated luminosities of $500\,\text{fb}^{-1}$, $1.5\,\text{ab}^{-1}$ and $3\,\text{ab}^{-1}$ are equally shared between $P(e^+, e^-) = (0, \pm 0.8)$ polarizations at centre-of-mass energies of $380\,\text{GeV}$, $1.5\,\text{TeV}$ and $3\,\text{TeV}$.

### 3.2 Lepton flavour violating processes and neutrino mass generation[33]

Many neutrino mass models contain NP scalars, neutral and/or (doubly-)charged. This is the case, for instance, in the left-right symmetric models [140–142] with type-I seesaw [143–147] or type-II seesaw [148–151]. Such new scalars might induce noticeable lepton flavour violating (LFV) signals in the charged lepton sector. In this section we estimate, in a model-independent way, the CLIC prospects for measuring such LFV couplings, both for the neutral and doubly-charged scalars [152, 153]. The CLIC measurements are largely complementary to the searches at the LHC and at the future high-luminosity/high-energy hadron colliders, as well as to the high-precision measurements at the low-energy experiments. Furthermore, the couplings of doubly-charged scalar to the charged leptons are also lepton number violating (LNV). These are probed through searches for neutrinoless double-beta decays ($0\nu\beta\beta$) [154–160] and the $e^-e^-$ scattering experiments such as MOLLER [161, 162]. The model-independent CLIC prospects for the LFV couplings can then be applied to some representative well-motivated models, like the type-II seesaw and the left-right extensions and the R-parity violating supersymmetric models [163–166]. In these models there is an intimate connection to the neutrino physics and complementarity with the other low and high energy experiments.

---

[33] Based on a contribution by Y. Zhang, R. Mohapatra and B. Dev.

### 3.2.1 Searching for a new neutral scalar

CLIC provides a "clean" environment to search for charged LFV (cLFV) scattering $e^+e^- \to \ell_\alpha^\pm \ell_\beta^\mp + X$, where $\alpha \neq \beta$ are the lepton flavour indices. In the SM such transitions are suppressed by neutrino masses and are vanishingly small. However, they need not be very suppressed in the presence of NP. Their discovery would thus unambiguously imply existence of NP. For instance, a new neutral scalar $\phi$, can have flavour violating couplings to leptons,

$$\mathcal{L}_Y = h_{\alpha\beta}(\bar{\ell}_{L\alpha}\ell_{R\beta,})\phi + \text{h.c.}. \qquad (75)$$

For simplicity we take the couplings $h_{\alpha\beta}$ to be real and symmetric in flavour indices, $\alpha, \beta$. In realistic models the scalar $\phi$ may be responsible for symmetry breaking and/or mass generation of other particles, and could be part of an electroweak singlet, doublet or triplet scalar field. For simplicity, again, we assume that it is CP-even and that its mixing with the SM Higgs is small.

**On-shell production:** If kinematically allowed, the neutral scalar $\phi$ can be singly produced at CLIC, $e^+e^- \to \ell_\alpha^\pm \ell_\beta^\mp \phi$, in association with a pair of charged leptons of differing flavours, $\alpha \neq \beta$. The $e^+e^- \to \ell_\alpha^\pm \ell_\beta^\mp \phi$ requires only a single LFV coupling $h_{\alpha\beta}$ to be nonzero. This is in contrast to most of the stringent low-energy cLFV constraints, such as $\mu \to eee$ and $\mu \to e\gamma$, which depend on the product $|h_{ee}^\dagger h_{e\mu}|$, and can always be made irrelevant by taking $h_{ee} \to 0$. Assuming only one nonzero LFV coupling thus significantly relaxes the low energy constraints. For the $h_{e\mu}$ coupling the relevant constraints are from muonium oscillation, the electron $g-2$ and the LEP $e^+e^- \to \mu^+\mu^-$ data [167], for $h_{e\tau}$, the limits are mainly from the electron $g-2$ and the LEP $e^+e^- \to \tau^+\tau^-$ data [167], while for $h_{\mu\tau}$ only the muon $g-2$ is relevant. The limits on the scalar mass, $m_\phi$, and the couplings $h_{\alpha\beta}$ are collected in Figure 40, with the shaded regions excluded (cf. also Ref. [152]). In the right panel of Figure 40, the gray region is excluded by the current muon $g-2$ data at the $5\sigma$ level.

In the future lepton colliders $\phi$ can also be produced from the laser "photon fusion", $\gamma\gamma \to \ell_\alpha^\pm \ell_\beta^\mp \phi$ [153]. Here the high luminosity photon beams are obtained from Compton backscattering of low energy, high intensity, laser beams off the high energy electron beams. The resulting effective photon luminosity distributions can be found, e.g., in [169–171]. Another option is to collide $e^\pm$ with the laser photons, which produces $\phi$ through the LFV process $e\gamma \to \ell\phi$, where $\ell = \mu, \tau$. For simplicity, we assume that $\phi$ has only one decay channel, $\phi \to \ell_\alpha^\pm \ell_\beta^\mp$. We apply nominal cuts $p_T(\ell) > 10$ GeV, $|\eta(\ell)| < 2.5$ and $\Delta R_{\ell\ell'} > 0.4$ using CalcHEP [172], and adopt a conservative efficiency of 60% for the $\tau$ lepton [173]. Figure 40 shows the projected reach for $h_{e\mu,e\tau,\mu\tau}$ couplings as functions of $m_\phi$ at 3 TeV CLIC with 2 ab$^{-1}$ of integrated luminosity using the $e^+e^-, \gamma\gamma \to \ell_\alpha^\pm \ell_\beta^\mp \phi$ or $e\gamma \to \ell\phi$ processes. All of the LFV processes are almost background free [152]. To denote the discovery potential lines in Figure 40 we used a simplistic rule of requiring a minimum of 10 signal events. It is clear that a large region of $m_\phi$ and $|h_{\alpha\beta}|$ can be probed at CLIC, extending the limits well beyond what is currently available. For instance, through $\gamma\gamma \to \mu^\pm \tau^\mp \phi$ production CLIC could directly test the possibility that the muon $g-2$ anomaly, $\Delta a_\mu$, is due to an exchange of a neutral scalar, $\phi$, that couples to muon and tau via LFV $h_{\mu\tau}$ couplings, see the right panel in Figure 40.

**Off-shell (and resonant) production:** The LFV signal $e^+e^- \to \ell_\alpha^\pm \ell_\beta^\mp$ could be due to an off-shell $\phi$ exchange, either in the $s$ or $t$ channel, depending on the specific combinations of the couplings $h_{\alpha\beta}$.

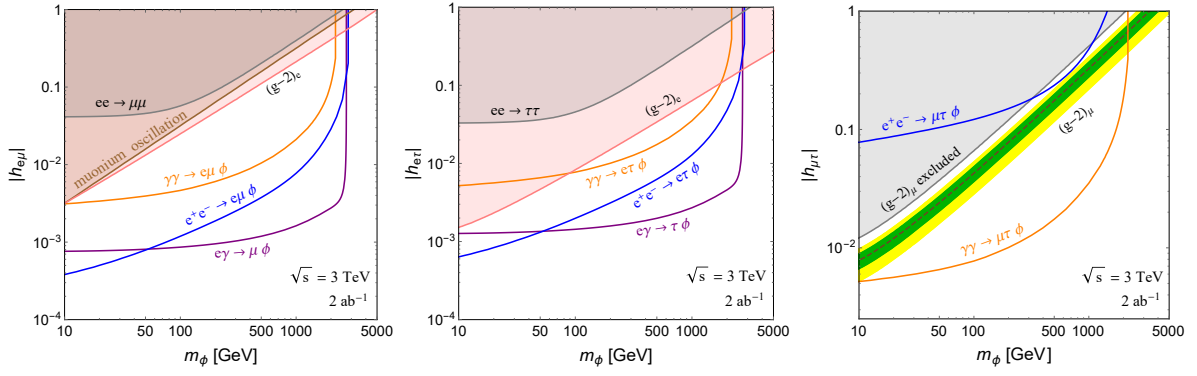

Figure 40: Prospects for probing the LFV couplings $h_{\alpha\beta}$ ($\alpha \neq \beta$) in the searches $e^+e^-, \gamma\gamma \to \ell_\alpha^\pm \ell_\beta^\mp \phi$ or $e\gamma \to \ell\phi$ at CLIC 3 TeV with a luminosity of 2 ab$^{-1}$, where the initial photons are laser photons. The projection lines correspond to 10 LFV signal events with $\phi$ decaying to lepton pairs, $\phi \to \ell_\alpha^\pm \ell_\beta^\mp$. The shaded regions are excluded by muonium oscillations, the electron and muon $g-2$ and the LEP $ee \to \ell\ell$ data [167], as indicated. In the right panel, the dashed brown line denotes the central value of $\Delta a_\mu$, with green and yellow bands denoting the $1\sigma$ and $2\sigma$ ranges of $\Delta a_\mu$ [168].

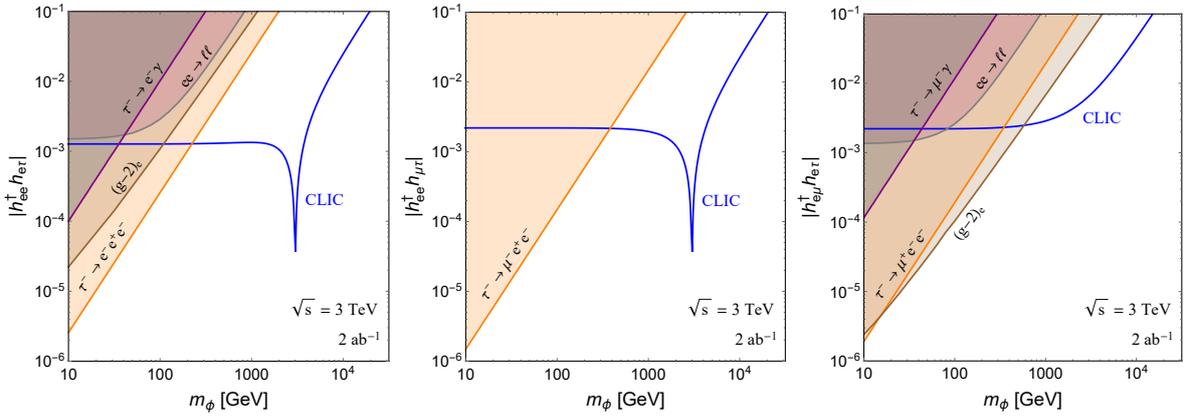

Figure 41: Prospects for probing $|h_{ee}^\dagger h_{e\tau}|$ (left), $|h_{ee}^\dagger h_{\mu\tau}|$ (middle) and $|h_{e\mu}^\dagger h_{e\tau}|$ (right) from $e^+e^- \to e^\pm \tau^\mp, \mu^\pm \tau^\mp$ searches at 3 TeV CLIC with a luminosity of 2 ab$^{-1}$. The projection lines assume 10 signal events. Also shown are the constraints from the LFV decays $\ell_\alpha \to \ell_\beta \ell_\gamma \ell_\delta$, $\ell_\alpha \to \ell_\beta \gamma$, electron $g-2$, and the LEP $ee \to \ell\ell$ data [167].

Different from the on-shell case, the off-shell production amplitudes depend quadratically on the Yukawa couplings (some of them might be flavour conserving), and are thus largely complementary to the on-shell LFV searches.

The amplitude for the $e^+e^- \to e^\pm \mu^\mp$ process is proportional to $h_{ee}^\dagger h_{e\mu}$, which is tightly constrained by $\mu \to eee$ bounds, and thus out of CLIC reach. The LFV couplings involving $\tau$ are much less constrained, so that the $e^+e^- \to e^\pm \tau^\mp, \mu^\pm \tau^\mp$ LFV signal is possible at CLIC. The $\mu\tau$ final state can arise from both the $s$ and $t$ channel $\phi$ exchanges, giving amplitudes proportional to $|h_{ee}^\dagger h_{\mu\tau}|$ and $|h_{e\mu}^\dagger h_{e\tau}|$, respectively. Figure 41 shows with blue curves the reach of 3 TeV CLIC with 2 ab$^{-1}$ for the relevant $|h^\dagger h|$ combinations assuming 10 signal events. The dips correspond to the resonance production at $m_\phi \simeq \sqrt{s} = 3$ TeV, where we set the $\phi$ width to $\Gamma_\phi = 30$ GeV. The relevant cLFV limits from the rare decays $\ell_\alpha \to \ell_\alpha \ell_\beta \ell_\delta$, $\ell_\alpha \to \ell_\beta \gamma$, electron $g-2$ and the LEP $ee \to \ell\ell$ data [167] are shown in Figure 41 as shaded regions. For light $\phi$, the CLIC reach is almost $m_\phi$ independent, while for heavy $\phi$ the production rate diminishes rapidly. The reach is then equivalent to bounding the four-fermion interaction $(\bar{e}e)(\bar{\ell}\ell)/m_\phi^2$. As shown in Figure 41, a broad range of $m_\phi$ and $|h^\dagger h|$ could be probed in both the $e\tau$ and $\mu\tau$ channels, and the effective cutoff scale $\Lambda \simeq m_\phi/|h|$ could go well above 10 TeV.

### 3.2.2 Doubly-charged scalar

LFV could also be due to a doubly-charged scalar, $\phi^{\pm\pm}$, that couples to leptons through Yukawa couplings. For simplicity we focus on the case of $\phi^{\pm\pm}$ that couples to the right-handed leptons,

$$\mathcal{L}_Y = f_{\alpha\beta}(\bar{\ell}_\alpha^c P_R \ell_\beta)\phi^{++} + \text{h.c.}, \tag{76}$$

while the main results in this section hold also for "left-handed" $\phi^{\pm\pm}$ (i.e., a $\phi^{\pm\pm}$ that couples to the left-handed leptons). Here $\alpha$, $\beta$ are the flavour indices and superscript $C$ denotes charge conjugation. Well-motivated candidates include both the left-handed $\phi^{\pm\pm}$ in the type-II seesaw [148–151] and the right-handed $\phi^{\pm\pm}$ from the left-right symmetric models [140–142, 174]. A "smoking-gun" signal of $\phi^{\pm\pm}$ are same-sign charged lepton pairs from the decay $\phi^{\pm\pm} \to \ell_\alpha^\pm \ell_\beta^\pm$. The current LHC dilepton limits exclude the right-handed (left-handed) $\phi^{\pm\pm}$ with mass up to roughly 650 GeV (800 GeV) [175, 176], depending on the specific lepton flavours. At 3 TeV CLIC the doubly-charged scalars could be probed to higher masses, as we show below. The present limits on the LFV Yukawa couplings $f_{\alpha\beta}$, $\alpha \neq \beta$, are mainly from LEP $e^+e^- \to \ell^+\ell^-$ data. A $t$-channel $\phi^{\pm\pm}$ exchange with the LFV couplings $f_{e\mu, e\tau}$ would change the $e^+e^- \to \mu^+\mu^-$, $\tau^+\tau^-$ cross sections. The absence of measured deviations excludes the shaded pink regions in Figure 42. The limits from electron and muon $g-2$ are highly suppressed by the charged lepton masses and are not shown in the plots. The vertical dashed gray lines indicate the current same-sign dilepton limits on the doubly-charged scalar mass from LHC [175, 176], assuming a BR($\phi^{\pm\pm} \to \ell_\alpha^\pm \ell_\beta^\pm$) = 100%, see Ref. [153] for further details.

At CLIC the doubly-charged scalar $\phi^{\pm\pm}$ can be explored through i) Drell-Yan pair production due to its couplings to photon and $Z$, ii) single production, and iii) the off-shell production. The latter two are sensitive to Yukawa couplings in (76). In single production, $e^+e^-, \gamma\gamma \to \phi^{\pm\pm}\ell_\alpha^\mp \ell_\beta^\mp$ [153, 177–180], or $e^\pm\gamma \to \phi^{\pm\pm}\ell_\alpha^\mp$ [177, 181–184], the $\phi^{\pm\pm}$ is emitted from an internal fermion line (here $\gamma$ is from the laser beam). The cross section is thus proportional to $|f_{e\mu, e\tau, \mu\tau}|^2$. In Figure 42 we show the corresponding projected sensitivities to $|f_{e\mu, e\tau, \mu\tau}|$ as functions of the $\phi^{\pm\pm}$ mass, $M_{\pm\pm}$, using the same cuts as in Section 3.2.1. For simplicity we assume that $\phi^{\pm\pm}$ decays predominantly to the same-sign charged leptons. Below $\sqrt{s}/2 \simeq 1.5$ TeV the $e^+e^- \to \phi^{\pm\pm}\ell_\alpha^\mp \ell_\beta^\mp$ production is dominated by the Drell-Yan pair production, $e^+e^- \to \phi^{++}\phi^{--}$, with one of the $\phi^{\pm\pm}$ decaying off-shell to $\ell_\alpha^\pm \ell_\beta^\pm$. This production mechanism is not sensitive to the Yukawa couplings, and thus we cut-off the projected limits at 1.5TeV.

The off-shell $t$-channel exchange of $\phi^{\pm\pm}$ generates $e^+e^- \to \ell_\alpha^\pm \ell_\beta^\mp$ and $e^\pm\gamma \to \ell_\alpha^\mp \ell_\beta^\pm \ell_\gamma^\pm$ transitions [153, 181–183], which depend quadratically on the product $|f^\dagger f|$. The sensitivity of LFV couplings from $e^\pm\gamma$ collision are weaker, because three-body phase space suppression in the final state. The stringent limit from $\mu \to eee$ also excludes any visible $ee \to e\mu$ signal at CLIC. We therefore focus on the reach of searches for $e^+e^- \to e^\pm\tau^\mp, \mu^\pm\tau^\mp$. These probe the combinations of couplings $|f_{ee}^\dagger f_{e\tau}|$ and $|f_{e\mu}^\dagger f_{e\tau}|$. The prospects at CLIC 3 TeV with a luminosity of 2 ab$^{-1}$ are shown as the blue curves in Figure 43. The most important existing constraints on the products $|f_{ee}^\dagger f_{e\tau}|$ and $|f_{e\mu}^\dagger f_{e\tau}|$ come from LFV decays $\ell_\alpha \to \ell_\beta \ell_\gamma \ell_\delta$, $\ell_\alpha \to \ell_\beta \gamma$, and the LEP $e^+e^- \to \ell^+\ell^-$ data, The corresponding exclusions are depicted as the shaded regions in Figure 43 [153].

Figures 42 and 43 show that both the on-shell and off-shell production of $\phi^{\pm\pm}$ at 3 TeV CLIC would probe a large part of allowed parameter space, both in terms of the mass $M_{\pm\pm}$ and the LFV couplings $f_{\alpha\beta}$. An integrated luminosity of 2 ab$^{-1}$ would allow for the couplings to be probed at the $10^{-3}$ level, corresponding to the reach in the effective cutoff scales, $\Lambda \simeq M_{\pm\pm}/\sqrt{|f^\dagger f|}$, in the 10s of TeV.

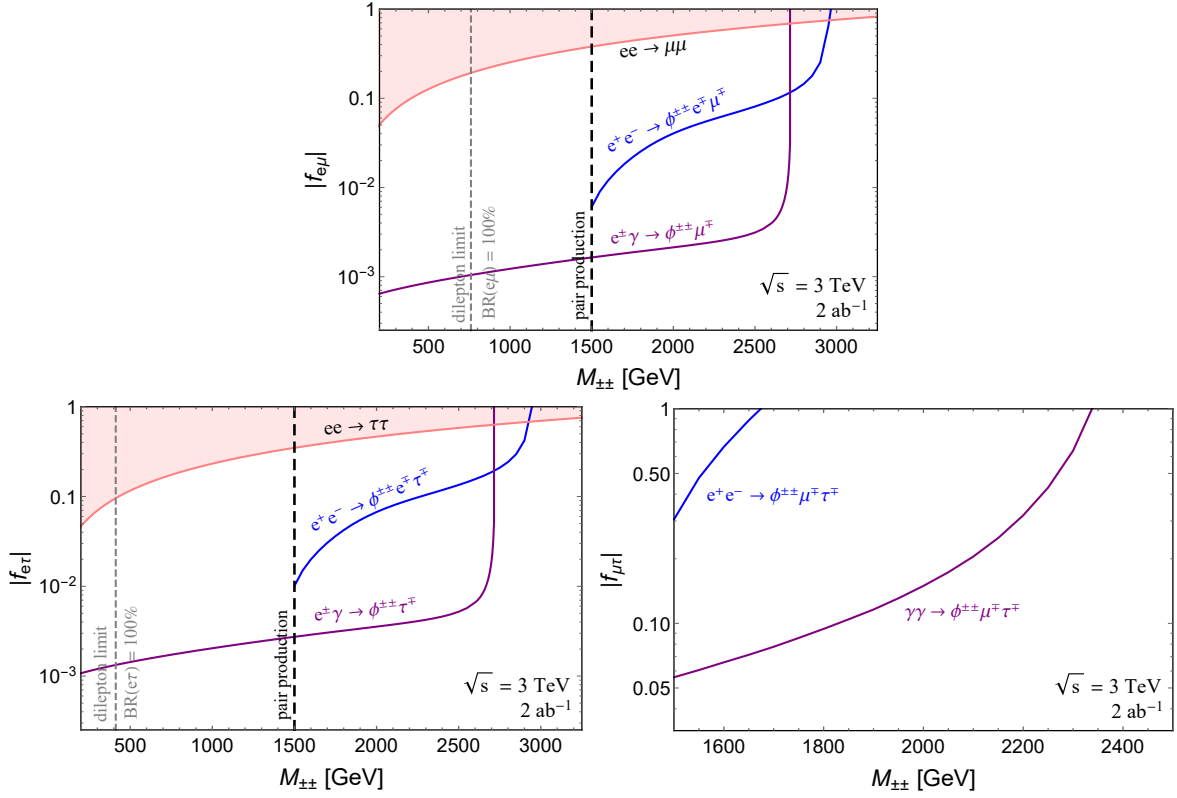

Figure 42: Prospects for probing the LFV couplings $f_{\alpha\beta}$, $\alpha \neq \beta$, using searches for $e^+e^- \to \phi^{++}\phi^{--}$, $e^+e^-, \gamma\gamma \to \ell_\alpha^\pm \ell_\beta^\pm \phi^{\mp\mp}$ or $e^\pm \gamma \to \ell^\mp \phi^{\pm\pm}$ at CLIC 3 TeV with 2 ab$^{-1}$ of integrated luminosity (here the initial photons are laser photon). The lines shown correspond to 10 LFV signal events, with $\phi^{\pm\pm}$ decaying to the lepton pairs, $\phi^{\pm\pm} \to \ell_\alpha^\pm \ell_\beta^\pm$. The shaded regions are excluded by the LEP $e^+eE- \to \ell^+\ell^-$ data [167]. The vertical dashed gray lines indicate the current LHC same-sign dilepton limits [175, 176], assuming BR($\phi^{\pm\pm} \to \ell_\alpha^\pm \ell_\beta^\pm$) = 100%.

### 3.2.3 Implications for specific models

We interpret next the CLIC reach for neutral and doubly-charged scalars, obtained in Sections 3.2.1 and 3.2.2, in the context of two specific models: the type-II see-saw mechanism [148–151] and the left-right symmetric models [140–142].

In type-II see-saw the new neutral and doubly-charged scalars are part of an electroweak triplet scalar $\Delta$ that couples to the left-handed leptons,

$$\mathcal{L}_Y = -f_{\alpha\beta} \bar{\psi}_{L,\alpha}^c \Delta \psi_{L,\beta} + \text{h.c.} \,. \tag{77}$$

Here $\psi_L = (\nu, \ell_L)^\mathsf{T}$ are the SM lepton doublets. Neglecting the mixing with the SM Higgs, the neutral scalar is $\phi = \text{Re}(\Delta^0)$. At tree-level it couples to the SM neutrinos, while coupling to charged leptons is induced at 1 loop through the trilinear $\phi\phi^{++}\phi^{--}$ term via triangle diagrams with doubly-charged scalar and charged leptons running in the loop. The effective $\phi\ell_\alpha^\pm \ell_\beta^\mp$ coupling is thus highly suppressed, both by the loop factor and by the ratio $m_\ell^2/v_{\text{EW}}^2$, where $m_\ell$ are the charged lepton masses and $v_{\text{EW}}$ is the electroweak VEV. This means that it will be very challenging to search for LFV signals induced by the type-II see-saw neutral scalar. First of all, the production cross section of $\phi$ at CLIC is expected to be very small. Furthermore, the branching fraction BR($\phi \to \ell_\alpha^\pm \ell_\beta^\mp$) is much smaller than for the neutrino channels BR($\phi \to \nu\bar\nu$).

The type-II see-saw doubly-charged scalar, $\phi^{\pm\pm}$, couples to same-sign left handed charged leptons ($\phi^{\pm\pm}\ell_\alpha^\mp \ell_\beta^\mp$), same-sign $W$ bosons ($\phi^{\pm\pm}W^\mp W^\mp$), as well as to the singly-charged scalar ($\phi^{\pm\pm}\phi^\mp\phi^\mp$ and $\phi^{\pm\pm}\phi^{\mp}W^{\mp}$, with $\phi^\pm = \Delta^\pm$. As a result of the current electroweak precision data [185], the mass splitting of the triplet scalars is required to be $\lesssim 60$ GeV, with a coupling that depends largely on the triplet VEV $\langle\Delta^0\rangle = v_\Delta$. In the limit of small $v_\Delta$, e.g., with a value smaller than roughly 0.1 MeV, $\phi^{\pm\pm}$ decays predominantly to same-sign dileptons [186]. In this case CLIC could detect a doubly-charged scalar $\phi^{\pm\pm}$ with mass up to $\sqrt{s}/2 \simeq 1.5$ TeV from Drell-Yan pair production, and also observe the LFV (and LNV) signal $\phi^{\pm\pm} \to \ell_\alpha^\pm \ell_\beta^\pm$, $\alpha \neq \beta$. In single production and in off-shell production of $\phi^{\pm\pm}$, the Yukawa couplings could be probed at the $\sim 10^{-3}$ level for a TeV-scale $\phi^{\pm\pm}$ [153] (see Figures 42 and 43), corresponding to $v_\Delta \lesssim 100$ eV. In this case the cascade decays $\phi^{\pm\pm} \to \phi^\pm W^\pm$, $\phi^\pm\phi^\pm$, $W^\pm W^\pm$ are completely negligible, and $\phi^{\pm\pm}$ decays are purely leptophilic. One should note that the BR($\phi^{\pm\pm} \to \ell_\alpha^\pm \ell_\beta^\pm$) are intimately correlated with the neutrino oscillation data, i.e., the mass squared differences, the neutrino mixing angles, the lightest neutrino mass and neutrino mass hierarchy, and the Dirac CP violating phase. The branching fractions BR($\phi^{\pm\pm} \to \ell_\alpha^\pm \ell_\beta^\pm$) could be used to set limits or even measure the low-energy parameters in the neutrino sector.

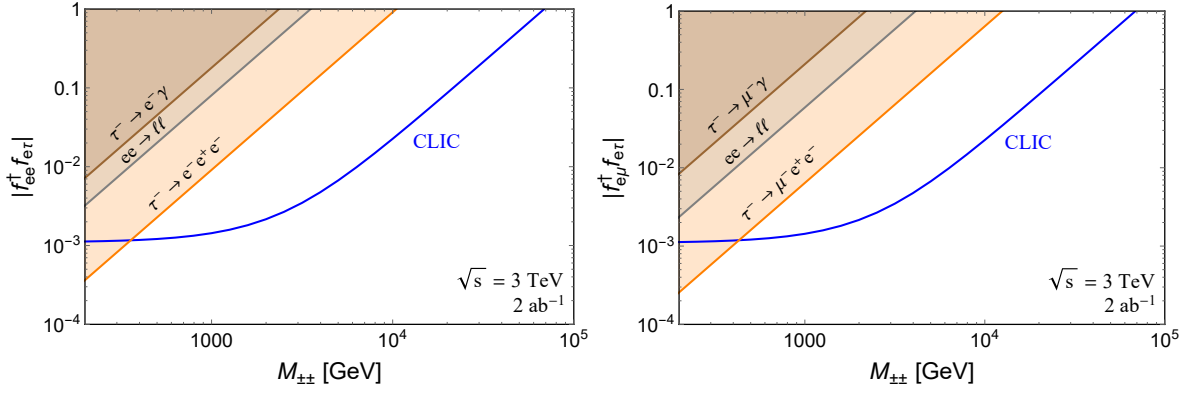

Figure 43: Prospects for probing the combinations of Yukawa couplings $|f_{ee}^\dagger f_{e\tau}|$ (left) and $|f_{e\mu}^\dagger f_{e\tau}|$ (right) of the doubly-charged scalar $\phi^{\pm\pm}$ from searches for the $e^+e^- \to e^\pm\tau^\mp$, $\mu^\pm\tau^\mp$ signals at the 3 TeV CLIC with a luminosity of 2 ab$^{-1}$ (we assume 10 signal events). Shaded regions show constraints from LFV decays $\ell_\alpha \to \ell_\beta\ell_\gamma\ell_\delta$, $\ell_\alpha \to \ell_\beta\gamma$, and from the LEP $e^+e^- \to \ell^+\ell^-$ data [167].

The scalar sector of the left-right symmetric models (LRSMs) consists of a bidoublet, $\Phi$, and the left- and right-handed triplets, $\Delta_{L,R}$,

$$\Phi = \begin{pmatrix} \phi_1^0 & \phi_2^+ \\ \phi_1^- & \phi_2^0 \end{pmatrix}, \qquad \Delta_{L,R} = \begin{pmatrix} \Delta_{L,R}^+/\sqrt{2} & \Delta_{L,R}^{++} \\ \Delta_{L,R}^0 & -\Delta_{L,R}^+/\sqrt{2} \end{pmatrix}. \qquad (78)$$

The left-handed triplet can be made to decouple from TeV-scale physics. This results in a simpler scalar sector that avoids unacceptably large contributions to the neutrino masses from the type-II see-saw [187]. In the limit of small scalar mixings the real component of $\phi_1^0$ is identified with the SM Higgs $h$. The CP-even ($H_1$) and CP-odd ($A_1$) neutral scalars are almost mass degenerate with the singly-charged scalar, $H_1^\pm = \phi_2^\pm$. The high-precision flavour data require their masses to be above $\gtrsim 10$ TeV, in order to avoid too large FCNCs [188]. Three degrees of freedom of $\Delta_R$ are eaten by the heavy $W_R$ and $Z_R$ bosons, leaving only a CP-even neutral scalar $\phi = \text{Re}(\Delta_R^0)$ and the right-handed doubly-charged scalar $\phi_R^{\pm\pm} = \Delta_R^{\pm\pm}$ [174].

The neutral scalar $\phi$ does not couple directly to the charged leptons. The LFV couplings $\phi\ell_\alpha^\pm \ell_\beta^\mp$ can be induced at tree-level from mixing with the heavy scalar $H_1$ [153], and at 1-loop from $W_R$ – right-handed neutrino loops and from doubly-charged scalar – charged lepton loops [189]. In the limit of the small mixing with the SM Higgs, the LFV decays $\phi \to \ell_\alpha^\pm \ell_\beta^\mp$ are the dominant decay modes. They are at least sizable, as long as the effective coupling in Eq. (75) is above $h_{\alpha\beta} \gtrsim 10^{-4}$. This can be achieved easily in the LRSM [153]. If LFV from the neutral scalar decay were to be found at CLIC, it would imply that the right-handed scale $v_R$ in the LRSM is not too far above the TeV, and that the heavy scalars from the bidoublet might be detectable at the future 100 TeV hadron colliders [174, 190] such as FCC-hh [191, 192] and SPPC [193].

If both $\Delta_L$ and $\Delta_R$ from the LRSM have TeV-scale masses, the left-handed doubly-charged scalar $\phi_L^{\pm\pm} = \Delta_L^{\pm\pm}$ behaves the same way as in the type-II see-saw, as long as the type-II see-saw contribution to neutrino masses dominate over the type-I see-saw [143–147]. The Yukawa couplings of $\phi_R^{\pm\pm}$ are given by

$$\mathcal{L}_Y = -(f_R)_{\alpha\beta} \bar{\psi}_{R,\alpha}^c \Delta_R \psi_{R,\beta} + \text{h.c.}, \qquad (79)$$

with $\psi_R = (N, \ell_R)^\mathsf{T}$ are the LRSM right-handed lepton doublets, with $N$ the heavy right-handed neutrino (RHN). The RHNs obtain their masses via the matrix $M_N = 2 f_R v_R$, where $v_R = \langle \Delta_R^0 \rangle$ is the right-handed VEV. The LFV couplings $(f_R)_{\alpha\beta}$ of $\phi_R^{\pm\pm}$ are closely related to the RHNs and their mass generation in the LRSM. If the $(f_R)_{\alpha\beta}$ were measured in the single production of $\phi_R^{\pm\pm}$ at CLIC, they would provide useful information on the RHN mixings as well as guidelines for direct searches for RHNs at future hadron and lepton colliders [194, 195], as well as for the indirect searches using low-energy high-precision experiments.

### 3.3 Implications of LFUV anomalies[34]

A specific feature of the SM is the universality of the gauge interactions in the lepton sector. This property leads to theoretically clean predictions for a series of observables that can be tested experimentally. Recent experimental results, from LHCb and from $B$-factories, are showing departures from the SM predictions at the $4\sigma$ level, suggesting possible hints of Lepton Flavour Universality Violation (LFUV) sources beyond the SM. These so called "flavour anomalies" can be classified into two sets:

- **LFUV in quark charged currents** $b \to c\tau\nu$: The anomalous observables are $R_D$ and $R_{D^*}$, the ratios of $B \to D^{(*)}\tau\nu$ to $B \to D^{(*)}\ell\nu$ branching fractions [196–201]. Global fits [202] put the combined statistical significance of the deviation just above the $4\sigma$ level. Assuming New Physics (NP), a good description of the data is obtained invoking NP in operators such as

$$\mathcal{L}_{\text{eff}} \supset -G_{\text{CC}} \left( \bar{c}_L \gamma^\mu b_L \right) \left( \bar{\tau}_L \gamma_\mu \nu_L \right), \qquad G_{\text{CC}}^{-1/2} \simeq 2.4 \, \text{TeV}. \qquad (80)$$

Other flavour data are compatible with such NP effects in processes involving the $\tau$ lepton (while analogous $B$-decays with electrons and muons in the final states exhibit a SM-like behaviour). What matters for CLIC is the large value of the associated NP Fermi constant, which leads to possibly observable effects both in direct and in indirect searches. In this brief study we present two strategies to explore the physics associated to the charged current anomalies at CLIC:

1. *Direct searches of the flavour mediator*
   The NP resonances cannot be much heavier than the NP scale hinted at by $G_{\text{CC}}$, Eq. (80). This means the flavour mediator could be produced on-shell at CLIC. We focus on the case of leptoquarks coupled with large couplings to the third family of quarks and leptons. This scenario is well motivated both theoretically and phenomenologically as shown in several papers [203]. We explore this possibility in Section 3.3.1.

2. *Indirect searches in $e^+e^- \to \tau^+\tau^-$*
   Once the effective operator (80) is embedded into operators respecting the $SU(2)_L \times U(1)_Y$ gauge symmetry, there are further phenomenological implications. The 1-loop contributions to the 4-lepton operators generate corrections to the $e^+e^- \to \tau^+\tau^-$ cross section that grow with energy and are at the level comparable with the CLIC sensitivity. This possibility is explored in Section 3.3.2.

---
[34] Based on a contribution by A. Greljo, M. Nardecchia, D. Marzocca and D. Buttazzo.

- **LFUV in quark neutral currents** $b \to s\ell\ell$: Another hint of LFUV beyond the SM is coming from the theoretically clean observables $R_K$ and $R_K^*$, the ratios of $b \to s\mu^+\mu^-$ to $b \to se^+e^-$ transitions [204, 205]. A departure from the SM would signal the presence of NP in muon and/or the electron modes. Assuming NP, a good description of the data is obtained with the NP operators such as

$$\mathcal{L}_{\text{eff}} \supset G_{\text{NC}} \left(\bar{s}_L \gamma^\mu b_L\right)\left(\bar{\ell}_L \gamma_\mu \ell_L\right), \qquad \ell = e, \mu, \qquad |G_{NC}|^{-1/2} \simeq 35 \text{ TeV}. \tag{81}$$

Note that $|G_{\text{NC}}| \ll |G_{\text{CC}}|$ so that in various explicit realizations the NP giving rise to $|G_{\text{NC}}|$ can be easily hidden from direct searches. The experimental data that goes beyond the $R_K$ and $R_K^*$ measurements, such as the angular observables in the $B \to K^*\mu\mu$ decay, favours the case of NP predominantly in the muon sector. However, these additional observables are subject to larger theoretical uncertainties. While disfavoured by these larger set of observables, NP with large couplings to the electron is still an open option. In this case CLIC could play a very important role, a possibility that we explore in Section 3.3.3.

Before proceeding, it is important to keep in mind that, even if the current experimental situation of the flavour anomalies were to change substantially with the future measurements at LHCb and Belle II, several aspects of the present study should remain useful. For example, even if the flavour anomalies are to disappear, the analysis in Section 3.3.1 can still be viewed as a representative case of searches at CLIC for leptoquarks decaying to third family fermions. On the other hand, were the evidence of flavour anomalies to get reinforced by the future data, it is still possible that these more precise measurements could suggest somewhat different signatures and search strategies at CLIC than the ones presented here. The presented studies should thus be viewed only as a rough guide on the CLIC capabilities to address the origin of flavour anomalies. In particular, we decided to focus on sample benchmarks that currently appear to us to be well motivated both at the theoretical and the phenomenological level.

### 3.3.1 Direct searches for leptoquarks mediators of the anomalies

Leptoquarks are a class of mediators that can explain the observed flavour anomalies. They can be realised in several different concrete models. We focus on two examples, where both the CC and the NC anomalies are explained simultaneously.

The first example is a mediator that is a vector leptoquark $U_\mu = (\mathbf{3}, \mathbf{1}, 2/3)$, a color triplet, electroweak singlet state with a hypercharge 2/3. This is the only example of a single mediator that can provide a good fit to both the $b \to c\tau\nu$ and the $b \to s\ell\ell$ flavour anomalies without further contributions from additional states. The leptoquark $U_\mu$ couples to the left-handed quarks and leptons through flavoured couplings $\beta_{i\alpha}$,

$$\begin{aligned}\mathcal{L}_{\text{vector}} = &-\frac{1}{2}U_{\mu\nu}^\dagger U^{\mu\nu} + M_U^2 U_\mu^\dagger U^\mu + g_U \beta_{i\alpha}(\bar{q}_L^i \gamma^\mu \ell_L^\alpha)\, U_\mu + \text{h.c.}+\\ &+ g_s \kappa_s U_\mu^\dagger T^a U_\nu G^{a\mu\nu} + g_Y \kappa_Y U_\mu^\dagger U_\nu B^{\mu\nu}\,,\end{aligned} \tag{82}$$

where $U_{\mu\nu} = D_\mu U_\nu - D_\nu U_\mu$. The last two terms are model-dependent non-minimal couplings to gluons and hypercharge, whose strengths are controlled by parameters $\kappa_{s,Y}$. Their values affect strongly the pair production cross section at the LHC and CLIC, respectively. A state with quantum numbers of $U_\mu$ appears naturally as a gauge vector in models based on the Pati-Salam group [206–210], and can be easily embedded as a composite resonance in models with a strongly interacting sector at the TeV scale [211–213].

A second class of models that are able to fit the observed anomalies contains two scalar leptoquarks of hypercharge 1/3, the electroweak singlet $S_1 = (\bar{\mathbf{3}}, \mathbf{1}, -1/3)$ and the triplet $S_3 = (\bar{\mathbf{3}}, \mathbf{3}, -1/3)$, [203, 214, 215],

$$\mathcal{L}_{\text{scalar}} = \mathcal{L}_{\text{kin}} + g_1 y_{i\alpha}^{(1)} (\bar{Q}_L^{ci}\epsilon L_L^\alpha) S_1 + g_3 y_{i\alpha}^{(3)} (\bar{q}_L^{ci}\epsilon \sigma^a \ell_L^\alpha) S_3^a + \text{h.c.}. \tag{83}$$

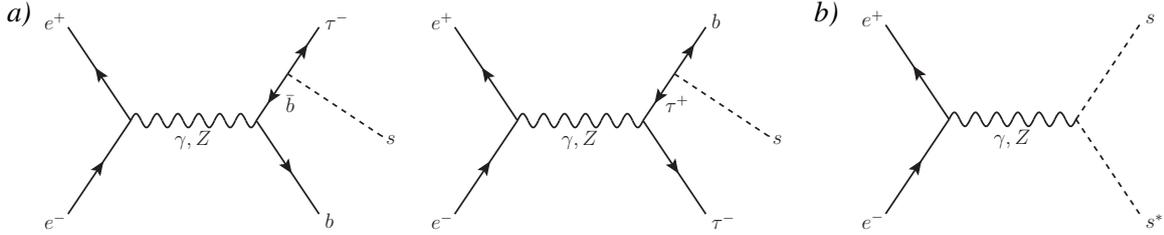

Figure 44: Single (a) and double (b) production of a scalar leptoquark $s_{4/3}$ with charge $4/3$. Analogous diagrams, but with $b \leftrightarrow \bar{b}$, contribute to production of vector leptoquarks.

Models have been proposed where the two leptoquark states are the pseudo-Goldstone bosons of a spontaneously broken symmetry, and are part of a strongly interacting sector together with a composite Higgs boson [215].

In the remainder of this section we investigate quantitatively the possibility of directly producing leptoquarks with a dominant coupling to third generation fermions at high-energy lepton colliders. Focusing on the above two cases of scalar leptoquarks and a vector leptoquark, in either case coupled to a $b - \tau$ current, we explore how well the leptoquarks with masses above 1.5 TeV can be searched for at CLIC. Lighter resonances will be probed by the LHC before the start of the high-luminosity program. For this reason, we consider as our only benchmark the final stage of the CLIC program, with a centre-of-mass energy of 3 TeV and an integrated luminosity of 3 ab$^{-1}$.

There are two main production modes for leptoquarks at the $e^+e^-$ colliders:

1. Single production through the coupling to the third-generation fermions, with a leptoquark radiated from a $b$-quark or a $\tau$ lepton, see Figure 44 (a). Similar processes involving top quarks and neutrinos are also possible, but for simplicity we restrict our analysis to the case of down-type fermions, which makes a very direct connection to the flavour anomalies.

2. Double production through the coupling to gauge vectors, as shown in Figure 44 (b), is relevant only for leptoquark masses $m < \sqrt{s}/2$. While the coupling of the scalar leptoquark to the photon is completely fixed by its electric charge, the vector leptoquarks can have in addition a non-minimal, model dependent, coupling to photons, see Eq. (82).

Both processes – single and double production – result in a $2b2\tau$ final state. We focus our numerical estimates for the CLIC reach on this final state. In the case of the scalar leptoquarks it has been shown that the $2b2\tau$ channel is the most sensitive one at the LHC [215], mainly because the component of $S_3$ with charge $Q_S = -4/3$ decays exclusively into $b\tau$. In the case of $U_\mu$, the other channel, $2t2\nu$, could be equally important, and a dedicated analysis for CLIC reach may be desired. Figure 45 shows the $e^+e^- \to 2b2\tau$ cross-section of the signal, calculated at tree-level with `MadGraph5_aMC@NLO` [31] using UFO model files from Ref. [216], for the two cases of vector and scalar leptoquarks, as a function of the leptoquark mass and of the coupling to fermions. Note that in the case of the vector leptoquark the relevant branching ratio is $Nr(U \to \bar{b}\tau^-) = 0.50$.

The irreducible SM background from EW $b\bar{b}\tau^+\tau^-$ production has a total cross-section of about 0.2 fb, mainly coming from $Zbb$ and $Z\tau\tau$. The $Z$ pole is eliminated by requiring the invariant masses of both $b$ and $\tau$ pairs to be greater than 200 GeV. This reduces the cross-section of the irreducible SM background to $3.4 \times 10^{-3}$ fb. The background is reduced further by imposing additional kinematic cuts: the decay products of the leptoquarks are required to have energies greater than $M^2/(2\sqrt{s})$, while the energies of the other two fermions need to be smaller than $(s - M^2)/(2\sqrt{s})$. Figure 46 shows the $b\tau$ invariant mass distribution of the SM background, $e^+e^- \to 2b2\tau$. The $b$-jet and $\tau$ lepton pair that is assumed to come from the decay of the leptoquark is chosen as the one that has the $b\tau$ invariant mass closest to the assumed leptoquark mass. In Figure 46 we also show the resonant signal from $U \to b\tau$, calculated at parton-level and applying a Gaussian smearing, assuming a 30% resolution on the resonance mass (at 95% CL) after tau reconstruction.

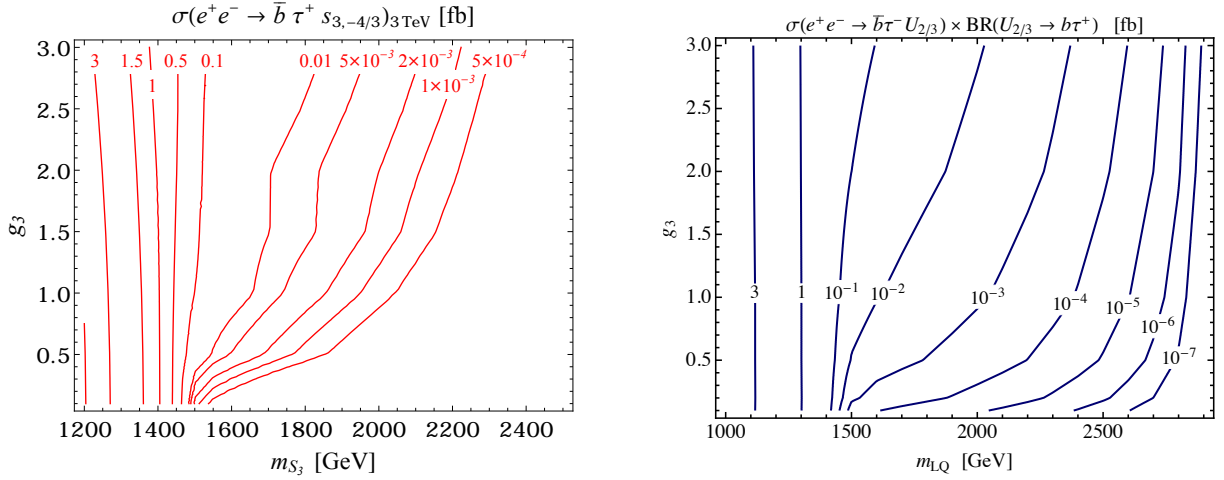

Figure 45: Total production cross section [fb] at 3 TeV as a function of leptoquark mass and coupling $g_3$ to third-generation fermions. Left: scalar leptoquark, $e^+e^- \to \bar{b}\tau^+ s_{3,-4/3}$. Right: vector leptoquark $e^+e^- \to b\tau^+ U_1$.

For light enough leptoquarks, below about 2 TeV for $g_{b\tau} = 1$, the background is negligible compared to the signal. We therefore estimate the reach of CLIC requiring the presence of at least 10 signal events. Given the strong dependence of the cross-section on the mass of the leptoquark, as shown in Figure 46 (right), relaxing this assumption will not have a significant impact on the reach.

In Figure 47 we compare the reach of CLIC for the scalar (left) and vector leptoquarks (right) with the one of LHC, showing the present exclusion and the expected reach from single and pair production after 3 ab$^{-1}$. In the same plots, we also show the values of masses and couplings that give the best fit to the flavour anomalies, under the assumption that the dominant coupling is to the third-generation fermions only, following the results of [203].

These results show that CLIC provides a unique opportunity to directly search for leptoquarks coupled to $b$ and $\tau$, in a range of masses and couplings interesting for the present flavour anomalies, but difficult to access at the LHC.

### 3.3.2 Indirect effects in $e^+e^- \to \tau^+\tau^-$

The combined solution to the $B$ anomalies requires at some high scale $\Lambda$ two semileptonic effective operators (see Ref. [203])

$$\mathcal{L}_{NP}^0(\Lambda) = \frac{1}{\Lambda^2}\left[C_S\left(\bar{q}_L^3\gamma^\mu q_L^3\right)\left(\bar{\ell}_L^3\gamma^\mu \ell_L^3\right) + C_T\left(\bar{q}_L^3\gamma^\mu\sigma^a q_L^3\right)\left(\bar{\ell}_L^3\gamma^\mu\sigma^a \ell_L^3\right)\right]. \tag{84}$$

In the following we assume that all the other operators with different flavour indices, either flavour-diagonal or flavour-violating, are suppressed by small CKM elements, as for example predicted by the approximate $U(2)_q \times U(2)_\ell$ symmetry, and are therefore negligible for the following analysis.

It has been pointed out in Refs. [217, 218] that at the loop level these operators contribute to the very well measured $Z\tau\tau$, $Z\nu\nu$ couplings, as well as to the $\tau$ decays. In the EFT these contributions are described by the renormalization group (RG) evolution of the two operators in Eq. (84) from the scale $\Lambda$ down to the $Z$ and $\tau$ mass scales. The RG evolution mixes the initial two operators into the operators that affect $Z$ couplings and contribute to $\tau$ decays:

$$\Delta g_{\nu_L}^{\tau\tau} = \frac{v^2}{16\pi^2\Lambda^2}\log\frac{\Lambda}{\mu}\left(\frac{1}{3}g_1^2 C_S - g_2^2 C_T + 3y_t^2(C_S + C_T)\right), \tag{85}$$

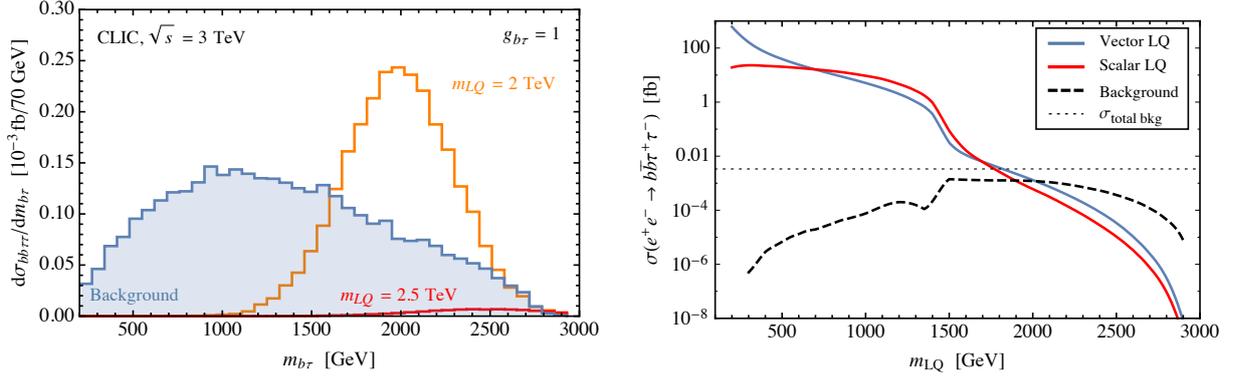

Figure 46: Left: Distribution for the $2b2\tau$ cross-section of the SM background in the invariant mass of the $b\tau$ pair (blue), at $\sqrt{s} = 3$ TeV, and two examples of a signal, a vector leptoquark with either $m_U = 2$ TeV (orange) and $m_U = 2.5$ TeV (red), taking $g_{b\tau} = 1$. Right: production cross-section times branching ratio into $b\tau$, as a function of the leptoquark mass, for the vector (blue) and scalar (red) leptoquarks, with $g_{b\tau} = 1$; the thick dashed line is the cross-section of the background after cuts, and the thin dotted line corresponds to $\sigma_{\text{bkg}} = 3.4 \times 10^{-3}$ fb ($Z$-pole cut only).

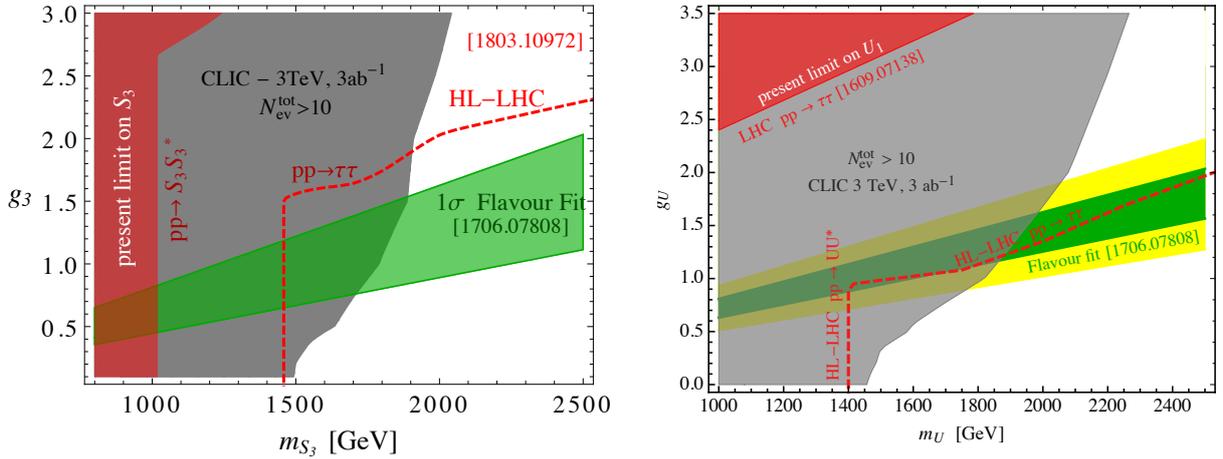

Figure 47: Comparison between the reach of CLIC and LHC in the $m_{\text{LQ}} - g_3$ plane. The gray region represents the parameter space where the expected total number of events for leptoquark production in the $b\tau$ channel will be larger than 10 at 3 TeV CLIC with $3\,\text{ab}^{-1}$ of luminosity; the red region is the present exclusion from LHC, combining pair production (vertical bound) and $\tau\tau$ searches; the red dashed line is the reach of HL-LHC. The green and yellow bands are the $1\sigma$ and $2\sigma$ preferred regions from the flavour fit of Ref. [203]. Left: scalar leptoquark $S_3$, from Ref. [215]. Right: vector leptoquark $U_1$, bounds quoted in Ref. [203].

$$\Delta g_{e_L}^{\tau\tau} = \frac{v^2}{16\pi^2\Lambda^2}\log\frac{\Lambda}{\mu}\left(\frac{1}{3}g_1^2 C_S + g_2^2 C_T + 3y_t^2(C_S - C_T)\right), \quad (86)$$

$$\Delta g_{u_L}^{33} = -\frac{v^2}{16\pi^2\Lambda^2}\log\frac{\Lambda}{\mu}\frac{1}{3}\left(g_1^2 C_S + g_2^2 C_T\right), \quad (87)$$

$$\Delta g_{d_L}^{33} = -\frac{v^2}{16\pi^2\Lambda^2}\log\frac{\Lambda}{\mu}\frac{1}{3}\left(g_1^2 C_S - g_2^2 C_T\right). \quad (88)$$

The renormalization group scale $\mu$ is to be identified with the energy at which the measurements are performed, either $\mu \sim m_Z$ at the $Z$ pole or $\mu \sim \sqrt{s}$, more generally.

While CLIC is not expected to improve substantially the measurements on these observables compared to LEP, it will instead measure very precisely the $e^+e^- \to f\bar{f}$ cross section at various energies $\sqrt{s}$. For values $\sqrt{s} \ll \Lambda$ one can use the EFT approach to describe the RG contributions from those two operators to the $Z$ couplings as well as to $e^+e^- f\bar{f}$ contact terms. While the former effect does not grow with the energy, the latter does and therefore is expected to be more important at large $\sqrt{s}$. The four-fermion contact interactions relevant for CLIC are (in the Warsaw basis)

$$\delta\mathcal{L}_L = \left[\left(\frac{2}{3}g_1^2 C_S + 2g_2^2 C_T\right)(\bar{\ell}_L^3\gamma^\mu\ell_L^3)(\bar{\ell}_L^1\gamma^\mu\ell_L^1) - 4g_2^2 C_T(\bar{\ell}_L^3\gamma^\mu\ell_L^1)(\bar{\ell}_L^1\gamma^\mu\ell_L^3) \right. \\ \left. +\frac{4}{3}g_1^2 C_S(\bar{\ell}_L^3\gamma^\mu\ell_L^3)(\bar{e}_R^1\gamma^\mu e_R^1)\right]\frac{1}{16\pi^2\Lambda^2}\log\frac{\Lambda}{\mu}, \quad (89)$$

$$\delta\mathcal{L}_{SL} \supset \left[-\frac{2}{3}g_1^2 C_S(\bar{q}_L^3\gamma^\mu q_L^3)(\bar{\ell}_L^1\gamma^\mu\ell_L^1) - \frac{2}{3}g_2^2 C_T(\bar{q}_L^3\gamma^\mu\sigma^a q_L^3)(\bar{\ell}_L^1\gamma^\mu\sigma^a\ell_L^1) \right. \\ \left. -\frac{4}{3}g_1^2 C_S(\bar{q}_L^3\gamma^\mu q_L^3)(\bar{e}_R^1\gamma^\mu e_R^1)\right]\frac{1}{16\pi^2\Lambda^2}\log\frac{\Lambda}{\mu}. \quad (90)$$

The estimate for the size of the effect due to these contact terms is

$$\frac{\delta\sigma}{\sigma_{\text{SM}}} \sim \frac{C_{S,T}}{16\pi^2}\frac{s}{\Lambda^2}\log\frac{\Lambda}{\sqrt{s}}. \quad (91)$$

A precise measurements of the $e^+e^- \to \tau^+\tau^-$, $b\bar{b}$, or $t\bar{t}$ cross sections at CLIC, with sub-percent precision, could test these effects.

For energies $\sqrt{s} \gtrsim \Lambda$ the EFT approach cannot be used, but the complete model with the explicit mediator inside the loop will give effects of similar size as the above EFT estimates.

### 3.3.3 Addressing $R(K^{(*)})$ with new physics in electrons

Anomalies in the $R(K^{(*)})$ ratios could well be due to NP that results in an excess of electrons rather than in a deficit of muons. Such NP can be studied at CLIC through the $e^-e^+ \to jj$ process. A good fit to data is obtained with the short-distance new physics contributions to

$$\mathcal{L} \supset \frac{1}{(36 \text{ TeV})^2}C_{XY}(\bar{s}\gamma_\mu P_X b)(\bar{e}\gamma^\mu P_Y e), \quad (92)$$

where the three best fit scenarios have been identified as $C_{LL}^{\text{BSM}} = 1.7$, $C_{LR}^{\text{BSM}} = -5.1$ or $C_{RR}^{\text{BSM}} = -5.6$ [219]. Here, $P_L$ and $P_R$ denote the left- and right-chirality projectors, respectively.

Such NP could well be beyond the mass reach of CLIC for on-shell particle production. However, indirect effects can still be observable as a deviation in the $e^+e^- \to jj$ cross section and angular distributions. The impact of new contact interactions on $e^+e^- \to b\bar{b}$ was studied in [11], where it was shown that with 1 ab$^{-1}$ of data at the $\sqrt{s} = 3$ TeV CLIC, a relative precision of $\epsilon(\sigma_{b\bar{b}}) \approx 1\%$ on the total cross section is achievable. The dominant correction to the cross section from the contact interaction is due to the interference with the SM contributions. The projected limit on the contact interaction $(\bar{b}\gamma_\mu P_L b)(\bar{e}\gamma^\mu P_L e)$ in Eq. (92) is thus $|C_{LL}^{bb}| \lesssim 1$ (0.1), depending on the polarisation choice of initial state electrons and positrons [11]. The precise limit depends slightly on the assumed systematic uncertainty. Similar limits are derived for other chiralities.

In a motivated class of models, in which the quark current $(\bar{s}\gamma_\mu P_L b)$ in (92) is due to an $SU(2)_L$ invariant operator with an MFV-like flavour structure, the pure third generation interactions, $(\bar{b}\gamma_\mu P_L b)$, are expected to be $1/V_{ts} \sim 25$ enhanced. In this class of models CLIC has an excellent potential to discover and characterise NP in $e^+e^- \to b\bar{b}$ process. In fact, in the models that predict $(\bar{s}\gamma_\mu P_X b)$ and $(\bar{b}\gamma_\mu P_X b)$ to be of a similar size, CLIC will still be a good probe.

CLIC could potentially probe directly the $(\bar{s}\gamma_\mu P_X b)(\bar{e}\gamma^\mu P_Y e)$ contact interaction. In contrast to $e^+e^- \to bb$, which arises at tree level in the SM, the SM background in $e^+e^- \to bs$ is generated only at one loop and is further suppressed by the small CKM element $V_{ts}$. A good estimate of the CLIC sensitivity can then be obtained by assuming that the transition is NP dominated. Naively rescaling previous results, we expect $|C_{LL}| \lesssim 30$, which is not enough to probe the best fit point directly using $e^+e^- \to bs$ production. This could potentially be improved with an excellent strange quark tagging technique.

Finally, CLIC has also an excellent opportunity to discover a potential signal in the $e^+e^- \to e^+e^-$ channel that arises in a class of models in which the $R(K^{(*)})$ anomalies are explained by a tree-level exchange of a $Z'$ boson. The main feature of these models is much stronger $(\bar{e}\gamma^\mu P_Y e)Z'_\mu$ interaction term compared to $(\bar{s}\gamma_\mu P_X b)Z'_\mu$, due to the stringent $B_s$ meson mixing constraint which typically requires $V_{ts}$ suppression in the later term. Previous discussion, coupled with the fact that $e^+e^-$ final state is cleaner than $b\bar{b}$, suggests good prospects at CLIC.

## 3.4 Exotic top decays

CLIC will produce a large set of clean $t\bar{t}$ events, which can be used to search for nonstandard top decays. In this section we cover both the expectations within common NP models, as well as the experimental projections for the searches at CLIC.

### 3.4.1 Sizes of $t \to ch, cZ, c\gamma, c+$MET in new physics models[35]

We first review the typical expectations, within various BSM models, for the sizes of flavour violating top decays, $t \to ch, t \to c\gamma, t \to c + \text{MET}$.

#### 3.4.1.1 The $t \to ch$ decay

The flavour violating $t \to ch$ and $t \to uh$ couplings are absent at tree level in the SM, but can be induced by NP effects. Below the scale of electroweak symmetry these couplings are described by a simple generic Lagrangian,

$$\mathcal{L} = y_{RL}(\bar{c}_R t_L)h + y_{LR}(\bar{c}_L t_R)h + \text{h.c.}, \tag{93}$$

and similarly for up quarks. The resulting $t \to ch$ partial decay width is [220]

$$\Gamma(t \to ch) = \frac{1}{32\pi}\left(|y_{LR}|^2 + |y_{RL}|^2\right) m_t \left(1 - m_h^2/m_t^2\right), \tag{94}$$

to be compared with the dominant $t \to bW$ partial decay width, which at leading order (LO) is given by,

$$\Gamma_t = \Gamma(t \to bW) = \frac{g^2}{64\pi}|V_{tb}|^2 \frac{m_t}{x_W^2}\left(1 - 3x_W^4 + 2x_W^6\right), \tag{95}$$

where $x_W \equiv m_W/m_t$. With the projected CLIC sensitivity of $\text{Br}(t \to ch) < 2.1(1.2) \times 10^{-4}$ for $500(1000)$ fb$^{-1}$ of data with an electron beam polarization of $-80\%$ at $\sqrt{s} = 380$ GeV, see Section 3.4.2, off-diagonal Yukawa couplings (93) larger than

$$\sqrt{|y_{LR}|^2 + |y_{RL}|^2} \gtrsim 0.021(0.015), \tag{96}$$

can be probed.

To appreciate the relevance of the above experimental reach it is worthwhile performing an analysis both within SM EFT as well as surveying a selection of NP models. We start with the SM EFT discussion. While the Higgs flavour violating interactions are absent in the renormalizable SM at the tree level, this is no longer true for SM EFT due to the dimension six operators. Starting from the Lagrangian

$$-\mathcal{L} = (\hat{y}_u)_{ij}\bar{q}_L^i \tilde{H} u_R^j + \frac{H^\dagger H}{\Lambda^2}(\hat{\epsilon}_u)_{ij}\bar{q}_L^i \tilde{H} u_R^j + \text{h.c.}, \tag{97}$$

---
[35]Based on a contribution by A. Azatov and A. Paul.

Table 25: Typical values for the flavour violating top decay branching ratios in several NP models, see the main text for details.

| | SM | MSSM | 2FHDM | CH | RS |
|---|---|---|---|---|---|
| Br$(t \to ch)$ | $3 \times 10^{-15}$ | $< 10^{-5}$ | $< 10^{-3}$ | $< 1.3 \times 10^{-5}$ | $< 2.6 \times 10^{-5}$ |
| Br$(t \to c\gamma)$ | $4.6 \times 10^{-14}$ | $< 5 \times 10^{-7}$ | $< 3.4 \times 10^{-6}$ | $< 5 \times 10^{-9}$ | $< 5 \times 10^{-9}$ |

the mass matrix and the Yukawa couplings are

$$(\hat{m}_u)_{ij} = \frac{v}{\sqrt{2}}(\hat{y}_u)_{ij} + \frac{v^3}{2\sqrt{2}\Lambda^2}(\hat{\epsilon}_u)_{ij}, \tag{98}$$

$$(\hat{Y}_u)_{ij} = \frac{(m_u)_{ij}}{v} + \frac{v^2}{\sqrt{2}\Lambda^2}(\hat{\epsilon}_u)_{ij}. \tag{99}$$

Since the $3 \times 3$ complex matrices $(\hat{\epsilon}_u)_{ij}, (\hat{y}_u)_{ij}$ have in general different origins, they typically cannot be simultaneously diagonalized. The misalignment between the two matrices induces the flavour violating $t \to ch$ decay. Its branching ratio depends on the size of off-diagonal elements of $(\hat{Y}_u)_{ij}$ in the mass eigenbasis, i.e., $y_{LR} = (\hat{Y}_u)_{23} \neq 0$ and $y_{RL} = (\hat{Y}_u)^*_{32} \neq 0$ in Eq. (93), and on the scale of NP, $\Lambda$, compared to the electroweak vev, $v$. The $t \to ch$ branching ratio vanishes for $\Lambda \to \infty$.

In Table 25 we report predictions for the $t \to ch$ branching ratio for a number of NP models:

- **SM and MSSM**: In the SM and MSSM the $t \to ch$ transition is generated at one loop (see [220] for the SM, and [221, 222] for the MSSM predictions). The resulting flavour violating branching ratio is therefore strongly suppressed, and is unobservable at CLIC.
- **Flavourful 2HDM**: The 2HDM of Type-III has by definition nonzero flavour off-diagonal Yukawa coupling. The non-holomorphic couplings of the Higgs to quarks then in general generates FCNCs in the quark sector. The resulting branching fractions are collected, e.g., in [223]. In particular, in the model proposed in [224, 225] the $y_{RL}$ coupling for $t \to ch$ transition can be significantly enhanced,

$$\mathcal{L} = -\frac{a}{2v_{SM}} m_t \sin\rho \left(h\bar{c}_R t_L + \text{h.c.}\right), \quad \text{where} \quad a = (\tan\beta + \cot\beta)\cos(\beta - \alpha). \tag{100}$$

Here $\rho$ is a parameter describing the mixing in the right-handed quark sector. Current constraints from direct and indirect searches only weakly bound $a, \rho$ parameters so that $a, \rho \sim \mathcal{O}(1)$ are still consistent with the experimental data and the branching ratios $Br(t \to ch) \lesssim 10^{-3}$ are allowed.
- **Composite Higgs**: In composite Higgs models with partial compositeness [15] the $t \to hc$ flavour violating Yukawa couplings in (93) are of the order of [226]

$$y_{RL} \sim \frac{g_*}{M_*} \frac{m_t}{M_*} \frac{m_c}{V_{cb}} \sim 5 \times 10^{-3} \left(\frac{\text{TeV}}{M_*}\right)\left(\frac{\text{TeV}}{M_*/g_*}\right), \tag{101}$$

$$y_{LR} \sim \frac{g_*}{M_*} \frac{m_t}{M_*} m_t V_{cb} \sim 1.1 \times 10^{-3} \left(\frac{\text{TeV}}{M_*}\right)\left(\frac{\text{TeV}}{M_*/g_*}\right), \tag{102}$$

where $g_*$ and $M_*$ are the typical coupling and mass of the composite resonances. For $Br(t \to hc)$ to be observable at CLIC requires a very low scale of compositeness. In this case there is a connection between the $t \to ch$ rate and the modification of the Higgs couplings to the vector bosons, $\Delta\kappa_V$,

$$\text{Br}(t \to ch) \simeq \frac{0.014}{g_*^2}(\Delta\kappa_V)^2. \tag{103}$$

Using presently allowed deviation $\Delta\kappa_V \sim 0.1$ gives Br$(t \to ch) \sim 10^{-5}$ for $g_* \sim 5$, out of CLIC reach unless $g_*$ is uncharacteristically small. Note that $\Delta\kappa_V$ is expected to be measured with per mille level precision at CLIC (see Section 2.1).

– **Randall-Sundrum warped models [227]**: In this class of models the estimates for $y_{RL}, y_{LR}$ are very similar to the composite Higgs models [228]

$$y_{RL} \sim \frac{g_*}{M_*}\frac{g_* v}{M_*}\frac{m_c}{V_{cb}} \sim 7 \times 10^{-3} \left(\frac{\text{TeV}}{M_*/g_*}\right)^2, \tag{104}$$

$$y_{LR} \sim \frac{g_*}{M_*}\frac{g_* v}{M_*} m_t V_{cb} \sim 1.7 \times 10^{-3} \left(\frac{\text{TeV}}{M_*/g_*}\right)^2. \tag{105}$$

Again, the possible observation of $t \to ch$ should be accompanied with stronger observation in the other channels, for example, a deviation of $hVV$ coupling from the SM expectations.

#### 3.4.1.2 The $t \to c\gamma$ transition

The $t \to c\gamma$ flavour violating decay can be generated by the effective Lagrangian [229]

$$\mathcal{L}_{\text{eff}} \supset A_{LR}(\bar{c}P_R\sigma_{\mu\nu}t)F^{\mu\nu} + A_{RL}(\bar{c}P_L\sigma_{\mu\nu}t)F^{\mu\nu} + \text{h.c.}, \tag{106}$$

giving the partial decay width

$$\Gamma(t \to c\gamma) = \frac{2}{\pi}\left(\frac{m_t^2 - m_c^2}{2m_t}\right)^3 \left(|A_{LR}|^2 + |A_{RL}|^2\right). \tag{107}$$

CLIC with 500 fb$^{-1}$ of data at $\sqrt{s} = 380$ GeV will be able to test this decay up to the precision $\text{Br}(t \to c\gamma) \lesssim 4.7 \times 10^{-5}$, cf. (114), which translates into

$$|A_{LR}| \lesssim 2 \times 10^{-5}\text{GeV}^{-1}. \tag{108}$$

In Table 25 we report the expectations for this branching ratio for various BSM scenarios:

– **SM and MSSM**: The $t \to c\gamma$ transition is generated at one loop, so that the resulting branching ratio is strongly suppressed and is unobservable at CLIC (for the SM values [220, 221]).
– **Randall-Sundrum and Composite Higgs**: For this class of models we can estimate the flavour violating dipole moments by multiplying with $1/(16\pi^2 v)$ the estimates in Eq. (104)

$$A_{LR} \sim 4.3 \times 10^{-8}\left(\frac{\text{TeV}g_*}{M_*}\right)^2 \text{GeV}^{-1}, \quad A_{RL} \sim 2 \times 10^{-7}\left(\frac{\text{TeV}g_*}{M_*}\right)^2 \text{GeV}^{-1}, \tag{109}$$

which makes this channel impossible to test at CLIC.
– **Flavourful 2HDM**: In the absence of full calculation we assume that the dipole loop is dominated by the SM Higgs exchange, which gives,

$$A_{RL} \sim 5.2 \times 10^{-6} a \sin\rho \Rightarrow Br(t \to c\gamma) \sim 3.4 \times 10^{-6} a^2 \sin^2\rho, \tag{110}$$

which is about an order of magnitude below CLIC reach. Note that $t \to c\gamma$ will always be accompanied by an even larger deviation in $t \to ch$.

#### 3.4.1.3 The $t \to c + $ MET transitions

A BSM scenario that leads to the $t \to c +$ MET signature is a flavour violating of light dark matter to quarks [230]. Taking as an example the effective dimension 6 operator,

$$\mathcal{L}_{\text{eff}} = \frac{1}{\Lambda^2}(\bar{\chi}\gamma_\mu\chi)(\bar{c}_R\gamma_\mu t_R) + \text{h.c.}, \tag{111}$$

the CLIC sensitivity of $Br(t \to c + \text{MET}) \lesssim (2-5) \times 10^{-4}$, see Figure 48 and [230], corresponds to

$$\frac{c}{\Lambda^2} \gtrsim 7 \times 10^{-6} \text{GeV}^{-2}. \tag{112}$$

If (111) is due to a tree level exchange of a $Z'$ that couples to DM current with strength $g_\chi$ and to the $\bar{c}_R \gamma_\mu t_R$ current with coupling $g_{tc}$, this translates into the bound

$$\frac{g_\chi g_{tc}}{M_{Z'}^2} \sim \frac{c}{\Lambda^2} \gtrsim 7 \times 10^{-6} \text{GeV}^{-2}. \tag{113}$$

As is commonly the case the direct searches for $Z'$ are more sensitive, e.g., for 800 GeV $Z'$ the LHC bound is $g_\chi g_{tc}/M_{Z'}^2 \gtrsim 4 \times 10^{-6}$. The two searches do have different model dependencies, though, with the interpretation of the direct search bound changing, if the $Z' \to tc$ and $Z' \to MET$ are not the only open channels.

### 3.4.2 Experimental prospects for FCNC top decays [36]

In the CLIC baseline staging scenario, the cross section of top-quark pair-production is highest at the forseen lowest energy stage at $\sqrt{s} = 380$ GeV with 723 fb corresponding to 346,000 top pair events with an integrated luminosity of $500\,\text{fb}^{-1}$. This makes the first energy stage particularly suitable for searching for rare decays of pair-produced top quarks. The experimental sensitivity of this search is dominated by the efficiency of reconstructing the rare decays and by the effectiveness of the background suppression. The reconstruction of events with $c$ quarks profits from the capability of the CLIC detector to identify jets likely to contain charm mesons. The studies below thus focus on FCNC production of $c$ quarks in top decays. (The $t \to u$ FCNC transitions can be probed well in top production at hadron colliders.) Signal events are defined as top pair production with a "signal top quark" decaying via FCNC and a "spectator top quark" decaying hadronically or leptonically according to the SM.

A detailed analysis of the prospects to search for FCNC top decays in $t\bar{t}$ events has been carried out in [10] for three channels: $t \to c\gamma$, $t \to cH$, and $t \to c+\text{MET}$. Full detector simulation of the signal and background processes as well as beam effects have been taken into account, as summarized below.

#### 3.4.2.1 The reach for $t \to c\gamma$ decay

The signal events were simulated with WHIZARD2.2.8 [55, 231]. For this analysis, only the hadronic decay channels of the spectator top quark are taken into account. We require a high-energy photon of at least 50 GeV in the event, leading to efficient background suppression. The background consists of 6-fermion final state processes compatible with top-quark pair production, 4-fermion production dominated by $W^+W^-$ pairs, and di-quark production of light flavours.

Jets are clustered using the VLC jet algorithm [99, 232] for a multiplicity of 4 jets (signal-like) and 6 jets (background-like). The kinematics are further evaluated using a $\chi^2$ measure to estimate their signal-likeness as well as to identify the two top-quark candidate jets. In order to further separate signal from background, a multivariate classifier using a Boosted Decision Tree (BDT) is trained on 42 input variables. The most discriminating variables include the properties of the photon, the reconstructed invariant mass of the signal top quark, the jet energies of the $b$ and $c$ candidates, the flavour tagging results and $\chi^2$ measures based on the signal and background hypotheses. A cut is applied on the BDT response to suppress background contributions. The remaining distribution of the BDT score is compared to the background-only BDT score distribution to extract the expected limit on $Br(t \to c\gamma)$. Using the $\text{CL}_s$ approach [233], the expected 95 % C.L. limit for the case that no signal is present in $500(1000)\,\text{fb}^{-1}$ of data at $\sqrt{s} = 380$ GeV, without electron beam polarization, is

$$\text{BR}(t \to c\gamma) < 4.7(2.6) \times 10^{-5}. \tag{114}$$

---

[36]Based on a contribution by N. van der Kolk and A. F. Żarnecki.

*3.4.2.2 The reach for $t \to ch$ decay*

The search for the FCNC decay $t \to ch$ is evaluated for $t\bar{t}$ production events where the signal top quark decays to $ch$. As the SM Higgs boson of mass 125 GeV has its highest branching fraction in the decay to $b\bar{b}$, only $h \to b\bar{b}$ decays are considered. The FCNC signal is produced using WHIZARD 2.2.8 with a 2HDM(III) model implemented in SARAH [234] and normalized to BR($t \to ch$)×BR($h \to b\bar{b}$) = $10^{-3}$, so that the change to top decay width can be ignored. Background contributions originate from 6-fermion top-quark pair production, 4-fermion final state processes dominated by intermediate weak boson pairs, and lighter flavour quark-pair production.

The selection is based on the kinematics of the final state jets and flavour tagging, where the "spectator top quark" can decay hadronically or leptonically. In a pre-selection step, events are selected with a topology of top-quark pair production. Tight requirements on the flavour tagging are applied to enhance the fraction of FCNC events which have at least 3 $b$ quarks and 1 $c$ quark in the final state, assigning two of the $b$-jets and one $c$-jet as the decay products of the FCNC $ch$ final state. The most likely jet configuration is determined using a $\chi^2$ measure. A BDT is trained to enhance the separation power between signal and background events. The training takes into account the following observables, among others: The kinematic $\chi^2$ value for the signal and background hypotheses, the reconstructed Higgs and $W$ boson masses, and the flavour tagging values of the jets. To select a signal-dominated sample for the extraction of the expected limit, a relatively tight selection cut on the BDT score is applied. Using the $CL_s$ approach, the BDT score distribution without signal is compared to the distribution expected with FCNC signal included. This yields the following expected 95 % C.L. limits for 500 (1000) fb$^{-1}$ of data with an electron beam polarization of $-80\,\%$ throughout the run at $\sqrt{s} = 380$ GeV:

$$\text{BR}(t \to ch) \times \text{BR}(h \to b\bar{b}) < 12\ (7.1) \times 10^{-5}. \tag{115}$$

The CLIC running scenarios foresee the luminosity of 1000 fb$^{-1}$ to be evenly split in the beam polarization modes $p(e^-) = -80\,\%$ and $p(e^-) = +80\,\%$, in which case the final limit is estimated to be

$$\text{BR}(t \to ch) \times \text{BR}(h \to b\bar{b}) < 9.1 \times 10^{-5}. \tag{116}$$

since the cross section for $e^+e^- \to t\bar{t}$ at 380 GeV is enhanced (reduced) by 34 % for the -80 % (+80 %) polarization configuration.

*3.4.2.3 The reach for $t \to c$+MET decay*

New stable particles with FCNC couplings to top quarks can result in a $c$+MET signature, if they escape detection. In order to facilitate the reconstruction of the mass of the heavy state, the analysis takes into account only the fully hadronic decay of the spectator top quark. Signal events are generated using WHIZARD 2.2.8 with a $t \to cX$ decay, where $X$ is a charge neutral scalar particle that is assumed to decay predominantly into the dark sector. The mass of $X$ is varied from 25 to 150 GeV. The background is dominated by processes with four fermions in the final state, primarily originating from $W^+W^-$ production.

Requiring the total transverse momentum of the hadronic final state to be above 20 GeV, the total invariant mass above 140 GeV, and the absolute value of the longitudinal momentum below 100 GeV, reduces large backgrounds due to 4-fermion and quark-pair production. Jets are clustered using the VLC algorithm with a required multiplicity of four jets. Cuts on the flavour tagging information are applied to enhance the fraction of events with one and only one $b$-jet. Among the four jets, the one with the highest $c$-tag value is assumed to originate from the $t \to c$+MET decay. The two remaining light jets are assigned to the $W$ boson from the spectator top decay. Energy–momentum conservation is used to determine the invariant mass of the signal top quark and the stable heavy particle. BDT classifiers are trained separately for low scalar masses up to 75 GeV and higher scalar masses ranging from 100 GeV to 150 GeV. The input variables include the reconstructed masses of the signal top-quark and the scalar, the total energy of the event, its total transverse momentum and invariant mass, the missing mass, sphericity and acoplanarity of the event, the minimum and maximum $y$ distance cuts for four-jet reconstruction with the VLC algorithm, $b$-jet energy and invariant mass, and the $\chi^2$ value calculated from the reconstructed mass of the $W$ boson and the two top quarks. For each of the mass points, the BDT distributions in a reconstructed mass window of $\pm 30$ GeV around the nominal mass are used to compare the background-only case to the one including signal using the $\text{CL}_s$ method. The resulting expected limits, as a function of the heavy scalar mass, for $500\,\text{fb}^{-1}$ of data collected at $\sqrt{s} = 380\,\text{GeV}$ with an electron beam polarization of $p(e^-) = -80\,\%$ are summarized in Figure 48. The limits scale approximately like $1/\sqrt{\mathcal{L}}$, such that the anticipated luminosity of $1000\,\text{fb}^{-1}$ at this stage results in limits which are better by a factor of about $1/\sqrt{2}$ than those reported in Figure 48.

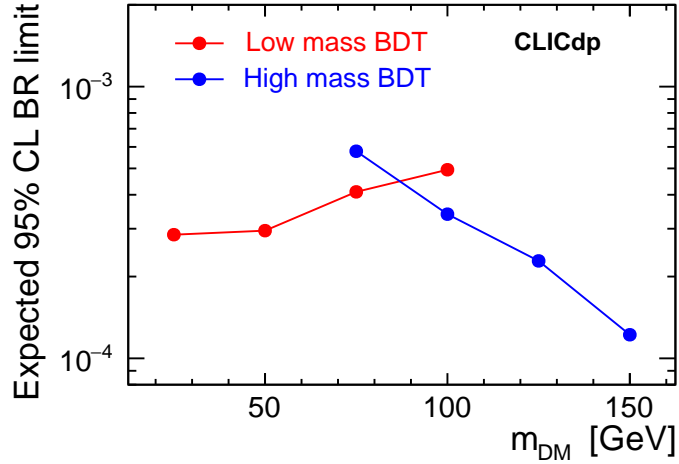

Figure 48: Expected limits from CLIC on the FCNC top decay $t \to c+\text{MET}$ for $500\,\text{fb}^{-1}$ of data collected at $\sqrt{s} = 380\,\text{GeV}$ with electron beam polarization $p(e^-) = -80\,\%$ in dependence of the mass of the invisible decay product, $m_{\text{DM}}$.

## 3.5 Determining Higgs couplings to light fermions[37]

### 3.5.1 New physics benchmarks for modified Higgs couplings

The Higgs couplings to the SM fermions, $f$, can differ from their SM values due to New Physics (NP). The size of the modification can be described using a generalized $\kappa$ framework,

$$\mathcal{L}_{\text{eff}} = -\kappa_{f_i} \frac{m_{f_i}}{v} h \bar{f}_i f_i + i\tilde{\kappa}_{f_i} \frac{m_{f_i}}{v} h \bar{f}_i \gamma_5 f_i - \left[ \left( \kappa_{f_i f_j} + i\tilde{\kappa}_{f_i f_j} \right) h \bar{f}_L^i f_R^j + \text{h.c.} \right]_{i \neq j}, \quad (117)$$

where a sum over fermion type $f = u, d, \ell$ and generations $i, j = 1, 2, 3$ is understood. The first two terms are flavour-diagonal with the first term CP-conserving and the second CP-violating. The terms in square brackets are flavour violating, where their real (imaginary) parts are CP conserving (violating). In the SM, we have $\kappa_{f_i} = 1$ while $\tilde{\kappa}_{f_i} = \kappa_{f_i f_j} = \tilde{\kappa}_{f_i f_j} = 0$.

The LHC measurements of the Higgs production and decay strengths constrain the flavour-diagonal CP-conserving Yukawa couplings to be [235–239] (taking couplings to be positive for simplicity, for future prospects see also [240–245])

$$\kappa_t = 1.09 \pm 0.14, \qquad \kappa_b = 1.04^{+0.27}_{-0.33}, \qquad \kappa_c \lesssim 6.2,$$
$$\kappa_s < 65, \qquad \kappa_d < 1.4 \cdot 10^3, \qquad \kappa_u < 3.0 \cdot 10^3, \quad (118)$$

---
[37]Based on a contribution by W. Altmannshofer, F. Bishara and M. Schlaffe.

$$\kappa_\tau = 1.01^{+0.17}_{-0.18}, \qquad \kappa_\mu = 0.2^{+1.2}_{-0.2}, \qquad \kappa_e \lesssim 630.$$

The above bounds on $\kappa_{t,b,c,s,d,u,\tau}$ were obtained by allowing BSM particles to modify the $h \to gg$ and $h \to \gamma\gamma$ couplings, i.e., $\delta\kappa_{g,\gamma}$ were floated, while not allowing any new decay channels, $\text{BR}_{\text{BSM}} = 0$. The upper bounds on $\kappa_{c,s,d,u}$ roughly correspond to the size of the SM bottom Yukawa coupling and are thus well above the SM value $\kappa_f = 1$. The upper bounds can be saturated only if one allows for large cancellations between the contribution to fermion masses from the Higgs vev and an equally large but opposite in sign contribution from NP. As we show below, in the models of NP motivated by the hierarchy problem, the effects of NP are generically well below these bounds.

The CP-violating flavour-diagonal Yukawa couplings, $\tilde\kappa_{f_i}$, are well constrained from bounds on the electric dipole moments (EDMs) [239, 246, 247] under the assumption of no cancellation with other contributions to EDMs beyond the Higgs contributions. The flavour violating Yukawa couplings are well constrained by the low-energy flavour-changing neutral current measurements [248–250]. A notable exception are the flavour-violating couplings involving a tau lepton. The strongest constraints on $\kappa_{\tau\mu}, \kappa_{\mu\tau}, \kappa_{\tau e}, \kappa_{e\tau}$ come from direct searches of flavour-violating Higgs decays at the LHC [251, 252].

The projected precision at CLIC with $\sqrt{s} = 3$ TeV and 2 ab$^{-1}$ of luminosity is [9]

$$\begin{aligned}\delta\kappa_t &\sim 4.1\%, & \delta\kappa_b &\sim 0.2\%, & \delta\kappa_c &\sim 1.7\%, \\ \delta\kappa_\tau &\sim 1.1\%, & \delta\kappa_\mu &\sim 7.8\%,\end{aligned} \qquad (119)$$

which is precise enough to probe well motivated NP modifications of the Yukawa couplings. Generically, the sizes of the deviations scale as $\mathcal{O}(v^2/\Lambda^2)$ where $v$ is the electroweak scale and $\Lambda$ the scale of NP. For $\Lambda = 1(2.5)$ TeV, one can naively expect deviations $\delta\kappa \sim 6(1)\%$. Of particular interest is the charm Yukawa which, e.g., can be $\mathcal{O}(1)$ enhanced in the modified Giudice-Lebedev model (see below). We will discuss the implication of CLIC for the charm Yukawa in Section 3.5.2.

In the rest of this subsection we review the expected sizes of $\kappa_{f_i}$ in popular models of weak scale NP, some of them motivated by the hierarchy problem.

### 3.5.1.1 Modified Yukawa couplings and electroweak new physics

Tables 26, 27, and 28, adapted from [253–257], summarize the predictions for the effective Yukawa couplings, $\kappa_f$, in the SM, multi-Higgs-doublet models (MHDM) with natural flavour conservation (NFC) [258, 259], a "flavourful" two-Higgs-doublet model beyond NFC (F2HDM) [260–263] the MSSM at tree level, a single Higgs doublet with a Froggat-Nielsen mechanism (FN) [264], the Giudice-Lebedev model of quark masses modified to 2HDM (GL2) [265], NP models with minimal flavour violation (MFV) [13], Randall-Sundrum models (RS) [227], and models with a composite Higgs where Higgs is a pseudo-Nambu-Goldstone boson (pNGB) [119, 266–268]. The flavour-violating couplings in the above set of NP models are collected in Tables 29 and 30. Next, we briefly discuss each of the above models, and show that the effects are either suppressed by $1/\Lambda^2$, where $\Lambda$ is the NP scale, or are proportional to the mixing angles between the Higgs and the extra scalars.

**Dimension-six operators with Minimal Flavor Violation (MFV):** We first assume that there is a mass gap between the SM and NP. Integrating out the NP states leads to dimension six operators (after absorbing the modifications of kinetic terms using equations of motion [269]), giving the effective Lagrangian in (97),

$$\mathcal{L}_{\text{EFT}} = \frac{\hat\epsilon_u}{\Lambda^2}\bar{Q}_L\tilde{H}u_R(H^\dagger H) + \frac{\hat\epsilon_d}{\Lambda^2}\bar{Q}_LHd_R(H^\dagger H) + \frac{\hat\epsilon_\ell}{\Lambda^2}\bar{L}_LH\ell_R(H^\dagger H) + \text{h.c.}, \qquad (120)$$

which correct the SM Yukawa interactions. The mismatch between the flavour structures of dimension 4 and dimension 6 contributions then results in the modifications of the Higgs Yukawa couplings at $\mathcal{O}(v^2/\Lambda^2)$, cf. Eq. (99).

Table 26: Predictions for the flavour-diagonal up-type Yukawa couplings in a sample of NP models (see text for details).

| Model | $\kappa_t$ | $\kappa_{c(u)}/\kappa_t$ | $\tilde{\kappa}_t/\kappa_t$ | $\tilde{\kappa}_{c(u)}/\kappa_t$ |
|---|---|---|---|---|
| SM | 1 | 1 | 0 | 0 |
| MFV | $1 + \frac{\Re(a_u v^2 + 2b_u m_t^2)}{\Lambda^2}$ | $1 - \frac{2\Re(b_u)m_t^2}{\Lambda^2}$ | $\frac{\Im(a_u v^2 + 2b_u m_t^2)}{\Lambda^2}$ | $\frac{\Im(a_u v^2)}{\Lambda^2}$ |
| NFC | $V_{hu}\, v/v_u$ | 1 | 0 | 0 |
| F2HDM | $\cos\alpha/\sin\beta$ | $-\tan\alpha/\tan\beta$ | $\mathcal{O}\left(\frac{m_c}{m_t}\frac{\cos(\beta-\alpha)}{\cos\alpha\cos\beta}\right)$ | $\mathcal{O}\left(\frac{m_{c(u)}^2}{m_t^2}\frac{\cos(\beta-\alpha)}{\cos\alpha\cos\beta}\right)$ |
| MSSM | $\cos\alpha/\sin\beta$ | 1 | 0 | 0 |
| FN | $1 + \mathcal{O}\left(\frac{v^2}{\Lambda^2}\right)$ | $1 + \mathcal{O}\left(\frac{v^2}{\Lambda^2}\right)$ | $\mathcal{O}\left(\frac{v^2}{\Lambda^2}\right)$ | $\mathcal{O}\left(\frac{v^2}{\Lambda^2}\right)$ |
| GL2 | $\cos\alpha/\sin\beta$ | $\simeq 3(7)$ | 0 | 0 |
| RS | $1 - \mathcal{O}\left(\frac{v^2}{m_{KK}^2}\bar{Y}^2\right)$ | $1 + \mathcal{O}\left(\frac{v^2}{m_{KK}^2}\bar{Y}^2\right)$ | $\mathcal{O}\left(\frac{v^2}{m_{KK}^2}\bar{Y}^2\right)$ | $\mathcal{O}\left(\frac{v^2}{m_{KK}^2}\bar{Y}^2\right)$ |
| pNGB | $1 + \mathcal{O}\left(\frac{v^2}{f^2}\right) + \mathcal{O}\left(y_*^2 \frac{\lambda^2 v^2}{M_*^2}\right)$ | $1 + \mathcal{O}\left(y_*^2 \frac{\lambda^2 v^2}{M_*^2}\right)$ | $\mathcal{O}\left(y_*^2 \lambda^2 \frac{v^2}{M_*^2}\right)$ | $\mathcal{O}\left(y_*^2 \lambda^2 \frac{v^2}{M_*^2}\right)$ |

Table 27: Same as Table 26 but for down-type Yukawa couplings.

| Model | $\kappa_b$ | $\kappa_{s(d)}/\kappa_b$ | $\tilde{\kappa}_b/\kappa_b$ | $\tilde{\kappa}_{s(d)}/\kappa_b$ |
|---|---|---|---|---|
| SM | 1 | 1 | 0 | 0 |
| MFV | $1 + \frac{\Re(a_d v^2 + 2c_d m_t^2)}{\Lambda^2}$ | $1 - \frac{2\Re(c_d)m_t^2}{\Lambda^2}$ | $\frac{\Im(a_d v^2 + 2c_d m_t^2)}{\Lambda^2}$ | $\frac{\Im(a_d v^2 + 2c_d |V_{ts(td)}|^2 m_t^2)}{\Lambda^2}$ |
| NFC | $V_{hd}\, v/v_d$ | 1 | 0 | 0 |
| F2HDM | $\cos\alpha/\sin\beta$ | $-\tan\alpha/\tan\beta$ | $\mathcal{O}\left(\frac{m_s}{m_b}\frac{\cos(\beta-\alpha)}{\cos\alpha\cos\beta}\right)$ | $\mathcal{O}\left(\frac{m_{s(d)}^2}{m_b^2}\frac{\cos(\beta-\alpha)}{\cos\alpha\cos\beta}\right)$ |
| MSSM | $-\sin\alpha/\cos\beta$ | 1 | 0 | 0 |
| FN | $1 + \mathcal{O}\left(\frac{v^2}{\Lambda^2}\right)$ | $1 + \mathcal{O}\left(\frac{v^2}{\Lambda^2}\right)$ | $\mathcal{O}\left(\frac{v^2}{\Lambda^2}\right)$ | $\mathcal{O}\left(\frac{v^2}{\Lambda^2}\right)$ |
| GL2 | $-\sin\alpha/\cos\beta$ | $\simeq 3(5)$ | 0 | 0 |
| RS | $1 - \mathcal{O}\left(\frac{v^2}{m_{KK}^2}\bar{Y}^2\right)$ | $1 + \mathcal{O}\left(\frac{v^2}{m_{KK}^2}\bar{Y}^2\right)$ | $\mathcal{O}\left(\frac{v^2}{m_{KK}^2}\bar{Y}^2\right)$ | $\mathcal{O}\left(\frac{v^2}{m_{KK}^2}\bar{Y}^2\right)$ |
| pNGB | $1 + \mathcal{O}\left(\frac{v^2}{f^2}\right) + \mathcal{O}\left(y_*^2 \lambda^2 \frac{v^2}{M_*^2}\right)$ | $1 + \mathcal{O}\left(y_*^2 \lambda^2 \frac{v^2}{M_*^2}\right)$ | $\mathcal{O}\left(y_*^2 \lambda^2 \frac{v^2}{M_*^2}\right)$ | $\mathcal{O}\left(y_*^2 \lambda^2 \frac{v^2}{M_*^2}\right)$ |

In Tables 26-30 we show the resulting $\kappa_f$ assuming MFV, i.e., that the flavour breaking in the NP sector is only due to the SM Yukawas [13, 270–275]. This gives $\hat{\epsilon}_u = a_u Y_u + b_u Y_u Y_u^\dagger Y_u + c_u Y_d Y_d^\dagger Y_u + \cdots$, and similarly for $\hat{\epsilon}_d$ with $u \leftrightarrow d$, while $a_q, b_q, c_q \sim \mathcal{O}(1)$ and are in general complex. For leptons we follow [256] and assume that the SM $Y_\ell$ is the only flavour-breaking spurion even for the neutrino mass matrix (see also [276]). Then $\hat{\epsilon}_\ell$ and $Y_\ell$ are diagonal in the same basis and there are no flavour-violating couplings. The flavour-diagonal $\kappa_\ell$ are given in Table 28.

**Multi-Higgs-doublet model with natural flavour conservation (NFC):** Natural flavour conservation in multi-Higgs-doublet models is an assumption that only one doublet, $H_u$, couples to the up-type quarks, only one Higgs doublet, $H_d$, couples to the down-type quarks, and only one doublet, $H_\ell$ couples to leptons (it is possible that any of these coincide, as in the SM where $H = H_u = H_d = H_\ell$) [258, 259]. The neutral scalar components of $H_i$ are $(v_i + h_i)/\sqrt{2}$, where $v^2 = \sum_i v_i^2$. The dynamical fields $h_i$ are a linear combination of the neutral Higgs mass eigenstates (and include $h_u$ and $h_d$). We thus have $h_i = V_{hi}h + \ldots$, where $V_{hi}$ are elements of the unitary matrix $V$ that diagonalizes the neutral-Higgs mass terms and we only write down the contribution of the lightest Higgs, $h$. NFC means that there are no tree-level Flavor Changing Neutral Currents (FCNCs) and no $CP$ violation in the Yukawa interactions, i.e., $\kappa_{qq'} = \tilde{\kappa}_{qq'} = 0$, $\tilde{\kappa}_q = 0$.

Table 28: Same as Table 26 but for lepton Yukawa couplings. NP effects in the pNGB model are negligible and therefore we do not report them here.

| Model | $\kappa_\tau$ | $\kappa_{\mu(e)}/\kappa_\tau$ | $\tilde{\kappa}_\tau/\kappa_\tau$ | $\tilde{\kappa}_{\mu(e)}/\kappa_\tau$ |
|---|---|---|---|---|
| SM | 1 | 1 | 0 | 0 |
| MFV | $1 + \frac{\Re(a_\ell)v^2}{\Lambda^2}$ | $1 - \frac{2\Re(b_\ell)m_\tau^2}{\Lambda^2}$ | $\frac{\Im(a_\ell)v^2}{\Lambda^2}$ | $\frac{\Im(a_\ell)v^2}{\Lambda^2}$ |
| NFC | $V_{h\ell}\, v/v_\ell$ | 1 | 0 | 0 |
| F2HDM | $\cos\alpha/\sin\beta$ | $-\tan\alpha/\tan\beta$ | $\mathcal{O}\left(\frac{m_\mu}{m_\tau}\frac{\cos(\beta-\alpha)}{\cos\alpha\cos\beta}\right)$ | $\mathcal{O}\left(\frac{m_{\mu(e)}^2}{m_\tau^2}\frac{\cos(\beta-\alpha)}{\cos\alpha\cos\beta}\right)$ |
| MSSM | $-\sin\alpha/\cos\beta$ | 1 | 0 | 0 |
| FN | $1 + \mathcal{O}\left(\frac{v^2}{\Lambda^2}\right)$ | $1 + \mathcal{O}\left(\frac{v^2}{\Lambda^2}\right)$ | $\mathcal{O}\left(\frac{v^2}{\Lambda^2}\right)$ | $\mathcal{O}\left(\frac{v^2}{\Lambda^2}\right)$ |
| GL2 | $-\sin\alpha/\cos\beta$ | $\simeq 3(5)$ | 0 | 0 |
| RS | $1 + \mathcal{O}\left(\bar{Y}^2 \frac{v^2}{m_{KK}^2}\right)$ | $1 + \mathcal{O}\left(\bar{Y}^2 \frac{v^2}{m_{KK}^2}\right)$ | $\mathcal{O}\left(\bar{Y}^2 \frac{v^2}{m_{KK}^2}\right)$ | $\mathcal{O}\left(\bar{Y}^2 \frac{v^2}{m_{KK}^2}\right)$ |

Table 29: Same as Table 26 but for flavour-violating up-type Yukawa couplings. In the SM, NFC and the tree-level MSSM the Higgs Yukawa couplings are flavour diagonal. The CP-violating $\tilde{\kappa}_{ff'}$ are obtained by replacing the real part, $\Re$, with the imaginary part, $\Im$. All the other models predict a zero contribution to these flavour changing couplings.

| Model | $\kappa_{ct(tc)}/\kappa_t$ | $\kappa_{ut(tu)}/\kappa_t$ | $\kappa_{uc(cu)}/\kappa_t$ |
|---|---|---|---|
| MFV | $\frac{\Re\left(c_u m_b^2 V_{cb}^{(*)}\right)}{\Lambda^2} \frac{\sqrt{2}m_{t(c)}}{v}$ | $\frac{\Re\left(c_u m_b^2 V_{ub}^{(*)}\right)}{\Lambda^2} \frac{\sqrt{2}m_{t(u)}}{v}$ | $\frac{\Re\left(c_u m_b^2 V_{ub(cb)} V_{cb(ub)}^*\right)}{\Lambda^2} \frac{\sqrt{2}m_{c(u)}}{v}$ |
| F2HDM | $\mathcal{O}\left(\frac{m_c}{m_t}\frac{\cos(\beta-\alpha)}{\cos\alpha\cos\beta}\right)$ | $\mathcal{O}\left(\frac{m_u}{m_t}\frac{\cos(\beta-\alpha)}{\cos\alpha\cos\beta}\right)$ | $\mathcal{O}\left(\frac{m_c m_u}{m_t^2}\frac{\cos(\beta-\alpha)}{\cos\alpha\cos\beta}\right)$ |
| FN | $\mathcal{O}\left(\frac{vm_{t(c)}}{\Lambda^2}|V_{cb}|^{\pm 1}\right)$ | $\mathcal{O}\left(\frac{vm_{t(u)}}{\Lambda^2}|V_{ub}|^{\pm 1}\right)$ | $\mathcal{O}\left(\frac{vm_{c(u)}}{\Lambda^2}|V_{us}|^{\pm 1}\right)$ |
| GL2 | $\epsilon(\epsilon^2)$ | $\epsilon(\epsilon^2)$ | $\epsilon^3$ |
| RS | $\sim \lambda^{(-)2}\frac{m_{t(c)}}{v}\bar{Y}^2\frac{v^2}{m_{KK}^2}$ | $\sim \lambda^{(-)3}\frac{m_{t(u)}}{v}\bar{Y}^2\frac{v^2}{m_{KK}^2}$ | $\sim \lambda^{(-)1}\frac{m_{c(u)}}{v}\bar{Y}^2\frac{v^2}{m_{KK}^2}$ |
| pNGB | $\mathcal{O}(y_*^2 \frac{m_t}{v}\frac{\lambda_{L(R),2}\lambda_{L(R),3}m_W^2}{M_*^2})$ | $\mathcal{O}(y_*^2 \frac{m_t}{v}\frac{\lambda_{L(R),1}\lambda_{L(R),3}m_W^2}{M_*^2})$ | $\mathcal{O}(y_*^2 \frac{m_c}{v}\frac{\lambda_{L(R),1}\lambda_{L(R),2}m_W^2}{M_*^2})$ |

In NFC there is a universal shift in all up-quark Yukawa couplings, $\kappa_u = \kappa_c = \kappa_t = V_{hu}v/v_u$. Similarly there is a (different) universal shift in all down-quark Yukawa couplings and in all lepton Yukawa couplings, see Tables 26 - 28.

**Higgs sector of the MSSM at tree level:** The MSSM tree-level Higgs potential and the couplings to quarks are the same as in the type-II two-Higgs-doublet model, see, e.g., [277]. This is an example of a 2HDM with natural flavour conservation in which $v_u = \sin\beta\, v$, $v_d = \cos\beta\, v$. The mixing of $h_{u,d}$ into the Higgs mass-eigenstates $h$ and $H$ is given by $h_u = \cos\alpha h + \sin\alpha H$, $h_d = -\sin\alpha h + \cos\alpha H$, where $h$ is the observed SM-like Higgs. The up-quark Yukawa couplings are rescaled universally, $\kappa_u = \kappa_c = \kappa_t = \cos\alpha/\sin\beta$, and similarly the down-quark Yukawas, $\kappa_d = \kappa_s = \kappa_b = -\sin\alpha/\cos\beta$.

Table 30: Same as Table 29 but for flavour-violating down-type Yukawa couplings.

| Model | $\kappa_{bs(sb)}/\kappa_b$ | $\kappa_{bd(db)}/\kappa_b$ | $\kappa_{sd(ds)}/\kappa_b$ |
|---|---|---|---|
| MFV | $\frac{\Re\left(c_d m_t^2 V_{ts}^{(*)}\right)}{\Lambda^2}\frac{\sqrt{2}m_{s(b)}}{v}$ | $\frac{\Re\left(c_d m_t^2 V_{td}^{(*)}\right)}{\Lambda^2}\frac{\sqrt{2}m_{d(b)}}{v}$ | $\frac{\Re\left(c_d m_t^2 V_{ts(td)}^* V_{td(ts)}\right)}{\Lambda^2}\frac{\sqrt{2}m_{s(d)}}{v}$ |
| F2HDM | $\mathcal{O}\left(\frac{m_s}{m_b}\frac{\cos(\beta-\alpha)}{\cos\alpha\cos\beta}\right)$ | $\mathcal{O}\left(\frac{m_d}{m_b}\frac{\cos(\beta-\alpha)}{\cos\alpha\cos\beta}\right)$ | $\mathcal{O}\left(\frac{m_s m_d}{m_b^2}\frac{\cos(\beta-\alpha)}{\cos\alpha\cos\beta}\right)$ |
| FN | $\mathcal{O}\left(\frac{v m_{b(s)}}{\Lambda^2}|V_{cb}|^{\pm 1}\right)$ | $\mathcal{O}\left(\frac{v m_{b(d)}}{\Lambda^2}|V_{ub}|^{\pm 1}\right)$ | $\mathcal{O}\left(\frac{v m_{s(d)}}{\Lambda^2}|V_{us}|^{\pm 1}\right)$ |
| GL2 | $\epsilon^2(\epsilon)$ | $\epsilon$ | $\epsilon^2(\epsilon^3)$ |
| RS | $\sim \lambda^{(-)2}\frac{m_{b(s)}}{v}\bar{Y}^2\frac{v^2}{m_{KK}^2}$ | $\sim \lambda^{(-)3}\frac{m_{b(d)}}{v}\bar{Y}^2\frac{v^2}{m_{KK}^2}$ | $\sim \lambda^{(-)1}\frac{m_{s(d)}}{v}\bar{Y}^2\frac{v^2}{m_{KK}^2}$ |
| pNGB | $\mathcal{O}(y_*^2\frac{m_b}{v}\frac{\lambda_{L(R),2}\lambda_{L(R),3}m_W^2}{M_*^2})$ | $\mathcal{O}(y_*^2\frac{m_b}{v}\frac{\lambda_{L(R),1}\lambda_{L(R),3}m_W^2}{M_*^2})$ | $\mathcal{O}(y_*^2\frac{m_s}{v}\frac{\lambda_{L(R),1}\lambda_{L(R),2}m_W^2}{M_*^2})$ |

Table 31: Same as Table 29 but for flavour-violating lepton Yukawa couplings.

| Model | $\kappa_{\tau\mu(\mu\tau)}/\kappa_\tau$ | $\kappa_{\tau e(e\tau)}/\kappa_\tau$ | $\kappa_{\mu e(e\mu)}/\kappa_\tau$ |
|---|---|---|---|
| F2HDM | $\mathcal{O}\left(\frac{m_\mu}{m_\tau}\frac{\cos(\beta-\alpha)}{\cos\alpha\cos\beta}\right)$ | $\mathcal{O}\left(\frac{m_e}{m_\tau}\frac{\cos(\beta-\alpha)}{\cos\alpha\cos\beta}\right)$ | $\mathcal{O}\left(\frac{m_\mu m_e}{m_\tau^2}\frac{\cos(\beta-\alpha)}{\cos\alpha\cos\beta}\right)$ |
| FN | $\mathcal{O}\left(\frac{v m_{\mu(\tau)}}{\Lambda^2}|U_{23}|^{\mp 1}\right)$ | $\mathcal{O}\left(\frac{v m_{e(\tau)}}{\Lambda^2}|U_{13}|^{\mp 1}\right)$ | $\mathcal{O}\left(\frac{v m_{e(\mu)}}{\Lambda^2}|U_{12}|^{\mp 1}\right)$ |
| GL2 | $\epsilon^2(\epsilon)$ | $\epsilon$ | $\epsilon^2(\epsilon^3)$ |
| RS | $\sim \sqrt{\frac{m_{\mu(\tau)}}{m_{\tau(\mu)}}}\bar{Y}^2\frac{v^2}{m_{KK}^2}$ | $\sim \sqrt{\frac{m_{e(\tau)}}{m_{\tau(e)}}}\bar{Y}^2\frac{v^2}{m_{KK}^2}$ | $\sim \sqrt{\frac{m_{e(\mu)}}{m_{\mu(e)}}}\bar{Y}^2\frac{v^2}{m_{KK}^2}$ |

The flavour-violating and CP-violating Yukawas are zero[38]. In Tables 26-28 we limit ourselves to the tree-level expectations, which are a good approximation for a large part of the MSSM parameter space.

In the alignment limit, $\beta - \alpha = \pi/2$ [279–285], the Yukawa couplings tend toward their SM value, $\kappa_i = 1$. The global fits to Higgs data in type-II 2HDM already constrain $\beta - \alpha$ to be not too far from $\pi/2$ [286–288] so that the couplings of the light Higgs are also constrained to be close to their SM values. Note that the decoupling limit of the 2HDM, where the heavy Higgs bosons become much heavier than the SM Higgs, implies the alignment limit while the reverse is not necessarily true [280].

**Flavorful two-Higgs-doublet model.** In [260] a 2HDM setup was introduced in which one Higgs doublet couples only to top, bottom and tau, and a second Higgs doublet couples to the remaining fermions (see also [289–292]). Such a 2HDM goes beyond NFC and therefore introduces FCNCs at tree level. However, the Yukawa couplings of the first Higgs doublet to the third generation fermions preserve a $U(2)^5$ flavour symmetry, only broken by the small couplings of the second Higgs doublet. This approximate $U(2)^5$ symmetry leads to a strong suppression of the most sensitive flavour violating transitions between the second and first generation.

The non-standard flavour structure of this "flavourful" 2HDM scenario leads to flavour non-universal modifications of all Higgs couplings. Neglecting corrections that are suppressed by ratios of second to third generation masses, one finds approximately $\kappa_t \neq \kappa_c \simeq \kappa_u$, $\kappa_b \neq \kappa_s \simeq \kappa_d$, $\kappa_\tau \neq \kappa_\mu \simeq \kappa_e$. CP violation in Higgs couplings can arise but is strongly suppressed by small fermion masses, see Tables 26 - 28. Also potentially sizable flavour violating Higgs couplings involving the third generation fermions arise, see Tables 29 - 31. As in all 2HDMs, the Higgs couplings approach their SM values in the alignment limit, $\beta - \alpha = \pi/2$.

---
[38]Note that beyond the tree level, in fine-tuned regions of parameter space the loops of sfermions and gauginos can lead to substantial corrections to these expressions [278].

**A single Higgs doublet with Froggatt-Nielsen mechanism (FN):** The Froggatt-Nielsen [264] mechanism provides a simple explanation of the size and hierarchy of the SM Yukawa couplings. In the simplest realization this is achieved by a $U(1)_H$ horizontal symmetry under which different generations of fermions carry different charges. The $U(1)_H$ is broken by a spurion, $\epsilon_H$. The entries of the SM Yukawa matrix are then parametrically suppressed by powers of $\epsilon_H$ as, for example, in the lepton sector

$$(Y_\ell)_{ij} \sim \epsilon_H^{H(L_i)-H(e_j)}, \tag{121}$$

where $H(e, L)$ are the FN charges of the right- and left-handed charged lepton, respectively. The dimension 6 operators in Eq. (120) due to electroweak NP have a similar flavour suppression, $(\hat{\epsilon}_\ell)_{ij} \sim \epsilon_H^{H(e_j)-H(L_i)} v^2/\Lambda^2$ [254, 256]. After rotating to the mass eigenbasis, the lepton masses and mixing angles are then given by [293, 294]

$$m_{\ell_i}/v \sim \epsilon_H^{|H(L_i)-H(e_i)|}, \quad |U_{ij}| \sim \epsilon_H^{|H(L_i)-H(L_j)|}, \tag{122}$$

giving the Higgs Yukawa couplings in Tables 28 and 31 in the row labeled 'FN' [254]. Similarly, for the quarks, after rotating to the mass eigenbasis, the masses and the mixings are given by [293]

$$m_{u_i(d_i)}/v \sim \epsilon_H^{|H(Q_i)-H(u_i(d_i))|}, \quad |V_{ij}| \sim \epsilon_H^{|H(Q_i)-H(Q_j)|}, \tag{123}$$

where $V$ is the Cabibbo-Kobayashi-Maskawa (CKM) mixing matrix and $H(u, d, Q)$ are the FN charges of the right-handed up and down and the left-handed quark fields, respectively.

**Higgs-dependent Yukawa couplings (GL2):** In the model of quark masses introduced by Giudice and Lebedev [265], the quark masses, apart from the top mass, are small because they arise from higher dimensional operators. The original GL proposal is ruled out by data, while the straightforward modification to a 2HDM (GL2) is not. It is given by an effective Lagrangian

$$\mathcal{L}_{\text{GL2}} = c_{ij}^u \left(\frac{H_1^\dagger H_1}{M^2}\right)^{n_{ij}^u} \bar{Q}_{L,i} u_{R,j} H_1 + c_{ij}^d \left(\frac{H_1^\dagger H_1}{M^2}\right)^{n_{ij}^d} \bar{Q}_{L,i} d_{R,j} H_2 + \\ c_{ij}^\ell \left(\frac{H_1^\dagger H_1}{M^2}\right)^{n_{ij}^\ell} \bar{L}_{L,i} e_{R,j} H_2 + \text{h.c.}, \tag{124}$$

where $M$ is the mass scale of the mediators. In the original GL model $H_2$ is identified with the SM Higgs, $H_2 = H$, while $H_1 = \tilde{H}$. Taking $c_{ij}^{u,d} \sim \mathcal{O}(1)$, the ansatz $n_{ij}^{u,d} = a_i + b_j^{u,d}$ with $a = (1, 1, 0)$, $b^d = (2, 1, 1)$, and $b^u = (2, 0, 0)$ then reproduces the hierarchies of the observed quark masses and mixing angles for $\epsilon \equiv v^2/M^2 \approx 1/60$. The Yukawa couplings are of the form $y_{ij}^{u,d} = (2n_{ij}^{u,d} + 1)(y_{ij}^{u,d})_{\text{SM}}$. The SM Yukawas are diagonal in the same basis as the quark masses, while the $y_{ij}^{u,d}$ are not. Because the bottom Yukawa is largely enhanced, $\kappa_b \simeq 3$, this simplest version of the GL model is already excluded by the Higgs data. Its modification, GL2, is still viable, though [253]. For $v_1/v_2 = \tan\beta \sim 1/\epsilon$ one can use the same ansatz for $n_{ij}^{u,d}$ as before, modifying only $b^d$, so that $b^d = (1, 0, 0)$, with the results shown in Tables 26-31. For leptons we use the same scalings as for right-handed quarks. Note that the $H_1^\dagger H_1$ is both a gauge singlet and a flavour singlet. From symmetry point of view it is easier to build flavour models, if $H_1 H_2$ acts as a spurion in (124), instead of $H_1^\dagger H_1$. This possibility is severely constrained phenomenologically, though [257, 295].

**Randall-Sundrum models (RS):** The Randall-Sundrum warped extra-dimensional model has been proposed to address the hierarchy problem and simultaneously explain the hierarchy of the SM fermion masses [227, 296–299]. Integrating out the Kaluza-Klein (KK) modes of mass $m_{KK}$, and working in the limit of a brane-localized Higgs, keeping only terms of leading order in $v^2/m_{KK}^2$, the SM quark mass matrices are given by [228] (see also [300–308], and Ref. [309] for a bulk Higgs scenario)

$$M_{ij}^{d(u)} = \left[F_q Y_{1(2)}^{5D} F_{d(u)}\right]_{ij} v. \tag{125}$$

The $F_{q,u,d}$ are $3 \times 3$ matrices of fermion wave-function overlaps with the Higgs and are diagonal and hierarchical. Assuming flavour anarchy, the 5D Yukawa matrices, $Y_{1,2}^{5D}$, are general $3 \times 3$ complex matrices with $\bar{Y} \sim \mathcal{O}(1)$ entries, but usually $\bar{Y} \lesssim 4$, see, e.g., [303]. At leading order in $v^2/m_{KK}^2$ the Higgs Yukawas are aligned with the quark masses, i.e., $M_{u,d} = y_{u,d} v/\sqrt{2} + \mathcal{O}(v^2/m_{KK}^2)$. The misalignments are generated by tree-level KK quark exchanges, giving

$$\left[y_{u(d)}\right]_{ij} - \frac{\sqrt{2}}{v}\left[M_{u,d}\right]_{ij} \sim -\frac{1}{3} F_{q_i} \bar{Y}^3 F_{u_j(d_j)} \frac{v^2}{m_{KK}^2}. \tag{126}$$

For the charged leptons, there are two choices for generating the hierarchy in the masses [228]. If left- and right-handed fermion profiles are both hierarchical (and taken to be similar) then the misalignment between the masses and Yukawas is $\sim \sqrt{m_i m_j/v^2} \times \mathcal{O}(\bar{Y}^2 v^2/m_{KK}^2)$. If only the right-handed profiles are hierarchical the misalignment is given by (see also Tables 28 and 31)

$$\left[y_\ell\right]_{ij} - \frac{\sqrt{2}}{v}\left[M_\ell\right]_{ij} \sim -\frac{1}{3} \bar{Y}^2 \frac{v^2}{m_{KK}^2} \frac{m_j^\ell}{v}. \tag{127}$$

The Higgs mediated FCNCs are suppressed by the same zero-mode wave-function overlaps that also suppress the quark masses, (125), giving rise to the RS GIM mechanism [310–312]. Using the fact that the CKM matrix elements are given by $V_{ij} \sim F_{q_i}/F_{q_j}$ for $i < j$, Eq. (126), one can rewrite the $\kappa_i$ as in Tables 26-30. The numerical analysis of Ref. [228] found that for diagonal Yukawas typically $\kappa_i < 1$, with deviations in $\kappa_{t(b)}$ up to $30\%(15\%)$, and in $\kappa_{s,c(u,d)}$ up to $\sim 5\%(1\%)$. For the charged leptons one obtains deviations in $\kappa_{\tau\mu(\mu\tau)} \sim 1(5) \times 10^{-5}$ [228]. These estimates were obtained fixing the mass of the first KK gluon excitation to 3.7 TeV, above the present ATLAS bound [313].

**Composite pseudo-Goldstone Higgs (pNGB):** Finally, we assume that the Higgs is a pseudo-Goldstone boson arising from the spontaneous breaking of a global symmetry in a strongly coupled sector, and couples to the composite sector with a typical coupling $y_*$ [119, 266–268] (for a review, see [16]). Assuming partial compositeness, the SM fermions couple linearly to composite operators $O_{L,R}$, giving the interactions $\lambda_{L,i}^q \bar{Q}_{L,i} O_R^i + \lambda_{R,j}^u \bar{u}_{R,j} O_L^j + h.c.$, where $i,j$ are flavour indices [314]. This is the 4D dual of fermion mass generation in 5D RS models. The SM masses and Yukawa couplings arise from expanding the two-point functions of the $O_{L,R}$ operators in powers of the Higgs field [315].

The new ingredient compared to the EFT analysis in (120) is that the shift symmetry due to the pNGB nature of the Higgs dictates the form of the higher-dimensional operators. The flavour structure and the composite Higgs coset structure completely factorize, if the SM fields couple to only one composite operator. The general decomposition of Higgs couplings then becomes [315] (see also [226, 316, 317])

$$Y_u \bar{Q}_L H u_R + \hat{\epsilon}_u \bar{Q}_L H u_R \frac{(H^\dagger H)}{\Lambda^2} + \ldots \quad \rightarrow \quad c_{ij}^u P(h/f) \bar{Q}_L^i H u_R^j, \tag{128}$$

and similarly for the down quarks. Here $f \gtrsim v$ is the equivalent of the pion decay constant, while $P(h/f) = a_0 + a_2(H^\dagger H/f^2) + \ldots$ is an analytic function whose form is fixed by the pattern of the spontaneous breaking and the embedding of the SM fields in the global symmetry of the strongly coupled sector. In (128) the flavour structure of $Y_u$ and $\hat{\epsilon}_u$ is the same. The resulting corrections to the quark Yukawa couplings are therefore strictly diagonal,

$$\kappa_q \sim 1 + \mathcal{O}(v^2/f^2). \tag{129}$$

Table 32: The assumed values for the model parameters used to produce Figure 49. See text for details.

| Model | $\kappa_c - 1$ | Assumptions |
|---|---|---|
| MFV | 6% | $\Lambda = 1$ TeV |
| FN | 6% | $\Lambda = 1$ TeV |
| GL2 | 3 | $\epsilon \sim 1/60$ |
| RS | 7% | $\bar{Y} = 4, m_{KK} = 2.2$ TeV |
| pNGB | 5% | $y_* = 4\pi, \lambda \sim V_{us}, M_* = 4\pi v$ |

For example, for the models based on the breaking of $SO(5)$ to $SO(4)$, the diagonal Yukawa couplings can be written as $\kappa_q = (1 + 2m - (1 + 2m + n)(v/f)^2)/\sqrt{1 - (v/f)^2}$, where $n, m$ are positive integers [318]. The MCHM4 model corresponds to $m = n = 0$, while MCHM5 is given by $m = 0, n = 1$.

The flavour-violating contributions to the quark Yukawa couplings arise only from corrections to the quark kinetic terms [315],

$$\bar{q}_L i \slashed{\partial} q_L \frac{H^\dagger H}{\Lambda^2}, \quad \bar{u}_R i \slashed{\partial} u_R \frac{H^\dagger H}{\Lambda^2}, \ldots, \tag{130}$$

due to the exchanges of composite vector resonances with typical mass $M_* \sim \Lambda$. After using the equations of motion these give (neglecting relative $\mathcal{O}(1)$ contributions in the sum) [226, 315, 319],

$$\kappa_{ij}^u \sim 2y_*^2 \frac{v^2}{M_*^2} \left( \lambda_{L,i}^q \lambda_{L,j}^q \frac{m_{u_j}}{v} + \lambda_{R,i}^u \lambda_{R,j}^u \frac{m_{u_i}}{v} \right), \tag{131}$$

and similarly for the down quarks. If the strong sector is CP violating, then $\tilde{\kappa}_{ij}^{u,d} \sim \kappa_{ij}^{u,d}$.

The exchange of composite vector resonances also contributes to the flavour-diagonal Yukawa couplings, shifting the estimate (129) by $\Delta \kappa_{q_i} \sim 2y_*^2 \frac{v^2}{M_*^2} \left[ \left(\lambda_{L,i}^q\right)^2 + \left(\lambda_{R,i}^u\right)^2 \right]$. This shift can be large for the quarks with a large composite component if the Higgs is strongly coupled to the vector resonances, $y_* \sim 4\pi$, and these resonances are relatively light, $M_* \sim 4\pi v \sim 3$ TeV. The left-handed top and bottom, as well as the right-handed top, are expected to be composite, explaining the large top mass (i.e., $\lambda_{L,3}^q \sim \lambda_{R,3}^u \sim 1$). In the anarchic flavour scenario, one expects the remaining quarks to be mostly elementary (so the remaining $\lambda_i \ll 1$). If there is some underlying flavour alignment, it is also possible that the light quarks are composite. This is most easily achieved in the right-handed sector [317, 320, 321].

In the case of the lepton sector, if we assume that there are no hierarchies in the composite sector [322] (see also [323–326]), then the NP effects in the flavour diagonal and off-diagonal Yukawas are negligible. For this reason, we do not report them in Tables 28 and 31.

### 3.5.2 The charm Yukawa at CLIC

As discussed in Section 3.5.1, CLIC has the potential to severely constrain deviations of the charm Yukawa obtainable in NP models of flavour. To illustrate this reach, Figure 49 depicts the size of the deviation of $\kappa_c$ from the SM prediction of $\kappa_c = 1$ along with the predicted reach of the three CLIC benchmark scenarios.

The deviations shown in Figure 49 were obtained using the current constraints from NP searches at the LHC and are shown in Table 32. For all the models considered here, CLIC is projected to constrain the allowed deviation by almost an order of magnitude, and thus severely constraining the parameter spaces of these models.

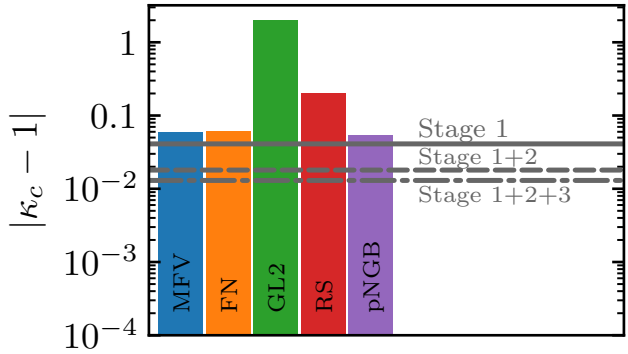

Figure 49: Deviations of the charm Yukawa from its SM value in the NP models discussed in Section 3.5.1.1 using the current bounds shown in Table 32. The predicted reach of the three CLIC scenarios [9] are shown with horizontal lines.

### 3.5.3 Strangeness tagging and $h \to s\bar{s}$

Typically, limits on the branching ratio of the Higgs into strange quarks are set by searches for exclusive Higgs decays, see e.g., Refs. [238, 240, 242, 327–329]. While these decays yield very clean final states, the searches are impeded by the reduced branching ratio to exclusive final states. Alternative approaches suggest using information from the event shape and other kinematic observables [330–333]. A new method to measure the branching ratio of $h \to s\bar{s}$ by tagging strange jets is discussed in Ref. [334]. Although it is challenging experimentally, it is not impossible, and was in fact used before in the context of $Z$ decays at DELPHI [335] and SLD [336].

In the following we summarize the setup of the strangeness tagger and its application to constraining the $h \to s\bar{s}$ branching ratio. While here the focus is on $h \to ss$ decays, the applicability of the tagger is not limited to this case but can be generalized for various other analyses where strange jets play a crucial role.

#### 3.5.3.1 Strangeness tagger

Tagging strange jets relies on the fact that strange quarks hadronize dominantly into kaons that are both prompt and carry a large fraction of the initial quark momentum. The tagger for strange jets is therefore based upon the following observables.

*Particle identification (PID).* The most important kaons for the tagger are the charged $K^\pm$ which, unlike neutral kaons, leave charged tracks in the detector before the decay. The $K^\pm$ account for about half of the kaons in a sample of strange jets. The specific energy loss $dE/dx$ can be used to distinguish kaon tracks from pion tracks. After reconstructing charged pions and kaons with a tracking efficiency of 95% we mimic PID as follows. We assume that the distributions of kaons and pions are two normal distributions separated by $1.5\sigma$ which, given the Bethe-Weizsäcker-formula, is in the achievable range for the planned resolution of 5%. After fixing a target efficiency for kaons, $\epsilon_{K^\pm}$, the fake rate for pions, $\epsilon_{\pi^\pm}$, follows directly. For instance, for $\epsilon_{K^\pm} = 0.95$ we obtain $\epsilon_{\pi^\pm} = 0.32$. Neutral kaons are only reconstructed if they decay within 1 m from the interaction point into charged pions. For the reconstruction of these, an efficiency of 85% is assumed.

In the two hemispheres defined by the jet axes in the Higgs rest frame, the hardest kaon pair is chosen for further analysis. For $h \to s\bar{s}$ study only the events with opposite charges for the selected kaons are kept. In principle, also one charged and one neutral or two neutral kaons would be possible, however, due to their lower abundance, these modes can only play a subleading role. Moreover, for neutral kaons the determination of the impact parameter is impossible and thus missing as important ingredient to the tagger.

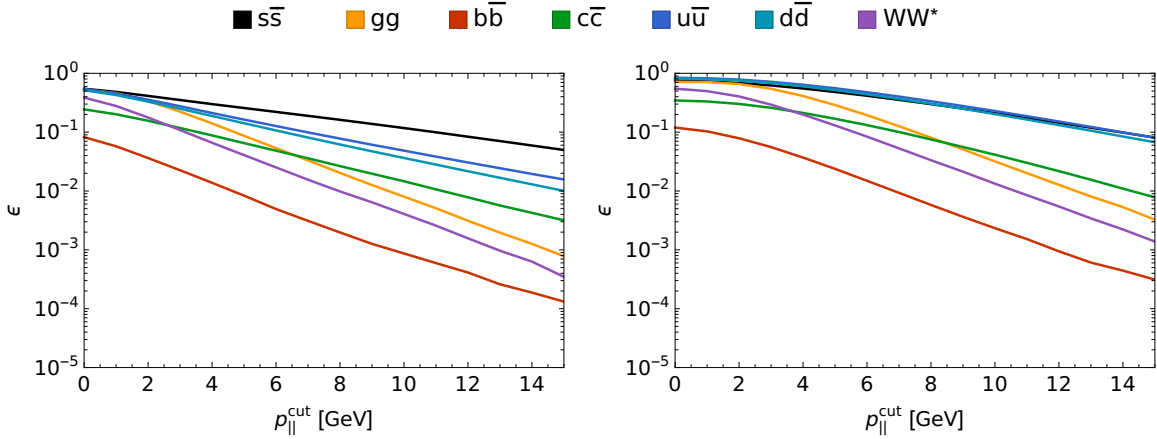

Figure 50: Efficiencies for tagging Higgs decays to the respective final states as a function of the cut on $p_{||}$. In the left plot the efficiency for kaon tagging is $\epsilon_{K^\pm} = 95\%$ whereas in the right plot no PID is assumed. The cut on the impact parameter is $d_0 < 16\ \mu$m.

*Impact parameter $d_0$.* The transverse impact parameter $d_0$ is a simple method to reject the background from heavy flavour jets due to the long lifetime of their mesons. The truth values of $d_0$ are smeared by a gaussian distribution with the width given by the anticipated resolution [9] as a function of the kaon momentum and direction. With this basic method, large fractions of the leading kaon in heavy flavour jets pass the displaced vertex cut even though they originate from the decay of a heavy flavour meson. Hence, by employing more sophisticated methods to find displaced vertices, a sizable improvement of the tagging performance can be expected. Despite this simple approach, good rejection efficiencies for charm and bottom jets are achieved.

*Large momentum fraction.* The hardest reconstructed kaon in a gluon jet tends to be softer than the corresponding kaon in a quark jet since gluons radiate more than quarks and thus their initial momentum is shared among more final states. Consequently, demanding that the momentum of the kaon projected onto the jet axis, $p_{||}$, is large, rejects the background from gluons. Similarly, this cut suppresses the background from fully hadronic $h \to WW^*$ that is falsely reconstructed as a dijet decay of the Higgs.

#### 3.5.3.2 Tagging $h \to s\bar{s}$

The efficiencies of the described tagger are evaluated using a sample of $e^+e^- \to hZ$ events generated by PYTHIA 8.219 [337, 338] at a centre-of-mass energy of $\sqrt{s} = 380$ GeV. In the analysis only the Higgs decay products are considered, the $Z$ decay products are ignored at truth level. The obtained efficiencies for a cut of $d_0 < 16\ \mu$m and $\epsilon_{K^\pm} = 0.95\%$ (no PID) are shown in the left (right) plot of Figure 50. For the $h \to WW^*$ decay mode it is assumed that both $W$ decay hadronically. The actual overall rejection for this channel is eventually greatly improved by the selection of a $h \to jj$ topography. The possible interplay between this selection and the flavour tagger is neglected. The effect of PID can be seen clearly by the lower efficiency for $h \to u\bar{u}, d\bar{d}$ compared to $h \to s\bar{s}$ in the left plot. While the first generation jets amount only for a negligible number of background events in a clean Higgs sample, these efficiencies are important to estimate the suppression of the non-Higgs background events.

Measurements of Higgs branching ratios into dijets at lepton colliders target different final states, according to the additional particles produced in the collision. While the $hqq$ channel profits from a large branching ratio of the $Z$ into two quarks, it is also the channel with the largest background from non-Higgs events. On the other hand, $h\ell\ell$ is a very clean channel which suffers from a very small branching ratio of the $Z$ into leptons and thus collects only a few events. The $h\nu\nu$ channel provides a good signal to background ratio combined with a sizable branching ratio of the $Z$ boson into neutrinos. In addition, the $W$ fusion process contributes significantly to this channel.

The performance of an analysis targeting dijet decays of the Higgs at CLIC with $\sqrt{s} = 350$ GeV in the three final states has been evaluated in Ref. [9]. Taking the detailed information on the abundance of the various flavours of the background jets from Ref. [339] and assuming that the change in the decomposition is negligible when running at $\sqrt{s} = 380$ GeV allows to estimate the limit that can be set by CLIC. With the amount of Higgs and background events reported for the $h\nu\nu$ channel in Ref. [9] an upper 95% CL on the signal strength $\mu \lesssim 50$ is obtained when neglecting systematic uncertainties. This is a significant improvement over the current bound, Eq. (118). The optimal cuts of the strangeness tagger for the given event numbers are relatively mild, requiring no PID, $p_{||} > 6$ GeV and $d_0 < 16$ $\mu$m. When the ratio of Higgs to non-Higgs events in the analysis is greater than $\mathcal{O}(1)$, the background from $WW$ dominates. In this case, the reduction of the background does not sufficiently compensate the loss of signal events unless the tagger cuts are rather loose. If the analysis that selects the $h \to jj$ topology is such that this background plays a less dominant role, the working point of the strangeness tagger can be chosen tighter such that it enhances significantly the signal over background ratio. In the ideal case with $10^4$ Higgses and only background from $h \to bb$, $cc$, and $gg$ the 95% CL upper limit is $\mu \lesssim 32$, obtained for a tagger without PID, $d_0 < 14$ $\mu$m, and $p_{||} > 10$ GeV. PID only plays a role if more than about $3 \times 10^4$ events pass the kinematic selection.

### 3.5.4 Flavour violating Higgs decays

In the Standard Model (SM), the tree level couplings of the Higgs boson are flavour diagonal. Therefore, in the SM flavour violating decays of the Higgs arise only at the loop level and have very small branching ratios. The kinematically allowed quark flavour violating decays $h \to bs$, $h \to bd$, $h \to sd$, and $h \to cu$, are strongly suppressed by small CKM matrix elements. Here and below, we use the notation $h \to ff' = h \to f\bar{f}' + \bar{f}f'$. The branching ratio of the largest of these decay modes can be estimated as $\mathrm{BR}(h \to bs)_{\mathrm{SM}} \sim (1/16\pi^2)^2 |V_{ts}^* V_{tb}|^2 \sim 10^{-7}$, far below expected experimental sensitivities. In the SM, the lepton flavour violating Higgs decays, $h \to \tau\mu$, $h \to \tau e$, and $h \to \mu e$ are suppressed by the tiny neutrino masses and thus below any imaginable experimental sensitivity. Observation of a flavour violating Higgs decay would therefore clearly indicate the presence of new physics.

New physics can generate appreciable flavour violating Higgs couplings. Using the generalized $\kappa$ framework notation, Eq. (117), the branching ratio for a flavour violating Higgs decay is

$$\mathrm{BR}(h \to f_i f_j) = \frac{m_h}{8\pi \Gamma_h} \left( \kappa_{f_i f_j}^2 + \tilde{\kappa}_{f_i f_j}^2 + \kappa_{f_j f_i}^2 + \tilde{\kappa}_{f_j f_i}^2 \right) , \qquad (132)$$

where $\Gamma_h$ is the total Higgs width. The flavour violating coefficients are expected to scale as $\kappa_{f_i f_j}, \tilde{\kappa}_{f_i f_j} \propto v^2/\Lambda^2$ where $v = 246$ GeV is the SM Higgs vev and $\Lambda$ is the scale of new physics.

Various indirect constraints exist on flavour violating couplings of the Higgs boson. Particularly strong bounds can be derived from kaon, $D$ meson, $B_d$ meson and $B_s$ meson oscillations. If Higgs has flavour violating quark couplings it contributes at tree level to these processes. The constraints from meson oscillations are generically weakest if flavour violating Higgs couplings to only one chirality of a particular quark flavour are non-zero. Barring accidental cancellations one finds in this case the following upper bounds (adapted from [249], see also [248])

$$\mathrm{BR}(h \to bs; bd; sd; cu)_{\mathrm{ind.}} \lesssim 8.2 \times 10^{-4};\ 3.7 \times 10^{-5};\ 8.2 \times 10^{-7};\ 5.7 \times 10^{-6}. \qquad (133)$$

The strongest constraints on lepton flavour violating Higgs decays come from bounds on radiative muon and tau decays $\tau \to \mu\gamma$, $\tau \to e\gamma$, and $\mu \to e\gamma$ that can be induced by Higgs exchange at the 1-loop and 2-loop level. Using current bounds, one finds (adapted from [248], see also [249])

$$\mathrm{BR}(h \to \tau\mu; \tau e; \mu e)_{\mathrm{ind.}} \lesssim 0.32;\ 0.24;\ 1.6 \times 10^{-8}, \qquad (134)$$

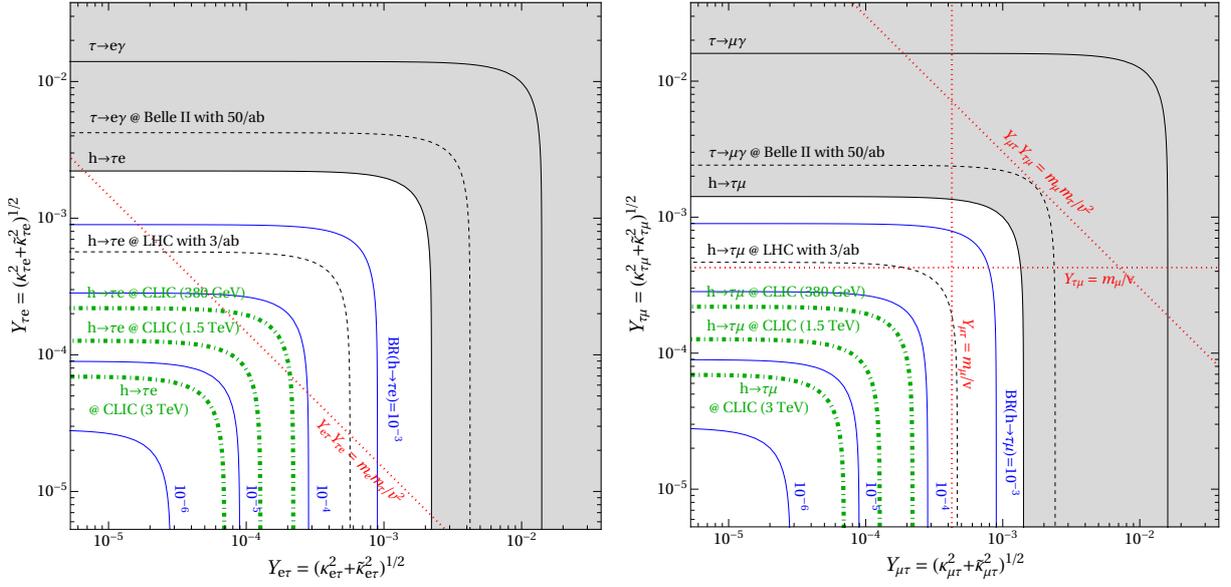

Figure 51: Current constraints (shaded regions) and expected sensitivities (dashed black lines) in the plane of lepton flavour violating Higgs couplings. The solid blue lines show the predicted values for the $h \to \tau\mu$ and $h \to \tau e$ branching ratios. The dotted red lines indicate theoretically motivated targets for the Higgs couplings.

and so large $h \to \tau\mu$ and $h \to \tau e$ branching ratios are still allowed. Direct searches for the decays $h \to \tau\mu$ and $h \to \tau e$ are currently being performed at the LHC[251, 252]. The current strongest bounds from the LHC are $\mathrm{BR}(h \to \tau\mu(e)) \lesssim 2.5(6.1) \times 10^{-3}$ [251, 252]. At the high luminosity LHC, assuming that the sensitivities scale with the square root of integrated luminosity the branching ratios down to few $\times 10^{-4}$ might be probed for both decay modes.

At CLIC, assuming that the backgrounds can be suppressed to a negligible level and using estimates for signal selection efficiency (including tau tagging) of 30% (15%) at 380 GeV (at higher energies) gives the projected 95% CL limit reach

$$\mathrm{BR}(h \to \tau\mu) \lesssim 6 \times 10^{-5}\,(380\,\mathrm{GeV});\quad 2 \times 10^{-5}\,(1.5\,\mathrm{TeV});\quad 6 \times 10^{-6}\,(3\,\mathrm{TeV}). \qquad (135)$$

The plots in Figure 51 show current constraints (shaded regions) and the expected sensitivities (dashed lines) in the plane of lepton flavour violating Higgs couplings. The solid blue lines show the predicted values for the $h \to \tau\mu$ and $h \to \tau e$ branching ratios. The dotted red lines indicate theoretically motivated targets for the Higgs couplings.

Flavour violating Higgs couplings are possible in new physics models where there is more than one term in the Lagrangian that contributes to the fermion masses. Additional mass contributions beyond the SM Yukawa couplings could for example arise (i) from dimension six operators containing three powers of the SM Higgs field (e.g. from integrating out heavy vector-like fermions or lepto-quarks), or (ii) from a new source of electroweak symmetry breaking, e.g., a second Higgs doublet or a technifermion condensate.

The simplest models with vector-like leptons or with lepto-quarks turn out to be strongly constrained by lepton flavour violating tau decays [340, 341]. Lepton flavour violating Higgs decays $h \to \tau\mu$ and $h \to \tau e$ at an observable level cannot be accommodated in these models. EFT arguments suggest that this is a generic property of models that do not contain a new source of electro-weak symmetry breaking [260]. The tau decay constraints can only be avoided in sufficiently complex new physics sectors where $h \to \tau\ell$ and $\tau \to \ell\gamma$ can be decoupled [342] or in models where $h \to \tau\ell$ is mimicked by a flavour conserving three-body decay of the form $h \to \tau \ell \phi$, with $\phi$ a flavourful mediator that decays into a dark sector [343].

In models with a new source of electroweak symmetry breaking, sizable branching ratios of lepton flavour violating Higgs decays are much more generic. In particular, models with extended Higgs sectors have been extensively studied in the literature [254, 260, 261, 278, 344–349]. It has been found that BR($h \to \tau\mu$) and BR($h \to \tau e$) up to the current direct bounds are possible in many of such models.

If a non-zero signal for either $h \to \tau\mu$ or $h \to \tau e$ is established by future searches, additional measurements can be made to identify some properties of the BSM physics that induces the flavour violating Higgs decay. For instance, lepton charge asymmetries can measure additional BSM sources of CP violation [350]. Measurements of the tau polarization could give insights on the chirality structure of the new physics couplings.

# 4 Electroweak symmetry breaking

## 4.1 Supersymmetry

Supersymmetry remains a greatly motivated extension of the Standard Model, which may provide a dynamical reason for the stabilization of the weak scale, as well as offering plausible dark matter candidates among the many new states that are necessary in supersymmetric models.

The present status of searches at the Large Hadron Collider puts the simplest models under a fair amount of pressure because of the absence so far of any hint of colored super-particles, e.g. scalar quarks and gluino. In full explorations of phenomenological models [351] it is found that super particles in the range of a few hundred GeV are generically not yet excluded. The light super-partners not yet excluded by current searches are usually not charged under strong interactions. However in some cases it is still possible to have such relatively light super-partners charged under strong interactions, e.g. because they give rise to experimental signatures that populate phase-space regions typical of Standard Model reactions, where backgrounds are large and searches are difficult in the busy $pp$ scattering events. A broad activity in searching for these states is in progress at the Large Hadron Collider and significant progress has been made as more data has been collected. Still, it is very hard to predict what the final coverage of the Large Hadron Collider will be on these challenging searches. Outside of these experimentally problematic regions of the parameters space one can find the expected reach of the Large Hadron Collider High-Luminosity phase in Refs. [352–354] from which one sees expected limits in the range of TeV masses for colored states, the precise bound depending on the details of the assumed spectrum of the supersymmetric states. For states that are not charged under strong interactions it is possible to have large regions of parameters space not explored by the Large Hadron Collider featuring several sub-TeV electroweak states, e.g. additional Higgs bosons, their supersymmetric fermionic partners, as well as partners of the electroweak gauge bosons.

The search for supersymmetric particles, and the measurement of their properties, is a topic of classic work at $e^+e^-$ and in particular at CLIC, where, thanks to the large centre-of-mass energy, one expects to produce several supersymmetric states and even to perform some tests of the underlying supersymmetric model and of its supersymmetry breaking pattern. A large body of work exists on this subject. In [355] the reader can find a detailed summary of studies on the discovery reach for supersymmetric particles at CLIC. The general conclusion of these searches is that superpartners can be discovered at CLIC when they are sufficiently light to be produced on-shell. In general their mass can also be measured with a few percent accuracy, opening interesting possibilities for testing the details of the underlying model. In the following we report some details of a sample of results about supersymmetric particles searches still not excluded by the Large Hadron Collider.

### 4.1.1 Higgs bosons

A recent evaluation of the search for additional Higgs bosons of the Minimal Supersymmetry Standard Model has appeared in CLIC CDR [5] as well as in previous works Refs. [356, 357] and Section 4.2 of [11], to which we refer the reader for details. Owing to the $SU(2)$ doublet nature of the Higgs bosons of the MSSM, these can be produced by gauge interactions and observed in decays to SM fermions such as in reactions

$$e^+e^- \to H^0 A^0 \to b\bar{b}b\bar{b}, b\bar{b}t\bar{t}, t\bar{t}t\bar{t}$$
$$e^+e^- \to H^+H^+ \to t\bar{b}t\bar{b}$$

The conclusion of these works is that Higgs bosons in the MSSM can be observed almost up to the kinematical threshold of CLIC at 1.5 TeV. Furthermore, these works have established the ability of CLIC to measure masses of the Higgs bosons of the MSSM to the percent level, and to distinguish the mass splitting among all of these states, which can be crucial for understanding the underlying model.

The indirect effects of these Higgs bosons in properties of the 125 GeV Higgs boson have been studied recently in Ref. [358], which finds sensitivity to Higgs bosons beyond the kinematic reach for some regions of the MSSM parameters space. Although these results consider the performance expected at the International Linear Collider, we expect similar results should hold for CLIC. In the following Section 4.3 we will present results from Ref. [359] that apply for more general Higgs sectors and confirm this expectation.

*4.1.2 Fermionic and bosonic superpartners*

The search for superpartners of the SM gauge bosons and fermions have been discussed at length in the body of literature on searches for supersymmetry at CLIC. In contrast with the case of Higgs bosons, in the search for superpartners one expects to produce partially undetectable final states, owing to the presence of neutral weakly interacting particles which do not interact with the detectors. These particles can appear at the bottom of the supersymmetric spectrum and therefore produce momentum imbalance between the initial and final state of each scattering event in which new particles are produced.

When the mass difference between these invisible states and the rest of the supersymmetric particles is not too small the imbalance of momentum can be leveraged to use distinguish new physics events from Standard Model events. In this case previous studies, summarized in [5] and [355], have shown that CLIC can produce and discover supersymmetric particles in the vast majority of the parameter space of the supersymmetric models.

Results for chargino and neutralinos have appeared in Ref. [360] for fully hadronic final states with four jets and missing transverse energy from the reactions

$$e^+e^- \to \chi^+\chi^- \to W^+W^-\chi_1^0\chi_1^0,$$
$$e^+e^- \to \chi_2^0\chi_2^0 \to ZZ\chi_1^0\chi_1^0.$$

The study was performed using full simulation and considering pileup from $\gamma\gamma \to$ hadrons. The discovery potential for charginos and neutralinos is established for masses of the lightest chargino 643.2 GeV and masses of the lightest and next-to-lightest neutralinos 340.3 GeV and 643.1 GeV, respectively. Masses and production rates of the supersymmetric states can be measured with few percent accuracy. These conclusions are expected to hold for most models with no small mass differences among the involved new states.

Detailed results for sleptons can be found in [361–363]. The final states studied include a two lepton final states and a final state with four jets and two leptons, which arise from reactions

$$e^+e^- \to \tilde{\ell}^+\tilde{\ell}^- \to \ell^+\ell^-\chi_1^0\chi_1^0$$
$$e^+e^- \to \tilde{\nu}\tilde{\nu} \to \ell^+\ell^-\chi^+\chi^-$$

plus a comprehensive list of backgrounds and underlying event activity. The results of these studies is that discovery and measurements of properties of the superpartners can be attained at CLIC, including percent and sub-percent mass determination.

Studies exists for the search for squarks. Ref. [364] has investigated the reach in the search for light-flavored squarks. This work considers four squark mass eigenstates (right-handed $u, d, s, c$ squarks) around 1 TeV and a lightest supersymmetric particle $\chi_1^0$ with mass 328 GeV, which give rise to the reaction

$$e^+e^- \to \tilde{q}\tilde{q}^* \to q\bar{q}\chi_1^0\chi_1^0.$$

A search in a final state with jets and missing momentum, including a comprehensive list of backgrounds and underlying $\gamma\gamma$ scattering activity, finds discovery potential for these squarks, which can safely be extrapolated to other models with non-compressed mass spectra and similar cross sections. Most such models are under heavy pressure from searches at the Large Hadron Collider, e.g. [365], and are expected to give rise to significant discrepancies with the SM prediction during the forthcoming runs of the Large Hadron Collider. As a consequence, the figure of merit of CLIC for the study of squarks lies in the capability to measure their properties. Ref. [364] finds a potential to measure the squarks masses with sub-percent accuracy and their production rate with a few percent accuracy. More detailed studies of the impact that squark properties determination can have in the understanding of electroweak symmetry breaking will be given in Section 4.4.1 for the case of stop and sbottom squarks and their impact on the calculation of the Higgs boson mass from the fundamental parameters of the MSSM.

### *4.1.3 Compressed spectra*

Especially after the experience of searches for superpartners at the Large Hadron Colliders it is very clear that challenging mass spectra are not uncommon, though not necessarily generic, in the MSSM and its extensions. The difficulty typically arises from either the small amount of momentum carried by invisible particles or the softness of visible final states in the reactions involving new physics states. These problems are quite generic when either there is little phase-space available for the new particles decay (due to mass degeneracy with the decay products), or the invisible particles are light, or both of these are true.

In all the above cases the detector activity expected in new physics reactions is very similar to that of copious Standard Model reactions, hence it is very hard to separate new physics from backgrounds. This is especially the case when the only handles to tag new physics events are soft objects, i.e. low $p_T$ tracks, jets, or leptons. The performance of the detector is crucial for assessing the potential for new physics giving rise to such challenging signatures. The potential of CLICdet [6] has not yet been evaluated on these difficult signals; still, we can have a very good impression of the capabilities of the CLICdet looking at the results obtained for the ILD detector in dedicated studies. We will review results from the ILC studies in this section, while in forthcoming Section 5.4 we will recast them for a specific case of interest for models of Dark Matter co-annihilation models.

In reference [366] the potential has been evaluated of a 500 GeV $e^+e^-$ collider to isolate new physics that gives rise to low visible particles activity events (e.g. characterized by low invariant mass for the recorded particles in the event) in a dataset corresponding to 0.5 ab$^{-1}$ integrated luminosity. In the case of first and second generation slepton and stau production

$$e^+e^- \to \tilde{\ell}^+\tilde{\ell}^- \to \ell^+\ell^-\chi_1^0\chi_1^0$$
$$e^+e^- \to \tilde{\tau}^+\tilde{\tau}^- \to \tau^+\tau^-\chi_1^0\chi_1^0$$

a detector level study[39] has been carried out taking into account beam conditions for the ILC. Taking into account relevant backgrounds it has been shown that an $e^+e^-$ collider has the potential for discovery for both sleptons and stau. In particular the result claimed is a discovery reach all the way up to few GeV from the kinematic limit of the machine. These results are expected to be qualitatively the same at different centre-of-mass energies, hence we deem that CLIC will have sensitivity to sleptons and staus in essentially the whole kinematically available mass range. Especially the results for $\tilde{\tau}$ are very encouraging for other cases of compressed spectra, as the tau leptons themselves give rise to either jetty or partially invisible final states, same challenges that need to be faced for the more complex squarks and electroweak-inos cases.

Strategies for compressed super-particles spectra involving Higgsinos have been studied for the ILC in Ref. [367], taking into account a comprehensive list of backgrounds and detector effects for the ILD detector. In this work mass spectra for Higgsinos are taken so that mass differences range between about 0.7 GeV and 2 GeV for Higgsino average mass around 165 GeV, which would be accessible at a 500 GeV $e^+e^-$ collider considered in this work. These Higgsino are discoverable and their cross sections and masses can also be measured. Ref. [367] finds that the mass differences can be measured to 40–300

---

[39]We refer to [366] for details on the simulation.

MeV, the absolute masses to 1.5–3.3 GeV, and the cross sections to 2–5% accuracy. These results are likely applicable at CLIC as well, as the ILD detector shares a good deal of similarities with ILD.

It should be remarked that the ability to discover almost degenerate Higgsinos is very important for the exploration of weak scale new physics. Additionally it is very important to be able to measure the properties of the discovered Higgsino as these are powerful probes of yet to be discovered new physics particles. The fact that searches at CLIC for moderately compressed spectra may lead to discovery at CLIC is of even greater importance, in view of the fact that a Higgsino around 1.1 TeV could be a candidate thermal relic dark matter particle [368, 369]. Indeed the search of a pure Higgsino at CLIC for even more compressed spectra will be discussed at length in Section 5.2 below. Here we anticipate the result that, owing to long-lived nature of the electrically charged Higgsino it is possible to search for short tracks in the inner tracker system, which disappear after few centimetres, yet have large momentum and are isolated from other activity in the event. The integrated luminosity and energy of the CLIC staging presently considered may allow exclusion of a pure Higgsino dark matter candidate even in the presence of some amount of background.

Furthermore, even when it is not a dark matter candidate, naturalness considerations tend to favour the Higgsino being the lightest supersymmetric particle, possibly much lighter than other electroweak-ino states. The mass splitting of the Higgsinos then becomes a very powerful diagnostic of the separation between the Higgsino mass scale and the heavier states that can split the Higgsino-like mass eigenstates through mixing with the bino and wino super-partners.

When all supersymmetric states are decoupled the Higgsino mass eigenstates are split by electroweak symmetry breaking effect at loop level and have a mass splitting predicted around 300 MeV [370]. For such small mass splittings it is hard to probe Higgsino at the Large Hadron Collider [371, 372]. However, when the bino and wino are light enough to have some component in the Higgsino-like mass eigenstates the mass splitting usually grows larger. If bino and wino are close enough to the Higgsino they can make the mass splitting large enough to make the model accessible at the Large Hadron Collider. The situation at at CLIC is quite different. In fact Higgsinos with large mass splitting can be probed by usual large-$p_T$ searches, e.g. [360], while small splittings can be addressed with specific techniques that benefit greatly from the clean $e^+e^-$ environment as discussed in Section 5.2 below.

All in all CLIC offers great chances to discover Higgsinos in very motivated mass ranges, as well as to measure their properties accurately, shedding lights on even higher mass scale physics. Remarkably, the typical mass splitting between the Higgsino multiplets states has an (inversely) linear dependence on the mass of heavy super-partners

$$m_{\chi^+} - m_{\chi^0} \propto \frac{v^2}{M_i}.$$

Therefore, it is important to interpret progress in the search for small mass splitting, even just a factor of a few smaller compared to the Large Hadron Collider, as an extension of the mass scales of heavy new physics probed by CLIC by roughly the same factor of a few [373].

## 4.2 Higgs plus singlet

In many extensions of the Standard Model the scalar sector is extended by new states that are not charged under the Standard Model gauge group. These states, usually referred as *singlet* state (from group theory terminology), do not interact with the Standard Model gauge bosons and fermions at tree level, but may acquire such interactions at loop level or because of mixing with the Higgs boson radial mode. We deal with loop level couplings in later Section 8.3 on so-called *axion-like* particles, here we present results on new states that couple to the Standard Model only though mixing with the Higgs boson.

The singlet state can have any mass, both below and above that of the 125 GeV Higgs boson. Indeed the mass of the singlet may or may not originate from the same dynamics responsible for the mass of the Higgs boson and the scale of weak interaction. In full generality we can parametrize the mass matrix of the Higgs plus singlet system by a symmetric 2-by-2 matrix, which in general has three independent real parameters. The diagonal elements of the matrix are always allowed by internal symmetries, whereas the off-diagonal one, which gives rise to the mixing, can be forbidden by internal symmetries. For the diagonal elements, however, there may be other reasons for suppression, e.g. a shift symmetry can be advocated to have a hierarchy between two diagonal elements. This argument may be used to make the singlet lighter than the 125 GeV Higgs boson or the vice-versa. In the two following subsections we will deal with both cases in turn.

For the mixing term, as said, it is always possible to assume that it is vanishing because of a selection rule stemming from an internal symmetry. For the largest possible value that the mixing term can attain more discussion is needed. In fact mixing terms tend to increase the mass difference between mass eigenstates and cannot be taken arbitrarily large for fixed diagonal terms, as they make the lightest state get a negative mass square. Parametrizing the mass matrix as

$$\begin{pmatrix} \mu^2 & \mu^2\rho \\ \mu^2\rho & M^2 \end{pmatrix}$$

in a generic basis $(h_1, h_2)$ we can derive the eigenvalues and eigenvectors of the matrix in the limit of large $M^2 \gg \rho\mu^2, \mu^2$

$$m_+ \simeq M^2 + \rho^2\mu^2\frac{\mu^2}{M^2}, \qquad s_+ \propto h_2 + \rho\frac{\mu^2}{M^2}h_1,$$
$$m_- \simeq \mu^2 - \rho^2\mu^2\frac{\mu^2}{M^2}, \qquad s_- \propto h_1 - \rho\frac{\mu^2}{M^2}h_2.$$

This means that if the mass scales that are linked to the Higgs boson and the singlet are sufficiently different from one another, i.e. $\mu \ll M$, and the mixing among these states is not forbidden by anything in the dynamics that generates the mass of the lightest state $h_1$, i.e. $\rho \simeq 1$, then the mixing angle is expected to be

$$\theta \lesssim \rho\mu^2/M^2 \simeq \left(\frac{m_-}{m_+}\right)^2, \qquad (136)$$

i.e. the mixing is dictated by the square of the ratio of the masses of the two states. This argument shows that for largely different masses of the two states one does not expect them to be largely mixed, the expected mixing scaling with the second power of the heavy mass for fixed light state mass. It should be noted that this argument made no use of "who is who"; that is to say, it holds both when the singlet is much lighter than the 125 GeV Higgs boson, or in the opposite case of a very heavy singlet.

Exceptions exist to this argument, simply because it has been derived under assumptions. For instance one can imagine a special mass matrix for the $h_1, h_2$ system

$$\begin{pmatrix} M & M \\ M & M \end{pmatrix}$$

which would have a very light state, indeed this matrix has a vanishing eigenvalue, and a maximal mixing angle $\theta \simeq \pi/4$. This would correspond to a large $\rho \simeq (M/\mu)^2$ for which indeed the expansion we have performed above breaks down.

A more moderate breaking of the above formulae and the resulting expectations arises when $\rho > 1$ for a systematic reason. This is the case of $\mu$ being originated from the breaking of a shift symmetry and the off diagonal term is generated at the same scale, e.g. $\mu^2 \simeq \left(\frac{v}{f}\right)^2 M^2$ and $\rho \simeq \frac{f}{v}$. In this case the expected mixing from formulae above is

$$\theta \lesssim \frac{v}{f} \simeq \frac{m_-}{m_+}. \qquad (137)$$

In this is case it is possible to have large mixings compared to the case discussed above; the mixing angle is saturated by a quantity that scales linearly with the mass of the heavier state for fixed mass of the light one.

These arguments, though they may fail in specific model because of peculiarities or tuning of the model, are quite a general guide to gauge the expected amount of mixing that it is interesting to probe in experiments that search for a mixed singlet state.

### *4.2.1 Heavy singlets* [40]

The motivations to consider extra singlet-like states at colliders are manifold. A relatively light scalar singlet is present in several extensions of the SM at the TeV scale, most notably the Next-to-Minimal Supersymmetric Standard Model (NMSSM, see e.g. [374]) and many realisations of the Twin Higgs idea [375]. The presence of a singlet can also modify the finite temperature potential of the SM Higgs inducing a first order electroweak phase transition [376–378] which is a necessary requirement for electroweak baryogenesis [379].

The capabilities of current experiments and possible upgrades of the LHC to test this kind of scenario have been addressed extensively in the literature, see e.g. [380]. It is therefore of high priority to study what are the possible exclusions attainable at a linear lepton collider such as CLIC, especially in multi-TeV stages where heavy singlets can be produced directly. In what follows we focus on the case where the extra singlet is heavier than the Higgs boson.

The following Lagrangian

$$\mathscr{L} = \mathscr{L}_{\text{SM}} + \frac{1}{2}(\partial_\mu S)^2 - \frac{1}{2}m_S^2 S^2 - a_S S|H|^2 - \frac{1}{2}\lambda_{HS} S^2 |H|^2 - V_S(S), \quad (138)$$

describes the most general renormalisable interactions of a real scalar singlet with the SM. Given the structure of the model it is very convenient to define the mixing angle $\gamma$ as the rotation angle needed to go from the interaction basis of eq.(138), where only the Higgs couples to fermions, to the mass basis. The angle $\gamma$ is for all intents and purposes the same as the angle $\theta$ discussed in the previous Section. In any event, we use the notation $\gamma$ for the mixing angle in this Section as we will map it to some explicit models in the following. For now the angle $\gamma$ can just be defined through the mass eigenstates definition

$$h = h_0 \cos\gamma + S \sin\gamma, \qquad \phi = S \cos\gamma - h_0 \sin\gamma, \quad (139)$$

where $h$ is the 125 GeV SM-like Higgs, and $\phi$ is the heavier singlet-like state with mass $M$. From this equation we see already that the phenomenology is mainly dictated by $\sin\gamma$, as it enters two main aspects of the model:

– Higgs signal strengths. They are universally rescaled by a factor $(1 - \sin^2\gamma)$. This implies that all the sensitivities at CLIC for the Higgs couplings can be immediately recast into a constraint on $\sin^2\gamma$.
– Single production of $\phi$. It corresponds to the production of a heavy SM Higgs with an overall rescaling given by $\sin^2\gamma$.

*Production of the singlet-like state*

We will consider both single and double production of the singlet. Single production is only sensitive to $\sin^2\gamma$ and the mass of the singlet-like state, while double production can in principle probe other parameters of the potential. A notable example is the case in which an internal symmetry, e.g. an exact $Z_2$ acting on the singlet, forbids the mixing with the Higgs. In this case double production of singlets is still allowed through the portal coupling $\lambda_{HS}$.

---

[40] Based on a contribution by D. Buttazzo, D. Redigolo, F. Sala and A. Tesi.

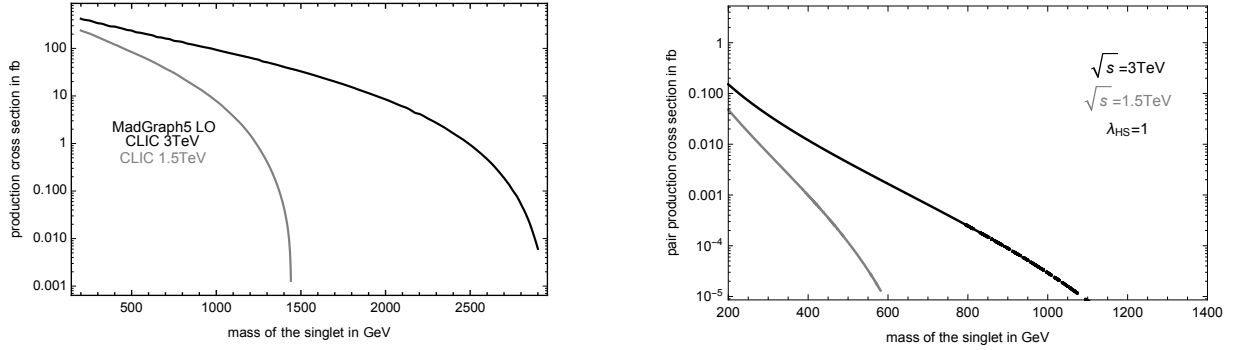

Figure 52: Left (right) panel: single (double) production of $\phi$ via $WW$-fusion, assuming $\sin^2\gamma = 1$ ($\sin^2\gamma = 0$).

At CLIC a few production channels for the singlet are available, whose relevance crucially depends on the centre-of-mass energy. By inspecting the behaviour at high energy, we see that the total rate for $\sigma(e^+e^- \to V^*V^*\nu\nu \to X\nu\nu)$ does not fall with energy in the case of $WW$-fusion for single and double production. The total rates can be computed

$$\sigma_{\nu\bar{\nu}\phi} = \frac{\sin^2\gamma}{v^2}\frac{g^4}{256\pi^3}\left[2\left(\frac{M^2}{s}-1\right)+\left(\frac{M^2}{s}+1\right)\log\left(\frac{s}{M^2}\right)\right] \approx \frac{\sin^2\gamma}{v^2}\frac{g^4 \log\left(\frac{s}{M^2}\right)}{256\pi^3}, \quad (140)$$

$$\sigma_{\nu\bar{\nu}\phi\phi} \approx \frac{g^4|\lambda_{HS}|^2}{49152\pi^5}\frac{(\log(s/M^2)-14/3)}{M^2}, \quad (141)$$

in the high energy limit and where $M$ is the physical mass of the singlet-like state. The full expressions are retained in the numerical calculations and have been also cross-checked with MADGRAPH 5 v2 at LO [31, 381]. The above formulas are extremely good approximations as long as the dominant contribution to the rates comes from kinematic configurations where $M_V^2/\hat{s} \ll 1$. The full computation is used to draw Figure 52, and we checked that it is reproduced to an excellent accuracy by the asymptotic expressions in Eqs. (140), (141). We remark that in what follows, we assume unpolarized electron beams, leaving possible optimization of the sensitivity that might result from enhanced polarized rates for future work. The cross section for Higgs-strahlung processes goes as $\sigma(e^+e^- \to Zh) \sim g^4/s$ at high energy (see also [382] for comparison). The same scaling holds for generic $\phi$-strahlung, therefore in the following we just consider $VV$-fusion processes for the production of scalar particles.

It is also interesting to analyze the different behaviour of single and double production with the parameters of the model. Two behaviours clearly emerge

- *Single production.* From the analytic expression it is evident that single production is a direct test of the mixing between the Higgs and the scalar. In the high-energy limit, the dependence is only logarithmic in the mass and linear in $\sin^2\gamma$. The limits will be not much sensitive to the mass of the scalar (for fixed mixing angle) until the kinematic threshold is reached. Within a concrete model, however, $\sin^2\gamma$ and the mass of the scalar are correlated. We will discuss later the behavior of single production in the different cases.
- *Double production.* It is a clean test of a combination of the quartic coupling and the mass of the singlet. This is a major difference with respect to the SM double Higgs production which is proportional both to $|\lambda_h|^2$ but also to $g^4$, with a numerically large coefficient in front of the latter contribution [58].

*Decay of the singlet-like state*

When the singlet is sufficiently heavy than the electroweak gauge bosons and the Higgs, the SO(4) symmetry of the SM Higgs potential and the equivalence theorem imply that

$$2\Gamma(\phi \to hh) = 2\Gamma(\phi \to ZZ) = \Gamma(h \to WW) = \sin^2\gamma \frac{M^3}{8\pi v^2} \qquad (142)$$

Other decay channels are $\Gamma(\phi \to t\bar{t}) = (3y_t^2)/(16\pi)\sin^2\gamma M$, which is subleading at high mass, and a possible decay into invisible or exotic decay products if other interactions are allowed beyond the minimal model in Eq. (138).

*Sensitivity at CLIC*

For single production we compute the significance for a "cut-and-count" experiment, that we define as

$$\text{significance} = \frac{S}{\sqrt{(S+B) + \alpha_{\text{sys.}}^2 B^2}}, \qquad (143)$$

where $B$ and $S$ are respectively the number of background and signal events. As systematic uncertainty on the background we choose $\alpha_{\text{sys.}} = 2\%$ in all the cases considered. We compute the single production rate as a function of the mass of the scalar analytically and numerically, and the results are shown in Figure 52. The value in the plot should be multiplied by the rescaling factor $\sin^2\gamma$ and by the branching fractions into a given final state. Given that, by virtue of the SO(4) symmetry of the SM, which is well approximated when $m_\phi \gg M_V$, we expect similar ratios into $ZZ$ and $hh$, it is evident that the individual largest rate is into $4b$.

*The decay channel $\phi \to VV$*

Given the large signal rate in $hh(4b)$, and given that the related backgrounds are expected to be smaller than those to $WW$ and $ZZ$, we do not perform a detailed study of the sensitivities from the $VV$ channels. The purpose of this Section is to rather to give a simple estimate of such sensitivity, which will then be shown to be a less powerful than the one from $hh(4b)$.

Following the discussion of the previous Section, we assume that the dominant backgrounds to $e^+e^- \to \nu\bar{\nu}\phi(VV)$ come from on-shell production of EW gauge bosons, and using MADGRAPH we simulate events for $e^+e^- \to \nu\bar{\nu}ZZ$ and $e^+e^- \to \nu\bar{\nu}WW$. We find a total cross section of 131 (52) fb and 57 (18) fb respectively at CLIC3 (CLIC1.5). We then assume that all $W(jj)$ and $Z(jj)$ will be told apart thanks to the excellent jet mass resolution of CLIC, so that we do not include backgrounds coming from the process $e^+\gamma \to e^+\nu W^-Z$ nor from its conjugate (for completeness we report the sum of their cross section is 330 fb at CLIC 3 TeV and 120 fb at CLIC 1.5 TeV). Finally, we assume that all backgrounds without neutrinos in the final state will become negligible upon imposing suitable missing energy cuts.

We consider the following four kinds of resonant signals: $ZZ(4\ell)$, $ZZ(2\ell 2j)$, $ZZ(4j)$ and $WW(4j)$. For each signal mass $m_\phi$, we select the simulated background events that satisfy $m_\phi - \Delta m_{\phi,X} < m_{VV} < m_\phi + \Delta m_{\phi,X}$, where $X = 4\ell, 2j2\ell, 4j$, and where we assume $\Delta m_{\phi,4\ell} = 5\%$, $\Delta m_{\phi,2j2\ell} = 10\%$, $\Delta m_{\phi,4j} = 15\%$. Using the total background cross sections, we then obtain the number of background events entering Eq. (143) at both CLIC stages. To obtain the number of signal events entering Eq. (143), we multiply the production cross section by the branching ratio of $\phi$ into $ZZ$, $WW$, and by the branching ratios of $Z$ and $W$ into the final state of interest. This procedure, quite optimistically for the estimate of the reach in the $VV$ channel, assumes that the cuts results in an efficiency of 100% for the signal. We show in Figure 53 the sensitivities obtained this way, for the exclusive channels $ZZ(4\ell)$ and $ZZ(2\ell 2j)$, and for the combination of the channels $ZZ(4\ell)$, $ZZ(2\ell 2j)$, $ZZ(4j)$ and $WW(4j)$. As it will be clear in the following, even these optimistic estimates, give worse sensitivity than other channels, to which we turn in the next Section.

*The decay channel $\phi \to hh$*

The decay of the singlet into two Higgs bosons is a very promising channel, which benefits from the large branching fraction $h \to b\bar{b}$. The dominant background is the irreducible contribution from $e^+e^- \to 2\nu 4b$, with a dominant component due to $ZZ(4b)$. A potentially large reducible contribution from $\gamma\gamma \to 4b$ is avoided imposing a cut on the transverse momentum of the $b$ quarks, and turns out to be negligible. The total cross section for $e^+e^- \to 2\nu 4b$ is computed with MADGRAPH to be 0.53 fb (1.65 fb) at the centre-of-mass energy of 1.5 TeV (3 TeV). We also compute the cross sections for the signal, $e^+e^- \to \phi(4b)\nu\bar{\nu}$, with MADGRAPH, after implementing the Lagrangian (138) in FEYNRULES 2.0 [383], and retaining the subdominant contribution from $\phi \to ZZ$. As detailed in Ref. [384] we use DELPHES3 [385] for detector simulation, using the CLICdet card [386], and applying the VLC jet reconstruction algorithm [232] with working point $R = 0.7$ and $N = 4$ (see also [99]).

In order to select the events we proceed with the following identification cuts:

1. $b$-tagging: we select events with four jets tagged as $b$, using a loose selection criterion in order not to excessively reduce the signal efficiency, and requiring each $b$-jet to have a $p_T$ of at least 20 GeV.

2. $h$ reconstruction: we identify the candidate Higgs bosons by choosing the pairing of the four reconstructed jets with an invariant mass that is closest to 125 GeV, and retaining the events having two distinct b-pairs with $m_{b\bar{b}}$ in the window of about [90, 130] GeV. The exact values of the invariant mass window are chosen for each $M_\phi$ hypothesis in order to maximise the significance of the signal.

3. $\phi$ reconstruction: we select the events with a total invariant mass of the $4b$ system in a window of about $[0.8, 1] M_\phi$ around the resonance peak, again optimising the exact size of the window for each signal hypothesis.

We tested the efficiencies $\epsilon_{S,B}$ for signal and background for steps $1, 2$ (i.e. after identification cuts) and we found no substantial differences varying the parameters of the jet reconstruction algorithm. Modulo detector effects, steps $1, 2$ appear to be effective in reducing the backgrounds down to the only contribution $e^+e^- \to 2\nu 2h$, indeed $\epsilon_B \lesssim 2\%$ for several sizes of $R$ in the VLC clustering algorithm, which is a factor of five smaller that the ratio between $\sigma(e^+e^- \to 2\nu 2h(4b))/\sigma(e^+e^- \to 2\nu 4b) \approx 10\%$. The efficiency on the signal after applying the cuts is $\epsilon_S \approx 30\%$. Notice that we simulated the signals assuming unit size coupling but we always worked in the narrow width approximation (NWA).

In Figure 53 we show the results for the CLIC sensitivity in $\sigma(e^+e^- \to \phi\nu\bar{\nu}) \times \mathrm{BR}(\phi \to hh)$ as a function of the mass of the singlet, and compare it with the reach in the two $\phi \to VV$ channels described above. Even though the $VV$ sensitivity is optimistic, and lacks a detailed simulation of both decay and detector effects, it can be seen in Figure 53 that the $hh$ channel results in being the best probe of scalar singlet production in this model over the entire mass range. One can also see that the sensitivity depends mildly on the collider energy and resonance mass, except for phase-space effects, which make heavier scalars completely inaccessible at lower centre-of-mass energies.

In Figure 54 we translate the projected bounds on the cross section into a limit on the mixing angle $\sin^2 \gamma$, which is simply obtained rescaling by the cross section of a SM Higgs with the same mass (we consider gluon fusion plus VBF plus VH production, and compute the first one with GGHIGGS v3.5 [387–389], and the second two with MADGRAPH). We compare the results with Higgs couplings measurements and direct searches at the LHC, showing the present exclusions [235, 390, 391][41] and projected Higgs couplings constraints from HL-LHC [21] as well as the projected direct search sensitivity at 300 fb$^{-1}$ and 3 ab$^{-1}$, that we update from ref. [380] using the procedure described therein, projecting the expected sensitivity of [390]. It can be seen that CLIC at 1.5 TeV and 3 TeV will be able to probe singlet masses up to about 1.3 TeV and 2.8 TeV, respectively. CLIC3TeV will be significantly more sensitive than the high-luminosity LHC over the full mass range. In particular if we use the arguments that lead to Eq (137) to estimate the maximal mixing that we expect in a given class of models, we should consider as theoretically motivated only the values of the mixing below the upper grey line in Figure 54. In this context the bounds from CLIC 3 TeV, reaching up to about 1.8 TeV, nearly doubles the mass of the singlet probed by the HL-LHC in theoretically motivated parameter space.

---

[41] For the Higgs couplings the 13 TeV best-fit signal strength from CMS [236] is larger than one by almost two sigma, so to be conservative we use the combined ATLAS and CMS 8 TeV fit [235].

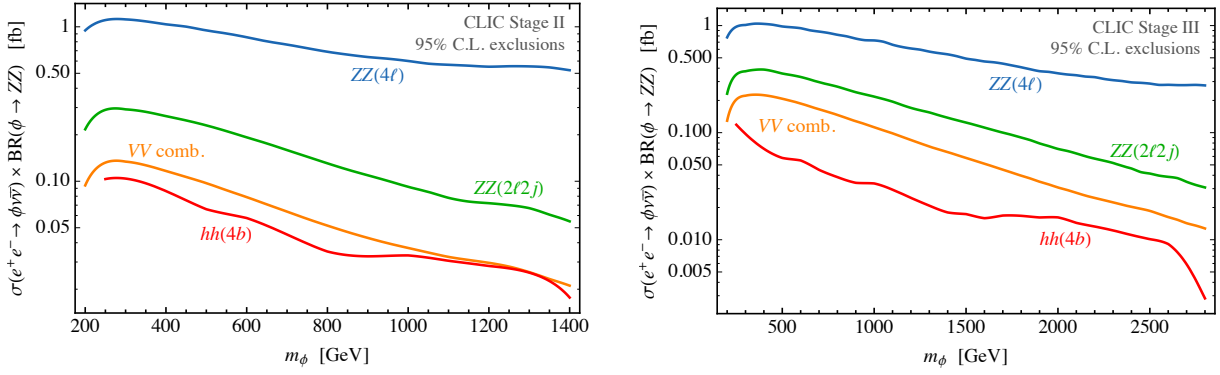

Figure 53: Comparison between the projected 95% C.L. exclusions in different channels on $\sigma(e^+e^- \to \phi\nu\bar{\nu}) \times \mathrm{BR}(\phi \to XX)$, with $X = h, Z$, at CLIC with $\sqrt{s} = 1.5\,\mathrm{TeV}$, $\mathcal{L} = 1.5\,\mathrm{ab}^{-1}$ (left-hand plot), and $\sqrt{s} = 3\,\mathrm{TeV}$, $\mathcal{L} = 3\,\mathrm{ab}^{-1}$ (right-hand plot). The limits, from top to bottom, come from $\phi \to ZZ \to 4\ell$ (blue), $\phi \to ZZ \to 2\ell 2j$ (green), combination of $\phi \to ZZ \to 4\ell, 2\ell 2j, 4j$ and $\phi \to WW \to 4j$ (orange), and finally $\phi \to hh \to 4b$ (red).

These bounds can be combined with those obtained by looking at indirect effects, especially Higgs boson couplings fits. In this model all the 125 GeV Higgs couplings are rescaled by a common factor, $\cos\gamma$, defined through Eq. (139). Applying the result of fitting the Higgs measurements of CLIC after the three energy stages from Section 2.1

$$\sin^2\gamma \leq 0.24\% \text{ at 95\% CL}$$

we can exclude all of the region of the plane above the horizontal dashed black line in Figure 54. We remark that there is significant overlap between the regions probed by the precision study of Higgs couplings, especially benefiting from the improvement from 3 TeV CLIC, and by the direct searches. Therefore a nice interplay, and cross-check in case of deviations from the SM, is expected between direct and indirect searches for heavy singlet states at the 3 TeV CLIC.

*Double production*

The double production rate is sensitive to the mass of the singlet, the quartic coupling $\lambda_{HS}$, and the higgs-singlet mixing $\sin^2\gamma$. However the contribution to the total rate from $\sin^2\gamma$ is small for the typical values of this coefficient allowed even by current constraints. The rate scales like $M^{-2}$ as in the right panel of Figure 52, where it is computed for $\sin\gamma = 0$. Numerically the cross section is tiny, therefore a relevant sensitivity could be retained only if the coupling $\lambda_{HS}$ is sizable.

As this process can be studied even for zero Higgs-singlet mixing it is interesting to consider generic decays of $\phi$, not necessarily the Higgs-like decays that we considered for single production. These decays are necessarily model dependent and we leave a comprehensive study for future work. As a guide to the potential reach of CLIC in probing the process

$$e^+e^- \to \phi\phi\nu\bar{\nu} \tag{144}$$

we draw iso-lines for 10 and 100 events at CLIC 3 TeV $3\mathrm{ab}^{-1}$ and 1.5 TeV $1.5\mathrm{ab}^{-1}$ in Figure 56. We remark that, even when $\phi$ decays only in Higgs-like modes, it gives rise to very structured events with many jets, e.g. $\phi\phi \to 4h \to 8b$ or $\phi\phi \to VV \to 8j$, and SM processes with such large multiplicity and such resonance structure are expected to be rare.

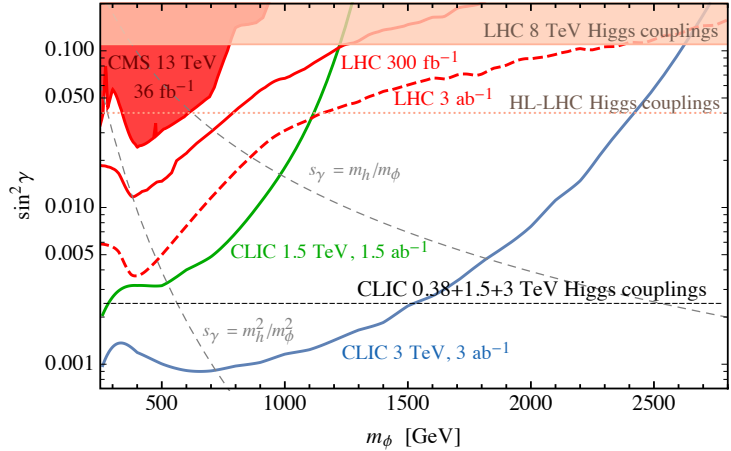

Figure 54: Constraints at 95% C.L. in the plane $(m_\phi, \sin^2\gamma)$. The shaded regions are the present constraints from LHC direct searches for $\phi \to ZZ$ (red) and Higgs couplings measurements (pink shaded) and projection for HL-LHC (pink dotted). The reach at CLIC Stage 2 (green) and Stage 3 (blue) is compared with the projections for LHC with a luminosity of 300 fb$^{-1}$ (solid red) and 3 ab$^{-1}$ (dashed red). We have fixed $\mathrm{BR}_{\phi \to hh} = \mathrm{BR}_{\phi \to ZZ} = 25\%$. The dashed gray lines represent the two upperbounds on the mixing as a function of the mass of the heavy singlet discussed in Eqs. (136) and (137).

*NMSSM interpretation*

The NMSSM, which is the MSSM augmented by a singlet chiral superfield with interactions $\mathcal{W} \supset \lambda H_u H_d S$, contains a scalar singlet in the spectrum. The presence of this singlet is known to alleviate the fine-tuning of the electro-weak scale in the case of a sizeable coupling $\lambda$ [392–395].

When the second Higgs doublet is decoupled, the SM-like Higgs is only coupled to the singlet and the phenomenology matches automatically to the one of eq. (138) discussed in the previous sections, with $\lambda_{HS} = |\lambda|^2$. Here we employ the convenient parametrization of the NMSSM Higgs sector proposed in [396, 397]. By exploiting the relation between the parameters of the model such as $\tan\beta$ and $\lambda$ (allowing also for loop corrections to the Higgs quartic, parametrized by the coefficient $\Delta_{hh}$), we can plot isolines of $\sin^2\gamma$ in the plane $(m_\phi, \tan\beta)$ as in Figure 55 (left). The shape of the contours is mostly determined by the upper bound on the Higgs mass, which is computed to be $m_h^2 \lesssim m_Z^2 c_{2\beta}^2 + \lambda^2 v^2 s_{2\beta}^2/2 + \Delta_{hh}^2$. For definiteness, we have fixed $\lambda = 1$, as it realises a most natural NMSSM region, and $\Delta_{hh} = 80$ GeV, which can be realised for stop masses and mixing in the range of 1-2 TeV.

In the NMSSM, already the second stage of CLIC would be more sensitive than the HL-LHC over the whole parameter space. The reach of the third CLIC stage underlines how powerful this machine would be in probing one of the most natural SUSY realisations. A smaller (larger) $\lambda$ would make all the exclusions and sensitivities weaker (stronger), while mild variations of $\Delta_{hh}$ do not significantly affect them. Varying other parameters has a very mild impact on the phenomenology we discuss. We conclude by noting that, unless $\lambda$ is pushed to the largest values allowed by perturbativity ($\simeq 2$), double singlet production does not play an important role in the NMSSM. We remark that the measurements of Higgs rates can also thoroughly probe this model. As shown by the pink iso-lines from Higgs couplings deviations, the NMSSM predicts several per mille deviations or larger in the bulk of the parameter space in which the singlet is at the TeV or lighter. Therefore we expect a very interesting interplay between direct and indirect searches in case of deviations from the Standard Model.

*Twin Higgs interpretation*

In the simplest realization of the Twin Higgs model the SM Higgs sector gets augmented by a single mirror copy [375]. The full scalar potential has an approximate SU(4) symmetry which is enforced at 1-loop by a discrete $Z_2$ between the SM and the mirror sector. The SU(4) symmetry gets spontaneously broken to SU(3) at a scale $f$ which is larger than $v$ after the breaking of the mirror symmetry is introduced. The radial mode of SU(4)/SU(3) is a singlet under the SM gauge group, which mixes with the SM Higgs and can be light enough to be accessible at colliders. In this context there are only two free parameters left after requiring the correct electro-weak vacuum and Higgs mass. These can be chosen to be the mass of the radial mode $m_\sigma$ and $f$ while the mixing with the SM Higgs is $\sin^2 \gamma \approx v^2/f^2$. We also assume the radial mode decays into the gauge bosons of the EW twin group for $m_\sigma \gtrsim m_W \times f/v$, and into the twin tops for $m_\sigma \gtrsim m_t \times f/v$. Such decays are effectively invisible for the analyses presented here.

The present constraints and future CLIC sensitivities are displayed in the Figure 55-right. In the case where these models are not too strongly coupled, they are expected to manifest themselves first via new diboson (longitudinal) resonances. On the contrary, their strong-coupling regime is expected to show up first in deviations in the Higgs couplings.

*Electroweak phase transition*

The Lagrangian in eq. (138) can induce a first order EW phase transition when it has a well approximated $Z_2$ symmetry as discussed in [377]. A viable scenario is realized when

$$a_S \langle S \rangle / m_S^2 \ll 1, \quad m_S^2 / \lambda_{HS} v^2 \ll 1,$$

therefore in a region where the singlet mass is mostly given by the Higgs vev and the mixing angle $\gamma$ is negligibly small (notice a factor of 2 difference in our definition of the quartic compared to the notation in [377]). We do not attempt to make a detailed numerical simulation, but we just show in Figure 56 what are the regions with a possible strong first order phase transition (a necessary condition for electroweak baryogenesis), that we take from [377]. Moreover we consider the case where the $Z_2$ is slightly broken, assuming it does not interfere with the dynamics of the finite temperature evolution, given the fact that the bounds attainable on $\sin^2 \gamma$ are so strong that the above condition can be easily satisfied with a decay of the singlet-like state within the detector. Therefore we plot in Figure 56 isolines of 10 and 100 number of events for $e^+e^- \to \phi\phi + \nu\nu$ at CLIC in the plane of $(M, \lambda_{HS})$. In the figure we also show the deviations from the SM in the Higgs trilinear couplings predicted in this model as well as the size of the predicted deviation in the overall coupling strength of linear Higgs couplings to the SM fields. This is denoted by $\kappa$ introduced in Section 2.1. In particular the lines $\Delta\kappa = 0.5\%$ and $\Delta\kappa = 0.1\%$ nearly corresponds to the 95% CL limits from the 380 GeV stage and all the stages up to 3 TeV CLIC on this models from Higgs couplings discussed in Section 2.1. We remark how this limit, especially after the 3 TeV stage of CLIC, can in principle probe almost all the region of interest for a strong first-order EWPT in this model.

For the particular model at hand the excellent sensitivity of CLIC to new physics in single Higgs couplings always overwhelms the bounds from the measurement of the Higgs trilinear coupling. However it should be remarked that the results of Section 2.1 are very model specific and minor modifications to the model may invalidate them. For instance if the Higgs boson had new decay channels, the bounds that would apply are those from Section 2.1, in which the total width of the Higgs is a free parameter. In that case the bounds on $\Delta\kappa$ would be in the 1% range, hence even looser than the line $\Delta\kappa = 0.5\%$ that we show in the figure. With such a loss of sensitivity from single Higgs couplings measurements is of utmost importance to have a precise measurement of the Higgs trilinear coupling, which becomes the most powerful probe of the parameter space where this model can deliver a strong first-order EWPT.

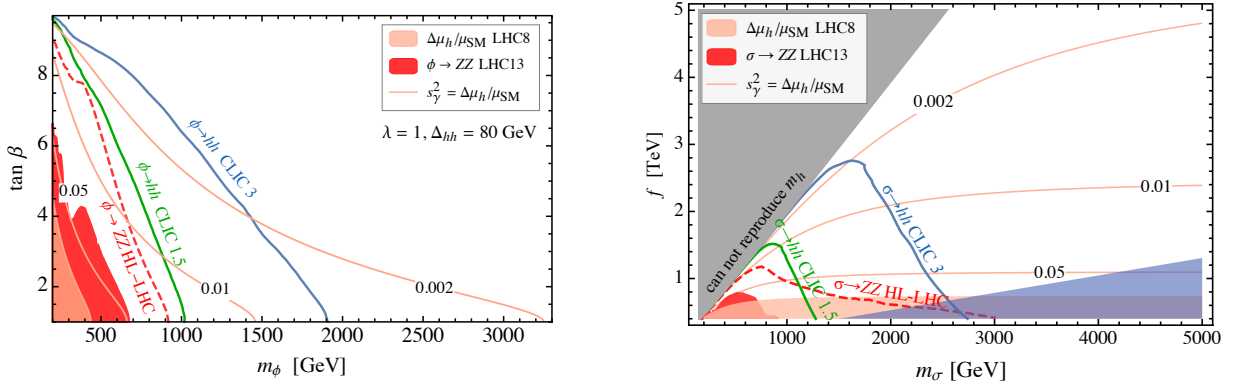

Figure 55: Left: NMSSM with couplings $\lambda = 1$ and with $\Delta_{hh} = 80$ GeV. Right: Twin Higgs models, where in the shaded area in the bottom-right corner one has $\Gamma_\sigma > m_\sigma$. See text for more details.

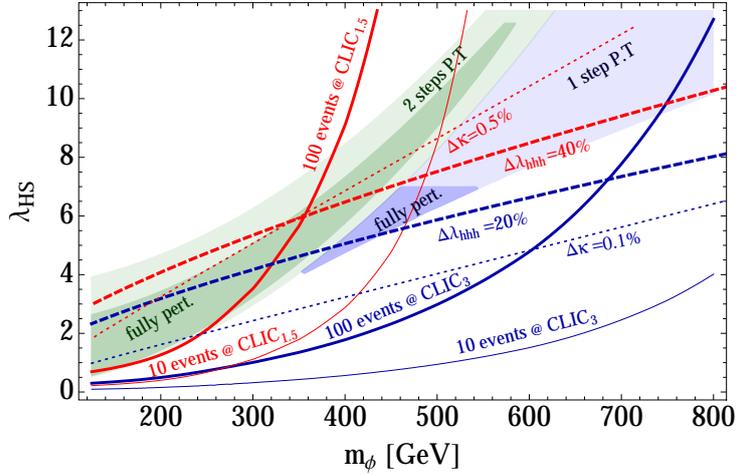

Figure 56: Iso-lines of total number of $\phi\phi\nu\bar{\nu}$ events at CLIC in the zero Higgs-singlet mixing limit. Red lines are for CLIC 1.5 TeV $1.5\text{ab}^{-1}$, blue lines are CLIC 3 TeV $3\text{ab}^{-1}$. Thin lines correspond to total number of double singlet production events $N_{\phi\phi} = 10$, thick lines to 100. The region with a possible first order electroweak phase transition is shaded in green (two-step transition) or blue (one-step transition) regions as discussed in the text. Darker shades corresponds to better perturbative control of the calculation of the strength of the phase transition. In addition we show iso-lines for the prediction of this model for the deviations in triple Higgs couplings and for the overall Higgs coupling strength modifier $\kappa$ defined in Section 2.1 which may be subject to constraints from Higgs physics studies.

### 4.2.2 Light singlets and relaxion [42]

Recently, a new mechanism [398] has been proposed that addresses the hierarchy problem in a way that goes beyond the conventional paradigm of symmetry-based solution to fine-tuning. This so-called relaxion mechanism belongs to the class of models where the solution is associated with the existence of a new and special kind of pseudo-Nambu-Goldstone boson (pNGB), the relaxion, which stabilizes the Higgs mass dynamically. The Higgs mass depends on the classical value of the relaxion field which evolves in time. Eventually, the relaxion stops its rolling in a special field value where the Higgs mass is much smaller than the theory's cutoff, hence addressing the fine tuning problem. Relaxion models do not require top, gauge or Higgs partners at the TeV scale, while a crucial role is played by the relaxion. The possible mass range for the relaxion is very broad, ranging from sub-eV to tens of GeV. Hence this scenario leads to interesting signatures associated with cosmology, the low-energy precision frontier, the intensity frontier, and at colliders.

---
[42]Based on a contribution by C. Frugiuele, E. Fuchs, G. Perez and M. Schlaffer.

We will briefly summarize the aspects of the relaxion mechanism that are relevant for the phenomenology at lepton colliders. The effective scalar potential of the theory depends both on the Higgs doublet $H$ and the relaxion $\phi$,

$$V(H,\phi) = \mu^2(\phi) H^\dagger H + \lambda (H^\dagger H)^2 + V_{\rm sr}(\phi) + V_{\rm br}(h,\phi)\,, \tag{145}$$

$$\mu^2(\phi) = -\Lambda^2 + g\Lambda\phi + \ldots\,, \tag{146}$$

where $\Lambda$ is the cutoff scale of a Higgs loop. The relaxion scans $\mu^2$ via the slow-roll potential

$$V_{\rm sr}(\phi) = rg\Lambda^3 \phi\,, \tag{147}$$

where $g$ is a small dimension-less coupling and $r > \frac{1}{16\pi^2}$ due to naturalness requirements. Once $\mu^2(\phi)$ becomes negative, the Higgs gets a vacuum expectation value (vev) $v^2(\phi) = -\frac{\mu^2(\phi)}{\lambda}$. The non-zero vev activates a periodic (model-dependent) backreaction potential $V_{\rm br}$ associated with the backreaction scale $\Lambda_{\rm br}$ that eventually stops the rolling of the relaxion at a value $\phi_0$, where $v(\phi_0) = 246\,{\rm GeV}$. Generically, the relaxion mechanism leads to $\mathcal{CP}$ violation and as a result, the relaxion $\phi$ mixes with the Higgs $h$ and inherits its couplings to SM fields [399, 400]. The relaxion mass $m_\phi$ and the mixing angle $\sin\theta$ can be expressed as

$$m_\phi \simeq \frac{\Lambda_{\rm br}^2}{f}\sqrt{c_0}\,, \tag{148}$$

$$\sin\theta \simeq 8\frac{\Lambda_{\rm br}^4}{v^3 f}s_0\,, \tag{149}$$

where $s_0 \equiv \sin\phi_0$, $c_0 \equiv \cos\phi_0$, and $f$ is the scale where the shift symmetry is broken. Combining Eqs. (148) and (149) with $4\Lambda_{\rm br}^2 s_0 < v^2\sqrt{c_0}$, which is fulfilled due to the suppressed $s_0$ at the endpoint of the rolling [400], the mixing angle as a function of the relaxion mass is approximately bounded by

$$\sin\theta \leq 2\frac{m_\phi}{v}\,. \tag{150}$$

Moreover, in the broken phase a trilinear relaxion-relaxion-Higgs coupling $c_{\phi\phi h}$ is generated [399, 401],

$$c_{\phi\phi h} = \frac{\Lambda_{\rm br}^4}{vf^2}c_0 c_\theta^3 - \frac{2\Lambda_{\rm br}^4}{v^2 f}s_0 c_\theta^2 s_\theta - \frac{\Lambda_{\rm br}^4}{2f^3}s_0 c_\theta^2 s_\theta - \frac{2\Lambda_{\rm br}^4}{vf^2}c_0 c_\theta s_\theta^2 + 3v\lambda c_\theta s_\theta^2 + \frac{\Lambda_{\rm br}^4}{v^2 f}s_0 s_\theta^3\,, \tag{151}$$

where $s_\theta \equiv \sin\theta$, $c_\theta \equiv \cos\theta$. In the limit of small mixing, the following approximation holds

$$c_{\phi\phi h} \simeq \frac{m_\phi^2}{v}\,. \tag{152}$$

Thus a bound on $c_{\phi\phi h}$ constrains the $(m_\phi, \sin\theta)$ parameter space, and in the above limit of small mixing it constrains directly the mass.

The most powerful strategy to exploit the sensitivity to the triple coupling $c_{\phi\phi h}$ is via the exotic Higgs decay $h \to \phi\phi$. There are two complementary ways to search for such a decay: looking directly for the decay products of the $\phi$ pair, or constraining this branching ratio by a global fit of Higgs couplings. Regarding the direct searches, CMS and ATLAS have performed studies for many of the relevant channels of the possible types $4f$, $2f2\gamma$ and $4\gamma$ [402–404].

Concerning the indirect bound, the measured rates of Higgs decays into SM particles $i$ allow for a global fit of the Higgs coupling modifiers, $\kappa_i$, and the branching ratio into new physics (NP),

BR($h \to$ NP), as an additional parameter. Different model assumptions enter the fit setup; for the case of the relaxion mixed with the Higgs, two fit parameters are applicable, namely a universal modifier of the Higgs to SM particles that can be identified as $\kappa \equiv \cos\theta$ (thus automatically requiring $\kappa_V \leq 1$, $V = W, Z$), and BR($h \to$ NP) that is realized by $h \to \phi\phi$. The total Higgs width is given by $\Gamma_h^{\text{tot}} = \cos^2\theta\,\Gamma_h^{\text{tot,SM}} + \Gamma_h^{\text{NP}}$. In general, the NP contribution to the Higgs width consists of $\Gamma_h^{\text{NP}} = \Gamma_h^{\text{inv}} + \Gamma_h^{\text{unt}}$, where $\Gamma_h^{\text{inv}}$ denotes the partial width into *invisible* particles and $\Gamma_h^{\text{unt}}$ denotes the partial width into *untagged* final states that are not necessarily undetectable, but were not accounted for in the data set included in the fit, see e.g. Refs. [405–407]. Both of these contributions could be accessed via the recoil of the $Z$ boson against the Higgs. In the relaxion case, we are interested in constraining $\Gamma_h^{\text{NP}} = \Gamma_h^{\text{unt}} = \Gamma(h \to \phi\phi)$, i.e. not the invisible width. For masses in the GeV range, the relaxion is short-lived and decays inside the detector even for small $\sin\theta$ (i.e. small couplings of the relaxion to SM particles), see Figure 1b of Ref. [399]. For the LHC, BR($h \to$ NP) has been constrained from Run-1 data and a projection for the HL-LHC has been worked out in Ref. [407].

At the LHC, the direct searches reach a lower sensitivity than the indirect bound via untagged Higgs decays. Therefore we present only the Run-1 limit and HL-LHC projection for BR($h \to$ NP) [407] as untagged decays in the left plot of Figure 57 with the potential to exclude a relaxion mass above 24 GeV.

Our goal is to show the unique capability of CLIC to probe a significant part of the so far not excluded mass window, below the HL-LHC projection. To our knowledge, a dedicated study of the sensitivity to the exotic Higgs decay into two light resonances and their decay products has not yet been performed for CLIC and would be highly desirable, both for the 380 GeV and the high energy runs. In the right plot of Figure 57 we show, in the approximation of small mixing, the mass-dependent branching ratios of the Higgs into 4 fermions, BR$(h \to \phi\phi \to 4f)$, where the BR into 4$b$ is dominant for masses of $m_\phi \gtrsim 9$ GeV. Branching ratios into two fermions and two photons or 4 photons are much smaller and therefore not shown in the plot.

Taking results for 240-250 GeV $e^+e^-$ colliders from Ref. [408] it is possible to estimate the reach of CLIC 380 GeV to search for such exotic Higgs decays. Our estimate is based on the assumption that the kinematics of the Higgs bosons produced at 380 GeV CLIC and at the 240-250 GeV colliders studied in Ref.[408] are sufficiently similar to make a reliable extrapolation. Our estimated bounds are then obtained by rescaling results from Figure 9 of Ref. [408] by the square root of the total number of Higgs events. As the bound from Ref. [408] is at the sub-permil level for the BR$(h \to \phi\phi \to 4b)$ it implies a sensitivity to per mille BR at CLIC. The result is shown as a yellow line in the $(m_\phi, \sin\theta)$ plane in Figure 57.

The results are largely driven by the number of produced Higgs bosons, therefore later stages of CLIC may have a very significant impact on this result. It should be remarked that the kinematics of the Higgs boson produced at the higher energies stages of CLIC, coming from the $W$ fusion process, is different than the kinematics of the $h$ strahlung process, which also provides a convenient $Z$ boson tag to identify the $Zh$ events. Therefore a careful dedicated assessment is needed to find out the sensitivity of later stages of CLIC to exotic Higgs decays. We leave for future work the assessment of the full potential of CLIC for the detection of exotic Higgs decays in the multi-TeV centre-of-mass energy phase[43].

In Ref. [409] it has been studied the process of production of a generic scalar in association with a $Z$ boson

$$e^+e^- \to \phi Z \tag{153}$$

Such $\phi$ strahlung process can be used to probe the relaxion model in a decay-independent manner. In

---

[43]For completeness we report that in following Section 8.3.1 we present results for Higgs boson decays to light pseudo-scalars, decaying to $\gamma\gamma$ or $ee$, that are relevant in the context of axion-like particles. Furthermore in Section 8.1 we present results for exotic Higgs boson decays to long lived particles. For other decay channels a good estimate can be obtained from ILC results contained in Ref. [408] after a rescaling of the bounds according to the square root of the ratio of the number of produced $Zh$ events at CLIC and ILC.

fact, by tagging the $Z$ boson it is possible to reconstruct the recoil mass, that is the invariant mass $(p_{e^-} + p_{e^+} - p_Z)^2$. Such quantity is expected to have a peak at the mass of the $\phi$ scalar, similarly to the case of Higgs strahlung used to produce the 125 GeV Higgs boson. The study has considered a 250 GeV ILC, but we can safely get estimates for CLIC 380 GeV with $1\text{ab}^{-1}$ as the kinematics of the Z boson produced are not largely different. In rescaling the sensitivity we take into account the reduction of cross sections, about a factor 2 for both signal and backgrounds, expected at a CLIC 380 GeV compared to a 240 GeV collider, as well as the luminosity. Figure 3 of Ref. [409] quotes limits for $\theta^2 \lesssim 0.01 - 0.001$ depending on the mass of the $\phi$ and on the requirement to observe the Higgs boson decay products or reconstruct it via the recoil mass. For CLIC 380 we estimate bounds a factor about 2 worse, therefore ranging from few per mille to few percent depending on the mass of $\phi$. These bounds are indicated as green double-dot horizontal line in Figure 57 [44]

Looking at the determination of the properties of the 125 GeV Higgs boson, we can set an approximate indirect bound on $c_{\phi\phi h}$ and thereby on $(m_\phi, \sin\theta)$ by requiring the branching ratio of the Higgs into a pair of relaxions to be less than the projected limit on untagged decays at CLIC 380 GeV for $1\text{ab}^{-1}$ integrated luminosity taken from Table 6 in Section 2.1:

$$\text{BR}(h \to \phi\phi) \lesssim \text{BR}(h \to \text{unt})_{\text{CLIC}} = 4.6\% \text{ at the 95\% confidence level} \qquad (154)$$

The resulting bound is shown in the left plot of Figure 57 as a dashed blue line. Bounds from later stages of CLIC are not shown because the result is heavily dependent on the direct determination of the width of the Higgs boson, that is performed best at the 380 GeV run and will not improve in the later stages of CLIC. In the same fit to the Higgs properties it is found that the overall coupling rescaling induced by the mixing with the relaxion is bound to be

$$\sin^2\theta \leq 2.3\% \text{ at 95\% confidence level}.$$

This bound excludes the region of the $(m_\phi, \sin\theta)$ plane above the dashed grey line in Figure 57, which does not improve on the region already probed by LEP.

The direct and indirect bounds from CLIC probe the parameter space below the theoretical upper bound of the mixing. CLIC will therefore explore relevant parameters of the model. While the HL-LHC may exclude a relaxion mass only above $24\,\text{GeV}$ for vanishing $\sin\theta$, CLIC will have the potential to exclude masses down to $20\,\text{GeV}$ with indirect searches and further down to 12 GeV with direct searches for $h \to 4b$ exotic decays.

Among the other searches sensitive to the relaxion, relaxion production via the rare $B$-decay $B \to K\phi$, $\phi \to \mu\mu$ at LHCb excludes $2m_\mu \leq m_\phi \lesssim 5\,\text{GeV}$ also for $\sin\theta$ smaller than shown in the left plot of Figure 57. In contrast, the bounds set by LEP1 via the 3-body $Z$-decay into $f\bar{f}\phi$ and by LEP2 via relaxion strahlung in $Z\phi$ production are sensitive only to mixing angles of $\sin^2\theta \gtrsim 10^{-2}$ and therefore constrain mostly parameter space above the theoretically motivated maximal mixing, or an area that is also excluded by untagged Higgs boson decays at the LHC.

As a conclusion, the higher precision reached at CLIC compared to the LHC—even in its high-luminosity run—is highly beneficial for constraining the relaxion parameter space down to small mixing angles and relaxion masses which are challenging at hadron colliders. The combination of direct and indirect search strategies allows to significantly reduce the region in parameter space in which the relaxion can live after bounds from heavy flavor factories, LHC and HL-LHC are taken into account.

## 4.3 Extended Higgs sectors [45]

The discussion of physics in future colliders has recently become a very important issue due to the absence of hints of New Physics beyond the Standard Model (BSM). In fact, although a Higgs boson has been discovered by the LHC experiments ATLAS [410] and CMS [411], no other solid hints of New Physics have been reported by the LHC collaborations until now. On the contrary, the LHC results point to a SM-like Higgs boson with couplings to the remaining SM particles well within the SM expectations.

---

[44] Due to the extrapolation from ILC and to the smoothing to a straight line of the result Ref. [409] the line in Figure 57 should be read as an indication of CLIC potential, possibly subject to variations from several effects.

[45] Based on a contribution by D. Azevedo, P. Ferreira, M. Mühlleitner, R. Santos and J. Wittbrodt.

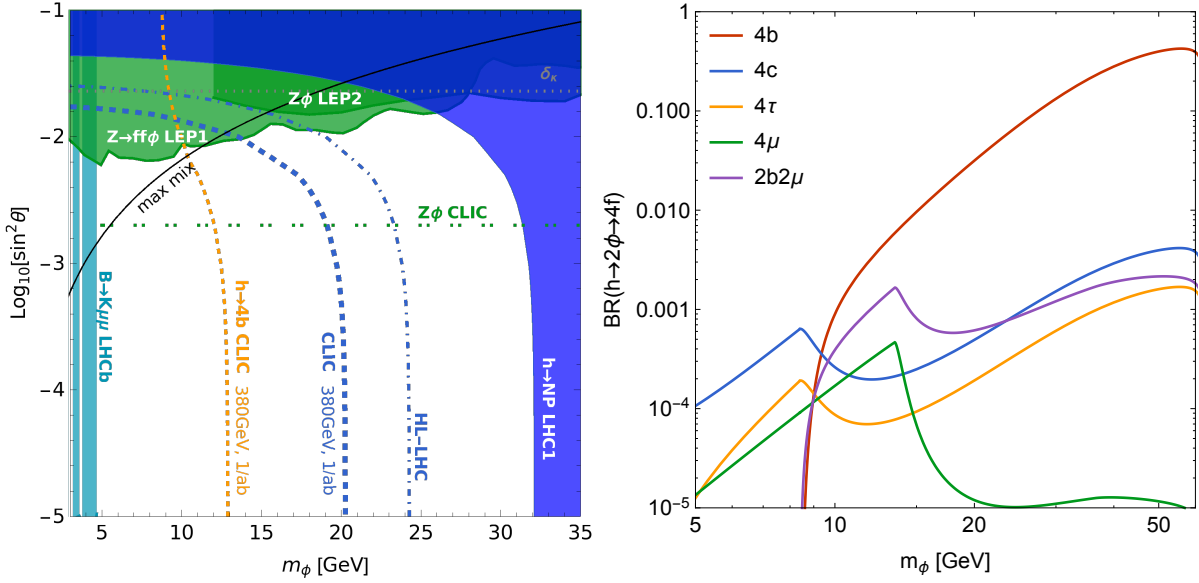

Figure 57: **Left:** Current and projected constraints on the relaxion mass $m_\phi$ and its mixing angle $\sin\theta$ with the Higgs. The bound from upper limits on the untagged branching ratio of the Higgs, here $h \to \phi\phi$: current [399] (blue area) and projected (blue, dash-dotted) exclusion from the HL-LHC at the 95% CL [407]; projection for CLIC at the 95% CL with $1\,\mathrm{ab}^{-1}$ running at $\sqrt{s} = 380\,\mathrm{GeV}$. The yellow lines represents the rescaling from Ref.[408] for $1\,\mathrm{ab}^{-1}$ at 380 GeV CLIC. The horizontal grey line represents the bound from the fit of the 125 GeV Higgs boson couplings from Table 6 and is very similar at the 3 stages of CLIC considered. The horizontal green double-dot line is an approximate the extrapolation of the ILC limits of Ref. [409] for direct $Z\phi$ production at the 380 GeV CLIC. We show further constraints from $B \to K\phi, \phi \to \mu\mu$ at LHCb (turquoise), the 3-body $Z$-decay into $f\bar{f}\phi$ at LEP1 and relaxion strahlung $\to Z\phi$ at LEP2 (green). The black line shows the upper bound on the mixing. For more details on the non-CLIC bounds see Ref. [401]. **Right:** Branching ratio of $h$ into 4 fermions via a pair of relaxions, showing the selection of the dominant channels.

Here we analyse several extensions of the SM: the SM extended by a complex singlet field (CxSM), the 2-Higgs-Doublet Model with a CP-conserving (2HDM) and a CP-violating (C2HDM) scalar sector, the singlet extension of the 2-Higgs-Doublet Model (N2HDM), and the Next-to-Minimal Supersymmetric SM extension (NMSSM). The models have in common the presence of at least three neutral bosons (one being the 125 GeV Higgs boson), which allow for the comparison of the production and decay rates of the other two scalars.

We focus mainly on two different issues. The first part of the work is about the nature of the discovered Higgs boson. The SM 125 GeV scalar originates from an $SU(2)$ doublet. When other fields are added to the SM content, mixing between fields from doublets and/or singlets takes place. The Higgs boson can acquire extra singlet or pseudoscalar components from the mixing. We will investigate what an electron-positron collider such as CLIC can tell us about the amount of mixing in the 125 GeV Higgs boson. The second part of the work focuses on the two non-125 GeV Higgs bosons and on the possibility to distinguish the different models if a new scalar is found. The issue addressed is whether we are able to disentangle the models based on Higgs rate measurements. We hope that we can shed some light on the relevance of a future electron-positron collider for BSM Higgs searches.

In Section 4.3.1 we briefly introduce the models and the scan over their parameter spaces. Section 4.3.2 is devoted to the nature of the 125 GeV Higgs boson after CLIC and in Section 4.3.3 we compare the signal rates of the two non-SM-like Higgs bosons within the different models. Our conclusions are given in Section 4.3.4.

*4.3.1 Short description of the models*

The models discussed in this work were introduced in detail in [412]. Here, we will only give their potentials, the particle spectrum and the independent parameters of the models.

- **Complex Singlet Extension of the SM (CxSM)**
  The model is an extension of the SM through the addition of a complex scalar singlet. The potential has a softly broken global $U(1)$ symmetry and is given by

  $$V = \frac{m^2}{2}H^\dagger H + \frac{\lambda}{4}(H^\dagger H)^2 + \frac{\delta_2}{2}H^\dagger H|\mathbb{S}|^2 + \frac{b_2}{2}|\mathbb{S}|^2 + \frac{d_2}{4}|\mathbb{S}|^4 + \left(\frac{b_1}{4}\mathbb{S}^2 + a_1\mathbb{S} + c.c.\right), \quad (155)$$

  where $\mathbb{S} = S + iA$ is a hypercharge zero scalar and the soft breaking terms are written in parenthesis. We further impose invariance under $\mathbb{S} \to \mathbb{S}^*$ (or $A \to -A$), and so $a_1$ and $b_1$ are real. We work in the broken phase where the three CP-even scalars mix. The mass eigenstates for these scalars are denoted by $H_i$ and are obtained from the gauge eigenstates via the rotation matrix $R$ that is fully defined in [413] and can be parametrised as

  $$R = \begin{pmatrix} c_1 c_2 & s_1 c_2 & s_2 \\ -(c_1 s_2 s_3 + s_1 c_3) & c_1 c_3 - s_1 s_2 s_3 & c_2 s_3 \\ -c_1 s_2 c_3 + s_1 s_3 & -(c_1 s_3 + s_1 s_2 c_3) & c_2 c_3 \end{pmatrix}, \quad (156)$$

  where we have defined $s_i \equiv \sin\alpha_i$ and $c_i \equiv \cos\alpha_i$, and

  $$-\frac{\pi}{2} \leq \alpha_i < \frac{\pi}{2}. \quad (157)$$

  The masses of the neutral Higgs bosons are ordered as $m_{H_1} \leq m_{H_2} \leq m_{H_3}$. We choose as input parameters the set

  $$\alpha_1, \quad \alpha_2, \quad \alpha_3, \quad v, \quad v_S, \quad m_{H_1} \quad \text{and} \quad m_{H_3}, \quad (158)$$

  and the remaining parameters are determined internally in `ScannerS` [414, 415] fulfilling the minimum conditions of the vacuum.
  All couplings of each Higgs boson $H_i$ to SM particles are rescaled by a common factor $R_{i1}$. Expressions for all couplings are available in [416] and the Higgs branching ratios, including the state-of-the art higher order QCD corrections and possible off-shell decays, can be obtained from `sHDECAY`[416][46] which implements the CxSM and also the RxSM both in their symmetric and broken phases in `HDECAY` [417, 418].

- **Two-Higgs Doublet Model - Real (2HDM) and Complex (C2HDM)**
  The model is an extension of the SM by a second scalar doublet. The potential is invariant under the softly broken $\mathbb{Z}_2$ transformation $\Phi_1 \to \Phi_1$ and $\Phi_2 \to -\Phi_2$ and can be written as

  $$\begin{aligned}V &= m_{11}^2|\Phi_1|^2 + m_{22}^2|\Phi_2|^2 - m_{12}^2(\Phi_1^\dagger\Phi_2 + h.c.) + \frac{\lambda_1}{2}(\Phi_1^\dagger\Phi_1)^2 + \frac{\lambda_2}{2}(\Phi_2^\dagger\Phi_2)^2 \\ &\quad + \lambda_3(\Phi_1^\dagger\Phi_1)(\Phi_2^\dagger\Phi_2) + \lambda_4(\Phi_1^\dagger\Phi_2)(\Phi_2^\dagger\Phi_1) + [\frac{\lambda_5}{2}(\Phi_1^\dagger\Phi_2)^2 + h.c.]\,.\end{aligned} \quad (159)$$

---

[46]The program sHDECAY can be downloaded from the url: itp.kit.edu/~maggie/sHDECAY.

The extension of the $\mathbb{Z}_2$ symmetry to the fermions guarantees that the model is free from tree-level flavour changing neutral currents (FCNC). The potential is CP-conserving and referred to as 2HDM if all parameters of the potential and the VEVs are real. The potential is CP-violating if the VEVs are real but $m_{12}^2$ and $\lambda_5$ are complex and we name it C2HDM [419]. Both models have three neutral scalars and two charged Higgs bosons. In the 2HDM the neutral scalars are $h$ and $H$, the lighter and the heavier CP-even states, while $A$ is the CP-odd state. In the C2HDM we define three Higgs mass eigenstates $H_i$ ($i = 1, 2, 3$) with no definite CP that are ordered as $m_{H_1} \leq m_{H_2} \leq m_{H_3}$. The rotation matrix $R$ that diagonalises the mass matrix is parametrised in Eqs. (156) and (157).

The 2HDM has 8 independent parameters while the C2HDM has 9 independent parameters. We define $v = \sqrt{v_1^2 + v_2^2} \approx 246$ GeV and $\tan\beta = v_2/v_1$ for both versions of the model. For the 2HDM we choose the independent parameters

$$v, \quad \tan\beta, \quad \alpha, \quad m_h, \quad m_H, \quad m_A, \quad m_{H^\pm} \quad \text{and} \quad m_{12}^2, \tag{160}$$

while for the C2HDM we use [420]

$$v, \quad \tan\beta, \quad \alpha_{1,2,3}, \quad m_{H_i}, \quad m_{H_j}, \quad m_{H^\pm} \quad \text{and} \quad Re(m_{12}^2), \tag{161}$$

where $m_{H_i}$ and $m_{H_j}$ denote any two of the three neutral Higgs bosons. The remaining mass is obtained from the other parameters [420].

All Higgs branching ratios, including the state-of-the art higher order QCD corrections and possible off-shell decays can be obtained from C2HDM_HDECAY[421][47] which is an implementation of the C2HDM in HDECAY [417, 418]. The complete set of Feynman rules for the C2HDM is available at:

porthos.tecnico.ulisboa.pt/arXiv/C2HDM/

The 2HDM branching ratios are part of the HDECAY release (see [418, 422] for details).

– **Next-to-Two-Higgs Doublet Model (N2HDM)**

The model [423] is an extension of the SM by a doublet and a real singlet. The potential is invariant under two discrete $\mathbb{Z}_2$ symmetries, $\Phi_1 \to \Phi_1, \Phi_2 \to -\Phi_2, \Phi_S \to \Phi_S$ (as in the 2HDM, to avoid tree-level FCNCs), softly broken by $m_{12}^2$, and $\Phi_1 \to \Phi_1, \Phi_2 \to \Phi_2, \Phi_S \to -\Phi_S$, which is not explicitly broken. The most general form of this scalar potential is

$$\begin{aligned} V &= m_{11}^2|\Phi_1|^2 + m_{22}^2|\Phi_2|^2 - m_{12}^2(\Phi_1^\dagger\Phi_2 + h.c.) + \frac{\lambda_1}{2}(\Phi_1^\dagger\Phi_1)^2 + \frac{\lambda_2}{2}(\Phi_2^\dagger\Phi_2)^2 \\ &+ \lambda_3(\Phi_1^\dagger\Phi_1)(\Phi_2^\dagger\Phi_2) + \lambda_4(\Phi_1^\dagger\Phi_2)(\Phi_2^\dagger\Phi_1) + \frac{\lambda_5}{2}[(\Phi_1^\dagger\Phi_2)^2 + h.c.] \\ &+ \frac{1}{2}m_S^2\Phi_S^2 + \frac{\lambda_6}{8}\Phi_S^4 + \frac{\lambda_7}{2}(\Phi_1^\dagger\Phi_1)\Phi_S^2 + \frac{\lambda_8}{2}(\Phi_2^\dagger\Phi_2)\Phi_S^2. \end{aligned} \tag{162}$$

This particular version of the N2HDM is CP-conserving and the particle spectrum consists of three CP-even scalars, one CP-odd scalar and a pair of charged Higgs bosons. The CP-even states are obtained from the gauge eigenstates via the same rotation matrix $R$ defined in Eqs. (156) and (157). These states are denoted by $H_1$, $H_2$ and $H_3$ and are ordered as $m_{H_1} < m_{H_2} < m_{H_3}$. The 12 independent parameters are

$$\alpha_1, \quad \alpha_2, \quad \alpha_3, \quad t_\beta, \quad v, \quad v_s, \quad m_{H_{1,2,3}}, \quad m_A, \quad m_{H^\pm}, \quad m_{12}^2. \tag{163}$$

All Higgs branching ratios, including the state-of-the art higher order QCD corrections and possible off-shell decays can be obtained from N2HDECAY[48] [424].

---
[47]The program C2HDM_HDECAY can be downloaded from itp.kit.edu/~maggie/C2HDM.
[48]The program N2HDECAY is available at gitlab.com/jonaswittbrodt/N2HDECAY and based on HDECAY [417, 418].

– **The Next-to-Minimal Supersymmetric Standard Model (NMSSM)**

The NMSSM is obtained by extending the two Higgs doublet superfields $\hat{H}_u$ and $\hat{H}_d$ in the Minimal Supersymmetric extension (MSSM) by a complex superfield $\hat{S}$. The NMSSM Higgs potential is derived from the superpotential, the soft SUSY breaking Lagrangian and the $D$-term contributions. In terms of the hatted superfields, the scale-invariant NMSSM superpotential is given by

$$\mathcal{W} = \lambda \hat{S} \hat{H}_u \hat{H}_d + \frac{\kappa}{3} \hat{S}^3 + h_t \hat{Q}_3 \hat{H}_u \hat{t}_R^c - h_b \hat{Q}_3 \hat{H}_d \hat{b}_R^c - h_\tau \hat{L}_3 \hat{H}_d \hat{\tau}_R^c ~, \tag{164}$$

where for simplicity only the third generation fermion superfields have been included. Here $\hat{Q}_3$ and $\hat{L}_3$ denote the left-handed doublet quark and lepton superfields, respectively, and $\hat{t}_R^c, \hat{b}_R^c$ and $\hat{\tau}_R^c$ the right-handed singlet quark and lepton superfields each. The soft SUSY breaking Lagrangian is given by the mass terms for the Higgs and the sfermion fields, built from the complex scalar components of the superfields,

$$\begin{aligned}
-\mathcal{L}_{\text{mass}} &= m_{H_u}^2 |H_u|^2 + m_{H_d}^2 |H_d|^2 + m_S^2 |S|^2 \\
&+ m_{\tilde{Q}_3}^2 |\tilde{Q}_3^2| + m_{\tilde{t}_R}^2 |\tilde{t}_R^2| + m_{\tilde{b}_R}^2 |\tilde{b}_R^2| + m_{\tilde{L}_3}^2 |\tilde{L}_3^2| + m_{\tilde{\tau}_R}^2 |\tilde{\tau}_R^2| ~,
\end{aligned} \tag{165}$$

the contribution from the trilinear soft SUSY breaking interactions between the sfermions and the Higgs fields

$$\begin{aligned}
-\mathcal{L}_{\text{tril}} &= \lambda A_\lambda H_u H_d S + \frac{1}{3} \kappa A_\kappa S^3 + h_t A_t \tilde{Q}_3 H_u \tilde{t}_R^c - h_b A_b \tilde{Q}_3 H_d \tilde{b}_R^c \\
&- h_\tau A_\tau \tilde{L}_3 H_d \tilde{\tau}_R^c + \text{h.c.} ~,
\end{aligned} \tag{166}$$

where the $A$'s denote the soft SUSY breaking trilinear couplings, and the contribution from the gaugino mass parameters $M_{1,2,3}$ of the bino ($\tilde{B}$), winos ($\tilde{W}$) and gluinos ($\tilde{G}$), respectively,

$$-\mathcal{L}_{\text{gauginos}} = \frac{1}{2} \left[ M_1 \tilde{B}\tilde{B} + M_2 \sum_{a=1}^{3} \tilde{W}^a \tilde{W}_a + M_3 \sum_{a=1}^{8} \tilde{G}^a \tilde{G}_a + \text{h.c.} \right] ~. \tag{167}$$

The soft terms are assumed to be non-universal at the GUT scale.

The particle spectrum of the NMSSM contains three CP-even Higgs mass eigenstates $H_i$ ($i = 1, 2, 3$), with $m_{H_1} \leq m_{H_2} \leq m_{H_3}$, two CP-odd mass eigenstates $A_1$ and $A_2$, with $m_{A_1} \leq m_{A_2}$, and a pair of charged Higgs bosons. Using the minimisation conditions we can parametrise the tree-level NMSSM Higgs sector by six independent parameters, chosen as

$$\lambda ~,~ \kappa ~,~ A_\lambda ~,~ A_\kappa ~,~ \tan\beta = v_u/v_d \quad \text{and} \quad \mu_{\text{eff}} = \lambda v_s / \sqrt{2} ~. \tag{168}$$

The sign conventions are such that $\lambda$ and $\tan\beta$ are positive, whereas $\kappa$, $A_\lambda$, $A_\kappa$ and $\mu_{\text{eff}}$ are allowed to have both signs. Due to the corrections to the SM-like Higgs boson mass, necessary to shift it to the measured 125 GeV, also the soft SUSY breaking mass terms for the scalars and the gauginos as well as the trilinear soft SUSY breaking couplings contribute to the Higgs sector. We use the `NMSSMTools` package [425–430] to calculate the Higgs masses and decay widths including the relevant higher order corrections. We have cross-checked these results against `NMSSMCALC` [431].

We have performed parameter scans in these models by varying the input parameters through the phenomenologically interesting ranges. Our scans take into account all applicable theoretical and experimental constraints. The parameter ranges and details on the applied constraints can be found in [412]. Note that the 125 GeV Higgs boson can be the lightest as well as a heavier scalar. This possibility is not excluded in any of the models.

## 4.3.2 The nature of the 125 GeV Higgs boson after CLIC

Over the last years, predictions for the measurement of the Higgs couplings to fermions and gauge bosons at CLIC were made for several benchmark energies and luminosities. We use results on the expected precision in the measurement of the Higgs couplings from [432] (see [355, 432] for details). The $\kappa_{Hii}$ are defined as

$$\kappa_{Hii} = \sqrt{\frac{\Gamma_{Hii}^{BSM}}{\Gamma_{Hii}^{SM}}} \, , \tag{169}$$

which at tree-level is just the ratio of the Higgs coupling in the BSM model and the corresponding SM Higgs coupling. We call the three benchmarks scenarios $Sc1$ (350 GeV), $Sc2$ (1.4 TeV) and $Sc3$ (3.0 TeV). With these predictions we now study the effect on the parameter space of each model presented in the previous Section. This will tell us how much an extra component from either a singlet (or more singlets) or a doublet contributes to the $h_{125}$ scalar boson. Clearly, if no new scalar is discovered one can only set bounds on the amount of mixing with extra fields. In the case of a CP-violating model it is possible to set a bound on the ratio of pseudoscalar to scalar Yukawa couplings, where there is an important interplay with the results from measurements of electric dipole moments (EDMs). The results presented in this Section assume that the measured central value is the SM expectation, meaning that all $\kappa_{Hii}$ extracted from [355, 432] have a central value of 1. If significant deviations from the SM predicted values are found the data has to be reinterpreted for each model.

### CxSM

Starting with the simplest extension, the CxSM, there are either one or two singlet components that mix with the real neutral part of the Higgs doublet. The admixture is given by the sum of the squared mixing matrix elements corresponding to the real and complex singlet parts, *i.e.*

$$\Sigma_i^{\text{CxSM}} = (R_{i2})^2 + (R_{i3})^2 \, , \tag{170}$$

with the matrix $R$ defined in Eq. (156)[49]. All Higgs couplings to SM particles are rescaled by a common factor. Therefore, we only need to consider the most accurate Higgs coupling measurement to get the best constraints on the Higgs admixture. The maximum allowed singlet admixture is given by the lower bound on the global signal strength $\mu$ which at present is

$$\Sigma_{\text{max LHC}}^{\text{CxSM}} \approx 1 - \mu_{\min} \approx 11\% \, . \tag{171}$$

In CLIC $Sc1$ the most accurate measurement is for the scaled coupling $\kappa_{HZZ}$, which would give

$$\Sigma_{\text{max CLIC@350GeV}}^{\text{CxSM}} \approx 0.85\% \, , \tag{172}$$

while for $Sc3$ one would obtain, from $\kappa_{HWW}$,

$$\Sigma_{\text{max CLIC@3TeV}}^{\text{CxSM}} \approx 0.22\% \, . \tag{173}$$

This implies, for this particular kind of extension, that the chances of finding a new scalar are reduced due to the orthogonality of the $R$ matrix. Note that in the limit of exact zero singlet component the singlet fields do not interact with the SM particles. The results for a real singlet are similar, with the bound being exactly the same but with a two by two orthogonal matrix replacing $R$. In this case it is exactly the value 0.22% that multiplies all production cross sections of the non-SM Higgs boson, after CLIC@3TeV.

---
[49] If a dark matter candidate is present one of the $R_{ij}, j = 2, 3$, is zero.

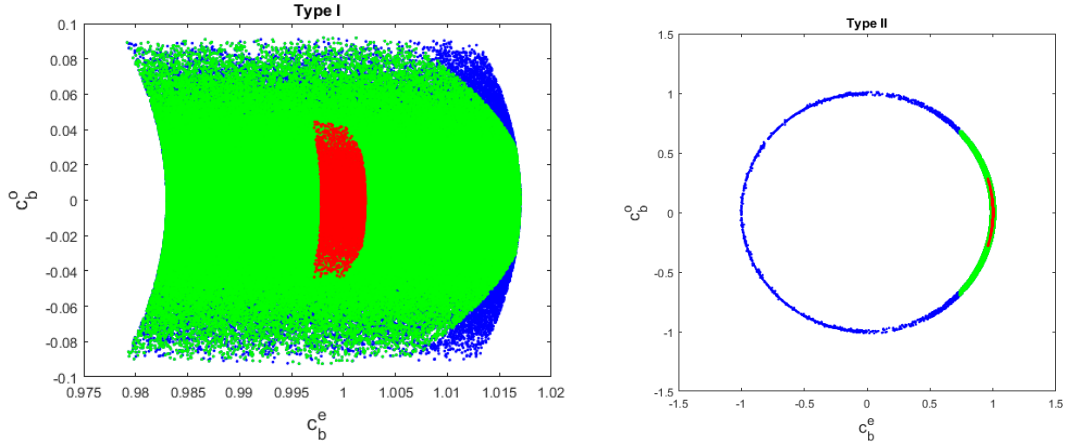

Figure 58: Yukawa couplings $c_b^o$ vs. $c_b^e$ (left) for the C2HDM Type I and Type II defined in detail in Ref. [412](right). The blue points are for $Sc1$ but without the constraints from $\kappa_{Hgg}$ and $\kappa_{H\gamma\gamma}$; the green points are for $Sc1$ including $\kappa_{Hgg}$ and the red points are for $Sc3$ including $\kappa_{Hgg}$ and $\kappa_{H\gamma\gamma}$.

*C2HDM*

This model with a CP-violating scalar shows a quite different behaviour in the four Yukawa versions of the model. In fact, the constraints act very differently in the four Yukawa versions of the model as shown in [421]. This is particularly true for the EDMs [421] - while for Type II the electron EDM constraint almost kills the pseudoscalar component of the $bbH$ coupling, the same is not true for the Flipped model and for the pseudoscalar component of the Higgs couplings to leptons in the Lepton Specific model. Since different Yukawa couplings enter the two-loop Barr-Zee diagrams, a small EDM can either be the result of small CP-violating Yukawa couplings or come from cancellations between diagrams. This can even allow for maximally CP-violating Yukawa couplings of the $h_{125}$ in some cases [421]. We will now study the indirect constraints from CLIC on CP-violating admixtures to the 125 GeV Higgs boson and compare them to direct constraints and constraints from EDMs.

*Type I*

In Figure 58 we show the pseudoscalar component of the $b$-quark Yukawa coupling $c_b^o$ versus its scalar component $c_b^e$. As all Yukawa couplings are equal in Type I, this plot is valid for all Type I Yukawa couplings. The blue points are for $Sc1$ but without the constraints from $\kappa_{Hgg}$ and $\kappa_{H\gamma\gamma}$. The green points are for $Sc1$ including $\kappa_{Hgg}$ ($\kappa_{H\gamma\gamma}$ is unconstrained by $Sc1$) and the red points are for $Sc3$ including $\kappa_{Hgg}$ and $\kappa_{H\gamma\gamma}$. Note that $\kappa_{Hgg}$ and $\kappa_{H\gamma\gamma}$ are the only measurements of couplings that can probe the interference between Yukawa couplings (in the case of $\kappa_{Hgg}$) and between Yukawa and Higgs gauge couplings (in the case of $\kappa_{H\gamma\gamma}$). We expect all pseudoscalar (scalar) Type I Yukawa couplings to be less than roughly 5% (0.5 %) away from the SM expectation by the end of the CLIC operation. We stress that this result assumes that experiments will not see deviations from the SM.

Recently, in [433] a study was performed for a 250 GeV electron-positron collider for Higgsstrahlung events in which the $Z$ boson decays into electrons, muons, or hadrons, and the Higgs boson decays into $\tau$ leptons, which subsequently decay into pions. The authors found that for an integrated luminosity of 2 ab$^{-1}$, the mixing angle between the CP-odd and CP-even components, defined as

$$\mathcal{L}_i = g\bar{\tau}\left[\cos\psi_{CP} + i\gamma_5 \sin\psi_{CP}\right]\tau H_i ,  \tag{174}$$

could be measured to a precision of $4.3^o$ which means that this is the best bound if the central measured value of the angle is zero. Their result is translated into our notation via

$$\tan\psi_{CP}^{\tau} = \frac{c^o(H_i\bar{\tau}\tau)}{c^e(H_i\bar{\tau}\tau)} \ . \tag{175}$$

Taking into account the values in Figure 58 we obtain bounds on $\psi_{CP}^{top} = \psi_{CP}^{bottom} = \psi_{CP}^{\tau}$, for Type I, that are of the order of $6^o$ for CLIC@350GeV and 3◦ for CLIC@3TeV. Therefore the indirect bounds are of the same order of magnitude as the direct ones.

*Type II*

For the Type II C2HDM we show in Figure 58 the pseudoscalar component of the $b$-quark Yukawa coupling $c_b^o$ vs. its scalar component $c_b^e$. The blue points are for $Sc1$ without the constraints from $\kappa_{Hgg}$ and $\kappa_{H\gamma\gamma}$. Whatever the constraint on the tree-level couplings is, the result will always be a ring in that plane that will become increasingly thinner with growing precision. The loop induced couplings, however, are sensitive to interference between Yukawa and Higgs gauge couplings. Even for CLIC@350GeV, including the constraint on $\kappa_{Hgg}$ reduces the ring to the green arch shown in the figure. By the end of the CLIC operation the arch will be further reduced to the red one. As discussed in previous works, a very precise measurement of $\kappa_{Hgg}$ or $\kappa_{H\gamma\gamma}$ will kill the wrong-sign limit[50], which corresponds in the figure to $c_b^e = -1$. Comparing these bounds with those from kinematic distributions of $h_{125} \to \tau^+\tau^-$ we see that direct investigations of the CP nature of $h_{125}$ are likely to be more powerful than the global analysis of rates. In fact the bound on $\psi_{CP}^{top}$ is the similar in all 2HDM types, as the top quark couples to the scalar sector in the same way. In Type II however we have $\psi_{CP}^{bottom} = \psi_{CP}^{\tau}$ that has different dependence on the mixing angles and $\tan\beta$ than $\psi_{CP}^{top}$. This different dependence results in looser constraints that can be seen in Figure 58 (right) for $\psi_{CP}^{bottom}$ and are of the order of $30°$ for CLIC@350GeV and $15°$ for CLIC@3TeV. Thus we conclude that for Type II the indirect bounds cannot compete with the direct ones. The EDM constraints also play a very important role in probing the CP-odd components of the couplings. In fact, in the particular scenario of the Type II C2HDM in which the lightest Higgs boson is the 125 GeV scalar, the bound is already constraining $\psi_{CP}^{bottom}$ to be below $20^o$ [421] clearly competing with the expectations for CLIC. These constraints may improve dramatically with the expected ACME II results [436].

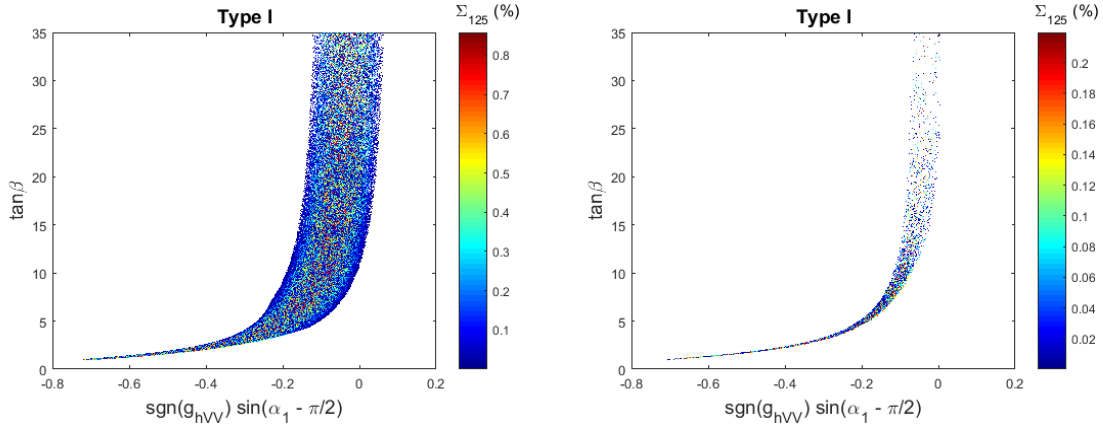

Figure 59: $\tan\beta$ as a function of $\sin(\alpha_1 - \frac{\pi}{2})$ for Type I in $Sc1$ (left) and $Sc3$ (right). The factor $-\frac{\pi}{2}$ is due to a different definition of the rotation angles relative to the 2HDM. Also shown in the colour code is the amount of singlet admixture present in $h_{125}$.

---

[50]The wrong sign limit refers to a Yukawa coupling that has a relative (to the coupling of the Higgs boson to the massive gauge bosons) minus sign to the corresponding SM coupling [434, 435].

*N2HDM and 2HDM*

The predictions for the N2HDM are very similar to the ones for the 2HDM and we will discuss them together. Although the N2HDM has an extra singlet field relative to the 2HDM, the couplings to gauge bosons and fermions are very similar. For instance, for the lightest Higgs boson the couplings to massive gauge bosons are related via $g_{hVV}^{N2HDM} = \sin\alpha_2\, g_{hVV}^{2HDM}$ which results in some extra freedom for the N2HDM parameter space. In Figure 59 we show $\tan\beta$ as a function of $\sin(\alpha_1 - \frac{\pi}{2})$ for Type I in $Sc1$ (left) and $Sc3$ (right) (the lepton-specific case behaves very similarly). The only notable difference between the N2HDM and the 2HDM is the colour bar where we show the percentage of the singlet component in the 125 GeV Higgs boson, $\Sigma_{125} = (R_{i3})^2$. In a previous work [437] we have shown that before the LHC run 2 the allowed admixture of the singlet was below 25% for Type I and the predictions for CLIC@350GeV and CLIC@3TeV are below 0.85% and 0.22%, respectively.

As expected, the allowed parameter space gets closer and closer to the SM line, that is the line $\sin(\beta - \alpha) = 1$ (alignment limit). Note that unless one detects a new particle there is no way to find the value of $\tan\beta$ if the models are in the alignment limit. In fact, if the lightest Higgs boson is the 125 GeV one and we are in the alignment limit, $\sin(\beta - \alpha) = 1$ in the 2HDM,[51] all couplings of the 125 GeV Higgs boson to SM particles are independent of the value of $\tan\beta$ (including the triple Higgs coupling). If the 125 GeV Higgs boson is not the lightest scalar in the model, the alignment limit corresponds to setting $\cos(\beta - \alpha) = 1$. In this limit the amount of mixing of a singlet or of a pseudoscalar is roughly the same as for the case where the 125 GeV Higgs is the lightest scalar. The main difference between the two scenarios is that the latter does not have a decoupling limit, making it in principle easier to probe, since there is a scalar lighter than 125 GeV in the model.

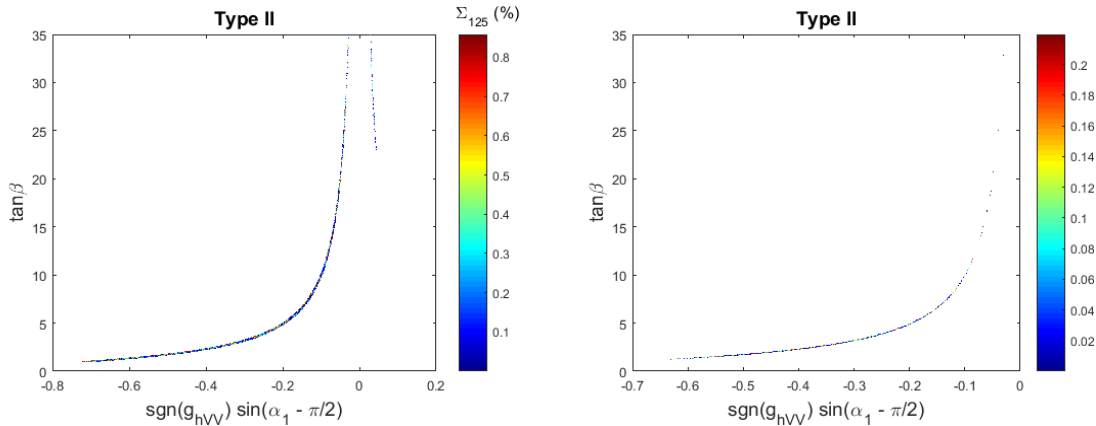

Figure 60: $\tan\beta$ as a function of $\sin(\alpha_1 - \frac{\pi}{2})$ for Type II in $Sc1$ (left) and $Sc3$ (right). The factor $-\frac{\pi}{2}$ is due to a different definition of the rotation angles relative to the 2HDM. Also shown in the colour code is the amount of singlet present in $h_{125}$.

In Figure 60 we show $\tan\beta$ as a function of $\sin(\alpha_1 - \frac{\pi}{2})$ for Type II in $Sc1$ (left) and $Sc3$ (right). These are typical plots not only for a Type II N2HDM but also for a Type II 2HDM (and similar plots are obtained for the Flipped versions of both models). As previously discussed we see that the right leg, corresponding to the wrong-sign limit, is very thin in the left plot, i.e. the parameter space is very reduced, and disappears in the right plot, signalling that this region of parameter space is not allowed. Again, this is true for both the 2HDM and the N2HDM. As for the percentage of the singlet component, it was constrained to 55% for Type II N2HDM at the end of run 1 [437] and the predictions for CLIC@350GeV and CLIC@3TeV are again below about 0.85% and 0.22%, respectively.

---

[51] In the N2HDM, the alignment limit is attained for $\cos(\beta-\alpha_1)\cos\alpha_2 = 1$ (where the $\cos(\beta-\alpha_1)$ appears due to a different definition of the angle $\alpha_1$ relative to the 2HDM). This means the N2HDM has SM-like couplings when $\cos(\beta - \alpha_1) = 1$ *and* $\cos\alpha_2 = 1$. In this limit the 125 GeV Higgs boson has no contribution from the singlet field.

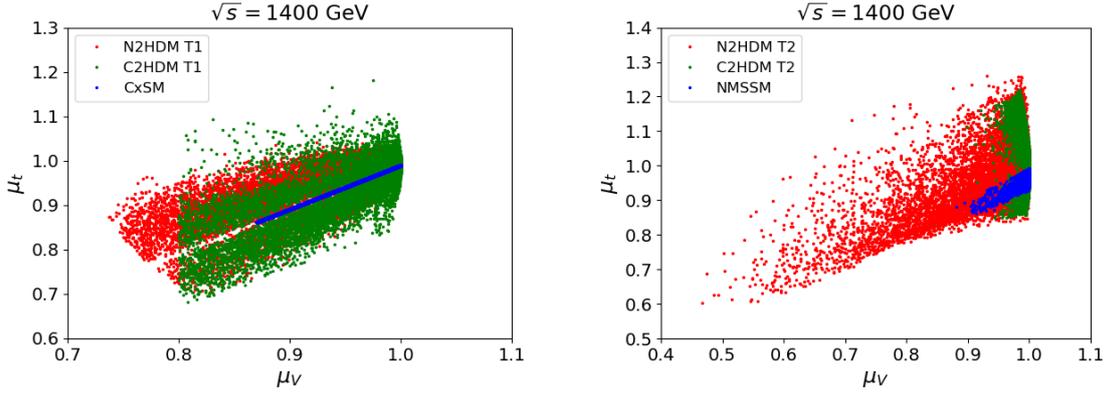

Figure 61: $\mu_t = \sigma^{BSM}_{\bar{t}th}/\sigma^{SM}_{\bar{t}th}$ as a function of $\mu_V = \sigma^{BSM}_{VVh}/\sigma^{SM}_{VVh} = \left(g^{BSM}_{VVh}/g^{SM}_{VVh}\right)^2$, where $V = W, Z$, for the 2HDM and N2HDM Type I and the CxSM (left) and for the 2HDM and N2HDM Type II and the NMSSM (right) for 1.4 TeV.

*tth vs. Zh*

We end this Section with a discussion on the correlations between different cross section measurements for the different models. In Figure 61 we present $\mu_t = \sigma^{BSM}_{\bar{t}th}/\sigma^{SM}_{\bar{t}th}$ as a function of $\mu_V = \sigma^{BSM}_{VVh}/\sigma^{SM}_{VVh} = \left(g^{BSM}_{VVh}/g^{SM}_{VVh}\right)^2$ for the 2HDM and N2HDM Type I and the CxSM (left) and for the 2HDM and N2HDM Type II and the NMSSM (right) for 1.4 TeV. The plots contain regions where precise measurements of deviations from the SM prediction could hint to a specific model. Take for instance the plot on the right and let us assume that the $\mu$'s could be measured with 5% precision. In this case a measurement $(\mu_t, \mu_V) = (1, 0.85)$ indicates that the model cannot be the C2HDM Type II nor the NMSSM. A measurement $(\mu_t, \mu_V) = (1.2, 1.0)$ excludes the NMSSM but not the remaining two models, in their Type II versions. Note that because $e^+e^- \to \bar{t}th$ (for which both Yukawa couplings and Higgs gauge couplings contribute) is not kinematically allowed for 350 GeV, the study of the correlations between this process and associated or $W$-fusion cross sections (for which only Higgs gauge couplings contribute) can only be performed for 1.4 TeV.

### 4.3.3 Signal rates of the non-SM-like Higgs bosons

In this section we present and compare the rates of the neutral non-SM-like Higgs bosons in the most relevant channels at a linear collider. We denote by $H_\downarrow$ the lighter and by $H_\uparrow$ the heavier of the two neutral non-$h_{125}$ Higgs bosons. All signal rates are obtained by multiplying the production cross section with the corresponding branching ratio obtained from sHDECAY, C2HDM_HDECAY, N2HDECAY and NMSSMCALC. For the particular processes presented in this Section, there is no distinction between particles with definite CP-numbers and CP-violating ones and they are therefore treated on equal footing. The main production processes for a Higgs boson at CLIC are associated production with a $Z$ boson, $e^+e^- \to ZH_i$, and $W$-boson fusion $e^+e^- \to \nu\bar{\nu}H_i$. We will be presenting results for two centre-of-mass energies, $\sqrt{s} = 350$ GeV and $\sqrt{s} = 1.4$ TeV. In the case of the former the cross sections are comparable in the mass range presented while for the latter the $W$-boson fusion cross section dominates in the entire Higgs boson mass range. In order to give some meaning to the event rates presented in this Section, we will use as a rough reference that at CLIC $10^{-1}$ fb for $Sc1$ correspond to 50 signal events and $10^{-2}$ fb for $Sc2$ correspond to 150 signal events.

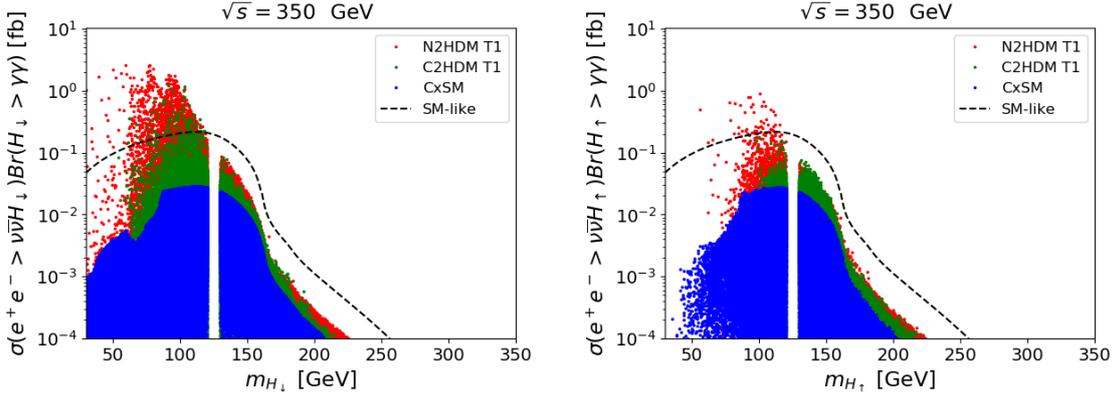

Figure 62: Total rate for $e^+e^- \to \nu\bar{\nu}H_i \to \nu\bar{\nu}\gamma\gamma$ as a function of the Higgs boson mass for $\sqrt{s} = 350$ GeV. The models presented are the CxSM and the Type I versions of the N2HDM and C2HDM. Also shown is the line for a SM-like Higgs boson. On the left panel we present the results for the lighter Higgs boson, $H_\downarrow$, and on the right we show the results for the heavier Higgs boson, $H_\uparrow$.

*4.3.3.1 The 350 GeV CLIC*

In Figure 62 we present the total rate for $e^+e^- \to \nu\bar{\nu}H_i \to \nu\bar{\nu}\gamma\gamma$ as a function of the Higgs boson mass for the CxSM and for the Type I versions of the N2HDM and C2HDM. Also shown is the line for a SM-like Higgs boson. The left panel contains the results for the lighter Higgs boson, $H_\downarrow$, and the right one for the heavier Higgs boson, $H_\uparrow$. The trend shown in the two plots is similar for all other final states. There is a hierarchy with the points of the N2HDM reaching the largest cross sections followed closely by the C2HDM and finally by the CxSM. This is easy to understand since the CxSM is the model with the least freedom - all couplings of the Higgs boson to SM particles are modified by the same factor - while the N2HDM is the least constrained model. This means that it is possible to distinguish between the singlet and the Type I doublet versions if a new scalar is found with a large enough rate. The $\gamma\gamma$ final state is one where the branching ratio decreases very fast with the mass. Still it is clear that there are regions of the parameter space that have large enough production rates to be promising for detection at the 350 GeV CLIC. The behaviour seen in the plots regarding the event rates for the lighter (left) and heavier (right) scalar is the same for the remaining final states and we will only show plots for the lighter Higgs boson in the remainder of this Section.

In Figure 63 we present the total rate for $e^+e^- \to ZH_\downarrow \to Zb\bar{b}$ (left) and for $e^+e^- \to \nu\bar{\nu}H_\downarrow \to \nu\bar{\nu}b\bar{b}$ (right) as a function of $m_{H_\downarrow}$ for $\sqrt{s} = 350$ GeV, for the NMSSM and for the Type II versions of the N2HDM and C2HDM. Clearly there is plenty of parameter space to be explored in the NMSSM and even more in the Type II N2HDM. For the Type II C2HDM, as discussed in a previous work [421], the constraints are such that points with masses below about 500 GeV are excluded. Again there are regions where the models can be distinguished but not if the cross sections are too small. As expected, for this centre-of-mass energy there is not much difference between the two production processes. For a 125 GeV scalar $\sigma(e^+e^- \to ZH_i) = \sigma(e^+e^- \to \nu\bar{\nu}H_i)$ for $\sqrt{s} \approx 400$ GeV. As the scalar mass grows so does the energy for which the values of the cross sections cross. We have also checked that the behaviour does not change significantly when the Higgs boson decays to other SM particles. That is, although the rates are much higher in $H_i \to b\bar{b}$ than in $H_i \to \gamma\gamma$, the overall behaviour is the same. The highest rates are obtained in all models for the final states $b\bar{b}$, $W^+W^-$, $ZZ$ and $\tau^+\tau^-$.

*4.3.3.2 The 1.4 TeV and 3 TeV CLIC stages*

As the centre-of-mass energy rises the $W$-fusion process becomes the dominant one. In Figure 64 we present the total rate for $e^+e^- \to \nu\bar{\nu}H_\downarrow \to \nu\bar{\nu}ZZ$ as a function of the lighter Higgs mass for $\sqrt{s} = 1.4$

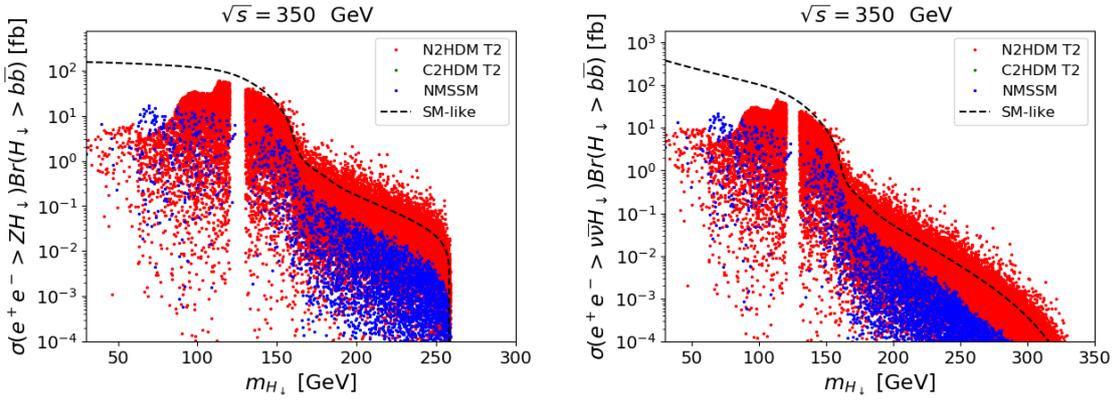

Figure 63: Total rate for $e^+e^- \to ZH_\downarrow \to Zb\bar{b}$ (left) and for $e^+e^- \to \nu\bar{\nu}H_\downarrow \to \nu\bar{\nu}b\bar{b}$ (right) as a function of $m_{H_\downarrow}$ for $\sqrt{s} = 350$ GeV. Plots are shown for the NMSSM and for the Type II versions of the N2HDM and C2HDM. Also shown is the line for a SM-like Higgs boson.

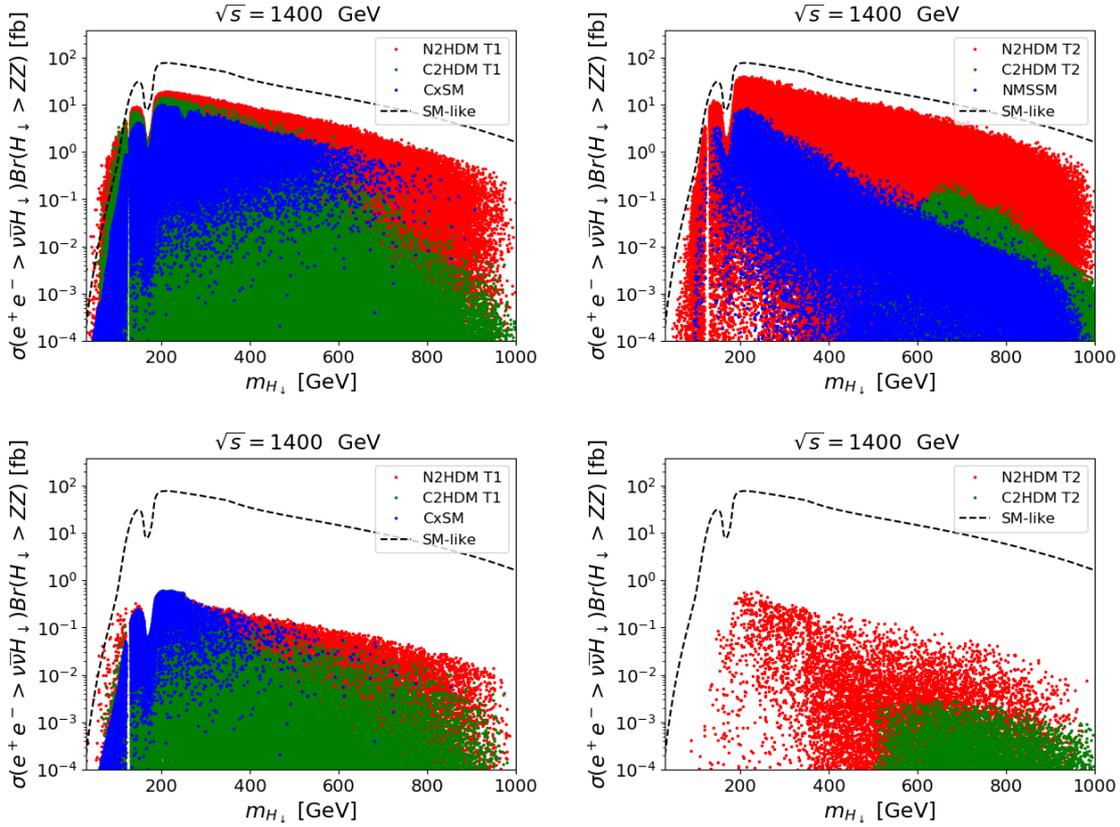

Figure 64: Total rate for $e^+e^- \to \nu\bar{\nu}H_\downarrow \to \nu\bar{\nu}ZZ$ as a function of the lighter Higgs boson mass for $\sqrt{s} = 1.4$ TeV. Left: models CxSM and Type I N2HDM and C2HDM; right: NMSSM and Type II N2HDM and C2HDM. Also shown is the line for a SM-like Higgs boson. In the bottom row the same quantities are shown after imposing the final results on the single Higgs boson measurements for the 350 GeV run.

TeV. In the left panel we show the rates for the CxSM and for the Type I N2HDM and C2HDM while in the right panel plots for the NMSSM and the Type II N2HDM and C2HDM are shown. We expect that total rates above roughly $10^{-2}$ fb can be explored at CLIC@1.4TeV. Hence, all models can be explored in a very large portion of their parameter space but the models are only distinguishable if large cross sections are observed. As previously discussed, the plots for the other final states do not differ much.

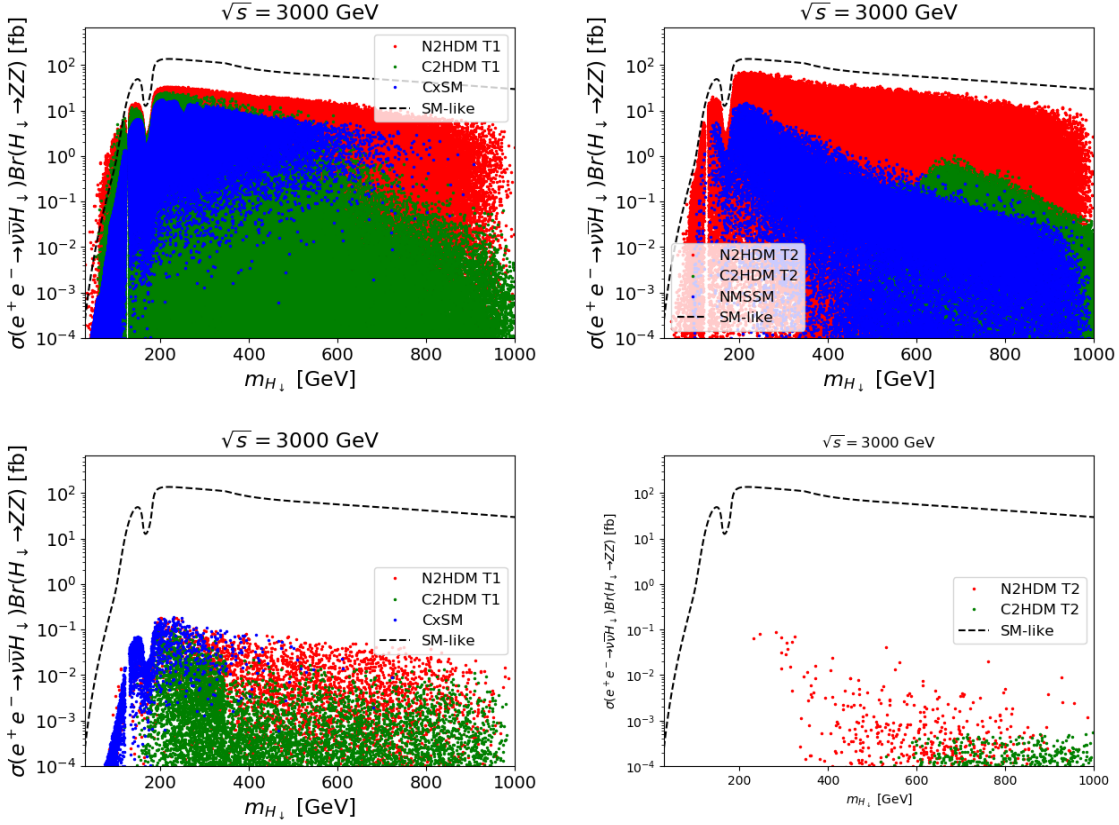

Figure 65: Total rate for $e^+e^- \to \nu\bar{\nu}H_\downarrow \to \nu\bar{\nu}ZZ$ as a function of the lighter Higgs boson mass for $\sqrt{s} = 3.0$ TeV. Left: models CxSM and Type I N2HDM and C2HDM; right: NMSSM and Type II N2HDM and C2HDM. Also shown is the line for a SM-like Higgs boson. In the bottom row the same quantities are shown after imposing the final results on the single Higgs boson measurements for the 1.4 TeV run.

However, once the 350 GeV run is complete, even if no new scalar is found, the measurement of the 125 GeV Higgs couplings will be more precise which reduces the parameter space of the models. In Figure 64 (bottom row) we present the total rate for $e^+e^- \to \nu\bar{\nu}H_\downarrow \to \nu\bar{\nu}ZZ$ as a function of the lighter Higgs boson mass for $\sqrt{s} = 1.4$ TeV including the predictions on the Higgs coupling measurements after the end of the 350 GeV run. We see that after imposing the constraints on the Higgs couplings the cross sections decrease by more than one order of magnitude. We find that the models can still be probed but are no longer distinguishable just by looking at the total rates to SM particles. Interestingly, all points from the NMMSM disappear when we impose the constraints from the 350 GeV run. This is of course related to the fact that we have used the SM central values for all predictions but it could very well be that at the end of this run we could be celebrating the discovery of a new NMSSM particle - or from any other model! Further analysis along the same direction is presented in Figure 65, which shows as well a remarkable overlap in sensitivity of direct searches of new scalars and deviations in Higgs couplings to be observed in previous states of CLIC operation.

*4.3.4 Conclusions*

We have investigated extensions of the SM scalar sector in several specific models: the CxSM, the 2HDM, C2HDM and N2HDM in the Type I and Type II versions as well as the NMSSM. The analysis is based on three CLIC benchmarks with centre-of-mass energies of 350 GeV, 1.4 TeV and 3 TeV. For each benchmark run, the precision in the measurement of the Higgs couplings was used to study possible deviations from the – CP-even and doublet-like – expected behaviour of the discovered Higgs boson. We concluded that the constraints on the admixtures of both a singlet and a pseudoscalar component to the 125 GeV Higgs boson, improve substantially from tens of percent to well below 1% when going from the LHC to the last stage of CLIC. In fact, as shown in [437], after the LHC Run 1 the constraints on the admixtures were as shown in Table 33, where $\Sigma$ stands for the singlet admixture and $\Psi$ is the pseudoscalar admixture. As noted in [437] the upper bound on $\Psi$ for the C2HDM type II is mainly due to the EDM constraints.

Table 33: Allowed singlet and pseudoscalar (for the C2HDM) admixtures.

| Model | CxSM | C2HDM II | C2HDM I | N2HDM II | N2HDM I | NMSSM |
|---|---|---|---|---|---|---|
| $(\Sigma \text{ or } \Psi)_{\text{allowed}}$ | 11% | 10% | 20% | 55% | 25% | 41% |

With the CLIC results the limits on the admixtures are completely dominated by the measurement of $\kappa_{HZZ}$ for $Sc1$ and by $\kappa_{HWW}$ for $Sc2$ and $Sc3$ through the unitarity relation

$$\kappa_{ZZ,WW}^2 + \Psi + \Sigma \leq 1. \tag{176}$$

Since this holds in all our models the constraints become independent of both model and Yukawa type and are given by

- $Sc1$: $\Sigma, \Psi < 0.85\%$ from $\kappa_{HZZ}$
- $Sc2$: $\Sigma, \Psi < 0.30\%$ from $\kappa_{HWW}$
- $Sc3$: $\Sigma, \Psi < 0.22\%$ from $\kappa_{HWW}$

In the second part of this work we investigated the potential to discover and study additional Higgs bosons at CLIC in $W$-boson fusion and Higgsstrahlung. We checked whether the models could be distinguished by a discovery in the first stage of CLIC. If no New Physics is found in the first stage of CLIC we discussed if the parameter space of the models still allows for large enough rates to be probed at the second stage.

- As expected the results are very similar for $W$-fusion and Higgsstrahlung for $\sqrt{s} = 350$ GeV. For the other two benchmark energies the $W$-fusion process dominates. Since the difference relative to the SM in both production processes is in the coupling $hVV$, $V = W, Z$, even for $\sqrt{s} = 350$ GeV, where the cross sections are of the same order, the two processes give the same information about the models.
- For $\sqrt{s} = 350$ GeV and for Type I models and CxSM, the latter is always the most constrained model as the couplings of the Higgs boson to SM particles are all modified by the same factor. Hence the Type I N2HDM and C2HDM, which in most cases are barely distinguishable, have rates that are always larger than the CxSM ones. For some final states the N2HDM rates are slightly above the C2HDM ones but always below the SM-like line, except for the $\gamma\gamma$ final states and only for Higgs boson masses below about 120 GeV. In these Type I models there are charged Higgs contributions in the $H_i \to \gamma\gamma$ loops and the charged Higgs mass is not as constrained as in the Type II models.

- For $\sqrt{s} = 350$ GeV and for Type II models and NMSSM, the C2HDM does not take part in the analysis due to the constraint on the non-125 GeV Higgs boson as previously explained. The Type II N2HDM has rates that are always above the corresponding NMSSM ones. So, it is possible to distinguish the two models in several regions of the parameter space which is expected since the N2HDM has more freedom.
- For $\sqrt{s} = 350$ GeV and for Type II models and NMSSM, the heavier neutral scalar can only be probed in the N2HDM where the rates can be up to two orders of magnitude above the SM line (these plots were not shown). CLIC can probe the lighter neutral scalar boson in both the NMSSM and the N2HDM and distinguishing the two models based on total rates alone may be possible.
- For $\sqrt{s} = 1400$ GeV the results are very similar in what regards the relative rates for the different processes. The main difference comes from imposing the predicted results for the 350 GeV run, if nothing is found and using the SM prediction as central value. This constrains the admixtures – and by unitarity the gauge couplings of the non-SM-like Higgs bosons – to tiny values identical in all models. Therefore, the models become harder to distinguish.

## 4.4 Discovering naturalness

In this section we want to study what CLIC can measure to test scenarios of new physics connected with the naturalness of the weak scale. In particular we study how well CLIC can probe particular predictions of these scenarios.

### 4.4.1 Testing the MSSM Higgs mass prediction [52]

Extending the Standard Model (SM) to its minimal supersymmetric extension (MSSM) makes the mass of the SM-like Higgs boson a prediction. In the decoupling limit, only the MSSM parameters $\mu$ and $\tan\beta$ fix $m_h$ at tree level. However, sizeable higher order corrections are expected, most importantly by contributions from the stops, the supersymmetric scalar partners of the top quark, and the gluino, the supersymmetric partner of the gluon. Though the Large Hadron Collider (LHC) may be able to observe the latter, the expected mass bounds on the former are significantly weaker and most importantly, a hadron collider is hardly sensitive to the mixing angle in the stop sector which is a crucial parameter to predict $m_h$ with a good accuracy. In contrast, the Compact Linear Collider (CLIC) [3–5, 8] can be expected to measure all relevant parameters of the stop sector with good precision and therefore — under the assumption that stops will be observed in this experiment — may be used to test the MSSM Higgs mass prediction.

Phenomenology studies of stop sector at high energy linear colliders have been performed before, see e.g. Refs. [438–440]. In many cases, they considered stop masses of the order $\mathcal{O}(\text{few } 100\,\text{GeV})$ which are becoming constrained by the LHC. Moreover, with the Higgs mass only being known since 2012, these studies often considered spectra which are incompatible with the observed value of $m_h \approx 125\,\text{GeV}$. This value requires large radiative corrections within the MSSM, which either requires a heavy stop sector or scenarios with large stop mixing.

In this work, we assess the expected CLIC sensitivity on a maximally mixed scenario, which allows for stops lighter than $1.5\,\text{TeV}$, by determining both masses and the mixing angle in the stop sector via polarised cross section measurements. We compare the results for scenarios where the heavier stop can be produced in a mixed $\tilde{t}_1\tilde{t}_2$ production and those where the mass of the $\tilde{t}_2$ is derived from observables in the sbottom sector. For that purpose, we discuss one illustrative benchmark which allows for an analysis in both scenarios and an overall combined analysis.

The determined sensitivity on all stop parameters is connected to the mass $m_h$ of the Higgs boson in two ways:

---

[52] Based on a contribution by D. Dercks, G. Moortgat-Pick and R. Rolbiecki.

- Assuming the gluino was observed at the high luminosity LHC phase, the Higgs mass can be predicted using the results of the CLIC stop parameter fit. Using the expected uncertainties on the determined stop parameters from the above analysis, we determine the expected accuracy of the Higgs mass prediction within the MSSM.
- Conversely, should the gluino be yet unobserved, we may use the known mass of the Higgs boson to determine lower and upper limits on $m_{\tilde{g}}$ via the MSSM mass formula. By showing two-dimensional confidence regions in various parameter planes we also discuss how additional information from kinematic mass measurements may improve the sensitivity to our fitted model parameters.

In all cases we obtain the value of the Higgs mass for given parameters from `FeynHiggs-2.14.2`[441–447]. For the time being such calculations are precise to the level of 1-2 GeV (see e.g. Ref. [447, 448] for a recent assessment of the several available calculations and their uncertainties). We work under the assumption that such calculations can be made more precise, both as a consequence of better known input SM parameters and better theoretical tools to perform this calculation.

This section is structured as follows: In Section 4.4.1.1 we summarise the most relevant analytical relations in the stop/sbottom sectors of the MSSM and discuss the predictability of the SM-like Higgs mass within the MSSM. In Section 4.4.1.2 we illustrate the advantages of cross section measurements at CLIC compared to the LHC and formulate minimum requirements on the MSSM stop/sbottom sectors to allow for an unambiguous determination of all theory parameters and a subsequent Higgs mass prediction at CLIC. An exemplary benchmark is defined and analysed in Sections 4.4.1.3/4.4.1.5 in the context of a known/unknown gluino mass, respectively. Finally we conclude in Section 4.4.1.6.

#### 4.4.1.1 The maximally mixed stops

**$\tilde{t}$-$\tilde{b}$ sector:** Within the MSSM, there are two scalar stop fields, $\tilde{t}_L, \tilde{t}_R$, and two scalar sbottom fields, $\tilde{b}_L, \tilde{b}_R$, whose gauge-eigenstate bases are fixed by the two electroweak parameters $\mu$, $\tan\beta$ and five soft SUSY-breaking parameters $M_{\tilde{Q}}^2, M_{\tilde{U}}^2, M_{\tilde{D}}^2, A_t$ and $A_b$. The mass matrix can be written as [439, 449]

$$\mathcal{M}_{\tilde{q}}^2 = \begin{pmatrix} m_{\tilde{q}_L}^2 & X_q m_q \\ X_q m_q & m_{\tilde{q}_R}^2 \end{pmatrix}, \tag{177}$$

with

$$m_{\tilde{q}_L}^2 \equiv M_{\tilde{Q}}^2 + m_Z^2 \cos 2\beta (T_q^3 - e_q \sin^2\theta_W) + m_q^2, \tag{178}$$

$$m_{\tilde{q}_R}^2 \equiv M_{\tilde{R}}^2 + e_q m_Z^2 \cos 2\beta \sin^2\theta_W + m_q^2, \tag{179}$$

$$X_t \equiv A_t - \mu \cot\beta, \tag{180}$$

$$X_b \equiv A_b - \mu \tan\beta, \tag{181}$$

where $e_q$ and $T_q^3$ are the charge and the third component of the weak isospin of the squark $\tilde{q}$, $M_{\tilde{R}} = M_{\tilde{U}}, M_{\tilde{D}}$ for $\tilde{q}_R = \tilde{t}_R, \tilde{b}_R$, respectively, and $m_q$ is the mass of the corresponding fermion. The two parameters, $\mu$ and $\tan\beta$, originate from the MSSM Higgs sector. Diagonalising the matrices $\mathcal{M}_{\tilde{t}}, \mathcal{M}_{\tilde{b}}$ yields the masses and mixings of the stop and sbottom sectors, respectively:

$$m_{\tilde{q}_{1,2}}^2 = \frac{m_{\tilde{q}_L}^2 + m_{\tilde{q}_R}^2}{2} \mp \sqrt{\frac{(m_{\tilde{q}_L}^2 - m_{\tilde{q}_R}^2)^2}{4} + X_q^2 m_q^2}, \tag{182}$$

$$\cos\theta_{\tilde{q}} = \frac{-X_q m_q}{\sqrt{(m_{\tilde{q}_L}^2 - m_{\tilde{q}_1}^2)^2 + X_q^2 m_q^2}}. \tag{183}$$

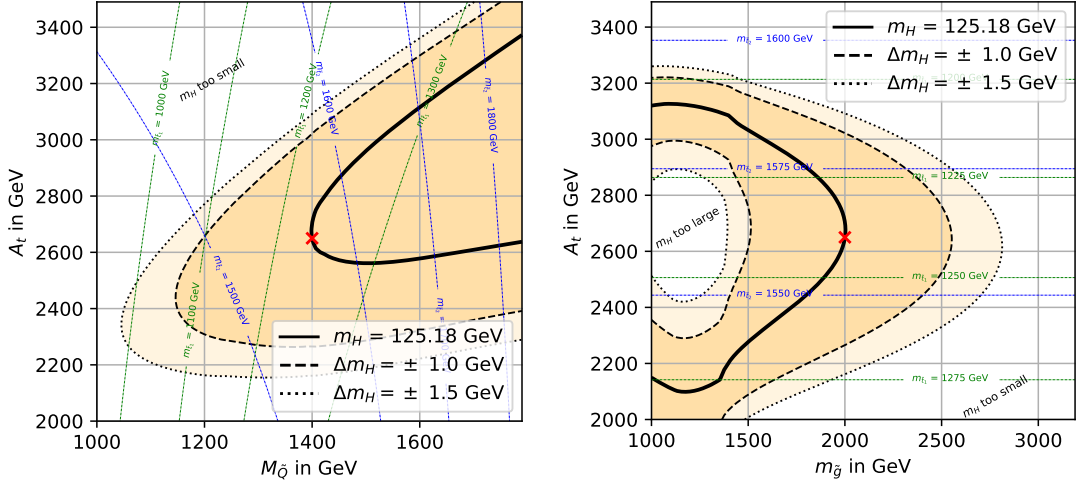

Figure 66: Higgs mass dependence determined with `FeynHiggs`. Unspecified MSSM parameters are chosen as in Section 4.4.1.3. The red cross denotes the benchmark point analysed within this work. (left) Dependence of $m_h$ on $M_{\tilde{Q}}$ and $A_t$. (right) Dependence of $m_h$ on $M_3 = m_{\tilde{g}}$ and $A_t$

Large values of the trilinear couplings $A_t$, $A_b$ result in unrealistic charge and colour-breaking vacuum configurations. This leads to the constraints [450–452]

$$A_b^2 < 3\Big(M_{\tilde{Q}}^2 + M_{\tilde{D}}^2 + M_A^2 + \frac{1}{2}M_Z^2\Big), \tag{184}$$

$$A_t^2 < 3\Big(M_{\tilde{Q}}^2 + M_{\tilde{U}}^2 - \frac{1}{2}M_Z^2\Big), \tag{185}$$

and hence restricts the theoretically allowed combinations of masses and mixing angles in Eqs. (182), (183).

Note that the stop and sbottom sectors are not independent due to the common SU(2)$_L$ doublet mass term $M_{\tilde{Q}}^2$ in both $\mathcal{M}_{\tilde{t}}^2$ and $\mathcal{M}_{\tilde{b}}^2$. Corresponding mass formulae can be derived, see e.g. Ref. [449],

$$\begin{aligned} m_{\tilde{t}_2}^2 \sin^2 \theta_{\tilde{t}} = {} & m_{\tilde{b}_2}^2 \sin^2 \theta_{\tilde{b}} + m_{\tilde{b}_1}^2 \cos^2 \theta_{\tilde{b}} - m_b^2 \\ & - m_{\tilde{t}_1}^2 \cos^2 \theta_{\tilde{t}} + m_t^2 + m_W^2 \cos 2\beta. \end{aligned} \tag{186}$$

This has the phenomenological consequence that the heavier stop mass, even if it lies beyond a collider's kinematic reach, can be determined if all other three stop/sbottom masses and both mixing angles are experimentally measured. This is particularly useful for the maximal mixing scenario analysed within this work which expects a significant mass splitting in the stop sector.

**Squark production at CLIC:** The squark production in electron-positron collisions, $e^+e^- \to \tilde{q}_i \tilde{q}_j^*$, proceeds via photon and $Z$ exchange. At CLIC centre-of-mass energies, $s \gg m_Z^2$, the unpolarised tree-level cross section is to a good approximation given by (see e.g. Refs. [438, 439]):

$$\sigma = \frac{\pi \alpha^2}{s} \beta_{ij}^3 \left( e_q^2 \delta_{ij} - e_q a_{ij} \delta_{ij} \frac{v_e}{8 c_W^2 s_W^2} + a_{ij}^2 \frac{a_e^2 + v_e^2}{256 s_W^4 c_W^4} \right), \tag{187}$$

with

$$\beta_{ij} = \sqrt{\left(1 - \frac{m_{\tilde{q}_i}^2}{s} - \frac{m_{\tilde{q}_j}^2}{s}\right)^2 - 4 \frac{m_{\tilde{q}_i}^2}{s} \frac{m_{\tilde{q}_j}^2}{s}}, \tag{188}$$

$$a_{11} = 4(T_q^3 \cos^2 \theta_{\tilde{q}} - s_W^2 e_q), \tag{189}$$

$$a_{22} = 4(T_q^3 \sin^2 \theta_{\tilde{q}} - s_W^2 e_q), \tag{190}$$

$$a_{12} = -2T_q^3 \sin 2\theta_{\tilde{q}}, \tag{191}$$

and $s_W \equiv \sin\theta_W$, the axial-vector and vector couplings of electron are given by $a_e = -1$ and $v_e = 4s_W^2 - 1$, respectively.

The first term of Eq. (187) is due to the photon exchange, the last term due to the $Z$-boson exchange, while the middle is the interference term. The photon exchange appears only for a diagonal pair production $\tilde{q}_i \tilde{q}_j^*$. The couplings of squarks to the $Z$ are given by factors $a_{ii}$ and $a_{12}$ for diagonal and non-diagonal pairs, respectively. These factors in turn depend on the mixing angle in the squark sector. Finally, the cross section depends on the produced squark masses only via the kinematic $\beta^3$ factor. Taking all the above information together one concludes that by measuring cross sections it is in principle possible to extract simultaneously squark masses and mixings.

An additional useful tool in studying of the production of squarks in $e^+e^-$ collisions is the possibility of having polarized beams, see e.g. Ref. [439]. The polarization will modify the couplings of the electron-positron current to $\gamma$ and $Z$, albeit in a different way. If one can measure the cross section at the same centre-of-mass energy for two different electron beam polarizations, $\sigma_L$ and $\sigma_R$, these two independent measurements typically yield mass and mixing angle with a good accuracy.

**Prediction of $m_h$ within the MSSM:** Whilst the mass of the Higgs boson is a free parameter within the Standard Model, it becomes a prediction within the MSSM. In the decoupling limit, $m_A \to \infty$, the leading order term for $m_h$ can be written as

$$m_{h,\text{tree}} \approx m_Z |\cos 2\beta|. \tag{192}$$

At higher orders in perturbation theory, $m_h$ receives sizeable corrections, which largely depend on the details of the remaining MSSM spectrum. Due to the large top Yukawa coupling, a significant contribution is arising from the two scalar top partners [453–456]. The 1-loop contribution can be written as

$$\Delta m_h^2(\tilde{t}) \approx \frac{3m_t^4}{4\pi^2 v^2} \left( \log\left(\frac{M_S^2}{m_t^2}\right) + \frac{X_t^2}{M_S^2} - \frac{X_t^4}{12 M_S^4} \right), \tag{193}$$

where $M_S^2 := (m_{\tilde{t}_1}^2 + m_{\tilde{t}_2}^2)/2$. As the leading order term is bounded from above by $m_Z^2$, large corrections are not only expected but also required in order to explain the experimentally observed value of $m_h \approx 125\,\text{GeV}$. If one wants to keep stops light, the $X_t$-dependent contribution should be maximised. Since the r.h.s. of Eq. (193) is maximal for $X_t = \sqrt{6} M_S$, such a scenario typically embeds a scalar top sector with a large $X_t$ and therefore, see Eqs. (182), (183), a nonvanishing mixing angle. Moreover, due to the common $SU(2)_L$ doublet mass parameter $M_{\tilde{Q}}^2$ in both $\mathcal{M}_{\tilde{t}}^2$ and $\mathcal{M}_{\tilde{b}}^2$ and due to the sbottom mixing being proportional to the small parameter $m_b^2$, at least one of the sbottom masses is expected to lie within the two scalar top masses. Note that the mass of the other scalar bottom is in principle still unconstrained as the $SU(2)_L$ singlet mass term $M_{\tilde{D}}$ in $\mathcal{M}_{\tilde{b}}^2$ allows one to arbitrarily rescale one eigenvalue in Eq. (182). Hence combining the above considerations, we expect the following mass hierarchy in the $\tilde{t}$–$\tilde{b}$ sector of a realistic MSSM scenario:

$$m_{\tilde{t}_1} < m_{\tilde{b}_1} < m_{\tilde{t}_2}; \quad m_{\tilde{b}_2} \text{ arbitrary}. \tag{194}$$

Another important quantum correction to the SM-like Higgs boson mass comes from the gluino, the SUSY Majorana partner of the SM gluon. Its contribution, $\Delta m_h^2(\tilde{g})$, cannot be given in a compact analytical form (see e.g. Refs. [442] for details on the calculation and numerical results). Instead, we show the dependence of the Higgs mass prediction on different model parameters as determined with FeynHiggs [441–447] in Figure 66, using the example benchmark scenario defined and analysed later in Section 4.4.1.3.

Figure 66 shows the dependence on the two stop parameters $M_{\tilde{Q}}$ and $A_t$ for a gluino mass of 2 TeV. We also show various iso-mass-contours for both stops. We observe that the Higgs mass indeed turns out to be too small unless $A_t$ and $M_{\tilde{Q}}$ are specifically chosen to maximise the stop contribution to $m_h$. With the given gluino mass, the correct Higgs mass value can only be reached for masses $m_{\tilde{t}_1} \gtrsim 1220\,\text{GeV}$, $m_{\tilde{t}_2} \gtrsim 1570\,\text{GeV}$. This scenario would indeed be testable at CLIC as we show below in Section 4.4.1.3. The dependence on the gluino mass is depicted in Figure 66 and we observe that larger values of $m_{\tilde{g}}$ tend to reduce the value of $m_h$ so that given values of the stop masses one obtains upper limits on the gluino mass. Therefore, if the gluino is not observed, a precise measurement of the stop parameters at CLIC will be expected to yield an upper bound on $m_{\tilde{g}}$ under the MSSM hypothesis. We discuss this approach below in Section 4.4.1.5.

*4.4.1.2 Testing the MSSM Higgs mass prediction at CLIC*

From our discussion in the previous section, it becomes clear that in order to predict $m_h$ within the MSSM, the gluino mass, both stop masses *and* the trilinear coupling $A_t$ — or alternatively[53] the mixing angle $\cos\theta_{\tilde{t}}$ — need to be known. Though the high luminosity LHC may be able to discover gluinos with multi-TeV masses [457, 458] and is sensitive to stop masses up to $\approx 1.2\,\text{TeV}$ [354, 459], it is unlikely that both stops could be observed and disentangled. Moreover, even if both stops should be observed, a precise measurement of the mixing angle may be very difficult as the LHC is mainly sensitive to the QCD production mode which is independent of the mixing angle, $\tilde{\theta}_q$.[54] Thus, the LHC may provide important contributions but is not expected to test the MSSM mass hypothesis completely.

At an $e^+e^-$ collider like CLIC, not only does the clean initial state allow for a far more precise final state analysis but there are also other important differences to the LHC which render the situation far more optimistic:

- At a lepton collider, scalar quark partners are predominantly produced via their electroweak interactions for which mixed production $e^+e^- \to \tilde{q}_i \tilde{q}_j^*$ can have a sizeable cross section. This is an important difference to the LHC where QCD-driven production does not yield such a mixed final state which turns out to provide important information to decipher the full stop/sbottom sector.
- The fully accessible target centre-of-mass energy of $\sqrt{s} = 3.0\,\text{TeV}$ at a lepton collider allows for a significantly higher kinematic reach up $m_{\tilde{t}_1} + m_{\tilde{t}_2} = 3\,\text{TeV}$, accessible via the mixed process $e^+e^- \to \tilde{t}_1 \tilde{t}_2^* + c.c.$ Note that this production mode has a lower threshold compared to the diagonal mode $e^+e^- \to \tilde{t}_2 \tilde{t}_2^*$ which requires $2 m_{\tilde{t}_2} \leq 3\,\text{TeV}$. Hence, the sensitivity to the mixed mode is particularly useful in our maximally mixed scenario as we require a large mass splitting which easily pushes the mass of the second stop beyond the threshold for direct $\tilde{t}_2 \tilde{t}_2^*$ production.
- Using spin-polarised electron beams and combining cross section measurements performed with two opposite beam polarisations provides two independent measurements and most importantly allows one to extract the mixing angles $\cos\theta_{\tilde{q}}$ with good accuracy.
- Not only is the measurement of the stop mixing angle $\cos\theta_{\tilde{t}}$ a necessary requirement to extract $A_t$ in Eq. (193), but the sensitivity to the sbottom mixing angle $\cos\theta_{\tilde{b}}$ allows one to measure values $m_{\tilde{t}_2}$ above the kinematic threshold by making use of the MSSM mass relation in Eq. (186).

Thus, should the gluino mass be known from previous LHC measurements, CLIC may be able to measure all remaining ingredients from the stop sector to predict $m_h$. Due to the relations between the stop and sbottom masses and mixing angles, not all parameters need to be measured directly for this purpose.

---

[53]The relation between stop mixing and $A_t$ requires the knowledge of the MSSM parameters $\mu$ and $\tan\beta$. For our discussion, we assume that these are known from an independent measurement, e.g. from precision electroweakino studies at CLIC.

[54]Attempts to extract the stop mixing angle from possible LHC measurements of branching ratios can be found in Ref. [460].

Table 34: Total production cross sections (including effects from NLO SUSY-QCD, beamstrahlung and ISR), the resulting expected event rates for the benchmark scenario discussed in the text. Event numbers are calculated using $N = \sigma \mathcal{L} \epsilon$, Eq. (209), and we assumed $\mathcal{L} = 2\,\text{ab}^{-1}/1\,\text{ab}^{-1}$ integrated luminosities for the left/right handed electron polarisation.

| channel | $P(e^-) = -0.8$ | $N(\epsilon_s = 10\%)$ | $N(\epsilon_s = 75\%)$ | $P(e^-) = +0.8$ | $N(\epsilon_s = 10\%)$ | $N(\epsilon_s = 75\%)$ |
|---|---|---|---|---|---|---|
| $\tilde{b}_1^*\tilde{b}_1$ | 0.070 fb | 13.9 | 105 | 0.010 fb | 1.0 | 7.8 |
| $\tilde{b}_1^*\tilde{b}_2$ + c.c. | 0.023 fb | 4.6 | 34.4 | 0.018 fb | 1.7 | 13.3 |
| $\tilde{b}_2^*\tilde{b}_2$ | 0.037 fb | 7.3 | 54.9 | 0.005 fb | 0.5 | 3.7 |
| $\tilde{t}_1^*\tilde{t}_1$ | 0.503 fb | 100.6 | 754.5 | 0.264 fb | 26.4 | 197.9 |
| $\tilde{t}_1^*\tilde{t}_2$ + c.c. | 0.022 fb | 4.4 | 33.7 | 0.017 fb | 1.7 | 13.1 |
| $\tilde{t}_2^*\tilde{t}_2$ | 0 fb | 0 | 0 | 0 fb | 0 | 0 |

Most importantly, information taken from measurements in the sbottom sector may be used to derive inaccessible observables in the stop sector. Regarding our expected MSSM hierarchy in Eq. (194), we can think about the following scenarios which would be testable at CLIC:

1. If both stops are kinematically accessible, i.e. $m_{\tilde{t}_1} + m_{\tilde{t}_2} \leq 3\,\text{TeV}$, *three* independent measurements would be sufficient to determine $m_{\tilde{t}_1}, m_{\tilde{t}_2}$ and $A_t$ required to evaluate Eq. (193). Note that for each production mode $\tilde{t}_i\tilde{t}_j^*$ we expect *two* independent measurements performed with opposite electron polarisations. Hence, accessibility of both $\tilde{t}_1\tilde{t}_1^*$ and $\tilde{t}_1\tilde{t}_2^*$ would be sufficient to predict $m_h$, even if $\tilde{t}_2\tilde{t}_2^*$ production is kinematically forbidden. The details of the $\tilde{b}$ sector would be completely irrelevant in this scenario.

2. Should the mass of $m_{\tilde{t}_2}$ in Eq. (194) be too heavy, a measurement of $\tilde{t}_1\tilde{t}_1^*|_{L/R}$ alone would leave us with an unconstrained degree of freedom in the scalar top sector. This may be fixed by making use of the sum rule in Eq. (186) *if* the sbottom sector can be fully determined experimentally. For this purpose, five independent measurements would be required in total and as $\tilde{t}_1\tilde{t}_1^*|_{L/R}$ and $\tilde{b}_1\tilde{b}_1^*|_{L/R}$ only provide four degrees of freedom, $m_{\tilde{b}_2}$ *must* be kinematically accessible, i.e. $m_{\tilde{b}_1} + m_{\tilde{b}_2} \leq 3\,\text{TeV}$, for this scenario to work.

3. In a lucky situation that both $\tilde{t}_2$ *and* $\tilde{b}_2$ are kinematically accessible, the $\tilde{t}/\tilde{b}$ system is in principle overconstrained. Measuring all kinematically accessible channels would then not only provide cross-checks if a single MSSM scenario can simultaneously fit all measurements but, as we show below, combining complementary information will result in a reduced uncertainty in the $m_h$ prediction. One could also envisage an experimental test of Eq. (186) and the consistency of the MSSM hypothesis in this regard.

At this stage we conclude: Should the MSSM be realised in Nature with $\tilde{t}_1$ and either $\tilde{t}_2$ or $\tilde{b}_2$ being within the CLIC kinematic reach, the MSSM stop sector can be fully determined from polarised measurements. If in addition the gluino mass is known, e.g. from high luminosity LHC results, results from CLIC may be used to predict $m_h$ and independently test if the MSSM Higgs mass prediction is in agreement with the observed value.

*4.4.1.3 A benchmark analysis*

Within this note, we use the following benchmark scenario to determine the CLIC sensitivity. The stop/sbottom parameters are chosen as

$$M_{\tilde{Q}_3} = M_{\tilde{U}_3} = M_{\tilde{D}_3} = 1.4\,\text{TeV}, \tag{195}$$

$$A_t = A_b = 2.65\,\text{TeV}. \tag{196}$$

For the gluino we choose

$$M_3 = 2.0 \,\text{TeV}, \tag{197}$$

and for all remaining parameters

$$M_A = 2.0 \,\text{TeV}, \tag{198}$$
$$\tan\beta = 20, \quad \mu = 1.0 \,\text{TeV}, \tag{199}$$
$$M_1 = 1.9 \,\text{TeV}, \quad M_2 = 4.0 \,\text{TeV}, \tag{200}$$
$$M_{\tilde{L}_{1\ldots3}} = M_{\tilde{E}_{1\ldots3}} = 4.0 \,\text{TeV}, \tag{201}$$
$$M_{\tilde{Q}_{1,2}} = M_{\tilde{U}_{1,2}} = M_{\tilde{D}_{1,2}} = 4.0 \,\text{TeV}, \tag{202}$$
$$A_{\tilde{\ell}_{1\ldots3}} = A_{\tilde{u}_{1\ldots2}} = A_{\tilde{d}_{1\ldots2}} = 2.65 \,\text{TeV}. \tag{203}$$

Parameters of the stop sector and the gluino mass are chosen such that we reproduce the current world average of the Higgs boson mass $m_h = 125.18 \,\text{GeV}$ [461], see Figure 66. The choice of $\mu = 1 \,\text{TeV}$ gives the mass of the lightest supersymmetric particle (LSP) $m_{\tilde{\chi}_1^0} = 997.7$ which is higgsino-like with another neutralino and the light chargino being almost mass degenerate. The LSP is a candidate for the cold dark matter, however with the relic density below the observed value. This choice has a profound impact on the decay patterns of stops and sbottoms and can affect expected experimental efficiency, which is however beyond the scope of the current study. In the following, we assume that the mass and character of the light higgsinos is accurately determined through different measurements. The gluino will decay via $\tilde{g} \to \tilde{t}_i t$ and $\tilde{g} \to \tilde{b}_i b$ and could evade a 5-$\sigma$ discovery at the HL-LHC though some hints can be expected. However, a detailed study is currently missing. With the above parameter choice for the stop/sbottom sectors, the phenomenologically relevant parameters[55] for our study are as follows:

$$m_{\tilde{t}_{1/2}} = 1.240 \,\text{TeV} \,/\, 1.561 \,\text{TeV}, \tag{204}$$
$$m_{\tilde{b}_{1/2}} = 1.379 \,\text{TeV} \,/\, 1.422 \,\text{TeV}, \tag{205}$$
$$\cos\theta_{\tilde{t}/\tilde{b}} = 0.7078 \,/\, 0.6987, \tag{206}$$
$$m_{\tilde{g}} = 2.000 \,\text{TeV}, \tag{207}$$
$$m_h = 125.18 \,\text{GeV}. \tag{208}$$

Note that this scenario is designed to contain a maximally mixed stop sector with the second stop being accessible via $\tilde{t}_1 \tilde{t}_2^*$ production. Note that because the parameters $M_{\tilde{Q}_3}, M_{\tilde{U}_3}$ and $M_{\tilde{D}_3}$ are equal $\tilde{b}_2$ is only slightly heavier than $\tilde{b}_1$ and also kinematically accessible. Hence, our benchmark corresponds to scenario 3 of the previous Section 4.4.1.2 and in principle yields more experimentally accessible channels than necessary to predict $m_h$. By simply ignoring the presence of $m_{\tilde{t}_2}$ or $m_{\tilde{b}_2}$, respectively, we may analyse the same benchmark also within the context of scenarios 1 and 2. Comparing the results of these approaches shows their complementarity and how much the results improve if information from both $\tilde{t}_2$ and $\tilde{b}_2$ measurements are combined.

The corresponding polarised cross sections for CLIC are provided in Table 34. Our calculations include higher order effects from NLO SUSY-QCD [438], which can be up to 10% of the tree-level cross section. The main motivation for taking them into account is however the dependence of the correction on the gluino mass. The beamstrahlung and initial state radiation (ISR) are determined with `Whizard-2.6.3` [55, 231], and they result in the cross section reduction by 40–70%, depending on the squark masses.[56] The number of signal events can be determined via

$$N(\epsilon_s) = \mathcal{L} \times \sigma \times \epsilon_s, \tag{209}$$

---

[55]All spectrum calculations have been performed with `FeynHiggs-2.14.2` [441–447] and the parameter definitions are to be understood in the on-shell scheme.

[56]Details on the cross section calculation will be provided in Ref. [462].

where $\mathcal{L}$ denotes the total integrated luminosity and $\epsilon_S$ the overall signal efficiency. It is assumed that the high luminosity CLIC phase will split runtimes with left- and right-handed electron beam polarisation with the ratio $2:1$ and use $\mathcal{L}[P(e^-) = -0.8/+0.8] = 2\,\mathrm{ab}^{-1}/1\,\mathrm{ab}^{-1}$.

The signal efficiency $\epsilon_s$ accounts for unspecified details in the event selection. While this will strongly depend on precise details of the entire MSSM SUSY spectrum, dedicated CLIC studies of various SUSY production processes achieve the selection efficiency in the range $\sim 30$–$100\,\%$ [364, 463, 464] using dedicated Boosted Decision Tree classifiers. In particular, stop production was studied in Ref. [465] achieving signal efficiency of $32\,\%$ and purity $50\,\%$. The resulting accuracy in cross section measurement is typically better than $10\,\%$ and is dominated by statistical uncertainty. Since the present study does not include a dedicated Monte Carlo analysis of various signals, we analyse and compare the results for efficiencies $10\,\%$ and a more optimistic value of $75\,\%$. This range should cover different expected signal efficiencies and purities, including possible contamination from SUSY backgrounds. We do not take into account further possible improvements owing to mass measurements using template fits or edges in differential distributions [466]. Once the signal is well-established, a further optimisation of selection method is warranted.

Due to the masses being close to the kinematic threshold, the expected event numbers are small. Still, there are sufficiently many channels which predict event rates large enough to potentially be associated to a statistically significant observation. Note that the null result for $\tilde{t}_2\tilde{t}_2^*$ is also considered an observation as it provides useful information to exclude MSSM parameter regions which predict $m_{\tilde{t}_2}$ significantly smaller than $\sqrt{s}/2$.

Similarly to the former studies in Refs. [439, 449], we assume that the predicted event numbers in Table 34 are exactly observed and determine how well CLIC could determine the underlying model parameters from a multidimensional fit.[57] With the best-fit-point being identical to the truth point by construction, we are interested in the $1\sigma$ confidence interval on the fitted model parameters $\vec{p}$ and determine those by using a likelihood-ratio test to quantify the compatibility of the predicted event rates $\mu_i(\vec{p})$ and the (pseudo)-observed event numbers $N_i$ of all channels according to a Poisson probability density

$$q(\vec{p}) = -2\log\left(\frac{L(\vec{p})}{\hat{L}(\vec{p})}\right)$$
$$= -2\sum_{\text{chan. } i}\left(N_i \log\left(\frac{\mu_i(\vec{p})}{N_i}\right) + N_i - \mu_i(\vec{p})\right). \qquad (210)$$

As our event rates are fairly low, we are dominated by statistical uncertainties and therefore ignore subleading effects from systematic uncertainties. In the large sample limit, the test statistics in Eq. (210) becomes $\chi^2$ distributed with minimum 0 and we determine $1\sigma$ contours on all parameters by scanning the $\tilde{t}/\tilde{b}$ parameter space.[58] We then find the $1\sigma$ single parameter confidence intervals corresponding to $\Delta q(\vec{p}) = q(\vec{p}) < 1$ and 2-dimensional $1\sigma$ contours by requiring $q(\vec{p}) < 2.3$.

*4.4.1.4 Results*

The general result of the above scan corresponds to a multidimensional confidence region and we focus on the most relevant phenomenological projections here. A table listing all determined $1\sigma$ confidence regions is provided in Table 35.[59] Figure 67 show the $1\sigma$ contours in various 2-dimensional parameter planes. Each plot compares the results using three different analyses, corresponding to the three scenarios listed in Section 4.4.1.2:

---

[57] As we fix observation equal to expectation we cannot perform a goodness-of-fit analysis here as e.g. the minimum $\chi^2$ is 0 by construction. We postpone this discussion to a future publication in Ref. [462].

[58] We constrain values of $A_t$, $A_b$ to those being in agreement with Eqs. (184), (185).

[59] More exhaustive results with 2-dimensional confidence regions on various model parameter combinations will be provided in Ref. [462].

Table 35: The $1\sigma$ parameter ranges determined in our analysis as described in Section 4.4.1.3. Numbers in square (rounded) brackets correspond to the signal efficiency of 10 (75) %. Note that these represent 1-dimensional confidence intervals and are therefore tighter than the 2-dimensional confidence regions shown in Figure 67.

| Parameter | "Only $\tilde{t}_1/\tilde{t}_2$ Observables" | | | "Only $\tilde{t}_1/\tilde{b}_1/\tilde{b}_2$ Observables" | | | "Full Combination" | | |
|---|---|---|---|---|---|---|---|---|---|
| $M_{\tilde{Q}}$ in GeV | [1359, | (1386, | 1415), | 1440] | [1387, | (1395, | 1405), | 1414] | [1387, | (1395, | 1405), | 1413] |
| $M_{\tilde{U}}$ in GeV | [1358, | (1386, | 1413), | 1440] | [1353, | (1383, | 1418), | 1455] | [1364, | (1387, | 1413), | 1438] |
| $M_{\tilde{D}}$ in GeV | | Not Analysed | | | [1381, | (1392, | 1406), | 1415] | [1381, | (1392, | 1406), | 1416] |
| $A_t$ in GeV | [2050, | (2442, | 2869), | 3221] | [2291, | (2519, | 2785), | 3052] | [2341, | (2537, | 2767), | 2973] |
| $A_b$ in GeV | | Not Analysed | | | [−7015, | (−5089, | 5285), | 7712] | [−6711, | (−5064, | 5232), | 9136] |
| $m_{\tilde{t}_1}$ in GeV | [1221, | (1233, | 1246), | 1259] | [1221, | (1234, | 1246), | 1258] | [1223, | (1234, | 1246), | 1257] |
| $m_{\tilde{t}_2}$ in GeV | [1499, | (1540, | 1583), | 1617] | [1521, | (1547, | 1577), | 1608] | [1529, | (1550, | 1574), | 1595] |
| $m_{\tilde{b}_1}$ in GeV | [1352, | (1371, | 1385), | 1392] | [1354, | (1365, | 1386), | 1396] | [1354, | (1364, | 1386), | 1398] |
| $m_{\tilde{b}_2}$ in GeV | [1408, | (1416, | 1431), | 1449] | [1406, | (1416, | 1433), | 1442] | [1406, | (1416, | 1432), | 1441] |
| $\cos\theta_{\tilde{t}}$ in GeV | [0.65, | (0.69, | 0.73), | 0.77] | [0.66, | (0.69, | 0.73), | 0.76] | [0.66, | (0.69, | 0.72), | 0.75] |
| $\cos\theta_{\tilde{b}}$ in GeV | [0.39, | (0.58, | 0.8), | 0.95] | [0.59, | (0.65, | 0.73), | 0.78] | [0.59, | (0.64, | 0.73), | 0.79] |
| $m_h$ in GeV | [123.47, | (124.85, | 125.25), | 125.32] | [124.45, | (125.03, | 125.23), | 125.32] | [124.59, | (125.06, | 125.22), | 125.29] |

1. *"Only $\tilde{t}_1/\tilde{t}_2$ Observables"* represents scenarios where the $\tilde{t}_2$ is kinematically accessible and no information from the sbottom sector is required to fully determine all stop parameters. All $\tilde{b}_i\tilde{b}_j^*$ measurements in Table 34 are ignored here.
2. *"Only $\tilde{t}_1/\tilde{b}_1/\tilde{b}_2$ Observables"* ignores any of the $\tilde{t}_1\tilde{t}_2^*$ and $\tilde{t}_2\tilde{t}_2^*$ measurements and yields sensitivities expected for a scenario where the heavier stop mass is beyond the CLIC kinematic reach.
3. *"Full Combination"* considers all measurements in Table 34 and represents the optimum scenario of an MSSM spectrum with all $\tilde{t}$ and $\tilde{b}$ being kinematically accessible.

To illustrate how strongly the results are limited by statistics we overlay the expected $1\sigma$ regions for both chosen benchmark signal efficiencies $\epsilon_s = 10$ and $75\,\%$.

In Figure 67 we show the $1\sigma$ contour of the determination of $m_{\tilde{t}_1}$ and $\cos\theta_t$. These two parameters can be determined simultaneously by measuring polarized cross sections of the $e^+e^- \to \tilde{t}_1\tilde{t}_1^*$ process, see also Refs. [439, 449, 467, 468]. This process has the largest expected event rates, see Table 34, and is always accessible if the stop sector is in the CLIC reach. As a consequence, the expected sensitivity on these two parameters are fairly independent of the three possible scenarios. As can be seen in Figure 67, all analyses determine these two parameters with positive correlation since the most sensitive, high-statistics process $\tilde{t}_1\tilde{t}_1^*|_L$ predicts the same number of events if mass and mixing angle are simultaneously increased within the given sensitivity range, see e.g. Eq. (187).

The situation is different for the determination of the combined $m_{\tilde{t}_1}$–$m_{\tilde{t}_2}$ confidence region, see in Figure 67. Here we observe that considering only the stop sector yields a significant anticorrelation between the two derived masses whilst considering the sbottom sector determines these two parameters with positive correlation. The anticorrelation in the stop measurement can be related to an ambiguity in the fitted event rate of the mixed $\tilde{t}_1\tilde{t}_2^*$ channel which stays constant if $m_{\tilde{t}_1}$ and $m_{\tilde{t}_2}$ are simultaneously rescaled in opposite directions. As a consequence, the overall upper/lower bound on the heavier stop mass is strongly connected to the corresponding lower/upper bound on the lighter stop mass derived from the $\tilde{t}_1\tilde{t}_1^*$ measurement seen in Figure 67. Note that regions which predict $m_{\tilde{t}_2} < 1.5\,\text{TeV}$ are constrained from the null-measurement of the $\tilde{t}_2\tilde{t}_2^*$ channel, explaining the sharp edge near the threshold in Figure 67.

Should the mass of the second stop be implicitly derived from measurements in the sbottom sector, a different ambiguity arises: determining all sbottom masses and the mixing angle explicitly from all $\tilde{b}_1/\tilde{b}_2$ measurements fixes the value of the MSSM parameter $M_{\tilde{Q}_3}$ within statistical uncertainty bands. Using this parameter to find relations between $m_{\tilde{t}_2}$, $m_{\tilde{t}_1}$ and $\cos\theta_{\tilde{t}}$ from Eqs. (182), (183) and using the strong positive correlation in the simultaneous determination of $m_{\tilde{t}_1}$ and $\cos\theta_{\tilde{t}}$, see above, results in a positive correlation between $m_{\tilde{t}_2}$ and $m_{\tilde{t}_1}$.

Though a combination of all channels does not significantly improve the results in the $m_{\tilde{t}_1}$–$\cos\theta_{\tilde{t}}$

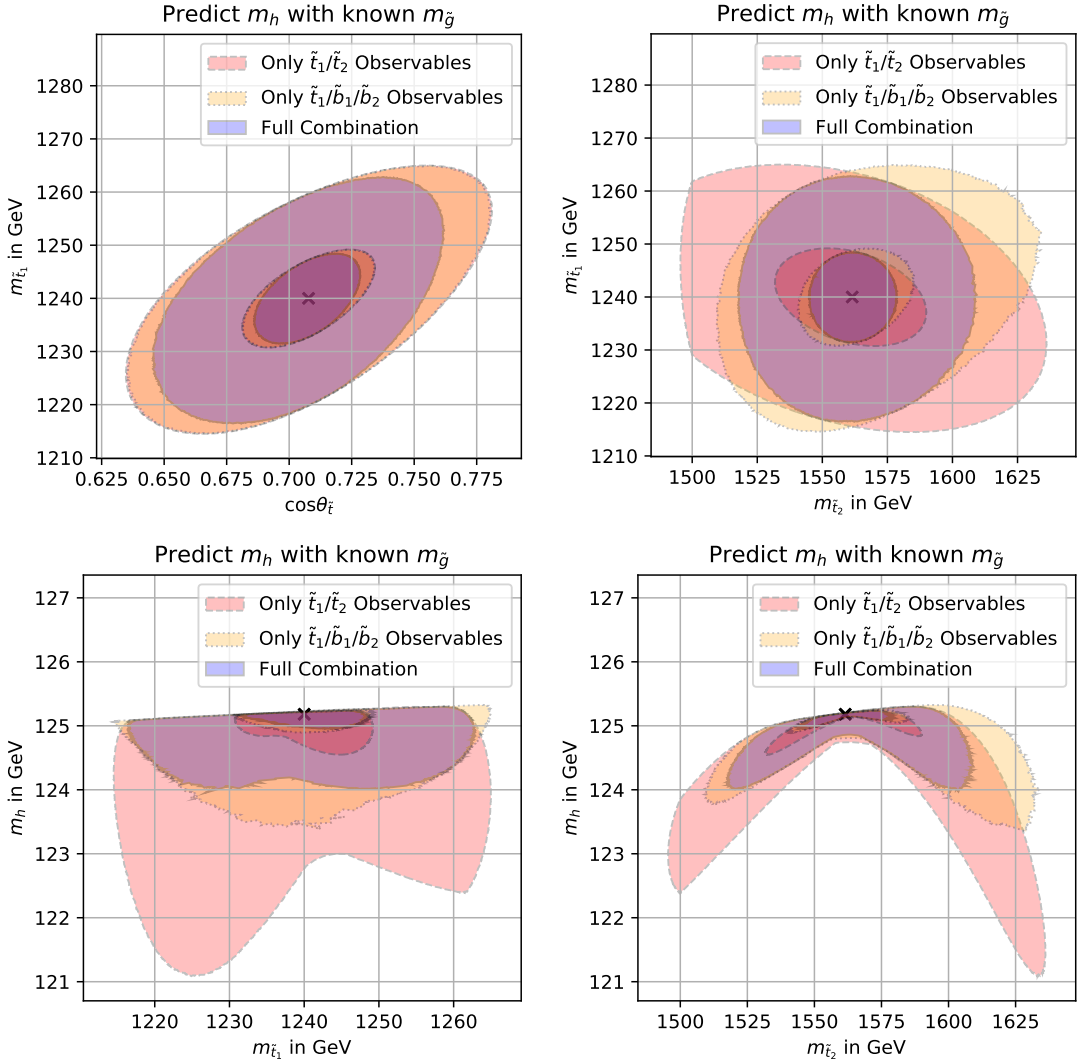

Figure 67: The 2-dimensional $1\sigma$ confidence regions after performing the model fits described in the text. The cross denotes the true point, see Section 4.4.1.3, which by construction is also the best fit point. Brighter/darker regions corresponds to signal efficiencies of $10\,\%/75\,\%$, respectively. (top-left) Measurement of mass and mixing angle. (top-right) Measurement of both stop masses. (bottom-left) Dependence of $m_h$ prediction on $m_{\tilde{t}_1}$. (bottom-right) Dependence of $m_h$ prediction on $m_{\tilde{t}_2}$.

plane, see Figure 67, it is useful when simultaneously determining the two masses of the scalar tops, see Figure 67. Here measurements in the stop and sbottom sectors provide partially complementary information and hence improves the result when considered in the full statistical combination.

With the confidence regions of the stop parameters at hand and the gluino mass assumed to be known at this stage, we can determine the $1\sigma$ interval of the corresponding predicted Higgs mass. Results are shown in Figure 67 and compare the dependence of the predicted Higgs mass on the determined masses of $\tilde{t}_1$, Figure 67, and $\tilde{t}_2$, 67, respectively. In general, we observe that the Higgs mass may be predicted within a few GeV uncertainty even if statistics are very poor. In the optimistic scenario with a combined measurement and a signal efficiency of $\epsilon_S = 75\,\%$, the Higgs mass uncertainty can be brought down to a few hundred MeV and therefore below the expected theory uncertainty on the Higgs mass calculation. For these scenarios, CLIC would indeed be capable of testing the MSSM Higgs mass hypothesis to a very good degree and predict the MSSM Higgs mass with such a good accuracy that any sizeable deviation of the best fit point from the observed value could lead to an exclusion of the MSSM.

Note that large deviations of $m_{\tilde{t}}$ from the assumed values introduce an ambiguity in the expected Higgs mass, especially in the case of $m_{\tilde{t}_2}$ but also in the case for $m_{\tilde{t}_1}$ in case of low sensitivity. This effect originates from the assumption of a maximally mixed stop sector. Here, significant deviations of the stop masses in any direction violates the maximal mixing assumption and therefore reduces the predicted Higgs mass as previously discussed. Therefore, to improve the overall lower bound on the predicted $m_h$ one needs to improve both lower *and* upper bound on the heavier stop mass.

As our analysis determines all masses only implicitly via their contribution to the total production cross section, it may be interesting to compare and/or combine this result with an independent determination of the mass via kinematic reconstruction. A study in Ref. [364], for example, quotes a sensitivity on the right-handed squark mass of approximately 0.5 %, corresponding to 6 GeV in our benchmark scenario, for a signal efficiency of $\approx 30\,\%$. This could improve the results on $\Delta m_{\tilde{t}_1}$ determined from our indirect measurement by about a factor of 2, however would hardly improve the resulting uncertainty on $m_h$.

*4.4.1.5 Inferring the unknown gluino mass from $m_h$*

In our previous analysis we used CLIC measurements to gain full information about the MSSM stop sector and combine it with a *known* gluino mass in order to predict $m_h$. This clearly assumes that the gluino must have been discovered prior to the discovery of the stops at CLIC. In the discussion of our chosen benchmark scenario in Section 4.4.1.3 we already commented on the dependence of this assumption on the details of the remaining MSSM spectrum, e.g. on the masses of the other squarks and electroweakinos. Though the gluino mass may be in the kinematic reach of the LHC, it may easily evade detection, for example because of only having decays with too soft decay products or having many competing decay topologies so none of the different LHC gluino searches will observe a striking signal. For our chosen benchmark it is therefore fair to also discuss the situation of an *unknown* gluino mass during CLIC operation.

Obviously, with one unknown ingredient the MSSM Higgs mass cannot be predicted as we did in the previous analysis. Therefore CLIC will not be able to test the MSSM Higgs mass hypothesis by checking if the observed Higgs mass is contained in the predicted $1\sigma$ MSSM contour. However, starting from the hypothesis "The MSSM is true", one can in turn use[60] $m_h = 125.18 \pm 1.00\,\text{GeV}$ as an additional auxiliary measurement and use it to determine the unknown parameter $M_3 = m_{\tilde{g}}$.

For this purpose, we follow the exact same procedure described at the end of Section 4.4.1.3, add $M_3$ as an additional scan parameter and add the Higgs mass measurement to the test statistic in Eq. (210):

$$q'(\vec{p}) = q(\vec{p}) + \left(\frac{m_h(\vec{p}) - 125.18\,\text{GeV}}{1.00\,\text{GeV}}\right)^2. \quad (211)$$

Note that changing the value of $m_{\tilde{g}}$ also slightly affects the expected $\tilde{t}/\tilde{b}$ production cross sections due to the NLO-QCD effect of the gluino.

As this analysis assumes that the gluino mass has not been observed directly, we have to exclude possibilities where the gluino would have been seen in clean two-body decays of the stops and sbottoms via $\tilde{q} \to \tilde{g}q$. Therefore, we constrain the gluino mass as follows:

$$m_{\tilde{g}} > m_{\tilde{q}_i} - m_{q_i}$$
$$\text{for all } \tilde{q}_i \text{ whose measurement has been considered in the scan.} \quad (212)$$

---

[60]We use the $\Delta m_h = 1\,\text{GeV}$ to describe the foreseen theory uncertainty on the Higgs mass calculation by the time CLIC will perform the analysis described here. Clearly, the precise value of this uncertainty cannot be predicted today. However, the impact of different values of $\Delta m_h$ can easily be inferred from our results.

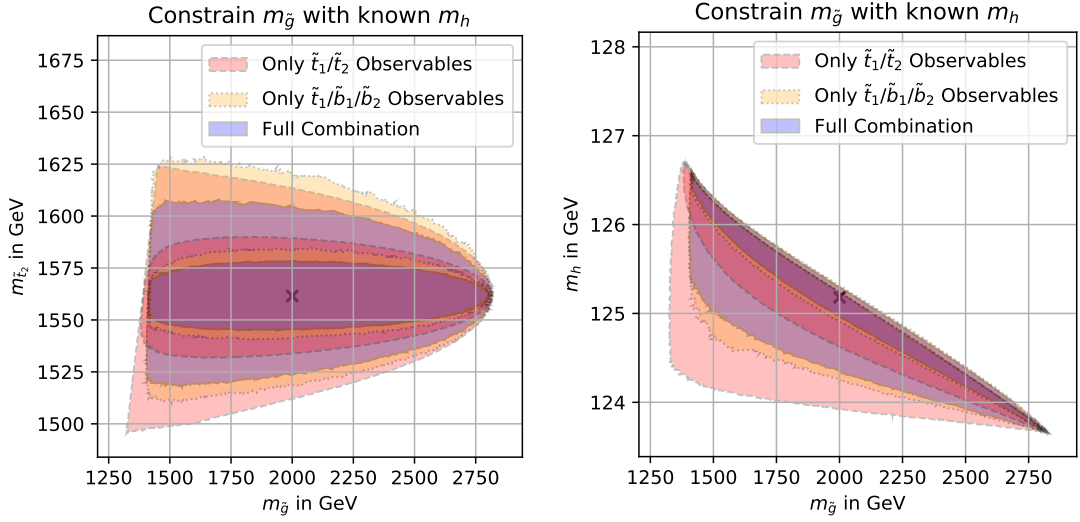

Figure 68: The 2-dimensional $1\sigma$ confidence regions after performing the model tests described in-text. The cross denotes the true point, see Section 4.4.1.3, which by construction is also the best fit point. Brighter/darker regions corresponds to signal efficiencies of 10 %/75 %, respectively. (left) Dependence on $m_{\tilde{g}}$ prediction on $m_{\tilde{t}_2}$. (right) Relation between fitted $m_{\tilde{g}}$ and observed $m_h$.

In any MSSM realisation where the above constraint is violated, the gluino mass could be directly measured much more precisely via kinematic reconstructions of stop/sbottom decays. As the results of our previous analysis, see Figure 67, already constrain $m_{\tilde{t}_1} \gtrsim 1.2\,\text{TeV}$, Eq. (212) requires $m_{\tilde{g}}$ always to lie above the 1 TeV threshold. Depending on which measurements of $\tilde{b}_1, \tilde{b}_2$ and $\tilde{t}_2$ are considered in the different scenarios, Eq. (212) may provide even tighter lower limits.

Remember that our analysis starts from the assumption that CLIC will observe the exact predictions as in Table 34. As a consequence, the best fit point of our pseudo-analysis will always be positioned at the true values of $m_{\tilde{t}_{1,2}}$, $m_h = 125.18\,\text{GeV}$, and is in perfect compatibility with all stop measurements (i.e. $q(\vec{p}_{\text{best fit}}) = 0$). For this particular parameter tuple, the lower and upper bound on the gluino mass can be trivially found by finding those values for which $m_H = 124.18\,\text{GeV}$ (upper bound) and $m_h = 126.18\,\text{GeV}$ (lower bound). From Figure 66, we determine

$$m_{\tilde{g}}(m_{\tilde{t}_i} = m_{\tilde{t}_i}^{\text{truth}}) \in [1.5\,\text{TeV}, 2.6\,\text{TeV}], \tag{213}$$

which is a $1\sigma$ 1-dimensional confidence interval. Interestingly, within our pseudo-analysis, the upper bound in Eq. (213) cannot be improved, regardless of the sensitivity in the stop searches. As can be seen in Figure 66, choosing $m_{\tilde{t}_i} = m_{\tilde{t}_i}^{\text{truth}}$ allows for the largest value of $m_{\tilde{g}}$ and as the truth point will be within the $1\sigma$-confidence region of any stop analysis, it creates an analysis-independent upper bound in our study. However, in a realistic experiment, all observed values will statistically vary around their truth prediction and therefore the best fit point most likely will have non-vanishing $q(\vec{p}_{\text{best fit}})$. In that case, the corresponding derived upper bound on $m_{\tilde{g}}$ will depend on the accuracy of the stop analysis.[61] It is therefore illuminating to not only derive the one-dimensional bound on $m_{\tilde{g}}$ but discuss the two-dimensional confidence regions in combination with other scan parameters.

The results of our analysis are shown in Table 36 (1-dimensional confidence intervals) and Figure 68 (2-dimensional confidence regions) using the same format as before. Figure 68 shows the combined sensitivity on the masses of the gluino and the heavier stop. As explained before, all studies result in the same upper bound on $m_{\tilde{g}}$ which appears for $m_{\tilde{t}_2} = m_{\tilde{t}_2}^{\text{truth}}$. However, it is interesting to observe how the gluino mass bound changes if $m_{\tilde{t}_2}$ is chosen differently ( for example if an actual measurement returned the best fit point at a different position). As observed in our previous analysis, more precisely in Figure 67, nearly any deviation of the stop masses from the true value results in a violation of the maximal mixing constraint and hence in a decreased prediction of $m_h$. As a consequence, the contribution of $m_{\tilde{g}}$ to $m_h$ needs to be reduced accordingly and as can be seen in Figure 66, this can only be accomplished by reducing the gluino mass. This is why in Figure 68 we observe that the *upper* gluino mass bound decreases if the mass of the second stop is larger *or* smaller than the value predicted by maximal mixing.

---

[61]This statement will be analysed in an upcoming study in Ref. [462].

Table 36: The $1\sigma$ parameter ranges determined in our analysis as described in Section 4.4.1.5. Numbers in square (rounded) brackets correspond to the signal efficiency of 10 (75) %.

| Parameter | "Only $\tilde{t}_1/\tilde{t}_2$ Observables" | | | | "Only $\tilde{t}_1/\tilde{b}_1/\tilde{b}_2$ Observables" | | | | "Full Combination" | | | |
|---|---|---|---|---|---|---|---|---|---|---|---|---|
| $M_{\tilde{Q}}$ in GeV | [1364, | (1387, | 1413), | 1436] | [1387, | (1395, | 1405), | 1413] | [1388, | (1396, | 1405), | 1412] |
| $M_{\tilde{U}}$ in GeV | [1365, | (1387, | 1413), | 1435] | [1359, | (1384, | 1417), | 1450] | [1367, | (1388, | 1412), | 1434] |
| $M_{\tilde{D}}$ in GeV | | Not Analysed | | | [1382, | (1392, | 1406), | 1413] | [1382, | (1392, | 1405), | 1413] |
| $A_t$ in GeV | [2143, | (2462, | 2840), | 3162] | [2337, | (2527, | 2781), | 3019] | [2372, | (2545, | 2760), | 2950] |
| $A_b$ in GeV | | Not Analysed | | | [-4886, | (-4874, | 4882), | 4894] | [-4881, | (-4869, | 4881), | 4892] |
| $m_{\tilde{t}_1}$ in GeV | [1222, | (1234, | 1246), | 1256] | [1223, | (1234, | 1246), | 1256] | [1224, | (1234, | 1246), | 1255] |
| $m_{\tilde{t}_2}$ in GeV | [1506, | (1542, | 1580), | 1611] | [1526, | (1548, | 1576), | 1605] | [1532, | (1550, | 1573), | 1592] |
| $m_{\tilde{b}_1}$ in GeV | | Not Analysed | | | [1356, | (1365, | 1386), | 1393] | [1356, | (1365, | 1386), | 1393] |
| $m_{\tilde{b}_2}$ in GeV | | Not Analysed | | | [1408, | (1416, | 1432), | 1440] | [1409, | (1416, | 1432), | 1439] |
| $\cos\theta_{\tilde{t}}$ | [0.66, | (0.69, | 0.73), | 0.76] | [0.66, | (0.69, | 0.72), | 0.76] | [0.67, | (0.69, | 0.72), | 0.74] |
| $\cos\theta_{\tilde{b}}$ | | Not Analysed | | | [0.6, | (0.64, | 0.73), | 0.77] | [0.6, | (0.65, | 0.73), | 0.77] |
| $m_{\tilde{g}}$ in GeV | [1335, | (1510, | 2552), | 2554] | [1462, | (1512, | 2546), | 2550] | [1486, | (1512, | 2547), | 2551] |

From the results of our previous analysis in Figure 67 one can read off the minimum possible Higgs mass achievable within each scenario while still being in agreement with all stop measurements. This value ranged between $m_h^{\min} = 122\,\text{GeV}$ (only stop measurements with $\epsilon_s = 10\,\%$) and $m_h^{\min} = 124.5\,\text{GeV}$ (combined analysis of all measurements with $\epsilon_s = 75\,\%$). These values, in turn, can be used to derive *lower* bounds on the gluino mass which minimise its negative contribution to $m_h$. Note that these bounds compete with the threshold constraint in Eq. (212) which results in a sharp cutoff on the left of the combined $m_{\tilde{g}}$–$m_{\tilde{t}_2}$ confidence region near $m_{\tilde{t}_2} = m_{\tilde{g}} - m_t$. For analyses with low sensitivity ($\tilde{t}$-only analysis with $\epsilon_S = 10\,\%$), it is the mass threshold which determines the absolute lower bound whilst high sensitivity searches (Combined analysis with $\epsilon_s = 75\,\%$) reach the best-case lower bound derived before in Eq. (213); see the $1\sigma$ 1-dimensional confidence intervals in the last row of Table 36.

In Figure 68 we show the sensitivity contours in the $m_{\tilde{g}}$–$m_h$ plane. As explained before, the 1-dimensional upper gluino mass bound corresponds to the best-case bound in Eq. (213) while the lower bound ranges between the corresponding lower constraint and the bound derived from Eq. (213); see Table 36. This figure shows that though the sensitivity of the stop analysis that is used hardly affects the overall lower and upper gluino mass bounds, a good sensitivity can be used to significantly reduce the ambiguity between gluino mass measurement and Higgs mass prediction. For example, should the gluino mass be observed to be 1.5 TeV, the low sensitivity study "Only $\tilde{t}_1/\tilde{t}_2$ observables" with $\epsilon_s = 10\,\%$ will predict MSSM Higgs masses with an uncertainty of $> \pm 1\,\text{GeV}$ whilst the high sensitivity study "Full Combination" with $\epsilon_S = 75\,\%$ would be able to determine the Higgs mass within $\pm 100\,\text{MeV}$. Note that this matches our conclusion from the previous analysis where we show that analyses with good sensitivity could predict $m_h$ with sub-GeV uncertainties if the gluino mass is known.

### 4.4.1.6 Conclusions

We studied a scenario within the MSSM with light, maximally mixed stops. Several, but not all, stop and sbottom states are kinematically accessible at CLIC. In the first step we tested the accuracy to which one can measure masses and mixing angles in the stop and sbottom sectors. Using these measurements it is possible to infer the values of the soft SUSY breaking parameters that enter the MSSM calculation of the Higgs boson masses through radiative corrections.

We considered several scenarios regarding both the accuracy of the measurements at CLIC and the accessibility of different states of the stop and sbottom sectors. The stop masses can be measured with the accuracy of 1.6–3.6 %, and the mixing angle to 7–21 %, depending on the scenario considered. Using the mass relations within the stop and sbottom sectors one can determine the masses of kinematically inaccessible states as well. Using the soft parameters as an input together with the gluino mass we determine that the Higgs boson mass can be predicted with the accuracy as good as 160 MeV, which provides a unique test of the consistency of the MSSM. Even in the case of a poor accuracy in the stop measurements, the information obtained can narrow the predicted Higgs boson mass down to 2 GeV.

Finally, we refine the analysis to include the *a priori* unknown gluino mass, which also enters the Higgs mass calculation at two loops. We demonstrate that using the measurements in the stop sector and the Higgs mass one can predict the gluino mass with the accuracy of about 1 TeV. This prediction is approximately independent of the assumed accuracy in the stop sector measurements.

In conclusion, we showed that CLIC can provide accurate measurements of the stop sector, which can be translated to the radiative contribution to the Higgs boson mass. This prediction can be used to test the consistency of the MSSM, if the gluino mass is known. For example, one can try to check if there are additional contributions to the Higgs boson mass, like in the Next-to-minimal SSM. Alternatively, the measurements can be used to estimate the gluino mass, guiding its searches at the HL-LHC.

### 4.4.2 *Measure* $\tan\beta$ *from ew-inos decays* [62]

#### 4.4.2.1 *Precision electroweakino physics at CLIC*

Even if the LHC discovers any new physics, most likely it will only provide a first glimpse of the full picture of beyond standard model physics. A future collider, either leptonic or hadronic collider, is needed to solidify the new standard model. A lot of studies have already been carried out, estimating the exclusion or discovery reach of new heavy particles at future colliders. Yet it is also important to investigate the power of a future collider in unraveling underlying new mechanisms and principles beyond the standard model. Given the discovery of a light 125 GeV Higgs, one such question we want LHC and future colliders to address is: what is the mechanism that generates the 125 GeV Higgs? More concretely, in the MSSM context, the Higgs mass is basically a function of the stop masses and $\tan\beta$. Could CLIC help measure those parameters to test whether MSSM is the right theory for Higgs mass?

There are two possibilities in MSSM to obtain the right Higgs mass. The first possibility is there is a large mixing between left- and right-handed stops via the parameter $X_t$, which allows for a 125 GeV Higgs boson with stops light enough to be directly studied at CLIC [453, 469, 470]. It may be possible to directly measure the stop-sector parameters [460, 471, 472], but we do not study this case in this note. A second possibility is (mini-)split SUSY with all the SUSY scalars heavy and the gaugino mass one loop below the scalar mass [473–478]. In the simplest case, only two parameters are needed to explain the Higgs mass: a universal SUSY scalar mass $m_0$ and $\tan\beta$. This simple scenario leads to the right Higgs mass when $m_0 \gtrsim \mathcal{O}(10)$ TeV at large $\tan\beta$. The one-loop mass hierarchy (factor of $\sim 100$) between the scalar and gauginos could naturally arise from anomaly-mediated SUSY breaking [479, 480] and some moduli mediation [481–484]. The spectrum automatically alleviates the SUSY flavor problem and solves the cosmological gravitino and moduli problems. While the scenario is meso-tuned in the sense that the little hierarchy problem is still unsolved, it stabilizes the large hierarchy between the Planck scale and $m_0$. It is possible to measure both $m_0$ and $\tan\beta$ through gluino and electroweakino decays at a future 100 TeV hadron collider [485]. In this note, we will study the possibility to measure these two quantities at CLIC. We focus on the region with $m_0 \subset (30 - 10^6)$ TeV and $\tan\beta \subset (2 - 4)$.

Precision measurement of the electroweakino sector, in particular, the decays of electroweakinos, can allow us to extract $\tan\beta$. The strategies depend on the relative ordering of electroweakino masses. As an example, we discuss one possible observable for $\tan\beta$ and present a collider study in the benchmark case with higgsino being the NLSP and bino being the LSP.

---
[62]Based on a contribution by J. Fan and M. Reece.

The case we focus on has the spectrum $M_2 > \mu > M_1$, which corresponds to the bino being the LSP and the Higgsino being the NLSP. One possible observable to probe $\tan\beta$, which we will study numerically below, is the decays of neutral higgsinos. In MSSM, Higgsinos are two doublet fermions $\widetilde{H}_u$ and $\widetilde{H}_d$ with hypercharge $+1/2$ and $-1/2$ respectively. The $\mu$ term gives a Dirac mass that may be interpreted as equal and opposite Majorana masses for the two neutral combinations $\widetilde{H}_\pm^0 = \frac{1}{\sqrt{2}}\left(\widetilde{H}_u^0 \pm \widetilde{H}_d^0\right)$. Mixing with the bino and wino splits the masses of neutral higgsinos, but the mass eigenstates remain approximately $\widetilde{H}_\pm^0$.

At an $e^+e^-$ collider, $\widetilde{H}_\pm^0$ could be produced in pairs since the $Z$ boson coupling to the neutral higgsinos is off-diagonal in the $\widetilde{H}_\pm^0$ basis:

$$i\widetilde{H}_u^\dagger \overline{\sigma}^\mu D_\mu \widetilde{H}_u + i\widetilde{H}_d^\dagger \overline{\sigma}^\mu D_\mu \widetilde{H}_d \supset \frac{g}{2\cos\theta_W} Z_\mu \left(\widetilde{H}_+^{0\dagger}\overline{\sigma}^\mu \widetilde{H}_-^0 + \widetilde{H}_-^{0\dagger}\overline{\sigma}^\mu \widetilde{H}_+^0\right). \tag{214}$$

The higgsinos mix with bino and wino through gauge-Yukawa couplings:

$$\mathcal{L} \supset \frac{\cos\beta}{2\sqrt{2}}(v+h)\left(g\widetilde{W}^0 - g'\widetilde{B}^0\right)\left[(1-\tan\beta)\widetilde{H}_+^0 - (1+\tan\beta)\widetilde{H}_-^0\right] + \text{h.c.}, \tag{215}$$

where we have used the replacement

$$H_u^0 \to \frac{1}{\sqrt{2}}(v+h)\sin\beta, \quad H_d^0 \to \frac{1}{\sqrt{2}}(v+h)\cos\beta, \tag{216}$$

which holds in the decoupling limit when all the other scalars are heavy. From Eq. 215, one could see that in the limit $\tan\beta = 1$, $\widetilde{H}_+^0$ does not mix with the bino or the wino or couple to the Higgs. As explained in Ref. [485, 486], this is due to a parity symmetry under which $\widetilde{H}_+^0$ and the $Z$ boson are odd, but all other neutralinos and the Higgs are even.

Decays of neutral higgsinos happen through the gauge-Yukawa couplings in Eq. 215. Thus, at $\tan\beta = 1$, $\widetilde{H}_+^0$ always decays to $Z + \widetilde{B}^0$ (through the mixing between $\widetilde{B}^0$ and $\widetilde{H}_-^0$ and the off-diagonal $Z$ coupling). On the other hand, decays of $\widetilde{H}_-^0$ all produce higgses. The final state of neutral higgsino pair production is thus always $Zh+$ missing particles. When $\tan\beta$ increases, each neutral higgsino produced could decay to both $Z$ and $h$ with different branching fractions. Thus, the number of $ZZ + \widetilde{B}^0\widetilde{B}^0$ events in the final states could be used to determine $\tan\beta$.

Needless to say, decays of neutral higgsinos in the spectrum with the higgsino in between the bino and wino is only one possible observable sensitive to $\tan\beta$. For example, in the same spectrum, an observable based on wino cascade decay, $\Gamma(\widetilde{W}^0 \to Zh\widetilde{B}^0)/(\Gamma(\widetilde{W}^0 \to ZZ\widetilde{B}^0) + \Gamma(\widetilde{W}^0 \to hh\widetilde{B}^0))$, also depends on $\tan\beta$. More details and other $\tan\beta$ dependent electroweakino observables could be found in Ref. [485].

We will present some preliminary collider studies here, focusing on the observable described in the previous section. In simulating events we use Madgraph 5 [31], showered by Pythia [338] with a decay table computed by SUSY-HIT [487]. Jets are clustered using FastJet [488, 489] with $\Delta R = 0.4$. For this preliminary analysis we have not used a detector simulation.

The benchmark spectrum has $M_1 = 200$ GeV, $M_2 = 800$ GeV, and $\mu = 400$ GeV. In this case, the lightest neutralino $\widetilde{\chi}_1^0$ is mostly bino while the heavier neutralinos $\widetilde{\chi}_2^0$ and $\widetilde{\chi}_3^0$ are almost degenerate in mass and are approximately $\widetilde{H}_\pm^0$. We search for $e^+e^- \to \widetilde{\chi}_2^0\widetilde{\chi}_3^0 \to ZZ\widetilde{\chi}_1^0\widetilde{\chi}_1^0$ in events with both $Z$'s decay hadronically, $Z \to q\bar{q}$. Decays of $\widetilde{\chi}_2^0$ and $\widetilde{\chi}_3^0$ at two different $\tan\beta$'s are shown in Table 37. Indeed $\widetilde{\chi}_2^0$ decays to Higgs dominantly while $\widetilde{\chi}_3^0$ mostly decays to the $Z$ boson. The branching fractions of the more rare decay channel depend on $\tan\beta$, as explained in the previous section. The final state has at least four jets and missing energy. We require each jet to have $p_T > 10$ GeV and the leading four jets to form two pairs with invariant masses in the $Z$ mass window 70 GeV $< m_{jj} <$ 110 GeV. The efficiencies of the cuts are 23% on signal events containing two neutral higgsinos decaying to two $Z$'s. We also simulate standard model backgrounds containing four jets plus two neutrinos. The background mis-tag rate is $6.2 \times 10^{-3}$. For $\sqrt{s} = 1.5$ TeV, the neutral higgsino pair production cross section is 19.7 fb. With 1.5 ab$^{-1}$ data, there will be 410 signal events after the cuts at $\tan\beta = 4$ and 142 signal events at $\tan\beta = 2$. The standard model background events are around 10 after the cuts and thus negligible. The estimated performance of the simple cut-and-count analysis is presented in Figure 69. The statistical uncertainty at $1\sigma$ is shown in orange, while a (somewhat arbitrarily chosen) 10% systematic uncertainty is displayed as a grey band.

Table 37: Branching fractions of $\widetilde{\chi}_2^0$ and $\widetilde{\chi}_3^0$ at $\tan\beta = 1.93$ and 4.04.

| $\tan\beta$ | Br($\widetilde{\chi}_2^0 \to \widetilde{\chi}_1^0 Z$) | Br($\widetilde{\chi}_2^0 \to \widetilde{\chi}_1^0 h$) | Br($\widetilde{\chi}_3^0 \to \widetilde{\chi}_1^0 Z$) | Br($\widetilde{\chi}_3^0 \to \widetilde{\chi}_1^0 h$) |
|---|---|---|---|---|
| 1.93 | $1.43 \times 10^{-2}$ | 0.986 | 0.993 | $6.70 \times 10^{-3}$ |
| 4.04 | $5.68 \times 10^{-2}$ | 0.943 | 0.974 | $2.59 \times 10^{-2}$ |

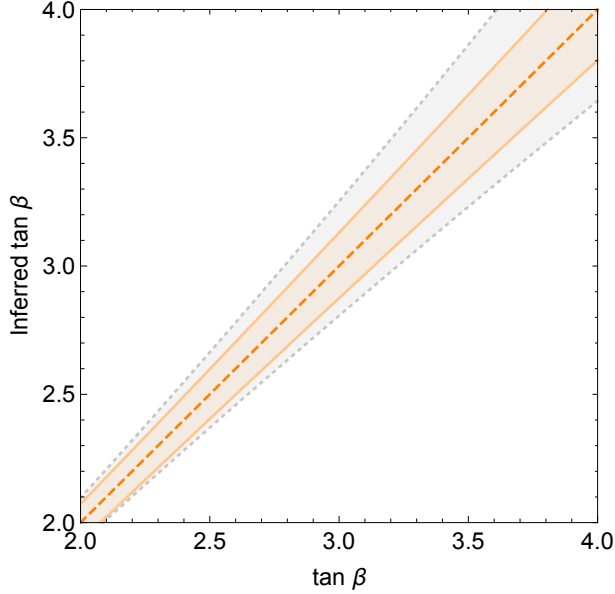

Figure 69: Inference of $\tan\beta$ from the measurement of the rate of $e^+e^- \to \widetilde{\chi}_2^0 \widetilde{\chi}_3^0 \to ZZ\widetilde{\chi}_1^0 \widetilde{\chi}_1^0$ decays. The parameters are $M_1 = 200$ GeV, $M_2 = 800$ GeV, and $\mu = 400$ GeV. The orange band shows the $1\sigma$ statistical uncertainty with $\sqrt{s} = 1.5$ TeV and 1.5 ab$^{-1}$ of data, while the grey band represents a 10% systematic uncertainty on cut efficiencies times cross section times luminosity.

In the (mini-)split SUSY scenario, scalars are too heavy to make directly at CLIC. Furthermore, the gluino is also likely to be out of reach, given current LHC constraints, and in any case can be produced only through loop processes with a small rate. This means that measuring a subdominant gluino branching ratio that is logarithmically sensitive to the heavy scalar mass scale is unlikely to be viable at CLIC as it might be at a future hadron collider [485, 490].

The best prospects for measuring the scalar mass scale at CLIC arise from a different RG effect: the gaugino–higgsino–Higgs Yukawa couplings in a SUSY theory are related to the gauge couplings, but once SUSY is broken they will run differently. Thus, deviations of these Yukawa couplings from the values predicted by tree-level SUSY are logarithmically sensitive to the scalar masses [491, 492]. For example, the Lagrangian contains a term (see [491] for details):

$$\frac{\widetilde{g}'_u}{\sqrt{2}} h^\dagger \widetilde{B}\widetilde{H}_u, \quad \text{where} \quad \widetilde{g}'_u \approx g' \sin\beta \left[ 1 + \frac{\log(m_{\text{scalar}}/\mu)}{16\pi^2} \left(-3y_t^2 + \mathcal{O}(g^2, g'^2)\right) \right]. \tag{217}$$

There are four Yukawa couplings, each involving the Higgs boson, either the bino or wino, and either the up-type or down-type Higgs boson. A leading estimate of $\tan\beta$ arises from either $\widetilde{g}_u/\widetilde{g}_d$ or $\widetilde{g}'_u/\widetilde{g}'_d$, but these estimates will be slightly different due to RG running, and the difference probes the scalar mass scale. From this we see that the scalar mass scale is intrinsically more difficult to measure from electroweakino physics than $\tan\beta$ itself. An early study of the prospects at $e^+e^-$ colliders reached optimistic conclusions [493], but explored light electroweakino parameter space that is already excluded by LHC data.

For our benchmark point, a measurement is much more challenging. For example, the production rate of $e^+e^- \to \widetilde{\chi}^0_1 \widetilde{\chi}^0_3$ at $\sqrt{s} = 1.5$ TeV is of order 1 fb (and mildly $\tan\beta$-dependent). This can be interpreted as associated production of one bino and one higgsino, and as such is a direct probe of the mixing angle determined by the Yukawa couplings. However, the final state is $Z + 2\widetilde{\chi}^0_1$: a mono-$Z$ recoiling against missing energy. This has a large Standard Model background of order 1 pb from the weak boson fusion process $e^+e^- \to \nu_e \bar{\nu}_e Z$. We simulate background with two neutrinos and two jets and find that the kinematics of the $Z$ is very similar in signal and background (with events peaking at large missing mass). As a result, measuring this signal directly is impossible.

While the Yukawa couplings can potentially be probed through other means than direct measurement of gaugino-higgsino pair production—for instance, through the mass splittings among the mostly-higgsino mass eigenstates—the RG running effect we seek to measure is always a subdominant effect compared to $\tan\beta$ itself. As such, it appears to be very difficult to measure at CLIC, for the benchmark spectrum we consider here.

There may be more promising observables in different spectra. For instance, if the higgsinos are heavier than both the winos and the bino, then ratios of branching fractions of a higgsino to a wino or to a bino, $\text{Br}(\widetilde{H}^0_i \to \widetilde{W}^0 h)/\text{Br}(\widetilde{H}^0_i \to \widetilde{B}^0 h)$ and $\text{Br}(\widetilde{H}^0_i \to \widetilde{W}^0 Z)/\text{Br}(\widetilde{H}^0_i \to \widetilde{B}^0 Z)$, depend on the ratio of Yukawa couplings $\widetilde{g}_i/\widetilde{g}'_i$ but not on $\tan\beta$, and hence are a more direct probe of the logarithmic running from the scalar mass scale. Assuming the neutral wino is heavier than the bino and decays to bino plus Higgs, one could try to compare the number of events from $e^+e^- \to \widetilde{\chi}^0_3 \widetilde{\chi}^0_4 \to 2\widetilde{\chi}^0_1 + Zh$ and that from $e^+e^- \to \widetilde{\chi}^0_3 \widetilde{\chi}^0_4 \to 2\widetilde{\chi}^0_2 + Zh \to 2\widetilde{\chi}^0_1 + Z + 3h$ with two additional Higgses. One challenge here is that, in addition to Standard Model background, there is SUSY background from decays of lighter winos. We leave a more detailed study for the future.

CLIC has excellent prospects for measuring properties of electroweakinos, and thus for probing key aspects of the physics of a (mini-)split SUSY scenario. We have shown in a benchmark scenario that a relatively simple analysis could give an accurate measurement of $\tan\beta$. However, obtaining an indirect hint of the mass scale of heavy scalars is much more challenging.

### 4.4.3 *Testing the scalar sector of the Twin Higgs model*

The hierarchy problem can be solved without any new colored particle in light spectrum of the theory that supersedes the Standard Model. In such theories the null results of the searches for new physics at the Large Hadron Collider are indeed what was to be expected. In these theories the mass of the Higgs mass is protected by a symmetry, as in SUSY, but the new particles associated with this symmetry are not charged under SM color. This makes probing the new states much more difficult, as it becomes a lot harder to produce them at a hadron collider. Several theories of this type have been proposed that stabilize the Higgs mass up to scales of order 5-10 TeV, the precision electroweak scale [375, 494–504]. The best known example of this class of models is the Mirror Twin Higgs (MTH) [375], in which the symmetry partners are neutral, not just under SM color, but under all the SM gauge groups.

In the MTH framework, the particle content of the SM is extended to include a mirror ("twin") copy of all the fields in the SM. A discrete $Z_2$ twin symmetry relates the particles and interactions in the SM and mirror sectors. The Higgs sector respects a larger global symmetry which, in the simplest incarnation of the model, is taken to be SU(4)×U(1). This global symmetry, like the discrete symmetry, is only approximate. The electroweak gauge symmetries of the SM and twin sectors are embedded inside the global symmetry. The fields that constitute the SM Higgs doublet are among the pseudo-Nambu-Goldstone bosons (pNGBs) associated with the spontaneous breaking of the global SU(4)×U(1) symmetry down to SU(3)×U(1) at the dynamical scale of the model, that we denote by $f$. Their mass is protected against one loop radiative corrections by the combination of the nonlinearly realized global symmetry and the discrete twin symmetry.

Recently an alternative class of Twin Higgs models, known as Fraternal Twin Higgs (FTH) models, has been proposed, in which the twin sector is more minimal than in the MTH, consisting of only those states that are required to address the hierarchy problem [505]. Specifically, the spectrum of light twin sector states includes only the third generation fermions, the electroweak gauge bosons, and the twin gluon. This framework naturally solves the cosmological problems of the MTH construction. It also leads to exotic collider signals since the lightest twin particles, the mirror glueballs, decay back to SM states, but with long lifetimes. Mirror glueballs can be produced in Higgs decays and will then decay far from the original interaction point, resulting in displaced vertices. The striking nature of these signals will allow the LHC to probe most of the preferred parameter space [506, 507].

The only communication between the visible and twin sectors that is required by the Twin Higgs framework is through the Higgs portal. After electroweak symmetry breaking the Higgs fields of the two sectors mix. The lighter mass eigenstate is identified with the 125 GeV Higgs particle. As a consequence of the mixing it has suppressed couplings to SM fields, resulting in a production cross section that is smaller than the SM prediction. This mixing also results in a contribution to the Higgs width from decays into invisible twin sector states. Unfortunately, while these signals are robust predictions of the MTH framework, they are not unique to it. They are expected to arise in any model in which the SM communicates with a light hidden sector through the Higgs portal.

If, however, the $Z_2$ symmetry is only softly broken, so that the Yukawa couplings in the two sectors are equal, the suppression in the Higgs production cross section and the Higgs invisible width are both determined by the mixing angle, leading to a prediction that can be tested by experiment [508]. This prediction does not apply to theories that exhibit hard breaking of $Z_2$, such as the FTH, or MTH models in which the Yukawa couplings in the two sectors are different. The prediction can be understood as a consequence of the mirror nature of the model. Since it does not depend on the enhanced global symmetry of the Higgs sector, this prediction is not specific to the MTH construction, but applies more generally to any mirror model [509, 510] in which the discrete $Z_2$ symmetry is only softly broken, so that the Yukawa couplings in the two sectors are equal.

If the breaking of the global symmetry is realized linearly, the radial mode in the Higgs potential is present in the spectrum and constitutes a second portal between the twin and SM sectors. We refer to this state as the twin sector Higgs. As we now explain, a study of the properties of this particle at colliders, when combined with precision measurements of the light Higgs, can be used to overdetermine the form of the scalar potential, thereby confirming that it possesses an enhanced global symmetry as dictated by the Twin Higgs mechanism.

In the case when the discrete $Z_2$ symmetry is only softly broken, the Higgs potential of the MTH model takes the form[63]

$$V = -\mu^2 \left( H_A^\dagger H_A + H_B^\dagger H_B \right) + \lambda \left( H_A^\dagger H_A + H_B^\dagger H_B \right)^2 \\ + m^2 \left( H_A^\dagger H_A - H_B^\dagger H_B \right) + \delta \left[ \left( H_A^\dagger H_A \right)^2 + \left( H_B^\dagger H_B \right)^2 \right]. \quad (218)$$

---

[63]We employ the notation of [494].

We distinguish the SM sector fields with the subscript $A$ and the twin sector fields with $B$. Overall the potential depends on four parameters, $\mu^2$ and $\lambda$, which respect the $Z_2$ symmetry, and $m^2$ and $\delta$, which break the $Z_2$ symmetry. We write the formula of the potential in two lines to highlight these symmetry breaking properties. In more detail, the terms in the top line of Eq. (218) respect both the global SU(4)×U(1) symmetry and the discrete $Z_2$ twin symmetry $A \leftrightarrow B$. The $m^2$ term explicitly breaks both the discrete and global symmetries, but only softly, and can naturally be smaller than $\mu^2$. The quartic term $\delta$ respects the $Z_2$ twin symmetry, but violates the global symmetry. In order to realize the light Higgs as a pNGB and thereby obtain a significant reduction in fine-tuning relative to the SM the parameter $\delta$ that violates the global symmetry must be much smaller than $\lambda$, which is invariant under SU(4)×U(1). Similarly, $m^2$ must be much smaller than $\mu^2$.

The parameters in the Higgs potential must reproduce the mass of the light Higgs and the electroweak vacuum expectation value (VEV). This fixes two combinations of the four parameters. Two additional measurements are then required to fully determine the potential. At a lepton collider the production cross section and invisible width of the light Higgs can be determined to a precision of order one part in a hundred [9]. This covers the entire range of interest for the MTH and fixes a third combination of the parameters. Finally, the discovery of the twin sector Higgs particle at a given mass would pin down all four parameters in the Higgs potential. Once the potential has been specified, in the absence of further $Z_2$ violation, the production cross section, width and branching ratios of the twin sector Higgs are all robustly predicted. Therefore, a measurement of the rate to any SM final state overdetermines the system, and constitutes a powerful consistency check on the form of the potential. These predictions remain true to a good approximation even in the presence of hard breaking of the $Z_2$ symmetry by the twin sector Yukawa couplings, provided that this breaking is not large enough to significantly alter the total width of the twin sector Higgs.

In the MTH framework, the breaking of the approximate global symmetry of the Higgs potential results in seven pNGBs. These include, in addition to the light Higgs, the longitudinal components of the $W^\pm$ and $Z$ bosons of both the SM and twin sector. It follows that in the limit that the global symmetry is exact, the couplings of the twin sector Higgs particle to all these seven states are the same, and are set by the SU(4)×U(1) invariant quartic term in the Higgs potential. In particular, the couplings of this state to the SM Higgs, $W^\pm$, and $Z$ are not suppressed by the mixing angle. In the limit that the twin sector Higgs particle is heavy, corresponding to the quartic term being large, its dominant decay modes are to these seven pNGBs. Furthermore, in the limit that the masses of the final state particles can be neglected, the branching ratio of the twin sector Higgs into each of these final states is the same. It follows that $WW$, $ZZ$ and di-Higgs are promising channels in which to search for the twin sector Higgs.

In Ref. [511] it was found that at the LHC, much of the range of parameter space in which the twin sector Higgs can be discovered is already disfavored by existing measurements of the couplings of the light Higgs. Here we summarize the results of the same reference about the high energy stage of CLIC, which can search for the twin sector Higgs boson as well as measure the invisible width of the light Higgs to percent level precision [9]. As we will show, combining measurements of the twin sector Higgs with precision studies of the couplings of the light Higgs results in much greater ability to confirm the MTH construction.

The discovery of the twin sector Higgs boson is expected to proceed in the reaction

$$e^+e^- \to H_+\nu\bar{\nu} \to hh\nu\bar{\nu} \to 4b\nu\bar{\nu}\,,$$

which has also been discussed in other studies for the discovery of new scalars at CLIC in Section 4.2.1. We refer to [511] for details on the collider analysis. From the analysis we are interested to the statistical precision of the measurement of the twin Higgs rate.

In order to quantify the confidence with which the Twin Higgs mechanism can be confirmed as follows. For a given parameter point, we calculate the uncertainty in the number of observed events after the cuts described above (due to Poisson statistics), and we also estimate the uncertainty in the expected number of events at that parameter point, where the leading contribution is the uncertainty in the value of the mixing between SM and Twin Higgs, namely $\sin^2(\vartheta - \theta)$ in the notation of [511], arising from Higgs coupling measurements. In particular, we assume that $\kappa_Z$, the multiplicative factor that measures the deviation of the Higgs coupling to the $Z$-boson, can be measured with a precision of 0.5% [512]. Combining the uncertainties in the number of expected and observed events, we arrive at the fractional uncertainty in the ratio of observed to expected events, centered around the value 1. The fractional uncertainty is plotted for CLIC ($\int dt \mathcal{L} = 1.5$ ab$^{-1}$) benchmarks in Figure 70 in the plane of the mass of the Twin Higgs, $m_+$, and $m_T$, the mass of the Twin sector fermions, that is a fixed function of the scale $f$ of the Twin Higgs in the (approximately) $Z_2$ symmetric limit that we have in mind. From the plot it is apparent that a test of Twin Higgs prediction for the rate of the new scalar, hence of the dynamical mechanism that stabilizes the weak scale, is feasible and CLIC may highlight a large deviation from the Twin Higgs prediction, or give a confirmation of the mechanism at the level of few tens of percent.

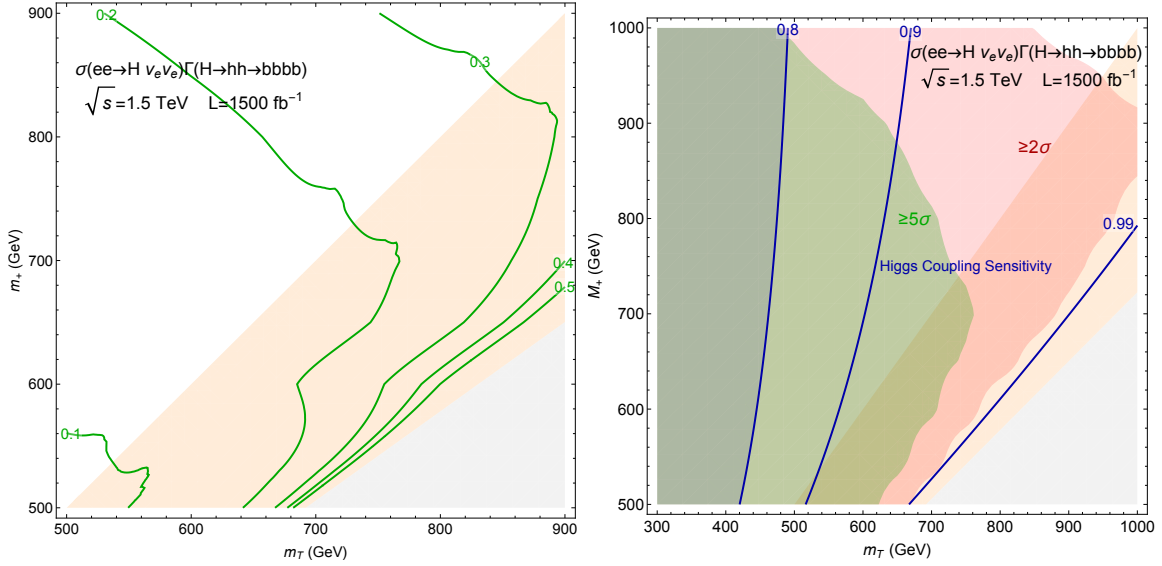

Figure 70: (left) Green lines display the uncertainty in the ratio of observed and expected events, centered around the value 1, for CLIC ($\int dt \mathcal{L} = 1.5$ ab$^{-1}$) 1.5 TeV stage as a function of $m_+$ and $m_T$. (right) Expected significance of $W$ fusion to heavy twin sector Higgs decaying to di-Higgs to 4 $b$. The blue contours indicate deviation in Higgs couplings, the region to the left of 0.8 is excluded by current measurements. Grey shaded region does not provide a stable vacuum. The orange region does not significantly improve tuning compared to the SM [511].

The Twin Higgs mechanism protects the mass of the Higgs against radiative corrections without requiring new particles charged under the SM gauge groups. In this framework, the light Higgs emerges as the pNGB associated with the breaking of a global symmetry, and its mass is protected against quantum effects by a combination of the global symmetry and a discrete $Z_2$ symmetry. If the breaking of the global symmetry is realized linearly, the radial mode of the Higgs potential, the twin sector Higgs, is present in the spectrum. This particle provides a new portal between the visible and twin sectors. We have shown that, if the discrete $Z_2$ symmetry is only softly broken, a measurement of the mass of the twin sector Higgs, when combined with precision measurements of the light Higgs, completely specifies the Higgs potential. The rates for twin sector Higgs events are then testable predictions of the Twin Higgs framework. This conclusion also applies to theories that exhibit hard breaking of the $Z_2$ symmetry by the twin sector Yukawa couplings, provided that this breaking is small enough that the correction to the overall width of the twin sector Higgs is small. While the high luminosity LHC can potentially discover the twin sector Higgs, CLIC has a much better precision and greater reach, allowing for a significant test of the principle of Twin Higgs stabilization of the weak scale.

# 5 Dark matter

The identification of Dark Matter is a vast problem. Depending on its interactions and production mechanism, viable dark matter candidates span a mass range of many orders of magnitude. The range of dark matter masses explorable at particle colliders lies in the GeV-TeV region and this makes a high-energy, multi-TeV $e^+e^-$ collider a very powerful tool to probe the existence of a dark matter candidate and, possibly, of a whole sector of dark states. In this section we consider possible dark matter candidates that can be directly produced at CLIC or that can affect Standard Model processes in a significant manner. We concentrate on the well defined case of dark matter being the lightest state of the dark sector, possibly accompanied by another state light enough to affect the dark matter phenomenology at colliders. We call the dark matter candidate the Lightest Dark Sector Particle (LDSP) and dub the Next to Lightest state as NLDSP. These are reminiscent of the Lightest Supersymmetric Particle and its Next-to-Lightest particle that are usually adopted in the discussion of supersymmetric dark matter scenarios. We want to stress, however, by using a different acronym, that the analysis of this chapter applies both to typical supersymmetric models and to non-supersymmetric models.

The mass difference between the NLDSP and LDSP states may range from zero to infinity. In the following we try to assess the reach of CLIC to explore this vast range of possibilities. Zero mass splittings lead to no energy released in the decay of the NLDSP to the LDSP and give rise to the same signals as very large (or infinite!) mass splitting, when the NDSLP is so heavy that it cannot be produced at CLIC. They all result in the production of invisible particles, which leads to apparent momentum non-conservation in the collisions. A measurable signal can be obtained by requesting the dark matter candidate to recoil against a hard photon, emitted from the electron or positron in the initial state, $e^+e^- \to \gamma + \text{invisible}$. Such processes are studied in Section 5.1.

For slightly larger mass splittings one can imagine that the NLDSP and the LDSP belong to the same weak multiplet. This explains very naturally why they are nearly degenerate in mass and is naturally realised in many extensions of the Standard Model in which the dark matter arises in well defined thermal scenarios for the history of the Universe. In this case it is possible to have short charged tracks left in the detector by the electrically charged members of the weak multiplet that accommodates the dark matter candidate. These may leave very distinctive exotic signals that we explore in Section 5.2. The role of this type of exotic searches with respect to more traditional $\gamma + \text{invisible}$ ones heavily depends on the performance of the tracking of the experiment, its design, which should favour a sensitive tracking detector as close as possible to the beam, and the background that naturally arises from a number of instrumental sources. Even with very conservative choices on these parameters it has been shown recently that these searches can be very powerful at CLIC [513]. We will show in the following how at CLIC, for reasonable and realistic running scenarios, these searches can be even more powerful, managing to exclude well motivated dark matter candidates, such as a thermal relic dark matter Dirac fermion weak doublet that may be identified with the Higgsino of supersymmetric models.

For larger mass splittings, about one order of magnitude below the mass of the dark matter, it is possible that the NLDSP and LDSP interact non-trivially in the early universe plasma, giving rise to a substantial change in the thermal relic abundance due to their so-called "co-annihilation". This scenario poses a well defined challenge to deal with soft objects, e.g. soft leptons and soft jets, as the energy released in the NLDSP decay may be tiny on the scale of the typical event at the collider. This non-minimal scenario of thermal history is studied in detail in Section 5.4, where we show how the clean environment of an $e^+e^-$ machine can play a key role in tagging these challenging events.

Even larger mass splitting can in general arise and give rise to more traditional signals, characterised by hard leptons, photons, and jets, and an apparent non-conservation of momentum. There is a vast literature on these signals and in general $e^+e^-$ machines do well on these signals, as is the case of homologous signals studied in the context of supersymmetry and summarized in Section 4.1.2.

A more challenging scenario is that of the Inert Doublet Model (IDM), a simple extension of the Standard Model that may provide a scalar dark matter candidate. At $e^+e^-$ colliders one can search for pair production of neutral or charged IDM scalars, with heavier scalars decaying into electroweak gauge bosons and the lightest scalar (dark matter candidate). Due to the small weak charge and spin multiplicity of this dark matter candidate, the signal cross-section for this model is the smallest among weakly charged dark matter candidates, which poses a challenge for discovery. Depending on the scalar mass splittings, a wide range of experimental signatures can be considered. We will see in Section 5.5 that already the first stage of CLIC can be sensitive to these scenarios, in spite of the relatively small expected signal rate for the considered channels.

All of the above studies are direct searches, in which a signal from the dark matter candidate or its partner is sought. By their nature, direct searches only probe the scenario they are designed for and it is thus very hard to have a robust full coverage of all possible signatures. However, if a weakly charged multiplet accommodates dark matter, it will induce a plethora of consequences on reactions involving SM particles at high momentum transfers. These may indirectly reveal the presence of new weakly charged states either through their virtual excitation or even by observing the effect of their production in loop corrections of SM processes. This path to discovery goes generically under the name of Electroweak Precision Tests and is largely explored in Chapter 2. Here in Section 5.3 we consider the effect of weakly charged states in the simplest reactions at $e^+e^-$ colliders, that is $e^+e^- \to f\bar{f}$ at high energy. These studies show interesting sensitivity to thermal dark matter candidates such as a Dirac fermion weak triplet around 2.0 TeV.

For more exotic, and necessarily more vaguely defined, dark sectors we refer the reader to the existing literature on linear colliders, e.g. [514, 515] and references therein, as well as 'parasitic' uses of the CLIC beam in fixed target experiments to search for light feebly interacting particles, possibly motivated by dark matter [516]. We also provide some discussion of relatively exotic signals that may be used in the search of weakly coupled dark sectors in Section 8.

## 5.1 Mono-photon [64]

An electron-positron collider allows the search for Dark Matter candidates using ISR photons in the reaction $e^+e^- \to \chi\chi\gamma$. This approach is complementary to mono-jet searches at hadron colliders, as the coupling to leptons is probed. In the following, a full simulation study is discussed for the case of higgsino pair production with an ISR photon at the 380 GeV stage of CLIC.

The event samples for the mono-photon analysis were generated in Whizard 2.3.1 [517] including ISR and the expected luminosity spectrum at 380 GeV CLIC. The full simulation and reconstruction software chain for the CLICdet detector model was used [2]. Particle identification, in particular for photons and electrons, was performed by the PandoraPFA [96, 518, 519] package. Pile-up from $\gamma\gamma \to$ hadrons interactions was overlaid to the physics events.

The following kinematic cuts on the final-state photons are imposed: $E_\gamma > 10\,\text{GeV}$ and $10° < \theta_\gamma < 170°$ to avoid fake photon candidates caused by electrons. The main SM background processes for this analysis are: $e^+e^- \to \nu\bar{\nu}\gamma$, which includes single-Z boson production with a hard ISR photon, and Bhabha scattering with an ISR photon, $e^+e^- \to e^+e^-\gamma$. The latter process is suppressed further by an electron veto in the forward calorimeters of the detector. The effective cross sections for the $e^+e^- \to \nu\bar{\nu}\gamma$ and $e^+e^- \to e^+e^-\gamma$ processes after selection are about 2500 fb and 100 fb, respectively. Other background processes are found to be negligible.

The upper cross section limit for the process $e^+e^- \to \chi\chi\gamma$ is calculated as a function of $m_\chi$ assuming an integrated luminosity of $1\,\text{ab}^{-1}$. For this purpose, photons in the energy range between 10 GeV and $E_{\gamma,max} = \sqrt{s}/2 - 2m_\chi^2/\sqrt{s}$ are considered, where $\sqrt{s}$ is the nominal centre-of-mass energy. The resulting limit at 95% C.L. is shown in Figure 71. A 380 GeV CLIC collider would be sensitive to the $e^+e^- \to \chi\chi\gamma$ process down to cross sections of about 2-3 fb in the mass range from the LEP limit of about 100 GeV almost up to 180 GeV. Higgsino pair production with an ISR photon could be excluded in the entire mass range.

---

[64]Based on a contribution by J.-J. Blaising.

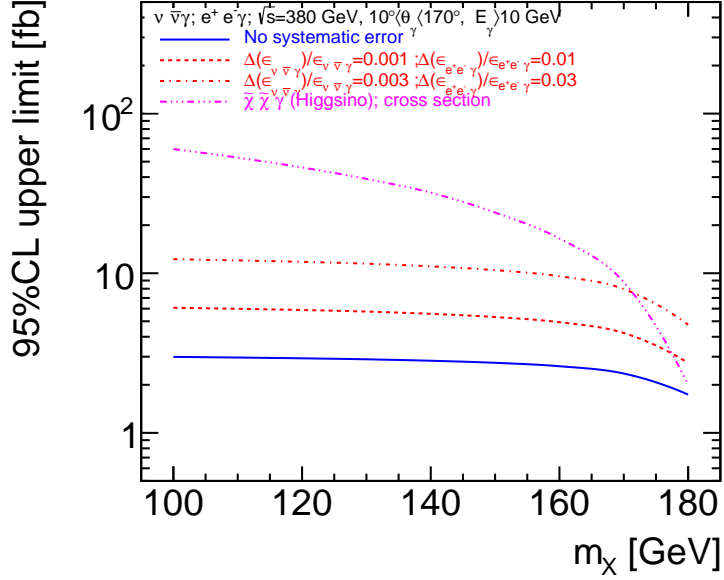

Figure 71: Upper limit (95% C.L.) on the $e^+e^- \to \chi\chi\gamma$ cross section at $\sqrt{s} = 380$ GeV as a function of $m_\chi$. The limit without including systematic uncertainties (blue) is compared to two different assumptions on the systematic uncertainty for the two main background processes (red). In addition, the cross section for higgsino pair production with an ISR photon is shown (magenta).

The measurement described here is sensitive to systematic uncertainties affecting the overall normalisation of the background processes. This is illustrated in Figure 71 for two example assumptions. The understanding of the electron tagging efficiency in the very forward direction is crucial in case of the $e^+e^- \to e^+e^-\gamma$ background. For this reason, larger systematic uncertainties are assumed compared to the $e^+e^- \to \nu\bar{\nu}\gamma$ process.

Once a signal is established, the photon energy distribution could be used to measure the mass of the Dark Matter candidate. For the case of a 120 GeV Higgsino particle, a precision on its mass of about 2 GeV is expected using the endpoint of the photon energy distribution.

## 5.2 Degenerate Higgsino DM stub tracks [65]

The higgsino is among the most compelling and elusive targets of searches for supersymmetric extensions of the Standard Model, being intimately connected to the naturalness of the weak scale, crucial for the success of gauge coupling unification, and an ideal WIMP dark matter candidate. Surprisingly few constraints exist on the higgsino when all other superpartners are decoupled, in which case the best limit from collider searches remains the combined LEP limit of $m_\chi > 92.4$ GeV. Naturalness considerations strongly motivate higgsinos in the range $m_\chi \lesssim$ TeV, while the observed dark matter relic abundance singles out $m_\chi \simeq 1.1$ TeV for thermal higgsino dark matter. High centre-of-mass lepton colliders such as CLIC provide one of the best avenues for probing higgsinos across this mass range.

In the limit that all other superpartners are decoupled, the higgsino multiplet consists of an SU(2)-doublet Dirac fermion. Electroweak symmetry breaking induces a small and calculable splitting between the charged and neutral components of the higgsino multiplet due to photon and $Z$ exchange, which renders the charged components $\chi^{\pm}$ slightly heavier than the neutral components $\chi^0_{1,2}$. This splitting takes the form [520]

$$\delta m = \frac{\alpha}{2} m_Z f(m_\chi^2/m_Z^2) \qquad (219)$$

where

$$f(x) = \frac{\sqrt{x}}{\pi} \int_0^1 dy (2-y) \log\left[1 + \frac{y}{x(1-y)^2}\right] \qquad (220)$$

and obtains the asymptotic value $\delta m = \frac{1}{2}\alpha m_Z$ in the limit $m_\chi^2 \gg m_Z^2$.

---

[65]*Based on a contribution by S. Alipour-Fard and N. Craig.*

When charged higgsinos $\chi^{\pm}$ are produced, they decay into the neutral $\chi^0$'s via charged current interactions, primarily via the two-body decay $\chi^{\pm} \to \chi^0 \pi^{\pm}$ with partial width

$$\Gamma(\chi^{\pm} \to \chi^0 \pi^{\pm}) = \frac{G_F^2}{\pi} \cos^2\theta_c f_\pi^2 \delta m^3 \sqrt{1 - \frac{m_\pi^2}{\delta m^2}}. \qquad (221)$$

where $\theta_c$ is the Cabibbo angle. The small splitting in Eq. (219) implies that the $\chi^{\pm}$ travel a macroscopic distance, of order one centimetre, before decaying into an invisible $\chi^0$ and soft Standard Model states. The charged higgsino lifetime makes it a particularly challenging target for LHC searches, as the "charged stub" left by the charged higgsino traversing the tracker is typically too short to be resolved by LHC detectors, leaving only a weak missing energy signature with unobservably soft decay products.

Here we study CLIC prospects for probing the pure higgsino, focusing on the production of charged higgsino pairs and the ensuing disappearing track signature. As a detailed treatment of backgrounds is beyond the scope of this study, we determine the signal efficiency for a variety of possible search strategies requiring one or more charged stubs per event, with or without hard photon ISR.

Our analysis strategy is loosely based on existing searches for pure higgsinos or winos at the LHC, requiring one or more charged stubs in each event with the further option of requiring hard photon ISR. We pursue two possible strategies involving only charged stub requirements: an optimistic strategy requiring at least one identifiable charged stub per signal event, and a more conservative strategy requiring two charged stubs. We also illustrate six possible strategies involving a hard ISR photon of energy $E_\gamma > 50, 100$, or $200$ GeV in addition to $\geq 1$ or exactly 2 charged stubs. For completeness, we study these strategies for three CLIC configurations: 500 fb$^{-1}$ at $\sqrt{s} = 380$ GeV, 1500 fb$^{-1}$ at $\sqrt{s} = 1.5$ TeV, and 3000 fb$^{-1}$ at $\sqrt{s} = 3$ TeV. We neglect beam polarization effects, which do not significantly impact the signal efficiency.

### 5.2.1 Charged stub-only analysis

The acceptance for a search requiring only one or more charged stubs can be determined analytically. To do so, we first compute the leading-order unpolarized differential cross section $d\sigma/d\cos\theta$ as a function of polar angle $\theta$ for $e^+e^- \to \chi^+\chi^-$ (see e.g. [521]). Once produced, each $\chi^{\pm}$ travels some distance before decaying, leading to a distinctive charged stub in the tracker. In order to be counted as a charged stub, the $\chi^{\pm}$ must traverse at least 4 layers of the CLIC tracker before decaying.

As such, we define $d_{\min}(\theta)$ as the minimum distance a single $\chi^{\pm}$ must travel in the detector before decaying in order to register 4 hits in the CLIC tracker, thereby enabling identification of the charged stub. This corresponds to the following requirements [522]:

$$d_{\min}(\theta) = \begin{cases} \frac{4.4\,\text{cm}}{\sin\theta} & 19° < \theta < 90° \\ \frac{22\,\text{cm}}{\cos\theta} & 13° < \theta < 19° \\ \frac{29\,\text{cm}}{\cos\theta} & 8° < \theta < 13° \end{cases}, \qquad (222)$$

with the usual symmetry about $\theta = 90°$. Particles produced at polar angles $\theta < 8°$ are assumed to exit the detector without registering hits.

The probability that a single $\chi^\pm$ of three-momentum $\vec{p}_\chi$ travels a distance $d_{\min}$ or greater in the detector is simply the survival probability

$$P_{\rm s}(d_{\min}) = e^{-m_\chi d_{\min} \Gamma_\chi / |\vec{p}_\chi|}\,, \tag{223}$$

where $\Gamma_\chi$ is given in Eq. (221). From this we can compute the number of events $N_{\rm evts}^{1-{\rm stub}}$ with *at least* one identifiable charged stub, or the number of events $N_{\rm evts}^{2-{\rm stub}}$ with exactly two identifiable charged stubs, at a given centre-of-mass energy $\sqrt{s}$ and integrated luminosity $\mathcal{L}_{\rm int}$ by integrating the differential cross section against the appropriate combination of survival probabilities over all polar angles:

$$N_{\rm evts}^{1-{\rm stub}} = \mathcal{L}_{\rm int} \times \int_{-1}^{1} \frac{d\sigma(e^+e^- \to \chi^+\chi^-)}{d\cos\theta} \left[2P_{\rm s}(d_{\min}) - P_{\rm s}(d_{\min})^2\right] d\cos\theta\,, \tag{224}$$

$$N_{\rm evts}^{2-{\rm stub}} = \mathcal{L}_{\rm int} \times \int_{-1}^{1} \frac{d\sigma(e^+e^- \to \chi^+\chi^-)}{d\cos\theta} P_{\rm s}(d_{\min})^2\, d\cos\theta\,. \tag{225}$$

This treatment does not account for possible additional efficiency factors associated with the identification of charged stubs beyond the requirement that the stub traverse 4 tracker layers before disappearing.

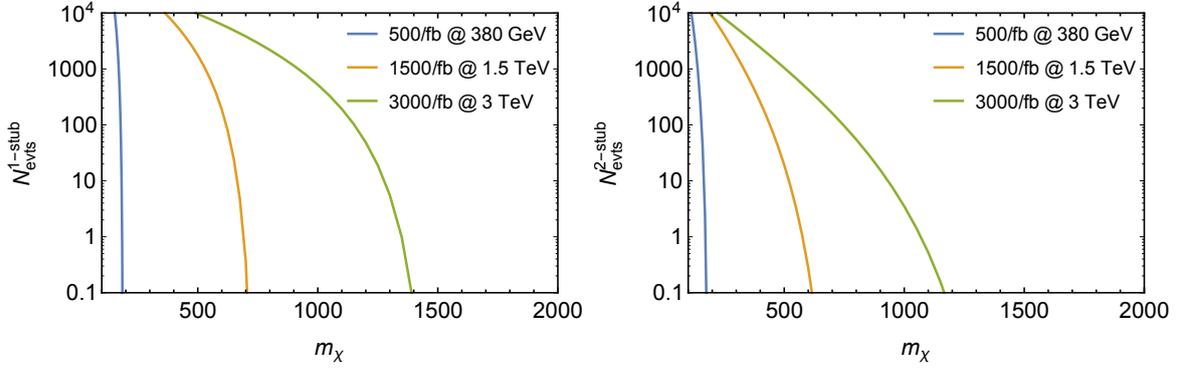

Figure 72: Number of expected signal events at $\sqrt{s} = 380$ GeV, 1.5 TeV, and 3 TeV for the $\geq 1$ charged stub selection (left) and $= 2$ charged stub selection (right) as a function of the charged higgsino mass $m_\chi$.

We validate our analytic treatment by simulating the process $e^+e^- \to \chi^+\chi^-$ at leading order at centre-of-mass energies $\sqrt{s} = 0.380, 1.5, 3.0$ TeV using MadGraph 5. We simulate 50,000 events at each of $m_\chi = 100 - 180$ GeV in 10 GeV intervals at $\sqrt{s} = 380$ GeV, $m_\chi = 100 - 800$ GeV in 100 GeV intervals at $\sqrt{s} = 1.5$ TeV, and $m_\chi = 100 - 1600$ GeV in 100 GeV intervals at $\sqrt{s} = 3$ TeV. Each $\chi^\pm$ is then decayed by drawing randomly from the appropriate distribution of lifetimes given by Eq. (221), and counted as a charged stub if it travels a distance greater than the corresponding $d_{\min}$ before decaying. The number of events with at least one charged stub, or with exactly two charged stubs, is then compared to the analytic expectation. We find excellent agreement between the analytic result and Monte Carlo simulation.

The results of the charged stub-only analysis are shown in Figure 72 for each of the three CLIC operating configurations.

### 5.2.2 *Charged stub + photon analysis*

Depending on the results of complete background characterization, background reduction may require the imposition of additional cuts beyond the appearance of charged stubs. One possible strategy is to require sufficiently hard initial state radiation (ISR) in conjunction with one or more charged stubs. To characterize the impact of possible cuts on the energy of a hard ISR photon, we extend the above analysis by considering $e^+e^- \to \chi^+\chi^- + \gamma$, counting signal events with a hard ISR photon of energy $E_\gamma > E_\gamma^{\min}$ in addition to one or more identifiable charged stubs.

We perform this analysis solely via Monte Carlo by simulating the process $e^+e^- \to \chi^+\chi^- + \gamma$ at leading order at center-of-mass energies $\sqrt{s} = 0.380, 1.5, 3.0$ TeV using MadGraph 5. We again simulate 50,000 events at each of $m_\chi = 100 - 180$ GeV in 10 GeV intervals at $\sqrt{s} = 380$ GeV, $m_\chi = 100 - 800$ GeV in 100 GeV intervals at $\sqrt{s} = 1.5$ TeV, and $m_\chi = 100 - 1600$ GeV in 100 GeV intervals at $\sqrt{s} = 3$ TeV. Each $\chi^\pm$ is then decayed by drawing randomly from the appropriate distribution of lifetimes given by Eq. (221), and counted as a charged stub if it travels a distance greater than the corresponding $d_{\min}$ before decaying. We then compute the number of events with at least one charged stub, or with exactly two charged stubs, *and* a photon of energy $E_\gamma > E_\gamma^{\min}$ produced in a range of polar angles $10° < \theta < 170°$. We illustrate the effects of three possible ISR cuts, corresponding to $E_\gamma^{\min} = 50, 100$, and 200 GeV, respectively, for a total of six charged stub + photon search strategies. As with the charged stub-only analysis, we do not account for possible additional efficiency factors associated with the identification of charged stubs beyond the requirement that the stub traverse 4 tracker layers before disappearing, nor do we account for the finite detection efficiency ($\sim 93\%$) of photons produced between $10° < \theta < 170°$ at CLIC.

The results of the charged stub + photon analysis are shown in Figure 73 for each of the three CLIC operating configurations and the six possible combinations of stub requirements and photon ISR energies.

*5.2.3 Discussion & conclusions*

Assuming a given set of selection requirements is sufficient to attain zero expected background in the signal region, the 95% exclusion limit can be obtained for each analysis by requiring $N_{\text{evts}} = 3$. The corresponding 95% exclusion reach is illustrated in Figure 74 for each of the eight analysis strategies discussed above, at each of the three CLIC operating configurations.

All analysis strategies are capable of covering a significant range of higgsino masses, well in excess of current collider limits. The most optimistic analysis strategies – namely, those requiring at least one charged stub, or at least one charged stub in conjunction with an ISR photon of energy $> 50$ GeV or $> 100$ GeV – are capable of covering higgsino masses up to the thermal dark matter target of $m_\chi \simeq 1.1$ TeV. This demonstrates the potential for CLIC to cover a highly motivated range of supersymmetric parameter space using a search for charged stubs, though detailed study of backgrounds is required in order to determine whether the zero-background assumption is justified in each analysis strategy.

In order to estimate the effect of non-negligible backgrounds we present also predictions for isolines of number of events in the acceptance discussed above for the strategy requiring at least one charged stub and for the strategy requiring two charged stubs. In addition we consider the effect of possible variations of the higgsino lifetime compared with the prediction of Eq. (219), which may arise if the higgsino is not exactly a pure state. We present iso-lines for 3 and 30 events in the plane lifetime vs. Higgsino mass in Figure 75. For the 1-stub strategy it seems possible to have up to about 30 events in the acceptance, hence CLIC 3 TeV should be able to probe the thermal relic higgsino dark matter even in the presence of some level of background.

### 5.3 Dark matter in loops: minimal dark matter and milli-charged [66]

Many new physics scenarios, motivated e.g. by dark matter (DM), feature new states charged under $SU(2)_L \times U(1)_Y$ in which the lightest particle in the multiplet is stable and neutral. If this is the case, direct searches at hadron colliders are notoriously difficult, and current LHC bounds, based on mono-X searches and disappearing tracks, remain limited to the 200–500 GeV domain, depending on the gauge quantum numbers of the new multiplets (see e.g. [523, 524] and Refs. therein). On the other hand, lepton colliders such as CLIC offer the possibility to indirectly probe the existence of those states via precision measurements, even if the beam energy is below the production threshold of the new particles.

---

[66]*Based on a contribution by L. Di Luzio, R. Gröber, and G. Panico.*

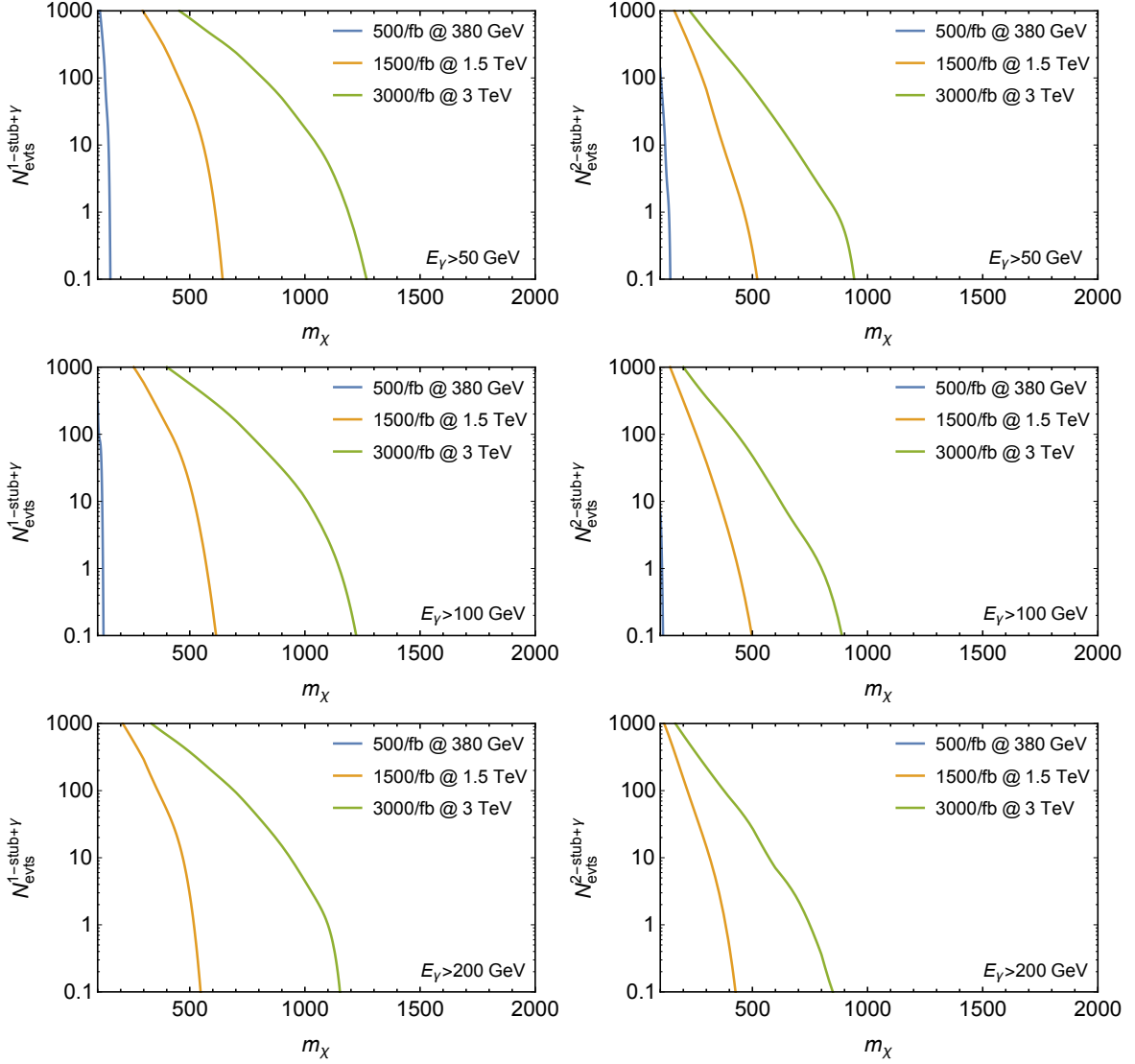

Figure 73: Number of expected signal events at $\sqrt{s} = 380$ GeV, 1.5 TeV, and 3 TeV for the $\geq 1$ charged stub + photon selection (left) and $= 2$ charged stub + photon selection (right) as a function of the charged higgsino mass $m_\chi$. The cut on photon ISR energy is $> 50$ GeV (top), $> 100$ GeV (middle), $> 200$ GeV (bottom).

After reviewing the physics case for new states with electroweak (EW) quantum numbers in Sections 5.3.1-5.3.3, we consider in Section 5.3.4 the universal loop corrections to the SM process $e^+e^- \to f\bar{f}$ (where $f$ denotes a standard model (SM) fermion), and provide in turn the projected exclusion limits of CLIC for various benchmark models. Our analysis follows closely the one of Ref. [525].

New states $\chi \sim (1, n, Y)$ charged under $SU(2)_L \times U(1)_Y$, whose lightest particle (LP) in the $n$-dimensional multiplet is stable and neutral appear in many motivated beyond-the-SM (BSM) scenarios. The EW sector of supersymmetry (SUSY) comprising the wino/higgsino system is maybe one of the most compelling cases for new EW multiplets, although there are also other frameworks (mainly related to DM) which motivate the existence of new particles with $n > 3$. In the following, we briefly review some of them.

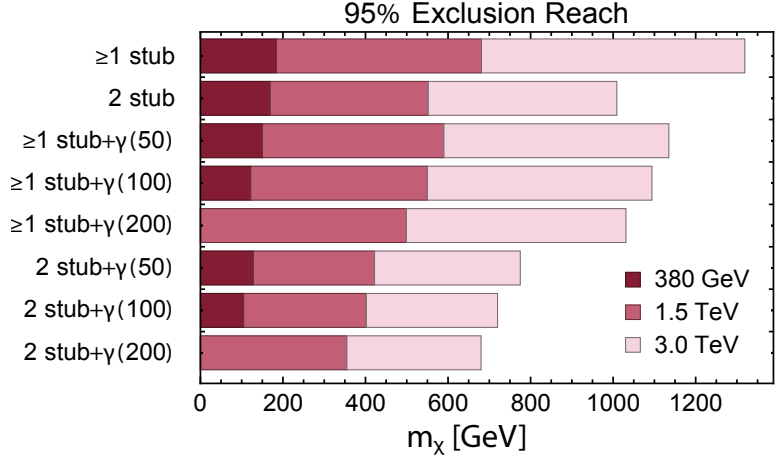

Figure 74: The 95% CLIC exclusion reach for pure higgsinos in each of the eight analysis strategies, assuming zero background in each analysis.

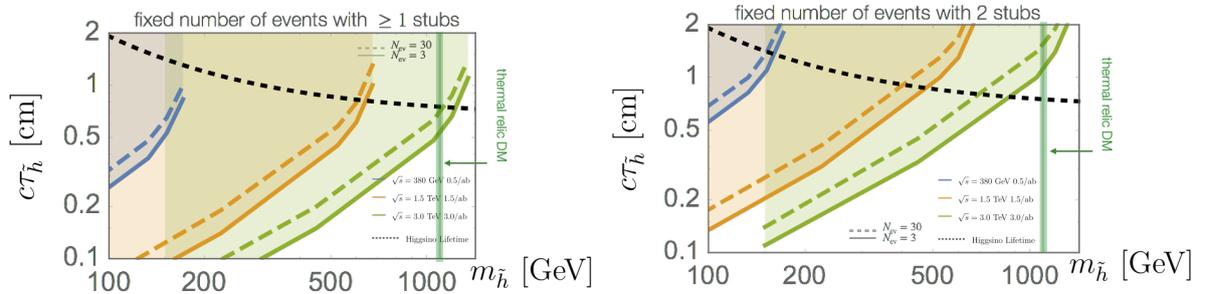

Figure 75: Contours in the place lifetime-mass for N=3 (solid) and N=30 (dashed) higgsino events in the acceptance defined by Eq. (222) at the three stages of CLIC: 380 GeV 0.5 ab$^{-1}$ (blue), 1.5 TeV 1.5 ab$^{-1}$ (yellow), and 3.0 TeV 3 ab$^{-1}$ (green).

### 5.3.1 Minimal (milli-charged) dark matter

The idea behind Minimal Dark Matter (MDM) [526] is to introduce a single EW multiplet $\chi$ which is accidentally stable at the renormalizable level due to the SM gauge symmetry. One further assumes $Y = 0$ (to avoid direct detection bounds from $Z$ exchange) and that the lightest particle (LP) in the multiplet is neutral. This is actually a prediction if the mass splitting is purely radiative as in the case of fermions, while scalars can receive a tree-level splitting from the scalar potential which is assumed to be sub-leading. The contribution to the relic density is then completely fixed by known EW gauge interactions and the mass of the new state $m_\chi$, thus making the framework extremely predictive. If one further requires that the theory remains weakly coupled up to the Planck scale and that $d < 6$ $\chi$-decay operators are not allowed (otherwise they would lead to a too fast $\chi$ decay, even with a Planck scale cutoff), this leads to one single option: the Majorana fermion representation $(1, 5, 0)_{\rm MF}$.[67] In the following, we use the labels RS, CS, MF, and DF to denote a real scalar, complex scalar, Majorana fermion, and Dirac fermion representation, respectively.

The MDM framework was extended in [528] to contemplate the possibility of a milli-charge $\epsilon \ll 1$. Bounds from DM direct detection imply $\epsilon \lesssim 10^{-9}$. The milli-charge has hence no bearing on collider physics, but it ensures the (exact) stability of the LP in the EW multiplet. The various MDM candidates (including for completeness also the wino-like DM $(1,3,0)_{\rm MF}$ which requires a stabilization mechanism beyond the SM gauge symmetry) are summarized in Table 38, together with their thermal mass saturating the DM relic density and the projected reach of the third CLIC stage ($\sqrt{s} = 3$ TeV and $L = 4$ ab$^{-1}$ for $P(e^-=-80\%)$. The details of the analysis are presented in Section 5.3.4.

---

[67]Originally also the real scalar representation $(1, 7, 0)_{\rm RS}$ was included in the list, but it was shown later in [527] that a previously overlooked $d = 5$ operator leads to a loop-induced decay of the neutral component in $\chi$, whose lifetime is shorter that the age of the Universe.

A notable feature of the milli-charged scenario is that the contribution of the complex multiplet to the relic density is doubled compared to the case of a single real component (thus making the thermal mass roughly a factor $\sqrt{2}$ smaller). On the other hand, the degrees of freedom are also doubled, thus improving the indirect testability of those scenarios via EW precision tests at lepton colliders. It turns out indeed that the hypothesis of $(1,3,\epsilon)_{\rm DF}$, see red-shaded line in Table 38, comprising the whole DM relic density can be fully tested at CLIC-3, while we find no sensitivity to the state $(1,3,\epsilon)_{\rm CS}$ for masses above the kinematical threshold of pair production.[68] For all the other cases the thermal mass lie well above the reach of CLIC Stage 3.

Table 38: MDM candidates, together with the corresponding masses saturating the DM relic density and the projected 95% CL exclusion limits from EW precision tests at the third CLIC stage ($\sqrt{s} = 3$ TeV, $L = 4$ ab$^{-1}$, $P_{e^-} = -80\%$, $P_{e^+} = 0$, and 0.3% systematic error). The exclusions refer only to the cases where $m_\chi > 1.5$ TeV. For masses below the threshold for pair production $m_\chi < \sqrt{s}/2$ the bound is characterized by a non-trivial profile – see Section 5.3.4 for details. The thermal masses are extracted from Ref. [528] ($\epsilon \neq 0$ cases) and Ref. [529] ($\epsilon = 0$ cases). A conservative 10% theoretical uncertainty is understood, originating from the inclusion of non-perturbative effects such as Sommerfeld enhancement and bound state formation.

| $\chi$ | $m_\chi^{\rm (DM)}$ [TeV] | $m_\chi^{\rm (3\,TeV CLIC)}$ [TeV] |
|---|---|---|
| $(1,2,1/2)_{\rm DF}$ | 1.1 | 1.5 |
| $(1,3,\epsilon)_{\rm CS}$ | 1.6 | - |
| $(1,3,\epsilon)_{\rm DF}$ | 2.0 | 2.0 |
| $(1,3,0)_{\rm MF}$ | 2.8 | 1.7 |
| $(1,5,\epsilon)_{\rm CS}$ | 6.6 | 1.6 |
| $(1,5,\epsilon)_{\rm DF}$ | 6.6 | 4.1 |
| $(1,5,0)_{\rm MF}$ | 11 | 3.0 |
| $(1,7,\epsilon)_{\rm CS}$ | 16 | 2.5 |
| $(1,7,\epsilon)_{\rm DF}$ | 16 | 6.7 |

#### 5.3.2 Accidental matter

From a more phenomenological point of view, one could ask the following question [527]: *Which extensions of the SM particle content with masses close to the EW scale (i) automatically preserve the accidental and approximate symmetry structure of the SM, (ii) are cosmologically viable, and (iii) form consistent EFTs with a cut-off scale as high as $10^{15}$ GeV (as suggested e.g. by neutrino masses)?* Those SM extensions are simply motivated by the fact that they can be discovered at high-energy particle colliders, without being constrained by other indirect probes such a flavour and baryon/lepton number violating process. A finite list of cases can be selected (see [527] for details), and among those a subset features a neutral LP in the EW multiplet: $(1,5,0)_{\rm RS}$, $(1,5,1)_{\rm CS}$, $(1,5,2)_{\rm CS}$, $(1,7,0)_{\rm RS}$, $(1,4,3/2)_{\rm DF}$, $(1,5,0)_{\rm MF}$, which are hence a natural target for our study. It turns out that the value of the hypercharge, unless exotically large, plays a subleading role for the extraction of the bound. Hence, instead of reporting explicitly the projected reach of CLIC for all the accidental matter candidates, we refer directly to Figure 76.

---

[68]Given a 10% uncertainty on the thermal masses, the DM hypothesis for the CS triplet can be potentially explored in direct searches at CLIC.

### 5.3.3 split-SUSY

A full analysis of EW precision tests of SUSY at lepton colliders goes beyond our scopes, since it would require the inclusion of non-universal corrections to SM fermion vertices (see e.g. [530] for a LEP analysis in this direction). On the other hand, in the motivated split-SUSY [473, 491] limit, where all the scalar partners are decoupled, the radiative corrections due to the gaugino/higgsino system are universal. In our analysis we neglect the mass splitting within the EW multiplets (namely we work in the regime $S \simeq T \simeq 0$), which is a good approximation in the high-energy limit probed by 3 TeV CLIC. Hence, our bounds can be eventually reinterpreted for the split-SUSY scenario as well.

### 5.3.4 Electroweak precision tests at CLIC

At lepton colliders one can study the modifications of the process $e^+e^- \to f\bar{f}$, where $f$ is a SM fermion, due to the presence of a new state $\chi \sim (1, n, Y)$ that modifies the EW gauge boson propagators at one loop. These effects can be parametrized via the inclusion of form factors in the effective Lagrangian [525, 531]

$$\mathcal{L}_{\text{eff}} = \mathcal{L}_{\text{SM}} + \frac{g^2 C_{WW}^{\text{eff}}}{8} W_{\mu\nu}^a \Pi(-D^2/m_\chi^2) W^{a\mu\nu} + \frac{g'^2 C_{BB}^{\text{eff}}}{8} B_{\mu\nu} \Pi(-\partial^2/m_\chi^2) B^{\mu\nu}, \quad (226)$$

where $C_{WW}^{\text{eff}} = \kappa(n^3 - n)/6$, $C_{BB}^{\text{eff}} = \kappa 2nY^2$, and $\kappa = 1/2, 1, 4, 8$, respectively for $\chi$ being a RS, CS, MF, DF. We further assume that $\chi$ does not interact at the renormalizable level with the SM matter fields and that the mass splitting within the $n$-plet is negligible.[69] If that is the case, $\chi$ only contributes to the transverse part of the gauge boson propagators and the renormalized form factors are (in the $\overline{\text{MS}}$ scheme and for the scale choice $\mu = m_\chi$) [532]

$$\Pi(x) = \begin{cases} -\dfrac{8(x-3) + 3x\left(\frac{x-4}{x}\right)^{3/2} \log\left(\frac{1}{2}\left(\left(\sqrt{\frac{x-4}{x}} - 1\right)x + 2\right)\right)}{144\pi^2 x} & \text{(scalars)} \\ -\dfrac{12 + 5x + 3\sqrt{\frac{x-4}{x}}(x+2) \log\left(\frac{1}{2}\left(\left(\sqrt{\frac{x-4}{x}} - 1\right)x + 2\right)\right)}{288\pi^2 x} & \text{(fermions)} \end{cases}, \quad (227)$$

where $x = s/m^2$, and $\sqrt{s}$ is the external momentum of the gauge boson propagator. In the effective field theory (EFT) limit, $x \ll 1$, the expanded form factor is $\Pi(x) \simeq -x/(480\pi^2)$, both for scalars and fermions. Since $\Pi(0) = 0$ there is no contribution to the oblique parameters $S$, $T$, $U$ [61], while $W$ and $Y$ [62], defined via the $d = 6$ operators $-\frac{W}{4m_W^2}\left(D_\rho W_{\mu\nu}^a\right)^2$ and $-\frac{Y}{4m_W^2}\left(\partial_\rho B_{\mu\nu}\right)^2$, are given by[70]

$$W = \frac{g^2 C_{WW}^{\text{eff}}}{960\pi^2} \frac{m_W^2}{m_\chi^2}, \qquad Y = \frac{g'^2 C_{BB}^{\text{eff}}}{960\pi^2} \frac{m_W^2}{m_\chi^2}. \quad (228)$$

For $x \simeq 1$ the EFT breaks down and hence the full momentum dependence of the form factor must be taken into account, while for $x = 4$ the form factor develops an imaginary part corresponding to the pair-production threshold. It should be stressed that for weakly coupled forms of new physics the energy reach of $W$ and $Y$ is screened by the weak coupling, so that it becomes important to include the full kinematical dependence of the form factors even below the pair-production threshold.

### 5.3.5 Description of the analysis

The $\chi$-induced corrections to the polarized SM amplitude $e^+e^- \to f\bar{f}$ can be obtained from Eq. (226). We refer to [525] for the relevant formulae. Note that since the radiative corrections are universal, the main effect is due to the interference with the SM amplitude. Following [525] we perform a binned likelihood analysis on the differential cross section of the process $e^+e^- \to f\bar{f}$ with respect to the cosine of the scattering angle $\theta$. In particular, we divide the latter into ten uniform intervals for $\cos\theta \in [-0.95, 0.95]$. For the final states we assume the following detection efficiencies: 100% for leptons, 80% for b-jets and 50% for c-jets. We then define a $\chi^2$ function

$$\chi^2 = \sum_{i=1}^{10} \frac{\left(N_i^{\text{SM+BSM}} - N_i^{\text{SM}}\right)^2}{N_i^{\text{SM}} + \left(\epsilon_i N_i^{\text{SM}}\right)^2}, \qquad (229)$$

where $N_i^{\text{SM+BSM}}$ ($N_i^{\text{SM}}$) is the expected number of events with (without) the $\chi$ contribution. The denominator of the $\chi^2$ includes both a statistical and a systematic error, the latter parametrized by $\epsilon_i$, which we assume to take values between 0 (pure statistical error) and 1%. The polarization of the incoming electrons and positrons can be used in order to enhance the cross section and effectively increase the integrated luminosity. The cross section of a generically polarized $e^+e^-$ beam is defined in terms of the polarization fractions $P_{e^-}$ and $P_{e^+}$, by

$$\begin{aligned}\sigma_{P_{e^-}P_{e^+}} = \frac{1}{4} [&(1+P_{e^-})(1+P_{e^+})\sigma_{RR} + (1-P_{e^-})(1-P_{e^+})\sigma_{LL} \\ &+ (1+P_{e^-})(1-P_{e^+})\sigma_{RL} + (1-P_{e^-})(1+P_{e^+})\sigma_{LR}], \end{aligned} \qquad (230)$$

where $\sigma_{LR}$ stands for instance for the cross section if the $e^-$-beam has complete left-handed polarization ($P_{e^-} = -1$) and the $e^+$-beam has complete right-handed polarization ($P_{e^+} = +1$). In the baseline CLIC design [3], the electron beam can be polarised up to $\pm 80\%$. There is also the possibility of positron polarisation at a lower level, although positron polarisation is not part of the baseline CLIC design.

---

[69]These assumptions are automatically satisfied for fermions with $n > 3$, while in the case of scalars they require that potential terms allowed by gauge invariance are subleading.

[70]See Section 2.6 for an interpretation of the result on W and Y in other models with flavor universal new physics effects.

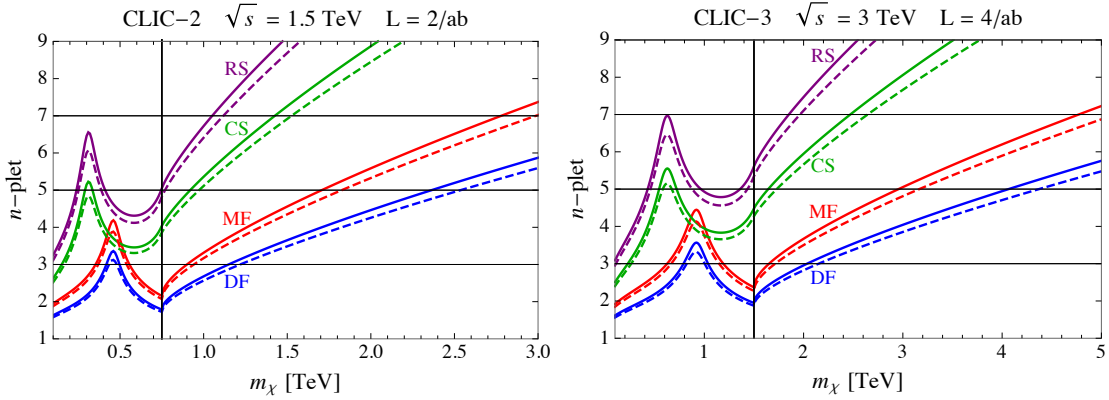

Figure 76: 95% CL exclusion limits for CLIC-2 (left panel) and CLIC-3 (right panel), obtained by combining the $e/\mu/b/c$ channels with 0.3% systematic error and polarization fractions $(P_{e^-}, P_{e^+}) = (-80\%, 0)$ [full lines] and $(P_{e^-}, P_{e^+}) = (-80\%, +30\%)$ [dashed lines].

### 5.3.6 Results

Our main results, see also Ref. [532], are displayed in Figure 76 where we show the 95% exclusion limits in the plane $(m_\chi, n)$ for different Lorentz representations (RS, CS, MF, DF) and for the two late stages of CLIC, denoted respectively CLIC-2 ($\sqrt{s} = 1.5$ TeV, $L = 2$ ab$^{-1}$) and CLIC-3 ($\sqrt{s} = 3$ TeV, $L = 4$ ab$^{-1}$). To obtain these exclusions we have combined the $e/\mu/b/c$ channels [71] assuming a systematic error of 0.3% (cf. Figure 77) and polarization fractions $P_{e^-} = -80\%$ and $P_{e^+} = 0$.

---

[71]For top-quark final states, results can be found in Ref. [10] and are considered in Section 2.6 as well.

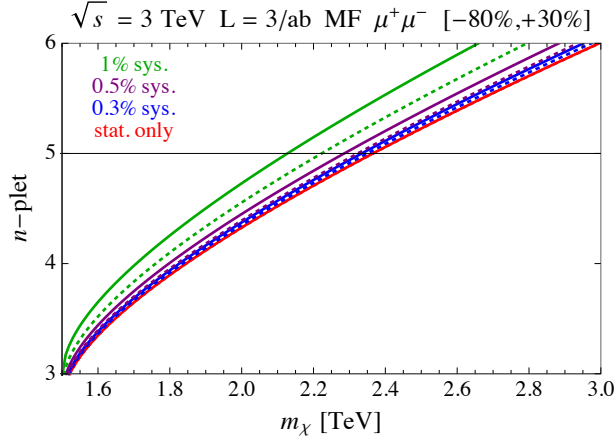

Figure 77: Impact of systematic error: this plot shows e.g. that the 0.3% systematic error line is almost indistinguishable from the "pure statistical" one. We also superimpose (dotted lines) the exclusions obtained by augmenting the number of bins from 10 to 20 (same colour code for the error treatment as before). We see that increasing the numbers of bins helps for larger systematic errors, but does not matter much for e.g. 0.3% systematics. Hence, in the following we stick to 0.1% systematics with 10 bins.

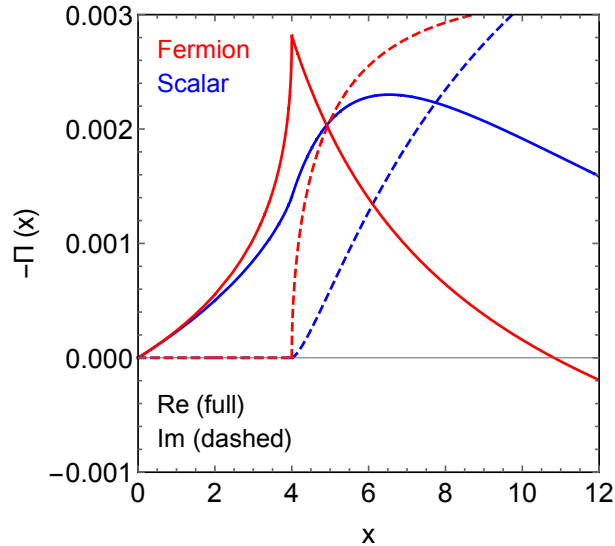

Figure 78: Profile of the form factors describing the modification of the electroweak gauge boson propagators (cf. Eq. (226)) due to $\chi$. The shape of the real parts (which gives the leading contribution to the interference term with the SM amplitude), are useful in order to understand the corresponding bounds in Figure 76. E.g. the cusp shape at threshold for the case of fermions and the resonance-like behaviour above threshold (both for fermions and scalars) corresponding to the zero of the real part of the form factor.

The vertical black line in both plots denotes the kinematical threshold for pair-production $\sqrt{s}/2$. In the region below threshold (on the right side of the vertical black line) the bound on the mass grows with the dimensionality of the multiplet and eventually enters the EFT regime for $m_\chi \gg \sqrt{s}/2$. The bounds in the region above threshold (on the left side of the vertical black line) have some non-trivial features which can be understood by following the shape of the real part of the form factor above threshold (cf. Figure 78).

## 5.4 Co-annihilation [72]

The nature of dark matter (DM) remains one of the most pressing issues in the field of high-energy physics. The thermal WIMP persists as one of the most studied solutions to this problem since it leads to detectable signatures at direct and indirect detection experiments. In the past decades there has been a significant experimental effort in searches for WIMP DM. These experiments have, however, not seen a conclusive signal, providing exclusion limits on models of DM.

At present there is a plethora of models beyond the Standard Model (SM) to explain DM. Consequently, in finding experimental constraints it would be ideal to be as model-independent as possible. This can be achieved by studying effective operators between the DM and the SM particles [533–546]. Nevertheless, the EFT approximation breaks down when studying collider signatures since the cut-off of the effective field theory may not be larger than the collider energy scale or the dark sector often requires a new mediator particle other than the DM which may dramatically alter the collider signature itself [547–549]. In this context the scientific community has proposed the use of *simplified models*, where the minimum particle content is assumed for the dark sector. In particular, it is common to introduce the mediator particle connecting visible and dark sectors in addition to the DM [550–553]. The simplified model approach also gives a framework to study the complementarity between different experiments [554, 555].

The simplified models with a mediator particle can be classified by its spin and quantum numbers, and they offer a rich phenomenology. However, not all features that may be present in more complete models are implemented within this framework. The primary example is the *coannihilation mechanism*, in which the DM ($\chi$) comes with an almost mass degenerate coannihilation partner (CAP, $\eta$) and the DM relic abundance is determined not by the $\chi$-$\chi$ scattering but mainly by the $\eta$-$\eta$ and $\eta$-$\chi$ scattering. This mechanism appears in various extensions of the SM, such as supersymmetric and extra dimensional models, and does not require a mediator particle. In particular, the stau-coannihilation ($\eta = \tilde{\tau}$) is often found in phenomenological scans of the MSSM parameter space [556, 557], since the lightest stau tends to be the next-to-the-lightest SUSY particle after the neutralino DM.

Phenomenology of the coannihilation mechanism is quite different from that in models with mediators. In the latter, the interaction dictating thermal freeze-out connects the DM and SM particles and severe constraints are placed by the direct/indirect detection experiments. On the other hand, if the coannihilation mechanism is operative, the thermal freeze-out is controlled by the interaction between the CAP and SM particles, and the direct/indirect detection constraints can easily be avoided. LHC phenomenology is also very different. Unlike mediator particles, the coannihilation partner decays into the DM and SM particles very softly, and the signal is easily swamped by the overwhelming background. Therefore, the LHC can do very little on the coannihilation DM models in general. The only exception is the extreme case where the mass splitting between the CAP and DM is smaller than the tau-lepton mass, 1.777 GeV. In such a case, the coannihilation partner may have a detector-scale lifetime and its production can be constrained at the LHC by looking for highly ionizing and/or slowly moving anomalous tracks. This possibility has recently been studied in Ref. [558].

In this report we construct DM simplified models with tau-philic coannihilation partners and study them in light of the future Compact Linear Collider (CLIC). We demonstrate that, unlike the LHC, CLIC can resolve soft tau-lepton signature and offers the ideal opportunity to explore this class of models. Even though CLIC provides clean final states for the signal, the soft tau background produced by bremsstrahlung photon collisions, $\gamma\gamma \to \tau^+\tau^-$, is significant. We take this effect into account and show how well CLIC can constrain the bulk of the model parameter space in each stage of the experiment.

---

[72]*Based on a contribution by A. Plascencia and K. Sakurai.*

Table 39: Simplified Models of DM with a colourless coannihilation partner (CAP)

| | **Model-1** | | |
|---|---|---|---|
| Component | Field | Charge | Interaction |
| DM | Majorana fermion ($\chi$) | $Y = 0$ | $\phi^*(\chi\tau_R) + \text{h.c.}$ |
| CAP | Complex scalar ($\phi$) | $Y = -1$ | |

| | **Model-2** | | |
|---|---|---|---|
| Component | Field | Charge | Interaction |
| DM | Real scalar ($S$) | $Y = 0$ | $S(\overline{\Psi}P_R\tau) + \text{h.c.}$ |
| CAP | Dirac fermion ($\Psi$) | $Y = -1$ | |

| | **Model-3** | | |
|---|---|---|---|
| Component | Field | Charge | Interaction |
| DM | Vector ($V_\mu$) | $Y = 0$ | $V_\mu(\overline{\Psi}\gamma^\mu P_R\tau) + \text{h.c.}$ |
| CAP | Dirac fermion ($\Psi$) | $Y = -1$ | |

### 5.4.1 Simplified models for tau-philic dark matter

Our simplified models consist of two new degrees of freedom: the gauge singlet DM particle, $\chi$, and the charged coannihilation partner (CAP), $\eta^{(\pm)}$. We assign these particles the odd $Z_2$ charge to ensure the stability of the DM. Inspired by the supersymmetric stau-coannihilation, we assume the DM and CAP interact together with a tau-lepton. The interaction term is given by

$$\mathcal{L} \supset g_{\text{DM}}\,\chi\,\eta\,\bar{\tau}_R + \text{h.c.}, \tag{231}$$

where $g_{\text{DM}}$ is the dark sector coupling which we take to be real. The gauge invariance forces $\eta$ to be singlet under $SU(3)_c$ and $SU(2)_L$ and have the hypercharge $-1$ as for the right-handed tau. Restricting the particles not to have spins higher than 1, we consider three possible spin assignments for the $(\chi, \eta)$ pair:[73] $(\frac{1}{2}, 0)$, $(0, \frac{1}{2})$ and $(1, \frac{1}{2})$. We refer to them as Model-1, 2 and 3, respectively. Those models together with our notation are summarised in Table 39.

*Coannihilation*: The DM annihilation channel in our simplified model is unique: $\chi\chi \to \tau^+\tau^-$. In Model-1 (-2) where the DM is a Majorana fermion $\chi$ (a real scalar $S$), this channel is suppressed. The initial state in both cases forms a spin-0 state (due to the Pauli blocking in the Majorana case). To conserve the angular momentum, the $\tau^+\tau^-$ pair in the final state must have the opposite chiralities in $s$-wave, rendering the contribution proportional to $m_\tau^2$ (chiral suppression). The dominant contribution then comes from the $p$-wave for a Majorana DM and $d$-wave for a scalar DM, which are suppressed by the factor $v^2$ and $v^4$, respectively, where $v$ is the average velocity of the annihilating DM particles.

As is well known, the DM relic abundance scales as

$$\Omega_{\text{DM}}h^2 \propto \langle\sigma_{\text{eff}}\,v\rangle^{-1}, \tag{232}$$

where $\langle\sigma_{\text{eff}}\,v\rangle$ is the thermal average of the effective annihilation cross-section that is given by [559]

$$\sigma_{\text{eff}}\,v = \frac{1}{(g_\chi + \bar{g}_\eta)^2}\left[g_\chi^2\cdot\sigma(\chi\chi \to \tau^+\tau^-) + \right.$$

---

[73] An additional potential assignment $(\frac{1}{2}, 1)$ leads to $\eta$ being an electrically charged vector boson which prevent us from finding an $SU(2)_L \times U(1)_Y$ invariant operator for Eq. (231). We will therefore not consider this option further.

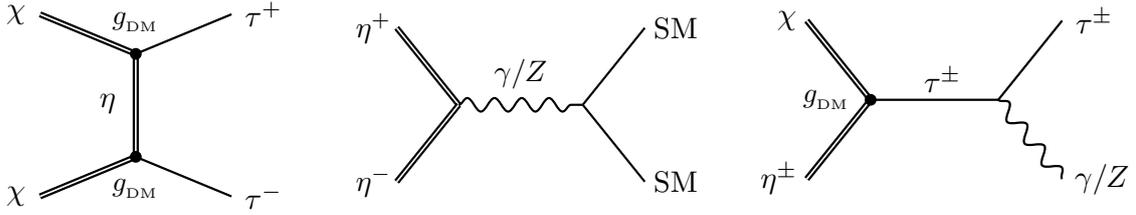

Figure 79: Feynman diagrams for the annihilation and coannihilation processes.

$$\left. \begin{array}{c} g_\chi \bar{g}_\eta \cdot \sigma(\chi\,\eta \to SM \text{ particles}) + \\ \bar{g}_\eta^2 \cdot \sigma(\eta\,\eta \to SM \text{ particles}) \end{array} \right] v\,, \qquad (233)$$

with

$$\bar{g}_\eta = g_\eta \Big(\frac{M_\eta}{m_\chi}\Big)^{3/2} \exp\Big(-\frac{\Delta M}{T}\Big)\,, \qquad (234)$$

where $g_\chi$ and $g_\eta$ denote the degrees of freedom of the fields $\chi$ and $\eta$, respectively, and should not be confused with the dark sector coupling $g_{DM}$. Their explicit values are given as $(g_S, g_\chi, g_\phi, g_{V_\mu}, g_\Psi) = (1, 2, 2, 3, 4)$. Each line of Eq. (233) corresponds to the contribution from a specific initial state illustrated in Figure 79. The latter two contributions are exponentially sensitive to the mass difference $\Delta M \equiv M_\eta - m_{DM}$ as can be seen in the expression of $\bar{g}_\eta$. Since the freeze-out occurs around $T \sim m_{DM}/25$, the above equations require $\Delta M \lesssim m_{DM}/25$ for the coannihilation mechanism to be operative. For our numerical study we compute the DM relic density using MicrOMEGAs 4.1.5 [560] implementing the simplified models with the help of FeynRules 2.0 [383] and LanHEP 3.2 [561].

*Experimental signatures*: The experimental signatures of the tau-philic simplified DM models have been studied in Ref. [558] and we summarise the result briefly here.

Direct detection experiments measure the nuclei recoil resulting from their interaction with the DM, but such interactions involving DM and quarks/gluons are absent at tree-level in our simplified models. Nonetheless, one-loop diagrams generate the relevant operators at dimension 6, which are suppressed by $1/M_\eta^2$. For the Majorana DM case, such an operator is given by the anapole moment operator $\mathcal{A}\,\bar{\chi}\gamma_\mu\gamma_5\chi\partial^\nu F_{\mu\nu}$. For $m_{DM} \simeq 500\,\text{GeV}$ and $\Delta M/m_\tau < 1$, the anapole moment is roughly given by $\mathcal{A}/g_{DM}^2 \sim 8 \cdot 10^{-7}\,[\mu_N\cdot\text{fm}]$ [562], which is more than one order of magnitude smaller than the current limit obtained by LUX [563] and also smaller than the projected sensitivity of LZ [564], even for the $g_{DM} = 1$ case.[74]

Regarding indirect detection experiments, the only relevant process is $\chi\chi \to \tau^+\tau^-$ as far as $2 \to 2$ processes are concerned. As mentioned above, for $\chi = \{\chi, S\}$ this process suffers from the chiral suppression, and the signal rate for indirect detection is below the experimental sensitivity. The chiral suppression is absent for $\chi = V_\mu$ (Model-3). Nevertheless, the annihilation rate is two orders of magnitude smaller than the current Fermi-LAT limit [565] even in this case. The $2 \to 3$ scattering, $\chi\chi \to \tau^+\tau^-\gamma$, may be more promising because it is free from the chiral suppression. In the case where $\chi = S$, the future Cherenkov Telescope Array (CTA) will be able to probe this annihilation channel in the region where $\Delta M > m_\chi/10$ [566].

At particle colliders the possibility arises of studying pair production of the charged coannihilation partners via an off-shell neutral gauge boson ($\gamma/Z$) exchange. The produced CAPs subsequently decay into the DM particle and a tau lepton. In the bulk of the viable parameter region, the mass splitting is small ($\Delta M \sim 20$ GeV) and the decay products of the CAP become very soft. In this region the LHC is hopeless to distinguish the signal from the overwhelming background.

---

[74] The limit mentioned here assumes the observed energy density of the DM. On the other hand, for $m_{DM} \simeq 500\,\text{GeV}$ and $g_{DM} \simeq 1$, all of our simplified models underproduce the DM. The actual constraints would therefore be even milder if this effect is taken into account.

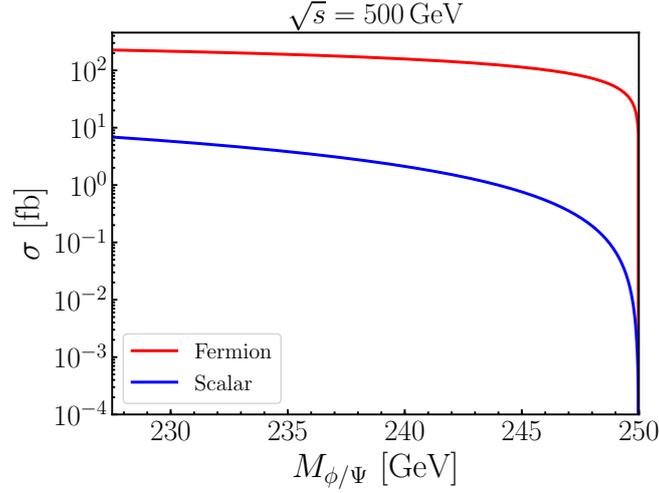

Figure 80: The cross sections for pair production of coannihilation partners. The scenario with Dirac fermion (complex scalar) as the coannihilation partner correspond to red (blue) curves.

The exceptional situation is the case where the mass splitting is smaller than the tau-lepton mass ($\Delta M < 1.777$ GeV) and the CAP has the detector-scale lifetime. This region can be explored at the LHC by looking for anomalous charged tracks, such as highly ionizing or slowly moving charged tracks. Those exotic signatures are intensively searched for by ATLAS [567, 568] and CMS [569, 570] and also can be investigated by the MoEDAL experiment [571]. Ref. [558] has shown that these searches have already excluded a part of the parameter region of the simplified models.

### 5.4.2 Expected sensitivity at CLIC

As mentioned above, the $e^+e^-$ collider can create pairs of coannihilation partners ($\eta$) via a neutral gauge boson exchange. The produced CAPs then decay into the DM particle $\chi$ and a tau lepton:

$$e^+e^- \to \eta^+\eta^- \to \tau^+\tau^-\chi\chi. \tag{235}$$

We focus our study on the signal coming from prompt decays of $\eta^\pm$ and hence we study the region of parameter space with $\Delta M > m_\tau$. The opposite case ($\Delta M \leq m_\tau$) may be probed at the LHC by looking for anomalous charged track signatures since $\eta$ can be long-lived in this region [558].

The production cross sections of scalar ($\phi$) and fermionic ($\Psi$) CAPs with $Y=-1$ are given by [572–574]

$$\sigma(e^+e^- \to \phi^+\phi^-) = \alpha^2 \pi s \cdot \mathcal{A} \cdot \frac{1}{6}\beta^3, \tag{236}$$

$$\sigma(e^+e^- \to \Psi^+\Psi^-) = \alpha^2 \pi s \cdot \mathcal{A} \cdot \beta\left(1 - \frac{1}{3}\beta^2\right), \tag{237}$$

with

$$\mathcal{A} = \frac{2}{s^2} + \frac{2}{s}\frac{(g_L + g_R)g_R}{(s - m_Z^2)} + \frac{(g_L^2 + g_R^2)g_R^2}{(s - m_Z^2)^2}, \tag{238}$$

$$g_L = \frac{-\frac{1}{2} + s_W^2}{s_W c_W}, \quad g_R = \frac{s_W^2}{s_W c_W}, \tag{239}$$

where $\alpha$ stands for the fine-structure constant, $g_L$ and $g_R$ correspond to the couplings between the $Z$ boson and the electron, and $\beta$ is the velocity of the outgoing $\eta$s

$$\beta = \sqrt{1 - \frac{4M_{\phi/\Psi}^2}{s}} \,. \tag{240}$$

These simple formulae neglect the subleading effects of the $Z$ boson width and the energy loss of incoming electrons due to bremsstrahlung photons.

Figure 80 shows the cross sections of scalar ($\phi$) and fermionic ($\Psi$) CAPs at the 500 GeV CLIC. In the formulae we can see that the cross section is proportional to $\beta$ for fermions while it is proportional to $\beta^3$ for scalars as $\beta \to 0$; therefore, the scalar production is significantly reduced as the mass gets closer to half of the $e^+e^-$ centre-of-mass energy. This feature is clearly seen in Figure 80. Moreover, we note that the production rate is independent of $g_{\rm DM}$.

We also comment on the vector boson ($\gamma/Z$) fusion (VBF) channel, $e^+e^- \to \eta^+\eta^-e^+e^-$.[75] Unlike the Drell-Yan process, the production rate of this channel is not proportional to $1/s$ and could potentially be important for large $s$. We have estimated the LO cross section of this process with MadGraph [31] requiring that out-going electrons have $p_T > 0.01$ GeV and $|\eta| < 7$ to avoid the $t$-channel singularity in the forward region. For $m_\eta = 300$ GeV we find the cross section of this process to be $\sigma_\phi^{\rm VBF} = 0.17$ fb and $\sigma_\psi^{\rm VBF} = 0.9$ fb, both of which are an order of magnitude smaller than the Drell-Yan processes of the corresponding models. We therefore do not include this process in our study.

In the region where the coannihilation mechanism works, the final state taus are very soft due to a small mass splitting between the CAP and DM. This region suffers from a large soft tau background produced by collisions of forward photons emitted by the incoming electrons: $\gamma\gamma \to \tau^+\tau^-$. This background can be suppressed by demanding a high energy ISR photon in the event. If such a photon is produced, one of the beam-remnant electrons will be deflected and detected, and the event can be safely rejected [575]. The efficiency of the analysis based on this technique in the case of the hadronic tau final state is studied in detail in Ref. [576]. The latter work provides the 95 % CL exclusion limit in the $(M_\eta, m_\chi)$ plane assuming a 500 GeV $e^+e^-$ collider with 500 fb$^{-1}$. We recast their result into our simplified models in the following way: along the exclusion contour, we calculate the required signal events, $N_{\max}(\Delta M)$, (before event selection) needed for exclusion for each value of $\Delta M$. For different collider energies $\sqrt{s}$, integrated luminosities $\mathcal{L}$ and spins $\phi/\Psi$, we demand the signal events before event selection not to exceed the corresponding upper limit:

$$\sigma_{\phi/\Psi}^{\sqrt{s}}(M_{\phi/\Psi}) \cdot \mathcal{L} \leq N_{\max}(\Delta M) \,. \tag{241}$$

This recasting method has been commonly used in the literature [556, 558, 577] and proved to work well empirically. At $e^+e^-$ colliders, where often cross sections for signal and background processes have the same short-distance behaviour, the assumption behind this estimate amounts to neglect radiation effects that may benefit more the background rate than the signal one and to assume similar efficiencies at the two points in $(M, \Delta M)$ space for both analyses carried out at the two different $\sqrt{s}$ values.

### *Result*

We present our results in Figure 81 where the projected sensitivities at 95 % CL are shown for various assumptions on the collider energy and luminosity. The blue bands show the region corresponding to the DM relic density observed by the Planck satellite mission [578] within $3\,\sigma$ for several values of $g_{\rm DM}$. The region above the blue band is excluded due to overproduction of DM, unless the thermal history of the Universe is modified. These plots illustrate the complementarity between the projected limits for CLIC and those for searches for long-lived charged particles at the LHC, the latter corresponding to the region coloured in magenta.

---

[75] In our simplified models the $W$-boson fusion channel, $e^+e^- \to \eta^+\eta^-\nu\bar{\nu}$, is absent, since the coannihilation partner is SU(2)$_L$ singlet and does not couple to the $W$-bosons.

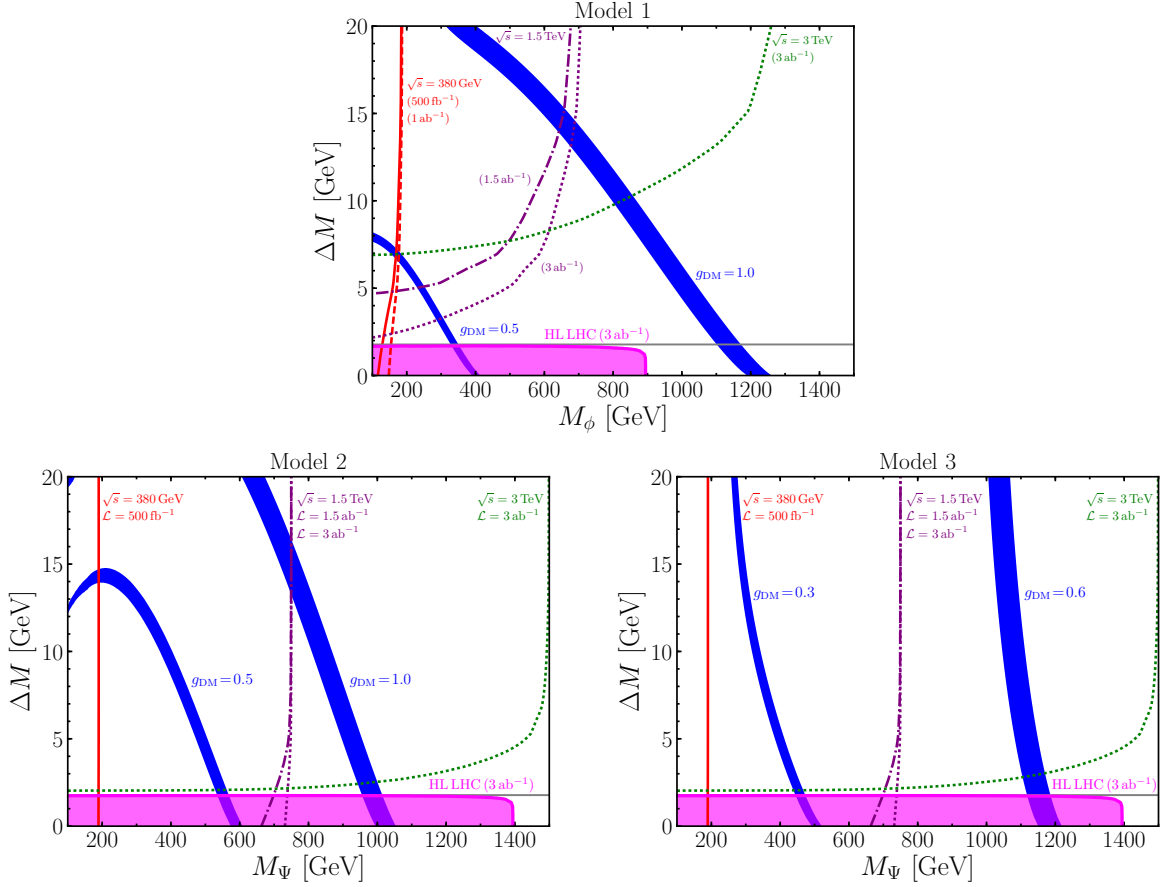

Figure 81: The DM coannihilation strip and the projected exclusion limits at CLIC for the three models presented in Table 39. Different colours correspond to different centre-of-mass energies $\sqrt{s}$ as shown in the plot. Solid, dashed, dot-dashed and dotted lines correspond to 500 fb$^{-1}$, 1 ab$^{-1}$, 1.5 ab$^{-1}$ and 3 ab$^{-1}$ for the integrated luminosities respectively. The region coloured in magenta corresponds to projected limits for long-lived charged particles searches at the high luminosity stage of the LHC; namely a centre-of-mass energy of 13 TeV and 3 ab$^{-1}$ [558]. The horizontal grey line indicates the mass of the $\tau$ lepton. The blue regions satisfy the correct dark matter relic abundance within $3\sigma$ for different values of the coupling $g_{\text{DM}}$.

The upper panel shows the projected sensitivity for Model-1 in which DM is a Majorana fermion. In this scenario the coannihilation partner is a complex scalar ($\phi$) and the production cross-section Eq. (236) gets suppressed by $\beta^3$ in the vicinity of the kinematic threshold; therefore, the exclusion limits on this scenario are weaker than those in the scenarios with a fermionic CAP ($\Psi$) (Model-2 and -3). Furthermore, the production rate gets smaller for larger $\sqrt{s}$ as can be seen in the expression of $\mathcal{A}$ in Eq. (238). Therefore, increasing collider energy does not help to explore the smaller $\Delta M$ region. In order to probe the coannihilation strip for $g_{\text{DM}} = 0.5$, increasing the luminosity from 1 to 3 ab$^{-1}$ represents a better improvement than increasing the centre-of-mass energy from 1.5 to 3 TeV.

The lower panel shows the exclusion limits on Model-2 and -3 corresponding to a scalar and a vector DM, respectively. The coannihilation partner is a charged Dirac fermion ($\Psi$) in both scenarios. For $\sqrt{s} = 380$ GeV with 500 fb$^{-1}$, the projected limits on these models are very close to the kinematic threshold ($M_\Psi = 190$ GeV). In Model-2, the DM overproduction constraint requires $M_\Psi$ to be smaller than 1 TeV for $g_{\rm DM} \leq 1$. This region can be explored by 3 TeV CLIC apart from a compressed mass region $\Delta M < 2.5$ GeV. Unlike Model-1 and -2, the DM density in Model-2 can easily be brought down to the allowed value without resorting small $\Delta M$ due to the absence of chiral suppression in the $\chi\chi \to \tau^+\tau^-$ mode. Thus, $M_\Psi$ can go higher than as 1.5 TeV for $g_{\rm DM} \gtrsim 0.7$, which exceeds the kinematical threshold of 3 TeV CLIC. On the other hand, almost the entire region with $g_{\rm DM} \lesssim 0.7$ can be explored by CLIC, as can be seen in the lower right panel of Figure 81.

*Summary*

We have studied the sensitivity of the future Compact Linear Collider to the tau-philic DM simplified models with a coannhilation partner. Three distinctive scenarios have been examined: **(i)** Majorana DM, **(ii)** Real scalar DM and **(iii)** Vector DM, where the CAP is a complex scalar in the first model, while it is a Dirac fermion in the latter two. We have found that CLIC has excellent sensitivity to these models. In particular, if the CAP is a Dirac fermion, almost the entire region allowed by the DM relic constraint can be explored by 3 TeV CLIC. If it is a complex scalar, the region with a small mass splitting $\Delta M < 10$ GeV may not be probed depending on the mass of the scalar. We found that larger luminosity helps greatly in exploring the small $\Delta M$ region even for lower energy stages ($\sqrt{s} = 380$ GeV and 1.5 TeV).

The models presented in this report are difficult to probe by direct and indirect DM detection experiments as well as by the LHC. Therefore, lepton colliders such as CLIC provide an almost unique opportunity to explore them. Consequently, a possible discovery of a new heavy electrically charged particle decaying into a $\tau$-lepton plus missing energy can provide information about one of the most pressing questions in high-energy physics; the nature of dark matter. In addition, this would present motivations to develop new techniques to explore models with compressed mass spectra at CLIC.

## 5.5 Inert doublet model [76]

Prospects for the discovery of Inert Doublet Model scalars at CLIC are described in detail in [579] and summarized in the following.

One of the simplest extensions of the SM which can provide a dark matter candidate is the Inert Doublet Model (IDM) [580–582]. In this model, the scalar sector is extended by a so-called inert or dark doublet $\Phi_D$ (the only field odd under $Z_2$ symmetry) in addition to the SM Higgs doublet $\Phi_S$. This results in five physical states after electroweak symmetry breaking: the SM Higgs boson $h$ and four dark scalars: two neutral, $H$ and $A$, and two charged, $H^\pm$. A discrete $Z_2$ symmetry prohibits the inert scalars from interacting with the SM fermions through Yukawa-type interactions and makes the lightest neutral scalar, chosen to be $H$ in this work, a good dark matter (DM) candidate. The free parameters of this model (after fixing the SM Higgs mass and vev) are the masses $m_H$, $m_A$, $m_{H^\pm}$, the Higgs-dark matter coupling $\lambda_{345}$, and the dark sector internal coupling $\lambda_2$.

In this study, the following discovery channels for the inert scalars are considered:

$$\begin{aligned} e^+e^- &\to A\,H, \\ e^+e^- &\to H^+H^-. \end{aligned} \tag{242}$$

Two sets of benchmark points in agreement with all theoretical and current experimental constraints were selected according to their respective kinematic accessibility at the initial energy Stage 1 (380 GeV) or at the higher-energy Stages 2 and 3 (1.5 and 3 TeV), as described in [583]. For these scenarios the dark scalar $A$ decays as $A \to Z^{(*)}H$, while the charged bosons $H^\pm$ decay predominantly as $H^\pm \to W^{\pm(*)}H$. Taking advantage of their clean signature, only leptonic decays of the $W^\pm$ and $Z$ bosons are considered. The DM candidate $H$ escapes detection, leading to a final state of two opposite-charge leptons (muons or electrons) and missing transverse energy. We here employ muon decay of the $Z$ boson from $AH$

---

[76]Based on a contribution by J. Kalinowski, W. Kotlarski, T. Robens, D. Sokołowska and A.F. Żarnecki.

production and different-flavor lepton pairs for $H^+H^-$ production:

$$e^+e^- \to \quad A\,H \to \quad Z^{(*)}HH \quad \to \mu^+\mu^- HH,$$
$$e^+e^- \to \quad H^+H^- \to HW^{+(*)}HW^{-(*)} \to H\ell^+\nu H\ell'^-\bar{\nu}' \quad (243)$$

The cross sections for the selected benchmark points according to their values of $m_A + m_H$ and $2m_{H^\pm}$ for the processes (242) are presented in Figure 82.

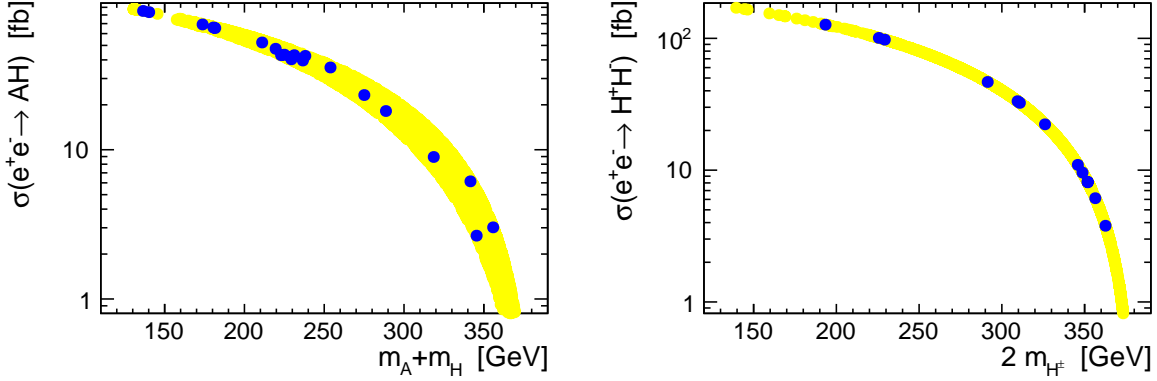

Figure 82: Leading-order cross sections for $AH$ (left) and $H^+H^-$ (right) inert scalar production at $\sqrt{s} = 380$ GeV. The yellow band represents all benchmark points obtained in the model scan [583], the blue points mark the scenarios selected for this study. Beam energy spectra are not included.

Event samples for the processes (243) are generated using WHIZARD2.2.8 [55]. In the simulations, all intermediate processes leading to $\ell^+ (\ell^-)' + \not{E}_\perp$ are considered. All major Standard Model background processes leading to the production of opposite-charge lepton pairs, including production and leptonic decays of tau leptons, with and without an associated pair of $\nu\bar{\nu}'$ are considered. Beam energy spectra and initial state radiation (ISR) are taken into account.

To select a signal enhanced sample of events, a multivariate analysis is performed using a Boosted Decision Tree (BDT) [584]. The following observables describing the kinematics of the dilepton final state are used as BDT input variables: the total energy of the lepton pair, the dilepton invariant mass and transverse momentum, the polar angle of the dilepton pair, angular observables of the leptons with respect to the beam and the dilepton pair boost direction, the Lorentz boost of the dilepton pair and the reconstructed missing recoil mass.

### 5.5.1 Neutral inert scalar pair production $e^+e^- \to AH$ at stage 1

The discovery prospects for inert scalars at the first stage of CLIC with $\sqrt{s} = 380$ GeV are determined assuming an integrated luminosity of 1 ab$^{-1}$.

The invariant mass of the lepton pair from (virtual) $Z$ or $\gamma$ decay depends on the mass splitting between $A$ and $H$ and is relatively small for most benchmark scenarios. Therefore the longitudinal boost and the invariant mass of the lepton pair are much smaller in the signal than in the background, particularly in the dominant background process $e^+e^- \to \mu^+\mu^-$. This allows to apply pre-selection cuts when generating the events: the di-muon invariant mass is required to be $M_{\mu\mu} < 100$ GeV and the di-muon longitudinal momentum $|p_Z^{\mu\mu}| < 140$ GeV. After these pre-selection cuts, the multivariate analysis is used to discriminate between signal and background events, and the final cut on the BDT response is optimised to obtain highest signal significance for each scenario. The resulting significances for the chosen low-energy benchmark points are shown in Figure 83.

A discovery, corresponding to $5\sigma$, at the initial stage of CLIC is expected for scenarios with the signal cross section (in the $\mu^+\mu^-$ channel, after pre-selection cuts) above about 0.5 fb, which corresponds to the neutral inert scalar mass sum below about 290 GeV. For the considered benchmark points we do not observe any sizable dependence of the expected significance on the mass splitting between the two neutral scalars, $m_A - m_H$ (indicated by the colour scale in Figure 83).

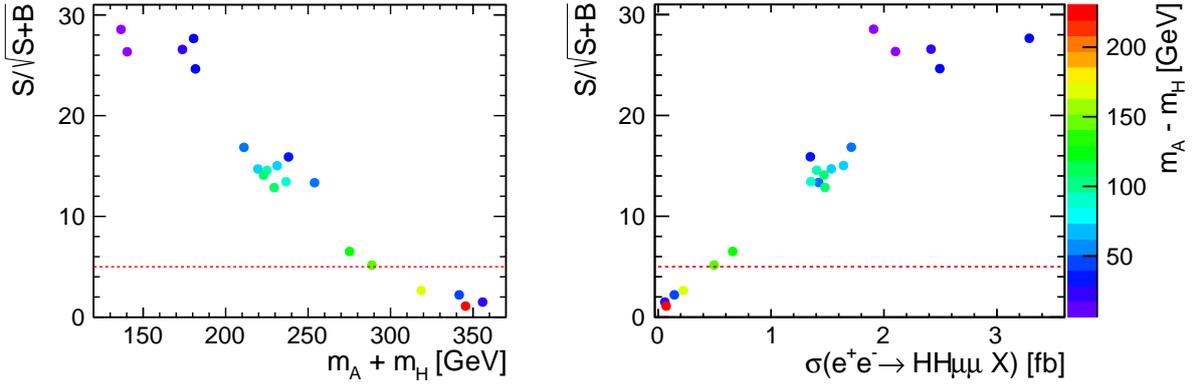

Figure 83: Expected significance of the deviations from the Standard Model predictions observed at 380 GeV CLIC for events with two muons in the final state ($\mu^+\mu^-$) as a function of the neutral inert scalar mass sum (left) and the production cross section for the considered signal channel, after pre-selection cuts (right), for the low-energy benchmark points. The colour indicates the mass splitting between the $A$ and $H$ scalars (right scale applies to both plots).

#### 5.5.2  Charged inert scalar pair production $e^+e^- \to H^+H^-$ at stage1

The selection of $H^+H^-$ production events is more challenging than for the $AH$ channel, as the two leptons in the final state no longer originate from a single (on- or off-shell) intermediate state. The large potential background of same-flavour lepton pair production $e^+e^- \to \ell^+\ell^-$ is suppressed by only considering different-flavour lepton pairs. A multivariate analysis, based on the same kinematical observables as for the neutral scalar production, exploits the fact that the kinematic space available for lepton pair production is reduced in the signal due to the massive intermediate scalars, which is not the case in the backgrounds.

As in the case of neutral scalar production, the expected discovery significance is mainly related to the production cross section. This is illustrated in Figure 84. Discovery at the initial stage of CLIC is only possible for scenarios with signal cross sections (in the electron-muon channel) above about 1 fb. This corresponds to charged scalar masses below roughly 150 GeV. We do not observe any sizable dependence of the expected significance on the mass splitting between the charged and neutral inert scalars, $m_{H^\pm} - m_H$ (indicated by colour scale in Figure 84), within the considered range of parameters.

#### 5.5.3  Inert scalars at the high-energy stages 2 and 3 of CLIC

The discovery prospects of the high-energy Stages 2 and 3 at 1.5 TeV and 3 TeV are determined assuming integrated luminosities of 2.5 ab$^{-1}$ and 5 ab$^{-1}$, respectively. In this case, benchmark points not accessible at the low-energy stage are added to the study. Similarly to the low-energy stage, generator-level acceptance cuts are applied, followed by a multivariate analysis to suppress background.

Figure 85 shows the expected discovery significances for the IDM signal in the $AH$ and $H^+H^-$ channels at the high-energy CLIC. In the $AH$ channel, the increase of collision energy and integrated luminosity leads to an increase of the sum of neutral inert scalar masses which can be probed from 290 GeV at Stage 1 to 550 GeV at Stage 2. The significance is mainly driven by the signal production cross section and is approximately proportional to the square-root of the integrated luminosity. For parameter points that are already accessible at Stage 1 the $AH$ production cross sections decrease with the collision energy much faster than most of the backgrounds and the significance of observation decreases at Stage 2. Only for points with $M_A + M_H \gtrsim 300\,\text{GeV}$, which are close to the production threshold at Stage 1, higher integrated luminosity and the production cross sections enhanced by up to a factor of 2 result in better sensitivity at the centre-of-mass energy of 1.5 TeV. Similarly, when going from 1.5 TeV to 3 TeV, the significance of observation increases only for scenarios with $M_A + M_H \gtrsim 1.2\,\text{TeV}$.

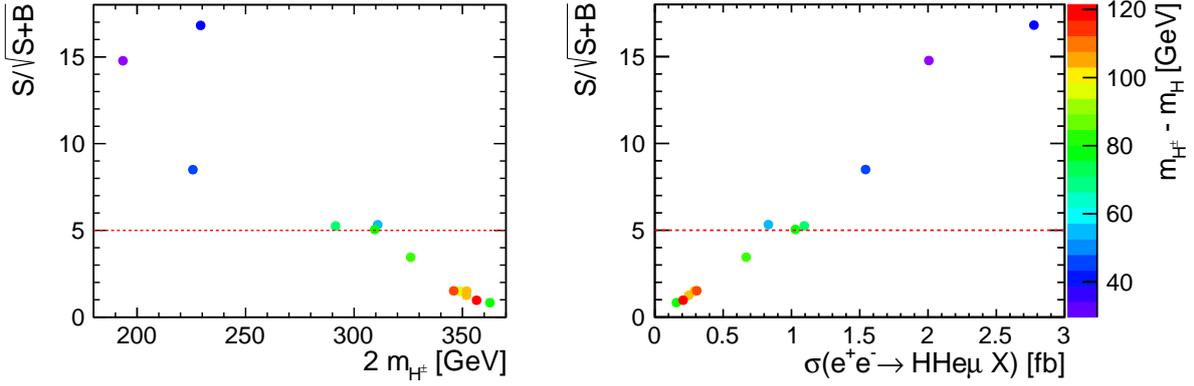

Figure 84: Expected significance of the deviations from the Standard Model predictions observed at 380 GeV CLIC for events with electron-muon pair in the final state ($e^+\mu^-$ or $\mu^+e^-$) as a function of $2m_{H^\pm}$ (left) and the production cross section for the considered signal channel (right), for different IDM benchmark points. The colour indicates the mass splitting between the $H^\pm$ and $H$ scalars (right scale applies to both plots).

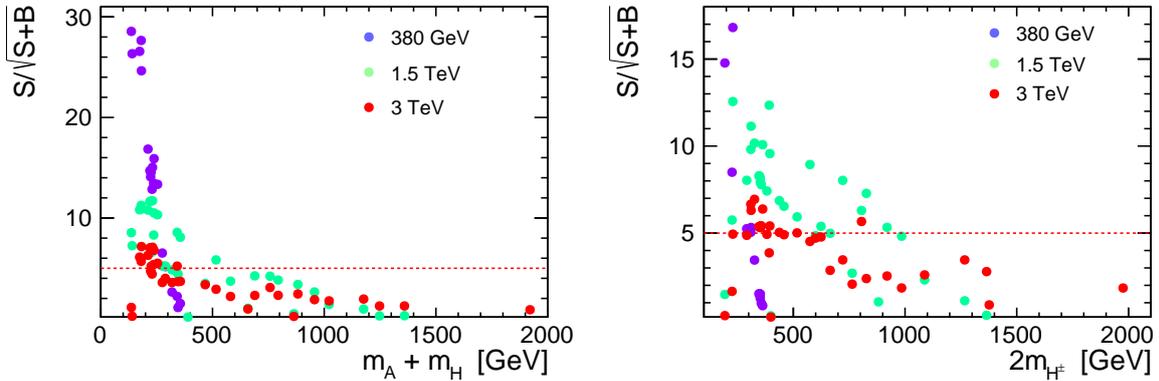

Figure 85: Significance of the deviations from the Standard Model predictions expected at the higher-energy CLIC stages for: (left) events with two muons in the final state ($\mu^+\mu^-$) as a function of the sum of neutral inert scalar masses and (right) events with an electron and a muon in the final state ($e^+\mu^-$ or $e^-\mu^+$) as a function of $2M_{H^\pm}$, for low and high energy benchmark points.

In the $H^+H^-$ channel, the production cross sections exhibit the same behaviour, with a general decrease with centre-of-mass energy, apart from parameter points close to the production threshold. Between the first and second energy stages, the discovery masses increase from about 150 GeV to 500 GeV.

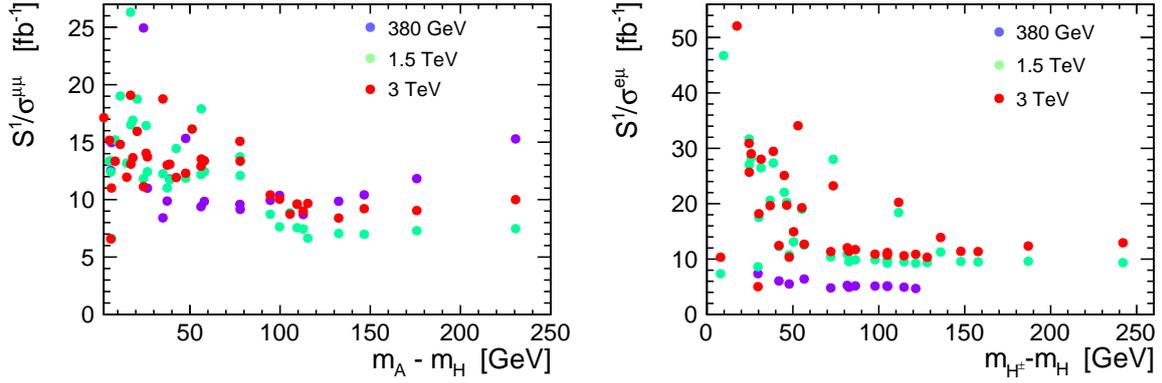

Figure 86: Ratio of the expected significance at the integrated luminosity of $1\,\text{ab}^{-1}$ to the signal cross section in the considered channel: (left) with two muons in the final state ($\mu^+\mu^-$) and (right) with electron-muon pair production ($e^+\mu^-$ or $e^-\mu^+$), as a function of the scalar mass differences, for different IDM benchmark points.

The dependence of the signal significance on the mass difference between neutral or charged inert scalar and the DM candidate is illustrated in Figure 86. It shows the ratio of the significance calculated for $1\,\text{ab}^{-1}$ to the respective cross section, and demonstrates that in both the $AH$ and the $H^+H^-$ channel the sensitivity at the higher-energy stages is better for low mass differences as this gives best kinematic separation of signal from background.

#### 5.5.4 Summary and outlook

We found that most of the considered low-mass benchmark scenarios can be observed with high significance already with $1\,\text{ab}^{-1}$ collected at $380\,\text{GeV}$ (the first stage of CLIC), provided that the sum of neutral inert scalar masses, $m_A + m_H < 290\,\text{GeV}$ or charged scalar mass, $m_{H^\pm} < 150\,\text{GeV}$. At $1.5\,\text{TeV}$ the discovery reach is extended to the sum of scalar masses of about $550\,\text{GeV}$ in the dimuon channel and for charged scalar masses up to about $500\,\text{GeV}$ in the $e^\pm\mu^\mp$ channels. For scenarios where the signal cross sections in the dilepton channel are too small, it might be worthwhile investigating semi-leptonic decays in the $H^+H^-$ production channel. Due to the much larger branching ratios the expected number of $H^+H^-$ signal events in the semi-leptonic final state is over an order of magnitude larger than for the electron-muon signature. As a similar scaling is expected for the background processes, we expect that the significance of the observation in the semi-leptonic channel should be increased by at least a factor of 3.

## 6 Baryogenesis

The observed asymmetry between baryons and anti-baryons in our Universe is one of the biggest mysteries in our understanding of nature. It is often cast in terms of the baryon-to-photon ratio [585]

$$Y_B = \frac{n_b}{s} = (8.59 \pm 0.11) \times 10^{-11} \quad (244)$$

where $n_B$ ($s$) is the baryon number (entropy) density. While this number is tiny, within the Standard Model it should be many orders of magnitude smaller. In principle such a non-vanishing $Y_B$ could originate in tuned initial conditions during the Big Bang, or from dynamics of grand unified theories at energy scales above $\sim 10^{16}$ GeV. However, in practice, the success of the inflationary paradigm implies that any matter-antimatter asymmetry created in either of these ways would have been washed away during inflation. Thus, it is likely that a later stage of the evolution of the Universe is responsible for the creation of the matter-antimatter asymmetry. Sakharov [586], over 40 years ago, formulated three key ingredients that must have been present in the early Universe in order to generate a non-vanishing $Y_B$: (I) baryon number violation, (II) violation of both C- and CP-invariance, and (III) either departure from thermal equilibrium dynamics or violation of CPT-invariance. The SM contains the first ingredient in the form of (B+L)-violating instanton transitions at high temperatures, but falls short with respect to the second and third condition. While a large number of possible mechanisms have been proposed to augment the Standard Model such that all three 'Sakharov' conditions are satisfied, the most theoretically attractive and experimentally testable are those that introduce new particles in the few-hundred GeV to TeV mass range.

The most thoroughly studied scenario is electroweak baryogenesis; for a review, see e.g. [379]. In electroweak baryogenesis, the Universe undergoes a first order phase transition during which the electroweak symmetry is broken. This can be achieved through the presence of additional scalar particles in the mass range of $\mathcal{O}(500)$ GeV, contributing to the finite-temperature effective potential. Thus, the search for additional scalar particles within the kinematic reach of CLIC is of crucial importance to test the viability of such models. In Section 6.1 we extend the Standard Model by a scalar particle $h_2$ that is a singlet under the Standard Model gauge group. This corresponds to the most minimal extension that allows the realisation of a strong first-order phase transition, and thus electroweak baryogenesis. We find that CLIC can perform very well in testing the parameter space of such a model, either by searching directly for the decay $h_2 \to hh$, by measuring the value of the trilinear coupling of the Higgs boson, or by measuring the single Higgs couplings. The latter is a particularly powerful way to test the minimal model we focus on, where the pattern of deviations of Higgs couplings is very constrained and one can apply the result in Table 6 in Section 2.1. However, these indirect limits can be relaxed in less minimal models while maintaining the ability to induce a strong first-order phase transition. The ability of CLIC to test this scenario in all the three ways outlined before, outperforming the HL-LHC in each of them, is thus very important to ensure that CLIC can probe a large class of models in which one or more of these three signals may be slightly different than in the specific model we study.

Another model which employs electroweak-scale particles that decay out of thermal equilibrium via B- and CP-violating interactions to generate the matter-antimatter asymmetry is the so-called WIMP baryogenesis mechanism, see Section 6.2. A striking phenomenological feature of such models is the predicted existence of long-lived particles. Such signatures can be exploited at CLIC to discover this well-motivated mechanism to generate a matter-antimatter asymmetry.

### 6.1 General Higgs plus singlet model [77]

Following [587], we discuss the reach of CLIC in searching for a heavy Higgs boson $H$ which decays to a pair of 125 GeV Higgs bosons $h$.[78] In addition, we assess the capability of CLIC heavy Higgs searches in probing the nature of the EW phase transition in the context of a general real singlet scalar extension of the SM [588–590]. This scenario can also capture the phenomenology of the Higgs sector in more complete theories beyond the SM such as the NMSSM (see [374] and references therein) or Twin Higgs theories [375], as discussed in Section 4.2.1. At the same time, the singlet scalar extension of the SM constitutes a paradigm for achieving a strongly first order EW phase transition that could generate the observed matter-antimatter asymmetry of the Universe.

---

[77] Based on a contribution by J. M. No.

[78] In [587] we have shown that CLIC's potential in improving on the HL-LHC's sensitivity in searches for $H$ decaying into gauge bosons is limited.

The three dominant processes contributing to Higgs boson production at a high-energy electron-positron collider are $e^+e^- \to HZ$, $e^+e^- \to H\nu\nu$ and $e^+e^- \to He^+e^-$. Assuming a heavy scalar $H$ with SM-like properties, we compute the production cross section[79] as a function of the scalar mass $m_H$ for each of the three processes and for $\sqrt{s} = 0.38, 1.4, 3$ TeV, shown in Figure 87. We show both the case of unpolarized electron and positron beams (solid lines) and the possibility of using beam polarization, which can constitute a strong advantage in searching for new physics [591], assuming for definiteness an electron-positron beam polarization $P_{e^-}$, $P_{e^+} = -80\%, +30\%$ (dashed lines)[80] to correspond to a CLIC operation scenario.

As highlighted in Figure 87, the dominant Higgs production mechanism for both $\sqrt{s} = 1.4$ and 3 TeV is the vector boson fusion (VBF) process $e^+e^- \to H\nu\nu$. We also emphasize that $\sqrt{s} = 380$ GeV does not allow to probe high values of $m_H$, and moreover it does not yield as many kinematical handles to disentangle the heavy scalar signal from SM backgrounds. In the rest of the section we then focus on $e^+e^- \to H\nu\nu$ as Higgs production mechanism in CLIC, considering $\sqrt{s} = 1.4$ and 3 TeV as c.o.m. energies. The respective projected integrated luminosities we consider are $\mathcal{L} = 1500$ fb$^{-1}$ and 2000 fb$^{-1}$ [8]. In all our subsequent analyses, we simulate CLIC production of the new scalar $H$ via $e^+e^- \to H\nu\nu$ using MADGRAPH_AMC@NLO [31] with a subsequent decay into the relevant final state, and assuming electron and positron polarized beams with $P_{e^-}$, $P_{e^+} = -80\%, +30\%$ in all our analyses. We then shower/hadronise our events with PYTHIA 8.2 [338] and use DELPHES [385] for a simulation of the detector performance with the Delphes Tune for CLIC studies [6] (see also [386]).

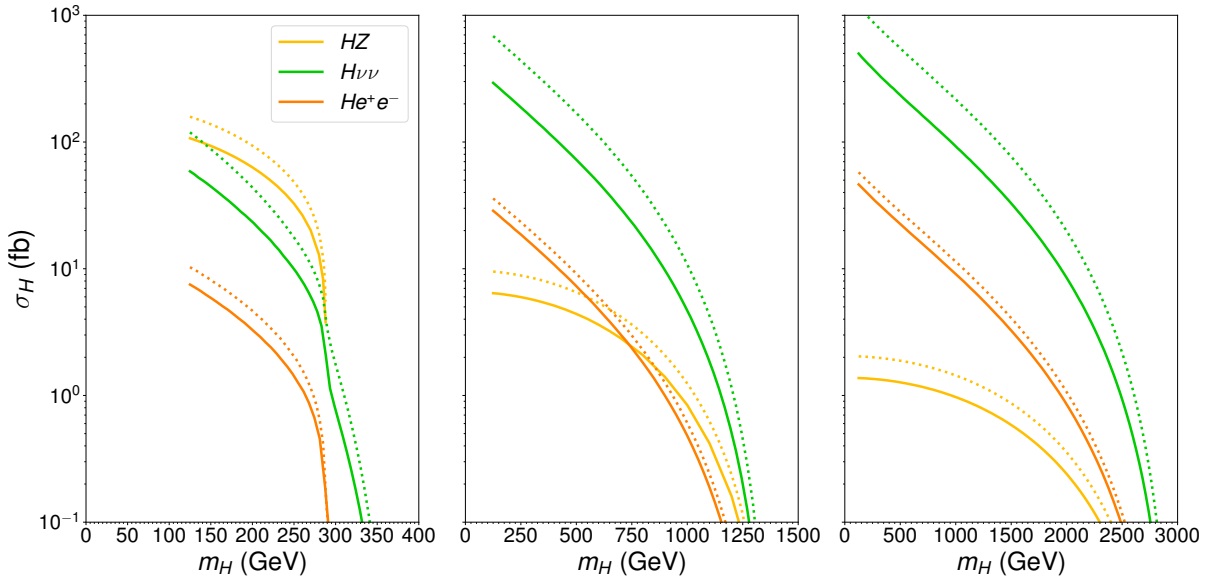

Figure 87: Higgs production cross sections $\sigma_H$ (in fb), assuming SM-like properties for $H$, as a function of $m_H$, for $\sqrt{s} = 380$ GeV (left), $\sqrt{s} = 1400$ GeV (middle) and $\sqrt{s} = 3000$ GeV (right), for unpolarized beams (solid) and $P_{e^-}$, $P_{e^+} = -80\%, +30\%$ (dashed).

---

[79]For $e^+e^- \to He^+e^-$, the outgoing electrons are required to satisfy $|\eta| < 5$, $P_T > 5$ GeV.
[80]Here, $-100\%$ corresponds to a fully left-handed polarized beam and $+100\%$ to a fully right-handed polarized beam.

We now explore the CLIC sensitivity to new scalars through resonant di-Higgs signatures $H \to hh$ in the VBF production channel. We focus on the $hh \to b\bar{b}b\bar{b}$ final state, which has the largest branching fraction while it not suffering from the very large QCD background one has to face in the LHC environment [592, 593]. We will show in what follows that resonant di-Higgs searches at CLIC constitute a very sensitive probe of the existence of new scalars that decay predominantly into $hh$. In Section 6.1.1 we analyse the $\sqrt{s} = 3$ TeV CLIC prospects, and discuss those for $\sqrt{s} = 1.4$ TeV in Section 6.1.2.

### 6.1.1 High-energy scenario $\sqrt{s} = 3$ TeV

The dominant backgrounds to the $e^+e^- \to H\nu\nu$ ($H \to hh \to 4b$) process at CLIC are from EW (including the SM non-resonant di-Higgs production contribution, on which we will comment in Section 6.1.3) and QCD processes yielding a $4b + 2\nu$ final state. We reconstruct jets (within DELPHES) with FASTJET [489], using the Valencia clustering algorithm [232] (particularly well-suited for jet reconstruction in high energy $e^+e^-$ colliders) in exclusive mode with $R = 0.7$ and $N = 4$ (number of jets). We perform our analysis for two different $b$-tagging working points within the CLIC Delphes Tune, corresponding to a 70% and 90% $b$-tagging efficiency, respectively[81]. In each case, we select events with 4 $b$-tagged jets, which are subsequently paired into two 125 GeV Higgs candidates by minimizing

$$\chi = \sqrt{\frac{(m_{b_1 b_2} - \overline{m_h})^2}{\Delta_h^2} + \frac{(m_{b_3 b_4} - \overline{m_h})^2}{\Delta_h^2}}, \quad (245)$$

where $\overline{m_h} = 102$ GeV and $\Delta_h = 30$ GeV are obtained from an approximate fit to the signal simulation (we note that the average Higgs mass $\overline{m_h}$ is somewhat lower than the truth value $m_h = 125$ GeV as a result of the jet reconstruction process). We then select events with two SM Higgs candidates by requiring $\chi < 1$. As a result, the only relevant SM background is from the EW processes.

Table 40: 3 TeV CLIC cross section (in fb) for signal (for $m_H = 300, 600, 900$ GeV respectively) and SM backgrounds for a $b$-tagging efficiency of 70% (90%), at different stages in the event selection and in the signal region (SR) for $m_H = 300, 600, 900$ GeV respectively (see text for details).

|              | $\sigma_S^{300}$ | $\sigma_S^{600}$ | $\sigma_S^{900}$ | $\sigma_B^{EW}$ | $\sigma_B^{QCD}$ |
|--------------|------------------|------------------|------------------|-----------------|------------------|
| Selection    | 12.85 (36.09)    | 8.52 (23.58)     | 5.19 (14.56)     | 0.407 (1.14)    | 0.048 (0.136)    |
| $\chi < 1$   | 9.26 (25.80)     | 5.29 (14.60)     | 3.52 (9.64)      | 0.146 (0.413)   | -                |
| SR$_{300}$   | 8.99 (25.80)     |                  |                  | 0.0444 (0.126)  | -                |
| SR$_{600}$   |                  | 4.80 (13.32)     |                  | 0.0236 (0.063)  | -                |
| SR$_{900}$   |                  |                  | 3.03 (8.25)      | 0.0098 (0.028)  | -                |

We define the Signal Region (SR) as

$$m_{4b} \in [C - \Delta, C + \Delta] \quad , \quad \begin{cases} C(m_H) = 0.96 \times m_H - 45\,\text{GeV} \\ \Delta(m_H) = 0.05 \times m_H + 40\,\text{GeV} \end{cases}, \quad (246)$$

with both $C(m_H)$ and $\Delta(m_H)$ extracted from a fit to the signal simulation. The cross section of three benchmark signal scenarios ($m_H = 300$ GeV, 600 GeV, 900 GeV) and the SM backgrounds at various stages in the selection process is shown in Table 40 for a $b$-tagging efficiency of 70% as well as for 90%.

---

[81]For the 90% $b$-tagging working point, the background contribution from events with $c$-jets which are mis-identified as $b$-jets ceases to be negligible and should be considered in a dedicated study. Nevertheless, the ratio of $b$-tagging efficiency to $c$-jet mistag rate is in this case $\sim 0.2$ (and backgrounds with mis-identified $c$-jets need to contain at least two of those), such that events with mis-identified jets are still subdominant, and we will not consider them here.

From the above analysis, we obtain the projected 95% C.L. sensitivity reach of $\sqrt{s} = 3$ TeV CLIC ($\mathcal{L} = 2000$ fb) for $H \to hh \to b\bar{b}b\bar{b}$ in the mass range $m_H \in [300\,\text{GeV}, 1\,\text{TeV}]$. Here the signal strength $\kappa$ is defined as $\kappa \equiv \sigma_S/\sigma_S^{\text{SM}} \times \text{BR}(H \to hh)$ (with $\sigma_S/\sigma_S^{\text{SM}}$ the ratio of the production cross section of $H$ to its SM value). The resuls of this section are summarized in Figure 88, and discussed in detail in the following Section 6.1.2 together with those obtained for $\sqrt{s} = 1.4$ TeV.

### 6.1.2 *Intermediate energy region $\sqrt{s} = 1.4$ TeV*

We now repeat the above analysis for a CLIC c.o.m. energy $\sqrt{s} = 1.4$ TeV with $\mathcal{L} = 1.5$ ab$^{-1}$. The cross sections for the signal (for $m_H = 300$ GeV, 600 GeV, 900 GeV) and the SM backgrounds are shown in Table 41, with the signal region being defined as in the analysis from Section 6.1.1 and given by Eq. (246).

Table 41: 1.4 TeV CLIC cross section (in fb) for signal (for $m_H = 300, 600, 900$ GeV respectively) and SM backgrounds for a $b$-tagging efficiency of 70% (90%), at different stages in the event selection and in the signal region (SR) for $m_H = 300, 600, 900$ GeV respectively (see text for details).

|  | $\sigma_S^{300}$ | $\sigma_S^{600}$ | $\sigma_S^{900}$ | $\sigma_B^{EW}$ | $\sigma_B^{QCD}$ |
|---|---|---|---|---|---|
| Selection | 6.18 (17.25) | 2.17 (5.88) | 0.456 (1.26) | 0.140 (0.385) | 0.039 (0.108) |
| $\chi < 1$ | 4.61 (12.85) | 1.36 (3.64) | 0.306 (0.843) | 0.052 (0.143) | - |
| SR$_{300}$ | 4.50 (12.51) |  |  | 0.022 (0.059) | - |
| SR$_{600}$ |  | 1.24 (3.32) |  | 0.022 (0.018) | - |
| SR$_{900}$ |  |  | 0.263 (0.725) | 0.0014 (0.0042) | - |

In Figure 88 we show the sensitivity of CLIC corresponding to $\sqrt{s} = 1.4$ TeV (blue) and $\sqrt{s} = 3$ TeV (orange) for 70% $b$-tagging (solid) and 90% $b$-tagging (dashed) efficiencies, together with the present limits from CMS $H \to hh \to b\bar{b}b\bar{b}$ searches [594] with $\mathcal{L} = 35.9$ fb$^{-1}$ (solid red) and the projected 95% C.L. sensitivity for HL-LHC with $\mathcal{L} = 3$ ab$^{-1}$ (dashed red) based on a $\sqrt{\mathcal{L}}$ scaling w.r.t. to the present expected exclusion sensitivity from [594]. As Figure 88 highlights, CLIC would greatly surpass the sensitivity of HL-LHC to resonant di-Higgs production: for a c.o.m. energy $\sqrt{s} = 1.4$ TeV the increase in sensitivity w.r.t. HL-LHC ranges from a factor $30 - 50$ for $m_2 \lesssim 400$ GeV, to roughly a factor 10 for $m_2 \sim 1$ TeV. For $\sqrt{s} = 3$ TeV the increase in sensitivity is a factor 50 or larger in the entire mass range $m_2 \in [250\,\text{GeV}, 1\,\text{TeV}]$, reaching two orders of magnitude sensitivity increase for $m_2 < 400$ GeV and $m_2 > 800$ GeV. At the same time, our results show that increasing the $b$-tagging efficiency above the 70% working point would benefit the reach of this search at CLIC substantially. In our work we specifically explore a 90% working point, but a less extreme increase of the $b$-tagging efficiency would display a comparable associated sensitivity increase.

*Altogether, the results of this section show that resonant di-Higgs production searches are a prominent and very sensitive probe of heavier Higgs bosons with CLIC. In the remainder of Section 6.1, we explore the sensitivity of these searches to the existence of a new singlet-like scalar interacting with the SM Higgs, and the implications for the properties of the EW phase transition in the early Universe.*

### 6.1.3 *Singlet scalar extension of the Standard Model*

The phenomenology of the SM extended by a real scalar singlet $S$ (SM + $S$) has been widely studied in the literature (see e.g. [380, 588–590, 595–604]), including the connection to the EW phase transition [588, 590, 595, 596, 600, 601, 603] (see also [377, 605]). We analyse here the sensitivity of CLIC to the parameter space leading to a first order EW phase transition by casting the results from the previous sections in terms of the SM + $S$ scenario. We also explore the complementarity of CLIC with other probes of the EW phase transition – favoured parameter space in this scenario from the HL-LHC [601, 603].

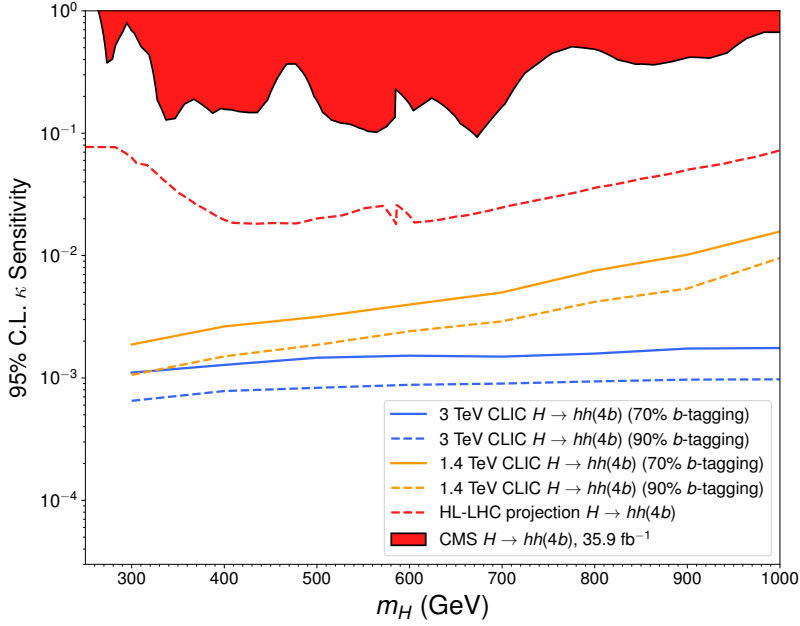

Figure 88: CLIC 95% C.L. sensitivity to $\kappa = \sigma_S/\sigma_S^{\rm SM} \times {\rm BR}(H \to hh)$ as a function of $m_H$ for $e^+e^- \to H\nu\nu$ ($H \to hh \to 4b$) at $\sqrt{s} = 1.4$ TeV with $\mathcal{L} = 1500$ fb$^{-1}$ (orange) and $\sqrt{s} = 3$ TeV with $\mathcal{L} = 2000$ fb$^{-1}$ (blue). In both cases the solid line corresponds to a 70% $b$-tagging efficiency and the dashed line to a 90% $b$-tagging efficiency. Shown for comparison are the LHC 95% C.L. excluded region from present CMS $H \to hh \to 4b$ searches [594] (red region) and the projected HL-LHC (13 TeV, $\mathcal{L} = 3$ ab$^{-1}$) expected 95% C.L. exclusion sensitivity (dashed red line).

We consider the most general form for the SM + $S$ scalar potential that depends on a Higgs doublet $\Phi$ and real singlet $S$ (see e.g. [588, 590]):

$$V(\Phi, S) = -\mu^2 \left(\Phi^\dagger \Phi\right) + \lambda \left(\Phi^\dagger \Phi\right)^2 + \frac{a_1}{2} \left(\Phi^\dagger \Phi\right) S$$
$$+ \frac{a_2}{2} \left(\Phi^\dagger \Phi\right) S^2 + b_1 S + \frac{b_2}{2} S^2 + \frac{b_3}{3} S^3 + \frac{b_4}{4} S^4. \quad (247)$$

Upon EW symmetry breaking, $\Phi \to (v + h)/\sqrt{2}$ with $v = 246$ GeV. We note that a shift in the singlet field $S + \delta S$ does not lead to any change in the physics, which may be used to choose a vanishing vev for the singlet field in the EW broken minimum by requiring $b_1 = -a_1 v^2/4$. This is the choice we adopt in the following. Once the EW symmetry is broken, the singlet $S$ and the SM Higgs $h$ mix in the presence of $a_1$, yielding two mass eigestates $h_1, h_2$. We identify $h_1$ with the 125 GeV Higgs boson, and $h_2$ with the heavy state $H$ discussed in the previous sections. The masses $m_1 = 125$ GeV, $m_2$ and the singlet-doublet mixing angle $\theta$ are related to the scalar potential parameters as

$$\begin{aligned} a_1 &= \frac{m_1^2 - m_2^2}{v} 2 \sin\theta \cos\theta, \\ b_2 + \frac{a_2 v^2}{2} &= m_1^2 \sin^2\theta + m_2^2 \cos^2\theta, \\ \lambda &= \frac{m_1^2 \cos^2\theta + m_2^2 \sin^2\theta}{2 v^2}, \end{aligned} \quad (248)$$

with $\mu^2 = \lambda v^2$. In the following we consider as independent parameters for our analysis the set

$\{v, m_1, m_2, \theta, a_2, b_3, b_4\}$.

In order to obtain a viable SM + $S$ scenario, we need to satisfy several theoretical constraints which we discuss below:

• *(Perturbative) unitarity and perturbativity*: The size of the quartic scalar couplings in Eq. (247) is constrained by perturbative unitarity of the partial wave expansion of scattering amplitudes. The bound $|a_0| \leq 0.5$ for the leading order term in the partial wave expansion of the $h_2 h_2 \to h_2 h_2$ scattering amplitude, $a_0(h_2 h_2 \to h_2 h_2) = 3b_4/(8\pi)$, yields $b_4 < 4\pi/3$ (see e.g. [604]). In addition, we require perturbative values for $a_2$ and $b_3/v$: $|a_2| < 4\pi$, $|b_3|/v < 4\pi$.

• *Boundedness from below of scalar potential*: We require the absence of runaway directions in the scalar potential (247) at large field values. Along the $h$ and $S$ directions, this leads respectively to the bounds $\lambda > 0$ and $b_4 > 0$. For $a_2 < 0$ we further require $a_2 > -2\sqrt{\lambda \, b_4}$ to ensure boundedness from below along an arbitrary field direction.

• *Absolute stability of EW vacuum*: First, the EW vacuum $(\langle h \rangle, \langle S \rangle) = (v, 0)$ must be a minimum. On one hand, this requires $b_2 > 0$, which by virtue of (248) yields an upper bound on the value of $a_2$

$$a_2 < \frac{2}{v^2}(m_1^2 \sin^2\theta + m_2^2 \cos^2\theta). \tag{249}$$

On the other hand, for $(v, 0)$ to be a minimum the determinant of the scalar squared-mass matrix has to be positive

$$\mathrm{Det} \begin{pmatrix} \partial^2 V/\partial h^2 & \partial^2 V/\partial h \partial S \\ \partial^2 V/\partial h \partial S & \partial^2 V/\partial S^2 \end{pmatrix} \bigg|_{(v,0)} \equiv \mathrm{Det}\mathcal{M}_S^2 = 2\lambda v^2 b_2 - \frac{a_1^2 v^2}{4} > 0. \tag{250}$$

In addition, we require that the EW vacuum is the absolute minimum of the potential. The conditions for this are discussed in detail in [590], and we summarise them here. It will prove convenient to define the quantities

$$\begin{aligned}
\overline{\lambda}^2 &\equiv \lambda b_4 - \frac{a_2^2}{4}, \\
m_* &\equiv \frac{\lambda b_3}{3} - \frac{a_2 a_1}{8}, \\
\mathcal{D}^2(S) &\equiv v^2 \left(1 - \frac{a_1 S}{2\lambda v^2} - \frac{a_2 S^2}{2\lambda v^2}\right),
\end{aligned} \tag{251}$$
$$\tag{252}$$

with $h^2 = \mathcal{D}^2(S)$ corresponding to the minimization condition $\partial V/\partial h = 0$ for values $h \neq 0$. From the analysis of [590], we immediately find that a sufficient (though not necessary) condition for the EW vacuum to be the absolute minimum of $V$ is given by

$$\overline{\lambda}^2 > \frac{m_*^2 v^2}{16 \, \mathrm{Det}\mathcal{M}_S^2}. \tag{253}$$

When (253) is not satisfied, there exists for $\overline{\lambda}^2 > 0$ a minimum $S = \omega$ along $\mathcal{D}^2(S)$ which is deeper than the EW vacuum, and in order for the EW vacuum to still be the absolute minimum of $V$, it is necessary that $\mathcal{D}^2(\omega) < 0$ (in order for this new minimum to be unphysical). In addition, in this case we also need to require that no new minimum exists along the $h = 0$ field direction which is deeper than the EW one. The extrema along this direction are given by the real solutions of the equation

$$b_4 S^3 + b_3 S^2 + b_2 S + b_1 = 0. \tag{254}$$

Finally, when $\overline{\lambda}^2 < 0$ a necessary and sufficient condition for the EW vacuum to be the absolute minimum of $V$ is the absence of a deeper minimum along the $h = 0$ field direction, which we have just discussed above.

In Figures 90–92, we show, for fixed values of $m_2 = 300$ GeV, $500$ GeV, $700$ GeV and $\sin\theta = 0.1$, $0.05$, the points that satisfy the above requirements in the plane $a_2$, $b_3/v$, with the parameter $b_4$ being scanned over, indicated by red circles. We find that, for a given choice of $(a_2, b_3/v)$, the requirements are generically satisfied more robustly as $b_4$ increases[82], and as such we demand that there is a value of $b_4 \in [0, 4\pi/3]$ above which the EW vacuum is the absolute minimum of the potential.

We note that for large values of $a_2$ and $b_3$ the 1-loop corrections may become important and might allow for new regions that fulfil the above stability/unitarity/perturbativity conditions (see the discussion in [603]). We leave an investigation of the impact of 1-loop corrections on the above theoretical constraints for the future. We also note that, as compared to [603], our analysis has a smaller range of allowed values for $b_4$ which is partially responsible (together with the different chosen range for $m_2$) for the different shape of the tree-level allowed region.

### 6.1.4 EW phase transition in the SM + S

The EW symmetry is (generally) restored at high temperatures $T \gg v$. EW symmetry breaking then occurs when the temperature of the Universe drops due to expansion, and it becomes energetically favorable for the Higgs field $\Phi$ to acquire a non-zero expectation value $\varphi_h = v_T \neq 0$. When there exists a potential barrier separating the symmetric vacuum $\varphi_h = 0$ from the broken one $v_T$, the EW phase transition is of first order. The temperature at which the two vacua become degenerate in energy is known as the critical temperature $T_c$, and the EW phase transition is considered to be strongly first order if[83] $v_T(T_c)/T_c \gtrsim 1$.

For the analysis of the EW phase transition in the SM + $S$ scenario, we adopt in the following a conservative strategy: It is known that including the 1-loop $T = 0$ (Coleman-Weinberg) contributions to the effective potential introduces a gauge-dependence[84] in the evaluation of various phase transition parameters, such as $T_c$ [606–608]. However for a singlet-driven first order EW phase transition as in the SM + $S$, the properties of the transition are dominantly determined by tree-level effects. It is then possible in a first approximation to perform the analysis of the phase transition using the tree-level potential (247) augmented by the $T^2$ terms from the high-$T$ expansion of the finite-temperature effective potential (see e.g. [590]):

$$V_{T^2} = \left( \frac{c_h}{2} h^2 + \frac{c_s}{2} S^2 + c_t S \right) T^2, \tag{255}$$

where

$$c_h = \frac{1}{48} \left( 9g^2 + 3g'^2 + 12 y_t^2 + 24\lambda + 2a_2 \right),$$
$$c_s = \frac{1}{12} \left( 2a_2 + 3b_4 \right),$$
$$c_t = \frac{1}{12} \left( a_1 + b_3 \right),$$

as these are manifestly gauge invariant[85]. This approach, which we take in the present work, nevertheless disregards 1-loop terms that could be numerically important in certain regions of parameter space, particularly for large values of $a_2$ and/or $b_3$, strengthening the phase transition in those regions. We believe the choice made here then provides a conservative prediction for a strongly first order EW phase transition.

---

[82] This is true except in certain regions of $a_2 < 0$, where "islands of stability" in the parameter $b_4$ exist (that is, a very narrow range of $b_4$ within $[0, 4\pi/3]$ where the EW vacuum is the absolute minimum of the potential. These regions are however not relevant for the subsequent EW phase transition discussion, and we disregard them in the following.

[83] A more accurate criterion can be obtained by considering the "nucleation" temperature $T_n$ at which the phase transition actually takes place, and requiring $v_T(T_n)/T_n \gtrsim 1$. It is nevertheless a reasonable approximation in general to consider $v_T(T_c)/T_c \gtrsim 1$ instead.

[84] This gauge-dependence arises from the Goldstone and gauge boson contribution to the Coleman-Weinberg potential, as well as to the cubic term of the finite-temperature potential in the high-$T$ expansion (see [606] for a detailed discussion).

[85] The last term in (255) is gauge invariant at 1-loop, but not necessarily at higher loop order [596, 603]. Still, we choose here to keep it in the analysis (in contrast to [596, 603], where such term is discarded).

In the following we use the numerical programme COSMOTRANSITIONS [609] (v2.0.2) to find the points in parameter space with a viable strongly first order EW phase transition, for fixed values of $m_2$ and $\sin\theta$ while scanning over $a_2$, $b_3$ and $b_4$. Specifically, for each scan point we evolve the effective potential (combining (247) and (255)) from $T = 0$ up and look for coexisting and degenerate phases at some temperature(s) $T_i^* = T_c$. We consider the point to have a strongly first order EW phase transition when at (any) such temperature there is coexistence of a phase with $\varphi_h = 0$ (irrespectively of the singlet vacuum expectation value) and a phase with $\varphi_h = v_T$, separated by a potential barrier and such that $v_T/T_c > 1$. The results of our EW phase transition scan are shown in Figures 90–92. We also overlay the projected sensitivities from CLIC, as well as those from HL-LHC, all discussed in the next section. Our EW phase transition scan shows that, as the mass $m_2$ increases, the values of $a_2$ and $b_3/v$ required to achieve a strongly first order transition also increase substantially, approaching the perturbativity limit (particularly for $a_2$) for $m_2 \sim 700 - 800$ GeV. This yields a clear target reach for high-energy colliders regarding a singlet-driven EW phase transition[86].

### 6.1.5 CLIC sensitivity to the SM + S: probing the EW phase transition

We analyse here the CLIC prospects for probing the parameter space leading to a strongly first order EW phase transition in the SM + $S$ scenario, based on the results from the previous sections. In addition, we discuss the complementarity with probes of this parameter space from the HL-LHC [601].

Let us start by pointing out that due to the singlet-doublet mixing, the couplings of $h_1$ ($h_2$) to SM gauge bosons and fermions are universally rescaled w.r.t. the corresponding SM Higgs coupling values by $\cos\theta$ ($\sin\theta$). In addition to these, the tri-scalar interactions play an important role in the discussion of both di-Higgs production at colliders and the nature of the EW phase transition. Specifically, we focus on the interactions $\lambda_{211} h_2 h_1 h_1$ and $\lambda_{111} h_1 h_1 h_1$, which follow from (247) after EWSB, with

$$\begin{aligned}\lambda_{211} &= \frac{1}{4}\left[a_1 c_\theta^3 + 4v(a_2 - 3\lambda) c_\theta^2 s_\theta - 2(a_1 - 2 b_3) c_\theta s_\theta^2 - 2 a_2 v s_\theta^3\right], \\ \lambda_{111} &= \lambda v c_\theta^3 + \frac{1}{4} a_1 c_\theta^2 s_\theta + \frac{1}{2} a_2 v c_\theta s_\theta^2 + \frac{b_3}{3} s_\theta^3, \end{aligned} \quad (256)$$

with $c_\theta \equiv \cos\theta$ and $s_\theta \equiv \sin\theta$. The coupling $\lambda_{211}$ controls the partial width of the decay $h_2 \to h_1 h_1$ for $m_2 > 250$ GeV, given by

$$\Gamma_{h_2 \to h_1 h_1} = \frac{\lambda_{211}^2 \sqrt{1 - 4 m_1^2/m_2^2}}{8\pi m_2}. \quad (257)$$

Denoting by $\Gamma^{\mathrm{SM}}(m_2)$ the total width of a SM-like Higgs with mass $m_2$ (as given *e.g.* in [405]), the branching fraction BR($h_2 \to h_1 h_1$) is simply given by

$$\mathrm{BR}(h_2 \to h_1 h_1) = \frac{\Gamma_{h_2 \to h_1 h_1}}{\sin^2\theta\, \Gamma^{\mathrm{SM}}(m_2) + \Gamma_{h_2 \to h_1 h_1}}. \quad (258)$$

In the limit of high $m_2$ masses, this branching fraction is expected to be fixed by the Equivalence Theorem[87], BR($h_2 \to h_1 h_1$) $\simeq 0.25$, but different values of $a_2$ and $b_3$ can lead to some departure from this expectation. We show in Figure 89 the values of BR($h_2 \to h_1 h_1$) for $m_2 = 500$, 700 GeV and $\sin\theta = 0.05$ for illustration. At the same time, the production cross section for $h_2$ normalized to the SM value (for a given mass $m_2$) takes in the case of the SM + $S$ scenario the very simple form $\sigma_S/\sigma_S^{\mathrm{SM}} = \sin^2\theta$, due to the universal rescaling discussed above.

---

[86] We emphasize again that the 1-loop Coleman-Weinberg and finite-$T$ terms of the effective potential disregarded here will have some impact on the precise shape of the parameter space region yielding a strongly first order EW phase transition, and the value of $m_2$ above which such a strong transition stops being feasible. Yet, the bound $m_2 \lesssim 700 - 800$ GeV will not be significantly modified.

[87] We are indebted to Andrea Tesi for reminding us of this.

With all these ingredients, we can readily interpret both the HL-LHC and CLIC sensitivities to the parameter space of the SM + $S$ scenario. First, we note that the projected HL-LHC sensitivity to the singlet-doublet mixing from a global fit to the measured 125 GeV Higgs signal strengths is $\sin\theta \simeq 0.18$ (assuming negligible theory uncertainties; taking into account the present theory uncertainties the projected value is $\sin\theta \simeq 0.25$). In the present work we have thus always considered $\sin\theta$ to be smaller than this value. The interpretation of the sensitivity of direct searches in CLIC in the context of the SM + $S$ scenario is shown in Figures 90–92 for $m_2 = 300, 500, 700$ GeV and $\sin\theta = 0.1, 0.05$: we show the resonant di-Higgs production sensitivity of CLIC with $\sqrt{s} = 1.4$ TeV (orange) and $\sqrt{s} = 3$ TeV (blue) for a respective $b$-tagging efficiency of 70% (solid) and 90% (dashed), with CLIC able to probe the region not contained within each pair of sensitivity lines. For the case $\sin\theta = 0.1$ (for $\sin\theta = 0.05$ there is no sensitivity) we also show the HL-LHC sensitivity to the process $pp \to h_2 \to ZZ$ as a shadowed yellow region.

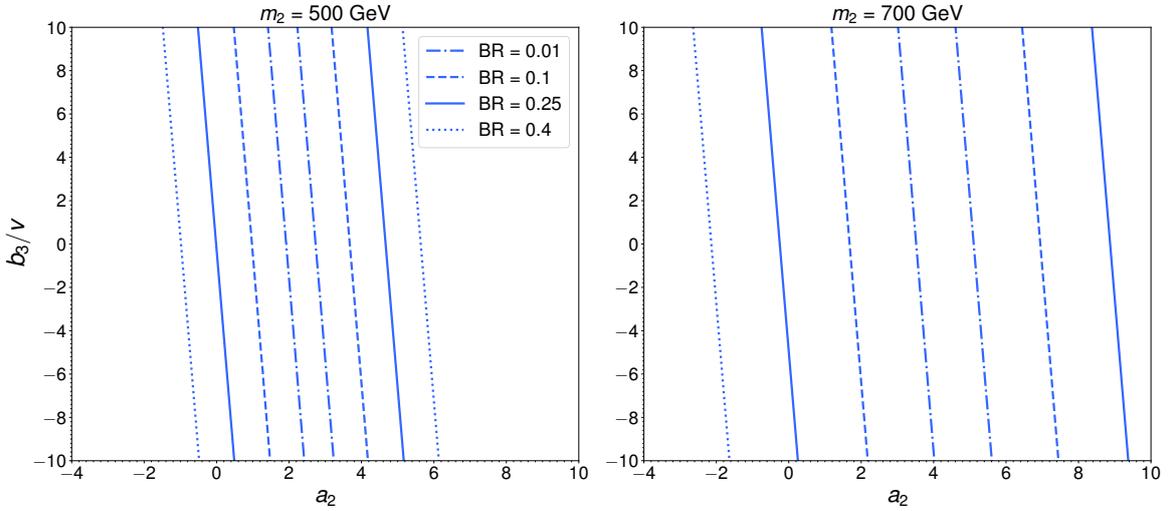

Figure 89: Branching fraction $\mathrm{BR}(h_2 \to h_1 h_1)$ values in the SM + $S$ scenario for $\sin\theta = 0.05$ and $m_2 = 500$ GeV (left), $m_2 = 700$ GeV (right) in the plane $(a_2, b_3/v)$.

In addition to the direct searches for $h_2$, we consider here two indirect collider probes of the SM + $S$ scenario:

*(i)* The measurement of the 125 GeV Higgs self-coupling $\lambda_{111}$. The projected sensitivity to the Higgs self-coupling at CLIC, combining the $\sqrt{1.4}$ TeV and $\sqrt{3}$ TeV runs is $\delta\lambda_{111} \equiv \left|\lambda_{111}^{\mathrm{SM}+S} - \lambda_{111}^{\mathrm{SM}}\right|/\lambda_{111}^{\mathrm{SM}} = 20\%$ (for a choice of beam polarization similar to the one considered in this work) [9], with $\lambda_{111}^{\mathrm{SM}} = \lambda\,v = 31.8$ GeV being the self-coupling value in the SM. For the Higgs self-coupling in the SM + $S$ scenario, we consider both the tree-level contribution from (256) and the 1-loop contribution computed to order $\sin\theta$ and given by [603] (note the different $\lambda_{111}$ normalization in our work w.r.t. [603]):

$$\Delta\lambda_{111}^{1-\mathrm{loop}} = \frac{1}{16\pi^2}\left(\frac{a_2^3\,v^3}{12\,m_2^2} + \frac{a_2^2\,b_3\,v^2}{2\,m_2^2}\sin\theta\right). \quad (259)$$

We then consider the region accessible to CLIC as $\left|(\lambda_{111} + \Delta\lambda_{111}^{1-\mathrm{loop}}) - \lambda_{111}^{\mathrm{SM}}\right|/\lambda_{111}^{\mathrm{SM}} = 0.20$ (the tree-level and 1-loop contributions given respectively by (256) and (259)), depicted in Figures 90–92 as a dashed-black curve. We nevertheless stress that it is not at all clear that the information on $\lambda_{111}^{\mathrm{SM}+S}$ from the non-resonant di-Higgs signal can be extracted from the data independently from the resonant di-Higgs contribution. In particular, since the non-resonant Higgs pair invariant mass distribution $m_{hh}$ peaks around $300 - 400$ GeV, for masses $m_2 \lesssim 500$ GeV disentangling the two contributions might be challenging.

*(ii) The measurement of the single-Higgs couplings at CLIC.* In this model all modifications to single Higgs processes follow from a universal rescaling of the Higgs interactions. We denote such rescaling as $\kappa$, as done in Section 2.1. Since such a correction cancels out in all branching ratios, the CLIC measurements of single Higgs processes can be interpreted as measurements of the production cross section via Higgsstrahlung or $W$ boson fusion. For a small singlet-doublet mixing (as we are considering here), the deviation in the Higgs production cross section with respect to its SM value is approximately given by (see e.g. [601, 603, 610]):

$$\delta\sigma_h(\approx 2\Delta\kappa) = \left| -\sin^2\theta + \frac{\lambda_{211}^2}{16\,\pi^2\,m_1^2}(1 - F(\tau)) \right|, \qquad (260)$$

where the first term is just the tree-level deviation and the second term corresponds to the leading 1-loop correction, with $\tau = m_1^2/(4m_2^2)$ and $F(\tau)$, $\lambda_{211}$ given by

$$F(\tau) = \frac{\text{Arcsin}(\sqrt{\tau})}{\sqrt{\tau(1-\tau)}}, \qquad (261)$$

$$\lambda_{221} = \frac{1}{2}\,a_2\,v\,c_\theta^3 + \left(b_3 - \frac{a_1}{2}\right)c_\theta^2 s_\theta + v(3\lambda - a_2)\,c_\theta s_\theta^2 + \frac{a_1}{4}\,s_\theta^3. \qquad (262)$$

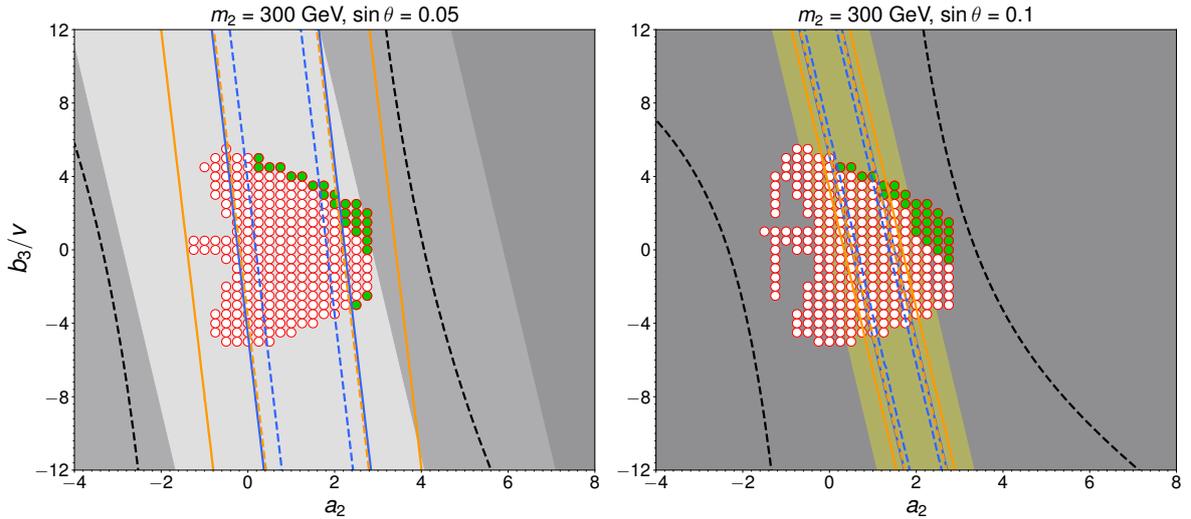

Figure 90: Region of parameter space in $(a_2, b_3/v)$ for $m_2 = 300$ GeV and $\sin\theta = 0.05$ (left), $\sin\theta = 0.1$ (right) within the 95% C.L. sensitivity reach of resonant di-Higgs production searches at CLIC with $\sqrt{s} = 1.4$ TeV (orange) and $\sqrt{s} = 3$ TeV (blue) for a $b$-tagging efficiency of 70% (solid) and 90% (dashed): CLIC sensitivity region is that not contained within each pair of (sensitivity) lines. The red circles indicate the region compatible with the requirements of unitary, perturbativity and absolute stability of the EW vacuum. The parameter $b_4$ has been scanned over (see text for details). Overlaid are the SM + S points compatible with unitary, perturbativity and absolute stability of the EW vacuum, and those yielding a strongly first order EW phase transition (green points). The dashed black lines correspond to the CLIC sensitivity to Higgs self-coupling deviations w.r.t. the SM $\delta\lambda_{111} = 0.20$. The yellow region (only for $\sin\theta = 0.1$) corresponds to the projected sensitivity of $pp \to h_2 \to ZZ$ searches at HL-LHC. The region within reach of a measurement of $\Delta\kappa$ at CLIC stage-1, stage-2 and stage3 are shown in dark, middle and light grey respectively. (See Table 6 in Section 2.1.)

In Figures 90–92 we show the indirect reach in the $(a_2, b_3/v)$ plane for fixed $m_2$ and $\sin\theta$ through the measurement of $\Delta\kappa$ at the different CLIC stages. For $\sin\theta = 0.1$, such a measurement of $\Delta\kappa$ at the first CLIC stage would already provide the most powerful constraint on the SM + S scenario, allowing to access the entire parameter space of the model. In contrast, for $\sin\theta = 0.05$ one needs to combine all three CLIC stages to be able to exclude the whole model parameter space. While in this particular model the resonant di-Higgs searches and the sensitivity to deviations in the Higgs self coupling are less constraining than the limits from single-Higgs coupling measurements for masses $m_2 \lesssim 500$ GeV, the latter could be significantly relaxed in extensions of the minimal scenario discussed here.

The yellow region corresponds to the projected sensitivity of $pp \to h_2 \to ZZ$ searches at the HL-LHC (these yield some sensitivity for $\sin\theta = 0.1$, but not for $\sin\theta = 0.05$). We note that LHC searches for $h_2 \to ZZ$ are most sensitive in the region of parameter space for $a_2$ and $b_3/v$ where the competing branching fraction $h_2 \to hh$ is smallest, as shown in Figure 89. LHC searches for $h_2 \to ZZ$ and resonant di-Higgs searches at CLIC, i.e. in the channel $e^+e^- \to v\bar{v}(h_2 \to hh)$, are then very much complementary (we note that, as shown in Figure 88, the sensitivity of resonant di-Higgs searches at the LHC is much worse than that of CLIC, and LHC resonant di-Higgs searches do not provide any meaningful constraint in our Figure).

The results from Figures 90–92 also highlight that it would be possible in many cases to simultaneously access via direct and indirect collider probes the region of parameter space yielding a strongly first order EW phase transition in the SM + S scenario. This would allow to correlate the information from the various probes towards providing a robust test of the nature of the EW phase transition.

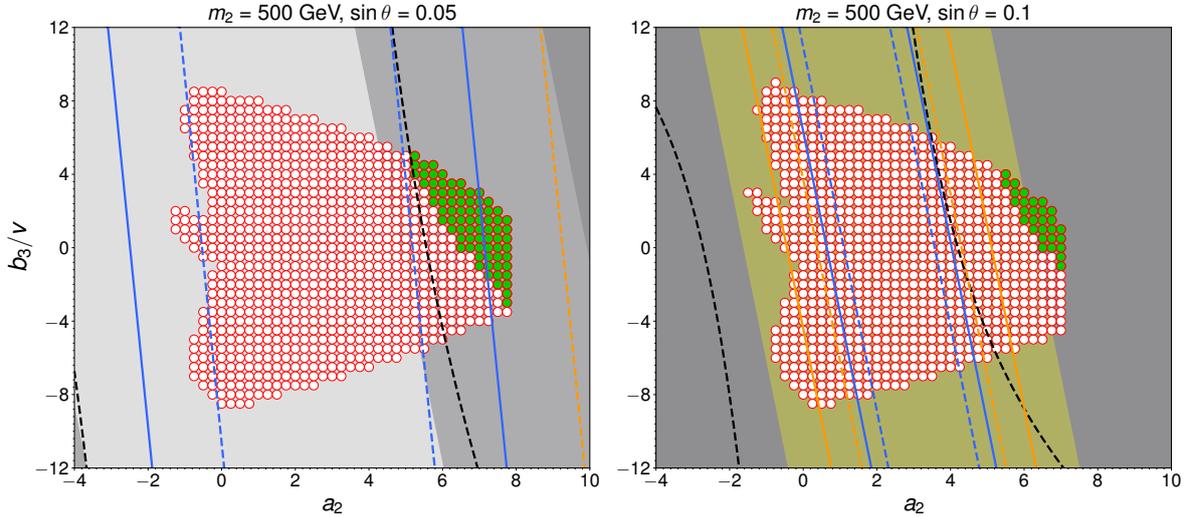

Figure 91: Same as Figure 90, but for $m_2 = 500$ GeV.

Before concluding, let us briefly comment on the limit of a vanishing singlet doublet mixing $\sin\theta \to 0$ (as is e.g. the case in the $Z_2$ symmetric limit of the SM + S scenario, discussed in Section 4.2). While the resonant di-Higgs signature vanishes in this limit, the indirect probes triple Higgs couplings and $\delta_\kappa$ at CLIC are still sensitive to the region yielding a strongly first order EW phase transition. This is shown in Figure 56 where it is shown that CLIC has enough sensitivity to probe the 1-loop corrections to triplet Higgs coupling and $\delta_\kappa$, which do not vanish as $\sin\theta \to 0$, and large regions of the parameter space for viable first-order phase transition can be excluded.

### *6.1.6 Conclusions*

Among the primary goals of future collider facilities is the precise analysis of the properties of the Higgs sector. We have shown in this work that a high-energy $e^+e^-$ machine like CLIC, when operating at multi-TeV c.o.m. energies, would yield very sensitive direct probes of the existence of new scalars, combining the energy reach with the clean environment of an electron-positron machine. In particular, resonant di-Higgs searches in the $4b$ final state at CLIC would surpass the reach of the HL-LHC by up to two orders of magnitude in the entire mass range $m_H \in [250\,\text{GeV}, 1\,\text{TeV}]$. At the same time, these searches provide a direct avenue to probe the nature of the EW phase transition for non-minimal scalar sectors, and the possible origin of the cosmic matter-antimatter asymmetry via EW baryogenesis.

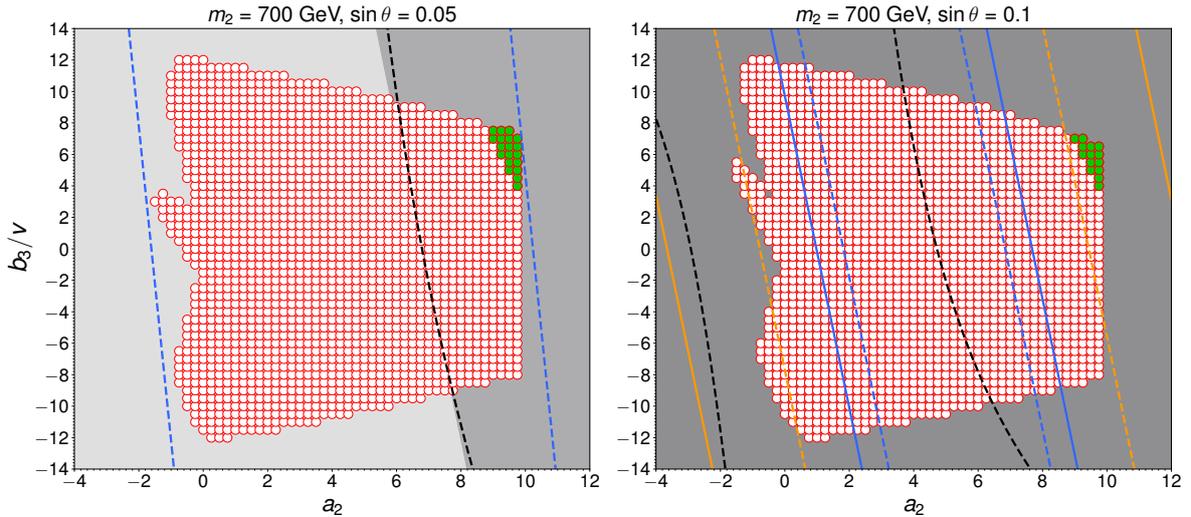

Figure 92: Same as Figure 90, but for $m_2 = 700$ GeV.

In the context of the extension of the SM by a real scalar singlet (SM + $S$, which could be viewed as a simple limit of the NMSSM or Twin Higgs theories), we have studied the sensitivity of CLIC to the parameter space where a strongly first order EW phase transition, as needed for successful baryogenesis, is realized. Our results show that there is a strong complementarity between direct searches for heavy Higgs bosons at CLIC via di-Higgs signatures, and indirect probes of BSM physics via measurements of the Higgs self-coupling $\lambda_{111}$ and the single Higgs measurements at CLIC. Combining the information from these searches could then allow to unravel the nature of EW symmetry breaking in the early Universe, and shed light on the origin of the baryon asymmetry of the Universe.

## 6.2 WIMP baryogenesis and displaced vertex [88]

### 6.2.1 Long-lived particles at the HL-LHC and CLIC: the case of WIMP baryogenesis

In this section, we highlight a class of baryogenesis models known as WIMP baryogenesis [611, 612] that can be *directly* tested with collider experiments, and we investigate the sensitivity of the high-energy colliders CLIC and the LHC to this scenario. The WIMP baryogenesis mechanism employs new EW-scale particles that decay out of thermal equilibrium via $B$- and $CP$-violating interactions. In such models, there is the exciting possibility of producing the parent particle(s) responsible for baryogenesis directly in a collider, and observing it decay through the same modes that generate the baryon asymmetry.

A particle decays out of equilibrium if its lifetime is longer than the Hubble time at a temperature comparable to its mass; $\tau_X > H^{-1}(T \sim M_X)$, where $M_X$ is the mass of the parent particle $X$ and $H(T)$ is the Hubble rate at temperature $T$. This gives a mass-dependent lower bound on the lifetime, generally predicting $c\tau_X \gtrsim 1$ cm for weak-scale $X$. Since the baryon asymmetry needs to be produced before the BBN, there is also an upper bound on this lifetime: $\tau_X \lesssim 1$ s, that is $c\tau_X \lesssim 10^8$ m in terms of proper decay length. If the primordial $X$ abundance decays prior to the EW phase transition, it may generate an asymmetry in either baryon or lepton number. However, if the lifetime is longer than $\sim 10^{-10}$ s, sphaleron processes are no longer active in the broken EW phase and hence $X$ must directly violate baryon number. Thus these baryogenesis models predict electroweak-scale, long-lived particles that decay hadronically.

---

[88]*Based on a contribution by Y. Cui, A. Joglekar, Z. Liu and B. Shuve.*

### 6.2.2 Model-independent features and phenomenology

We now summarize more quantitatively the model-independent features of baryogenesis in this scenario. We consider a weak scale particle $X$ that decays after its thermal freeze-out and triggers baryogenesis. A particle freezes out when its annihilation rate falls below the Hubble expansion rate. The temperature at freeze out, $T_{\text{fo}}$, depends only logarithmically on the annihilation cross section, such that $T_{\text{fo}} \sim M_X/20$ for annihilation cross sections $\sim$ fb.

The cosmological condition that $X$ decay out of equilibrium is

$$c\tau_X \gtrsim 1 \text{ cm} \left(\frac{100 \text{ GeV}}{M_X}\right)^2. \tag{263}$$

Scattering with the SM may keep $X$ in thermal equilibrium down to $T_{\text{fo}}$, in which case the decay length should be somewhat longer. If $X$ decays after freeze out, then the final baryon asymmetry is proportional to its would-be relic abundance if it does not decay (for details see [611]). In any case, it is clear that WIMP baryogenesis models predict new particles that can decay in various components of a detector at the LHC (e.g. ATLAS or CMS) or CLIC, but typically in the displaced vertex regime (or out of detector as missing energy) due to the above cosmological condition.

If the decay temperature, denoted by $T_d$, is less than the freeze-out temperature, $T_{\text{fo}} > T_d > T_{\text{BBN}}$, and assuming that we can neglect washout processes, the baryon asymmetry is given by

$$\Delta_B = \epsilon_{\text{CP}} n_X(T_{\text{fo}}), \tag{264}$$

where $\epsilon_{\text{CP}} < 1$ is a measure of $CP$ violation in the decays that can be generated by interference between tree-level and loop-level decay diagrams [611] or alternatives such as in [613]. Directly measuring such a CP violation effect tied to baryogenesis at collider experiments is exciting yet generally challenging. Therefore, we focus on displaced decay signals tied to the out-of-equilibrium condition for baryogenesis [586] recalled at the beginning of this Chapter as one of the necessary conditions to generate the baryon number of Universe.

It is important to note that the lifetime of the parent particle $X$ can be naturally very different from the couplings that lead to its production at the LHC. For example, if an approximately conserved $Z_2$ symmetry is responsible for the long lifetime, $X$ particles can still be produced in pairs via $Z_2$ conserving interactions but decay slowly through interactions that violate the symmetry. These particles could be copiously produced at the LHC and CLIC. An earlier study proposed simplified models for WIMP baryogenesis mechanisms along with studies of sensitivity to these models in some searches at ATLAS and CMS [614].

### 6.2.3 Example models

We briefly describe two benchmark models that have been proposed as concrete examples realizing the general idea of WIMP baryogenesis. These models utilize $B$-violating, out-of-equilibrium decay of a weak scale particle in order to satisfy the Sakharov conditions for baryogenesis, and thus generically predict displaced vertex signals at the LHC and CLIC.

**Model 1: Split SUSY with $R$-Parity Violation** Ref. [615] proposes a model that embeds the WIMP baryogenesis mechanism in mini-split SUSY with $R$-parity violation (RPV). The decays of the bino are responsible for generating the baryon asymmetry, and $CP$ violation arises from interference in the decays of the bino with intermediate wino states. Due to the very high mass of sfermions in such models, the rate of direct production of binos is negligible at the LHC. The dominant production of the new states responsible for baryogenesis is of the (nearly) pure wino, which has a lifetime comparable to the bino and is therefore also long-lived. The relevant effective interaction is:

$$\mathcal{L}_{\text{eff}} \supset \frac{\sqrt{2} g_2 \lambda''_{ijk}}{3 m_{\text{sq}}^2} T^a \tilde{W}^a \bar{u}_i \bar{d}_j \bar{d}_k + \text{h.c.}, \tag{265}$$

where heavy squarks with mass $m_{\text{sq}}$ are integrated out. The winos are pair produced via their gauge interactions.

**Model 2: Higgs-portal singlet** The model proposed in Ref. [611] incorporates new, meta-stable weak-scale particles $\chi$ (the parent responsible for baryogenesis) and $\psi$ (a heavier particle needed to violate $CP$) that couple to the SM via the Higgs portal and that decay to SM quarks through $B$-violating operators. This model can be embedded in the NMSSM with RPV couplings. The $\chi$ decay to SM quarks is mediated by a di-quark scalar $\phi$, leading to decays like $\chi \to u_i \phi^{(*)}$, $\phi^{(*)} \to d_j d_k$, where $i, j, k$ are flavor indices. The $\chi$ particles can annihilate to SM particles via an intermediate scalar $S$ that mixes with the Higgs boson. To avoid flavor constraints, $\chi$ decays preferentially to heavy-flavor quarks, and $\phi$ is typically heavy to evade dijet constraints. This model realizes $CP$ violation and the baryon asymmetry is predominantly generated in the decays of $\chi$ [611].

The LHC phenomenology can be derived by considering a low-energy effective description of the full UV model:

$$\mathcal{L}_{\text{eff}} \supset \frac{c_H}{\Lambda_H} \chi^2 |H|^2 + \frac{c_q}{\Lambda_q} g_{ijk} \chi u_i d_j d_k. \tag{266}$$

The LLP is produced via the Higgs portal $h\chi\chi$ interaction, either via an on-shell or off-shell $h$ depending on $m_\chi$ being above or below $m_h/2$, and decays to three (anti-)quarks. Note that the $\chi$ decay may have an on- or off-shell mediator $\phi$, which would somewhat change the $\chi$ decay kinematics.

### 6.2.4 Analysis

The phenomenology of the two models from Section 6.2.3 are quite distinct. In the split-SUSY model described above, the gauge charge of the wino give an irreducible pair production cross section for the winos via the weak interactions. Existing constraints require $M_X \gtrsim 700$ GeV (see Ref. [614] and Figure 93 below). Therefore, the LLPs in the split-SUSY scenario will be heavy, giving rise to signatures with large transverse momenta. In the second model where the LLP couples to the SM via the Higgs portal, the dominant production is via on- or off-shell Higgs decays. The cross section declines precipitously for $2M_X > 125$ GeV, and so the phenomenologically accessible parameter space is for low-mass LLPs, $M_X \lesssim 200$ GeV.

#### 6.2.4.1 RPV split-SUSY (wino) model

In WIMP baryogenesis models, the LLP often decays at least partially hadronically. Thus, searches for displaced jets or high-track-multiplicity vertices constrain this scenario. We recast a CMS search for displaced dijets [616], which provided the dominant constraints on LLP signatures in RPV-SUSY models [614, 617]; similar results can be obtained by recasting the ATLAS multi-track vertex search [618] and the CMS displaced multijet search [619]. Our recast is based upon the simulation methods of Ref. [617], and more details on the simulation framework can be found there.

To give an example of the sensitivity of the HL-LHC, we show the projected sensitivity for long-lived RPV wino pair production at HL-LHC along with the constraints from earlier LHC runs. In Figure 93 we show the results for RPV wino from the split-SUSY model. In the left panel, we show the displaced dijet efficiencies for various masses of the RPV wino at 13 TeV LHC, for masses between 100 to 1500 GeV, as a function of the proper lifetime of the LLP. The maximal signal effciency approaches 50% for heavy winos with $c\tau \sim 0.1$ m, where the majority of the decays take place before or in the earlier part of the pixel tracking system, satisfying track reconstruction quality and impact parameter requirements for tracks in the displaced jets. For lower mass winos, the signal efficiency suffers from the low jet $p_T$ that makes it difficult to satisfy the $p_T > 60$ GeV requirement for signal selection; at higher masses, the signal is fully efficient for the $p_T$ selection.

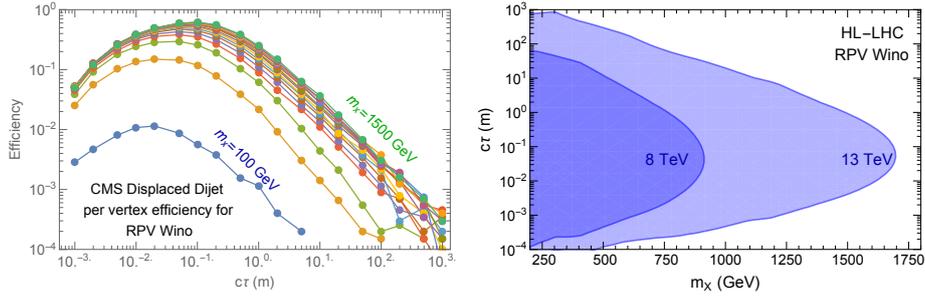

Figure 93: Left panel: the signal selection efficiency by the CMS displaced dijet search for the RPV wino signals as a function of their lifetimes and masses. Right panel: the recasted current and HL-LHC ($3\ \mathrm{ab}^{-1}$) projected 95% C.L. exclusion limit for the RPV wino signal as the function of wino mass and its lifetime.

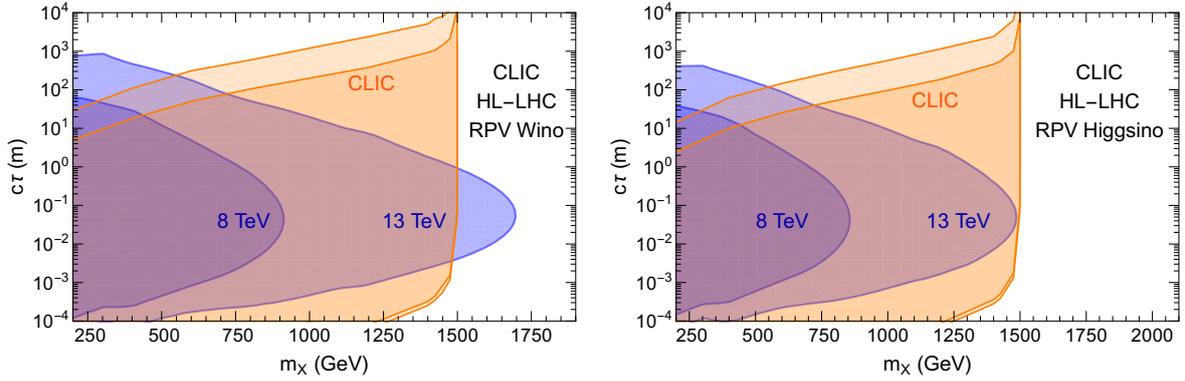

Figure 94: Event rates and exclusions for the wino and higgsino signal in the lifetime vs. mass plane. Orange: darker region corresponds to $N > 30$ events in the CLIC acceptance, lighter orange regions corresponds to $N > 3$ events and correspond to a projected 95% C.L. exclusion limit for zero expected background. The left (right) panel refers to the RPV wino (higgsino) signal. Blue region: same as in Figure 93

In the right panel of Figure 93, we show the current and projected sensitivity at the LHC for this RPV wino benchmark. We can see that the HL-LHC improves the current limit from around 800 GeV up to 1650 GeV for $c\tau \sim 0.1$ m. Note that, despite the small efficiency at low masses, the large wino production cross section still leads to the model being excluded. The coverage extends to long and short lifetime as well, covering 0.1 millimetres to 500 metres for a 500 GeV wino. These pair-produced winos have low boost factors and therefore move slowly. Further development in using the precision timing for LLPs at the LHC, similar to the GMSB Higgsino benchmark study in Ref. [620], could improve the HL-LHC sensitivity significantly, especially for the long lifetime regime.

As we can see from Figure 93, the advantage of high collision energy enables the LHC to cover wino mass up to 1650 GeV in the most sensitive $c\tau$ range ($\sim 10$ cm). 3 TeV CLIC thus cannot compete with LHC in terms of the mass reach of wino in general. However, Figure 93 also shows that there is ample parameter space in $c\tau$ at masses below 1.5 TeV that HL-LHC are not sensitive to. This is due to both the large QCD background at the LHC and the current limit in vertex reconstruction efficiency. In contrast, CLIC, as a $e^+e^-$ collider, provides a much cleaner environment for these searches, with almost full coverage for electroweak states below 1.5 TeV mass. With much lower background (in particular for the hadronic channel) and improved vertex reconstruction techniques, CLIC has the great potential to close up the region that HL-LHC is not capable of effectively probing, which is illustrated in Figure 94.

In Figure 94, the projected exclusion limit for 3 TeV CLIC at 95% C.L. for the luminosity of $3\,\text{ab}^{-1}$ is indicated by the orange region in the wino mass and $c\tau$ parameter space, overlaid on the blue regions showing the LHC sensitivity copied from Figure 93. Here we simulated pair production of wino-like charginos at 3 TeV. The charginos almost exclusively decay to wino-like neutralino and a soft pion, since the couplings to bino-like neutralino states are heavily suppressed by large $\mu$ term. The wino-like neutralino decays hadronically via RPV couplings. We make a simplified assumption for charginos: $c\tau_{\chi^\pm \to \chi^0} \ll c\tau_{\chi^\pm(\text{RPV})}$, so that tracks contributing to DVs come entirely from wino-like neutralinos and the track coming from the soft pion emitted by chargino decay is not associated with any vertex. For the analysis we assume a nearly perfect vertex reconstruction efficiency in the $c\tau$ range of $0.3 - 100$ mm as suggested in e.g. [621]. It is evident that CLIC at 3 TeV with $3\,\text{ab}^{-1}$ luminosity is sensitive to the large parts of parameter space in Figure 93 that LHC is not, below the kinematic limit of wino production of 1500 GeV. It almost entirely covers the uncovered parameter space in the Figure 93 for $c\tau > 1$ cm and $m_\chi < 1500$ GeV. For lower $c\tau$, CLIC can offer up to an order of magnitude improvement in terms of the reach in $c\tau$. Similar considerations can be applied to the case in which the displaced decays is coming from an SU(2) doublet particle, such as an Higgsino with RPV decays. As in the wino case, the tree level chargino production is mediated through photon and Z boson. The coupling between the higgsino-like charginos to Z boson is significantly reduced compared to the coupling between wino-like charginos and Z boson. Therefore, the cross-section for higgsino pair production is reduced compared to the wino pair production. This significantly affects the mass reach at the LHC, while leaving the mass reach at CLIC almost unchanged. This makes CLIC competitive with the LHC in terms of the mass reach as evident from the right panel of Figure 93. All in all we conclude that CLIC will significantly extend the reach of previous hadron colliders for searching long lived electroweak charged particles and will explore previously uncharted territory in the parameters space of models for the generation of the net baryon number of Universe.

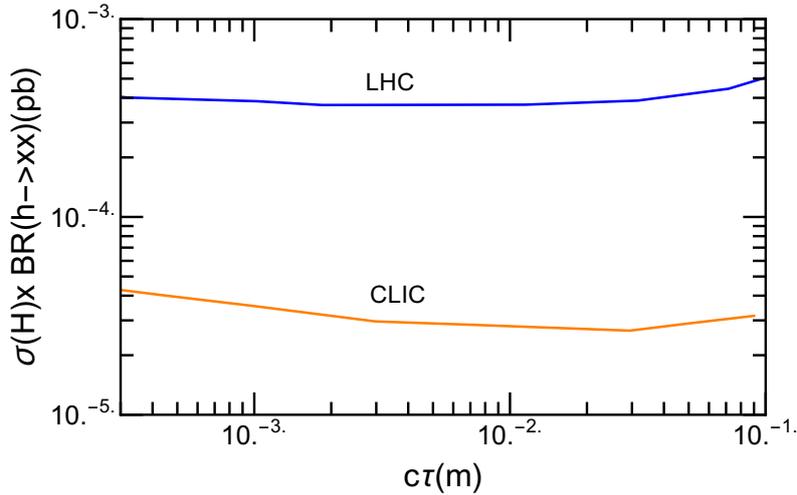

Figure 95: Blue line: HL-LHC projected 95% C.L. exclusion limit for the Higgs portal singlet model as the function of $c\tau_\chi$ for $m_\chi = 30$ GeV. Orange line: projection for CLIC with the same model.

*6.2.4.2 Higgs-portal singlet model*

The Higgs-portal singlet models like the one proposed in Ref. [611] have haronically decaying weak scale mass singlet ($\chi$) particles with $c\tau \gtrsim \mathcal{O}(\text{mm})$ that are motivated by WIMP baryogenesis scenario. In Figure 95, we compare the 95% C.L. reach of HL-LHC with CLIC for this class of models. LHC sensitivity of various classes of Higgs portal models is studied in [622, 623]. These include Twin Higgs models, Folded SUSY models, quirky little Higgs models etc. Similar LHC sensitivity is obtained for the Higgs portal singlet model embedded in RPV-NMSSM that decays to SM quarks via RPV couplings as shown by the blue line in the Figure 95 for $m_\chi = 30$ GeV.

At CLIC, the dominant mode of Higgs production is via W fusion, which has the cross section an order of magnitude lower than that at the LHC. Since we are dealing with on-shell production of light states, the cleaner environments and vertex reconstruction efficiencies can enable CLIC to have better $c\tau_\chi$ coverage than HL-LHC for a given mass of the exotic particle similar to the heavier case of RPV wino discussed before. This can be observed in Figure 95. CLIC sensitivity to $h \to \chi\chi$ at 3 TeV for 95% C.L. is projected as indicated by the orange line using the sensitivity given for the Hidden valley models in [621]. CLIC will clearly have an order of magnitude better reach in the $c\tau_\chi$ range favored by the WIMP baryogenesis models with light singlets ($< 100$ GeV).

## 7 New neutrinos and see-saw mediators

The Standard Model, despite of its many major successes, suffers from its inability to explain the observed light neutrino masses and mixings. In contrast to the other particles of the SM, the three neutrinos have extremely small $\sim$ eV masses. The measured solar and atmospheric mass-square differences from neutrino oscillation experiments are about $10^{-5}$ eV$^2$ and $10^{-3}$eV$^2$, and the mixing angles are approximately 34-degrees, 42-degrees and 8-degrees [624]. Recent bounds from cosmology have further constrained the SM neutrino masses to be less than about 1 eV [578]. An explanation of these measurements requires beyond the Standard Model physics and is one of the outstanding goals of the high-energy physics program.

Over the past decades, there have been several important developments at the theory and experimental frontiers to address the key questions of how neutrino masses are generated. The most widely adopted approach to explain small neutrino masses is the so-called seesaw mechanism [625] where the light neutrinos receive Majorana masses from a dimension-5 operator. Depending on the envisioned underlying high-scale model, the simplest seesaw mechanisms can be categorised into Type-I [143, 145, 146], Type-II and Type-III [148, 626] seesaw scenarios. The Type-I and Type-II models can further be embedded into the Left-Right Symmetric Model [140–142]. The model extends the Standard Model particle content by three generations of singlet heavy neutrinos $N_R$, that are introduced as the parity gauge partners of the corresponding left-handed neutrino fields, a triplet scalar field under $SU(2)_L$, a triplet scalar field under $SU(2)_R$ and gauge bosons related to the spontaneously broken gauge groups $SU(2)_R$ and $U(1)_{B-L}$. Further, the Higgs field becomes a bi-doublet under $SU(2)_L \times SU(2)_R$. This model can naturally explain small neutrino masses through the Type-I seesaw (RH neutrinos) and Type-II seesaw ($SU(2)$-triplet scalars) mechanisms.

Thus, models addressing the generation of neutrino masses often predict additional gauge bosons and scalar particles which can be searched for at CLIC.

In Section 7.1 we evaluate CLIC's sensitivity in probing low-scale seesaw models using the $W^+W^-H$ production process. While this analysis is performed for the inverse seesaw mechanism, we expect the results to hold for other low-scale seesaw models as well. This process is a particularly interesting probe for models with diagonal and real Yukawa couplings which are otherwise difficult to test.

Double charged scalars are a striking signal for new physics, often associated with seesaw models, where they can arise as SU(2) triplets or singlets. Doubly charged Higgs scalars descending from a triplet representation are studied in Section 7.2, where they manifest themselves as resonance peaks in invariant mass distributions, and the $SU(2)$-singlet scenario in Section 7.3, where searches in s- and t-channel processes are proposed.

The Zee model [627] is one of the simplest neutrino–mass generation scenarios, where light neutrino masses are generated at one–loop level. The model consists of an extra scalar doublet and a charged singlet scalar field, which can be embedded in Left-Right Symmetric extensions. Due to the relatively large coupling of the charged Higgs boson to leptons, its production cross-section at $e^+e^-$ colliders is strongly enhanced compared to hadron colliders. This feature makes it an interesting scenario to study at CLIC, see Section 7.4.

### 7.1 New neutrinos and large mixing see-saw models in the $W^+W^-H$ process [89]

Low-scale seesaw models are attractive scenarios to account for the tiny masses of the three light active neutrinos. They introduce new fermionic gauge singlets that can naturally have large Yukawa couplings in the presence of a nearly conserved lepton number symmetry [628, 629], opening new search strategies involving that particle (see for example [630–647]). This section presents a novel way to probe these neutrino mass models at lepton colliders, using the $W^+W^-H$ production process. As an illustrative example, we present a study in the inverse seesaw (ISS) but we expect our results to hold for other low-seesaw models. More details can be found in the original study [648].

---
[89]Based on a contribution by J. Baglio and C. Weiland.

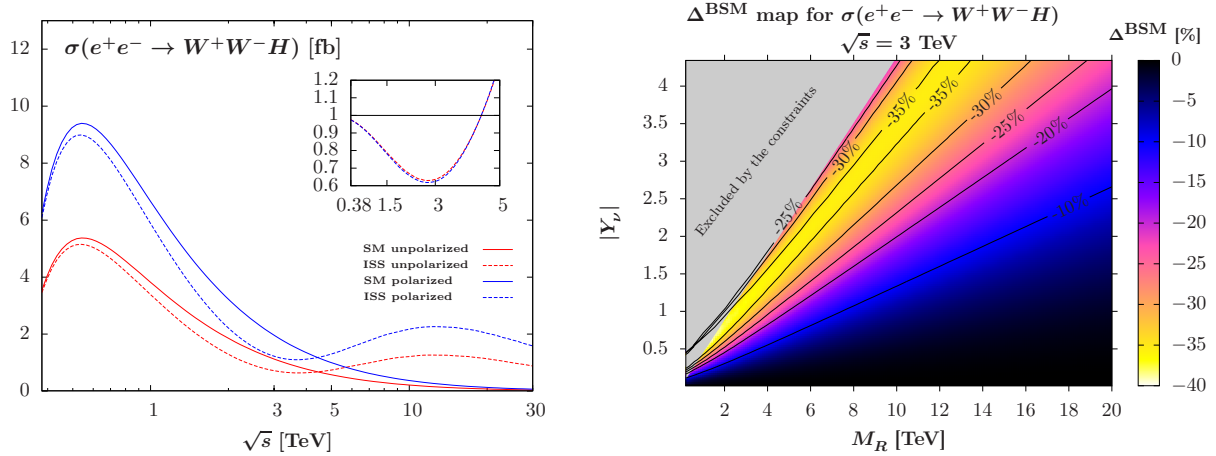

Figure 96: Left: LO total cross-section $\sigma(e^+e^- \to W^+W^-H)$ as a function of the collider energy $\sqrt{s}$. The solid curves are for the SM predictions while the dashed curves stand for the ISS predictions using a benchmark scenario defined in the text. The red (blue) curves are for an unpolarized ($-80\%$ polarized electron beam) cross-section. The ratio of the ISS prediction with respect to the SM cross-section is displayed in the insert up to 5 TeV. Right: Contour map of the neutrino corrections $\Delta^{\text{BSM}}$ at the 3 TeV CLIC, using a $-80\%$ polarized electron beam, as a function of the seesaw scale $M_R$ and $|Y_\nu|$. The grey area is excluded by the constraints.

In these models, the heavy neutrinos form pseudo-Dirac pairs that couple to the Standard Model (SM) particles through their potentially large mixing with SM fields. The calculation of the total cross-section $\sigma(e^+e^- \to W^+W^-H)$ is performed at leading order (LO)[649], comparing the case with unpolarized beams to the case with a polarized electron beam using $P_{e^-} = -80\%$, based on the Compact Linear Collider (CLIC) baseline [8]. The SM Higgs boson mass is fixed to $M_H = 125$ GeV while the other SM parameters are fixed to their Particle Data Group values [168]. Low-energy neutrino data from the global fit NuFIT 3.0 [650] are used as input of the $\mu_X$-parametrization [637], that was extended in [645]. Experimental constraints are dominated by the global fit to electroweak precision observables and low-energy data [651], see Ref. [648] for the complete list including theoretical constraints as well.

Our results are presented in Figure 96 in terms of deviations with respect to the SM prediction,

$$\Delta^{\text{BSM}} = \frac{\sigma^{\text{ISS}} - \sigma^{\text{SM}}}{\sigma^{\text{SM}}}, \qquad (267)$$

with $\sigma^{\text{ISS}}$ being the cross-section calculated in the ISS model. To illustrate the heavy neutrino effects in the ISS, a diagonal Yukawa texture $Y_\nu = |Y_\nu|I_3$ is used as well as a hierarchical heavy neutrino mass matrix with $M_{R_1} = 1.51 M_R$, $M_{R_2} = 3.59 M_R$, and $M_{R_3} = M_R$. Figure 96 (left) presents the variation of the total production cross-section $\sigma(e^+e^- \to W^+W^-H)$ as a function of the collider energy $\sqrt{s}$, using a benchmark scenario with $|Y_\nu| = 1$ and $M_R = 2.4$ TeV. We have extended here our previous results to 30 TeV, in order to provide predictions that could be relevant to future lepton colliders such as the ALEGRO project or the Low EMittance Muon Accelerator (LEMMA) project (see e.g. [652]. We emphasize that our results can be directly translated to a muon collider by doing the exchanging $M_{R_1}$ and $M_{R_2}$ values in our calculation, leading to larger deviations at lower collider energies. First, our results demonstrate the gain in cross-section from using polarized beams. Second, we see that below $M_{R_1} \approx 3.6$ TeV which is the mass of the t-channel heavy neutrino, negative interferences decrease the total cross-section, leading to a maximum correction of $-38\%$ at an energy close to 3 TeV, while for larger masses the correction leads to an enhanced cross-section, giving a second maximum above 10 TeV that is absent in the SM.

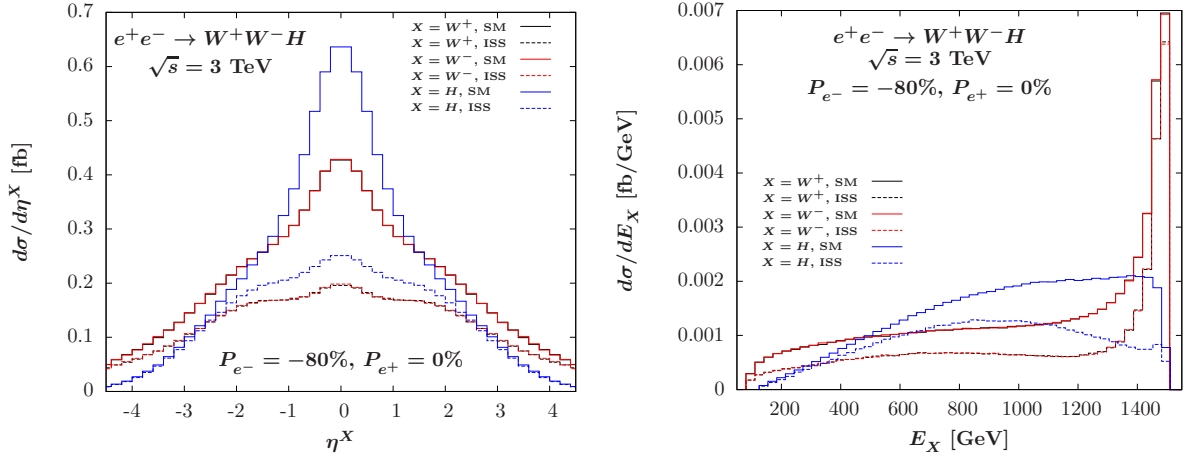

Figure 97: Pseudo-rapidity (left) and energy (right) distributions of the $W^+$ (black), $W^-$ (red) and Higgs (blue) bosons in the process $e^+e^- \to W^+W^-H$ at $\sqrt{s} = 3$ TeV, using a $-80\%$ polarized electron beam. The solid curves stand for the SM predictions, the dashed curves stand for the ISS predictions using the benchmark scenario described in the text.

Figure 96 (right) presents the contour map of the heavy neutrino corrections as a function of the two parameters $|Y_\nu|$ and $M_R$ at the 3 TeV CLIC with a -80% polarized electron beam. The size of the grey area excluded by the constraints is mostly driven by the global fit [651]. Sizable deviation of at least $-20\%$ are allowed for a large fraction of the parameter space, much larger than the similar map obtained with the study of the triple Higgs coupling in [645]. The largest effect in the ISS reaches a $-38\%$ deviation for Yukawa couplings $|Y_\nu| \sim 1$ and a seesaw scale of a few TeV. These results can be approximated within 1% by the following formula for $M_R > 3$ TeV,

$$\begin{aligned}
\mathcal{A}_{\text{approx}}^{\text{ISS}} &= \frac{(1\,\text{TeV})^2}{M_R^2} \text{Tr}(Y_\nu Y_\nu^\dagger) \left( 17.07 - \frac{19.79\,\text{TeV}^2}{M_R^2} \right), \\
\Delta_{\text{approx}}^{\text{BSM}} &= (\mathcal{A}_{\text{approx}}^{\text{ISS}})^2 - 11.94\, \mathcal{A}_{\text{approx}}^{\text{ISS}}.
\end{aligned} \quad (268)$$

As can be seen from the kinematic distribution presented in Figure 97, the shape of the SM and ISS are clearly distinguishable, with a marked difference in the central region and for boosted Higgs bosons. As discussed in [648], the deviation from the ISS can thus be enhanced with a simple choice of cuts. It was found that with $|\eta_{H/W^\pm}| < 1$ and $E_H > 1$ TeV it is possible to reach corrections down to $-66\%$ without decreasing the cross-section by more than an order of magnitude. The deviations $\Delta^{\text{BSM}}$ are found to be large enough in a significant fraction of the parameter space to trigger a detailed sensitivity analysis of the process $\ell^+\ell^- \to W^+W^-H$ at lepton colliders, $\ell = e/\mu$. This makes this process a new probe of neutrino mass models, allowing testing of regimes with diagonal and real Yukawa couplings difficult to access otherwise, which is highly complementary to other existing probes such as lepton flavour violating processes in the $\mathcal{O}(10)$ TeV range of the heavy neutrino masses.

## 7.2 Doubly-charged mediators from type-2 see-saw models [90]

The most characteristic feature of the Type II seesaw model is the presence of the doubly-charged Higgs boson $H^{\pm\pm}$, that can decay into same-sign leptonic or bosonic states and gives unique signatures at high energy colliders. The pair-production of $H^{\pm\pm}H^{\mp\mp}$ occurs via $s$-channel $Z$, and $\gamma$ mediated diagrams,

---

[90]*Based on a contribution by M. Mitra.*

where the $H^{\pm\pm}H^{\mp\mp}V$ vertices ($V$ being $Z$ and $\gamma$) depend on gauge couplings and are independent of the triplet vev $v_\Delta$. As shown in Figure 98, the production cross-section is $\sigma_p \sim 400$ fb - 0.1 fb for a wide range of doubly charged Higgs mass, and different centre-of-mass energies (c.m.energies). The produced $H^{\pm\pm}$ can decay to leptonic final states, i.e., $H^{\pm\pm} \to l^\pm l^\pm$, and to same-sign gauge bosons $H^{\pm\pm} \to W^\pm W^\pm$. For the choice $M_{H^{\pm\pm}} < M_{H^\pm}$, the decay mode into singly-charged Higgs $H^{\pm\pm} \to H^\pm W^*$ is absent. The branching ratios of $H^{\pm\pm}$ into different decay modes depend crucially on the triplet vev $v_\Delta$. For smaller triplet vev, $v_\Delta \lesssim 10^{-4}$ GeV, the $H^{\pm\pm}$ predominantly decays into same-sign leptons $H^{\pm\pm} \to l^\pm l^\pm$, whereas for larger $v_\Delta$, the gauge boson mode $H^{\pm\pm} \to W^\pm W^\pm$ becomes dominant. The dependency of the branching ratios on the triplet vev can be seen from the right panel of Figure 98. Therefore, identifying the dominant decay of $H^{\pm\pm}$ at colliders can possibly hint towards the possibilities $v_\Delta \gtrsim 10^{-4}$ GeV, or $v_\Delta \lesssim 10^{-4}$ GeV.

So far, a number of searches have been performed at LEP-II and LHC to discover $H^{\pm\pm}$ in the decay mode $H^{\pm\pm} \to l^\pm l^\pm$, which constrains the small $v_\Delta$ region. The CMS collaboration looked for different leptonic flavors including $ee, e\mu, e\tau, \mu\mu, \mu\tau$ and $\tau\tau$. In addition, the CMS searches also include the associated production $pp \to H^{\pm\pm}H^\mp$ and the subsequent decays, $H^\pm \to l^\pm \nu$. This combined channel of pair-production and associated production gives the stringent constraint $M_{H^{\pm\pm}} > 820$ GeV [176] at 95% C.L for $e, \mu$ flavor. The constraint from ATLAS searches is $M_{H^{\pm\pm}} > 870$ GeV at 95% C.L [653]. Searches have also been performed at LEP-II that results in $M_{H^{\pm\pm}} > 97.3$ GeV [654] at 95% C.L. Search of $H^{\pm\pm}$ in vector boson fusion channel [655, 656] gives a very relaxed constraint, although this is relevant for larger $v_\Delta$. Note that the VBF cross-section scales quadratically with the triplet vev, and hence increases for a very large vev. However, the range of $v_\Delta \sim 10^{-4} - 10^{-1}$ GeV cannot be probed at the 13 TeV LHC in VBF channel, as the cross section becomes extremely small in this range. Additionally, for large mass of the doubly-charged Higgs, the LHC cross section becomes significantly low, as shown in Figure 98. On the other hand, the fall in the cross section at a $e^+e^-$ collider is relatively smaller. This motivates to exploration of the signatures of doubly-charged Higgs at a lepton collider [657], where the cross section remains larger for heavy charged Higgs masses.

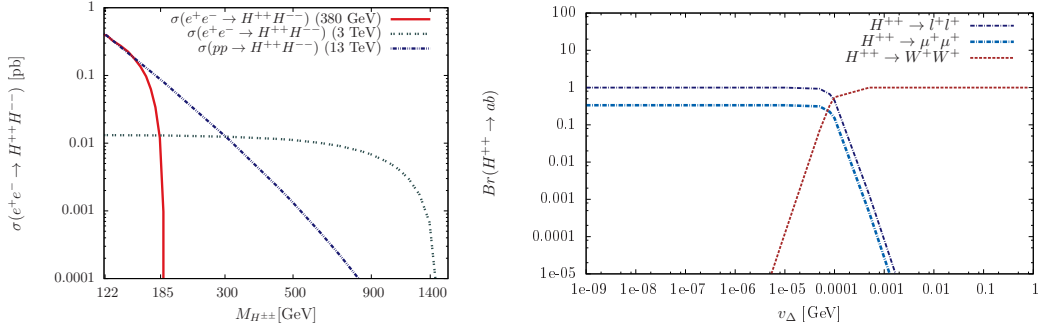

Figure 98: Left panel: Production cross section at $e^+e^-$ collider. The c.m.energies are $\sqrt{s} = 380$ GeV and 3 TeV. For comparison, the cross section at 13 TeV LHC has also been shown. The pair-production cross section increases by a factor of two, if CLIC uses 80%, and 30% beam polarization for electron and positron beam. Right panel: Branching ratio of $H^{\pm\pm}$ into leptonic and gauge boson decay modes.

The results presented in [657] correspond to the simulation of all hadronic final states at CLIC, assuming two different c.m.energies $\sqrt{s} = 380$ GeV and 3 TeV. For definiteness, the large triplet vev $v_\Delta = 10^{-2}$ GeV has been considered. The results are however valid for a wide range of $v_\Delta \sim 10^{-4} - 1$ GeV. For the choice of $v_\Delta$, the produced $H^{\pm\pm}$ will decay into $W^\pm W^\pm$ gauge bosons with almost 100% branching ratio. The hadronic final states from $W^\pm$ decays have been analysed in detail. The discovery prospect of two distinct mass regions have been explored - A) low mass $H^{\pm\pm}$ at $\sqrt{s} = 380$ GeV and B) heavy $H^{\pm\pm}$ at $\sqrt{s} = 3$ TeV.

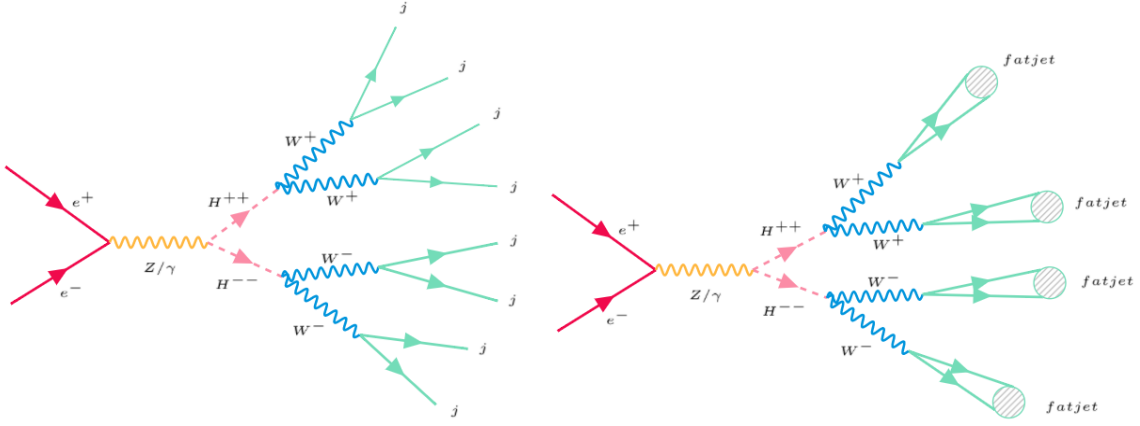

Figure 99: Left panel: The Feynman diagram for $H^{++}H^{--}$ pair-production and its subsequent decays into gauge bosons, relevant for 380 GeV analysis. Right panel: Same production mode of $H^{\pm\pm}$ and its hadronic decays leading to 4 fat-jets, relevant for 3 TeV analysis.

**A) Low mass $H^{\pm\pm}$ at $\sqrt{s} = 380$ GeV:**

Lighter Higgs states with mass up to $M_{H^{\pm\pm}} \sim 190$ GeV can be probed at 380 GeV c.m. energy. The model signature in this scenario is $e^+e^- \to H^{\pm\pm}H^{\mp\mp} \to 4W \geq 7j$ for $M_{H^{\pm\pm}} \gtrsim 2M_W$ (on-shell decay of $H^{\pm\pm} \to W^+W^+$), and $e^+e^- \to H^{\pm\pm}H^{\mp\mp} \to W^\pm jj W^\mp jj \geq 7j$ for $M_{H^{\pm\pm}} < 2M_W$ (off-shell decay of $H^{\pm\pm}$ to $W^+jj$).

Table 42: The cross sections for the signal and background for the fully hadronic final states, arising from $e^+e^- \to H^{\pm\pm}H^{\mp\mp}$. $\sigma_p$ refers to the partonic cross section. $\sigma_d$ is the cross section after taking into account detector effects. The last column represents the cross section with $b$-veto. The c.m.energy is $\sqrt{s} = 380$ GeV.

| | $e^+e^- \to H^{++}H^{--} \to N_j \geq 7j$ | | |
|---|---|---|---|
| Mass (GeV) | $\sigma_p$ (fb) | $\sigma_d(N_j \geq 7j)$ (fb) | $\sigma_d(N_j \geq 7j + b\,\text{veto})$ (fb) |
| 121 | 0.80 | 0.30 | 0.20 |
| 137 | 2.08 | 0.94 | 0.66 |
| 159 | 5.45 | 2.58 | 1.82 |
| 172 | 5.04 | 2.48 | 1.74 |
| 184 | 1.11 | 0.53 | 0.38 |

| | Backgrounds | | |
|---|---|---|---|
| Processes | $\sigma_p$ (fb) $\times 10^{-2}$ | $\sigma_d(N_j \geq 7j)$ (fb)$\times 10^{-2}$ | $\sigma_d(N_j \geq 7j + b\,\text{veto})$ (fb) $\times 10^{-2}$ |
| $e^+e^- \to t\bar{t} \to 6j$ | 10341.0 | 338.0 | 36.0 |
| $W^+W^-3j, W^\pm \to 2j$ | 8.89 | 1.18 | 0.88 |
| $ZZ + 3j, Z \to 2j$ | 0.98 | 0.13 | 0.10 |
| $7j$ | 30.32 | 1.13 | 0.88 |
| $W^\pm + 5j, W^\pm \to jj$ | 30.18 | 4.64 | 3.54 |
| $Z + 5j, Z \to jj$ | 18.32 | 2.15 | 1.61 |

The event simulation has been carried out using `FeynRules` [383]-`MadGraph5_aMC@NLO` [31]-`Pythia 6` [658]. The detector simulation has been taken into account by `Delphes-3.3.0` [385], with the ILD card. Jets have been formed using inclusive anti-$k_t$ jet clustering algorithm [659]. For the partonic event generation, the following sets of cuts at MadGraph level both for the signal and backgrounds have been implemented: the transverse momentum of light jets $p_T(j_i) > 20$ GeV for all the final state partons, the pseudo-rapidity $|\eta| < 5.0$, and the separation between the light jets $\Delta R(j_i, j_j) > 0.4$. A number of backgrounds have been considered in the analysis, including $6j$ from $t\bar{t}$; $W^+W^- + 3j$, $W^{\pm} \to 2j$; $ZZ + 3j$, $Z \to 2j$; $7j$; $W^{\pm} + 5j$, $W^{\pm} \to jj$; and $Z + 5j$, $Z \to jj$. Among the different backgrounds, $e^+e^- \to 7j$ includes diagrams of coupling order $\alpha_{EW}^2 \alpha_S^5$ with quarks and gluons as intermediate particles.

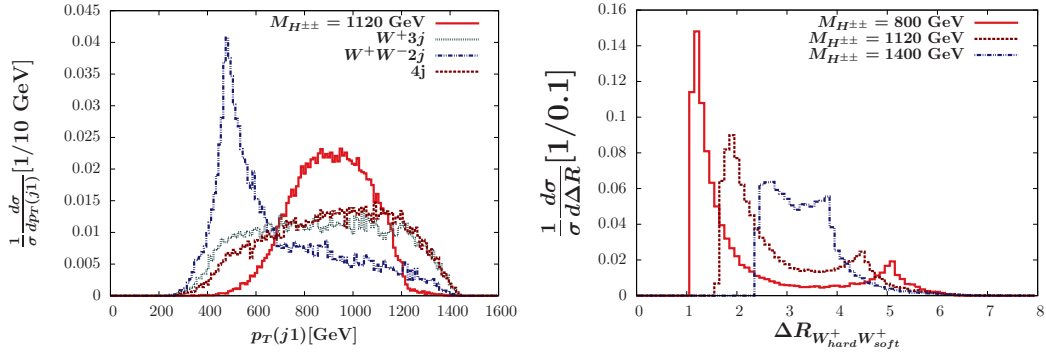

Figure 100: Left panel: $p_T$ distribution of the leading fat-jet. Right panel: $\Delta R$ separation of the $W^+W^+$ pair.

The details of the signal and background cross sections and the cut-flow are given in Table 42 for several illustrative mass points between $M_{H^{\pm\pm}} \sim 121$ GeV and the kinematic threshold $M_{H^{\pm\pm}} \sim 184$ GeV. Since the signal comprises higher jet multiplicity, $N_j \geq 7j$ has been demanded. This reduces the background to $\sigma_d \sim 3$ fb. In Table 43, the statistical significance $n_s$, and the required luminosity to achieve a $5\sigma$ significance, have been shown. Other than the extreme low and high mass ranges $M_{H^{\pm\pm}} = 121$ and 184 GeV, all other mass points have a large discovery prospect with 124 fb$^{-1}$ of data. The doubly-charged Higgs boson with intermediate mass of 159 GeV (172 GeV) can be discovered with $5\sigma$ significance with only $\mathcal{L} \sim 22$ (24) fb$^{-1}$. This further improves to $\mathcal{L} \sim 16$ (17) fb$^{-1}$ after applying a $b$-veto ($50 - 60\%$ efficiency and $1\%$ mis-tag efficiency), whic helps in reducing the dominant top-quark pair background [657].

Table 43: The statistical significance $n_s$ for $\mathcal{L} = 100$ fb$^{-1}$. The third column displays the luminosity required to achieve $5\sigma$ significance. The c.m.energy is $\sqrt{s} = 380$ GeV.

| $e^+e^- \to H^{++}H^{--} \to N_j \geq 7j$ | | |
| --- | --- | --- |
| Mass (GeV) | $n_s$ | $\mathcal{L}$ (fb$^{-1}$) |
| 121 | 1.54 | 1054.14 |
| 137 | 4.48 | 124.56 |
| 159 | 10.48 | 22.76 |
| 172 | 10.15 | 24.26 |
| 184 | 2.69 | 345.48 |

**B) Heavy $H^{\pm\pm}$ at $\sqrt{s} = 3$ TeV:**

Heavier $H^{\pm\pm}$ with mass $M_{H^{\pm\pm}} \sim 800$ GeV to 1.4 TeV can ideally be probed at CLIC, operating with c.m.energy $\sqrt{s} = 3$ TeV. For such a heavy Higgs, the produced $W^{\pm\pm}$ boson will have large transverse momentum, and hence will be boosted. The final hadronic decay products are therefore highly collimated, and can be reconstructed as fat-jets, see Figure 99. Therefore, the model signature in this

Table 44: The cut-flow for the signal and backgrounds. The cross sections are in fb. $\sigma_p$ refers to the partonic cross section. In the backgrounds the decays of the $W^\pm$ boson and top quark to jets are included. Here MD refers to Mass-drop. See text for details.

| $e^+e^- \to H^{++}H^{--} \to W^+W^+W^-W^- \to Nj_{\text{fat}}$ | | | | | | | |
|---|---|---|---|---|---|---|---|
| Masses (GeV) | $\sigma_p$ (ab) | $4j_{\text{fat}}$ (> 120 GeV) | 4 MD | 1 tagged | 2 tagged | 3-tagged | 4-tagged |
| 800 | 1250 | 812.9 | 758.0 | 757.9 | 748.9 | 671.8 | 389.0 |
| 1000 | 850.6 | 527.0 | 492.5 | 492.3 | 486.1 | 436.6 | 258.9 |
| 1120 | 670.0 | 380.0 | 358.4 | 358.3 | 354.2 | 321.9 | 193.1 |
| 1350 | 167.1 | 80.4 | 75.54 | 75.52 | 74.88 | 68.2 | 42.0 |
| 1400 | 94.36 | 45.54 | 42.85 | 42.84 | 42.42 | 38.6 | 24.0 |

| Backgrounds | | | | | | | |
|---|---|---|---|---|---|---|---|
| Processes | $\sigma_p$ (ab) | $4j$ (> 120 GeV) | 4 MD | 1 tagged | 2 tagged | 3-tagged | 4-tagged |
| $4j$ | 6900.0 | 1310.0 | 895.0 | 360.0 | 68.0 | 5.5 | 0.0 |
| $W^+3j$ & $W^-3j$ | 1900.0 | 320.0 | 220.0 | 166.0 | 44.0 | 4.8 | $1.52 \times 10^{-1}$ |
| $W^+W^-2j$ | 190.0 | 25.6 | 17.7 | 15.6 | 8.3 | 1.23 | $5.7 \times 10^{-2}$ |
| $W^+W^-Zjj$ | 4.23 | - | - | - | - | - | - |
| $t\bar{t}$ | 42 | - | - | - | - | - | - |

scenario is [657]
$$e^+e^- \to H^{\pm\pm}H^{\mp\mp} \to W^\pm W^\mp W^\pm W^\mp \to 4 \text{ fat} - \text{jet}.$$

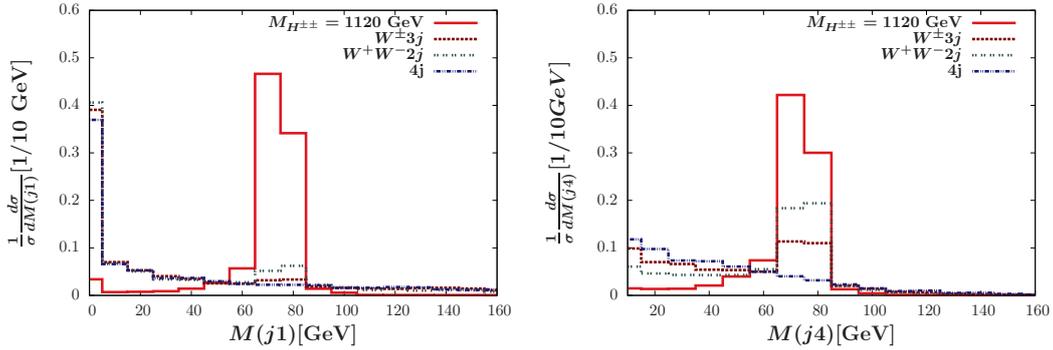

Figure 101: The invariant mass of the fat-jet(leading and 4th leading) constructed using sub-jets four momentum. For signal $M_{H^{\pm\pm}} = 1120$ GeV has been considered as illustrative point.

Event simulation for the signal and background have been carried out using the same tool-chain as for 380 GeV, except for the use of Delphes. The output of Pythia 8 [660] (in HepMC [661] format) has been analyzed and fat-jets have been reclustered using Cambridge-Aachen algorithm [662] in FastJet-3.0.0 [489] with radius parameter $R = 1.0$. A number of backgrounds can lead to the final states with multiple fat-jets. These are: $4j$ (includes both the QED and QCD contributions), $W^+W^-2j$, and $W^+/W^-3j$, $W^+W^-Zjj$ and $t\bar{t}$, with subsequent decays of $W$ boson and the top quark into jets. See [657] for detail discussion of the backgrounds, pre-selection and selection cuts. The partonic cross sections of the signal and backgrounds are shown in Table 44. The cross sections for $W^+W^-Zjj$ and $t\bar{t}$ are small compared to other backgrounds. Therefore, a detailed analysis of these backgrounds have not been included.

In Figure 100, the transverse momentum of the leading fat-jet $j_1$ and the $\Delta R$ distribution of the two same-sign $W^+W^+$ has been shown. As expected, the $\Delta R$ separation reduces for relatively lower mass 800 GeV, resulting in more collimated jets. The produced jets have a very high $p_T$. This motivates implementation of the following selection cuts: a) the number of fat-jets $N_{j_{\text{fat}}} = 4$, b) $p_{T_{j_{\text{fat}}}} > 120$ GeV for all the fat-jets. The background further reduces after reconstruction of the $W$ bosons using the mass-drop tagger [663], that indicates if the fat-jet was initiated by a $W$ boson or a parton. For the signal, the subjets inside a fat-jet are generated from the $W$. Therefore, as shown in Figure 101, the distribution of the invariant mass of the sub-jets peaks around the $W$ mass. The largest background from Table 44) is $e^+e^- \to 4j$ events with the partonic cross section $\sigma_p(4j) \sim 6.9\,\text{fb} \gg \sigma_p(\text{signal})$. The higher transverse momentum cut on jet $p_T$ and demanding that 4 fat-jets have a non-trivial substructure (referred to as mass-drop MD in Table 44) makes the background negligible. The required luminosity ( see Table 45) to achieve a discovery for 800-1120 GeV doubly-charged Higgs boson is $\mathcal{L} =$39 - 95 $\text{fb}^{-1}$, with at least 2 fat-jets tagged as W-bosons. However, for higher masses, such as 1.4 TeV a minimum 3 tagged jets will be required.

Table 45: The statistical significance $n_s$ for $\mathcal{L} = 500\,\text{fb}^{-1}$ and the required luminosity to achieve $5\sigma$ significance. The c.m.energy is $\sqrt{s} = 3$ TeV. In the 2nd column, to derive significance, we consider 2 tagged events for 800-1120 GeV mass range and 3 tagged events for the higher mass range. Here 2-tag implies two or more than two fat-jet masses are within the window of 60-100 GeV, and the fat-jets are tagged as $W$ jets. Similar criteria applies for 3-tagged jets.

| $e^+e^- \to H^{++}H^{--} \to W^+W^+W^-W^- \to Nj_{\text{fat}}$ | | |
|---|---|---|
| Masses (GeV) | $n_s$ (2, 3-tagged $\mathcal{L} = 500\,\text{fb}^{-1}$) | $\mathcal{L}(\text{fb}^{-1})$( with 2,3-tagged) |
| 800 | 17.96(2-tag) | 38.75 |
| 1000 | 13.95(2-tag) | 64.23 |
| 1120 | 11.49(2-tag) | 94.68 |
| 1350 | 5.48(3-tag) | 416.24 |
| 1400 | 3.95(3-tag) | 801.15 |

### 7.3 SU(2) singlets and neutrino mass from higher dimensional operators [91]

Doubly charged scalars (DCSs) are hypothetical particles [148, 149, 664–667] introduced in many beyond-Standard-Model (BSM) realisations, often in connection with radiative generation of neutrino masses [668–675].

From the phenomenological point of view, a DCS can produce intriguing leptonic signatures at low- and high-energy, such as lepton-number and lepton-flavour violating decays, same-sign boosted lepton pairs in high-energy collisions [676–678]. Hence, this option has to be adequately investigated in present and future experiments.

In this Section, we review the phenomenology of a DCS which is singlet under the $SU(2)$ weak symmetry of the SM, with a mass lying in the reach of future collider energies. This setup is realised by adding the following set of operators to the SM Lagrangian:

$$\mathcal{L}_{\text{UV}} = \mathcal{L}_{\text{SM}} + \left(D_\mu S^{++}\right)^\dagger \left(D^\mu S^{++}\right) + \left(\lambda_{ab} \overline{(\ell_R)}^c_a (\ell_R)_b\, S^{++} + \text{h.c.}\right) + \\ + \lambda_2 \left(H^\dagger H\right) \left(S^{--}S^{++}\right) + \lambda_4 \left(S^{--}S^{++}\right)^2 + [\ldots], \quad (269)$$

where $a$ and $b$ are flavour indices and $\lambda_{ab}$ is a complex coupling matrix in the flavour space. In general, this Lagrangian introduces 16 parameters: the mass of the DCS $m_S$, six complex Yukawa parameters $\lambda_{ab}$, a coupling to the Higgs sector $\lambda_2$, the $\lambda_4$ quartic self-coupling, and the DCS width $\Gamma_S$. No specific assumption on the origin of $m_S$ is made, therefore $\lambda_2$ and $m_S$ are understood to be unconstrained by the electroweak-symmetry-breaking (EWSB) mechanism. Any form of new physics contributing to the value of $m_S$ and $\Gamma_S$ (*e.g.*, heavier degrees of freedom dynamically contributing to $m_S$, or undetected decay modes affecting $\Gamma_S$ not directly incorporated in Eq. (269)) is intended to be represented by the ellipsis.

---
[91] Based on a contribution by M. Ghezzi, L. Panizzi and G.M. Pruna.

The Lagrangian in Eq. (269) allows for a plethora of phenomenological consequences. In this Section we focus on the lepton sector and two peculiar signatures at CLIC:

– *opposite-sign di-lepton final states*: such final state is produced through a DCS exchange in the $t$-channel [679] and the CLIC will display a unique power to explore this effect (even for very high DCS mass scales);
– *on-shell single production of the DCS*: this channel is open when the centre-of-mass energies are in the reach of the DCS mass: the DCS is produced in association with same-sign leptons with same or different flavours and subsequently decays into boosted same-sign leptons.

In our study, we will consider both of these characteristic signatures and, for the latter, we will include finite width effects.

### 7.3.1 Neutrino mass generation and low-energy observables

As previously mentioned, DCSs can provide a natural explanation for neutrino masses. In fact, under the assumption that the $SU(2)$-singlet doubly charged scalar emerges as an accidental low-energy degree of freedom of a Grand-Unified Theory (GUT), then also effective dimension-seven operators are potentially produced at the GUT scale. Furthermore, if the underlying theory allows for a dimension-seven operator like

$$O^{(7)} \equiv \frac{C^{(7)}}{\Lambda^3} S^{--} \left[(D_\mu H)^T i\sigma_2 H\right]^2 + \text{h.c.}, \qquad (270)$$

the genuine two-loop diagram in Figure 102 produces neutrino masses just below the GUT scale.

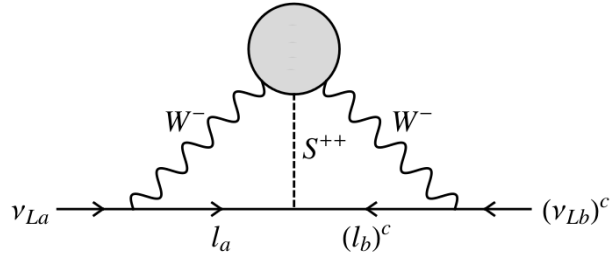

Figure 102: Genuine two-loop diagram for the neutrino Majorana mass terms triggered by an $SU(2)$-singlet doubly charged scalar and an effective dimension-seven operator involving the doubly charged scalar and $W$-bosons.

Of course, the extensive computation of the multi-loop Majorana neutrino mass generated in the framework of a dimension-seven theory with the inclusion of a $SU(2)$-singlet doubly charged scalar at the GUT scale and its evolution down to the neutrino mass scale goes beyond the scope of the present report. However, at the radiative level, it will produce an effective Majorana mass term proportional to

$$m^\nu_{ab} \propto C^{(7)} \lambda_{ab} \frac{m^l_a m^l_b}{\Lambda} \mathcal{F}(m_W, m_S, \Lambda), \qquad (271)$$

where $m^l$ indicates the lepton mass, $a$ and $b$ are flavour indices, $\Lambda$ is the GUT scale, and $\mathcal{F}$ is a dimensionless function that depends on the mass of $m_S$, $\Lambda$ and the $W$-boson mass $m_W$. From Eq. (271) one can infer that even large couplings $\lambda_{ab} \sim 1$ are compatible with the neutrino mass scale, provided that an adequate perturbative suppression is generated by the effective coupling $C^{(7)}$, the GUT scale $\Lambda$ and the perturbative function $\mathcal{F}$ combined.

Since the Pontecorvo–Maki–Nakagawa–Sakata matrix does not display any strong hierarchy among its entries [650], we focus on a scenario where such realisation is caused by a sufficiently anarchic behaviour in the $m^\nu$ mass matrix. Formally, this can be obtained by choosing $\lambda_{ab} \sim \left(y_a^l y_b^l\right)^{-1}$, where $y^l$ indicates the lepton Yukawa couplings. Therefore, in this framework the most sizeable couplings are $\lambda_{11}$ and $\lambda_{12}$. Hence, in the following we will pay exclusive attention to the impact of a DCS in the phenomenology of electrons and muons.

With regard to low-energy observables, the most important constraint comes from the lepton-flavour violating muon three-body decay, i.e. $\mu \to 3e$. The current limit is set by the SINDRUM collaboration [680] (BR$\leq 10^{-12}$) and will be improved by the MU3E collaboration [681] (BR$\leq 5 \cdot 10^{-15}$).

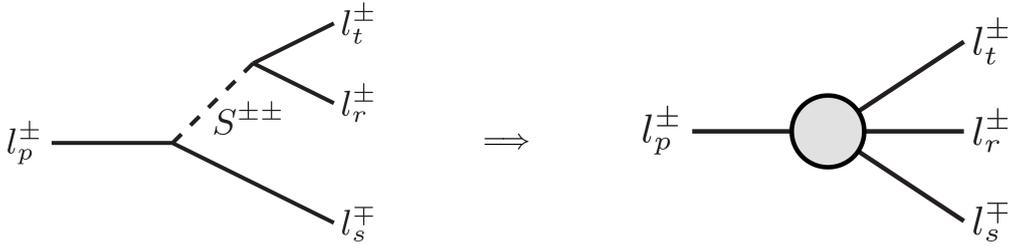

Figure 103: Tree-level diagrams for the lepton three-body decay triggered by a doubly charged scalar (left panel) and after such particle is integrated out at the lepton mass scale (right panel).

In Figure 103 we show the tree-level diagrams for the lepton three-body decay. The related branching ratio is

$$\mathrm{BR}(l_p^\pm \to l_r^\pm l_s^\mp l_t^\pm) \simeq \frac{m_p^5 |\lambda_{ps}|^2 |\lambda_{rt}|^2}{s_{rt} 6(4\pi)^3 m_S^4 \Gamma_p}, \qquad (272)$$

where the symmetry factor reads $s_{rt} = 1 + \delta_{rt}$. Eq. (272) allows strong experimental bounds to be obtained on the quantity $\lambda_{11}\lambda_{12}/m_S^2$ and the parameter space involved in future collider searches to be constrained.

### 7.3.2 Phenomenology at the CLIC

CLIC [3, 5] would be an ideal machine to search for BSM physics coupled to the leptonic sector. For our purposes, CLIC would be essential to accurately probe the Yukawa couplings of the $SU(2)$-singlet DCS.

In order to explore the aforementioned DCS $t$-channel exchange and resonant single production, we implemented the Lagrangian of Eq. (269) in FeynRules v2.3 [383] and extracted a model file for CalcHEP v3.6.30 [172].

We performed our numerical simulations by taking into account the initial-state radiation and beamstrahlung effects, implemented in CalcHEP according to [682, 683]. For the latter, we adopted the parameters listed in the CLIC Conceptual Design Report [3]. Full details on our calculations can be found in [684].

With regard to the centre-of-mass energy and luminosity, we consider several experimental stages (summarised in Table 46). All the runs are performed with standard acceptance cuts for the charged-lepton final state:

$$E \geq 10\,\mathrm{GeV}, \qquad |\cos(\theta)| \leq 0.95, \qquad (273)$$

where $E$ is the energy of the charged lepton, and $\theta$ represents its scattering angle.

Table 46: Centre-of-mass energies and integrated luminosities of the CLIC prototype at different operational stages.

| Stage | Ia | Ib | II | III |
|---|---|---|---|---|
| $\sqrt{s}$ | 350 GeV | 380 GeV | 1500 GeV | 3000 GeV |
| $\mathcal{L}$ | 100 fb$^{-1}$ | 500 fb$^{-1}$ | 1500 fb$^{-1}$ | 3000 fb$^{-1}$ |

Both low-energy facilities and hadronic machines can only be sensitive to combinations of DCS Yukawa couplings. Instead, the CLIC could explore both individual $\lambda$-couplings and combinations via lepton pair-production processes with a DCS exchanged in the $t$-channel at the tree level.

Since the DCS of Eq. (269) only couples to right-handed currents, an adequate polarisation of the beams would enhance the production cross sections. This option is available in the CLIC [3], where the electron beam can be polarised up to $P_{e^-} = \pm 80\%$.

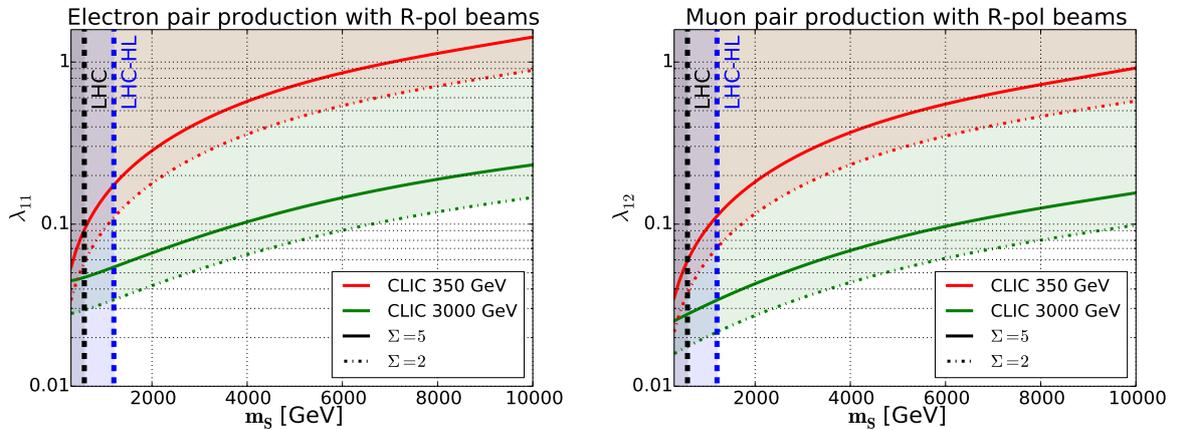

Figure 104: Significance contours for the processes $e^+e^- \to e^+e^-$ (left) and $e^+e^- \to \mu^+\mu^-$ (right) plotted in the $\{\lambda, m_S\}$ plane. The initial-state electron is right-handed polarised. For the electron-positron pair production, the restriction $|\cos\theta| \leq 0.5$ is also applied. Limits from the current LHC data (black-dashed line) and the future HL phase (blue-dashed line) are displayed.

### 7.3.3 Opposite-sign di-lepton channel

In Figure 104, the significance[92] contours for discovery, $\Sigma = 5$, and exclusion, $\Sigma = 2$, are shown as functions of the DCS mass and couplings, at various CLIC operational stages and their related luminosities, for the channels $e^+e^- \to e^+e^-$ (left panel) and $e^+e^- \to \mu^+\mu^-$ (right panel). We applied the cuts of Eq. (273) on the integrated cross sections, plus the stronger cut $|\cos\theta| \leq 0.5$ in case of electron final states to control the large SM background (as suggested in [679]). Limits from the current Large Hadron Collider (LHC) data and future high-luminosity (HL) phase [93] are also plotted. The main results are summarised in Table 47, where we show the minimum values of the couplings $\lambda_{11}$ and $\lambda_{12}$ for which the CLIC can make a significant observation of the particle at the five-sigma level, together with the minimum values which can be excluded, for a choice of centre-of-mass energies and DCS masses.

---

[92] We adopted a definition of the significance, $\Sigma \equiv S/\sqrt{S+B} = \sqrt{\mathcal{L}}\, \sigma_S/\sqrt{\sigma_S + \sigma_B}$, that does not include systematic errors. Although advisable for a better quantitative estimate of the limits, their inclusion should not change the qualitative outcome of the present document.

[93] The LHC limits have been obtained by recasting the 13 TeV CMS search [176] for pair production of a doubly charged scalar decaying into same-sign leptons and considering results for the $S^{\pm\pm} \to 2e^\pm$ decay channel, with the inclusion of both the $q\bar{q}$- and $\gamma\gamma$-initiated processes. The limits are weakly dependent on $\Gamma_S$ due to the specific cuts of the CMS search, and especially to the requirement of having same-sign leptons with an invariant mass within a small window around $m_S$. Limits for the $2\mu^\pm$ and the mixed $e^\pm\mu^\pm$ decay channels are estimated to be similar to the $2e^\pm$ case.

Table 47: CLIC sensitivity to $\lambda_{11}$ and $\lambda_{12}$ at various operational stages for several choices of $m_S$ at the five-sigma level.

| $\lambda_{11}$ | $m_S = 500$ GeV | $m_S = 1$ TeV | $m_S = 2$ TeV |
|---|---|---|---|
| CLIC 380 | $5.3 \times 10^{-2}$ | $9.5 \times 10^{-2}$ | $1.91 \times 10^{-1}$ |
| CLIC 1500 | $4.3 \times 10^{-2}$ | $5.5 \times 10^{-2}$ | $8.5 \times 10^{-2}$ |
| CLIC 3000 | $4.8 \times 10^{-2}$ | $5.2 \times 10^{-2}$ | $6.7 \times 10^{-2}$ |

| $\lambda_{12}$ | $m_S = 500$ GeV | $m_S = 1$ TeV | $m_S = 2$ TeV |
|---|---|---|---|
| CLIC 380 | $4.7 \times 10^{-2}$ | $8.5 \times 10^{-2}$ | $1.63 \times 10^{-1}$ |
| CLIC 1500 | $3.5 \times 10^{-2}$ | $4.7 \times 10^{-2}$ | $7.3 \times 10^{-2}$ |
| CLIC 3000 | $3.7 \times 10^{-2}$ | $4.4 \times 10^{-2}$ | $5.8 \times 10^{-2}$ |

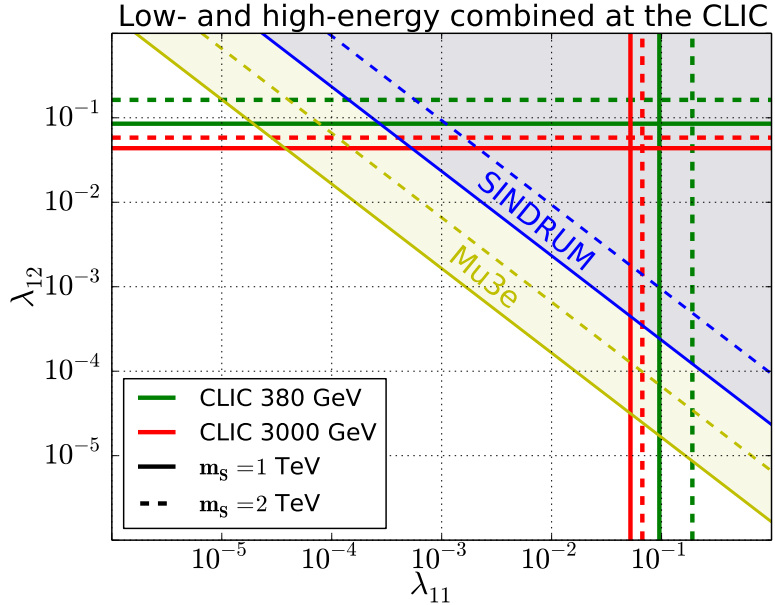

Figure 105: Current and future limits from the SINDRUM and the MU3E experiment, respectively, and discovery power of the CLIC experiment for $m_S = 1$ TeV and $m_S = 2$ TeV.

The next step is to include the actual sensitivity of low-energy experiments in the previous analysis. A combined limit on $\lambda_{11}$ and $\lambda_{12}$ comes from the SINDRUM experiment [680] and will be improved by the MU3E experiment [681]. Instead, CLIC will explore $\lambda_{11}$ and $\lambda_{12}$ independently. We combined the analysis in Figure 105 to show the strong complementarity between the low- and high-energy investigations.

### 7.3.4 Resonant single production

In the framework of Eq. (269), CLIC allows exploration of a production channel that is only available at lepton-lepton colliders: the single production above energy threshold of a DCS associated with two same-sign uncorrelated leptons.

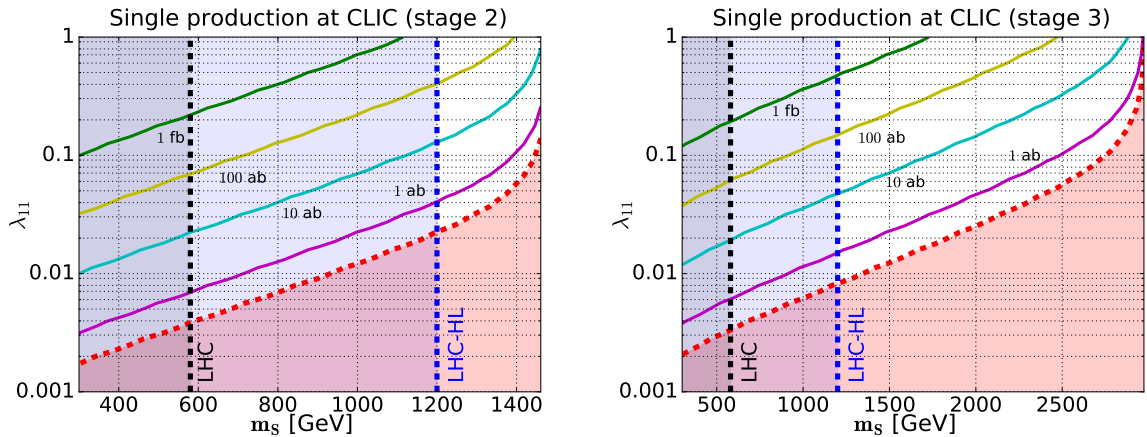

Figure 106: Contours for $e^+e^- \to S^{++}e^-e^-$ cross section in the $\{\lambda_{11}, m_S\}$ plane for the CLIC in the operational stage 2 (left panel) and stage 3 (right panel). The red-dashed line marks the threshold for the production of a single event. The expected bounds from LHC and HL-LHC are also reported.

Such a process occurs through a combination of boson fusion and strahlung of the DCS from initial or final states. These sub-channels strongly interfere and cannot be treated separately. In Figure 106, we show the production cross sections for the second ($\sqrt{s} = 1.5$ TeV) and third ($\sqrt{s} = 3$ TeV) operational stages of the CLIC; the cross sections for $e^+e^- \to S^{++}e^-e^-$ are given as a function of the coupling $\lambda_{11}$ and the mass $m_S$ of the DCS. As one can infer from the plots, there is a considerable portion of the parameter space that can be explored at the latest stages of the CLIC programme. This is possible because for values of the same-sign lepton invariant mass higher than $M_{SSL} \sim 500$ GeV the SM background is around $\sim 10$ ab, i.e. the signal extracted from the differential cross section with respect to the same-sign lepton invariant mass is essentially background-free around the peaks. Hence, with adequate selection cuts (that we are not applying in our simulation), a bunch of events around the production peak is sufficient to discover the particle and possibly study its properties, provided it lives in the parameter space that we have roughly identified with the white region of Figure 106 (allowed by both current and future LHC searches and by the single-event production threshold). Of course, a more accurate analysis would require the inclusion of detector effects, a precise estimate of high-order corrections and the possibility that more BSM channels are open at the same time, therefore calling for a combined analysis. However, our qualitative conclusion should not change.

Furthermore, CLIC is a powerful tool to profile the DCS. Its unique capability to determine the line shape of the particle is displayed in Figure 107. Here, we show the normalised cross section (signal-minus-background) for the $e^+e^- \to S^{++}e^-e^- \to e^-e^+e^-e^-$ process binned with respect to the invariant mass of the same-sign leptons. The resolution of 30 GeV per bin has been roughly borrowed from the $Z'$ studies [685]. For illustrative purposes, we consider the DCS width $\Gamma_S$ as a free parameter and focus on the third operational stage of the CLIC. We investigate the range of masses $m_S = [0.8 - 2.8]$ TeV and set $\Gamma_S/m_S = 5(10)\%$ in the left(right) panel. Then, the coupling is fixed to $\lambda_{11} = 1$ and each distribution is tagged with its value for the integrated cross section. Of course, such values can be naively rescaled to account for different values of the coupling. Recalling that above $M_{SSL} \sim 500$ GeV the SM background is almost negligible, we stress again the cleanliness of the DCS line shape determination at the CLIC, even in presence of a small number of events (i.e. small values of the $\lambda$ couplings).

To conclude, in this section we presented the unique sensitivity of the Compact Linear Collider to explore the phenomenology of a doubly charged $SU(2)$-singlet scalar in di-lepton final states via its exchange in $t$-channel and through its resonant single production.

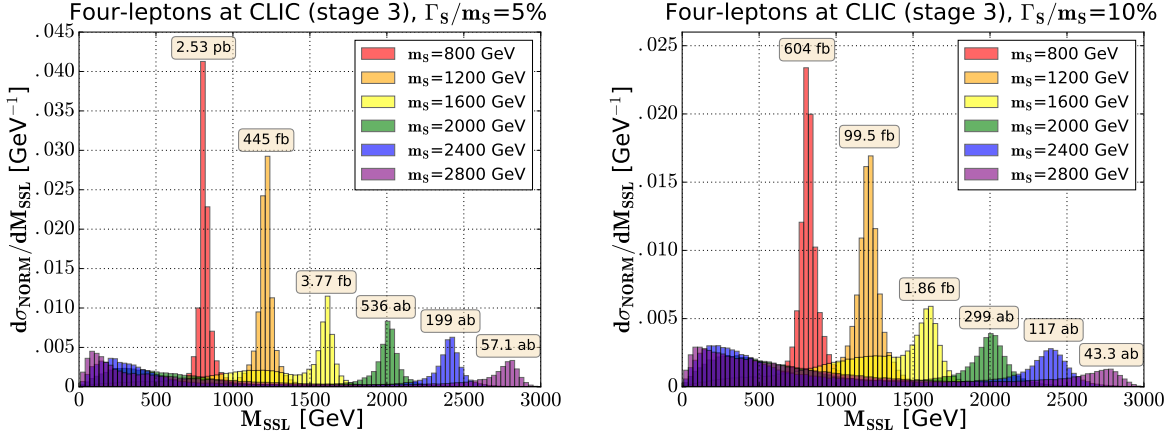

Figure 107: Normalised distributions of the invariant mass of same-sign lepton pairs. $\Gamma_S/m_S$ is assumed to be $5(10)\%$ in the left(right) panel. Total cross sections are shown upon the peak of each distribution.

### 7.4 Zee left-right model [94]

The Zee model [627] is one of the simplest scenarios where light neutrino masses are generated at one–loop. The model consists of an extra scalar doublet and a charged singlet scalar field. Although the simplest form of the Zee model is ruled out by experimental neutrino data [686–688], its Left-Right Symmetric extensions are still viable [689–691]. The extended Zee model in a Left-Right Symmetric (LRS) framework [141, 142, 692] is well-fitted to match with neutrino oscillation data as well as has very promising collider implications for the charged singlet Higgs discovery at $e^+e^-$ machine [691]. The other interesting aspect of this model is the presence of light right–handed (RH) neutrinos of mass from MeV down to eV. The RH neutrinos get their masses, in contrast with the Minimal Left-Right Symmetric Model, through one loop, and hence, can naturally be light. The light neutrino masses are generated via a combination of loop–induced processes and seesaw mechanism. Owing to the extra interaction of the charged Higgs with the RH neutrinos and for relatively larger Yukawas, the cross section at $e^+e^-$ collider is enormous compared to the LHC.

**Model and Charged Higgs** : The gauge group of the model is $SU(3)_C \times SU(2)_L \times SU(2)_R \times U(1)_{B-L}$, with the charge of a particle $Q$: $Q = I_{3L} + I_{3R} + (B-L)/2$. The quarks and leptons consist of left–handed and right–handed doublet fields. The RH neutrinos, therefore, are naturally embedded in this framework. The Higgs sector consists of two doublets $H_R(1,1,2,1)$ and $H_L(1,2,1,1)$, one bi-doublet $\Phi(1,2,2,0)$ and the charged singlet scalar $\delta(1,1,1,2)$. The Yukawa Lagrangian involving lepton doublets and charged Higgs $\delta^{\pm}$ is given below:

$$\mathcal{L}_Y = \lambda_{L_{ij}} l_{Li}^T i\tau_2 l_{Lj}\delta^+ + \lambda_{R_{ij}} l_{Ri}^T i\tau_2 l_{Rj}\delta^+ + \text{h.c.}, \qquad (274)$$

with $\lambda$ being the Yukawa couplings. See [689–691] for the details of the model, symmetry breaking, and neutrino mass generation. The Yukawa $\lambda_{L,R}$ contribute in neutrino masses, as well as, the pair-production of the charged Higgs at lepton collider. In the charged Higgs sector, there are three physical charged Higgses, that are linear combinations of $\phi_2^{\pm}$, $H_L^{\pm}$ and $\delta^{\pm}$. Among the three, flavor constraints indirectly force $\phi_2^{\pm}$ to be more than 15 TeV, and hence de-coupled from the theory. Therefore, $\delta^{\pm}$ primarily mixes only with $H_L^{\pm}$. Two potentially interesting scenarios are: $i$) the lightest charged Higgs $H_1^{\pm}$ consists almost entirely of the charged singlet field $\delta^{\pm}$, and $ii$) the lightest physical state $H_1^{\pm}$ is almost equal admixture of $\delta^{\pm}$ and $H_L^{\pm}$. These two scenarios are dubbed as "Zero" and "Half", respectively, in the display of their respective results in the following. A list of different charged Higgs eigenstates considered for detail analysis has been presented in Table 48.

---

[94]*Based on a contribution by M. Mitra.*

Table 48: Lightest charged Higgs boson $H_1^\pm$. Two possible scenarios are zero-mixing ($H_1^\pm$ has major contribution from $\delta^\pm$) and half-mixing ($H_1^\pm$ has significant contribution both from $\delta^\pm$ and $H_L^\pm$).

| Mass | Composition |
|---|---|
| 473.32 | $0.002\phi_2^+ + 0.999\delta^+$ |
| 1000.7 | $0.002\phi_2^+ + 0.999\delta^+$ |
| 432.58 | $0.03{\phi_1^-}^* - 0.006\phi_2^+ + 0.72H_L^+ + 0.69\delta^+$ |
| 1000.9 | $0.03{\phi_1^-}^* - 0.006\phi_2^+ + 0.76H_L^+ + 0.65\delta^+$ |

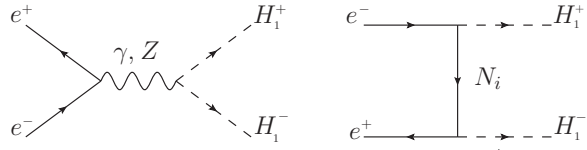

Figure 108: Feynman diagram for the production of $H_1^+ H_1^-$ at $e^+e^-$ collider.

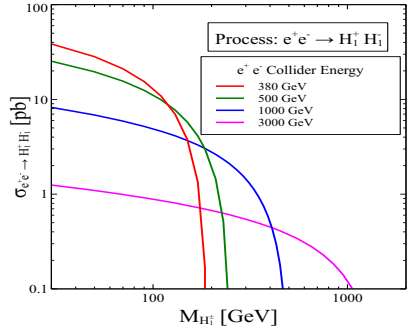

Figure 109: Pair–production cross section of $H_1^+ H_1^-$ at $e^+e^-$ collider for different centre-of-mass energies.

**Collider Signature**: The pair-production of the charged Higgs $H_1^\pm$ at $e^+e^-$ collider occurs through the s-channel process mediated by $\gamma$, $Z$ and $Z_R$ bosons, and an additional t-channel process mediated by the RH neutrinos $N_i$ (see Figure 108). Due to the large couplings of the charged singlet with the right-handed leptons and the small masses of the right-handed neutrinos in this model, this t-channel process gives the major contribution in the cross section. The $Z_R$ mass is in few TeV, and hence gives very little contribution. Figure 109 shows the pair-production cross section of the charged singlet Higgs as a function of its mass for four different centre-of-mass energies (c.m.energies) at lepton colliders. The cross section varies between 0.1-10 pb, for higher c.m.energies $\sqrt{s} = 1$ [173, 693–696] and 3 TeV, while the latter is relevant for CLIC [5, 11, 355, 697]. At LHC, the t-channel contribution is absent. Hence, compared to the LHC, the cross section at lepton collider is larger. The produced charged Higgs $H_1^\pm$ decays into a charged lepton and RH neutrinos, which gives a final state of dileptons ($l^+$ and $l^-$) and missing transverse energy (MET). The RH neutrinos of this model being $M_{N_1} \sim$ eV-MeV is very light, and hence, their decay occurs outside the detector. The model signature at $e^+e^-$ collider therefore looks like

$$e^+e^- \to H_1^+ H_1^- \to l^+l^- \not{E}_T + X, \tag{275}$$

where $l^\pm$ is $e^\pm$, $\mu^\pm$ and $\tau^\pm$. The $\tau$ will decay further and can give rise to leptonic or hadronic final states. The signature analysed in [691] corresponds to the all leptonic final states. The different SM backgrounds in this scenario are $e^+e^- \to l^+l^- Z(\to \nu_l \bar{\nu}_l)$, $e^+e^- \to W^+W^- \to l^+l^-\nu_l\bar{\nu}_l$, and $e^+e^- \to t(\to b\,l^+\nu_l)\,\bar{t}(\to \bar{b}\,l^-\bar{\nu}_l)$, with $W^+W^-$ being the most dominant.

Table 49: cross sections corresponding to the SM backgrounds for different c.m.energies.

|  | SM Backgrounds | | Eff Cross sec after applying cuts |
|---|---|---|---|
|  | Channels | cross section | $\sigma_b$ (fb) |
| $\sqrt{s}=1$ TeV | $l^+l^- Z(\to \nu_l \bar{\nu}_l)$ | 18.68 | 1.67 |
|  | $W^+(\to l^+\nu_l)\,W^-(\to l^-\bar{\nu}_l)$ | 126.88 | 7.05 |
|  | $t(\to bl^+\nu_l)\,\bar{t}(\to \bar{b}l^-\bar{\nu}_l)$ | 13.96 | 0.05 |
|  | Total Backgrounds |  | 8.77 |
| $\sqrt{s}=3$ TeV | $l^+l^- Z(\to \nu_l \bar{\nu}_l)$ | 6.33 | 0.44 |
|  | $W^+(\to l^+\nu_l)\,W^-(\to l^-\bar{\nu}_l)$ | 13.85 | 1.13 |
|  | $t(\to bl^+\nu_l)\,\bar{t}(\to \bar{b}l^-\bar{\nu}_l)$ | 1.76 | 0.002 |
|  | Total Backgrounds |  | 1.57 |

The relevant vertices that are required for collider analysis have been included in FeynRules [383]. The tool-chain MadGraph[31, 381] - Pythia[337], and Delphes[385, 698, 699] have been used for event generation (with the ILD card), hadronization, and detector simulation, respectively. See [691] for details of the various kinematical cuts that have been used in the study. Table 49 gives the background cross section for 1 TeV and 3 TeV c.m.energies after all the selection cuts. The backgrounds have become quite small ($\sigma \sim 1-8$ fb) after all the cuts. The signal cross sections ($\sigma \sim 5-53$ fb after cut) and their statistical significance ($\mathcal{S}$) over the background are given in Table 50. Clearly the case with zero mixing in the Higgs state is the most optimistic scenario. In particular, only 1 fb$^{-1}$ luminosity is required for zero mixing to discover the charged Higgs $H_1^\pm$ with mass range 473 GeV - 1 TeV. For the relatively less optimistic scenario of half-mixing, 3 fb$^{-1}$ will be required to claim discovery. Therefore, the charged Higgs with higher masses can be discovered with much less data at the high energy run of CLIC with c.m.energy $\sqrt{s}=3$ TeV.

Table 50: Signal cross sections, and the statistical significance for 1 TeV, and 3 TeV c.m.energies.

|  | Signal at $e^+e^-$ Collider | | | Cross sec after cuts | Stat Significance ($\mathcal{S}$) | |
|---|---|---|---|---|---|---|
| c.m.energies | Mass (GeV) | Mixing | CS (fb) | $\sigma_s$ (fb) | $\mathcal{L}=1$ fb$^{-1}$ | $\mathcal{L}=3$ fb$^{-1}$ |
| $\sqrt{s}=1$ TeV | 473.32 | Zero | 192.67 | 53.63 | 11.73 | 20.32 |
|  | 432.58 | Half | 49.50 | 12.51 | 3.56 | 6.174 |
| $\sqrt{s}=3$ TeV | 1000.70 | Zero | 100.31 | 27.07 | 10.78 | 18.67 |
|  | 1000.92 | Half | 17.86 | 4.99 | 2.96 | 5.13 |

# 8 Feebly interacting particles

To explain phenomena like Dark Matter, the matter-antimatter asymmetry or neutrino masses new particles or interactions are needed. The absence of a discovery of such particles can either mean that they are very heavy, i.e. out of the kinematic reach of high-energy experiments, or very weakly coupled to Standard Model particles. In this section we explore the latter case in a model independent way, i.e. using a simplified model [700], that is useful to cast our result onto a large class of models.

A standard benchmark in this context is provided by the study of Hidden Valley models [701], which we discuss in Section 8.1 quoting full simulation results from a CLICdp analysis [621] focused on displaced vertex signatures. We leverage these results and show a possible recast, in which the search for a heavy Higgs bosons is pursued looking for its decay into long-lived particles final states. This result applies to some theories of the so-called "neutral naturalness" already presented e.g. in Sections 4.2.1 and 4.4.3 and clearly shows the power of a clean $e^+e^-$ machine to go after subtle signal from new physics. We also recall the results in Section 6.2 on WIMP baryogenesis. This is yet another example of recasting the CLICdp analysis [621], which reinforces the strength and enlarges the scope of exotic signal searches that can be conducted in the clean $e^+e^-$ environment.

In addition we study the reach for generic pseudo-scalar particles. Because of the pattern of couplings we pick for these particles, it is possible to imagine that they arise as a pseudo Nambu-Goldstone boson. These can be coupled feebly to the SM fields and give rise both to prompt and displaced signatures. In the following we pursue two variants of couplings patterns motivated by different possible UV origins of these light pseudo-scalars in Section 8.3. These are the so called Axion-like particle, studied in Section 8.3.1, and its photo-phobic variant in which the axion-like particle is assumed to not have a sizable coupling to photons, which is studied in Section 8.3.2.

## 8.1 Hidden valley searches in Higgs boson decays [95]

The experimental prospects of searches for Hidden Valley particles are described in [621] and summarized here. The study is based on models with a hidden gauge sector coupling to SM particles at high energies [701, 702]. These models contain new massive Long-Lived Particles (LLP) with a measurable flight distance in the detector. The search for such LLPs relies on the reconstruction of displaced vertices (DV) by the CLIC tracking system.

As the search benefits from a large sample of produced Higgs bosons, the dominant channel is the production of $h\nu_e\bar{\nu}_e$ at $\sqrt{s} = 3$ TeV. The LLPs decay predominantly into $b$ quarks. Therefore, the experimental signature studied here is the process $h \to \pi_v^0 \pi_v^0$ in the final state of 4 $b$ quarks. Signal and background MC event samples are produced with full detector simulation for the CLIC_ILD detector model [5] using WHIZARD 1.95 [55] and PYTHIA6 [337], configured to produce Hidden Valley processes. Signal event samples are generated for various combinations of $\pi_v^0$ lifetimes from 1 to 300 ps and masses between 25 and 50 GeV/c$^2$. The main backgrounds are $e^+e^- \to q\bar{q}$, $e^+e^- \to q\bar{q}\nu\bar{\nu}$, $e^+e^- \to q\bar{q}q\bar{q}$, and $e^+e^- \to q\bar{q}q\bar{q}\nu\bar{\nu}$. Beam-induced backgrounds are taken into account.

The search is based on the displacement of the decay vertex of the $\pi_v^0 \to b\bar{b}$ with respect to the primary vertex and the beam axis, which depend on the lifetime of the $\pi_v^0$ as illustrated in Figure 110. The simulated events are processed in the following way: particles are reconstructed with the PANDORA Particle Flow Analysis package [96], applying tight requirements on timing and track transverse momentum. All particles in the event are combined into four jets using the longitudinally invariant $k_t$ algorithm [703]. Based on these, the LCFI+ [100] tool identifies the primary and secondary vertices of the event and the jet clustering is redone using this information. Flavor tagging based on a Boosted Decision Tree (BDT) [704] is performed to assign each jet a $b$- and $c$-tag probability. A jet with a $b$-tag probability of more than 0.95 is considered $b$-tagged.

Events are required to contain at least two displaced vertices and four $b$-tagged jets. Two $b$-tagged jets are assigned to the respective displaced vertex with which they share the most charged tracks. These events are then passed to a multivariate analysis based on a BDT using seven variables related to the track and displaced vertex multiplicities, invariant masses of two and four jets, and measures of the jet clustering multiplicity. A selection cut on the BDT response is used to separate signals from the background.

---

[95] Based on a contribution by M. Kucharczyk and T. Wojton.

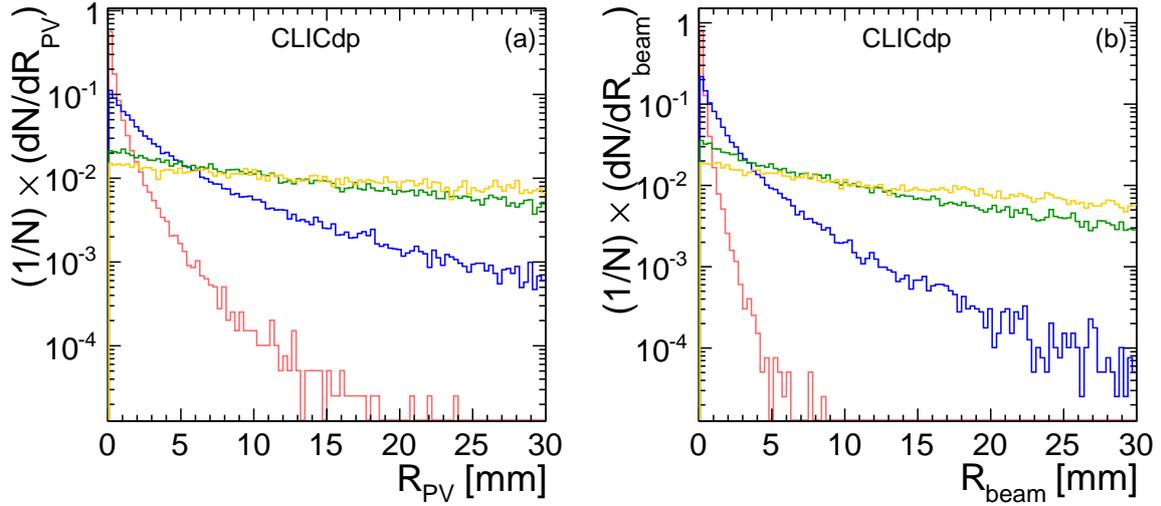

Figure 110: Distance of the $\pi_v^0$ to (a) the primary vertex (PV) and (b) its radial distance to the beam axis for $\pi_v^0$s generated with a mass of 50 GeV/c² and with four different lifetimes: 1 ps (red line), 10 ps (blue line), 100 ps (green line) and 300 ps (yellow line).

Based on this selection, the sensitivity to observe $\pi_v^0$ particles through the Higgs boson decay $h \to \pi_v^0 \pi_v^0$ has been estimated in dependence of the $\pi_v^0$ mass and lifetimes according to model [702] for an integrated luminosity of 3 ab$^{-1}$. The prospective upper limits on the production cross-section and Branching Ratio of the Higgs decay to LLPs are evaluated using the CL$_s$ method [233, 705]. The resulting expectation for upper limits at 95 % C.L. on the cross section $\sigma(h) \times BR(h \to \pi_v^0 \pi_v^0)$ in the case of absence of the signal for the assumption of 100 % Branching Fraction of $\pi_v^0 \to b\bar{b}$ is determined according to the mass and lifetime as presented in Figure 111.

## 8.2 Long-lived particles from heavy Higgs decays [96]

Exotic decays of the Standard Model-like Higgs into long-lived particles (LLPs) are a smoking-gun signature of new physics and arise in a variety of models addressing the nature of dark matter, the origin of the baryon asymmetry, or the electroweak hierarchy problem. Higgs decays into LLPs lead to a range of novel signatures at colliders, including the appearance of one or more displaced vertices (DVs) away from the primary vertex. Although a sophisticated suite of searches for these exotic decays has been developed at the LHC, their reach is largely limited by trigger thresholds.[97] This leaves significant room for probing Higgs decays into LLPs at CLIC.

An even richer set of signatures is possible if the observed Standard Model-like Higgs is part of an extended Higgs sector. In this case, the same underlying physics that leads to decays of the 125 GeV SM-like Higgs $h(125)$ into long-lived particles can also lead to a significant rate for long-lived particle production via the decays of heavier states in the Higgs sector. These decays are often kinematically more distinctive than exotic decays of the $h(125)$, and may provide a leading channel for the discovery of additional Higgs states. With its high centre-of-mass energy, CLIC is particularly well-suited to probing these latter signatures.

---

[96]Based on a contribution by S. Alipour-Fard and N. Craig.
[97]For a recent review of theoretical motivations and LHC searches, see [706].

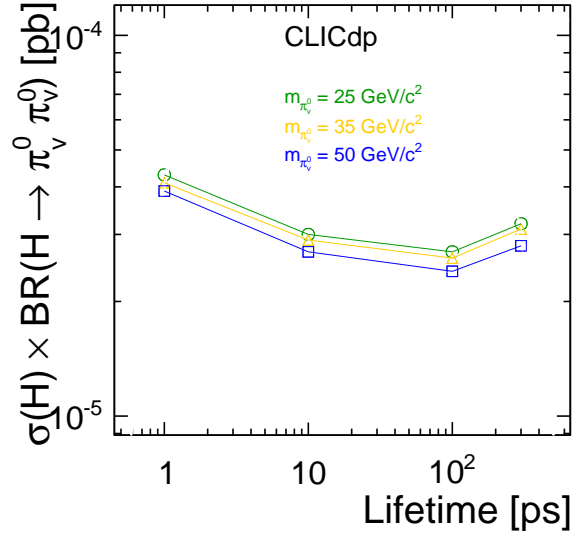

Figure 111: Observed 95 %C.L. upper limits on the cross section and branching ratio $\sigma(h) \times BR(h \to \pi_v^0 \pi_v^0)$ for three different masses of $\pi_v^0$: 25 eV/c$^2$ (green), 35 GeV/c$^2$ (yellow), 50 GeV/c$^2$ (blue), as a function of $\pi_v^0$ lifetime.

Prototypical theories exhibiting decays of both the SM-like Higgs and heavier Higgs states into LLPs are the folded SUSY [707], the fraternal Twin Higgs [505], and the Hyperbolic Higgs [503], all of which feature both an extended Higgs sector and a confining hidden sector whose bound states are produced in Higgs decays and are LLPs that subsequently decay back to the Standard Model via mixing with the Higgs. The minimal branching ratio for $h(125)$ decays into pairs of LLPs in these scenarios is typically of the order of $10\% \times (v/f)^4$, where $f$ is a symmetry-breaking scale expected to be $\mathcal{O}(\text{TeV})$. In the fraternal Twin Higgs and Hyperbolic Higgs scenarios, Higgs coupling deviations are proportional to $(v/f)^2$, leading to a current limit of $f \gtrsim 3v$ from LHC Higgs data and allowing branching ratios for Higgs decays into LLPs as large as $0.1\%$; direct limits on new particles in folded SUSY lead to similar rates. Both the mass and lifetime of the LLPs in the confining hidden sector can vary in these scenarios, but are typically in the mass range of $10 - 100$ GeV with decay lengths ranging from $\mu$m-km.

In addition, these theories feature at least one heavier CP-even Higgs state – the radial mode of spontaneous symmetry breaking associated with the scale $f$ in the case of the fraternal Twin Higgs and Hyperbolic Higgs, and the heavy CP-even Higgs of a supersymmetric Higgs sector in the case of folded SUSY. For definiteness, we will focus on the former case here. The mass of the radial mode in the fraternal Twin Higgs and Hyperbolic Higgs is $\mathcal{O}(1) \times f$, and it mixes with the SM-like Higgs with a mixing angle proportional to $v/f$. Once produced, it decays with roughly equal branching ratios into the Standard Model and the hidden sector, with the majority of hidden sector decays ending in two or more LLPs. For example, the rate for the production of long-lived particles via the heavy radial mode in the fraternal Twin Higgs model for the best-case scenario $f = 3v$ [505] in $e^+e^-$ collisions at $\sqrt{s} = 3$ TeV is illustrated in Figure 112. This provides a motivated theory target for characterizing the performance of CLIC searches for Higgs decays into LLPs, although the motivation for SM-like and heavy Higgs decays into LLPs is quite general [706].

In this work we explore the potential for CLIC to observe or exclude decays of either $h(125)$ or additional CP-even Higgs scalars into long-lived particles that decay hadronically within the tracker volume. To do so, we work in terms of a simplified model involving a Higgs-like CP-even scalar $H$ of mass $m_H$, and a long-lived CP-even scalar $X$ of mass $m_X$ and width $\Gamma_X$. The couplings of $H$ to the Standard Model include (at least) Higgs-like couplings to $W$ and $Z$ bosons, giving rise to production via $WW$ fusion at CLIC. The couplings of $X$ allow it to decay back to pairs of Standard Model quarks. For definiteness we take $\text{Br}(X \to b\bar{b}) = 80\%$, loosely corresponding to decays induced by mixing with the Higgs. The $H$ and $X$ bosons are coupled via a cubic interaction $HXX$, allowing for the decay $H \to XX$ when kinematically available. The signal process of interest consists of $e^+e^- \to \nu\bar{\nu}(H \to XX \to (q\bar{q})(q\bar{q}))$. The finite lifetime of the $X$ bosons leads to displaced dijet pairs. While CLIC has the potential to probe these signatures across all operating energies, we focus on $\sqrt{s} = 3$ TeV where the production rate for both $h(125)$ and heavier CP-even Higgses via $WW$ fusion is maximized.

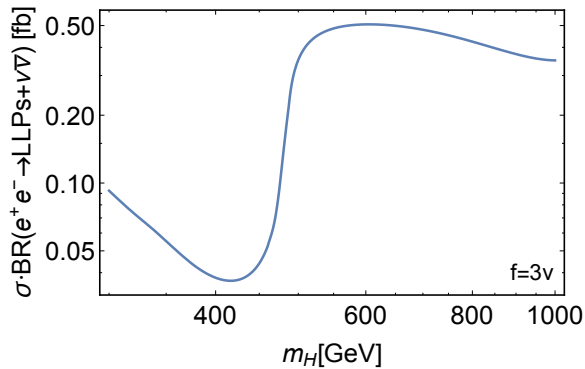

Figure 112: The rate for production of long-lived particles (hidden glueballs) via the heavy radial Higgs in the fraternal Twin Higgs model as a function of the heavy Higgs mass $m_H$ for $f = 3v$. Here the rate includes all decays of the radial Higgs mode into twin bottom quarks, including via production of intermediate twin top quarks and twin $W, Z$ bosons. These decay modes subsequently lead to the production of multiple LLPs from twin confinement. The increase in rate around 500 GeV is due to kinematic thresholds associated with the twin top quark and twin $W, Z$ bosons.

As a detailed treatment of backgrounds to long-lived particle searches is well beyond the scope of this study, we determine the signal efficiency for a pair of analysis strategies aimed at achieving a zero-background selection. We perform our analysis at parton level, imposing a variety of cuts and efficiency factors at parton level to emulate the effects of more detailed track-level requirements. Our choice of cuts and efficiency factors is motivated in part by the detailed study of [621].

### 8.2.1 Simulation & analysis

We simulate $e^+e^- \to \nu\bar{\nu}H$ with $H \to XX$ at $\sqrt{s} = 3$ TeV in Madgraph 5, followed by the parton-level displaced decay $X \to q\bar{q}$ in Pythia 8, which correctly propagates the secondary vertex of the $X$ decay. We consider the benchmark $H$ masses $m_H = 125, 200, 400, 600, 800, 1000$ GeV, $X$ masses $m_X = 25, 50, 100$ GeV, and $X$ decay lengths ranging from $c\tau = 10^{-2} - 10^3$ cm. Our analysis is aimed at determining the efficiency for identifying one or both of the displaced vertices coming from the decay of the two $X$ bosons, including conservative cuts aimed at reducing Standard Model backgrounds.

To resolve and identify a displaced vertex (DV), we impose the following requirements, motivated in part by [621]:

- In order to mimic the role of the track impact parameter significance variable, IPS = IP / $\sigma_{\text{IP}}$, in identifying secondary vertices distinct from the primary vertex, we require the two $b$ quarks in the displaced vertex to satisfy IPS > 16. The impact parameter resolution $\sigma_{\text{IP}}$ is taken from [5].
- In order to ensure that tracks originating from a displaced vertex are identifiable as such, we require the trajectory of each of the $b$ quarks to intersect at least 5 tracker layers before exiting the tracker, using the tracker geometry detailed in [6].
- To suppress backgrounds, we require two resolved $b$-jets from the displaced vertex, which we emulate by requiring $\Delta R > 0.5$ between the two $b$ quarks and imposing an additional flat efficiency of 90% per tag corresponding to the 'loose' working point.
- As noted in [621], significant background discrimination is possible by requiring each displaced vertex to contain 5 or more tracks. Based on the track multiplicity of the signal model in [621] with $m_H = 125$ GeV and $m_X = 50$ GeV, we assign a flat efficiency factor of 0.5 to each displaced vertex to emulate the impact of a track multiplicity cut. We do not account for possible variation of the track multiplicity distribution with $m_X$.

We study two possible analysis strategies. The first strategy ("1DV") requires the identification of one or more displaced vertex per signal event. The second strategy ("2DV") requires the identification of exactly two displaced vertices per signal event. For each analysis strategy we obtain the 95% CL exclusion on the cross section times branching ratio $\sigma(e^+e^- \to \nu\bar{\nu}H) \cdot \mathrm{Br}(H \to XX)$ at $\sqrt{s} = 3$ TeV assuming 3000/fb of integrated luminosity and zero background events. The zero background assumption for the 2DV analysis with $m_H = 125$ GeV is justified by the background study in [621], and likely holds for higher values of $m_H$. Reaching zero background for the 1DV analysis will require improved background discrimination.

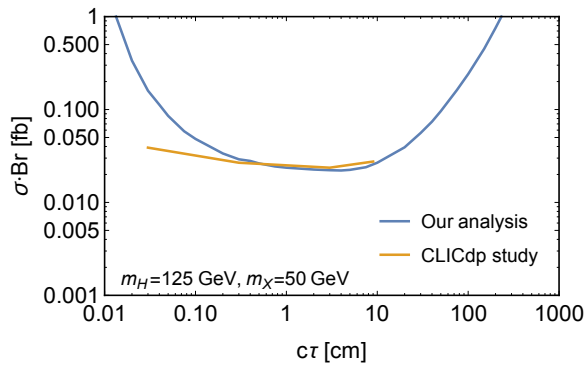

Figure 113: Comparison between this work and the CLICdp study [621] of the 95% CL limit from the 2DV analysis on $\sigma(e^+e^- \to \nu\bar{\nu}H) \cdot \mathrm{Br}(H \to XX)$ at $\sqrt{s} = 3$ TeV, $\mathcal{L}_{\mathrm{int}} = 3000$/fb for $m_H = 125$ GeV, $m_X = 50$ GeV.

We validate our analysis strategy by comparison with the 2DV analysis performed for $m_H = 126$ GeV and $m_X = 50$ GeV in [621], obtaining good agreement over the primary range of interest for a tracker-based search, $c\tau \sim 1 - 10$ cm, as illustrated in Figure 113. Our analysis is somewhat more conservative at lower lifetimes but captures the expected falloff in the limit.

### 8.2.2 Discussion & conclusions

The 95% exclusion on $\sigma(e^+e^- \to \nu\bar{\nu}H) \cdot \mathrm{Br}(H \to XX)$ is shown in Figures 114, 115, and 116 as a function of various choices of $m_H, m_X$, and $c\tau$. Several salient features are apparent. As is clear in all three figures, optimal sensitivity is obtained for decay lengths between 0.1-10cm, for which the LLPs have the highest likelihood of satisfying both the IPS selection (distinguishing the secondary vertex) and the track hit selection.

As is evident in Figure 114, sensitivity falls off at higher $m_H$ at fixed $m_X$ due to the significant boost of the LLPs, which collimates their decay products; for $m_X = 50$ GeV the typical separation between the two bottom quarks in the $X$ decay falls below the angular separation cut of $R = 0.5$ for $m_H \gtrsim 400$ GeV. This likewise explains the improved sensitivity to higher values of $m_X$ among the heavier $m_H$ benchmarks in Figure 115.

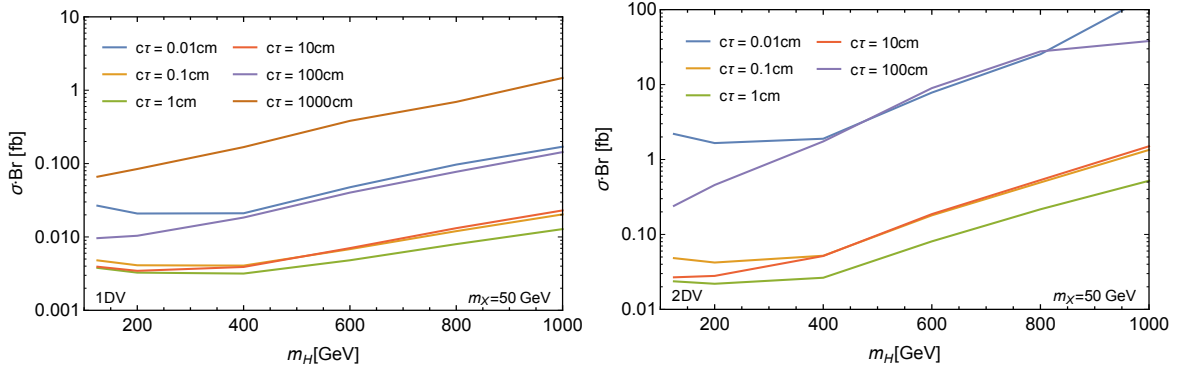

Figure 114: 95% CL limit from the 1DV analysis (left) and the 2DV analysis (right) on $\sigma(e^+e^- \to \nu\bar{\nu}H) \cdot \text{Br}(H \to XX)$ at $\sqrt{s} = 3$ TeV, $\mathcal{L}_{\text{int}} = 3000/\text{fb}$ as a function of $m_H$ for various values of $c\tau$. Here $m_X = 50$ GeV.

The sensitivity of the 2DV analysis is typically one order of magnitude weaker in $\sigma \cdot \text{Br}$ compared to the 1DV analysis, but sensitivity falls much more rapidly in regions where the collimation of the decay products or the average decay length of the LLP make it unlikely for both DVs in the event to pass selection cuts. Peak sensitivity for the 1DV analysis approaches the attobarn level for many combinations of $m_H, m_X$, while peak sensitivity for the 2DV analysis approaches the tens of attobarns.

For a Standard-Model-like Higgs at $m_H = 125$ GeV, the 1DV analysis approaches a sensitivity of $\sigma \cdot \text{Br} \lesssim 4$ ab for LLP decay lengths between $0.1 - 10$ cm, while the 2DV analysis approaches a sensitivity of $\sigma \cdot \text{Br} \lesssim 20 - 40$ ab for 50 GeV $> m_X >$ 25 GeV over the same range of decay lengths. This corresponds to an exclusion reach on the exotic branching ratio of the SM-like Higgs to LLPs of $\text{Br}(H \to XX) \lesssim 10^{-5}$ for the 1DV analysis and $\text{Br}(H \to XX) \lesssim 10^{-4}$ for the 2DV analysis, demonstrating considerable improvement over current [616] (and likely future [706]) LHC reach for the same scenario. Sensitivity at the level of $\text{Br}(H \to XX) \lesssim 10^{-5}$ would probe fraternal Twin Higgs and Hyperbolic Higgs scenarios to the level of $f \gtrsim 9v$, corresponding to sub-percent tuning of the electroweak scale.

As for the coverage of heavy Higgses decaying to LLPs, for a heavy Higgs of $m_H = 1$ TeV the 1DV analysis approaches a sensitivity of $\sigma \cdot \text{Br} \lesssim 2 - 10$ ab for LLP decay lengths between $0.1 - 10$ cm and LLP masses between $50 - 100$ GeV, while the 2DV analysis approaches a sensitivity of $\sigma \cdot \text{Br} \lesssim 30 - 600$ ab over the same range of LLP decay lengths and masses, easily reaching motivated parameter space.

Both analysis strategies exhibit significant sensitivity to decays of either $h(125)$ or potential heavy Higgses into long-lived particles with lifetimes between $0.1 - 10$cm. This demonstrates the potential for LLP searches at CLIC to cover a compelling range of parameter space beyond that of the LHC, motivating further detailed study of backgrounds and refinement of analysis strategies.

## 8.3 Axion-like particles

Pseudoscalar particles appear in several extensions of the Standard Model for instance motivated by the solution of strong CP problem of the Standard Model, like the QCD axion[708–711], or the hierarchy problem of the weak scale [398]. These particles may also be related to the Dark Matter of the Universe. Therefore they are interesting targets for searches at colliders. In a wide set of cases pseudo-scalar particles can emerge as pseudo Nambu-Goldstone bosons (pNGB), hence it is meaningful to consider them as light degrees of freedom of a theory that extends the Standard Model, and possibly the only new degrees of freedom accessible at colliders. In the following we consider a generic pseudo-scalar particle that interacts like a pNGB of a field shift symmetry allowing the mass of the pNGB boson to be a free parameter. We consider in Section 8.3.1 a general case for the interaction of the pNGB motivated by couplings similar to those of an axion, hence named axion-like particle, and a more specific case in which the axion-like particle does not have a coupling to photons, that we discuss in Section 8.3.2.

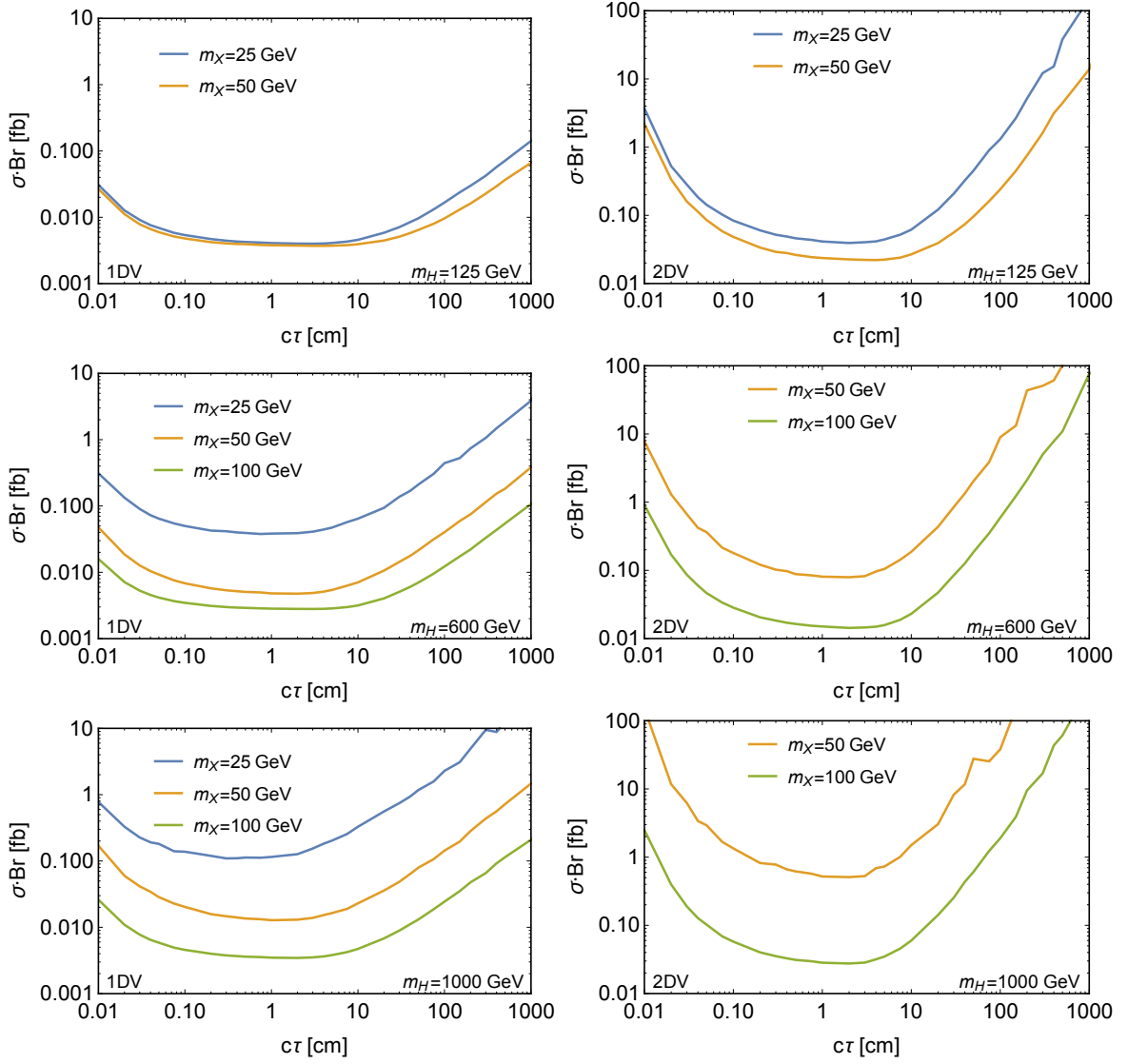

Figure 115: 95% CL limit from the 1DV analysis (left column) and the 2DV analysis (right column) on $\sigma(e^+e^- \to \nu\bar{\nu}H) \cdot \mathrm{Br}(H \to XX)$ at $\sqrt{s} = 3$ TeV, $\mathcal{L}_{\mathrm{int}} = 3000/\mathrm{fb}$ as a function of $c\tau$ for various values of $m_X$. From top to bottom, $m_H = 125$ GeV, $m_H = 600$ GeV, $m_H = 1000$ GeV.

### 8.3.1 General axion-like particles [98]

Assuming that the ALP respects a shift symmetry apart from a soft breaking through an explicit mass term, its interactions with Standard Model (SM) particles are described by dimension-5 operators or higher

$$\mathcal{L}_{\mathrm{eff}}^{D \leq 5} = \frac{1}{2}(\partial_\mu a)(\partial^\mu a) - \frac{m_{a,0}^2}{2}a^2 + \sum_\psi \frac{c_\psi}{2}\frac{\partial^\mu a}{f}\bar{\psi}\gamma_\mu\gamma_5\psi + \frac{c_3\alpha_s}{4\pi}\frac{a}{f}G^A_{\mu\nu}\tilde{G}^{\mu\nu,A} \\ + \frac{c_2\alpha_2}{4\pi}\frac{a}{f}W^A_{\mu\nu}\tilde{W}^{\mu\nu,A} + \frac{c_1\alpha_1}{4\pi}\frac{a}{f}B_{\mu\nu}\tilde{B}^{\mu\nu}, \quad (276)$$

---
[98]Based on a contribution by M. Bauer, M. Neubert, and A. Thamm.

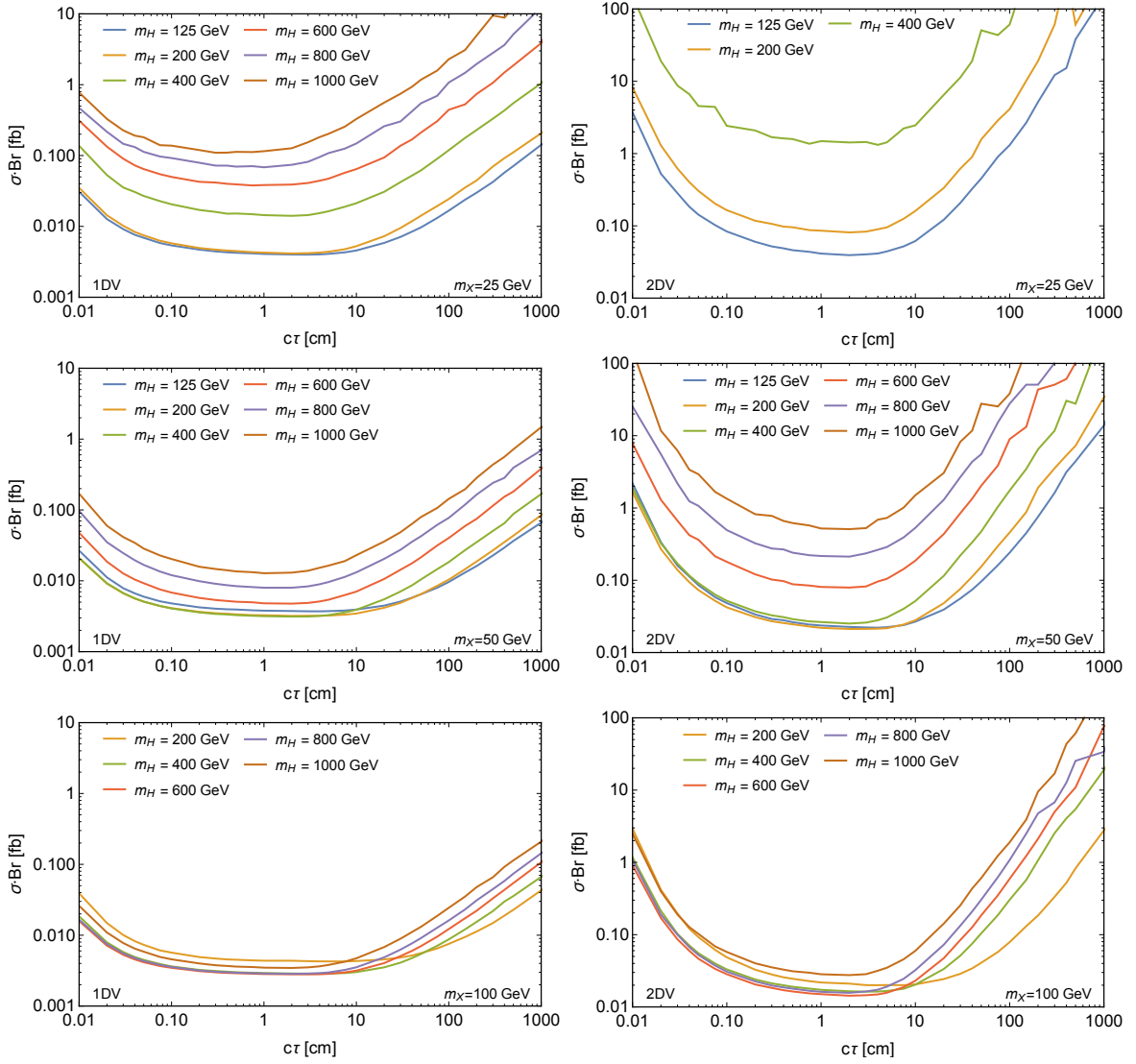

Figure 116: 95% CL limit from the 1DV analysis (left column) and the 2DV analysis (right column) on $\sigma(e^+e^- \to \nu\bar{\nu}H) \cdot \mathrm{Br}(H \to XX)$ at $\sqrt{s} = 3$ TeV, $\mathcal{L}_{\mathrm{int}} = 3000/\mathrm{fb}$ as a function of $c\tau$ for various values of $m_H$. From top to bottom, $m_X = 25$ GeV, $m_X = 50$ GeV, $m_X = 100$ GeV.

where $m_{a,0}$ is the explicit symmetry breaking mass term and $\alpha_1 = e^2/4\pi c_w^2$ and $\alpha_2 = e^2/4\pi s_w^2$, where $c_w$ and $s_w$ are the cosine and sine of the weak mixing angle, respectively. Interactions with the Higgs boson, $\phi$, are described by the dimension-6 and 7 operators

$$\mathcal{L}_{\mathrm{eff}}^{D \geq 6} = \frac{c_{ah}}{f^2} (\partial_\mu a)(\partial^\mu a)\, \phi^\dagger \phi + \frac{c_{Zh}}{f^3} (\partial^\mu a) \left( \phi^\dagger i D_\mu \phi + \mathrm{h.c.} \right) \phi^\dagger \phi + \ldots, \qquad (277)$$

where the first operator mediates the decay $h \to aa$, while the second one is responsible for $h \to Za$. Note that a possible dimension-5 operator coupling the ALP to the Higgs current is redundant unless it is introduced by integrating out a heavy new particle which acquires most of its mass through electroweak symmetry breaking [712–715].

The relevant partial widths for this study are the decays into photons and leptons. For the derivation and one-loop contributions we refer the reader to [715]

$$\Gamma(a \to \gamma\gamma) = \frac{\alpha^2 m_a^3}{(4\pi)^3 f^2} \left| c_1 + c_2 \right|^2, \qquad (278)$$

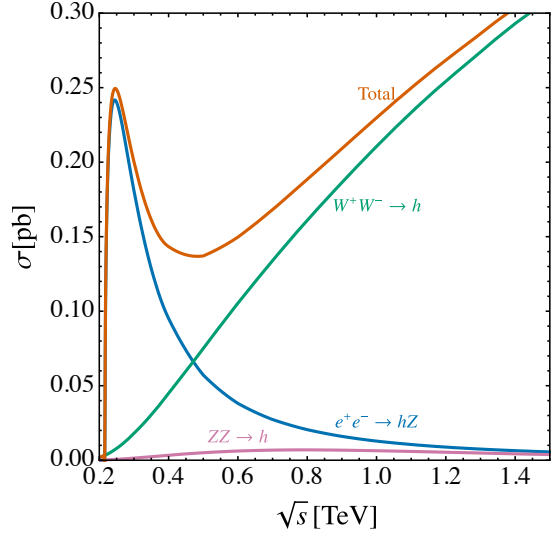

Figure 117: Higgs production cross-sections at $e^+e^-$ colliders as a function of the centre-of-mass energy, $\sqrt{s}$.

$$\Gamma(a \to \ell^+\ell^-) = \frac{m_a m_\ell^2}{8\pi f^2} |c_{\ell\ell}^{\text{eff}}|^2 \sqrt{1 - \frac{4m_\ell^2}{m_a^2}}. \tag{279}$$

In the following we will consider the processes $h \to Za$ and $h \to aa$ with the corresponding partial widths

$$\Gamma(h \to Za) = \frac{m_h^3}{16\pi f^2} |c_{Zh}^{\text{eff}}|^2 \lambda^{3/2}\left(\frac{m_Z^2}{m_h^2}, \frac{m_a^2}{m_h^2}\right), \tag{280}$$

$$\Gamma(h \to aa) = \frac{m_h^3 v^2}{32\pi f^4} |c_{ah}|^2 \left(1 - \frac{2m_a^2}{m_h^2}\right)^2 \sqrt{1 - \frac{4m_a^2}{m_h^2}}, \tag{281}$$

where $\lambda(x,y) = (1 - x - y)^2 - 4xy$ and we define $c_{Zh}^{\text{eff}} = c_{Zh}^{(5)} + c_{Zh} v^2/2f^2$ to take into account possible contributions from a dimension-5 operator which originates from integrating out chiral heavy new physics.

At CLIC, ALPs can be produced directly in a Drell-Yan process as an s-channel resonance, $e^+e^- \to a$. Furthermore, the ALP can be produced via exotic decays of the Higgs, $h \to Za$ and $h \to aa$ [714, 715] or in association with photons, jets or Z bosons, e.g. $e^+e^- \to a\gamma$, or Higgs bosons $e^+e^- \to ha$ [716–719]. In the following we will discuss these three production modes. For more details on these production modes see [720].

*Resonant ALP production*

Direct Drell-Yan production, $e^+e^- \to a$, is strongly suppressed by the electron mass and is therefore not the prevailing production mode. The cross-section is given by

$$\sigma(e^+e^- \to a) \stackrel{s \approx m_a^2}{=} \frac{4\pi \Gamma_a}{(s - m_a^2)^2 + m_a^2 \Gamma_a^2} \frac{\sqrt{s} m_e^2}{8\pi f^2} |c_{ee}^{\text{eff}}|^2, \tag{282}$$

where we set $m_e^2/s \to 0$ and $\Gamma_a$ denotes the total decay width of the ALP. This cross-section is highly suppressed by the electron mass and this process is therefore not the dominant production mode for an ALP.

*ALPs in visible exotic Higgs decays*

Exotic decays are interesting because very small couplings can lead to appreciable branching ratios. Light or weakly coupled ALPs can be long-lived and thus only a small fraction of them decay inside the detector. Here we focus on ALPs decaying inside the detector. The average decay length of the ALPs perpendicular to the beam axis is given by

$$L_a^\perp(\theta) = \frac{\sqrt{\gamma_a^2 - 1}}{\Gamma_a} \sin\theta, \tag{283}$$

where $\gamma_a$ denotes the relativistic boost factor given by

$$\gamma_a = \begin{cases} \dfrac{m_h^2 - m_Z^2 + m_a^2}{2 m_a m_h}, & \text{for } h \to Za, \\ \dfrac{m_h}{2 m_a}, & \text{for } h \to aa. \end{cases} \tag{284}$$

The fraction of ALPs decaying before they have travelled a certain distance related to the size of the relevant detector component, $L_{\text{det}}$, is then given by

$$\begin{aligned} f_{\text{dec}}^{a} &= \int_0^{\pi/2} d\theta \, \sin\theta \left(1 - e^{-L_{\text{det}}/L_a^\perp(\theta)}\right), \\ f_{\text{dec}}^{aa} &= \int_0^{\pi/2} d\theta \, \sin\theta \left(1 - e^{-L_{\text{det}}/L_a^\perp(\theta)}\right)^2, \end{aligned} \tag{285}$$

where $f_{\text{dec}}^{a}$ is relevant for $h \to Za$ decays and $f_{\text{dec}}^{aa}$ for $h \to aa$ decays. For simplicity in the following we make the assumption that the ALPs are produced at maximum scattering angle in the laboratory frame, corresponding to $\sin\theta = 1$ in (283). This is a somewhat optimistic assumption, but serves the purpose of getting the estimates we intend to compute. For further details on the angular distribution of the ALP we refer to [720] and references therein.

Here we focus on ALP decays into photons and leptons to illustrate our results but ALP decays into jets, heavy flavours or missing energy final states also lead to interesting signatures. For prompt ALP decays, we demand all final state particles to be detected in order to reconstruct the decaying SM particle with standard techniques. For the decay into photons we require the ALP to decay before the electromagnetic calorimeter which we take to be at $L_{\text{det}} = 1.5\,\text{m}$ for LHC detectors as well as for CLIC. Analogously, the ALP should decay before the inner tracker for an $e^+e^-$ final state to be detected, thus we use $L_{\text{det}} = 0.6\,\text{m}$ for CLIC [6]. We also require $L_{\text{det}} = 0.6\,\text{m}$ for muon and tau final states in order to take full advantage of the tracker information in reconstructing these events. For LHC experiments we use instead $L_{\text{det}} = 0.02\,\text{m}$.

We define the effective branching ratios

$$\text{Br}(h \to Za \to Y\bar{Y} + X\bar{X})\big|_{\text{eff}} = \text{Br}(h \to Za)\,\text{Br}(a \to X\bar{X})\,f_{\text{dec}}^{a}\,\text{Br}(Z \to Y\bar{Y}), \tag{286}$$

$$\text{Br}(h \to aa \to X\bar{X} + X\bar{X})\big|_{\text{eff}} = \text{Br}(h \to aa)\,\text{Br}(a \to X\bar{X})^2\,f_{\text{dec}}^{aa}, \tag{287}$$

where $X = \gamma, e, \mu, \tau, \text{jet}$ and $Y = \ell, \text{hadrons}$. Multiplying the effective branching ratios by the production cross section and luminosity allows us to derive results at a specific collider. As shown in Figure 117, at CLIC, with a centre-of-mass energy of 380 GeV, 1.5 and 3 TeV, $\sigma(pp \to h) = 0.15, 0.32, 0.50$ pb. We do not distinguish displaced from prompt decays and derive the reach for a certain number of signal events. While we required 100 events at the LHC, as this is what is typically needed to suppress backgrounds in new physics searches with prompt Higgs decays (see also [715] for further discussion), we consider 4 events to be sufficient for reconstruction at CLIC due to the significantly cleaner collision environment and smaller backgrounds. We do not distinguish between vector-boson fusion or associated Higgs production.

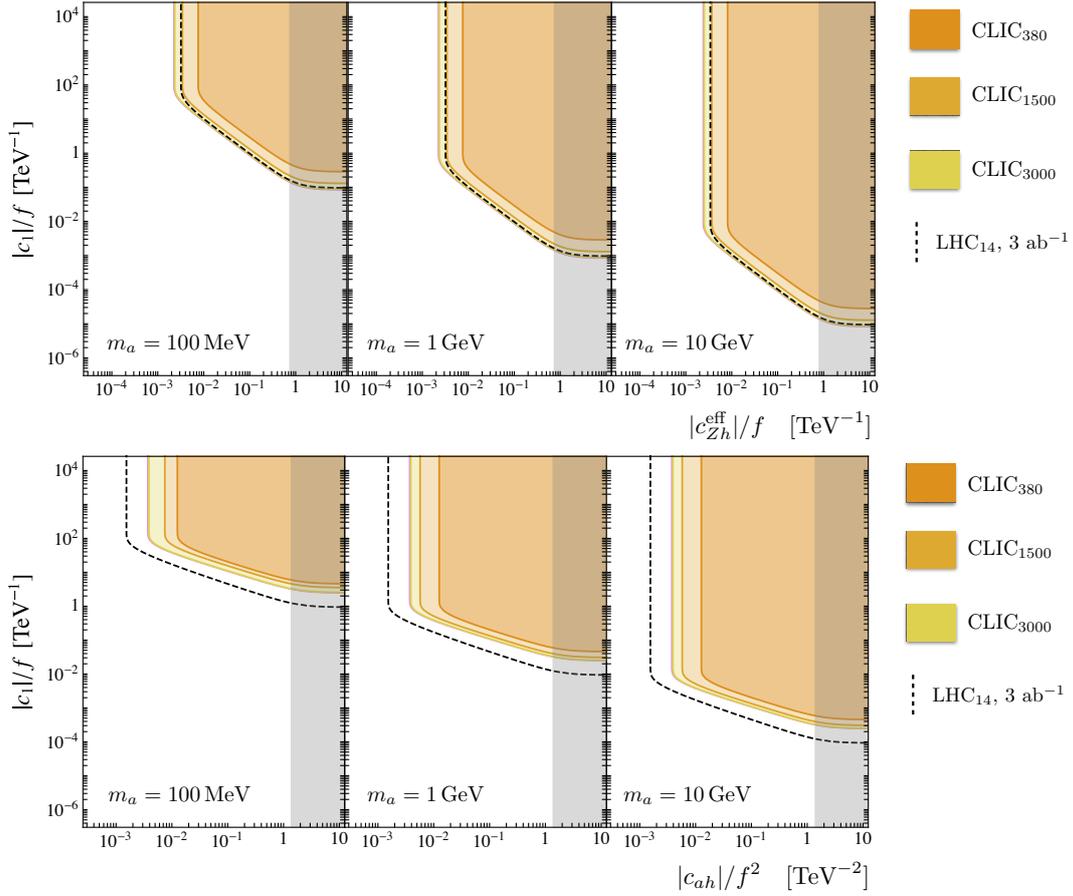

Figure 118: Parameter regions which can be probed in the decay $h \to Za$ (top) and $h \to aa$ (bottom) with $a \to \gamma\gamma$ at CLIC$_{3000}$ (yellow), CLIC$_{1500}$ (light orange), CLIC$_{380}$ (dark orange). The projected bounds for the LHC are shown in dashed black lines. The grey shaded area is excluded by LHC Higgs coupling measurements.

ALPs can be searched for in the exotic Higgs decays $h \to Za$ and $h \to aa$ at CLIC. The Higgs production cross section at lepton colliders is typically at least one order of magnitude smaller compared to the LHC. This implies that lepton colliders are most powerful for light ALPs with dominant decay channels for which backgrounds at hadron colliders are large. In Figure 118 we show the expected reach of the $h \to Za$ and $h \to aa$ processes at CLIC with $\sqrt{s} = 3, 1.5$ TeV and $380$ GeV with $3, 1.5$ and $0.5$ ab$^{-1}$ respectively, denoted by CLIC$_{3000}$, CLIC$_{1500}$ and CLIC$_{380}$, in yellow, light orange and dark orange in the coupling plane $|c_{Zh}^{\text{eff}}|$ vs $|c_1|$ and $|c_{ah}^{\text{eff}}|$ vs $|c_1|$ for $c_2 = 0$ for three different ALP masses $m_a = 100$ MeV, $1$ GeV and $10$ GeV. (violet regions). The black dashed contour shows the projected reach for the LHC with $3$ ab$^{-1}$. Since the reach in searches for exotic Higgs decays is directly proportional to the number of Higgses produced, the higher the luminosity the greater the sensitivity. The reach in $h \to Za$ is larger than the one in $h \to aa$. This is due to the fact that hadronic decays of the Z can be easily reconstructed at an electron-positron collider. In order to reconstruct the Higgs, we demand the Z from the Higgs decay as well as all ALPs to decay into visible final states with Br($Z \to$ visible) $= 0.8$ and Br($a \to \gamma\gamma$) $= 1$. This condition can be relaxed if the electrons in ZZ-fusion or the additional Z in associated Higgs production are detected. Note that Higgs coupling measurements set an upper limit on BR(h $\to$ BSM) $< 0.34$ [235] which constrains the coefficient $|c_{Zh}| < 0.72\,(f/\text{TeV})$ depicted by the grey region. The same limit leads to the constraint $|c_{ah}| < 1.34\,(f/\text{TeV})^2$. For leptonic ALP decays, the analogous plots are shown in Figure 119.

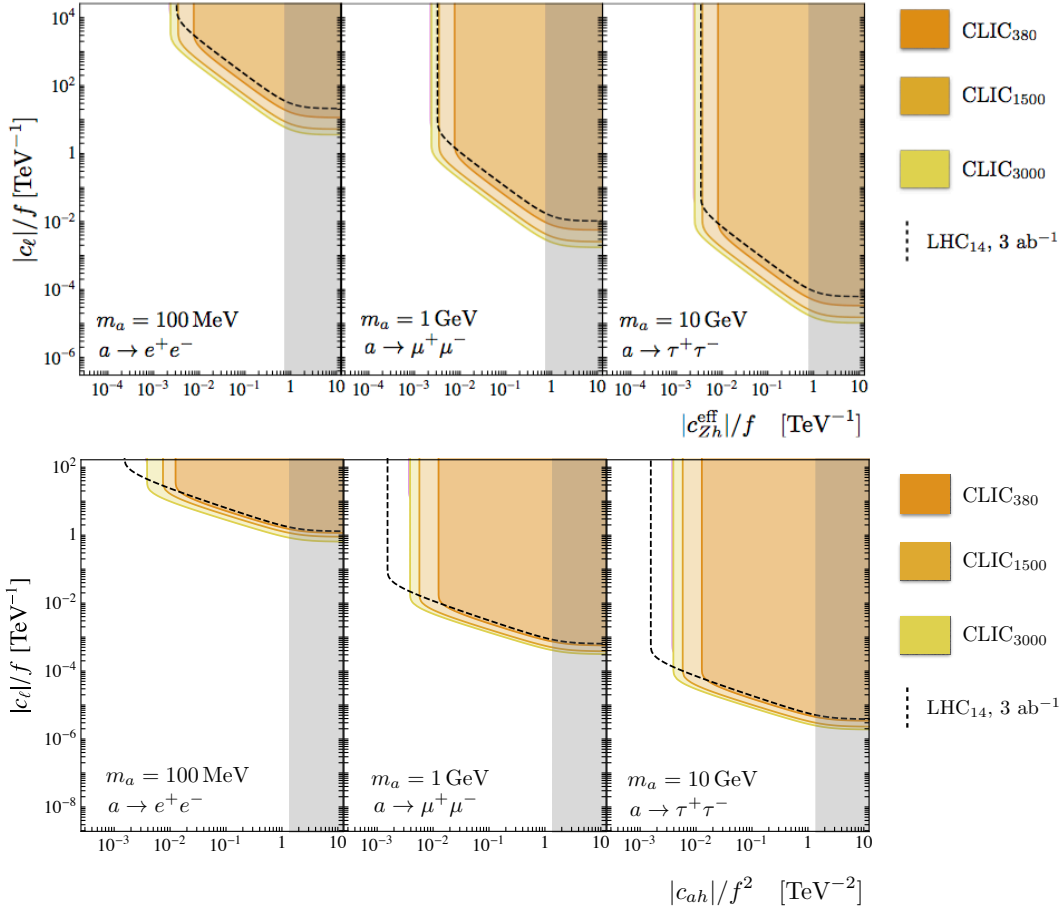

Figure 119: Parameter regions which can be probed in the decay $h \to Za$ (top) and $h \to aa$ (bottom) with $a \to \ell^+\ell^-$ CLIC$_{3000}$ (yellow), CLIC$_{1500}$ (light orange), CLIC$_{380}$ (dark orange). The projected bounds for the LHC are shown in dashed black lines. The grey shaded area is excluded by LHC Higgs coupling measurements.

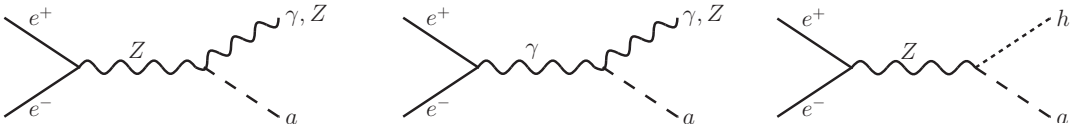

Figure 120: The left and middle Feynman diagrams contribute to the processes $e^+e^- \to \gamma a$ and $e^+e^- \to Za$, while the right diagram contributes to $e^+e^- \to ha$.

### ALP production in association with a Higgs, a Z-boson or a photon

ALPs can be radiated from a $Z$ boson or a photon and thereby be produced in association with a $\gamma$, a $Z$ or a Higgs boson. The relevant Feynman diagrams are shown in Figure 120. Additional t-channel diagrams with an electron as a mediator are suppressed by $m_e^2/\Lambda^2$ and hence neglected here. The differential cross sections are given by

$$\frac{d\sigma(e^+e^- \to \gamma a)}{d\Omega} = 2\pi\alpha\alpha^2(s)\frac{s^2}{f^2}\left(1 - \frac{m_a^2}{s}\right)^3 (1 + \cos^2\theta)\left(|V_\gamma(s)|^2 + |A_\gamma(s)|^2\right), \quad (288)$$

$$\frac{d\sigma(e^+e^- \to Za)}{d\Omega} = 2\pi\alpha\alpha^2(s)\frac{s^2}{f^2}\lambda^{\frac{3}{2}}\left(\frac{m_a^2}{s}, \frac{m_Z^2}{s}\right)(1 + \cos^2\theta)\left(|V_Z(s)|^2 + |A_Z(s)|^2\right), \quad (289)$$

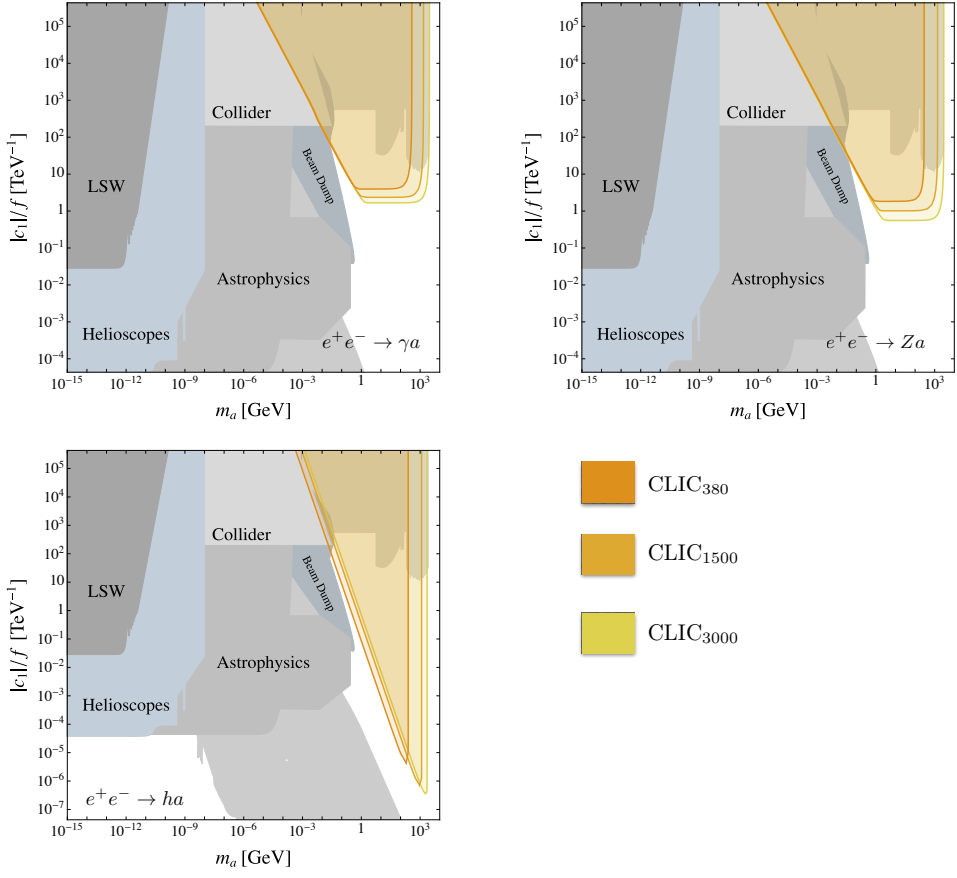

Figure 121: Projected exclusion contours for searches for $e^+e^- \to \gamma a \to 3\gamma$ (top left), $e^+e^- \to Za \to Z\gamma\gamma$ (top right) and $e^+e^- \to ha \to \bar{b}b\gamma\gamma$ (bottom left) for CLIC$_{380}$ (dark orange), CLIC$_{1500}$ (light orange), CLIC$_{3000}$ (yellow) assuming $c_2 = 0$. The constraints from other experiments are in grey in the background. For more details see [715, 720].

$$\frac{d\sigma(e^+e^- \to ha)}{d\Omega} = \frac{\alpha}{128\pi\, c_w^2 s_w^2} \frac{|c_{Zh}|^2}{f^2} \frac{s\, m_Z^2}{(s - m_Z^2)^2} \lambda^{\frac{3}{2}}\left(\frac{m_a^2}{s}, \frac{m_h^2}{s}\right) \sin^2\theta \left(g_V^2 + g_A^2\right), \quad (290)$$

where

$$V_\gamma(s) = \frac{C_{\gamma\gamma}}{s} + \frac{g_V}{2c_w^2 s_w^2} \frac{C_{\gamma Z}}{s - m_Z^2 + im_Z \Gamma_Z}, \qquad A_\gamma(s) = \frac{g_A}{2c_w^2 s_w^2} \frac{C_{\gamma Z}}{s - m_Z^2 + im_Z \Gamma_Z}, \quad (291)$$

$$V_Z(s) = \frac{1}{c_w s_w} \frac{C_{\gamma Z}}{s} + \frac{g_V}{2c_w^3 s_w^3} \frac{C_{ZZ}}{s - m_Z^2 + im_Z \Gamma_Z}, \qquad A_Z(s) = \frac{g_A}{2c_w^3 s_w^3} \frac{C_{ZZ}}{s - m_Z^2 + im_Z \Gamma_Z}, \quad (292)$$

and $g_V = 2s_w^2 - 1/2$ and $g_A = -1/2$. Here $C_{\gamma\gamma} = (c_1 + c_2)/(4\pi)^2$ and $C_{\gamma Z} = (c_w^2 c_2 - s_w^2 c_1)/(4\pi)^2$. Note that the cross sections with a gauge boson in the final state become independent of $s$ for $m_a^2, m_Z^2 \ll s$, while the cross section decreases as $1/s$ for $e^+e^- \to ha$.

In order to obtain the total cross section, we integrate the differential distributions (288) with the lifetime dependent factor in (285),

$$\sigma(e^+e^- \to Xa) = \int d\Omega\, \frac{d\sigma(e^+e^- \to aX)}{d\Omega}\left(1 - e^{-L_{\text{det}}/L_a^\perp(\theta)}\right). \quad (293)$$

In contrast to hadron colliders, $e^+e^-$-machines offer a much cleaner detector environment allowing us to identify ALPs produced in association with a $Z$-boson, a photon or a Higgs boson. For $e^+e^- \to$ $\gamma a \to 3\gamma$ and $e^+e^- \to Za \to Z\gamma\gamma$, the process only depends on the photon coupling. We assume that only a coupling to the hypercharge gauge boson is present, in which case the process depends only on $c_1/f$. We show the projections for these two channels at CLIC$_{380}$ (dark orange), CLIC$_{1500}$ (light orange) and CLIC$_{3000}$ (yellow) for the photon final state in Figure 121 and for leptonic final states in Figure 122. The parameter space corresponds to at least 4 expected signal events and we consider only visible decays of the $Z$-boson, Br($Z \to$ visible)=0.8. We also impose the constraint $|C_{\gamma Z}| < 1.48\, f/\text{TeV}$ from the LEP measurement of the total width of the $Z$ boson. In both processes, a higher centre-of-mass energy at CLIC allows us to access larger ALP masses. The reach in the ALP-photon coupling is mostly determined by the integrated luminosity. Higher energy stages at CLIC are expected to accumulate more luminosity which is why they are more sensitive to smaller $c_1$.

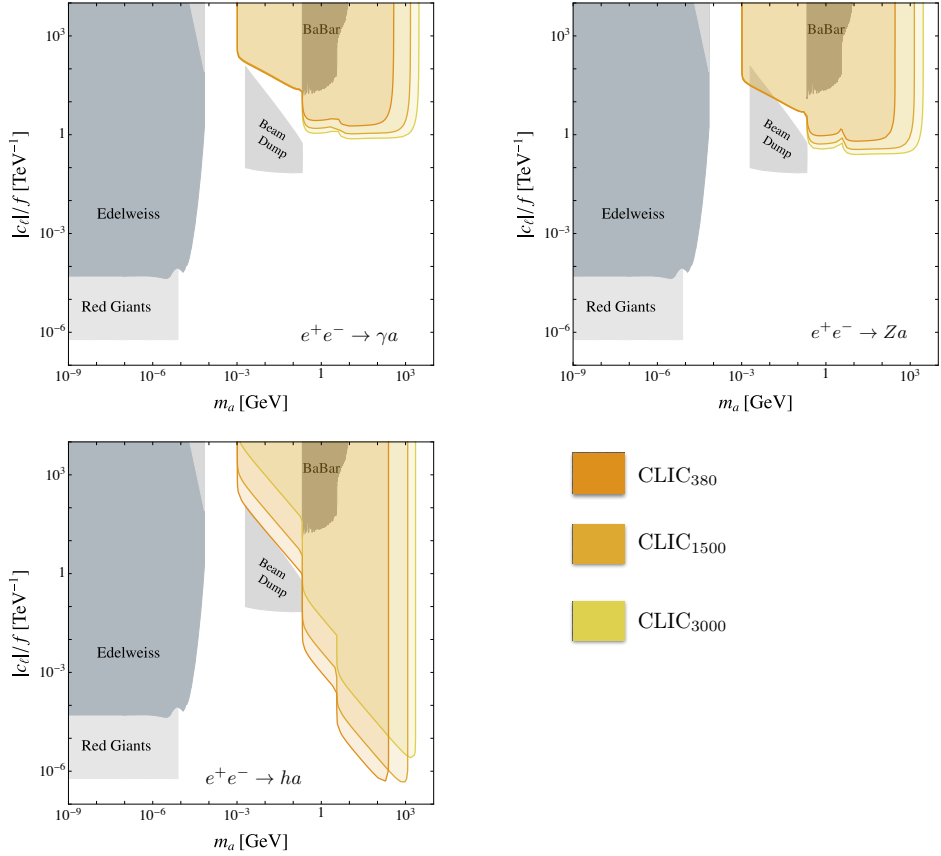

Figure 122: Projected exclusion contours for searches for $e^+e^- \to \gamma a \to \gamma \ell^+\ell^-$ (top left), $e^+e^- \to Za \to Z\ell^+\ell^-$ (top right) and $e^+e^- \to ha \to \bar{b}b\ell^+\ell^-$ (bottom left) for CLIC$_{380}$ (dark orange), CLIC$_{1500}$ (light orange), CLIC$_{3000}$ (yellow). The constraints from other experiments are in grey in the background. For more details see [715, 720].

For $e^+e^- \to ha \to b\bar{b}\gamma\gamma$ with BR($h \to b\bar{b}$) = 0.58, gauge invariance does not impose any relation between the coefficients. The reach in the ALP-photon and ALP-mass plane is shown in the bottom row of Figures 121 and 122. The projections for CLIC$_{380}$ (dark orange), CLIC$_{1500}$ (light orange) and CLIC$_{3000}$ (yellow) are shown. All the contours assume $|c_{Zh}|/f = 0.72\,\text{TeV}^{-1}$. Figure 123 shows the dependence on $|c_{Zh}|/f$. Since there is no relation between the two couplings entering this process, smaller values of $|c_1|/f$ can be probed.

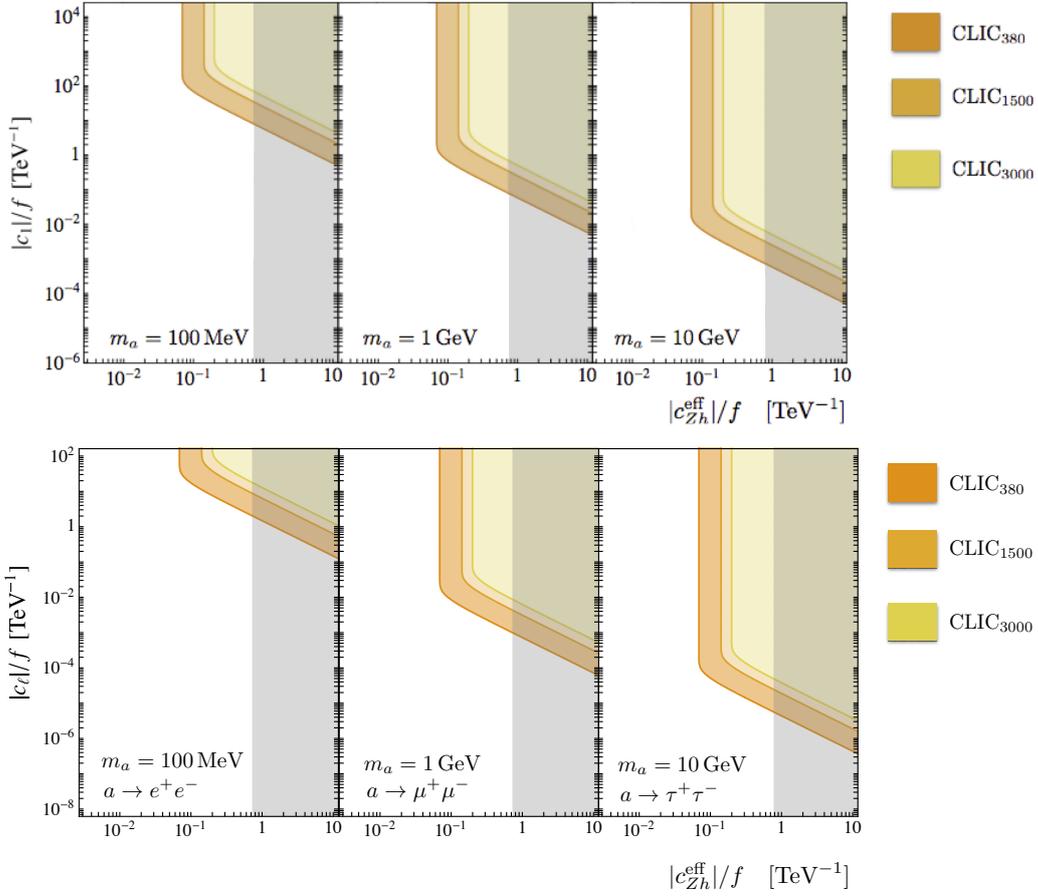

Figure 123: Projected exclusion contours for searches for $e^+e^- \to ha \to b\bar{b}\gamma\gamma$ (top) and $e^+e^- \to ha \to b\bar{b}\ell^+\ell^-$ (bottom) for CLIC$_{380}$ (dark orange), CLIC$_{1500}$ (light orange), CLIC$_{3000}$ (yellow). The constraints from LHC Higgs coupling measurements are shown in grey (for more details see [715, 720]).

### 8.3.2 *Photo-phobic axion-like particles* [99]

Axion-like particles or ALPs arise in a wide range of scenarios of physics beyond the Standard Model, ranging from solutions to the Strong CP and dark matter problems to scenarios of composite dynamics. Their effective Lagrangian presents a well defined, minimal description of an exotic, SM-singlet, pseudo-scalar state whose interactions are protected by an approximate shift symmetry, characterized by a preference for derivative interactions with SM particles. In general, its leading interactions with SM particles arise at canonical dimension-5 [721] and can be written as

$$\mathcal{L}_a = -\frac{c_1 \alpha_1}{4\pi}\frac{a}{f_a} B_{\mu\nu}\widetilde{B}^{\mu\nu} - \frac{c_2 \alpha_2}{4\pi}\frac{a}{f_a} W^I_{\mu\nu}\widetilde{W}^{\mu\nu}_I - \frac{c_3 \alpha_3}{4\pi}\frac{a}{f_a} G^A_{\mu\nu}\widetilde{G}^{\mu\nu}_A + \frac{\partial_\mu a}{f_a}\sum_\psi \bar{\psi}_i \gamma_\mu c^{ij}_\psi \psi_j, \quad (294)$$

where the sum over $\psi$ denotes one over all chiral fermion representations and the ALP kinetic and mass terms are left understood. It is well known that for light ALPs below the MeV scale, a host of cosmological and astrophysical measurements place very tight bounds on the couplings to photons and electrons of order $10^{-8}$ TeV$^{-1}$ (see, *e.g.* [722] for a review). For heavier ALPs, the limits are less stringent. In particular, once the ALP mass goes beyond the $B$-meson mass scale, the only relevant constraints that can be obtained are from direct production at beam dumps and collider experiments. There has been a renewed interest in this type of scenario in recent years, with a number of studies on ALP phenomenology at both $e^+e^-$ and $pp$ colliders considering the sensitivity of direct production processes to the ALP-gauge boson couplings [714–717, 719, 723–726]. The focus of this study is the set of couplings to electroweak (EW) gauge bosons, namely the photon, $W$ and $Z$ bosons, given after EW symmetry breaking by

$$\mathcal{L}_{EW} = - c_{\gamma\gamma} \frac{e^2}{4\pi^2} \frac{a}{f_a} F_{\mu\nu}\tilde{F}^{\mu\nu} - c_{ZZ} \frac{e^2}{4\pi^2} \frac{a}{f_a} Z_{\mu\nu}\tilde{Z}^{\mu\nu} - c_{Z\gamma} \frac{e^2}{4\pi^2} \frac{a}{f_a} F_{\mu\nu}\tilde{Z}^{\mu\nu} - c_2 \frac{g_2^2}{4\pi^2} \frac{a}{f_a} W^+_{\mu\nu}\tilde{W}^{\mu\nu}_-. \tag{295}$$

---

[99] Based on a contribution by K. Mimasu.

The $\gamma\gamma$, $Z\gamma$ and $ZZ$ couplings are expressed in terms of the original parameters $c_2$ and $c_1$ as well as the weak mixing angle $\theta_W$ as

$$c_{\gamma\gamma} = \frac{1}{4}(c_1 + c_2), \tag{296}$$

$$c_{ZZ} = \frac{1}{4}\left(c_1 \tan^2\theta_W + \frac{c_2}{\tan^2\theta_W}\right), \tag{297}$$

$$c_{Z\gamma} = \frac{1}{2}\left(c_1 \tan\theta_W - \frac{c_2}{\tan\theta_W}\right). \tag{298}$$

While considerable attention has been devoted to $c_{\gamma\gamma}$, the couplings to other weak gauge bosons have received somewhat less interest. It is important to explore the sensitivity of collider experiments to this orthogonal direction in parameter space, where the ALP couplings to photons is absent or at least suppressed [719]. This is partly motivated by the fact that, as mentioned, light ALPs coupling to photons are extremely well constrained, certainly beyond the reach of existing or future collider experiments. This 'photophobic' scenario corresponds to the case where $c_1 \simeq -c_2$ such that

$$c_{\gamma\gamma} \simeq 0, \tag{299}$$

$$c_{ZZ} \simeq \frac{\cos 2\theta_W}{\sin^2 2\theta_W} c_2, \tag{300}$$

$$c_{Z\gamma} \simeq -\frac{c_2}{\sin 2\theta_W}. \tag{301}$$

It was recently shown that the photophobic ALP is not necessarily a consequence of extreme parameter tuning, even though the ALP gauge bosons are not shift symmetry preserving and will in general mix among themselves due to renormalisation group evolution [727]. This particular relation among the couplings, in the massless case, possesses a remnant shift symmetry that protects $c_{\gamma\gamma} = 0$ from such RGE effects and is only explicitly broken by the ALP mass term. One therefore obtains, in the case of a massive ALP, a predictive model for a quasi-photophobic ALP with a single coupling parameter to the EW gauge bosons and an irreducible, albeit loop-suppressed coupling to photons. The aforementioned astrophysical and cosmological bounds for low masses are somewhat alleviated but still remain well below the expected sensitivity of collider experiments that may probe this scenario in the 'mono-X' channels where the ALP is produced in association with a gauge boson but is sufficiently long-lived to result in a missing energy signature. For more massive, unstable ALPs, $e^+e^-$ machines are a promising place to study the couplings to the EW sector, with a number of possibilities for production and decay modes that are detailed in the next section. In general, the relatively clean environment allows for excellent discrimination between signal and background.

In the following we first give a general discussion of the possible ALP production modes at $e^+e^-$ colliders involving the EW gauge boson coupling as well as the final state signatures that can be expected assuming decays into EW gauge bosons in the photophobic case. Then, a concrete study is presented that aims to determine the CLIC sensitivity to heavier, photophobic ALPs above 100 GeV. The reach in $(c_2/f_a, m_a)$ plane is investigated assuming the proposed staging of the experiment in Ref. [6], consisting of $\sqrt{s} =$ 380 GeV, 1.5 TeV and 3 TeV and integrated luminosities of 0.5, 1.5 and 3 ab$^{-1}$ respectively.

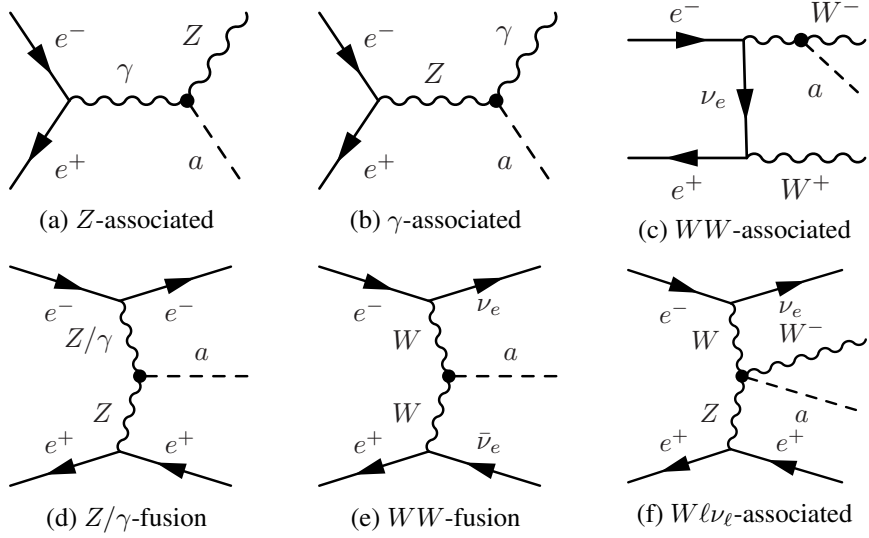

(a) $Z$-associated   (b) $\gamma$-associated   (c) $WW$-associated

(d) $Z/\gamma$-fusion   (e) $WW$-fusion   (f) $W\ell\nu_\ell$-associated

Figure 124: Representative Feynman diagrams for ALP production processes at $e^+e^-$ colliders.

*Electroweak production & decay of the photophobic ALP*

Since even the 'photophobic' ALP possesses an irreducible coupling to photons, the main difference between this and a generic ALP scenario is the relative size of this coupling to the others. From a collider perspective, the dominant ALP production processes involving the EW gauge boson couplings are in association with a $Z$-boson or a photon. The photophobic scenario leads to a specific predictions for the relative abundances of these two processes. Other potentially relevant modes are $WW$-fusion, $WW$-associated production, $Z/\gamma$-fusion and $W\ell\nu$-associated production. Representative Feynman diagrams for the main production modes are shown in Figure 124 and cross section predictions computed with `MadGraph5_aMC@NLO` [728] at the three CLIC centre-of-mass energies are shown in Figure 125.

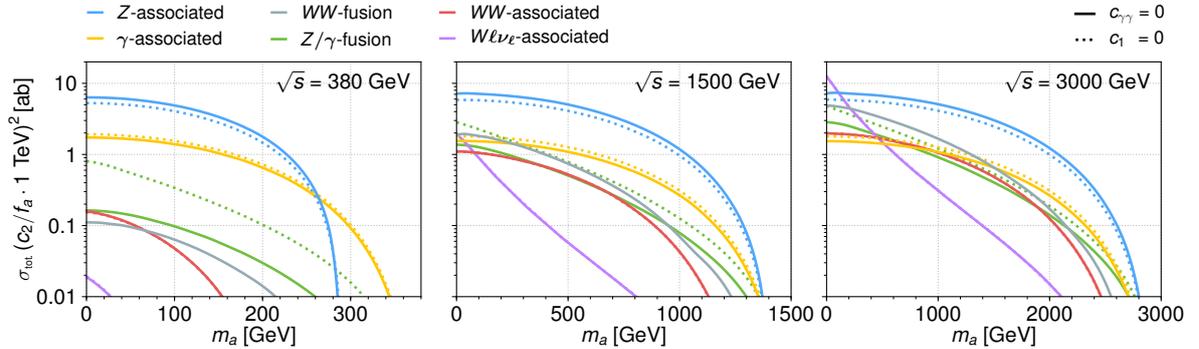

Figure 125: Inclusive cross section for ALP production processes as a function of ALP mass, $m_a$, at an $e^+e^-$ collider with 380, 1500 and 3000 GeV of center of mass energy, corresponding to the three stages of CLIC. The predictions are given in units of attobarn and $(c_2/f_a \cdot 1\,\text{TeV})^2$, such that they correspond to the cross section for $c_2/f_a = 1\,\text{TeV}^{-1}$. The rates were computed with a $p_T$ cut of 5 GeV on any final state leptons and photons along with a $\Delta R$ separation of 0.4 between any two such particles. Solid lines denote the prediction in the photophobic scenario of interest, while dotted lines show for comparison the rates for $c_1 = 0$.

One can observe that the $Z$- and $\gamma$-associated modes are independent of the centre-of-mass energy in the massless limit, due to the momentum dependence of the dimension-5 ALP-gauge boson interaction. Although $Z$-associated production always gives the largest cross section, one should bear in mind that a reduction in rate will always be paid through the branching fraction of the $Z$ into the specific final state searched for. If one considers leptonic decay modes for the $Z$, the effective rate becomes smaller than that of $\gamma$-associated production. It can also be seen that, at 380 GeV, the remaining production modes are very subdominant but that they start to become competitive with increasing centre-of-mass energy, with $WW$-fusion and $WW$-associated production becoming larger than $\gamma$-associated production for low ALP masses, although the latter will also have a reduced effective cross section due to the final state $W$-decays. Figure 125 also includes for comparison, the corresponding production cross sections as a function of $c_2$ with $c_1$ set to zero, as an example of a general non-photophobic scenario. This has a large impact on the $Z/\gamma$-fusion production, particularly at low centre of mass energies where an increase of nearly an order of magnitude is induced by the appearance of the $\gamma\gamma-$fusion channel. The other modes are not significantly affected beyond slight modifications to the $Z$- and $\gamma$-associated production rates.

Another major phenomenological consequence of the absence/reduction of ALP-photon couplings is the relative suppression of its decay mode into photons, which is the typical channel for searching for such objects. For the photophobic ALP, this decay mode is only relevant for ALP masses below $2m_e$. Above the fermion thresholds, decays into these dominate and once decays to EW gauge bosons are available, these immediately become the primary modes. Above the EW gauge boson thresholds, the photophobic ALP predicts a specific pattern of branching rations into the three possible modes, $WW$, $Z\gamma$ and $ZZ$, independent of $c_2/f_a$ and tending to 65%, 20% and 15% respectively once kinematic threshold effects are no longer relevant. The branching fractions into these three modes for masses above 100 GeV, are shown in Figure 126, including off-shell effects. Some three-body bosonic decay modes can become relevant at high masses but remain at the order of a few percent and are neglected.

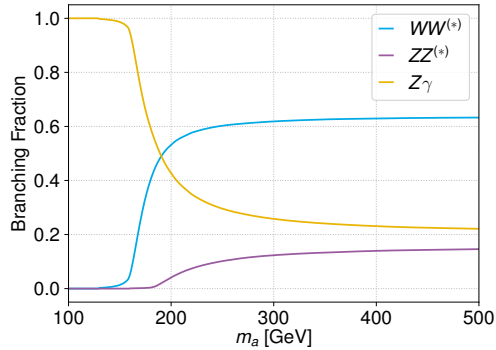

Figure 126: Branching fractions of the photophobic ALP into two gauge bosons as a function of ALP mass, $m_a$, including of-shell effects. At high masses, the $WW$, $Z\gamma$ and $ZZ$ modes asymptote to 65%, 20% and 15% respectively.

The combination of different production and decay modes lead to a rich set of possibilities for ALP searches in EW gauge boson final states at $e^+e^-$ colliders. These are summarised in Table 51, which shows that the final states between different production and decay modes can even overlap, although the resonant kinematical structure of the final states will, of course, differ between them. In this exploratory study, leptonic ($e$ & $\mu$) decays for the final state gauge bosons are considered. This motivates considering the $Z\gamma$ decay mode for the ALP owing to the lack of missing energy in the decay final state as well as the penalty of only one $Z$-boson leptonic branching fraction. One can then expect that the resonant $\gamma\ell^+\ell^-$ system can be cleanly reconstructed and distinguished over the background continuum. The CLIC reach of the $\gamma$-associated and $WW$-fusion production modes will be investigated as examples of a canonical production mechanism as well as an important alternative production mode which also predicts missing energy in the final state. In reality, the $WW$-fusion mode corresponds to the associated production of and ALP with two neutrinos. This final state can also be produced by the $Z$-associated mode in which the $Z$ decays into a neutrino anti-neutrino pair. This component is included together with $WW$-fusion under $\nu\bar{\nu}$-associated production. The reach of this mode at 380 GeV was not considered due to the comparatively small $WW$-fusion rate at this center of mass energy, as shown in Figure 125, although the $Z$-associated component of the signal may be worth investigating at this energy, an exercise left for future work.

Table 51: Predicted final states of each production mode combined with each decay mode of the ALP at $e^+e^-$ colliders.

| Decay | $Z$-assoc. | $\gamma$-assoc. | $WW$-fusion | $WW$-assoc. | $Z\gamma$-fusion | $W\ell\nu$-assoc. |
|---|---|---|---|---|---|---|
| $WW$ | $ZWW$ | $\gamma WW$ | $WW + \not{E}$ | $WWWW$ | $eeWW$ | $We\nu WW$ |
| $Z\gamma$ | $ZZ\gamma$ | $\gamma Z\gamma$ | $Z\gamma + \not{E}$ | $WWZ\gamma$ | $eeZ\gamma$ | $We\nu Z\gamma$ |
| $ZZ$ | $ZZZ$ | $\gamma ZZ$ | $ZZ + \not{E}$ | $WWZZ$ | $eeZZ$ | $We\nu ZZ$ |

*Phenomenological study of $a \to Z\gamma$ signals*

In order to estimate the CLIC reach in the $(c_2/f_a, m_a)$ plane for this scenario, a kinematic selection will be performed for the ALP signature against dominant sources of SM background, coming from multiboson final states. The FeynRules [729] model for the ALP effective Lagrangian presented in [719] was adapted to the photophobic limit. The signal and background samples were generated at leading order with `MadGraph5_aMC@NLO`, for a range of mass points and a coupling value of $c_2/f_a = 1$ TeV$^{-1}$ at the three energy stages of CLIC, and subsequently decayed into $Z\gamma$ with `MadSpin` [730]. For this coupling value, the ALP remains narrow, with its width not exceeding $\sim$10% of its mass over the majority of the accessible mass range. This is important to ensure the validity of the narrow width approximation used in determining the signal cross section predictions. No interference effects between the signal and background are considered. The samples were then showered with `PYTHIA8` [337] and passed through a fast simulation of the CLIC detector response provided by `Delphes` [385], using CLIC-specific configuration cards [386], which approximate the detector resolution and acceptance for the three energy stages. A simple event selection is performed, which will be further detailed in the next two sections, to obtain event samples of the targeted final states in which the leptonic $Z$-boson and ALP candidates are reconstructed.

Given the clean resonant structure of the signal, a simple cut and count selection is likely to yield acceptable background discrimination. However, since this is a sensitivity study, it is worthwhile to extract the maximum possible reach by exploiting as much kinematical information as possible. To this end, basic multivariate methods were employed in the form of a Neural Network classifier to distinguish between signal and background, which makes use of the full inter-correlated kinematic information to learn to optimally perform said task. The classifier is implemented thanks to `Keras` [731] with the `TensorFlow` [732] backend. Further details are not important for the purposes of this study, in which the tool was mainly used to streamline the analysis procedure and possibly obtain moderate gains in discrimination power. Once trained, the classifier was evaluated on validation samples to determine the signal and background selection efficiencies as function of a cut on the discriminant. From this, one can determine the predicted number of signal and background events and obtain a Poissonian likelihood ratio between the signal and background only hypotheses assuming an observation of the latter. Finally, 95% confidence level exclusions on $c_2/f_a$ were computed as a function of $m_a$ using the CLs method.

*$\gamma$-associated analysis*

This search is for two photons, one of which is highly energetic, and a leptonically decaying $Z$-boson that, along with the (usually) less energetic photon, reconstructs the resonance mass. The dominant irreducible background for this signature is the SM $Z\gamma\gamma$ process, and is considered in this study as the only background source. The visible cross section is defined by that which produced events containing at least two reconstructed photons and two same flavor, oppositely charged leptons, meaning they have satisfied the acceptance and efficiency requirements of the `Delphes` configuration cards for the given CLIC energy stage. These are summarised in Table 52 for the mass hypotheses tested in this study along with the $Z\gamma\gamma$ background.

Table 52: Total and visible cross sections in zb as defined in the text for $\gamma$-associated ALP production times its $Z\gamma$ branching fraction for the simulated mass points as well as that of the dominant SM $Z\gamma\gamma$ background at the three CLIC energy stages. The signal cross sections correspond to the parameter point $c_2/f_a = 1$ TeV$^{-1}$.

| $\sqrt{s}$ = 380 GeV | | | $\sqrt{s}$ = 1.5 TeV | | | $\sqrt{s}$ = 3 TeV | | |
|---|---|---|---|---|---|---|---|---|
| $m_a$ [GeV] | $\sigma_{\text{tot.}}$ | $\sigma \cdot \mathcal{A} \cdot \varepsilon$ | $m_a$ [GeV] | $\sigma_{\text{tot.}}$ | $\sigma \cdot \mathcal{A} \cdot \varepsilon$ | $m_a$ [GeV] | $\sigma_{\text{tot.}}$ | $\sigma \cdot \mathcal{A} \cdot \varepsilon$ |
| 100 | 94 | 43 | 100 | 103 | 4.3 | 100 | 103 | 0.52 |
| 120 | 85 | 45 | 300 | 24 | 11 | 300 | 26 | 3.2 |
| 140 | 74 | 43 | 500 | 16 | 8.2 | 600 | 19 | 5.5 |
| 160 | 61 | 37 | 800 | 7.7 | 3.6 | 900 | 16 | 4.6 |
| 200 | 19 | 10 | 1000 | 3.5 | 1.4 | 1200 | 12 | 3.1 |
| 240 | 7.8 | 5.0 | 1200 | 0.91 | 0.30 | 1600 | 7.6 | 1.4 |
| 280 | 3.1 | 1.9 | 1400 | 0.045 | 0.011 | 2000 | 3.5 | 0.46 |
| 320 | 0.76 | 0.44 | | | | 2400 | 0.91 | 0.12 |
| | | | | | | 2800 | 0.045 | 0.003 |
| $Z\gamma\gamma$ | 26.7 fb | 7.8 fb | $Z\gamma\gamma$ | 4.5 fb | 0.63 fb | $Z\gamma\gamma$ | 1.7 fb | 0.11 fb |

For the signal, we see the drop in cross section with increasing mass also reflected in Figure 125. In the case of the background, the cross section decreases with increasing centre-of-mass energy. The acceptance and reconstruction efficiency is also reduced, although less dramatically, for the signal. When $m_a$ becomes small with respect to the collider energy, we also observe a reduction in the effective cross section due to the collimation of the boosted ALP decay products resulting in a reduced identification efficiency of the photon and lepton candidates. Such an effect could be mitigated with substructure techniques that lie beyond the scope of this work.

The pair of leptons that most closely reconstructed the $Z$ mass is retained to form the $Z$-boson candidate. No cut on the invariant mass is performed although this would likely occur in a real analysis to suppress other, reducible backgrounds. The photon that, along with the $Z$-candidate, best reconstructed the hypothesis mass is retained to form the ALP-candidate. The remaining photon is almost always the most energetic one in the signal samples since it was the associated photon recoiling against the ALP and has a fixed energy at parton level as a function of $m_a$ and $\sqrt{s}$. The resonance mass and the recoil photon energy are by far the most effective variables to distinguish signal from background. The kinematic properties of the leptons and photons as well as the $Z$- and ALP-candidates are fed as input into the Neural Network to train a binary classification discriminant. These include the energies, absolute and relative polar/azimuthal angles and invariant masses for the $Z$- and ALP-candidates. For each mass hypothesis, a cut on the discriminant output evaluated on the validation sample is made that maximises the 95% exclusion on $c_2/f_a$. The limits are shown as solid lines in Figure 127. Table 53 summarises, for the mass hypotheses tested, the expected number of events for signal and background after the discriminant cut is applied as well as the derived limit on $c_2/f_a$. Comparing with Table 52 shows that percent background efficiency is generally achieved and that the kinematic selection improves $S/\sqrt{S+B}$ by orders of magnitude.

Table 53: Expected number of events at the three CLIC stages for the $\gamma$-associated signal with $c_2/f_a = 10$ TeV$^{-1}$ and the SM background after a cut on the Neural Network discriminant is applied to events satisfying the visible cross section requirements. Also shown is the derived 95% exclusion on $c_2/f_a$ in TeV$^{-1}$ obtained with the CLs method assuming an observation of the background only hypothesis.

| $\sqrt{s} = 380$ GeV, $\mathcal{L}^{\text{int.}} = 0.5$ ab$^{-1}$ | | | | $\sqrt{s} = 1.5$ TeV, $\mathcal{L}^{\text{int.}} = 1.5$ ab$^{-1}$ | | | | $\sqrt{s} = 3$ TeV, $\mathcal{L}^{\text{int.}} = 2$ ab$^{-1}$ | | | |
|---|---|---|---|---|---|---|---|---|---|---|---|
| $m_a$ [TeV] | $N_{\text{sig.}}$ | $N_{\text{bkg.}}$ | $c_2/f_a$ | $m_a$ [GeV] | $N_{\text{sig.}}$ | $N_{\text{bkg.}}$ | $c_2/f_a$ | $m_a$ [GeV] | $N_{\text{sig.}}$ | $N_{\text{bkg.}}$ | $c_2/f_a$ |
| 100 | 1.68 | 13.3 | 23 | 100 | 0.49 | 2.8 | 33 | 100 | 0.060 | 0.22 | 71 |
| 120 | 1.39 | 19.4 | 27 | 300 | 1.30 | 13.2 | 26 | 300 | 0.56 | 3.5 | 30 |
| 140 | 1.13 | 15.3 | 29 | 500 | 1.07 | 10.5 | 27 | 600 | 0.91 | 2.3 | 23 |
| 160 | 1.11 | 16.8 | 30 | 800 | 0.46 | 9.3 | 41 | 900 | 0.81 | 2.4 | 24 |
| 200 | 0.41 | 45.4 | 60 | 1000 | 0.18 | 9.4 | 64 | 1200 | 0.51 | 1.4 | 28 |
| 240 | 0.17 | 24.0 | 83 | 1200 | 0.036 | 9.3 | 146 | 1600 | 0.21 | 1.4 | 44 |
| 280 | 0.078 | 91.3 | 162 | 1400 | 0.0012 | 9.3 | 805 | 2000 | 0.075 | 1.5 | 73 |
| 320 | 0.015 | 40.2 | 308 | | | | | 2400 | 0.020 | 2.4 | 151 |
| | | | | | | | | 2800 | 0.0006 | 5.5 | 1032 |

*$\nu\bar{\nu}$-associated analysis*

The second production mode leads to an ALP final state decaying to $Z\gamma$ in association with missing energy. The dominant irreducible background in this case is the SM $Z\gamma\nu\bar{\nu}$ process. There is another possible background in the form of the $Z\gamma\gamma$ process where one of the two photons is out of the calorimeter acceptance. The visible cross section is defined by requiring the presence of at least one photon and at least two leptons, vetoing events containing a further photon of energy greater than 15 GeV. No requirement on the missing energy is made, although this may be included in an experimental search to reduce other background sources.

The visible cross section predictions for the mass hypotheses and two backgrounds are summarised in Table 54. At 380 GeV, $Z\gamma\gamma$ is more important than $Z\gamma\nu\bar{\nu}$, even after requiring that one photon be lost and the two backgrounds show opposite behaviour with increasing centre of mass energy, with the former decreasing and the latter increasing. Although the full $Z\gamma\gamma$ cross section is comparable to the $Z\gamma\nu\bar{\nu}$ rate, the requirement that one photon be lost reduces it to a subdominant concern, particularly at $\sqrt{s} = 3$ TeV. It is nonetheless included in the training sample for the Neural Network classifier. Conversely, the main SM $Z\gamma\nu\bar{\nu}$ cross section increases slightly between 1.5 and 3 TeV. The main difference between this production mode and the $\gamma$-associated one is the absence of a loss in reconstruction efficiency for low $m_a$ relative to $\sqrt{s}$, this indicated that the $WW$-fusion component is dominant in the visible cross section in this case, as it will not necessarily produce a highly boosted ALP in the final state as opposed to the $Z$ or $\gamma$-associated production always produces an ALP with an energy close to $\sqrt{s}/2$. This can be verified in Figure 125, where the $WW$-fusion rates at 1.5 and 3 TeV are greater than the $Z$-associated production at low masses when factoring in the 20% invisible $Z$ branching fraction.

The $Z$-candidate is identified as for the $\gamma$-associated production analysis and the ALP-candidate is reconstructed from the leading photon and the $Z$-boson candidate. The ALP candidate four momentum is effectively equal to the missing momentum in each event. The same kinematical properties of the final state and reconstructed particles are fed into the Neural Network to construct the signal discriminant. For this analysis involving a two-component background with very different kinematical properties, a multi-class discriminant was trained to identify signal and the two backgrounds independently. The discriminant outputs three values: $P_{\nu\bar{\nu}a}$, $P_{Z\gamma\nu\bar{\nu}}$ and $P_{Z\gamma\gamma}$, which can be interpreted as the probability of a given event belonging to the signal, $Z\gamma\nu\bar{\nu}$ and $Z\gamma\gamma$ processes, respectively. From these, the combined discriminants

$$P_A = \frac{P_{\nu\bar{\nu}a}}{P_{\nu\bar{\nu}a} + P_{Z\gamma\nu\bar{\nu}}} \quad \text{and} \quad P_B = \frac{P_{\nu\bar{\nu}a}}{P_{\nu\bar{\nu}a} + P_{Z\gamma\gamma}} \tag{302}$$

are constructed to independently distinguish the signal from the two backgrounds. A two dimensional cut on $p_A$ and $P_B$ is made to determine the signal and background efficiencies and eventually the 95% confidence level exclusion on $c_2/f_a$ for each $m_a$, as represented by the dashed lines in Figure 127. Table 55, details the expected number of events for the signal hypotheses tested along with the two background components, as well as the derived limit on $c_2/f_a$.

Table 54: Total and visible cross sections in zb as defined in the text for $\nu\bar\nu$-associated ALP production times its $Z\gamma$ branching fraction for the simulated mass points as well as that of the two main SM backgrounds, $Z\gamma\nu\bar\nu$ and $Z\gamma\gamma$ where a photon is out of acceptance, at the three CLIC energy stages. The signal cross sections correspond to the parameter point $c_2/f_a = 1~\text{TeV}^{-1}$.

| $\sqrt{s} = 380$ GeV | | | $\sqrt{s} = 1.5$ TeV | | | $\sqrt{s} = 3$ TeV | | |
|---|---|---|---|---|---|---|---|---|
| $m_a$ [GeV] | $\sigma_{\text{tot.}}$ | $\sigma \cdot \mathcal{A} \cdot \varepsilon$ | $m_a$ [GeV] | $\sigma_{\text{tot.}}$ | $\sigma \cdot \mathcal{A} \cdot \varepsilon$ | $m_a$ [GeV] | $\sigma_{\text{tot.}}$ | $\sigma \cdot \mathcal{A} \cdot \varepsilon$ |
| 100 | 71 | 31 | 100 | 224 | 33 | 100 | 422 | 39 |
| 120 | 62 | 37 | 300 | 45 | 26 | 300 | 96 | 38 |
| 140 | 52 | 34 | 500 | 27 | 16 | 600 | 59 | 27 |
| 160 | 41 | 28 | 800 | 10 | 5.7 | 900 | 41 | 17 |
| 200 | 10 | 7.4 | 1000 | 4.3 | 2.0 | 1200 | 28 | 8.9 |
| 240 | 2.9 | 2.1 | 1200 | 0.95 | 0.34 | 1600 | 14 | 3.1 |
| 280 | 0.18 | 0.13 | 1400 | 0.0046 | 0.0013 | 2000 | 5.7 | 0.86 |
| 320 | 0.0016 | 0.0012 | | | | 2400 | 1.3 | 0.14 |
| | | | | | | 2800 | 0.037 | 0.0030 |
| $Z\gamma\nu\bar\nu$ | 1.6 fb | 0.73 fb | $Z\gamma\nu\bar\nu$ | 8.3 fb | 3.3 fb | $Z\gamma\nu\bar\nu$ | 17.3 fb | 5.9 fb |
| $Z\gamma(\gamma)$ | 26.7 fb | 3.0 fb | $Z\gamma(\gamma)$ | 4.5 fb | 0.44 fb | $Z\gamma(\gamma)$ | 1.7 fb | 0.10 fb |

Table 55: Expected number of events at the three CLIC stages for the $\nu\bar\nu$-associated signal with $c_2/f_a = 10~\text{TeV}^{-1}$ and the SM background after a cut on the two-dimensional Neural Network discriminant is applied to events satisfying the visible cross section requirements. Also shown is the derived 95% exclusion on $c_2/f_a$ in $[\text{TeV}^{-1}]$ obtained with the CLs method assuming an observation of the background only hypothesis.

| $\sqrt{s} = 380$ GeV, $\mathcal{L}^{\text{int.}} = 0.5~\text{ab}^{-1}$ | | | | $\sqrt{s} = 1.5$ TeV, $\mathcal{L}^{\text{int.}} = 1.5~\text{ab}^{-1}$ | | | | $\sqrt{s} = 3$ TeV, $\mathcal{L}^{\text{int.}} = 2~\text{ab}^{-1}$ | | | |
|---|---|---|---|---|---|---|---|---|---|---|---|
| $m_a$ [GeV] | $N_{\text{sig.}}$ | $N_{\text{bkg.}}$ | $c_2/f_a$ | $m_a$ [GeV] | $N_{\text{sig.}}$ | $N_{\text{bkg.}}$ | $c_2/f_a$ | $m_a$ [GeV] | $N_{\text{sig.}}$ | $N_{\text{bkg.}}$ | $c_2/f_a$ |
| 100 | 0.85 | 0.49 | 19 | 100 | 2.20 | 11.0 | 19 | 100 | 3.55 | 21.4 | 17 |
| 120 | 1.28 | 4.4 | 21 | 300 | 1.96 | 5.3 | 18 | 300 | 3.46 | 20.3 | 17 |
| 140 | 1.09 | 3.5 | 22 | 500 | 1.58 | 10.3 | 22 | 600 | 3.94 | 34.4 | 18 |
| 160 | 0.99 | 4.5 | 24 | 800 | 0.56 | 4.4 | 32 | 900 | 1.99 | 11.3 | 20 |
| 200 | 0.21 | 2.5 | 47 | 1000 | 0.25 | 19.5 | 64 | 1200 | 1.21 | 6.4 | 23 |
| 240 | 0.076 | 3.5 | 83 | 1200 | 0.037 | 6.4 | 133 | 1600 | 0.40 | 3.5 | 36 |
| 280 | 0.0056 | 2.4 | 288 | 1400 | 0.00015 | 6.1 | 2104 | 2000 | 0.14 | 3.4 | 62 |
| 320 | $5 \times 10^{-5}$ | 5.5 | 3641 | | | | | 2400 | 0.022 | 2.5 | 147 |
| | | | | | | | | 2800 | 0.00041 | 2.5 | 1068 |

*Results & conclusions*

The projected sensitivities to $c_2/f_a$ for a range of ALP masses in the two production and decay $a \to Z\gamma$ are summarised in Figure 127. For reference, some recently published LHC limits on this scenario are included that come from reinterpretations of triboson cross section measurements at $\sqrt{s} = 8$ TeV [727]. Overall, it can be seen that the CLIC projections significantly extend the current sensitivity in the $(c_2/f_a, m_a)$ plane of the photophobic ALP scenario. Due to the absence of the loss in reconstruction efficiency from boosted ALP production, the $\nu\bar\nu$-associated mode outperforms the $\gamma$-associated one at low $m_a$. The coupling reach is of order 20 TeV$^{-1}$ and is similar between $\sqrt{s} = 1.5$ and 3 TeV. This is likely because of the increase in cross section for the dominant background at 3 TeV as seen in Table 54. The increasing centre-of-mass energy evidently pushes farther out in $m_a$ into the multi-TeV range, although the coupling sensitivity at the kinematic edges of the collider reach is such that the narrow width approximation used in the signal cross section predictions breaks down and eventually the ALP width eventually grows beyond $m_a$.

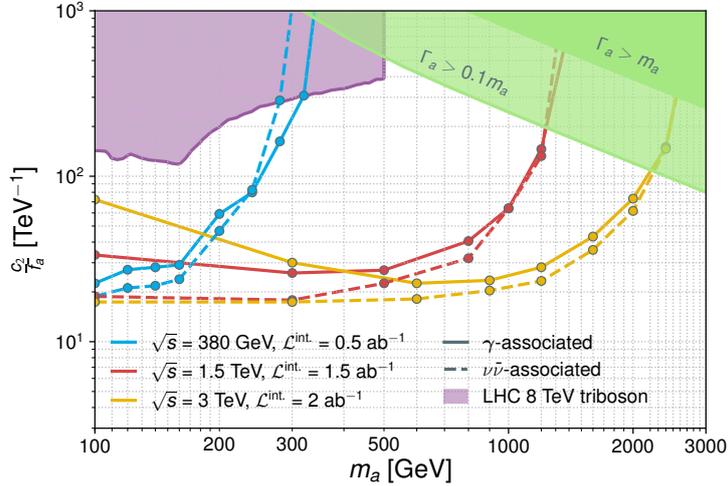

Figure 127: Projected exclusions at 95% Confidence Level on the photophobic ALP obtainable at the three CLIC stages via the $\gamma$-associated and $WW$-fusion production modes and decaying into the $Z\gamma$ final state. Also shown are the LHC limits obtained in Reference [727] from reinterpretations of triboson searches at $\sqrt{s} = 8$ TeV. The regions in which the ALP width is above 10% of its mass and is equal to its mass are shaded in green.

Due to the striking resonant signature, discrimination between the signal and background is excellent and one is almost always able to obtain better than 1% background rejection while keeping nearly all of the signal. This is likely to be possible either in a cut-based or multivariate analysis. The coupling reach is maximised for a typically small ($< 20$) number of expected background events, such that the observation of only a few events would constitute evidence for new physics. This also motivates, for the future, the consideration of additional backgrounds, e.g., coming from fakes that could contribute a handful of events in the final selection.

It should be remarked that an ALP can also be searched for in the decay to WW and ZZ, which gives rise to a similar signature to that considered for the search for new singlet particles investigated in Section 4.2.1. In this case we can recast the results for the search of a singlet in $VV$ final state of Section 4.2.1 for the search of an ALP governed by the Lagrangian Eq. (295). In Figure 128 we show the reach in the plane $(m_a, c_2/f_a)$ of the combined $VV$ channel at CLIC 3 TeV 3 ab$^{-1}$. For this result we also show extra lines useful for interpreting the search in models in which the heavy ALP is a pNGB originating from symmetry breaking at scale $f$ in a UV sector characterized by a coupling $g_*$. Due to the imagined UV origin [384] of the $c_2$ coefficient we take $c_2 = (4\pi/g_*)^2$. For two choices of the coupling $g_*$ we show i) lines corresponding to expected masses $m_f$ of the lowest lying new states of the UV completion of the ALP ii) the line at which $m_a = m_f$, that is the maximal $m_a$ for which the effective theory we have imagined is still valid. Considering only the portion of the plane to the left of these lines $m_a = m_f$ the figure shows that for strongly coupled theories the reach of searches for the ALP is inferior to that of direct searches for the states in the UV completion of the ALP effective theory, while for more weakly coupled theories the search for ALP can probe larger values of $f_a$ than the search for the states in the UV completion. Complementarity is expected in the pursuit of both these types of searches.

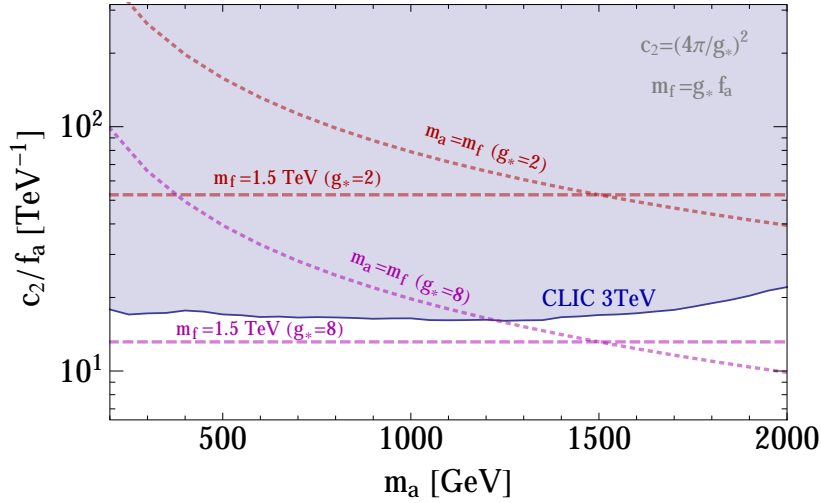

Figure 128: Projected exclusions at 95% Confidence Level in the plane $c_2/f_a$ vs. $m_a$ from the search $e^+e^- \to a\nu\bar{\nu} \to VV\nu\bar{\nu}$ adapter from Ref. [384].

In conclusion we have presented an exploratory study on the CLIC sensitivity to heavy photophobic ALPs focusing on the $\gamma$-associated and $\nu\bar{\nu}$-associated modes considering both decays $a \to Z\gamma$ and $a \to VV$ and taking into account some detector effects via Delphes. The individual projected reach for each analysis extends significantly beyond existing limits in the mass-coupling plane set by LHC measurements of multiboson processes. CLIC will be able not only to extend the mass reach to such objects beyond 2 TeV, but also to improve the coupling sensitivity by around an order of magnitude. It was observed that, although the cross section predictions for the $\gamma$-associated modes are important at all energy stages, the $WW$-fusion mechanism eventually outperforms the former at low masses due to the absence of a reduction in reconstruction efficiency from the fact that the $\gamma$-associated mode produces significantly more boosted ALPs. Furthermore, the sensitivity of $\nu\bar{\nu}$-associated production is similar to that of $\gamma$-associated production even at high masses due to the presence of the $Z$-associated component, whose rate evolves similarly to $\gamma$-associated production and does not deteriorate as fast as $WW$-fusion with increasing $m_a$.

Accompanying the study is a general discussion on the various production mechanisms and decay modes that test ALP couplings to EW gauge bosons at $e^+e^-$ colliders. A great deal more possibilities exist that have yet to be studied and a combination of various searches in different decay modes is likely to increase the final CLIC reach in this scenario. Of particular interest would be considering hadronic gauge boson decay modes in the $a \to Z\gamma$ decay to increase the signal cross section. These are left for future work. It is also worth noting that most of what has been studied here can be applied to the non-photophobic case where $c_{\tilde{B}}$ is left free, the main difference being the appearance of the $\gamma\gamma$ decay mode. These or any future results from studies of other production/decay modes could safely be recast into the non-photophobic case by appropriate rescalings of the cross section times branching fraction, assuming that the kinematics of the production are not significantly altered by the appearance of the ALP-photon interaction.

## Acknowledgements

R. Franceschini was supported by Programma per Giovani Ricercatori "Rita Levi Montalcini" granted by Ministero dell'Istruzione, dell'Università e della Ricerca (MIUR). T. Vantalon was supported by the "Severo Ochoa" excellence program of MINECO (grant SO-2012-0234). F. Riva, D. M. Lombardo and M. Riembau acknowledge support from the Swiss National Science Foundation grant no. PP00P2-170578. C. Grojean is supported by the Helmholtz Association through the recruitment initiative program. A. Mitov is supported by the European Research Council Consolidator Grant NNLOforLHC2 and by the UK STFC grants ST/L002760/1 and ST/K004883/1. J. Reuter acknowledges partial support from the EU COST Network "VBScan" (COST Action CA16108) and the Collaborative Research Unit SFB676, Project B1 of the German Research Association (DFG). O. Matsedonskyi acknowledges the IASH postdoctoral fellowship for foreign researchers. C. Zhang is supported by IHEP under Contract No. Y7515540U1 and by the United States Department of Energy under Grant Contracts de-sc0012704. The research of Ignacio García, M. Perelló and M. Vos is supported by the Spanish national program for particle physics under project number FPA2015-65652-C4-3-R (MINECO/FEDER-UE), the "Severo Ochoa" excellence program under grant number SEV-2014-0398-05 (ref. BES-2015-072974) and the Generalitat Valenciana under grant PROMETEO/2018/060. The research of J. Kalinowski, W. Kotlarski, T. Robens, D. Sokołowska andA. F. Żarnecki was supported by the National Science Centre, Poland, the HARMONIA project under contract UMO-2015/18/M/ST2/00518 (2016-2019) and the OPUS project under contract UMO-2017/25/B/ST2/00496 (2018-2020). The work of T. Robens was supported by the National Science Foundation under grant number 1519045, by the National Research, Development and Innovation Fund, Hungary under grant number K 125105 and by the European Union through the European Regional Development Fund - the Competitiveness and Cohesion Operational Programme (KK.01.1.1.06). The work of W. Kotlarski was supported by the German Research Foundation (DFG) under grants number STO 876/4-1 and STO 876/2-2. The work of J. Kalinowski was supported by the DFG under grant number SFB 676. J. Zupan acknowledges support in part by the DOE grant de-sc0011784. T. M. P. Tait acknowledges support in part by NSF Grant No. PHY-1620638. W. Altmannshofer acknowledges support in part by NSF Grant No. PHY-1720252. D. Buttazzo is supported by the INFN grant "FLAVOR". G. Durieux is supported at the Technion by a fellowship from the Lady Davis Foundation. The research of C. Kilic is supported by the National Science Foundation Grant Number PHY-1620610. Z. Chacko is supported in part by the National Science Foundation under Grant Number PHY-1620074, the Fermilab Intensity Frontier Fellowship and the Visiting Scholars Award #17-S-02 from the Universities Research Association. C. B. Verhaaren is supported in part by Department of Energy Grant number DE-SC-000999. S. Najjari is supported by Vrije Universiteit Brussel through the Strategic Research Program "High Energy Physics" and also supported by FWO under the EOS-be.h project n. 30820817. K. Mimasu is supported by a Marie Sklodowska-Curie Individual Fellowship of the European Commission's Horizon 2020 Programme under contract number 707983. The work of A. D. Plascencia is supported by CONACYT. The work of K. Sakurai is partially supported by the National Science Centre, Poland, under research grants 2017/26/E/ST2/00135 and DEC-2016/23/G/ST2/04301. Y. Cui is supported in part by the US Department of Energy (DOE) grant DE-SC0008541. A. Joglekar is supported in part by the US Department of Energy (DOE) grant DE-SC0008541. Z. Liu has been supported by Fermi Research Alliance, LLC under Contract No. DE-AC02-07CH11359 with the U.S. Department of Energy, Office of Science, Office of High Energy Physics, and in part by the NSF under Grant No. PHY1620074 and by the Maryland Center for Fundamental Physics. The work of B. Shuve is supported by the National Science Foundation under Grant No. PHY-1820770. M. Ghezzi is partially supported by the SNSF under contract 200021_160156. G. M. Pruna is partially supported by the SNSF under contract 200021_160156. The work of L. Di Luzio was supported by the ERC grant NEO-NAT. R. Gröber was partially supported by a COFUND/Durham Junior Research Fellowship under the EU grant number 609412. The work of G. Moortgat-Pick and D. Dercks has been supported by the Deutsche Forschungsgemeinschaft through funds of the SFB 676 Particles, Strings and the Early Universe. K. Rolbiecki is supported by the National Science Centre, Poland, under research grants DEC-2016/23/G/ST2/04301 and 2015/19/D/ST2/03136. The work of S. Alipour-Fard and N. Craig has been supported by the Department of Energy grant DE-SC0014129. J.M.No was supported by the Programa Atraccion de Talento de la Comunidad de Madrid under grant 2017-T1/TIC-5202, and also acknowledges support from the Spanish MINECO's "Centro de Excelencia Severo Ochoa" Programme under grant SEV-2012-0249. J.Baglio acknowledges the support from the German Research Foundation (DFG) through the Grant JA 1954/1, as well as the support from the Carl-Zeiss foundation. C. Weiland acknowledges the support from the Seventh Framework Programme (FP/2007-2013)/ERC Grant NuMass Agreement No. 617143 and partial support from the European Union's Horizon 2020 research and innovation programme under the Marie Sklodowska-Curie grant agreements No. 690575 and No. 674896. G. Perez acknowledges support from BSF, ERC-COG, ISF and Minerva. D. Azevedo, P. Ferreira and R. Santos were supported in part by the National Science Centre, Poland, the HARMONIA project under contract UMO-2015/18/M/ST2/00518 and by CERN fund grant CERN/FIS-PAR/0002/2017. M. Mühlleitner acknowledges the grant GRK1694 "Elementary particle physics at highest energy and highest precision". M. Reece was supported by the DOE Grant de-sc0013607. J. Fan is supported by the DOE grant DE-SC- 0010010 and NASA grant 80NSSC18K1010. F. Sala is supported in part by a Pier Seed Project funding (Project ID PIF-2017-72). A. Tesi is supported by the INFN grant "STRONG".